\documentclass[useAMS,usenatbib,longnamesfirst,nonamebreak]{mn2e}
\usepackage{graphicx}
\usepackage{epstopdf}
\usepackage{longtable}
\usepackage{dcolumn}
\usepackage{color}
\usepackage{amssymb}
\usepackage{subfig}
\usepackage{float}
\usepackage{afterpage}
\usepackage{setspace}
\usepackage{here}
\usepackage{morefloats}
\providecommand{\e}[1]{\ensuremath{\times 10^{#1}}}
\newcommand{\lsim}{\mathrel{\hbox{\rlap{\lower.55ex \hbox{$\sim$}} \kern-.3em \raise.4ex \hbox{$<$}}}}

\title[NAL variability in SDSS Quasars]{Narrow absorption line variability in repeat quasar observations from the Sloan Digital Sky Survey}

\author[T. L. Hacker et al.]{Troy L. Hacker,$^{1}$ Robert J. Brunner,$^{1}$\thanks{E-mail: bigdog@illinois.edu} Britt F. Lundgren,$^{2,3}$ and Donald G. York$^{4,5}$\\
$^{1}$Department of Astronomy, University of Illinois at Urbana-Champaign, 1002 West Green Street, Urbana, IL 61801, USA\\
$^{2}$NSF Astronomy and Astrophysics Postdoctoral Fellow\\
$^{3}$Department of Astronomy, University of Wisconsin, 475 N Charter Street, Madison, WI 53706, USA\\
$^{4}$Department of Astronomy and Astrophysics, University of Chicago, 5640 S Ellis Ave, Chicago, IL 60637, USA\\
$^{5}$Enrico Fermi Institute, University of Chicago, Chicago, IL 60637, USA}
\date{Accepted 2013 June 05. Received 2013 May 10; in original form 2013 March 10}
\begin{document}
 
\pagerange{1--77} \pubyear{2013}

\maketitle

\label{firstpage}

\begin{abstract}

We present the results from a time domain study of absorption lines detected in quasar spectra with repeat observations from the Sloan Digital Sky Survey Data Release 7 (SDSS DR7).  Beginning with over 4500 unique time separation baselines of various absorption line species identified in the SDSS DR7 quasar spectra, we create a catalogue of 2522 quasar absorption line systems with two to eight repeat observations, representing the largest collection of unbiased and homogeneous multi-epoch absorption systems ever published.  To investigate these systems for time variability of narrow absorption lines, we refine this sample based on the reliability of the system detection, the proximity of pixels with bright sky contamination to individual absorption lines, and the quality of the continuum fit.  Variability measurements of this sub-sample based on the absorption line equivalent widths yield a total of 33 systems with indications of significantly variable absorption strengths on time-scales ranging from one day to several years in the rest frame of the absorption system.  Of these, at least 10 are from a class known as intervening absorption systems caused by foreground galaxies along the line of sight to the background quasar.  This is the first evidence of possible absorption line variability detected in intervening systems, and their short time-scale variations suggest that small-scale structures ($\sim$10-100 au) are likely to exist in their host foreground galaxies.

\end{abstract}


\section{INTRODUCTION} \label{secint}

Absorption lines in the spectra of quasars have proven to be sensitive and unbiased probes of the gaseous content of the Universe from the formation of the first collapsed structures to the present day (see \citealt{mei09} for a recent comprehensive review).  Quasar absorption lines (QALs) provide unique insight on the temperatures, ionization states, densities, metallicities, and kinematics of extragalactic gas that would otherwise be invisible.  QALs are thought to originate either in material associated with the quasar or host galaxy or in intervening galaxies along the line of sight \citep{wey79}, and have been observed to contain a wide range of ions from neutral hydrogen to highly ionized metals such as \mbox{N\,{\sc v}} and \mbox{O\,{\sc vi}} \citep[e.g.,][]{fol83,and87,ham04,yor13}.  Typically, QALs are classified into three categories based on line width: broad absorption lines (BALs), narrow absorption lines (NALs), and intermediate, mini-broad absorption lines (mini-BALs).  BALs are particularly broad lines created by outflowing material from the central region of a quasar \citep[e.g.,][]{wey91} while mini-BALs are somewhat smaller features also thought to reside in quasar outflows \citep*[e.g.,][]{ham97a,chu99,ham04,nar04,mis07,rod11}. 

NALs, with widths small enough to resolve strong absorption doublets such as \mbox{Mg\,{\sc ii}} and \mbox{C\,{\sc iv}}, are further classified into three categories based on their proximity to the background quasar: intrinsic absorption lines generated by gas physically tied to the quasar environment, associated absorption lines (AALs) formed in the quasar host galaxy or galaxy cluster, and intervening absorption lines (IALs) located at cosmological distances from the background source \citep{wey79}.  The dividing line between intervening and associated NALs is ambiguous due to the numerous competing mechanisms that affect the observed absorption lines.  Previous studies of narrow AALs have found that these absorption systems primarily have redshifts ($z_{\scriptsize{\textrm{abs}}}$) corresponding to velocities of $\sim$0.02--0.04$c$ in the quasar rest frame \citep*[][]{fol86,nes08,van08,wil08}.  However, BALs and mini-BALs are routinely detected with velocities approaching, and sometimes even exceeding, $0.2c$ \citep*{fol83,jan96,ham97a,rod11} meaning that, in some cases, absorption systems with redshifts much smaller that the quasar emission redshift ($z_{\scriptsize{\textrm{qso}}}$) that appear to be intervening may actually be associated with the background quasar.  

Another proposed method to distinguish between different types of NALs is time variability \citep*{bar97,ham97b,nar04}.  Variations in the radiative output of quasars are known to exist \citep[e.g.,][]{van04,wil05}, so any absorbing gas physically associated with the quasar should also exhibit variability.  Absorption line variability can be caused by bulk motion of the absorbing gas or changes in the gas properties such as ionization states and column densities \citep[see, e.g.,][]{nar04}.  The quasar environment is prone to dramatic, short-term changes in ionizing flux and gas dynamics, and such conditions could result in variable absorption lines.  Conversely, such mechanisms for changing the properties of the absorbing gas on short time-scales are thought to be unfeasible in intervening systems due to the large sizes and low densities of intervening cloud structures, so time variability has become one of the strongest indicators that an absorption system is intrinsic to the quasar \citep[see][and references therein]{ham95,nar04}.  

Time variability of BALs and mini-BALs in quasar spectra is well documented \citep*[e.g.,][]{bar94,nar04,lun07,mis07,gib08,gib10,ham08,kro10,cap11,cap12,cap13,hal11,rod11,fil12,viv12}, but NAL variability has only been detected in a few studies of a small number of objects \citep{ham95,ham11,bar97,ham97c,gan01,nar04,wis04,mis05}.  The results from these studies, summarized in Table \ref{tabprev}, amount to a total of 48 systems with repeat spectroscopic observations that were analysed for absorption line variability.  For the 18 systems where NAL variability was observed, the cause was attributed to gas changes in high-velocity outflows intrinsic to the quasar environment.  However, temporal variations in absorption line profiles have been detected in very small-scale structures ($\sim$10-100 au) within the Milky Way interstellar medium \citep{lau00,lau03,wel07} and a damped Ly$\alpha$ quasar absorption system at high redshift \citep{kan01}.  Since similar structures must also exist in other galaxies, small time-scale variations could arise from sampling different regions of a foreground cloud structure as it moves relative to the background quasar \citep{van08}.  Thus time variability of QALs by itself may not always be a conclusive indicator of intrinsic absorption.

\begin{table*}
\begin{center}
\caption[Summary of previous NAL variability studies]{Summary of the previous studies that detected NAL variability in repeat observations. \label{tabprev}}
\begin{tabular}{@{}ccccccl@{}}
\hline
\\[-2mm]
\multicolumn{2}{c}{Number of systems} & Number of & Variable absorption lines & Range of $\beta$ values & Range of time baselines & \multicolumn{1}{c}{Reference} \\
\cline{1-2}
Total & Variable & observations & & & (days in the observed frame) \\
\hline
1 &1 &10 &\mbox{N\,{\sc v}} $\lambda \lambda$1239, 1243 &0.0051 &$\sim$30--3300 &H95$^a$ \\
& & &\mbox{C\,{\sc iv}} $\lambda \lambda$1548, 1551 \\
1 &1 &3 &\mbox{N\,{\sc v}} $\lambda \lambda$1239, 1243 &0.08 &$\sim$400--4000 &H97$^b$ \\
& & &\mbox{Si\,{\sc iv}} $\lambda \lambda$1394, 1403 \\
& & &\mbox{C\,{\sc iv}} $\lambda \lambda$1548, 1551 \\
1 &1 &2 &\mbox{C\,{\sc iv}} $\lambda \lambda$1548, 1551 &0.0026 &$\sim$700 &B97$^c$ \\
1 &1 &2 &\mbox{C\,{\sc iv}} $\lambda$1548 &0.0035 &1556 &G01$^d$ \\
13 &4 &2--4 &\mbox{N\,{\sc v}} $\lambda \lambda$1239, 1243 &0.0029--0.11 &$\sim$150--6270 &N04$^e$ \\
& & &\mbox{C\,{\sc iv}} $\lambda \lambda$1548, 1551 \\
19 &4 &2 &\mbox{H\,{\sc i}} $\lambda$1216 &--0.0016--0.000 89 &1556--3070 &W04$^f$ \\
& & &\mbox{C\,{\sc iv}} $\lambda \lambda$1548, 1551 \\
4 &1 &2 &\mbox{C\,{\sc iv}} $\lambda$1548 &0.032 &$\sim$500 &M05$^g$ \\
8 &5 &2 &\mbox{C\,{\sc iv}} $\lambda \lambda$1548, 1551 &0.032--0.047 &742 &H11$^h$ \\
\hline
\multicolumn{7}{l}{{$^a$} \citet{ham95}; {$^b$} \citet{ham97a}; {$^c$} \citet{bar97}; {$^d$} \citet{gan01}; {$^e$} \citet{nar04};} \\
\multicolumn{7}{l}{{$^f$} \citet{wis04}; {$^g$} \citet{mis05}; {$^h$} \citet{ham11}} \\
\end{tabular}
\end{center}
\end{table*}

To verify this conclusion, we can estimate the range of time-scales needed to detect absorption variations in intervening clouds on the order of 10-100 au in diameter.  Since ionization state changes are unlikely in these environments \citep[see, e.g.,][]{nar04}, we consider absorption line variability caused by bulk motion transverse to the quasar line of sight.  Assuming that the variability time-scale is set by the time needed to cross the quasar beam, small-scale absorbing gas clouds moving at typical orbital velocities in normal galaxies ($\sim$200 km sec$^{-1}$) should move a distance equal to the cloud diameter in $\sim$0.25--2.5 yr in the rest frame of the absorption system.  If the cloud and quasar beam have similar sizes, variations in QAL systems on time-scales of months to years should be observable.  Such variations are well within the range of current observations (see, e.g., Table \ref{tabprev}) depending on the distribution and frequency of small-scale clouds in intervening galaxies.  

Although the existence of QAL variability has been clearly established, no large, systematic study of NAL variability in QAL systems has yet been conducted.  Such a study would provide a more definitive understanding of the connections between variable NALs and the properties and environmental conditions of absorbing gas clouds.  Repeat spectroscopic observations from the Sloan Digital Sky Survey \citep[SDSS; ][]{yor00} provide a remarkable opportunity to search for variability in a large, unbiased, and homogeneous population of quasar absorption systems.  In this paper, we present the results from our systematic study of NAL variability drawn from a sample of over 2500 absorption systems with at least two SDSS observations.  In Section \ref{secobs}, we explain our procedure for identifying quasars in SDSS DR7 with repeat spectroscopic observations and creating a catalogue of NAL systems detected in their spectra.  Section \ref{secda} describes our data analysis methods to refine the sample of NAL systems, measure absorption line variations between observations, and quantify the probability of false variations in our sample.  Next is Section \ref{secdis} which discusses the global properties of the variable systems we detect, our attempts to identify any systematic biases in this sample, and the implications of the measured variations on the sizes and velocities of the absorbing gas clouds.  Finally, we summarize our results from this study and present ideas for future work in Section \ref{seccon}.

\section{OBSERVATIONS AND DATA REDUCTION} \label{secobs}

\subsection{The Sloan Digital Sky Survey} \label{secsdss}

The data for this paper are part of the Seventh Data Release (DR7) of the SDSS.  Marking the completion of the phase known as SDSS-II, DR7 includes over 11,000 deg$^{2}$ of imaging data with follow-up spectroscopy for more than 930,000 galaxies and 120,000 quasars over 9380 deg$^{2}$ \citep{aba09}.  DR7 also includes all objects that were observed more than once during the course of the survey.  Although not a primary aim of the survey, repeat observations were taken for a number of reasons including calibrations of the spectroscopic data processing pipeline, re-observations of entire plates due to poor seeing conditions, and repeated objects on overlapping plates \citep[see, e.g.,][]{wil05}.  There are thousands of objects with multiple repeat spectroscopic observations and time separations ranging from days to years making the SDSS one of the largest collections of multi-epoch spectra ever published.  

The SDSS was designed to select candidate galaxies \citep{eis01,str02}, quasars \citep{ric02}, and stars \citep{sto02} for follow-up spectroscopy based on their photometric signatures.  In a process known as tiling, targeted objects were mapped on to the sky to determine the locations of the 0.2 mm diameter holes (3 arcsec on the sky) drilled into the 3$\degr$ diameter aluminum plates \citep{yor00,bla03}.  Each plate contained 640 holes corresponding to the object positions and, during spectroscopic observations, the plates were placed in the imaging plane of the SDSS telescope and plugged with optical fibres to transmit the light from each object to twin spectrographs.  The resulting spectra, obtained by co-adding three or four 15-min exposures, have a resolution $R \equiv \lambda/\Delta \lambda$ of about 1800 in the wavelength range of 3800--9200 \textrm{\AA} \citep{sto02}.  Initial calibration occurs in the \texttt{spectro2d} data processing pipeline where the spectra are flat-fielded and flux calibrated.  A second pipeline, \texttt{spectro1d}, identifies spectral features and classifies objects by spectral type \citep{sub02}.  Flexible Image Transport System (FITS) files containing the output spectra from both pipelines for all observed objects are available for download from the SDSS Data Archive Server (DAS).\footnote{http://das.sdss.org/}

\begin{figure*}
\begin{center}
\includegraphics[width=84mm]{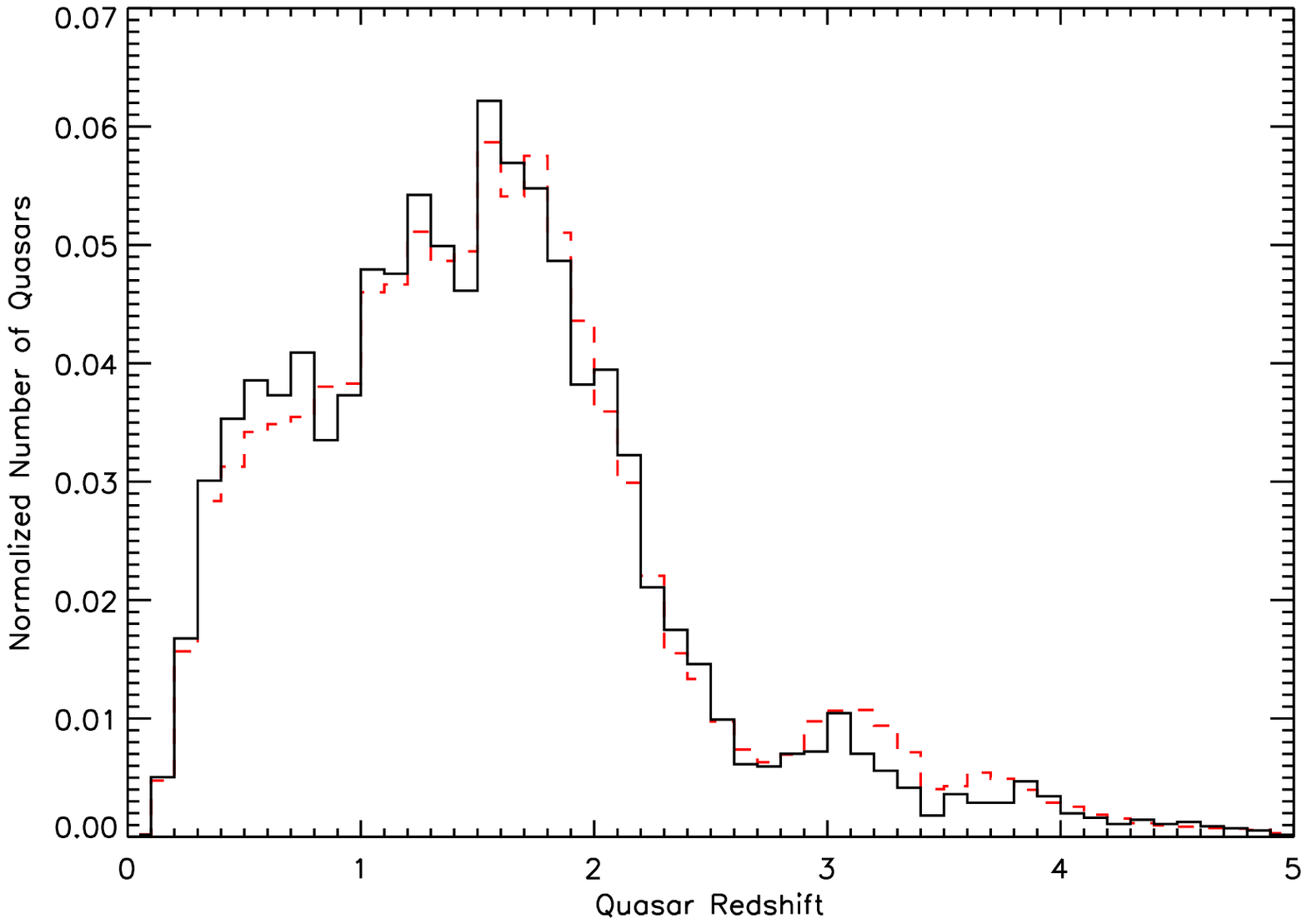}
\includegraphics[width=84mm]{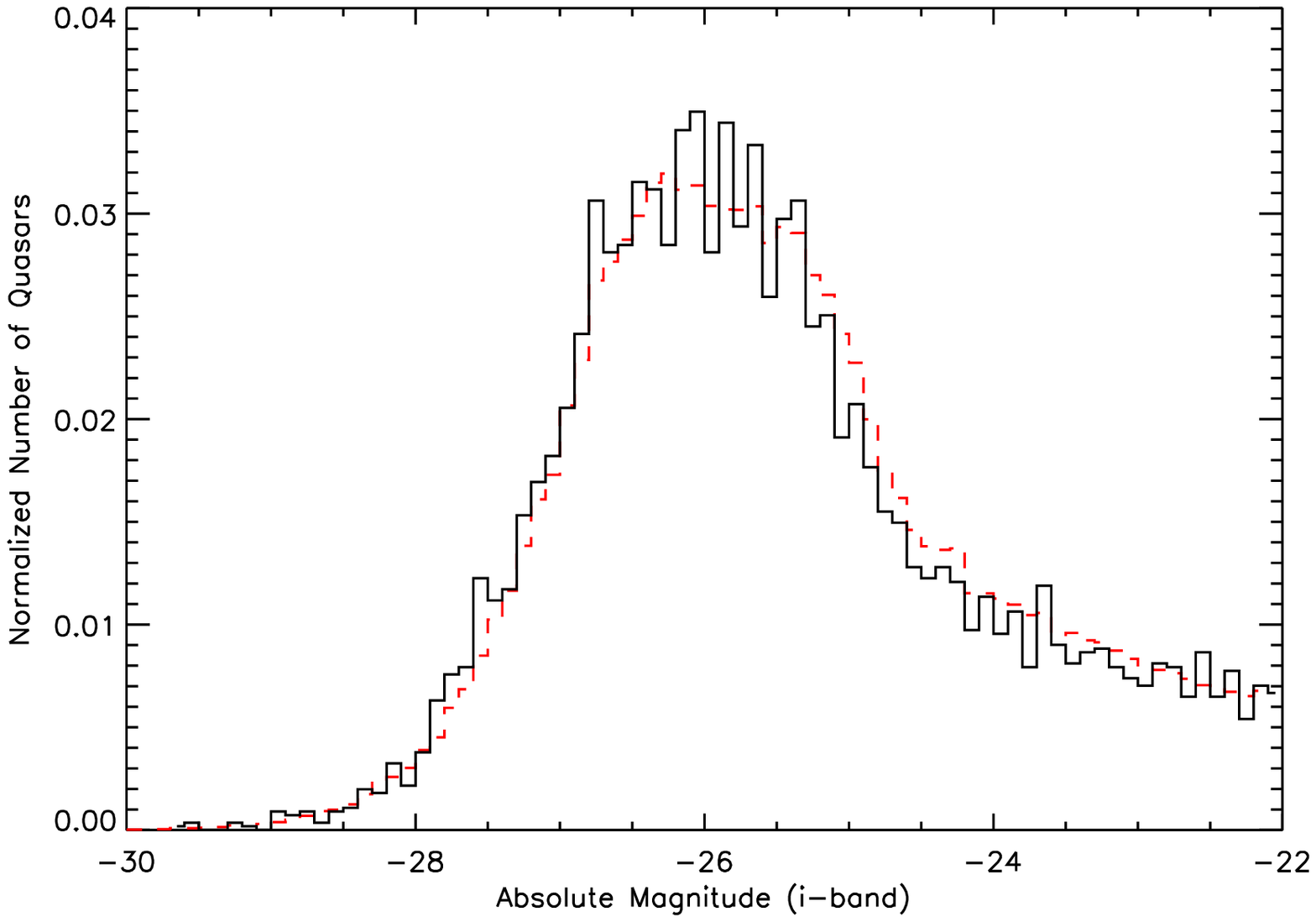}
\caption[Number distributions of SDSS DR7 quasars with repeat observations as a function of $z_{\scriptsize{\textrm{qso}}}$ and $M_{i}$]{Number distributions of the 5,550 SDSS DR7 quasars with repeat observations as a function of quasar redshift (left-hand panel: $z_{\scriptsize{\textrm{qso}}}$) and $i$-band absolute magnitude (right-hand: $M_{i}$).  The bin size is 0.1, and the y-axis is normalized to enable comparison with the entire DR7 quasar catalogue \citep[dashed red line;][]{sch10}.  The shape of both histograms closely matches, suggesting that the sample of quasars with repeat observations is unbiased with respect to the full SDSS DR7 quasar sample. \label{fignzqr}}
\end{center}
\end{figure*}

\subsection{QAL systems catalogue} \label{seccat}

The SDSS DR7 Catalogue Archive Server \citep[CAS\footnote{http://cas.sdss.org/}; see][]{tha08} contains 6,081 quasars with repeat observations (see Appendix \ref{seczid} for the query we used to identify these quasars).  After comparison with the \citet{sch10} DR7 quasar catalogue, 485 objects were removed that were incorrectly identified as quasars by the SDSS \texttt{spectro1d} pipeline.  These objects were visually inspected and rejected as quasars because either their luminosities, emission line widths, or spectral characteristics did not satisfy the quasar criteria \citep{sch10}.  The quasar redshifts were also modified using the \citet{hew10} catalogue to minimize systematic biases that are present in the published SDSS quasar redshifts.  The remaining 5596 quasars yield 9067 unique quasar/time-separation pairings where some were observed more than once on the same or back-to-back nights while others spanned the duration of SDSS-I/II operations.  In some cases, spectra from multiple observations of the same object are co-added by the \texttt{spectro2d} pipeline when the time between observations is small (less than $\sim$30 d), the same plate is re-observed, and the plate has not been re-plugged \citep{wil05}.  The FITS header for each plate (spPlate file) lists the Modified Julian Day (MJD) for all observations of a plate that are co-added.  Checking all of the quasar/time--separation pairings reveals that the second observation of Plate 2516 was co-added with the first observation on the previous day.  Since none of the 46 quasars on this plate have additional observations, all of them were removed from the repeat observation sample.  Of the remaining 5550 quasars, over 83\% have only two observations and, when combined with the comparatively few quasars with more observations, yield a total of 12 471 spectra.

\begin{figure}
\begin{center}
\includegraphics[width=84mm]{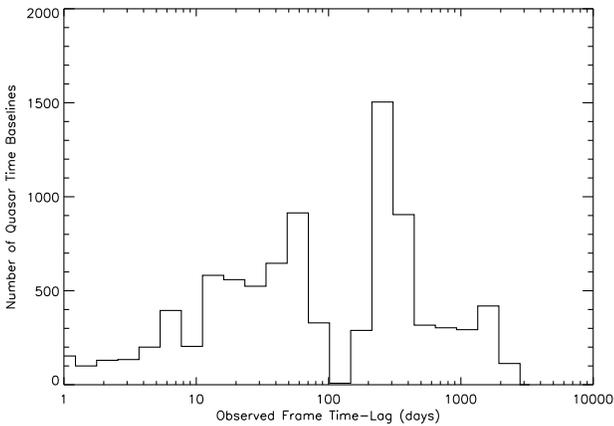}
\caption[Number distribution of SDSS DR7 quasars with repeat observations as a function of $\Delta t_{obs}$]{Number distribution of the 9021 time baselines for the 5550 SDSS DR7 quasars with repeat observations as a function of observed frame time between observations ($\Delta t_{obs}$).  A logarithmic scale is used for the \textit{x}-axis due to the large range of time separations spanning days to years.  Although 30\% of the repeat observations are separated by less than one month, the FITS file headers indicate that none of the spectra have been co-added. \label{figntqr}}
\end{center}
\end{figure}

The redshift ($z_{\scriptsize{\textrm{qso}}}$), $i$-band absolute magnitude ($M_{i}$), and observed frame time between observations ($\Delta t_{obs}$) distributions of the 5550 quasars with repeat observations are plotted in Figs \ref{fignzqr} and \ref{figntqr}.  Also plotted in Fig. \ref{fignzqr} are the same distributions for the entire SDSS DR7 quasar catalogue \citep[cf.][Fig. 5]{sch10}.  Since the general shape of the histograms shows only slight deviations, there are no obvious systematic biases in the quasars that were observed multiple times \citep{sch10}.  Fig. \ref{figntqr} presents the distribution of the remaining 9021 unique quasar/time-separation pairings where none have been co-added even though there are 2596 instances where the time between observations is less than a month.  Thus, an arbitrary minimum time between observations is unwarranted and variations in the properties of these quasars can be studied on both short and long time-scales.

The SDSS DR7 repeat quasar spectra were individually analysed by the automated QAL detection pipeline developed by York et al. (in preparation; hereafter Y13) to produce a sample of 2522 metal absorption systems in 1,641 quasars with reliable detection of either \mbox{Mg\,{\sc ii}} or \mbox{C\,{\sc iv}} in at least one observation over the redshift range of 0.36 $< z_{\scriptsize{\textrm{abs}}} <$ 3.6 (see Appendix \ref{seczid} for descriptions of the Y13 QAL pipeline and confidence grades to quantify reliability).  Given that many of the quasars have multiple absorption systems with varying levels of confidence, we compared the measured system redshifts between repeat observations of the same quasar.  When the difference was small ($\Delta z \leq $ 0.001 or, equivalently, $\Delta v \leq$ 300 km $\textrm{s}^{-1}$), these were considered to be the same system.  This offset is based on the nominal $\sim$2 $\textrm{\AA}$ width required to resolve strong lines such as \mbox{Mg\,{\sc ii}} and \mbox{C\,{\sc iv}} that are used to identify QAL systems \citep{lun09}.  Once $\Delta z > $ 0.001, absorption systems with different redshifts are likely distinct and were treated separately.  In addition, 500 of these systems have between three and eight observations, and including each unique time separation gives a total of 4454 combinations of two-epoch quasar spectra.  

\begin{table*}
\begin{center}
\caption[Catalogue of QAL systems detected in SDSS DR7 quasars with repeat observations]{Catalogue of 2522 QAL systems detected in SDSS DR7 quasars with repeat observations.  The object identifiers, position coordinates, and plate-MJD-fibre designations are taken from the \texttt{SpecObjAll} table in the SDSS Catalogue Archive Server (CAS) while the quasar redshifts ($z_{\scriptsize{\textrm{qso}}}$) are from \citet*{hew10}.  The absorption system redshift ($z_{\scriptsize{\textrm{abs}}}$), system grade, and detected lines are outputs of the Y QAL detection pipeline.  Some absorption lines are flagged based on alternate identifications ($a$), proximity of masked pixels ($b$), and/or questionable continuum fits ($c$).  The full table is available as an ancillary file in the astro-ph link for this paper. \label{tabsys}}
\begin{tabular}{@{}lrrccrrcrcl@{}}
\hline
\multicolumn{1}{c}{SDSS J} & \multicolumn{1}{c}{RA} & \multicolumn{1}{c}{Dec.} & $z_{\scriptsize{\textrm{qso}}}$ & $z_{\scriptsize{\textrm{abs}}}$ & \multicolumn{1}{c}{\phantom{0}$\beta$} & \multicolumn{1}{c}{Plate} & MJD & \multicolumn{1}{c}{Fibre} & Grade & Detected Lines\\
\hline
000042.89+005539.5&  0.178 80&  0.927 69&0.9528&0.4963& 0.2601& 387&51791&531  &B &\mbox{Fe\,{\sc ii}} $\lambda$2587, $\lambda$2600;  \\
  &  &  &  &  &  &  &  &  & &\mbox{Mg\,{\sc ii}} $\lambda \lambda$2796, 2804  \\
  &  &  &  &  &  & 685&52203&501  &B &\mbox{Fe\,{\sc ii}} $\lambda$2600; \mbox{Mg\,{\sc ii}}  \\
  &  &  &  &  &  &  &  &  & &$\lambda \lambda$2796, 2804; \mbox{Mn\,{\sc ii}} $\lambda$2577  \\
000433.93-004844.7&  1.141 35& --0.812 45&1.2975&1.2929& 0.0020& 388&51793&289  &B &\mbox{Mg\,{\sc i}} $\lambda$2853; \mbox{Mg\,{\sc ii}}  \\
  &  &  &  &  &  &  &  &  & &$\lambda \lambda$2796, 2804  \\
  &  &  &  &  &  & 686&52519&286  &N & \\
000654.10-001533.4&  1.725 47& --0.259 25&1.7224&1.1317& 0.2398& 388&51793&234  &A &\mbox{Cr\,{\sc i}} $\lambda$3595; \mbox{Fe\,{\sc ii}}  \\
  &  &  &  &  &  &  &  &  & &$\lambda$2344$^{a}$, $\lambda$2383$^{a}$, $\lambda$2587$^{a}$,  \\
  &  &  &  &  &  &  &  &  & &$\lambda$2600$^{a}$; \mbox{Mg\,{\sc i}} $\lambda$2853;  \\
  &  &  &  &  &  &  &  &  & &\mbox{Mg\,{\sc ii}} $\lambda \lambda$2796, 2804  \\
  &  &  &  &  &  & 686&52519&182  &A &\mbox{Fe\,{\sc ii}} $\lambda$2344, $\lambda$2383, $\lambda$2587,  \\
  &  &  &  &  &  &  &  &  & &$\lambda$2600; \mbox{Mg\,{\sc ii}} $\lambda \lambda$2796$^{a}$,  \\
  &  &  &  &  &  &  &  &  & &2804$^{a}$  \\
000725.18+003645.0&  1.854 96&  0.612 49&2.0359&2.0355& 0.0001& 388&51793&463  &C &\mbox{C\,{\sc iv}} $\lambda \lambda$1548, 1551;  \\
  &  &  &  &  &  &  &  &  & &\mbox{Fe\,{\sc ii}} $\lambda$2261$^{b}$, $\lambda$2600$^{bc}$;  \\
  &  &  &  &  &  &  &  &  & &\mbox{Si\,{\sc iv}} $\lambda$1394  \\
  &  &  &  &  &  & 686&52519&483  &B &\mbox{Al\,{\sc ii}} $\lambda$1671; \mbox{C\,{\sc i}}  \\
  &  &  &  &  &  &  &  &  & &$\lambda$1657; \mbox{C\,{\sc iv}} $\lambda \lambda$1548,  \\
  &  &  &  &  &  &  &  &  & &1551; \mbox{Fe\,{\sc i}} $\lambda$2524; \mbox{Ni\,{\sc ii}}  \\
  &  &  &  &  &  &  &  &  & &$\lambda$1317$^{a}$, $\lambda$1370$^{a}$; \mbox{S\,{\sc ii}}  \\
  &  &  &  &  &  &  &  &  & &$\lambda$1254$^{bc}$; \mbox{Si\,{\sc ii}} $\lambda$1304;  \\
  &  &  &  &  &  &  &  &  & &\mbox{Si\,{\sc iv}} $\lambda$1394  \\
\hline
\end{tabular}
\end{center}
\end{table*}

\begin{figure*}
\begin{center}
\includegraphics[width=84mm]{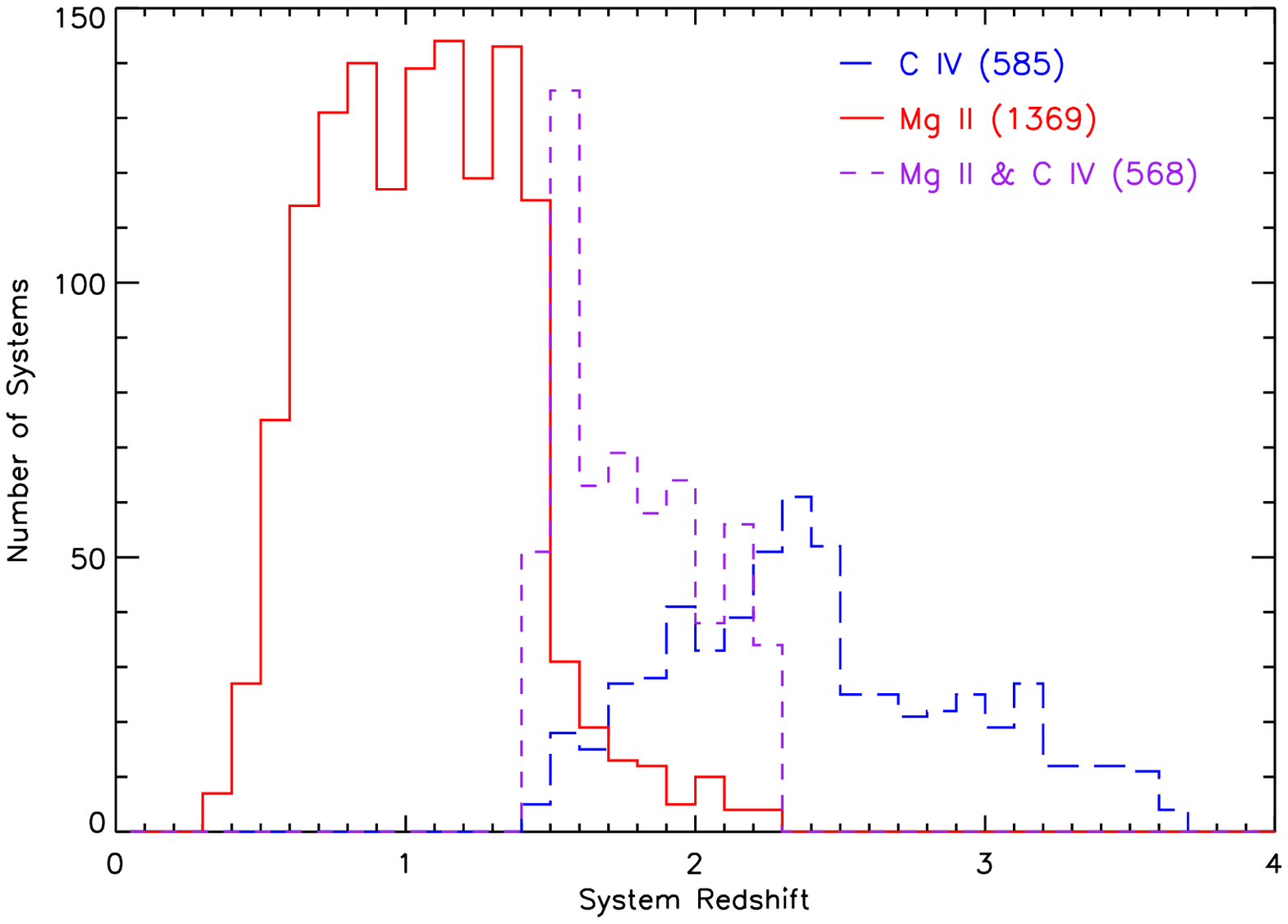}
\includegraphics[width=84mm]{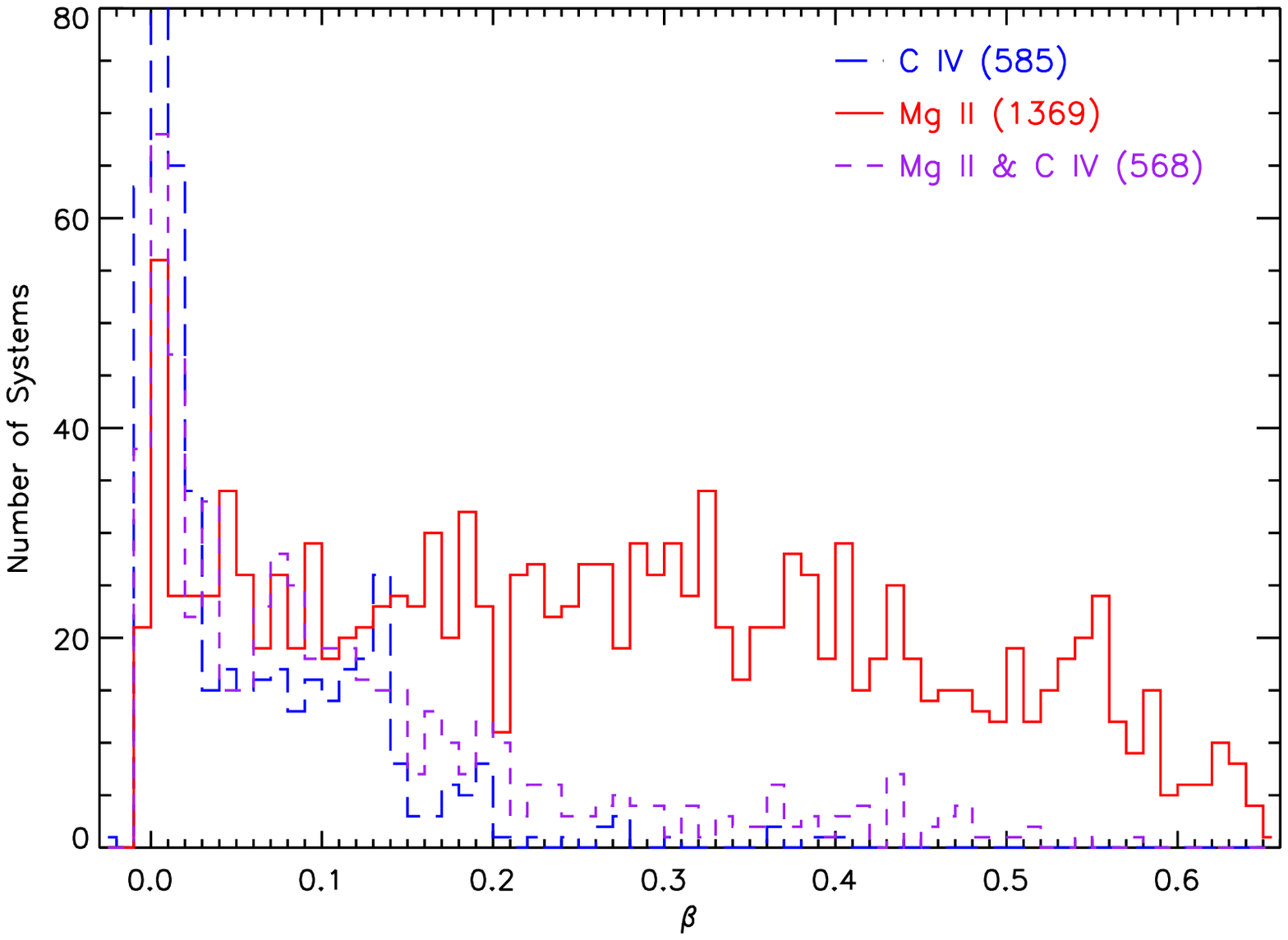}
\caption[Number distributions of SDSS DR7 QAL systems with repeat observations as a function of $z_{\scriptsize{\textrm{abs}}}$ and $\beta$]{The absorption system redshift (left-hand panel: $z_{\scriptsize{\textrm{abs}}}$) and $\beta$ (right-hand) distributions of the 2522 SDSS DR7 QAL systems with repeat observations.  The bin sizes are 0.1 and 0.01 respectively, and the \textit{y}-axis range on the right-hand panel is truncated to more clearly show the trends for all system classifications (the peak of the \mbox{C\,{\sc iv}} line is 191 systems at 0 $\leq \beta <$ 0.01).  The \mbox{Mg\,{\sc ii}} and \mbox{C\,{\sc iv}} system redshift distributions are determined by the quasar redshift distribution and the redshift ranges where these doublets are longward of the Ly$\alpha$ forest and shifted to optical wavelengths.  The $\beta$ distributions reflect the typical environments where different ionization species are found. \label{fignzsy}}
\end{center}
\end{figure*}

The system properties and detected lines are listed in Table \ref{tabsys}, and their redshift and quasar rest-frame velocity\footnote{Redshift differences and quasar rest-frame velocity separation between quasars and absorption systems are related with the dimensionless parameter $\beta$, where
\begin{equation}
\beta = \frac{v}{c} = \frac{\left [ (1+z_{\scriptsize{\textrm{qso}}})/(1+z_{\scriptsize{\textrm{abs}}}) \right]^{2}-1}{\left [ (1+z_{\scriptsize{\textrm{qso}}})/(1+z_{\scriptsize{\textrm{abs}}}) \right]^{2}+1}.
\end{equation}} ($\beta$) distributions are plotted in Fig. \ref{fignzsy}.  All of these systems are classified by the Y13 QAL pipeline based on the detection of the \mbox{Mg\,{\sc ii}} or \mbox{C\,{\sc iv}} doublets because systems without at least one of these ions are most likely false.  A \mbox{Mg\,{\sc ii}} system is defined as a system with a reported detection of at least one of the lines in the \mbox{Mg\,{\sc ii}} $\lambda \lambda$2796, 2804 doublet and no detection of the \mbox{C\,{\sc iv}} $\lambda \lambda$1548, 1551 lines.  \mbox{C\,{\sc iv}} systems are similarly defined (at least one \mbox{C\,{\sc iv}} line and no \mbox{Mg\,{\sc ii}} lines), and systems with detections of lines from both doublets are also specified.  The redshift distributions of the \mbox{Mg\,{\sc ii}} and \mbox{C\,{\sc iv}} systems reflect the higher number of quasars observed by the SDSS at $z_{\scriptsize{\textrm{qso}}} \leq 2$ (see Fig. \ref{fignzqr}) and the redshift intervals where their rest wavelengths are both longward of the Ly$\alpha$ forest and shifted into the observed range of the SDSS spectrographs.  Their $\beta$ distributions indicate that higher ionization ions such as \mbox{C\,{\sc iv}} are more prevalent in close proximity to the background quasar (i.e., intrinsic and associated systems) due to the extreme luminosity of the central accretion disc while lower ionization ions like \mbox{Mg\,{\sc ii}} are found over a broader range of galactic environments \citep[see][and references therein]{lun09}.

\subsection{QAL error flags} \label{secnflg}

Absorption line measurements in some of the QAL systems are likely erroneous due to various mitigating factors.  These lines are flagged when any of the following conditions are present.
\begin{enumerate}
\renewcommand{\labelenumi}{($\alph{enumi}$)}
\item Alternate line identification for a different QAL system detected in the same quasar spectrum.
\item Masked or multiple negative flux density pixels within seven pixels of the line centre wavelength.
\item The continuum fit near the line is suspect.
\end{enumerate}  
The first condition is provided by the Y13 QAL pipeline.  In the course of searching for QAL absorption systems, the pipeline routinely detects multiple systems at various redshifts in a given SDSS quasar spectrum.  On occasion, certain absorption line species identified in one system correspond to different species in another system.  The lines flagged with multiple ion/redshift possibilities require more careful analysis to determine the most likely line identification.

The last two conditions reflect the reliability of the line detection.  The proximity of masked pixels could significantly alter the absorption line profile (see Appendix \ref{secmsk} for more details).  Similarly, a poor continuum fit resulting either from regions with multiple emission and/or absorption lines or from nearing the wavelength limits of the SDSS spectrograph yield line measurements that are likely incorrect.  In addition to the catalogue of QAL systems in SDSS repeat spectra, the Y13 QAL pipeline produces normalized spectra for all quasars even when no systems are detected.  The continuum fitting algorithm uses 30-pixel windows for regions longward of the Ly$\alpha$ emission line.  Based on this window size, we set the minimum distance between an absorption line trough and a masked pixel such that the line and nearby continuum are free of masked pixels over half of a continuum window at a minimum.  To evaluate the quality of the continuum fit near an absorption line, we calculated the mean of the normalized pixel flux density over a region spanning two continuum windows.  However, this region is likely to contain other absorption lines and other deviant pixels so we use the \texttt{RESISTANT\_MEAN}\footnote{http://idlastro.gsfc.nasa.gov/} \textrm{\scriptsize{IDL}} routine to trim away outliers based on deviations from the median.  Since this is a rather crude estimate of the continuum mean flux density, we only flagged lines with values exceeding 1.0 $\pm$ 0.1.

Of the 5795 total system observations in Table \ref{tabsys}, 3857, or 67\%, have at least one flagged absorption line.  Breaking this down by system grade gives the following fractions of observations with flagged lines: 80\% for grade A observations (1966 out of 2453), 82\% (1315/1610) for grade B, and 33\% (576/1732) for grade C and below.  This trend suggests that although grade A and B systems are deemed reliable and accurate QAL systems, each system generally has more detected lines than systems with lower grades and are thus more likely to have an absorption line trigger one of the error conditions.  To determine the probability of having an absorption line with one of the error conditions, we compared the number of flagged lines to the total number of detected lines in each system observation.  Overall, 40\% of the absorption lines in a QAL system listed are flagged with 35\% in grade A observations, 47\% in grade B observations, and 44\% in grade C and below observations.  Given these results, additional care must be taken to accurately analyse absorption lines in a manner that does not introduce false artefacts when comparing lines from different observations of the same QAL system.

\section{DATA ANALYSIS} \label{secda}

Our primary reason for creating the catalogue of QAL systems in Table \ref{tabsys} with repeat SDSS observations was to conduct a systematic search for NAL time variability in a large, homogenous, and unbiased sample of QAL systems.  The Y13 QAL pipeline independently analyses each quasar observation for absorption lines from many different ion species that are prevalent in SDSS spectra \citep[see, e.g.,][]{lun09}, and variations in any one of these lines could signify short-term changes in the properties or dynamics of the absorbing gas clouds.  The redshift calculated for an absorption system often has slight deviations between multiple observations of the same object.  Since we are ultimately measuring the variability of these absorption features, we must apply the same system redshift to each observation.  When $\Delta z \leq $ 0.001 for the different observations (see Section \ref{seccat}), we averaged the different redshifts for the same system to get a single result that is used for all observations of the same absorption system.  

\subsection{NAL system refinement} \label{secnref}

Before these QAL systems can be analysed for NAL variability, the catalogue must be further refined by selecting only the highest quality spectra to minimize the occurrence of false variability due to noise fluctuations or errors in the data.  First, we chose the most reliable and accurate systems by requiring an `A' grade in at least one observation (see the electronic version of Table \ref{tabsys} for definitions of all the confidence grade categories).  Additionally, there is a known systematic error in the Y13 QAL pipeline that results in poor continuum fits near emission features \citep{lun09}.  Since the quality of the continuum fit is critical to measuring and comparing the absorption lines, we rejected systems with $\beta < 0.01$.  Finally, we only considered combinations of two quasar spectra with common absorption lines detected in both observations.  Together, these criteria removed 1501 systems.

The reliability of the individual absorption lines detected in the remaining 1084 QAL systems must also be assessed.  We automatically threw out all line detections shortward of Ly$\alpha$ because the Ly$\alpha$ forest makes any attempt to measure variations of absorption lines in this region extremely challenging and likely inaccurate due primarily to the difficulty in obtaining an accurate fit to the continuum flux level and distinguishing between hydrogen and metal absorption lines.  Of the three error conditions for which NALs are flagged in Table \ref{tabsys}, those in close proximity to masked pixels or a poor continuum fit are highly suspect and were not included in our variability measurements (flags $b$ and $c$ in Section \ref{secnflg}).  Absorption lines with an alternate identification were not removed because virtually all lines detected in a grade A system actually belong to that system.  However, there are exceptions to this rule so variable absorption lines must be individually inspected to determine the correct line identification.

\subsection{NAL variability} \label{secnvar}

Previous studies of small time-scale absorption line variability focused primarily on changes in the line rest frame equivalent width (EW) between observations since normalizing the pixel flux with the measured continuum distinguishes between absorption and continuum variability.  In addition to the absorption system redshift and confidence grade described in Appendix \ref{seczid}, the Y13 QAL pipeline calculates an EW and error from the normalized spectrum for each detected absorption line.  Any significant changes in the EW of an absorption feature between two different observations could be the result of variations in the absorbing gas.  Since the short-term variability of BALs and mini-BALs in quasar spectra has already been confirmed \citep[see, e.g.,][and references therein]{lun07,gib08}, we confined our search to NALs with $\textrm{EW} < 2$ $\textrm{\AA}$ in the rest frame of the absorption system.  After calculating $\Delta \textrm{EW} = | \textrm{EW}_1 - \textrm{EW}_2 |$ for all common absorption lines between two observations of a NAL system, we required a minimum $\Delta \textrm{EW}/\sigma_{\Delta \scriptsize{\textrm{EW}}} \geq 4$ for one line or $\geq 3$ for two lines to identify the strongest cases of absorption line variability.  These cuts yielded 69 absorption systems with lines that exhibit significant EW differences between observation epochs. 

Although the $\beta <$ 0.01 cut described in Section \ref{secnref} removed many systems in close proximity to the background quasar emission lines, absorption lines in foreground systems at higher $\beta$ values can also be superimposed on these regions of the quasar spectrum.  Due to the questionable continuum fits in these cases, systems with significantly variable absorption lines in close proximity to emission lines need to be excluded.  We calculated the quasar rest-frame wavelengths for all of the variable lines and compared them to a composite quasar spectrum created from SDSS observations \citep{van01}.  Prominent emission features in this spectrum that are coincident with the significantly variable absorption lines we detected include Ly$\alpha$, \mbox{N\,{\sc iv}}, \mbox{Si\,{\sc iv}}, \mbox{C\,{\sc iv}}, and \mbox{Mg\,{\sc ii}}.   Variable lines located within areas equal to twice the measured widths of these emission lines were rejected, which resulted in the removal of 21 absorption systems from our variable sample. 

The spectra of the common absorption lines in the remaining 48 candidate variable NAL systems were all visually inspected to assess the accuracy of the measured EW variations.  We were somewhat hindered in this process by the resolution of SDSS spectra that causes multiple individual components to blend together into a single line.  We primarily searched for significant changes in the line profile and pixel flux values (i.e., pixel-to-pixel flux variations greater than the flux errors).  Ion doublets such as \mbox{Mg\,{\sc ii}}, \mbox{C\,{\sc iv}}, \mbox{Al\,{\sc iii}}, \mbox{Ca\,{\sc ii}}, \mbox{N\,{\sc v}}, and \mbox{Si\,{\sc iv}} are particularly useful because detection of both lines reduces the likelihood of a valid alternate identification, and variations in the weaker, redder line should also be seen in the stronger, bluer line unless the line is highly saturated.  We also examined lines for variations consistent with their relative line strengths as described in Appendix \ref{secvar}.

A total of 15 systems were rejected after visual inspection because their significantly variable absorption lines displayed one or more of the following indicators of false variability.
\begin{enumerate}
\item Negligible flux variations outside the error bars.
\item Poor continuum fit.
\item Erroneous line detection or identification.
\item Unrealistic line variations.
\end{enumerate}
The first two conditions were found in nine of the rejected systems, mainly due to blending with nearby absorption lines and/or a high concentration of lines in a small wavelength range.  Five systems were rejected based on the last two conditions while one was removed because the significantly variable line in quasar SDSS J131347.68+294201.3 had alternate identifications from two grade A systems at different redshifts (\mbox{Fe\,{\sc ii}} $\lambda$2383 at $z_{\scriptsize{\textrm{abs}}}$ = 1.5157 and \mbox{Fe\,{\sc ii}} $\lambda$2600 at $z_{\scriptsize{\textrm{abs}}}$ = 1.3057).  After visual inspection of all absorption lines in these two systems, the latter identification was discarded because the former system showed more consistent variations in lines with similar strengths.  

\subsection{False variability} \label{secfvar}

Although all of the significantly variable absorption lines detected in each system were visually inspected for indicators of false variability, there still remains the possibility that the observed line variations between observation epochs were caused by noise fluctuations in the SDSS spectra.  Even with the stringent NAL variability criteria in Section \ref{secnvar}, we might expect some false variations simply from the statistical noise.  EW changes resulting from such variations would not be visually distinguishable from those that are real.  To test if statistical fluctuations in a large catalogue of NAL systems cause the absorption line variations we measured in this sample, we simulated multiple observations of each line in each system using the EW measurements from each real observation. 

In this Monte Carlo simulation, our main premise was that the EW of a given absorption line did not change between observations.  Thus, any variations detected in the simulated EWs would be the result of statistical fluctuations.  Under this assumption of no intrinsic variation, we simulated two independent observations of the same underlying absorption line based on the Gaussian nature of the noise.  Specifically, we generated two new EWs by randomly sampling Gaussian distributions with the same mean, specified by one of the measured EWs, but the two standard deviations were given by the two measured EW errors.  We simulated all lines within a system for each pair of epochs and accumulated any resultant statistical fluctuations in terms of the simulated EW differences ($\Delta \textrm{EW}_{\scriptsize{\textrm{sim}}} = | \textrm{EW}_{1,\scriptsize{\textrm{sim}}} - \textrm{EW}_{2,\scriptsize{\textrm{sim}}} |$) and the measured EW errors added in quadrature.  This process was repeated 10 000 times for each system, and we counted the number of occurrences where a system exhibited significant EW variations caused by statistical fluctuations ($\Delta \textrm{EW}_{\scriptsize{\textrm{sim}}}/\sigma_{\Delta \scriptsize{\textrm{EW}}} \geq 4$ for one line in a system or $\geq 3$ for two lines; see Section \ref{secnvar}).

\begin{table*}
\begin{center}
\caption{Catalogue format for the variable NAL systems detected in SDSS DR7 quasars with repeat observations.  See Table \ref{tabsys}, Appendix \ref{seczid}, and Section \ref{secnflg} for descriptions of these data.  The full table is available as an ancillary file in the astro-ph link for this paper.  \label{tabvar}}
\begin{tabular}{@{}cll@{}}
\hline
Column & Label & Explanation \\
\hline
1&             System&          System number index \\
2&             SDSS J&          SDSS DR7 object designation \\
3&             RA&              Right ascension in decimal degrees (J2000) \\
4&             Dec.&             Declination in decimal degrees (J2000) \\
5&             $z_{\scriptsize{\textrm{qso}}}$&           Quasar redshift \\
6&             $z_{\scriptsize{\textrm{abs}}}$&           NAL system redshift \\
7&             $\beta$&            NAL system velocity (v/c) in the QSO rest frame \\
8&             $\textrm{Plate}_1$&         SDSS plate ID - first observation \\
9&             $\textrm{Fibre}_1$&         SDSS fibre ID - first observation \\
10&            $\textrm{MJD}_1$&           SDSS MJD - first observation \\
11&            $\textrm{Plate}_2$&         SDSS plate ID - second observation \\
12&            $\textrm{Fibre}_2$&         SDSS fibre ID - second observation \\
13&            $\textrm{MJD}_2$&           SDSS MJD - second observation \\
14&            $\Delta t_{\scriptsize{\textrm{r}}}$&          Time between observation in the QSO rest frame (d) \\
15&            Ion&             Absorption line ion \\
16&            $\lambda_0$&     Absorption line rest wavelength (A) \\
17&            $\textrm{EW}_1$&            Absorption line equivalent width - first observation (A) \\
18&            $\sigma_{\scriptsize{\textrm{EW}_1}}$&          Absorption line equivalent width error - first observation (A) \\
19&            $\textrm{flag}_1$&          Absorption line flag - first observation \\
20&            $\textrm{EW}_2$&            Absorption line equivalent width - second observation (A) \\
21&            $\sigma_{\scriptsize{\textrm{EW}_2}}$&          Absorption line equivalent width error - second observation (A) \\
22&            $\textrm{flag}_2$&          Absorption line flag - second observation \\
\hline
\end{tabular}
\end{center}
\end{table*}

To quantify the probability of false variability due to statistical fluctuations, we took multiple approaches in this simulation.  First, we used the catalogue of 1084 NAL systems from Section \ref{secnref} but did not make any of the rest wavelength, error condition, or EW cuts that were applied to the real data.  This is the most conservative option because it includes systems in the simulation that would otherwise be removed with these cuts, thereby providing an upper limit on the false variability probability given by equation \ref{eqnfvp}
\begin{equation}
P_{\scriptsize{\textrm{false}}} = \frac {\sum \limits_{i=1}^{N_{\tiny{\textrm{sys}}}} \sum \limits_{j=1}^{N_{\tiny{\textrm{sim}}}} N_{\scriptsize{\textrm{false}}}}{N_{\scriptsize{\textrm{sys}}} N_{\scriptsize{\textrm{sim}}}} \label{eqnfvp}
\end{equation} 
where $P_{\scriptsize{\textrm{false}}}$ is the probability of false EW variations in our sample of NAL systems, $N_{\scriptsize{\textrm{false}}}$ is the number of trials where a significant false variation occurred (4$\sigma$ in one line or 3$\sigma$ in two or more lines), $N_{\scriptsize{\textrm{sys}}}$ is the number of systems, and $N_{\scriptsize{\textrm{sim}}}$ is the number of trials.  The resulting probability of false variations from our simulation was 8.3\e{-4} using the EW measured in the first observation of each system to generate the two simulated EWs in each trial, or roughly one system with false variations in a sample of 1200 with two observations.  The same probability is obtained to two significant Figs when using the EW measured in the second observation of each system.  If we apply the rest wavelength, error condition, or EW cuts from Sections \ref{secnref} and \ref{secnvar}, the probability of false variations drops to 4.5\e{-4}, or one system in 2200 with two observations.  Therefore, based on the results of this simulations, we conclude that the absorption line EW variations observed in the real data are not caused by statistical noise fluctuations.

\section{DISCUSSION} \label{secdis}

The system properties and absorption line EW measurements of the remaining 33 NAL systems with significant EW variations are presented in Table \ref{tabvar} with a detailed discussion in Appendix \ref{secvar}.  This population of variable NAL systems is the largest ever assembled, so we now have a sufficiently large sample to characterize the global properties of these systems.  We apply a cut at $\beta = 0.22$ to divide the population into absorption systems that are most likely intervening and those that are a mix of intervening, associated, and intrinsic systems (see Section \ref{secgp}).  The 10 NAL systems with $\beta > 0.22$ represent the first plausible evidence of variable intervening systems in quasar spectra, and we use this sample to estimate the size of small-scale structures in the foreground galaxies at cosmological distances from the background quasars.  The remaining 23 NAL systems with $\beta < 0.22$ are either high-velocity quasar outflows or gas clouds similar to those in intervening systems but located in galaxies much closer to the background quasar.  Comparing these low-$\beta$ systems to the 18 variable NAL systems from previous studies in Table \ref{tabprev}, we detect a similar number of variable NAL systems but have an initial sample size of low-$\beta$ systems that is over a magnitude larger (607 of the 1084 systems we searched for NAL variability have $\beta < 0.22$).  Additionally, our low-$\beta$ variable NAL systems have a wider range of variable absorption lines, higher $\beta$ values, and shorter time baselines than those in Table \ref{tabprev}.  These discrepancies can be attributed to the differences between the targeted, high-resolution spectroscopic observations of known NAL systems in past studies and the large, medium-resolution, systematic survey of previously unknown quasars in the SDSS.  In all likelihood, there are more variable NAL systems in the quasars observed by the SDSS, but we are limited by the fact that these observations were not optimized for detailed absorption line studies.

\subsection{Global properties} \label{secgp}

\begin{figure}
\begin{center}
\includegraphics[width=84mm]{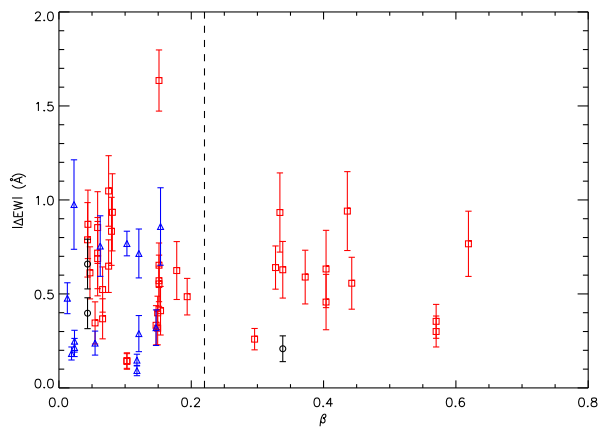}
\caption{The absolute change in EW ($|\Delta$EW$|$) versus $\beta$ for our sample of significantly variable absorption lines.  Different lines with the same $\beta$ are from the same NAL system.  The 13 lines from 10 intervening systems with $\beta > 0.22$ are all low-ionization lines since any high-ionization lines are shifted into the Ly$\alpha$ forest at these $\beta$ values.  The 41 lines from the remaining 23 NAL systems have a mix of ionization states due to the extreme environment close to the background quasar and the higher redshifts of these systems (see Fig. \ref{figdewzs}). \label{figdewb}}
\end{center}
\end{figure}

Table \ref{tabval} presents a breakdown by ion of the 54 absorption lines with significant EW changes in the variable NAL systems.  Given the energetic environment surrounding the background quasars, we first look for a relationship between the amount of absorption line variability and proximity to the quasar as measured by the $\beta$ parameter in Fig. \ref{figdewb}.  Error bars for $\beta$ are not included because the quasar and absorption system redshifts are known to very high precision \citep[see, e.g.,][]{hew10}, and they would, therefore, be smaller than the size of the points.  The points in these plots span a wide range of $|\Delta$EW$|$ and $\beta$ values, indicating no obvious dependence between variability and proximity to the quasar in our sample.  This result may seem surprising since variable absorption systems in close proximity to the quasar are exposed to the highly dynamic environment of quasar outflows that would certainly cause greater absorption line variability compared to systems at higher $\beta$ values.  However, we removed the majority of these systems with our cut of $\beta > 0.01$ (see Section \ref{secnref}).

\begin{table}
\begin{center}
\caption[Significantly variable absorption lines for the population of variable NAL systems]{The significantly variable absorption lines in the population of variable 33 NAL systems in Table \ref{tabvar}.  The colour/symbol combinations are used in Figs \ref{figdewb}--\ref{figrewmei}. \label{tabval}}
\begin{tabular}{@{}cccc@{}}
\hline
Symbol & Ion & Absorption line & \multicolumn{1}{c}{Number of lines} \\
\hline
\rule{0pt}{3ex} {\large \color{red}$\square$}& \mbox{Al\,{\sc ii}}& $\lambda$1671& 3\\
& \mbox{C\,{\sc ii}}& $\lambda$1335& 1\\
& \mbox{Fe\,{\sc ii}}& $\lambda$2344& 17\\
& & $\lambda$2374\\
& & $\lambda$2383\\
& & $\lambda$2587\\
& & $\lambda$2600\\
& \mbox{Mg\,{\sc i}}& $\lambda$2853& 1\\
& \mbox{Mg\,{\sc ii}}& $\lambda \lambda$2796, 2804& 13\\
& \mbox{Si\,{\sc ii}}& $\lambda$1527& 2\\
{\large \color{blue}$\bigtriangleup$}& \mbox{Al\,{\sc iii}}& $\lambda$1855& 1\\
& \mbox{C\,{\sc iv}}& $\lambda \lambda$1548, 1551& 11\\
& \mbox{Si\,{\sc iv}}& $\lambda \lambda$1394, 1403& 2\\
$\bigcirc$& \mbox{C\,{\sc i}}& $\lambda$1277& 2\\
& & $\lambda$1657\\
& \mbox{Ni\,{\sc ii}}& $\lambda$1742& 1\\
\hline
\end{tabular}
\end{center}
\end{table}

\begin{figure}
\begin{center}
\includegraphics[width=84mm]{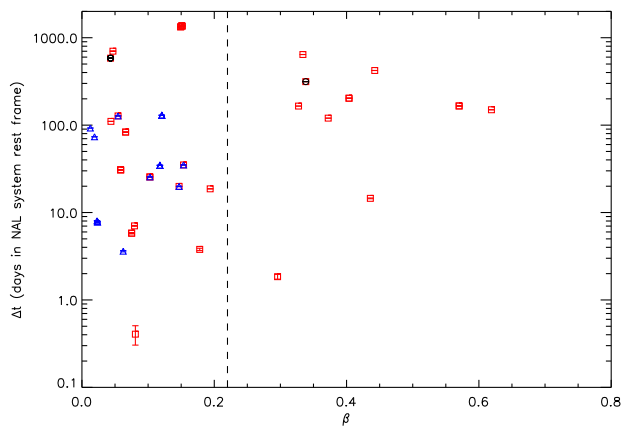}
\caption[$\Delta t_{\scriptsize{\textrm{r}}}$ versus $\beta$ for the population of variable NAL systems (colour-coded by ion)]{The rest-frame time between observations ($\Delta t_{\scriptsize{\textrm{r}}}$) versus $\beta$ for the significantly variable absorption lines in our sample.  A logarithmic scale is used for the \textit{y}-axis due to the large range of time separations spanning days to years, and we divide our sample into intervening and low-$\beta$ systems at $\beta = 0.22$.  The points are scattered in both the high and low-$\beta$ regimes meaning that there is no obvious dependence between $\Delta t_{\scriptsize{\textrm{r}}}$ and $\beta$. \label{figdtbe}}
\end{center}
\end{figure}

\begin{figure*}
\begin{center}
\includegraphics[width=84mm]{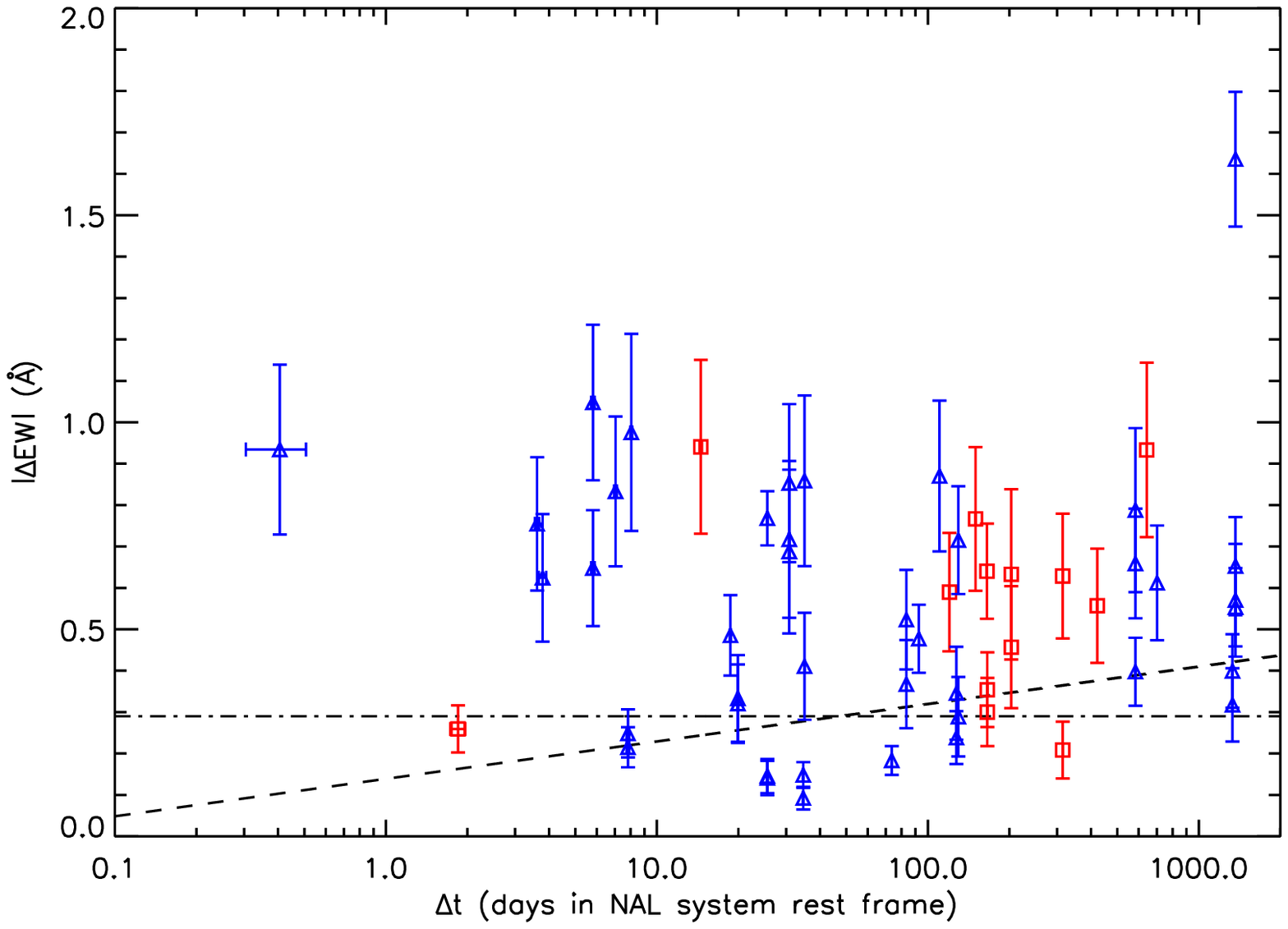}
\includegraphics[width=84mm]{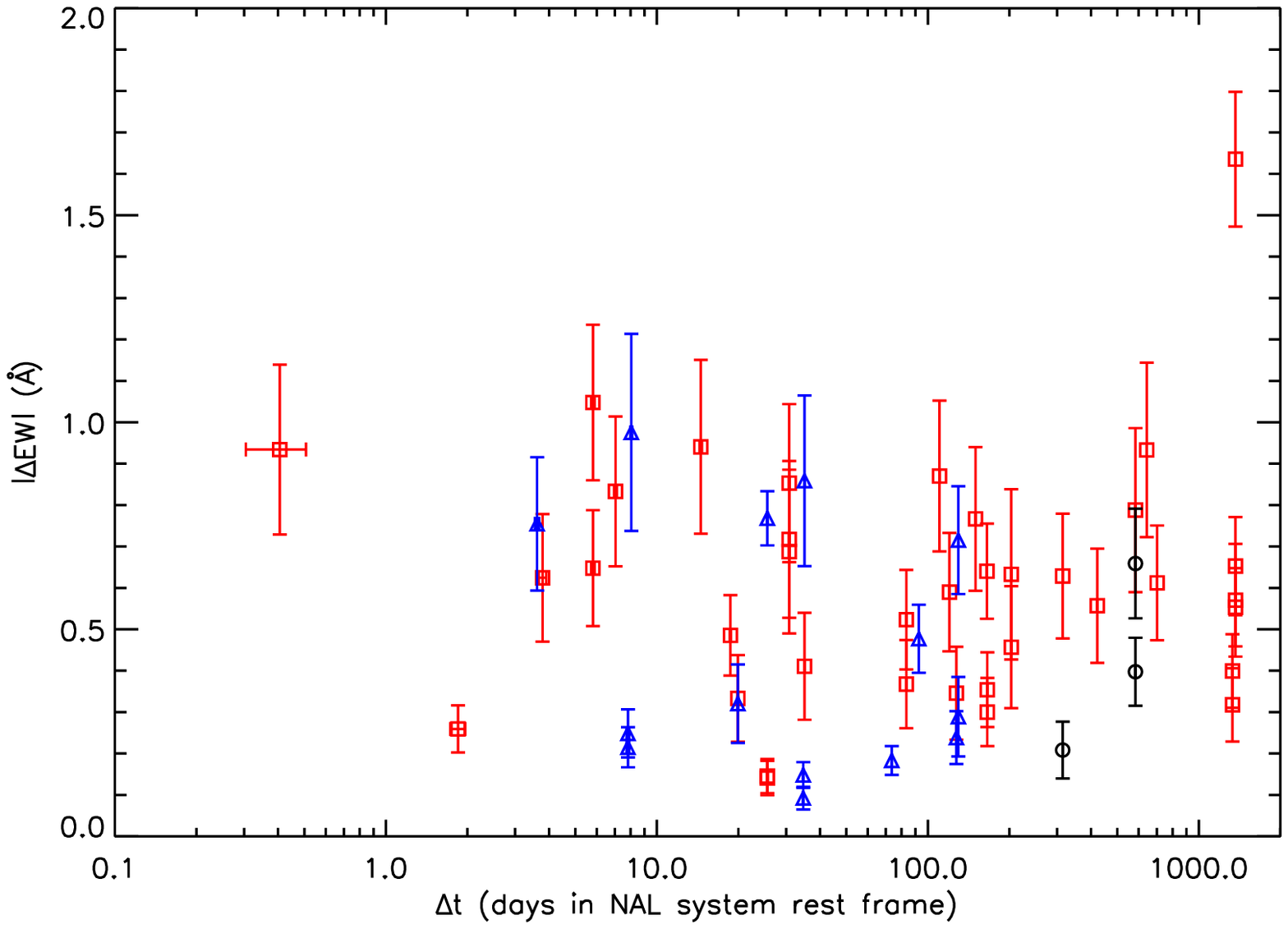}
\caption{The absolute change in EW ($|\Delta$EW$|$) versus the rest-frame time between observations ($\Delta t_{\scriptsize{\textrm{r}}}$) for the significantly variable absorption lines in our sample.  A logarithmic scale is used for the \textit{x}-axis due to the large range of time separations spanning days to years.  In the left-hand panel, the points are colour-coded by $\beta$ (high-$\beta$: red squares; low-$\beta$: blue triangles) with log-linear (dashed) and zero-slope (dot-dashed) best-fitting lines to all points.  These fits indicate no dependence between $\Delta$EW and $\Delta t_{\scriptsize{\textrm{r}}}$, which suggests that some of the absorbing gas clouds may be comprised of small-scale structures.  In the right-hand panel, the lines from a wide range of ions are scattered at all values of $\Delta$EW and $\Delta t_{\scriptsize{\textrm{r}}}$, which again suggests that the sample is unbiased. \label{figdewt}}
\end{center}
\end{figure*}

\begin{figure*}
\begin{center}
\includegraphics[width=84mm]{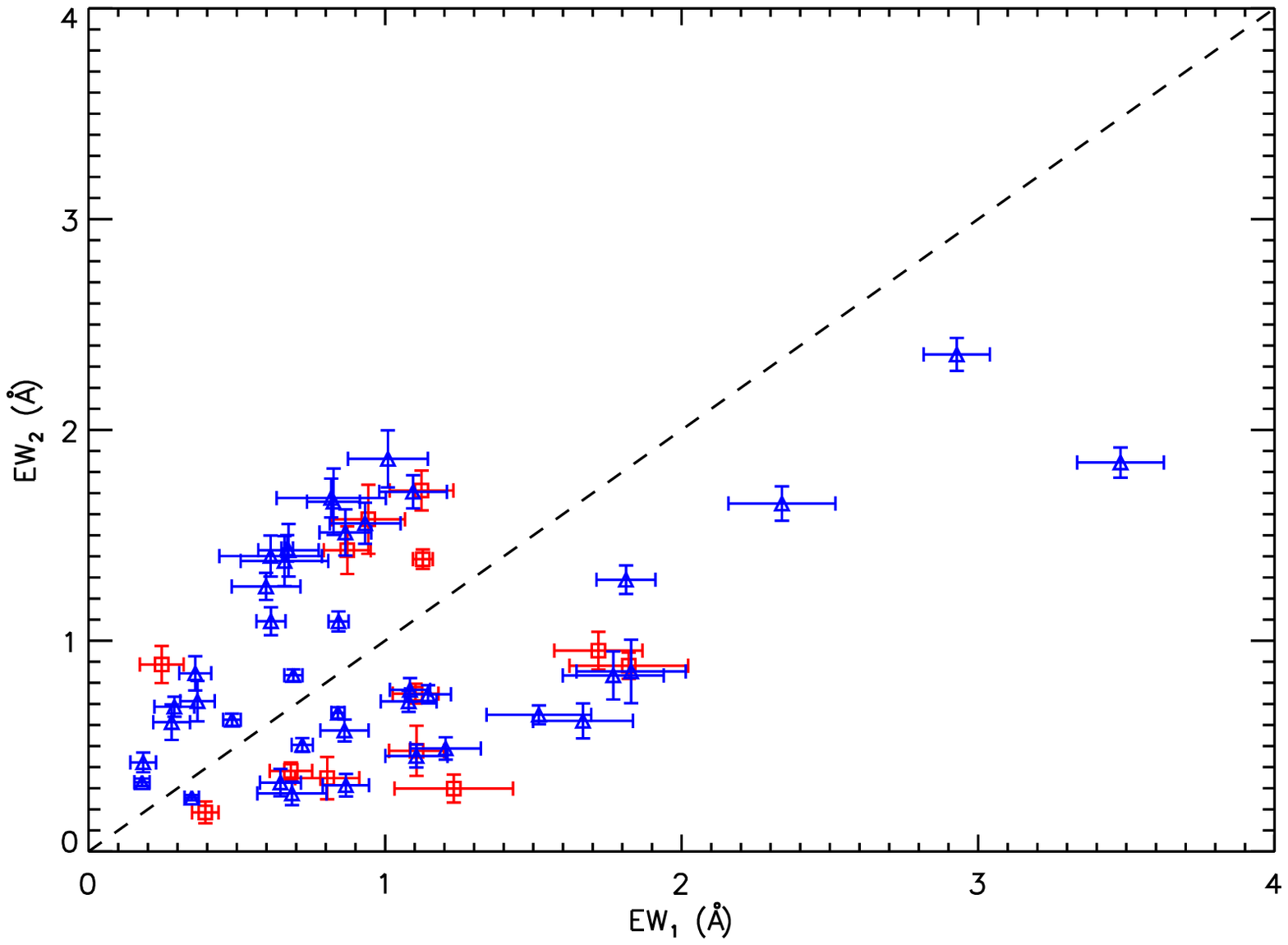}
\includegraphics[width=84mm]{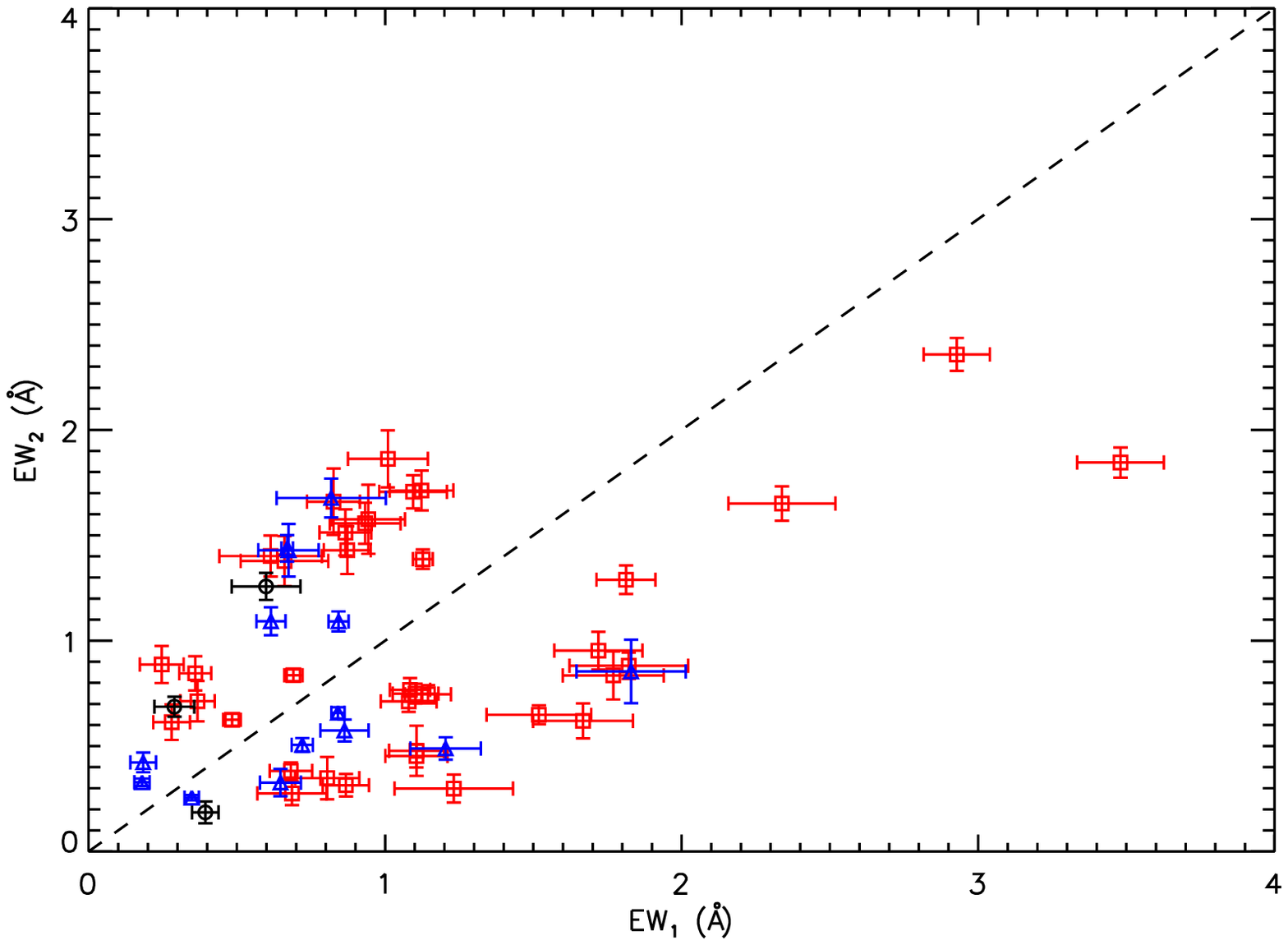}
\caption{The first (EW$_1$) and second (EW$_2$) epoch EW measurements for the significantly variable absorption lines in our sample.  In the left-hand panel, both the intervening ($\beta > 0.22$; red squares) and low-$\beta$ systems (blue triangles) have similar numbers of lines above (26) and below (28) the EW$_1$ = EW$_2$ line, which suggests that an absorption line is just as likely to get weaker as it is to get stronger between observations.  In the right-hand panel, all absorption lines are scattered around the EW$_1$ = EW$_2$ line.  The exception is \mbox{Mg\,{\sc ii}} with three outlying points with EW$_1$ $>$ EW$_2$ that are biased due to saturation. \label{figew12}}
\end{center}
\end{figure*}

\begin{figure*}
\begin{center}
\includegraphics[width=84mm]{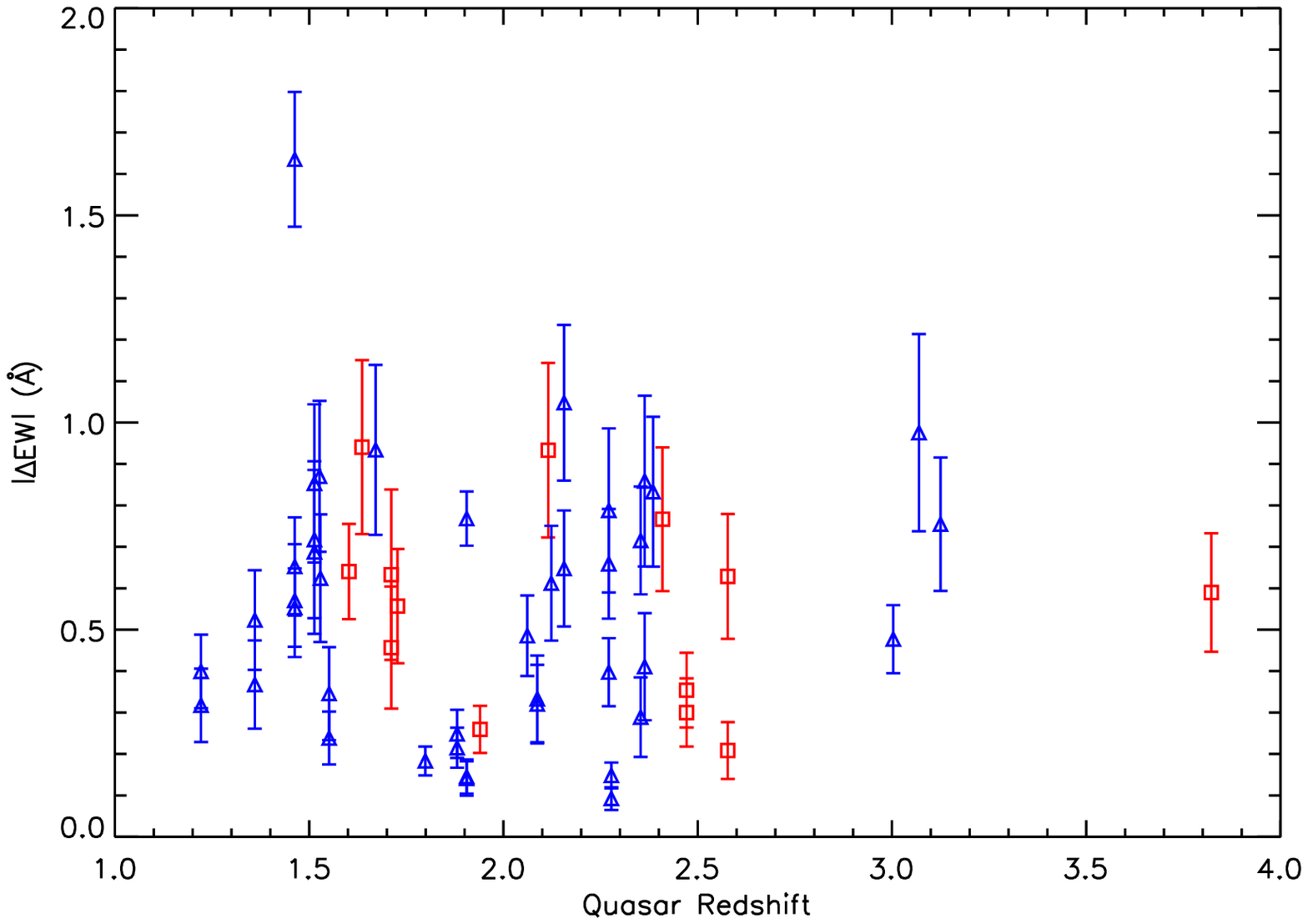}
\includegraphics[width=84mm]{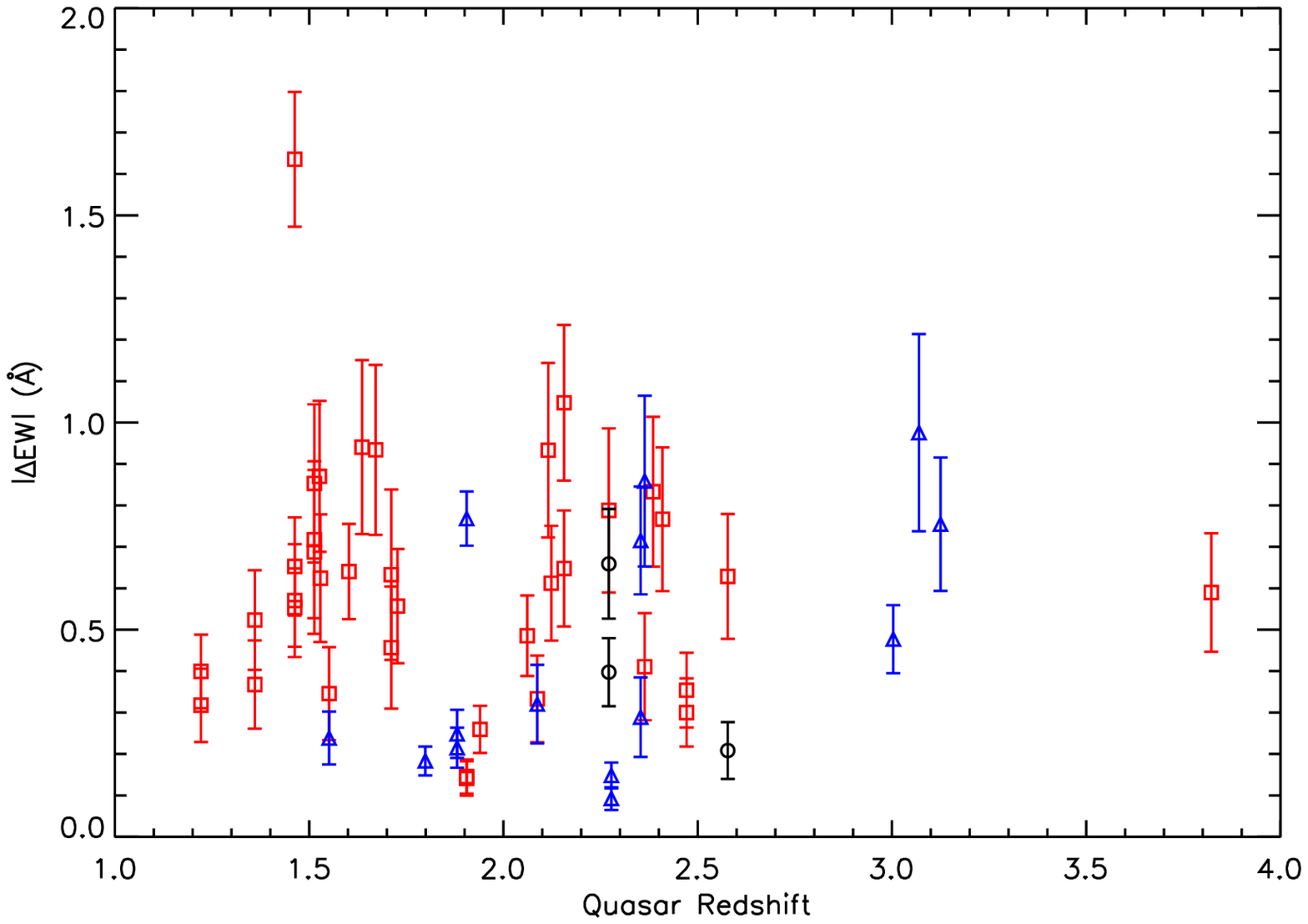}
\caption{The absolute change in EW ($|\Delta$EW$|$) versus quasar redshift ($z_{\scriptsize{\textrm{qso}}}$) for the significantly variable absorption lines in our sample.  Although the points in the left-hand panel (high-$\beta$ systems: red squares; low-$\beta$ systems: blue triangles) are scattered indicating that absorption line variability does not depend on $z_{\scriptsize{\textrm{qso}}}$, the overall distribution is slightly biased towards higher redshift compared to all SDSS DR7 quasars with repeat observations (see Fig. \ref{fignzqr}) because the UV absorption lines in quasars at lower $z_{\scriptsize{\textrm{qso}}}$ are not shifted into the optical range of the SDSS spectrographs.  In the right-hand panel, the high-ionization and minor lines have a higher average $z_{\scriptsize{\textrm{qso}}}$ compared to the low-ionization lines because their rest wavelengths require higher $z_{\scriptsize{\textrm{qso}}}$ for detection by the SDSS spectrographs. \label{figdewzq}}
\end{center}
\end{figure*}

\begin{figure*}
\begin{center}
\includegraphics[width=84mm]{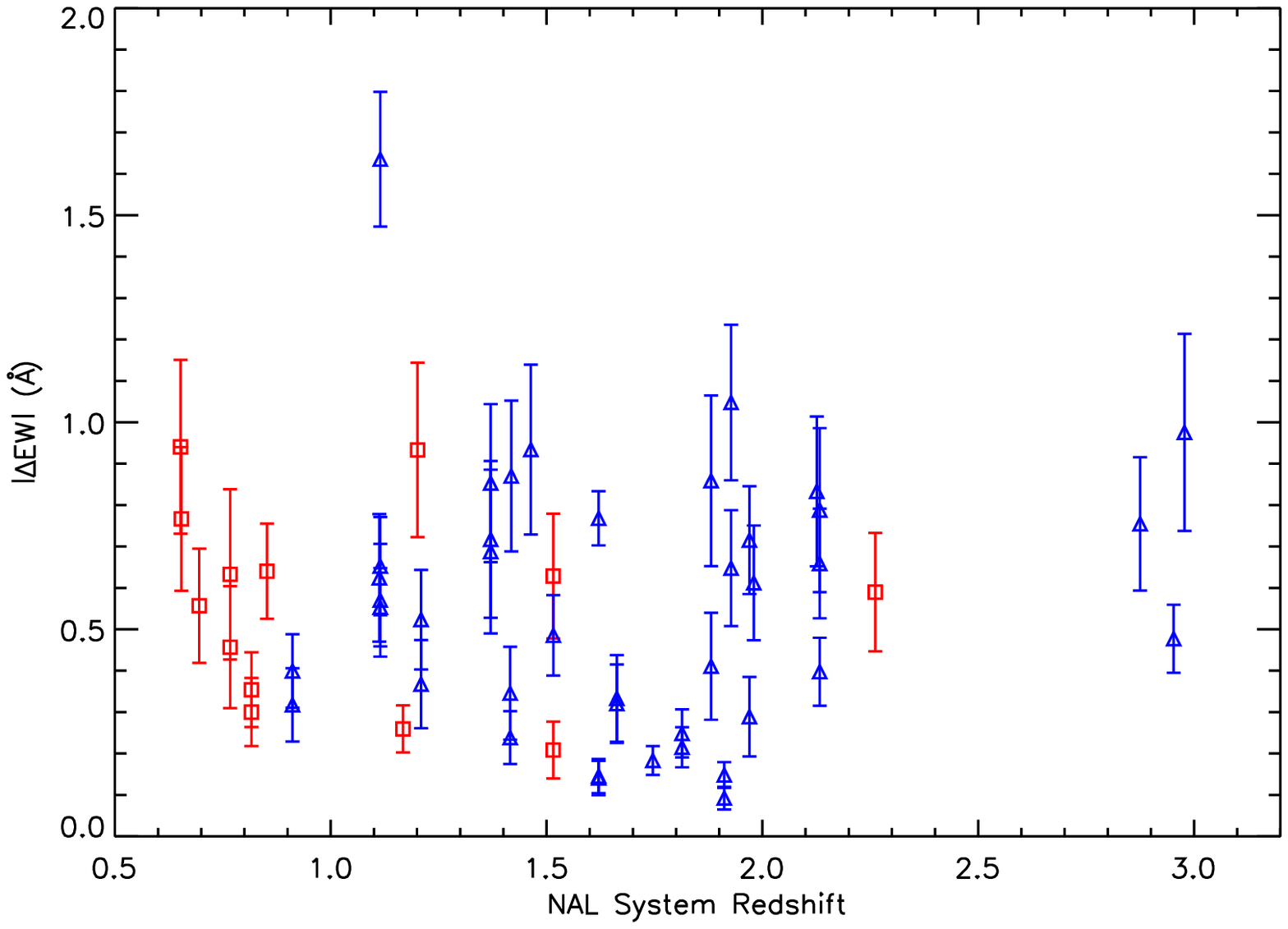}
\includegraphics[width=84mm]{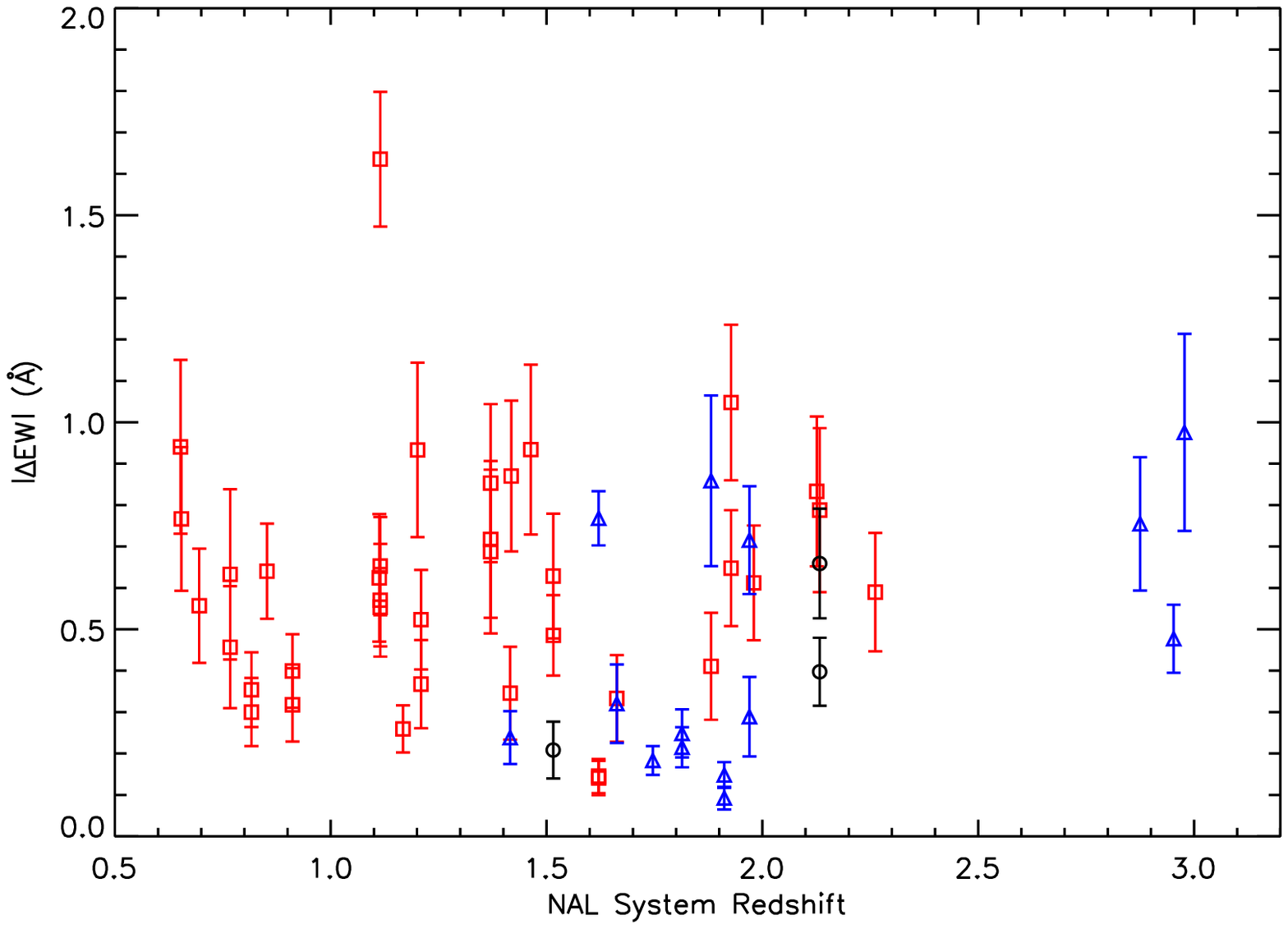}
\caption{The absolute change in EW ($|\Delta$EW$|$) versus NAL system redshift ($z_{\scriptsize{\textrm{abs}}}$) for the significantly variable absorption lines in our sample.  The points in the left-hand panel (high-$\beta$ systems: red squares; low-$\beta$ systems: blue triangles) are scattered, which confirms that there is no clear dependence between $|\Delta$EW$|$ and $z_{\scriptsize{\textrm{abs}}}$.  In the right-hand panel, there is a higher occurrence of low-ionization lines at lower $z_{\scriptsize{\textrm{abs}}}$ and high-ionization and minor lines at higher $z_{\scriptsize{\textrm{abs}}}$ due to the observable redshift window in SDSS quasar spectra. \label{figdewzs}}
\end{center}
\end{figure*}

\begin{figure*}
\begin{center}
\includegraphics[width=84mm]{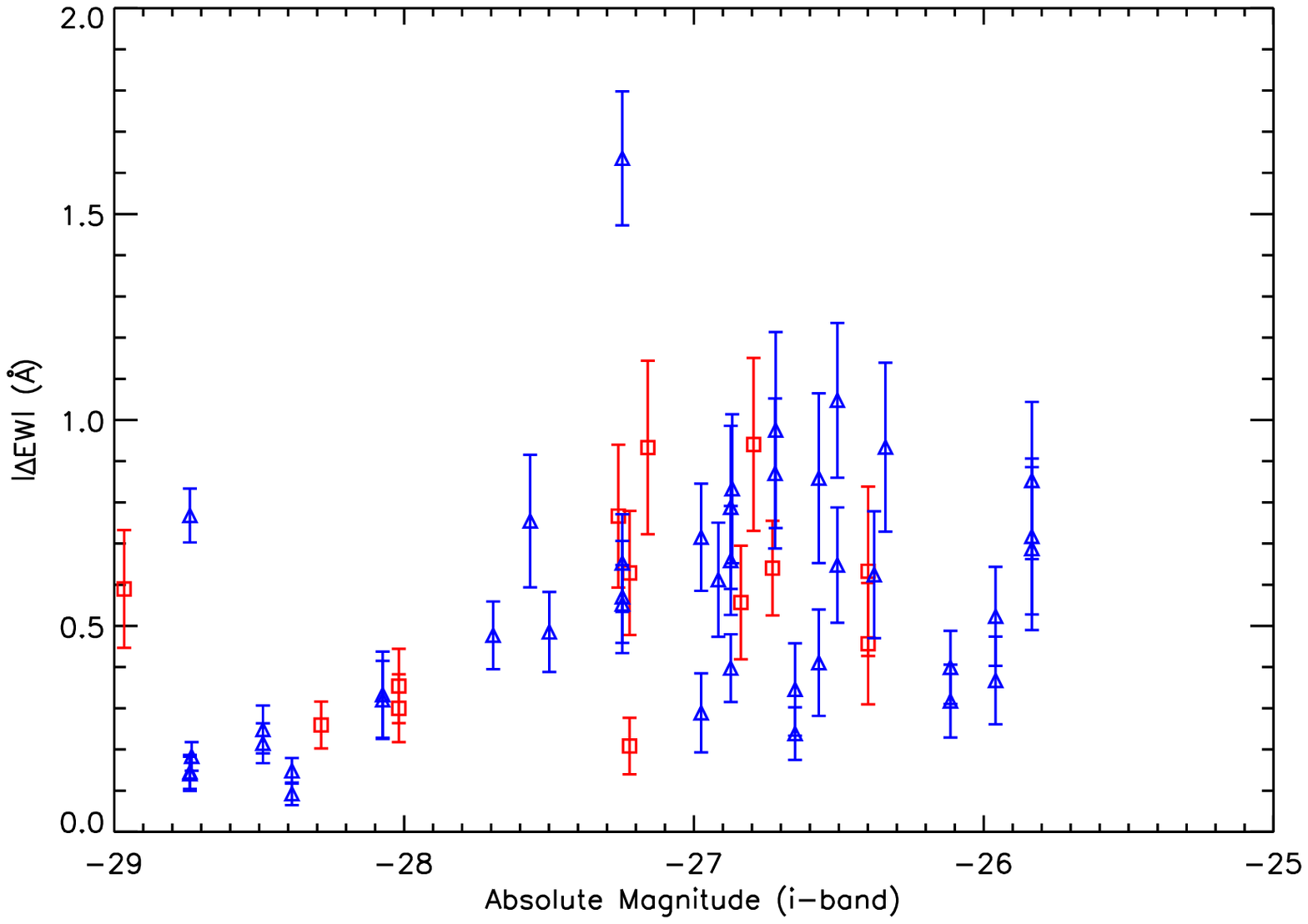}
\includegraphics[width=84mm]{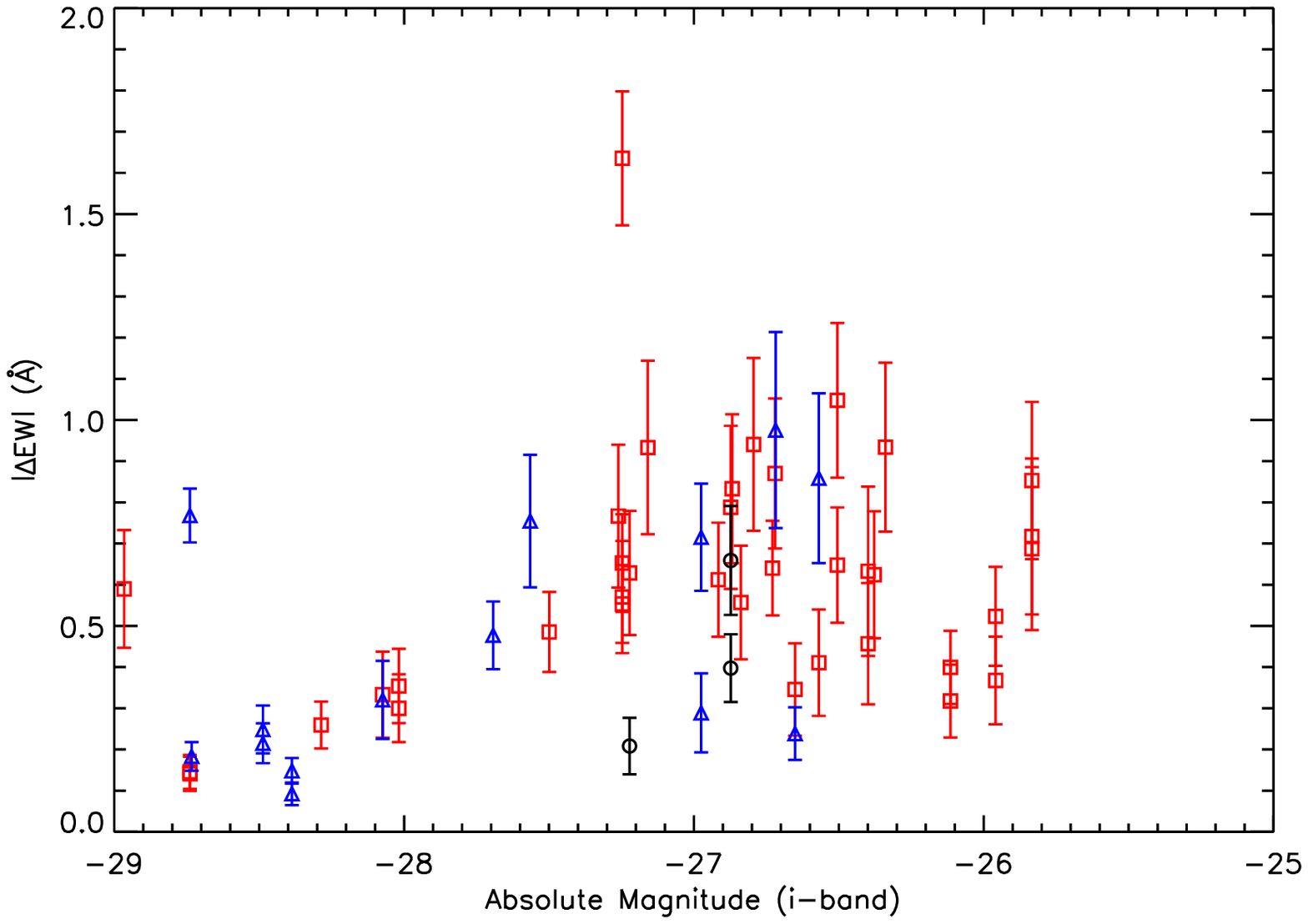}
\caption[$|\Delta$EW$|$ versus $M_i$ for the population of variable NAL systems (colour-coded by $\beta$)]{The absolute change in EW ($|\Delta$EW$|$) versus $i$-band absolute magnitude ($M_i$) for the significantly variable absorption lines in our sample.  For systems with $M_i$ $>$ --28, the points in the left-hand panel are scattered with no dependence between variability, quasar luminosity, or $\beta$ (high-$\beta$ systems: red squares; low-$\beta$ systems: blue triangles).  The apparent bias of points towards lower quasar luminosity is a selection effect caused by the peak in the distribution of SDSS DR7 quasars with repeat observations at --27 $<$ $M_i$ $<$ --25 (see Fig. \ref{fignzqr}).  In the right-hand panel, the high-ionization and minor lines are found in the brighter quasar spectra compared to low-ionization lines for two reasons: these lines must be at higher redshift to be shifted to the optical range of the SDSS spectrographs, and higher redshift quasars are preferentially brighter due to the flux limit of the survey. \label{figdewmi}}
\end{center}
\end{figure*}

The current highest velocity QAL system with absorption line variability was identified as a BAL at $\beta = 0.22$, which is marked by a dashed line in Figs \ref{figdewb} and \ref{figdtbe} \citep{fol83}.  Based on this result, we attribute systems with $\beta > 0.22$ values to intervening foreground galaxies while those at $\beta < 0.22$ are either intrinsic, associated, or intervening systems.  In Figs \ref{figdewb}--\ref{figrewmei} and Table \ref{tabval}, the absorption lines in the variable NAL systems are colour-coded based on common low-ionization (neutral and singly ionized: red squares), high-ionization (doubly and triply ionized: blue triangles), and minor (black circles) species.  The 13 lines from 10 intervening systems with $\beta > 0.22$ are all low-ionization lines as expected since high-ionization lines with shorter rest wavelengths are shifted into the Ly$\alpha$ forest at these high $\beta$ values.  The global properties of these systems are discussed in Section \ref{secial}.  The 41 lines from the remaining 23 NAL systems, described in Section \ref{secaal}, have a mix of ionization states due to the extreme environment close to the background quasar and the higher overall redshifts of these systems (see Fig. \ref{figdewzs}).   

Since the absorption line variability observed in our sample does not have a clear dependence on proximity to the background quasar, we searched for connections between these parameters and the rest-frame time between observations ($\Delta t_{\scriptsize{\textrm{r}}}$).  Fig. \ref{figdtbe} plots $\Delta t_{\scriptsize{\textrm{r}}}$ and $\beta$ where the $\Delta t_{\scriptsize{\textrm{r}}}$ error bars are $\pm$6 h shifted to the rest frame of the absorption system.  This estimate is used because the quasar spectra are identified with the MJD of the last observation, and multiple spectra of each object are taken during an observation window that are co-added to give the final spectrum \citep[see, e.g., Section \ref{secsdss} \&][]{wil05}.  Most SDSS spectroscopic observations of a given object are completed on a single night so we based our $\Delta t_{\scriptsize{\textrm{r}}}$ error estimate on a nominal 12-h observing window.  The scatter of points once again shows that there is no clear dependence between quasar proximity and time between observations (other than the normal seasonal imprint).  Also, the distribution of points on both sides of the $\beta = 0.22$ line shows that we did not introduce any bias by dividing our sample into high and low-$\beta$ systems.  We do not find any variable absorption lines in systems with high-$\beta$ and small $\Delta t_{\scriptsize{\textrm{r}}}$ (lower-right region in Fig. \ref{figdtbe}).  Although this result matches the distribution of high-$\beta$ systems in the overall NAL system sample, it is also physically plausible that this is related to the physical sizes of small-scale structures in the absorbing gas clouds (see Section \ref{secmod}).  However, we do not have a sufficient number of high-$\beta$ and small $\Delta t_{\scriptsize{\textrm{r}}}$ repeat observations to distinguish between these two possibilities.

We observe a similar result in the comparison of $|\Delta$EW$|$ and $\Delta t_{\scriptsize{\textrm{r}}}$ in Fig. \ref{figdewt}.  To confirm the lack of dependence between these parameters, we plot the weighted least-squares log-linear (dashed) and zero-slope (dot-dashed) best-fitting lines to all points in the left-hand panel of Fig. \ref{figdewt}.  The reduced chi-squared values are over eight for both fits, which indicates that $|\Delta$EW$|$ and $\Delta t_{\scriptsize{\textrm{r}}}$ are independent with no obvious systematic biases in our sample of variable NAL systems.  The observed EW variations suggest that the absorbing gas clouds could be comprised of small-scale structures moving relative to the quasar line of sight, especially in the high-$\beta$ systems where short time-scale changes in ionizing flux are thought to be unlikely (see Section \ref{secmod}).  Furthermore, variations in BAL systems on similar time-scales have been recently reported \citep{cap11,cap13}.  Most of the high-$\beta$ lines (red squares) have $\Delta t_{\scriptsize{\textrm{r}}}$ $>$ 100 d, but this could simply be the result of small numbers in the sample.  Finally, the scattered distribution of points by ion in the right-hand panel of Fig. \ref{figdewt} further implies that the 33 variable NAL systems do not suffer from any systematic effects, which means that there are likely multiple scales in the physical sizes of the absorbing gas clouds.

Given that variable absorption lines are identified based on rest-frame EW changes between observations, the first observation EW (EW$_1$) is compared to the second observation (EW$_2$) in Fig. \ref{figew12} for the 54 absorption lines with EW variations greater than the 3$\sigma$ level.  Lines with EW $>$ 2 $\textrm{\AA}$ are included from systems with another variable line below this limit.  There is an exclusion zone around EW$_1$ = EW$_2$ because all of these lines have significant differences between EW$_1$ and EW$_2$ as defined in Section \ref{secnvar}.  In the left-hand panel, there are similar numbers of lines above (26) and below (28) the EW$_1$ = EW$_2$ line, which suggests that an absorption line is just as likely to get weaker as it is to get stronger between observations regardless of whether it belongs to a high (red squares) or low-$\beta$ (blue triangles) NAL system.  However, additional observations of these systems could possibly reveal how the trend of these changes depends on time (e.g., at what point does a line that strengthens from EW$_1$ to EW$_2$ start to weaken or disappear as the absorbing gas cloud moves in front of the quasar beam).  In the right-hand panel of Fig. \ref{figew12}, all of the points are scattered around the EW$_1$ = EW$_2$ line with the exception of three outlying \mbox{Mg\,{\sc ii}} lines where EW$_1$ $>$ EW$_2$.  The measured EWs for all of these lines are biased due to saturation, so they do not accurately represent the conditions of the absorbing gas.  Thus, Fig. \ref{figew12} confirms that there is no obvious systematic bias in the measured variable absorption line EWs.

Next, we examine the absorption line variability measured in these NAL systems relative to the quasar ($z_{\scriptsize{\textrm{qso}}}$) and system ($z_{\scriptsize{\textrm{abs}}}$) redshifts.  In Fig. \ref{figdewzq}, we examine the absolute change in EW and $z_{\scriptsize{\textrm{qso}}}$.  In the left-hand panel, the points are scattered indicating that variability does not depend on $z_{\scriptsize{\textrm{qso}}}$.  However, their distribution is slightly biased towards higher redshifts compared to the distribution for all SDSS DR7 quasars with repeat observations (see Fig. \ref{fignzqr}) because the UV absorption lines in quasar spectra at lower $z_{\scriptsize{\textrm{qso}}}$ are not shifted into the optical range of the SDSS spectrographs.  There is a higher average $z_{\scriptsize{\textrm{qso}}}$ of 2.21 for high-$\beta$ systems (red squares) versus 1.94 for low-$\beta$ systems (blue triangles) due to the greater difference between $z_{\scriptsize{\textrm{qso}}}$ and $z_{\scriptsize{\textrm{abs}}}$ in high-$\beta$ systems.  This selection effect caused by the SDSS spectrographs also affects the distribution of points with respect to ion type as seen in the right-hand panel of Fig. \ref{figdewzq}.  As expected, the high-ionization and minor lines (blue triangles and black circles, see Table \ref{tabval}) have a higher average $z_{\scriptsize{\textrm{qso}}}$ of 2.33 compared to 1.97 for the low-ionization lines (red squares) because their shorter rest wavelengths require higher $z_{\scriptsize{\textrm{qso}}}$ for detection by the SDSS spectrographs.

The absolute change in EW is compared to $z_{\scriptsize{\textrm{abs}}}$ in Fig. \ref{figdewzs}.  Once again, the points are scattered with respect to $|\Delta \textrm{EW}|$, which confirms that there is no obvious dependence between variable EWs and $z_{\scriptsize{\textrm{abs}}}$.  In the left-hand panel, the high-$\beta$ lines (red squares) are biased towards low redshift because $z_{\scriptsize{\textrm{qso}}}$ $\gg$ $z_{\scriptsize{\textrm{abs}}}$ because any high-ionization lines in these systems are shifted into the Ly$\alpha$ forest.  The distribution of low-$\beta$ lines (blue triangles) more closely matches the overall QAL $\beta$ distribution in Fig. \ref{fignzsy}.  A similar result is evident in the right-hand panel of Fig. \ref{figdewzs} where the lines are highlighted by ion type.  There is a higher occurrence of low-ionization lines (red squares, see Table \ref{tabval}) at lower $z_{\scriptsize{\textrm{abs}}}$ and high-ionization and minor lines (blue triangles and black circles) at higher $z_{\scriptsize{\textrm{abs}}}$ due again to the observable redshift window in SDSS quasar spectra.

Finally, we consider the relationship between absorption line variability in these NAL systems and quasar luminosity.  Fig. \ref{figdewmi} compares the absolute change is EW with the $i$-band absolute magnitude ($M_i$).  For systems with $M_i$ $>$ --28, the points are scattered with no discernible dependence between variability and quasar luminosity.  The distribution of the points seems to suggest that variable absorption systems are preferentially found in lower luminosity quasars.  However, this apparent bias is likely a selection effect caused by the peak in the distribution of SDSS DR7 quasars with repeat observations at --27 $<$ $M_i$ $<$ --25 (see Fig. \ref{fignzqr}).  In the right-hand panel of Fig. \ref{figdewmi}, the high-ionization and minor lines (blue triangles and black circles, see Table \ref{tabval}) are found in the brighter quasar spectra compared to low-ionization lines (red squares).  There are two reasons for this result: these lines must be at higher redshift to be shifted into the optical range of the SDSS spectrographs, and higher redshift quasars are preferentially brighter due to the flux limit of the survey.

After looking at the EWs, redshifts, and quasar luminosities in the population of 33 variable NAL systems in Table \ref{tabvar}, we conclude that there are no obvious systematic biases in the systems that were identified.  Fig. \ref{figew12} shows that a variable absorption line is just as likely to get weaker as it is to get stronger between observations.  The distributions of variable lines versus the quasar and system redshifts in Figs \ref{figdewzq} and \ref{figdewzs} match the expected results based on the overall quasar and absorption system distributions, the occurrence of various ionization states based on the quasar and absorption system redshifts, and the properties of the SDSS spectrographs.  A similar result is seen when comparing variability to quasar luminosity in Fig. \ref{figdewmi}.  Having gained confidence that the absorption variability detected in these NAL systems is likely not the result of systematic biases in the sample, we next explore their properties to discover potential new insights into the conditions of absorbing gas clouds that might cause short time-scale variability.

\subsection{Intervening NAL systems} \label{secial}

Although there have been no previously reported detections of intervening NAL variability in quasar spectra, there was a claim to the discovery of \mbox{Fe\,{\sc ii}} and \mbox{Mg\,{\sc ii}} absorption variability in an intervening system detected in multi-epoch spectra of GRB060206 \citep{hao07}.  Following the discovery that the incidence of \mbox{Mg\,{\sc ii}} absorption systems along gamma-ray burst (GRB) sight lines is higher compared to quasars \citep{pro06}, a geometric solution was proposed to explain this result based on small-scale absorbing cloud structures ($\lsim$10$^{16}$ cm) and different beam sizes for these background sources \citep{fra07}.  A direct consequence of this model is the prediction that the absorption lines should vary in strength on short time-scales as the GRB evolves and the beam size changes.  However, the reported intervening absorption variations in GRB060206 were subsequently refuted by independent observations of the same GRB with higher resolution and better signal-to-noise ratio spectra that displayed no evidence of absorption line variability \citep{tho08,aok09}.  Furthermore, a recent investigation of four intervening NAL systems in high-resolution spectra of GRB080319B failed to find any evidence of absorption line variability at high confidence \citep{del10}.  Thus, the 10 systems we identified from the catalogue in Table \ref{tabsys} with significant EW changes represent the first detection of plausible absorption line variability in intervening NAL systems.

\begin{figure}
\begin{center}
\includegraphics[width=84mm]{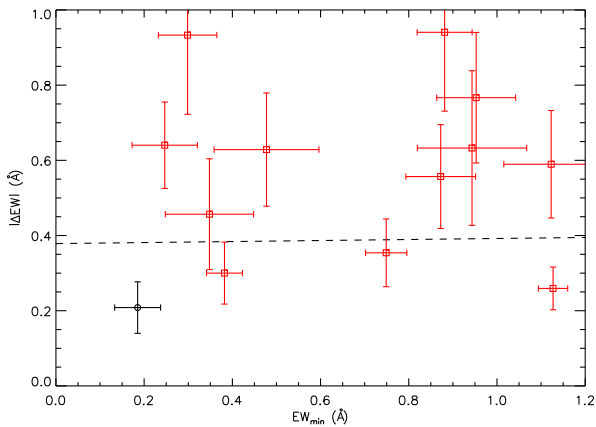}
\caption[$|\Delta$EW$|$ versus EW$_{\scriptsize{\textrm{min}}}$ for the population of variable intervening NAL systems (colour-coded by ion)]{The absolute change in EW ($|\Delta$EW$|$) versus the minimum EW (EW$_{\scriptsize{\textrm{min}}}$) for our sample of significantly variable IALs.  The best-fitting line (dashed black) implies that there is no strong dependence between variability and line strength. \label{figdewmei}}
\end{center}
\end{figure}

We first look for a relationship between the amount of variability in each IAL and the minimum EW.  \citet{nes05} showed that intervening \mbox{Mg\,{\sc ii}} absorption systems in SDSS quasar display the known dependence between absorption cross-section ($\partial N/\partial z$) and line strength, namely that $\partial N/\partial z$ is much higher in weak versus strong systems.  Thus, we might expect to see increased variability with decreasing line strength.  Fig. \ref{figdewmei} plots $|\Delta$EW$|$ versus EW$_{\scriptsize{\textrm{min}}}$, broken down by ion (see Table \ref{tabval} for colours/symbols).  The scatter of the points suggests that the measured variability in our sample of intervening NAL systems has very little dependence on line strength.  To confirm this, we perform a weighted linear least-squares fit to the points and included this result in Fig. \ref{figdewmei}.  The slope of this line is 0.01 $\pm$ 0.08, which implies that there is no strong dependence between the amount of variability and line strength in the intervening NAL system sample.  This outcome appears to dispute our predicted relationship based on the higher occurrence of weak absorption systems, but it has been observed in previous variable BAL studies \citep{bar94,gib08,cap11}.  

\begin{figure}
\begin{center}
\includegraphics[width=84mm]{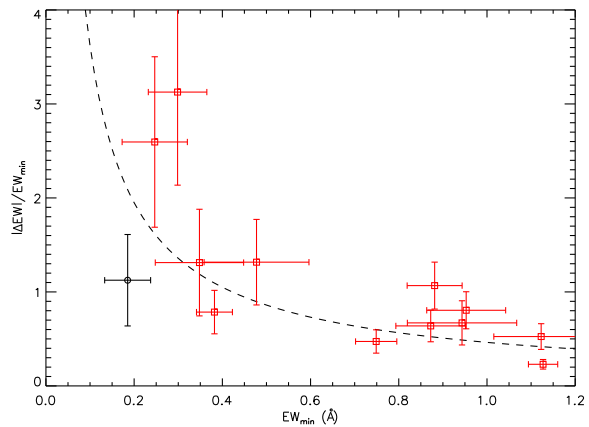}
\caption[$|\Delta$EW$|$/EW$_{\scriptsize{\textrm{min}}}$ versus EW$_{\scriptsize{\textrm{min}}}$ for the population of variable intervening NAL systems (colour-coded by ion)]{The absolute fractional change in EW ($|\Delta$EW$|$/EW$_{\scriptsize{\textrm{min}}}$) versus the minimum EW (EW$_{\scriptsize{\textrm{min}}}$) for our sample of significantly variable IALs.  The best-fitting line (dashed black) shows that weaker lines have higher fractional variability in agreement with previous BAL variability studies. \label{figfewmei}}
\end{center}
\end{figure}

\begin{figure}
\begin{center}
\includegraphics[width=84mm]{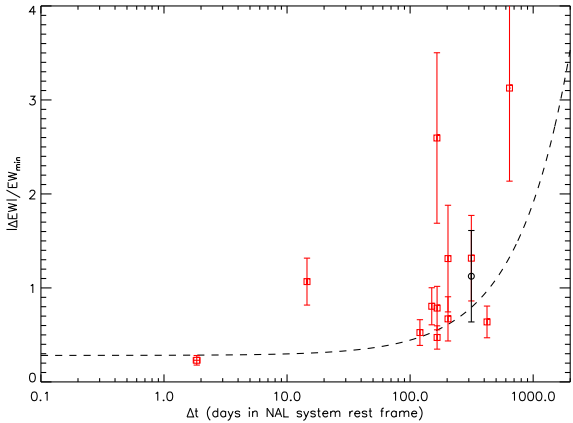}
\caption[$|\Delta$EW$|$/EW$_{\scriptsize{\textrm{min}}}$ versus $\Delta t_{\scriptsize{\textrm{r}}}$ for the population of variable intervening NAL systems (colour-coded by ion)]{The absolute fractional change in EW ($|\Delta$EW$|$/EW$_{\scriptsize{\textrm{min}}}$) versus the rest-frame time between observations ($\Delta t_{\scriptsize{\textrm{r}}}$) for our sample of significantly variable IALs.  A logarithmic scale is used for the x-axis due to the large range of time separations spanning days to years.  The slope of the best-fitting line (dashed black) indicates that IALs have bigger fractional changes over longer time periods. \label{figfewtei}}
\end{center}
\end{figure}

An alternative measure of absorption line variability is the fractional change in EW ($|\Delta$EW$|$/EW$_{\scriptsize{\textrm{min}}}$).  Fig. \ref{figfewmei} shows the fractional change as a function of the minimum EW.  Although EW changes in our sample of IALs have very little dependence on line strength, the fractional change clearly increases in weaker lines.  This conclusion is affirmed by the weighted least-squares fit to the points (dashed black line) that gives a power law index of -0.9 $\pm$ 0.5.  This result is in agreement with the expected increase in variability as line strength decreases, and these same trends are seen in previous variable BAL studies even though they used different definitions of fractional variability \citep{bar94,lun07,gib08,cap11}.  Furthermore, these BAL studies all noted a tendency towards increasing fractional change in EW with increasing time between observations in the rest frame of the absorption system.  Fig. \ref{figfewtei}, with a weighted least-squares linear best-fitting line slope of (16 $\pm$ 3) $\times$ 10$^{-4}$ (dashed black line), shows that this relationship also holds for the intervening NAL system sample.  This result is expected because BALs and NALs are both comprised of clouds of absorbing gas that have similar physical sizes and metal compositions even though they have vastly different origins and absorption line properties.  

\begin{figure}
\begin{center}
\includegraphics[width=84mm]{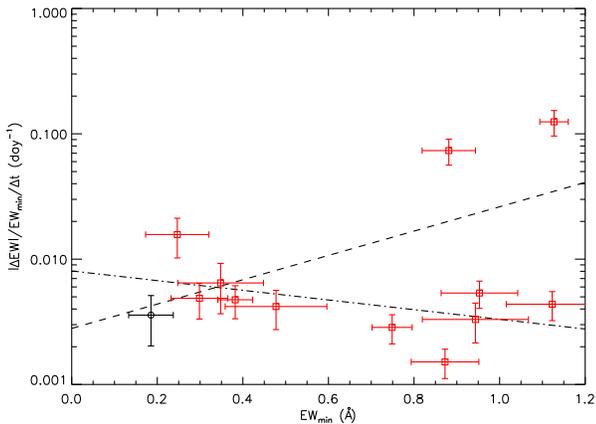}
\caption[$|\Delta$EW$|$/EW$_{\scriptsize{\textrm{min}}}$/$\Delta t_{\scriptsize{\textrm{r}}}$ versus EW$_{\scriptsize{\textrm{min}}}$ for the population of variable intervening NAL systems (colour-coded by ion)]{The absolute rate of fractional change in EW ($|\Delta$EW$|$/EW$_{\scriptsize{\textrm{min}}}$/$\Delta t_{\scriptsize{\textrm{r}}}$) versus the minimum EW (EW$_{\scriptsize{\textrm{min}}}$) for our sample of significantly variable IALs.  A logarithmic scale is used for the y-axis due to the large range of time separations spanning days to years.  Excluding the two lines with the highest rate of change gives an improved fit (dot-dashed black line) to the remaining 11 points and appears to show that weaker variable absorption lines have a faster rate of variability. \label{figrewmei}}
\end{center}
\end{figure}

The last relationship we examine for the variable intervening NAL system sample was the rate of fractional absorption line variability ($|\Delta$EW$|$/EW$_{\scriptsize{\textrm{min}}}$/$\Delta t_{\scriptsize{\textrm{r}}}$) and the minimum line strength in Fig. \ref{figrewmei}.  With the exception of two outliers, all of the lines have $|\Delta$EW$|$/EW$_{\scriptsize{\textrm{min}}}$/$\Delta t_{\scriptsize{\textrm{r}}}$ $<$ 0.02 d$^{-1}$ independent of line strength.  From Fig. \ref{figfewmei}, we might presume that weaker variable absorption lines in our sample should have a faster rate of variability than stronger lines.  The weighted least-squares log-linear fit (dashed line) to all the points is contrary to this prediction with a slope of 0.4 $\pm$ 3.0, but this fit is biased by the two lines with the highest rate of change.  Excluding these outliers gives a better fit to the remaining 11 points with a slope of --0.4 $\pm$ 2.7, implying that weaker absorption lines may have a faster rate of variability as expected.  However, these results are inconclusive so a larger sample size is needed to confirm the relationship between variability, line strength, and time between observations for variable intervening systems.  Finally, the outliers in Fig. \ref{figrewmei} have nearly an order of magnitude faster rate of variability, which could suggest the existence of sub-au scale absorbing gas clouds, small-scale density fluctuations, or limitations of our simple model.

\subsection{Interpretation of variable intervening systems} \label{secmod}

Since evidence of EW variability in intervening NAL systems has not previously been reported in quasar spectra, we construct a simple model to quantify the cloud sizes implied by absorption line variability on short time-scales.  We consider bulk motion of an absorbing gas cloud as the cause of the observed variability in these systems since changes in ionization state are unlikely to occur on short time-scales in foreground galaxies \citep[e.g.,][]{nar04}.  Assuming a spherical cloud of uniform density that moves transverse to our line of sight, the absorption variability time-scale is determined by the time it takes the cloud to cross the quasar beam.  The distance it travels is equal to the sum of the cloud and beam diameters and is related to the crossing time as follows:
\begin{equation}
R_{\textrm{\scriptsize{b}}} + R_{\textrm{\scriptsize{c}}} = \textstyle \frac{1}{2} \displaystyle v\Delta t \simeq \textrm{10--100} \textrm{ au} \left ( \frac{v}{200 \textrm{ km sec}^{-1}} \right ), \label{eqnsize}
\end{equation}
where $R_{\textrm{\scriptsize{b}}}$ is the quasar beam radius, $R_{\textrm{\scriptsize{c}}}$ is the cloud radius, $v$ is the orbital velocity, and the last expression is calculated from the main concentration of points with $\Delta t_{\scriptsize{\textrm{r}}} > 100$ d in Fig. \ref{figfewtei}.  

Although the results in equation \ref{eqnsize} are based on the variability time-scales for the high-$\beta$ intervening systems described in Section \ref{secial}, deviations from the underlying assumptions could have considerable impact on the cloud-size estimates.  For example, a significant velocity differential between the individual components that are blended into a single absorption line would require a more detailed modelling of their motion relative to one another and the line of sight to the background quasar.  Additionally, the size of the quasar beam must be known before the cloud size can be determined.  The beam size depends on the size of the accretion disc emitting region, the disc orientation with respect to the line of sight, and the distance from the quasar to the absorption system.  Estimates of the emitting region size range from $\sim$10--100 au with shorter wavelengths (i.e., X-rays) emitted at smaller radii and UV and optical wavelengths at larger radii \citep{dai10}.  If we assume a flat cosmology with $\Omega_{\textrm{\scriptsize{M}}}$ = 0.29, $\Omega_\Lambda$ = 0.71, and $h$ = 1.0, noticeable quasar beam size changes of no more than a factor of 1.2 occur solely in the high-$\beta$ quasar and absorption system redshift ranges in Table \ref{tabvar} \citep{wri06}.  Finally, the accretion disc orientation can reduce the beam size by up to an order or magnitude.  However, all of these factors are, at most, of the same order as the uncertainty in the size of the quasar emission region, so we can safely assume that they are incorporated into the size ranges given in equation \ref{eqnsize}.  These results suggest cloud sizes ranging from $\sim$1 to 100 au in accordance with the findings from small-scale structures in the Milky Way interstellar medium \citep{lau00,lau03,wel07} and halo \citep{mey99,ric03,nas12}.  Outside our Galaxy, the only known discovery of short time-scale variations attributed to small-scale structures occurred in the 21-cm absorption line profile of a damped Ly$\alpha$ quasar absorption system at high redshift \citep{kan01}.  More recent studies have also used measurements of 21-cm absorption to resolve parsec-scale cloud structures in quasar--galaxy pairs \citep{bor10,gup12,sri13}.

Given that none of the variable NAL systems in Table \ref{tabvar} completely crosses the background quasar beam (i.e., our sample does not include an `on/off' system), we next modify our model to account for partial covering factors between the cloud and quasar beam.  The EW of an absorption line is directly proportional to optical depth which, in turn, depends on the column density of the absorbing gas.  Thus, the ratio of EWs between observations (EW$_{\scriptsize{\textrm{min}}}$/EW$_{\scriptsize{\textrm{max}}}$) can be equated to the ratio of area of the quasar beam covered by the cloud to the total beam area ($A_{\scriptsize{\textrm{min}}}$/$A_{\scriptsize{\textrm{max}}}$):
\begin{equation} 
\frac{\textrm{EW}_{\scriptsize{\textrm{min}}}}{\textrm{EW}_{\scriptsize{\textrm{max}}}} = \frac{A_{\scriptsize{\textrm{min}}}}{A_{\scriptsize{\textrm{max}}}} \simeq \frac{(R-\textstyle \frac{1}{2}\displaystyle d)^2}{R^2} \label{eqncov}
\end{equation}
where $R$ is the beam radius, $d$ is the distance travelled by the cloud, and the area where the beam and cloud overlap is approximated as a circle.  We are also assuming that the maximum line strength (EW$_{\scriptsize{\textrm{max}}}$) corresponds to complete coverage of the background quasar.  Solving equation \ref{eqncov} for $d$ gives the following expression:
\begin{equation}
d \simeq 2R \left (1-\sqrt{\frac{\textrm{EW}_{\scriptsize{\textrm{min}}}}{\textrm{EW}_{\scriptsize{\textrm{max}}}}} \right ) \simeq \textrm{10--50} \textrm{ au} \left ( \frac{2R}{100 \textrm{ au}} \right ) \label{eqnd}
\end{equation}
where the EWs of the variable IALs are used to estimate the distance travelled by the absorbing gas cloud.  Once again, this result is consistent with the lower limit of cloud sizes reported in previous studies, but this estimate is likely more accurate because it is calculated using the measured strengths of the variable lines rather than assuming that the cloud completely crossed the quasar beam between observations.

Now that we have a better estimate of the distance travelled by the absorbing gas clouds in the variable intervening NAL systems, we can calculate their velocities from the rest-frame time differences between observations.  Dividing equation \ref{eqnd} by the range of $\Delta t_{\scriptsize{\textrm{r}}}$ values greater than 100 d gives the following result:
\begin{equation}
v \simeq \frac{2R}{\Delta t} \left (1-\sqrt{\frac{\textrm{EW}_{\scriptsize{\textrm{min}}}}{\textrm{EW}_{\scriptsize{\textrm{max}}}}} \right ) \simeq \textrm{50--500 km sec}^{-1} \left ( \frac{2R}{100 \textrm{ au}} \right ) \label{eqnvel}
\end{equation}
where 2$R$ is the quasar beam size.  This range of velocities agrees with typical peculiar velocities in the Milky Way \citep{car07} and the Canis Major dwarf galaxy \citep{mar04,din05} on the lower end, high-velocity clouds in the middle \citep{wak97}, and orbital velocities in larger galaxies on the upper end \citep{spa07}.  The two outliers imply velocities of $\sim$10$^3$--10$^4$ km sec$^{-1}$, where the lower limit is on the order of speeds measured in starburst-driven winds \citep{hec03} and velocity dispersions in massive galaxy clusters \citep{spa07} while velocities near the upper limit are found only in outflows from AGNs (see Section \ref{secaal}) or GRBs \citep[see, e.g.,][]{cuc09}.  These calculations are for this simple model, and they indicate that the measured variations in these NAL systems are reasonable when compared to known properties of typical host galaxies.

\begin{figure*}
\begin{center}
\includegraphics[width=84mm]{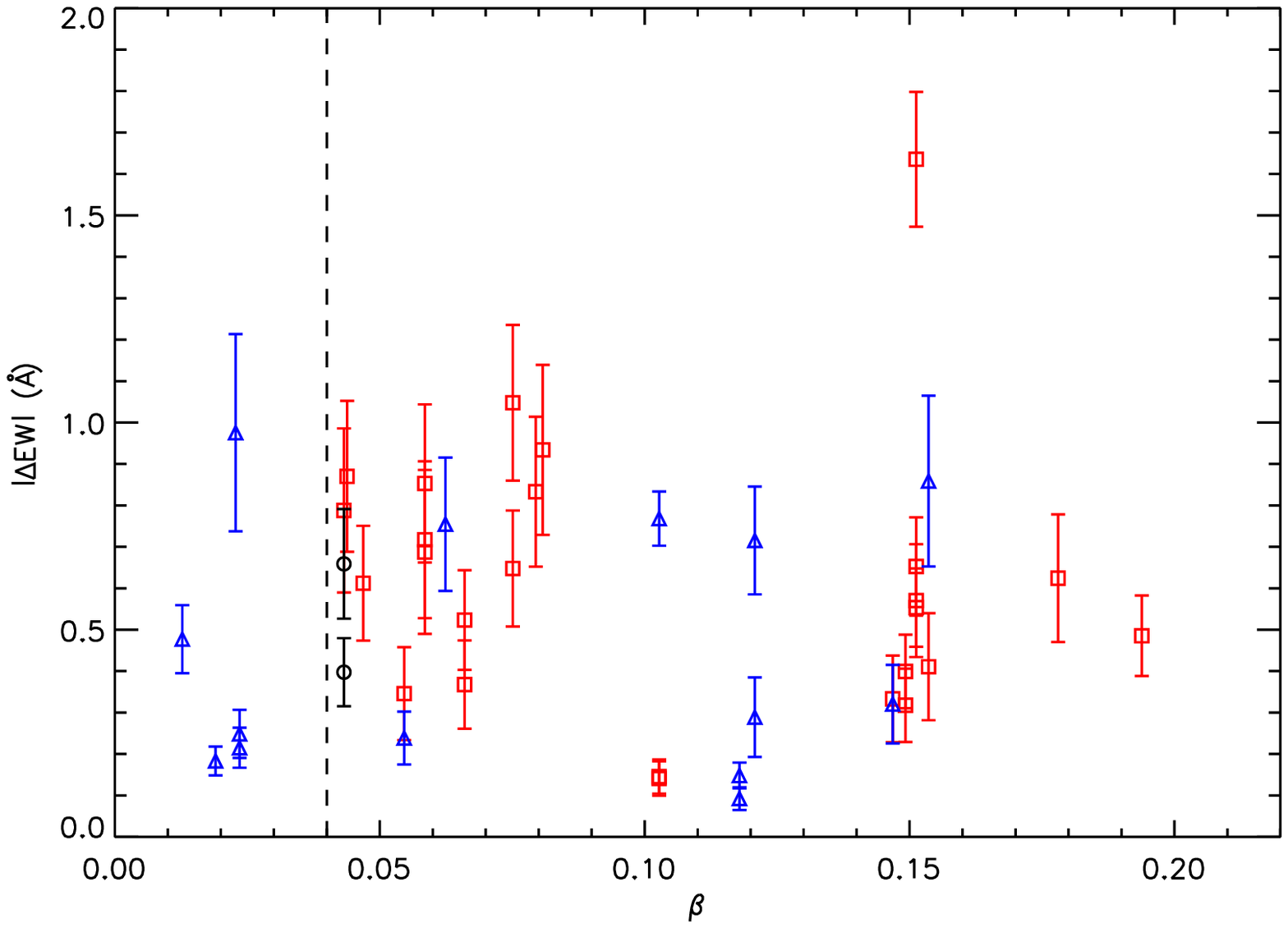}
\includegraphics[width=84mm]{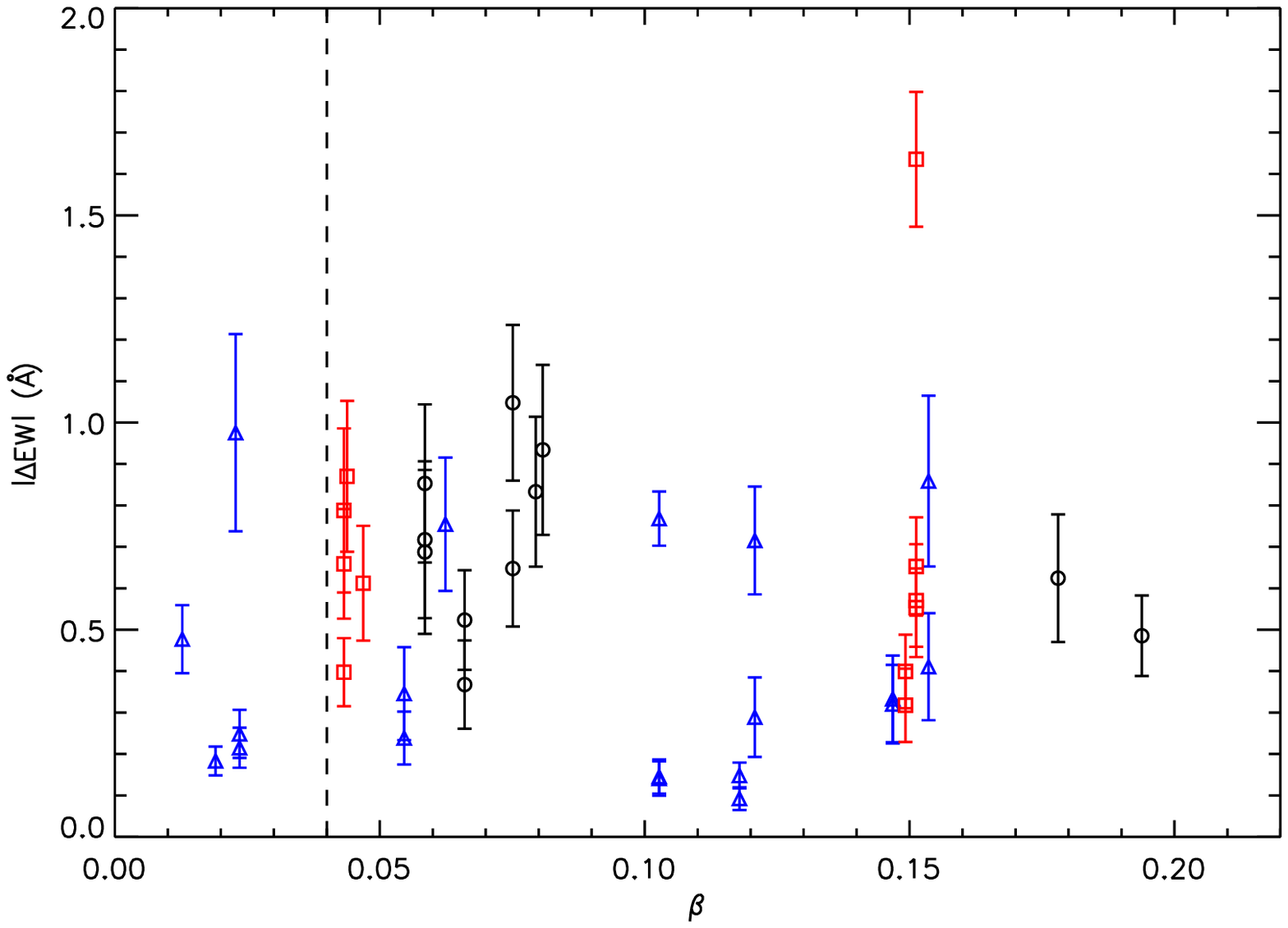}
\caption[$|\Delta$EW$|$ versus $\beta$ for the population of low-$\beta$ variable NAL systems (colour-coded by ion and low-$\beta$ sub-samples)]{The absolute change in EW ($|\Delta$EW$|$) versus $\beta$ for our sample of significantly variable low-$\beta$ absorption lines.  Different absorption lines with the same $\beta$ are from the same NAL system.  The dashed line at $\beta < 0.04$ is the nominal cut-off between associated and intrinsic/intervening NAL systems \citep{wil08}.  However, the existence of variable high-ionization lines in the left-hand panel at a wide range of $\beta$ values suggests that a simple $\beta$ cut is insufficient to distinguish between intrinsic, associated, and intervening NAL systems.  In the right-hand panel and Figs \ref{figdewmea}--\ref{figrewmea}, all of the systems with doubly and triply-ionized lines are grouped in a sub-sample of intrinsic NAL systems (blue triangles).  The other two groups are comprised of systems with variable low-ionization lines and are divided into sub-samples with $\Delta t_{\scriptsize{\textrm{r}}}$ $>$ 100 d (red squares) and $\Delta t_{\scriptsize{\textrm{r}}}$ $<$ 100 d (black circles). \label{figdewbia}}
\end{center}
\end{figure*}

Having gained confidence that the properties of absorbing gas clouds in variable NAL intervening systems are in agreement with the results from independent and unrelated observations, it would also be useful to estimate their neutral hydrogen column densities from the EWs of the variable lines.  Unfortunately, even a rough approximation of the hydrogen column density is not possible with the data used in this study due primarily to the saturation of prominent metal lines in SDSS spectra \citep{nes05}.  Alternative methods to calculate the column density include measuring the Ly$\alpha$ absorption line \citep{rao03} or estimating abundances from the weakest metal lines \citep{meir09}.  However, most of the variable NAL systems are not at high enough redshifts and do not include the Ly$\alpha$ line, many of the SDSS spectra do not have sufficient resolution to measure the Ly$\alpha$ line or signal-to-noise ratio to detect weak lines, and we have no knowledge of what the depletion would be in these systems.  Another technique uses spectra with only one QAL system and the dust to gas ratio in the Small Magellanic Cloud to derive the extinction to the background quasar, but this technique is not applicable to our sample because multiple absorption systems are detected in the spectra of our variable systems \citep{yor06}.  Additional targeted observations with improved resolution and signal-to-noise ratio are needed to determine neutral hydrogen column densities and system classifications, but we can conclude that most of the variable NAL systems have values ranging from $10^{17}$ to $10^{20}$ cm$^{-2}$ based on the statistics of QAL systems and the prominence of metal lines in SDSS spectra \citep{yor06}.  

Overall, this series of simplified models indicates that absorption line variability in intervening NAL systems is a reasonable consequence of the interaction between the background quasar beam and typical foreground galaxies.  The cloud sizes predicted by the measured EWs and the rest-frame time differences between observations are consistent with size estimates reported in other studies of variable absorption.  Moreover, the velocities implied by these sizes and variability time-scales are also consistent with a wide range of astrophysical phenomena.  Clearly, intervening NAL variability on short time-scales is rare; however, as we have shown, the presence of variability by itself may be insufficient to distinguish between intrinsic and intervening systems.  Of the 477 NAL systems classified as intervening based on the standard $\beta > 0.22$ cut in the refined sample of 1,084 systems with repeat observations (see Section \ref{secnref}), we find approximately 2.5\% of the intervening systems display absorption line EW variations.  This fraction is likely a lower limit due to the large number of quality cuts that we imposed on the sample, but given the quality of our data we cannot make a more quantitative assessment of this fraction without a targeted follow-up campaign \citep[cf.][]{daw13}.

\subsection{Low-$\beta$ NAL systems} \label{secaal}

In contrast to absorption line variability in intervening NAL systems, a number of previous studies have detected short time-scale variability in low-$\beta$ systems that were all attributed to high-velocity outflows intrinsic to the quasar environment (see Table \ref{tabprev}).  However, the existence of intervening NAL variability in our sample suggests that time variability alone may not always support the conclusion that a NAL system with variable absorption lines is intrinsic.  In Fig. \ref{figdewbia}, we take a closer look at the 41 variable lines in 23 NAL systems with $\beta < 0.22$ in an attempt to distinguish between intrinsic, associated, and intervening systems.  The dashed line at $\beta < 0.04$, based on the cut-off for associated NAL systems from \citet{wil08}, gives five lines in four systems.  All of these lines are all variable \mbox{C\,{\sc iv}} lines, which suggests that they are likely not associated systems but instead are high-velocity intrinsic systems.  Furthermore, \citet{ric99} reported that a significant fraction of \mbox{C\,{\sc iv}} NALs with $0.02 < \beta < 0.25$ may be intrinsic due to their correlation with the steepness of the quasar spectrum and the radio loudness of the background quasar.  Thus, it appears that simple $\beta$ cuts are insufficient to distinguish between intrinsic, associated, and intervening systems.  

Given that absorption line variability in high-ionization lines (blue triangles in the left-hand panel of Fig. \ref{figdewbia}) is a strong indicator of systems in quasar outflows, we decided to group all systems with at least one doubly or triply ionized variable line into a sub-sample of NAL systems that are likely intrinsic.  These criteria yield 11 systems comprised of 19 lines (blue triangles in the right-hand panel of Fig. \ref{figdewbia}).  Five of these systems have $\beta$ values that exceed the highest velocity intrinsic NAL system previously confirmed through absorption variability at $\beta = 0.08$ \citep[][unconfirmed variable systems push this number slightly higher; see Table \ref{tabprev}]{ham97a}.  Although some of these may be intervening systems, most are probably intrinsic due to their time variability and thus provide supporting evidence of extremely high-velocity quasar outflows ($\beta > 0.1$) that have only been previously detected in variable BAL and mini-BAL systems \citep{fol83,nar04,mis07,cap11,rod11}.

The remaining 22 lines in 12 systems all have variable low-ionization lines (red squares and black circles in the left-hand panel of Fig. \ref{figdewbia}).  To determine the origin of these systems, we recall from Fig. \ref{figfewtei} that most of the intervening NAL systems vary on time-scales longer than 100 d.  Thus we make another cut at $\Delta t_{\scriptsize{\textrm{r}}}$ = 100 d where systems that vary on smaller time-scales are probably intrinsic while those with $\Delta t_{\scriptsize{\textrm{r}}}$ $>$ 100 d have a higher likelihood of being intervening systems.  The outliers in Fig. \ref{figfewtei} suggest that we may be including some intervening systems in the intrinsic sub-sample and vice versa, but the vast majority of the lines with $\Delta t_{\scriptsize{\textrm{r}}}$ $<$ 100 d have $\beta < 0.22$, so they are most likely intrinsic (see Fig. \ref{figdewt}).  This cut breaks the low-ionization sample into two sub-samples: 11 lines in 7 systems (black circles in the right-hand panel of Fig. \ref{figdewbia}) with $\Delta t_{\scriptsize{\textrm{r}}}$ $<$ 100 d and 11 lines in 5 systems (red squares) with $\Delta t_{\scriptsize{\textrm{r}}}$ $>$ 100 d.

\begin{figure}
\begin{center}
\includegraphics[width=84mm]{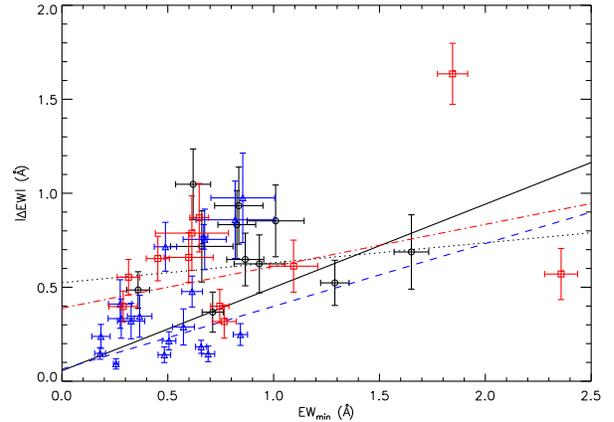}
\caption[$|\Delta$EW$|$ versus EW$_{\scriptsize{\textrm{min}}}$ for the population of low-$\beta$ variable NAL systems (colour-coded by low-$\beta$ sub-samples)]{The absolute change in EW ($|\Delta$EW$|$) versus the minimum EW (EW$_{\scriptsize{\textrm{min}}}$) for our sample of significantly variable low-$\beta$ absorption lines.  The slope of the best-fitting line (solid black) to all points indicates that the amount of absorption line variability increases with line strength.  Fits to the three sub-samples imply that they all are increasingly contaminated by intervening systems. \label{figdewmea}}
\end{center}
\end{figure}

\begin{figure}
\begin{center}
\includegraphics[width=84mm]{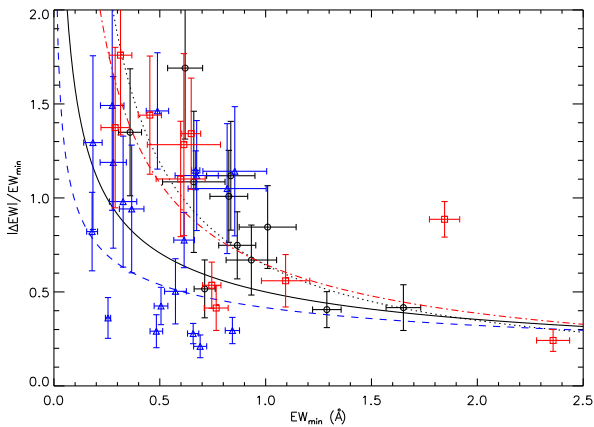}
\caption[$|\Delta$EW$|$/EW$_{\scriptsize{\textrm{min}}}$ versus EW$_{\scriptsize{\textrm{min}}}$ for the population of low-$\beta$ variable NAL systems (colour-coded by low-$\beta$ sub-samples)]{The absolute fractional change in EW ($|\Delta$EW$|$/EW$_{\scriptsize{\textrm{min}}}$) versus the minimum EW (EW$_{\scriptsize{\textrm{min}}}$) for our sample of significantly variable low-$\beta$ absorption lines.  As showed by the over-plotted best-fitting lines for the entire sample (solid black line) and each sub-sample, these points appear to show increasing fractional changes in weaker lines although the pattern is less pronounced compared to the high-$\beta$ systems (see Fig. \ref{figfewmei}). \label{figfewmea}}
\end{center}
\end{figure}

Having defined low-$\beta$ sub-samples based on the properties of previously reported intrinsic and intervening NAL systems, we analyse their properties using the same approach as the intervening systems in Section \ref{secial}.  First, we compare the amount of variability ($|\Delta$EW$|$) with the minimum line strength (EW$_{\scriptsize{\textrm{min}}}$) in Fig. \ref{figdewmea}.  Also plotted in this Fig. are the weighted linear least-squares best-fitting lines for the entire sample of low-$\beta$ NALs as well as each sub-sample.  Overall, the slope of 0.44 $\pm$ 0.04 indicates that the amount of absorption line variability increases with line strength (solid black line in Fig. \ref{figdewmea}).  Since most of the lines in the main low-$\beta$ sample are intrinsic, we use a comparison between this fit and the fit for the intervening sample in Fig. \ref{figdewmei} to evaluate the system classifications for the three sub-samples.  The fit to the high-ionization intrinsic sub-sample points (0.33 $\pm$ 0.06: dashed blue line) implies that most of the systems in this sub-sample are intrinsic, but some are probably intervening.  The low-ionization intervening sub-sample (0.22 $\pm$ 0.06: dot-dashed red line) appears to have a higher contamination of intrinsic systems.  Finally, the low-ionization intrinsic sub-sample (0.11 $\pm$ 0.13: dotted black line) has no dependence between $|\Delta$EW$|$ and EW$_{\scriptsize{\textrm{min}}}$, which matches the result for the high-$\beta$ intervening system sub-sample.  Although this result could signify a bias in this sub-sample, it may also indicate that there are intervening systems with short time-scale variability ($\Delta t_{\scriptsize{\textrm{r}}}$ $<$ 100 d) missing in the variable NAL systems at high-$\beta$.

\begin{figure}
\begin{center}
\includegraphics[width=84mm]{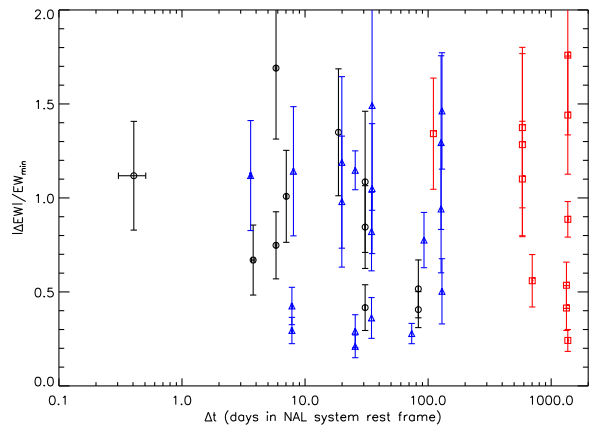}
\caption[$|\Delta$EW$|$/EW$_{\scriptsize{\textrm{min}}}$ versus $\Delta t_{\scriptsize{\textrm{r}}}$ for the population of low-$\beta$ variable NAL systems (colour-coded by low-$\beta$ sub-samples)]{The absolute fractional change in EW ($|\Delta$EW$|$/EW$_{\scriptsize{\textrm{min}}}$) versus the rest-frame time between observations ($\Delta t_{\scriptsize{\textrm{r}}}$) for our sample of significantly variable low-$\beta$ absorption lines.  A logarithmic scale is used for the \textit{x}-axis due to the large range of time separations spanning days to years.  Although there is one outlier, the general pattern of the low-$\beta$ absorption lines shows almost no dependence between fractional change and time between observations. \label{figfewtea}}
\end{center}
\end{figure}

Figs \ref{figfewmea} and \ref{figfewtea} plot the fractional change in EW ($|\Delta$EW$|$/EW$_{\scriptsize{\textrm{min}}}$) as a function of the minimum line strength (EW$_{\scriptsize{\textrm{min}}}$) and the rest-frame time between observations ($\Delta t_{\scriptsize{\textrm{r}}}$).  In the population of high-$\beta$ intervening systems described in Section \ref{secial}, we note a strong tendency towards fractional EW changes in weaker lines (see Fig. \ref{figfewmei}).  Although this pattern in repeated in Fig. \ref{figfewmea}, it is less pronounced for all three low-$\beta$ sub-samples due to the increased scatter in the points.  This result may suggest that absorption line variability is independent of line strength for the weakest lines below a certain EW threshold (e.g., EW$_{\scriptsize{\textrm{min}}}$ $<$ $\sim$1 $\textrm{\AA}$), but more detailed theoretical modelling is needed to determine this relationship.  The weighted least-squares fit to all the points (solid black line) gives a power law index of --0.5 $\pm$ 0.4, which is in agreement with the index for the high-$\beta$ variable IALs in Fig. \ref{figfewmei}.  Power-law fits to the points of the three sub-samples also agree with the high-$\beta$ lines, but the high-ionization intrinsic sub-sample (--0.4 $\pm$ 0.8: dashed blue line) has the lowest index while the low-ionization sub-samples ($\Delta t_{\scriptsize{\textrm{r}}}$ $<$ 100 d, --0.9 $\pm$ 0.8: dotted black line; $\Delta t_{\scriptsize{\textrm{r}}}$ $>$ 100 d, --0.7 $\pm$ 0.7: dot-dashed red line) indicate slightly stronger dependencies between $|\Delta$EW$|$/EW$_{\scriptsize{\textrm{min}}}$ and EW$_{\scriptsize{\textrm{min}}}$.  However, due to the small sample sizes, it is unclear whether the different power-law indices can be used to distinguish between intrinsic and intervening systems in these sub-samples.  

Differences between the high-$\beta$ intervening sample and the low-$\beta$ sub-samples are again evident in Fig. \ref{figfewtea}.  The general pattern of the low-$\beta$ absorption lines in this Fig. shows almost no dependence between fractional change and time between observations while the high-$\beta$ lines displayed a trend towards increasing $|\Delta$EW$|$/EW$_{\scriptsize{\textrm{min}}}$ with time (see Fig. \ref{figfewtei}).  This discrepancy could be the result of mixing of intrinsic, associated, and intervening systems in the low-$\beta$ sample, and it appears that a $\Delta t_{\scriptsize{\textrm{r}}}$ = 100 d cut for the low-ionization systems does not cleanly distinguish between these different system classifications.  However, given the apparent relationship between variability and time for the intervening systems in Fig. \ref{figdewt}, perhaps future detections of additional variable systems or improved theoretical modelling will uncover the true time dependencies for variable absorption in intrinsic, associated, and intervening NAL systems.

\begin{figure}
\begin{center}
\includegraphics[width=84mm]{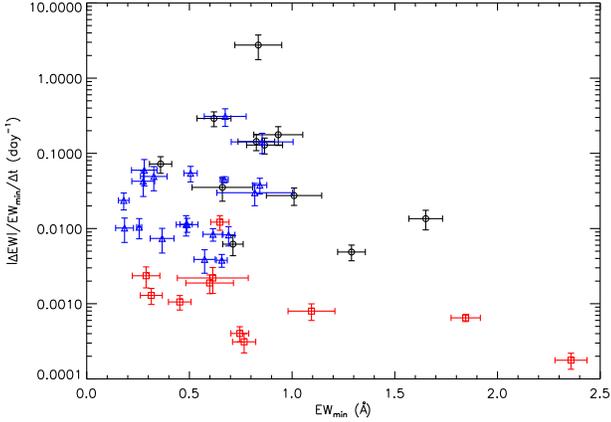}
\caption[$|\Delta$EW$|$/EW$_{\scriptsize{\textrm{min}}}$/$\Delta t_{\scriptsize{\textrm{r}}}$ versus EW$_{\scriptsize{\textrm{min}}}$ for the population of low-$\beta$ variable NAL systems (colour-coded by low-$\beta$ sub-samples)]{The absolute rate of fractional change in EW ($|\Delta$EW$|$/EW$_{\scriptsize{\textrm{min}}}$/$\Delta t_{\scriptsize{\textrm{r}}}$) versus the minimum EW (EW$_{\scriptsize{\textrm{min}}}$) for our sample of significantly variable low-$\beta$ absorption lines.  A logarithmic scale is used for the \textit{y}-axis due to the large range of time separations spanning days to years.  Most of the lines have a small rate of change with no obvious dependence on EW$_{\scriptsize{\textrm{min}}}$. \label{figrewmea}}
\end{center}
\end{figure}

Lastly, Fig. \ref{figrewmea} shows the relationship between the EW rate of change ($|\Delta$EW$|$/EW$_{\scriptsize{\textrm{min}}}$/$\Delta t_{\scriptsize{\textrm{r}}}$) and the minimum line strength (EW$_{\scriptsize{\textrm{min}}}$).  Most of the lines have a small rate of change with no obvious dependence on EW$_{\scriptsize{\textrm{min}}}$.  This result provides further evidence that the low-$\beta$ system sub-samples contain an uncertain combination of intrinsic, associated, and intervening NAL systems because the variable high-$\beta$ systems displayed a faster rate of variability for stronger lines in Fig. \ref{figrewmei}.  Since the low-ionization sub-samples do contain some intervening NAL systems, we again estimate the cloud size using the average $\Delta t_{\scriptsize{\textrm{r}}}$ and assuming bulk transverse motion of the absorbing gas clouds to the line of sight.  The result of $\sim$50 au is consistent with equations \ref{eqnsize} and \ref{eqncov} estimated from the population of high-$\beta$ intervening systems.  A similar estimate for the high-ionization intrinsic sub-sample would be incorrect due to the increased likelihood that the observed NAL variability is caused by both ionization state changes and bulk motion \citep{ham95,ham97c,ham11,nar04,wis04}.

\section{CONCLUSIONS} \label{seccon}

In this paper, we present our results from the largest ever study of quasar NAL systems in the time domain.  The global properties of the 33 NAL systems with measured variations in their absorption line EWs selected from the refined sample of 1,084 systems (see Section \ref{secnref}) with repeat SDSS observations contain no obvious systematics, suggesting that this sample is unbiased.  The 10 variable intervening NAL systems we discovered imply that approximately 200 systems must be observed to find a single variable intervening system.  Since all the previous studies of NAL variability combined amount to less than 50 systems (see Table \ref{tabprev}), it is not surprising that this effect has not been observed in previous studies.  Moreover, the variability fraction nearly doubles from 2.5\% (12/477) for the high-$\beta$ systems to 4.9\% (30/607) for the low-$\beta$ systems.  Therefore, our results show that, in most cases, time variability of NALs indicates that the system is intrinsic to the background quasar.  However, we found evidence of NAL systems where simple variability, $\beta$, and time separation cuts do not cleanly distinguish between intrinsic, associated, and intervening systems.  

We also constructed a basic model of an intervening absorbing gas cloud with variability caused by bulk motion of the cloud transverse to the background quasar beam.  The resulting sizes and velocities are in agreement with typical values found in normal foreground galaxies, which confirms that intervening NAL variability should be seen if a large enough sample of NAL systems is observed.  Additionally, the samples of low-$\beta$ systems include indications of absorption variability in NAL systems with velocities higher than previously reported, which could provide unique insight into the dynamics of quasar outflows.  Two of these systems (12 and 16 in Table \ref{tabvar}) have radio-loud background quasars which has been shown to correlate with the rate of incidence of absorption systems attributed to quasar outflows \citep{ber11,cha12}.  Ultimately, more observations of quasar NAL systems are needed, preferably with higher resolution and signal-to-noise ratio, to definitively confirm the existence of absorption line variability in these extreme cases.  One promising opportunity is our ancillary project with the Baryon Oscillation Spectroscopic Survey (BOSS) to obtain repeat observations of quasars detected in SDSS I/II that span a wide range of quasar and absorption system properties \citep{daw13}.  These observations will likely place additional constraints on the sizes of the absorbing gas clouds and the environmental conditions that lead to short-term absorption variability.

\section*{ACKNOWLEDGEMENTS}

The authors would like to thank Brian Fields for his genuine interest in our work and his many helpful suggestions for interpreting the results.  We also thank the anonymous referee for helpful comments on the manuscript.  Funding for the SDSS and SDSS-II has been provided by the Alfred P. Sloan Foundation, the Participating Institutions, the National Science Foundation, the US Department of Energy, the National Aeronautics and Space Administration, the Japanese Monbukagakusho, the Max Planck Society, and the Higher Education Funding Council for England. The SDSS Web Site is http://www.sdss.org/.

The SDSS is managed by the Astrophysical Research Consortium for the Participating Institutions. The Participating Institutions are the American Museum of Natural History, Astrophysical Institute Potsdam, University of Basel, University of Cambridge, Case Western Reserve University, University of Chicago, Drexel University, Fermilab, the Institute for Advanced Study, the Japan Participation Group, Johns Hopkins University, the Joint Institute for Nuclear Astrophysics, the Kavli Institute for Particle Astrophysics and Cosmology, the Korean Scientist Group, the Chinese Academy of Sciences (LAMOST), Los Alamos National Laboratory, the Max-Planck-Institute for Astronomy (MPIA), the Max-Planck-Institute for Astrophysics (MPA), New Mexico State University, Ohio State University, University of Pittsburgh, University of Portsmouth, Princeton University, the United States Naval Observatory, and the University of Washington.

\appendix

\section{ADDITIONAL DATA REDUCTION}

\subsection{NAL system identification} \label{seczid}

From the SDSS DR7 release, we identified all quasars (spectra classes 3 and 4 in the \texttt{SpecObjAll} table on the SDSS CAS) using the following SQL query:
\begin{description}
\tt
\item \noindent SELECT first.specObjID, first.mjd, first.plate,
\item first.fiberID, first.z, first.specClass, 
\item first.ra, first.dec, other.specObjID, 
\item other.mjd, other.plate, other.fiberID, 
\item other.z, other.specClass, other.ra, other.dec, 
\item COUNT(DISTINCT other.mjd) + COUNT(DISTINCT 
\item first.mjd) AS nightsObserved into 
\item mydb.qsorepeats 
\item \noindent FROM SpecObjAll first
\item JOIN SpecObjAll other ON 
\item first.bestObjID=other.bestObjID
\item JOIN PlateX AS firstPlate ON 
\item firstPlate.plate=first.plate
\item JOIN PlateX AS otherPlate ON 
\item otherPlate.plate=other.plate
\item \noindent WHERE first.scienceprimary=1 AND 
\item other.scienceprimary=0 AND other.bestObjID$>$0 
\item AND first.specClass=dbo.fSpecClass(`QSO') or 
\item first.specClass=dbo.fSpecClass(`HIZ$\_$QSO')
\item \noindent GROUP BY first.plate, other.plate, first.fiberID, 
\item other.fiberID, first.mjd, other.mjd 
\item \noindent ORDER BY nightsObserved DESC, first.plate, 
\item other.plate
\end{description}
After downloading all of the FITS files for these spectra from the \texttt{spectro1d} pipeline, they were each examined by the Y13 QAL detection pipeline to produce a catalogue of all metal absorption line systems identified in these spectra.  With the aim of sampling a wide range of galactic environments, the pipeline searches for over 50 different ions listed in Table \ref{tabmet} that span a wide range of elements, ionization states, and line strengths.  The details of the automated algorithm to detect QAL systems in SDSS quasar spectra are the subject of \citet{yor13}, but the following paragraph provides a brief summary of this process.

\begin{table*}
\begin{center}
\caption[Quasar absorption line data]{Narrow QALs analysed in this study \citep[reproduced from][]{mor03}.  \label{tabmet}}
\begin{tabular}{@{}lccccclccc@{}}
\cline{1-4}
\cline{7-10}
Ion & Rest wavelength & $\log f\lambda_{0}$ + 12.00 & Spectra label & & & Ion & Rest wavelength & $\log f\lambda_{0}$ + 12.00 & Spectra label \\ & $\lambda_{0}$ $(\textrm{\AA})$ & + $\log (N/N_{H})$ & & & & & $\lambda_{0}$ $(\textrm{\AA})$ & + $\log (N/N_{H})$ & \\
\cline{1-4}
\cline{7-10}
\mbox{Al\,{\sc ii}}&  1670.79&  9.95&  Al 1671& & & \mbox{Mg\,{\sc i}}&  2026.48&  9.94&  Mg 2026 \\
\mbox{Al\,{\sc iii}}&  1854.72&  9.51&  Al 1855& & & &  2852.96& 11.30&  Mg 2853 \\
&  1862.79&  9.20&   Al 1863& & & \mbox{Mg\,{\sc ii}}&  2796.35& 10.82&  Mg 2796 \\
\mbox{C\,{\sc i}}&  1277.24& 10.56&  C 1277& & & &  2803.53& 10.51&  Mg 2804 \\
&  1280.14& 10.05&  C 1280& & & \mbox{Mn\,{\sc ii}}&  2576.88&  8.50&  Mn 2577 \\
&  1328.83& 10.52&  C 1329& & & &  2594.50&  8.39&  Mn 2594 \\
&  1560.31& 10.60&  C 1560& & & &  2606.46&  8.24&  Mn 2606 \\
&  1656.92& 10.91&  C 1657& & & \mbox{N\,{\sc v}}&  1238.82& 10.24&  N 1239 \\
\mbox{C\,{\sc ii}}&  1334.53& 10.75&  C 1335& & & &  1242.80&  9.94&  N 1243 \\
\mbox{C\,{\sc iv}}&  1548.20& 10.99&  C 1548& & & \mbox{Ni\,{\sc ii}}&  1317.22&  8.26&  Ni 1317 \\
&  1550.78& 10.69&  C 1551& & & &  1370.13&  8.27&  Ni 1370 \\
\mbox{Ca\,{\sc ii}}&  3934.78&  9.75&  Ca 3935& & & &  1709.60&  7.99&  Ni 1710 \\
&  3969.59&  9.45&  Ca 3970& & & &  1741.55&  8.12&  Ni 1742 \\
\mbox{Cr\,{\sc i}}&  3579.71&  8.81&  Cr 3580& & & \mbox{O\,{\sc i}}&  1302.17& 10.53&  O 1302 \\
\mbox{Cr\,{\sc ii}}&  2056.26&  8.02&  Cr 2056& & & \mbox{S\,{\sc ii}}&  1250.57&  8.03&  S 1251 \\
&  2062.24&  7.88&  Cr 2062& & & &  1253.81&  8.34&  S 1254 \\
&  2066.16&  7.71&  Cr 2066& & & &  1259.52&  8.52&  S 1260 \\
\mbox{Fe\,{\sc i}}&  2484.02& 10.63&  Fe 2484& & & \mbox{Si\,{\sc i}}&  2515.07& 10.28&  Si 2515 \\
&  2523.61& 10.21&  Fe 2524& & & \mbox{Si\,{\sc ii}}&  1260.42& 10.73&  Si 1260 \\
&  3861.01&  9.42&  Fe 3861& & & &  1304.37&  9.61&  Si 1304 \\
\mbox{Fe\,{\sc ii}}&  1608.45&  9.47&  Fe 1608& & & &  1526.71&  9.87&  Si 1527 \\
&  2249.88&  8.11&  Fe 2250& & & &  1808.01&  8.14&  Si 1808 \\
&  2260.78&  8.24&  Fe 2261& & & \mbox{Si\,{\sc iv}}&  1393.76& 10.41&  Si 1394 \\
&  2344.21&  9.93&  Fe 2344& & & &  1402.77& 10.11&  Si 1403 \\
&  2374.46&  9.37&  Fe 2374& & & \mbox{Ti\,{\sc ii}}&  3242.92&  7.82&  Ti 3243 \\
&  2382.77& 10.38&  Fe 2383& & & &  3384.73&  8.02&  Ti 3385 \\
&  2586.65&  9.75&  Fe 2587& & & \mbox{Zn\,{\sc i}}&  2139.24&  8.17&  Zn 2139 \\
&  2600.17& 10.29&  Fe 2600& & & \mbox{Zn\,{\sc ii}}&  2026.13&  7.68&  Zn 2026 \\
& & & & & & &  2062.66&  7.38&  Zn 2063 \\
\cline{1-4}
\cline{7-10}
\end{tabular}
\end{center}
\end{table*}

The Y13 QAL pipeline is comprised of an absorption line-finding algorithm followed by a code that sorts detected lines into systems at the same redshift \citep{lun09}.  The first code performs a fit to the continuum flux that is used to normalize the spectrum.  Next, the line EWs are calculated using the method outlined in \citet{sch93} that was developed for the \textit{Hubble Space Telescope} Quasar Absorption Line Key Project \citep{bah93}.  This routine first generates an EW spectrum by convolving the pixels across the entire quasar spectrum with the point spread function of the telescope.  The identified absorption lines in the new spectrum are fitted with one or more Gaussian profiles to calculate the final EWs.  The absorption features with EW $\geq 3\sigma_{\textrm{\tiny{EW}}}$ are recorded and passed into a second code that matches different ions at the same redshift.  This code starts by trying to find the easily observable ion doublets of \mbox{C\,{\sc iv}} and \mbox{Mg\,{\sc ii}} and applies various selection criteria, including proper wavelength separations and reasonable doublet ratios, to ensure that the detected lines are real.  A redshift is determined for these positively identified doublets that is used to search the remaining list of detected lines for other ions in the same absorption system.  The last step is to assign a confidence grade based on the number of lines with EW $\geq 4\sigma_{\textrm{\tiny{EW}}}$.  Grade `A' systems have at least four lines with confident detections, grade `B' systems have three, and grade `C' systems have two from either the \mbox{Mg\,{\sc ii}} or \mbox{C\,{\sc iv}} doublets.  The accuracy of grade A systems is greater than 99\% while grade B systems are slightly less accurate but still considered to be reliable detections, and lower grades are successively less reliable \citep[see][and the electronic version of Table \ref{tabsys} for definitions of all the confidence grade categories]{yor13}.  

\subsection{Spectroscopic mask bits} \label{secmsk}

Nearly half of the repeat quasar absorption spectra contain pixels with unusually bright background sky flux and/or cosmic rays that adversely affect our absorption line measurements by producing artificial features in the line and/or nearby continuum profiles.  The spectra FITS files contain error mask arrays created during the \texttt{spectro2d} pipeline processing as described in table 11 of \citet{sto02}.  There are 13 different mask conditions that can be flagged for each pixel, and two mask values are given for every pixel: conditions present in all 15-min exposures are flagged in the \texttt{AND}-mask while those present in any of the individual exposures are flagged in the \texttt{OR}-mask \citep{sto02}.  

There is no clear recipe for mapping combinations of mask bits that represent highly contaminated pixels where the reported object flux and error should not be trusted.  Pixels flagged \texttt{FULLREJECT}, \texttt{NOSKY}, and \texttt{NODATA} already have their flux errors set identically to zero to indicate that these pixels are most likely corrupted.  For other error conditions, there is a \texttt{SKYMASK}\footnote{http://spectro.princeton.edu/} \textrm{\scriptsize{IDL}} routine in the SDSS \texttt{spec2d} spectroscopic pipeline utilities that masks regions where sky-subtraction errors are expected to dominate.  This routine masks pixels flagged \texttt{BRIGHTSKY} in the \texttt{AND}-mask or \texttt{BADSKYCHI} or \texttt{REDMONSTER} in the \texttt{OR}-mask.  The first flag indicates a region of the spectrum where the sky emission dominates the target flux while the latter two flags signify pixels where the sky subtraction is likely incorrect due to a poor fit to the sky flux.  The \texttt{SKYMASK} routine can also mask a user-defined number of neighbouring pixels that do not meet flag criteria but are likely contaminated.

For the quasar spectra with pixels flagged in the mask arrays, we compared the sky flux of pixels flagged \texttt{BRIGHTSKY} in the \texttt{AND}-mask to neighbouring unflagged pixels to quantify how they were affected by the saturated pixels.  In the entire sample, the average ratio of the pixel sky flux to median sky of the spectrum drops by $\sim$65\% in the nearest pixel and $\sim$82\% in the next pixel.  Similarly, the average fractional change in pixel flux errors between the flagged and unflagged pixels decreases by $\sim$44\% for the nearest pixel and $\sim$65\% for the next pixel.  Based on these results, we decided to mask a total of three pixels consisting of the flagged and two closest neighbouring pixels to balance the trade-off between keeping as many pixels as possible while accounting for bright sky contamination.

\section{VARIABLE NAL SYSTEMS} \label{secvar}

The comparison spectra for the 33 NAL systems with significant absorption line EW variability are presented in Figs B1--B33.  For each system, the top panel shows the normalized pixel flux values (first observations are red and second are blue), and the bottom panel plots the difference spectrum of the two observation epochs.  To facilitate visual comparison, 1$\sigma$ error bars are included for each pixel flux value, and shaded backgrounds identify masked pixels that are not included in our search for absorption line variability due to contamination from bright pixels (see Section \ref{secmsk}).  Line identifications for significantly variable absorption lines not flagged for proximity to masked pixels or poor continuum fits are italicized, while labels in bold font highlight lines detected in both observation epochs by the Y13 QAL pipeline (see Table \ref{tabmet} for abbreviated ion labels).  Undetected absorption lines in the vicinity of detected lines are labelled in regular font, and unlabelled absorption lines belong to other QAL systems in the quasar spectrum. 

We used the product of the absorption line rest wavelength ($\lambda_{0}$), oscillator strength ($f$), and metal cosmic abundance ($N/N_{H}$) as shown in Table \ref{tabmet} to estimate the optical depth of the various ions \citep{mor88}:
\begin{equation}
\tau_{\lambda} \propto (N/N_{H})f\lambda.
\end{equation}
The range of optical depth values determines the relative strengths of the different lines where strong lines have high optical depths and vice versa.  Variations in weaker lines typically occur in the line core while stronger lines, with a higher tendency towards saturation in the line core, show variability in the line wings.  We also compared lines with similar strengths to see if they displayed comparable and consistent variations.  Following is a discussion of the observed variations in each of the 33 variable NAL systems, sorted from high to low-$\beta$.  

\paragraph*{\it J082033.97+432751.8 (System 1, $z_{\scriptsize{\textrm{abs}}} = 0.6540$, $\beta = 0.6190$):} This system has the highest $\beta$ value of any of the variable NAL systems we detected.  The \mbox{Fe\,{\sc ii}} $\lambda$2587 ($\Delta$EW = --0.77 $\pm$ 0.17 $\textrm{\AA}$) line becomes about 40\% narrower in the line core as expected for a weak line.  Other \mbox{Fe\,{\sc ii}} lines with similar strengths could not be checked for comparable variations due to the poor continuum fits and significant noise fluctuations near the $\lambda$2344 and $\lambda$2374 lines.  The \mbox{Mg\,{\sc ii}} doublet appears to vary in the wings between the two lines consistent with expected variations in strong lines.  Although the \mbox{Fe\,{\sc ii}} $\lambda$2484 line also appears to vary, it is likely not part of this system because it has an alternate identification as a \mbox{Si\,{\sc iv}} $\lambda$1394 line in a $z_{\scriptsize{\textrm{abs}}}$ = 1.9478 grade C system, it is located near to two unidentified lines with similar variations, and it becomes wider and shallower which is inconsistent with the changes in the $\lambda$2587 line.

\paragraph*{\it J140323.39--000606.9 (System 2, $z_{\scriptsize{\textrm{abs}}} = 0.8163$, $\beta = 0.5701$):} Both the \mbox{Fe\,{\sc ii}} $\lambda$2383 ($\Delta$EW = --0.30 $\pm$ 0.08 $\textrm{\AA}$) and \mbox{Mg\,{\sc ii}} $\lambda$2804 ($\Delta$EW = --0.35 $\pm$ 0.09 $\textrm{\AA}$) lines have similar line strengths and showed comparable and consistent variations as the lines narrowed between observations.  Although the \mbox{Mg\,{\sc ii}} $\lambda$2796 line does not vary, the \mbox{Mg\,{\sc ii}} doublet ratio is approximately one indicating that this line is highly saturated.  The \mbox{Fe\,{\sc i}} $\lambda$2524 line is redshifted from the system average and shows slight flux variations along with the \mbox{Fe\,{\sc ii}} $\lambda$2587 line.  Although the \mbox{Fe\,{\sc ii}} $\lambda$2600 line is visible in both spectra and also appears to vary, it did not meet the Y13 QAL pipeline detection criteria in the first observation (see Section \ref{seczid}).

\paragraph*{\it J151652.69--005834.8 (System 3, $z_{\scriptsize{\textrm{abs}}} = 0.6953$, $\beta = 0.4426$):} The widening between observations of the \mbox{Mg\,{\sc ii}} $\lambda$2804 ($\Delta$EW = 0.56 $\pm$ 0.14 $\textrm{\AA}$) can also been seen in the $\lambda$2796 line even though a comparable variation is not reflected in the EW measurements.  Another strong line in this system, \mbox{Fe\,{\sc ii}} $\lambda$2383, also appears to widen in a manner consistent with the observed \mbox{Mg\,{\sc ii}} variations.  Finally, the \mbox{Fe\,{\sc ii}} $\lambda$2344 and $\lambda$2587 lines, which are weaker by a factor of $\sim$3, become deeper in the line core as expected for weaker lines.

\paragraph*{\it J215421.13--074430.0 (System 4, $z_{\scriptsize{\textrm{abs}}} = 0.6521$, $\beta = 0.4359$):} This system has the lowest redshift of all the variable NAL systems in Table 3.  The narrowing of the \mbox{Mg\,{\sc ii}} $\lambda$2796 ($\Delta$EW = --0.94 $\pm$ 0.21 $\textrm{\AA}$) line between observations is also visible in the \mbox{Fe\,{\sc ii}} $\lambda$2383 and $\lambda$2600 lines with similar line strengths.  Perhaps this is a result of changes in the amount of component blending that seems to be occurring in the \mbox{Mg\,{\sc ii}} $\lambda$2804 line.  In addition, flux variations are present in a few of the \mbox{Mg\,{\sc i}} $\lambda$2853 line pixels.  Finally, the \mbox{Fe\,{\sc ii}} $\lambda$2374 line is readily apparent in the first observation at $\lambda_{obs}$ = 3924 $\textrm{\AA}$ but has disappeared in the second observation.

\paragraph*{\it J025743.72+011144.5 (System 5, $z_{\scriptsize{\textrm{abs}}} = 0.7670$, $\beta = 0.4037$):} Both of the \mbox{Mg\,{\sc ii}} $\lambda \lambda$2796, 2804 ($\Delta$EW = 0.63 $\pm$ 0.21 $\textrm{\AA}$) lines show comparable and consistent widening between observations.  However, the \mbox{Mg\,{\sc i}} $\lambda$2853 ($\Delta$EW = --0.46 $\pm$ 0.15 $\textrm{\AA}$) line varies in an opposite manner, possibly signifying variations in both the bulk motion and the gas ionization state of this system.  There are slight flux differences in the \mbox{Fe\,{\sc ii}} $\lambda$2383 line but the remaining \mbox{Fe\,{\sc ii}} lines are unchanged.

\paragraph*{\it J012403.77+004432.6 (System 6, $z_{\scriptsize{\textrm{abs}}} = 2.2614$, $\beta = 0.3722$):} This system has the highest redshift of any of the high-$\beta$ systems ($\beta > 0.22$).  There are 20 different absorption lines detected in both observations of this system but many of the shorter wavelength lines have poor continuum fits.  The only line that shows any variability is the \mbox{Mg\,{\sc ii}} $\lambda$2796 ($\Delta$EW = 0.59 $\pm$ 0.14 $\textrm{\AA}$) line as it significantly widens between observations.  Other strong lines such as \mbox{Fe\,{\sc ii}} $\lambda$2383 and $\lambda$2600 could not be checked for similar variations because of their proximity to masked pixels.

\paragraph*{\it J092655.98+254830.5 (System 7, $z_{\scriptsize{\textrm{abs}}} = 1.5153$, $\beta = 0.3383$):} The \mbox{Fe\,{\sc ii}} $\lambda$2374 ($\Delta$EW = --0.63 $\pm$ 0.15 $\textrm{\AA}$) line displays variations in the line core as it becomes both narrower and shallower between observations.  This line has an \mbox{Fe\,{\sc ii}} $\lambda$2383 alternate identification in a grade C system at $z_{\scriptsize{\textrm{abs}}}$ = 1.4736.  The \mbox{Fe\,{\sc ii}} $\lambda$2587 line shows a similar trend but changes by a smaller amount.  Although there is one nearby bright pixel, significant variations are also seen in the \mbox{Ni\,{\sc ii}} $\lambda$1742 ($\Delta$EW = --0.21 $\pm$ 0.07 $\textrm{\AA}$) line which is weaker by an order of magnitude and becomes narrower with a shift away from the system redshift.  Pixel flux variations are visible in the \mbox{C\,{\sc iv}} $\lambda$1548 while the \mbox{Al\,{\sc iii}} doublet appears to vary in both lines even though the $\lambda$1863 line was not detected in the second observation.

\paragraph*{\it J075105.17+272116.8 (System 8, $z_{\scriptsize{\textrm{abs}}} = 1.2012$, $\beta = 0.3340$):} The \mbox{Fe\,{\sc ii}} $\lambda$2587 ($\Delta$EW = --0.93 $\pm$ 0.21 $\textrm{\AA}$) line narrows with a noticeable redshift difference between observations.  Flux and redshift variations are also visible in the \mbox{Fe\,{\sc ii}} $\lambda$2383 line although the line troughs have an opposite offset compared to the $\lambda$2587 line.

\paragraph*{\it J024603.68--003211.7 (System 9, $z_{\scriptsize{\textrm{abs}}} = 0.8524$, $\beta = 0.3274$):} The \mbox{Fe\,{\sc ii}} $\lambda$2344 ($\Delta$EW = 0.64 $\pm$ 0.12 $\textrm{\AA}$) line becomes narrower and deeper in the line core which translates to an overall EW increase between observations.  The \mbox{Fe\,{\sc ii}} $\lambda$2374 and $\lambda$2587 lines, with comparable strengths to the $\lambda$2344 line, display the same trend in their EW changes.  Pixel flux differences are also evident in the red edges of both \mbox{Mg\,{\sc ii}} lines suggesting variations occurred in the line wings as expected for stronger lines.

\paragraph*{\it J171748.76+275532.5 (System 10, $z_{\scriptsize{\textrm{abs}}} = 1.1674$, $\beta = 0.2957$):} The observations for this system are separated by only one day in the system rest frame.  The \mbox{Mg\,{\sc ii}} $\lambda$2804 line ($\Delta$EW = 0.26 $\pm$ 0.06 $\textrm{\AA}$) widens between observations with a consistent and comparable variation in the blue line ($\lambda$2796) of this doublet.  This system is blended with another grade A system at $z_{\scriptsize{\textrm{abs}}}$ = 1.1706 so the observed variations are likely caused by changes in the amount of blending between these systems.

\paragraph*{\it J131347.68+294201.3 (System 11, $z_{\scriptsize{\textrm{abs}}} = 1.5157$, $\beta$ = 0.1938):} The EW of the \mbox{Fe\,{\sc ii}} $\lambda$2383 ($\Delta$EW = 0.49 $\pm$ 0.10 $\textrm{\AA}$) line more than doubles as the profile becomes wider and deeper between observations.  This line has an \mbox{Fe\,{\sc ii}} $\lambda$2600 alternate identification in a grade A system at $z_{\scriptsize{\textrm{abs}}}$ = 1.3057.  Pixel flux variations are visible in the following lines: \mbox{Al\,{\sc ii}} $\lambda$1671, \mbox{Fe\,{\sc ii}} $\lambda$2600, and \mbox{Si\,{\sc ii}} $\lambda$1527. 

\paragraph*{\it J024534.07+010813.7 (System 12, $z_{\scriptsize{\textrm{abs}}} = 1.1125$, $\beta$ = 0.1780):} The observations for this system are separated by only three days in the system rest frame.  The \mbox{Fe\,{\sc ii}} $\lambda$2383 ($\Delta$EW = 0.62 $\pm$ 0.15 $\textrm{\AA}$) line becomes wider and deeper between observations.  Seven of the ten remaining lines are flagged for at least one of the three error conditions (see Section \ref{seccat}), and no variations are evident in the three unflagged lines.

\paragraph*{\it J004806.05+004623.6 (System 13, $z_{\scriptsize{\textrm{abs}}} = 1.8811$, $\beta$ = 0.1536):} The \mbox{C\,{\sc iv}} $\lambda$1551 ($\Delta$EW = 0.86 $\pm$ 0.21 $\textrm{\AA}$) line doubles in width as two distinct components in the first observation are blended together in the second.  A similar but smaller widening is visible in the \mbox{C\,{\sc iv}} $\lambda$1548 line.  The second significantly variable line, \mbox{Fe\,{\sc ii}} $\lambda$2344 ($\Delta$EW = --0.41 $\pm$ 0.13 $\textrm{\AA}$), has an opposite change of the same magnitude as the \mbox{C\,{\sc iv}} $\lambda$1551 line where it becomes more narrow and shallow between observations.  This line has a \mbox{Mg\,{\sc i}} $\lambda$2026 alternate identification in a grade D system not considered at $z_{\scriptsize{\textrm{abs}}}$ = 2.3309.  The remaining eight absorption lines all have error flags.  The relative changes of these two lines could indicate a change in ionization state in this system.  

\paragraph*{\it J152555.81+010835.4 (System 14, $z_{\scriptsize{\textrm{abs}}} = 1.1147$, $\beta$ = 0.1513):} At nearly four years between observations in the system rest frame, this system has the longest time separation of all the variable NAL systems.  It also has the most significantly variable lines that all decrease in EW between observations: \mbox{Mg\,{\sc ii}} $\lambda$2796: $\Delta$EW = --0.57 $\pm$ 0.14 $\textrm{\AA}$, $\lambda$2804: $\Delta$EW = --1.64 $\pm$ 0.16 $\textrm{\AA}$; \mbox{Fe\,{\sc ii}} $\lambda$2344: $\Delta$EW = --0.55 $\pm$ 0.10 $\textrm{\AA}$, $\lambda$2587: $\Delta$EW = --0.65 $\pm$ 0.12 $\textrm{\AA}$. The last line has a \mbox{Ca\,{\sc ii}} $\lambda$3935 alternate identification in a grade B system at $z_{\scriptsize{\textrm{abs}}}$ = 0.3902.  All of the lines detected in this system are blended with lines from the same ion species in a separate system at $z_{\scriptsize{\textrm{abs}}}$ = 1.1160, and the observed variability is caused by changes in the amount of blending between these two systems.

\paragraph*{\it J143826.73+642859.8 (System 15, $z_{\scriptsize{\textrm{abs}}} = 0.9114$, $\beta$ = 0.1492):} Both of the \mbox{Mg\,{\sc ii}} $\lambda \lambda$2796, 2804 ($\Delta$EW = --0.32 $\pm$ 0.09 $\textrm{\AA}$, $\Delta$EW = --0.40 $\pm$ 0.09 $\textrm{\AA}$) lines show comparable and consistent narrowing between observations while all the other detected lines are generally unchanged.

\paragraph*{\it J143229.24--010616.0 (System 16, $z_{\scriptsize{\textrm{abs}}} = 1.6627$, $\beta$ = 0.1468):} The \mbox{C\,{\sc iv}} $\lambda$1548 ($\Delta$EW = --0.32 $\pm$ 0.10 $\textrm{\AA}$) line decreases in width by almost a factor of 2 as the amount of blending of the individual components in this line changes between observations.  The $\lambda$1551 line shows similar flux variations which implies that it also varies even though the measured EWs do not change.  The second significantly variable line in this system, \mbox{Si\,{\sc ii}} $\lambda$1527 ($\Delta$EW = 0.33 $\pm$ 0.10 $\textrm{\AA}$), has an opposite change of the same magnitude as the line width more than doubles between observations.  This line has a \mbox{Mg\,{\sc ii}} $\lambda$2796 alternate identification in a grade D system at $z_{\scriptsize{\textrm{abs}}}$ = 0.4536.  The relative changes of these two lines could signify a change in ionization state in this system.  

\paragraph*{\it J022632.51--003841.4 (System 17, $z_{\scriptsize{\textrm{abs}}} = 1.9699$, $\beta$ = 0.1208):} Both of the \mbox{C\,{\sc iv}} $\lambda \lambda$1548, 1551 ($\Delta$EW = --0.29 $\pm$ 0.10 $\textrm{\AA}$, $\Delta$EW = --0.72 $\pm$ 0.13 $\textrm{\AA}$) lines show comparable and consistent narrowing between observations as reflected by the changes in the line blending.  A similar variation is noted in the \mbox{Al\,{\sc ii}} $\lambda$1671 line.

\paragraph*{\it J005202.40+010129.2 (System 18, $z_{\scriptsize{\textrm{abs}}} = 1.9116$, $\beta$ = 0.1179):} The \mbox{Si\,{\sc iv}} $\lambda \lambda$1394, 1403 ($\Delta$EW = --0.09 $\pm$ 0.03 $\textrm{\AA}$, $\Delta$EW = 0.15 $\pm$ 0.03 $\textrm{\AA}$) lines vary in the opposite manner that could be the result of changes in the line blending.  The red line has a \mbox{C\,{\sc i}} $\lambda$1280 alternate identification in a grade B system at $z_{\scriptsize{\textrm{abs}}}$ = 2.1916.  None of the other lines in this system show any variations.

\paragraph*{\it J004023.76+140807.3 (System 19, $z_{\scriptsize{\textrm{abs}}} = 1.6207$, $\beta$ = 0.1027):} This system has three significantly variable lines that all increase between observations, possibly indicating bulk motion as the cause of the observed variability.  The \mbox{C\,{\sc iv}} $\lambda \lambda$1548, 1551 ($\Delta$EW = 0.77 $\pm$ 0.07 $\textrm{\AA}$) doublet is blended together and could suggest the formation of a mini-BAL.  Both the \mbox{Al\,{\sc ii}} $\lambda$1671 ($\Delta$EW = 0.14 $\pm$ 0.04 $\textrm{\AA}$) and \mbox{Fe\,{\sc ii}} $\lambda$2383 ($\Delta$EW = 0.15 $\pm$ 0.04 $\textrm{\AA}$) lines become wider by the same amount between observations.  The \mbox{Al\,{\sc ii}} line has a \mbox{Mg\,{\sc ii}} $\lambda$2796 alternate identification in a grade E system at $z_{\scriptsize{\textrm{abs}}}$ = 0.5658.  No other lines varied in this system.

\paragraph*{\it J023620.79--003342.2 (System 20, $z_{\scriptsize{\textrm{abs}}} = 1.4635$, $\beta$ = 0.0808):} The observations of this system are separated by only one MJD.  There is a clear line blending change in the \mbox{Mg\,{\sc ii}} $\lambda$2796 ($\Delta$EW = --0.93 $\pm$ 0.21 $\textrm{\AA}$) line corresponding to an EW decrease of one-half.  Line blending changes are also visible in the \mbox{Mg\,{\sc ii}} $\lambda$2804 and \mbox{Fe\,{\sc ii}} $\lambda$2600 lines. 

\paragraph*{\it J024154.42--004757.6 (System 21, $z_{\scriptsize{\textrm{abs}}} = 2.1259$, $\beta$ = 0.0794):} The \mbox{Fe\,{\sc ii}} $\lambda$2383 ($\Delta$EW = 0.83 $\pm$ 0.18 $\textrm{\AA}$) line doubles in width between observations.  A similar trend is evident in the \mbox{Fe\,{\sc ii}} $\lambda$2600 line variations as expected for lines with comparable strengths.  Overall, this spectrum is quite noisy with error conditions flagged for 10 of the remaining 13 lines.

\paragraph*{\it J131712.01+020224.9 (System 22, $z_{\scriptsize{\textrm{abs}}} = 1.9274$, $\beta$ = 0.0751):} The variations in the \mbox{Si\,{\sc ii}} $\lambda$1527 ($\Delta$EW = --1.05 $\pm$ 0.19 $\textrm{\AA}$) and \mbox{Al\,{\sc ii}} $\lambda$1671($\Delta$EW = 0.65 $\pm$ 0.14 $\textrm{\AA}$) lines are caused by changes in the line blending making the EWs suspect.  The alternate identification for the \mbox{Al\,{\sc ii}} line is incorrectly flagged by the Y13 QAL pipeline.  Pixel flux variations are visible in the \mbox{C\,{\sc ii}} $\lambda$1335; \mbox{C\,{\sc iv}} $\lambda \lambda$1548, 1551; and \mbox{Si\,{\sc iv}} $\lambda \lambda$1394, 1403 lines although their low $S/N$ precludes a more precise variability measurement.

\paragraph*{\it J171244.11+555949.7 (System 23, $z_{\scriptsize{\textrm{abs}}} = 1.2092$, $\beta$ = 0.0660):} The \mbox{Fe\,{\sc ii}} $\lambda$2374 ($\Delta$EW = --0.37 $\pm$ 0.11 $\textrm{\AA}$) and \mbox{Mg\,{\sc ii}} $\lambda$2804 ($\Delta$EW = --0.52 $\pm$ 0.12 $\textrm{\AA}$) lines both become narrower between observations.  Comparable and consistent variations are noted in the following lines: \mbox{Mg\,{\sc ii}} $\lambda$2796; \mbox{Fe\,{\sc ii}} $\lambda$2344, $\lambda$2587.  

\paragraph*{\it J102046.62+282707.1 (System 24, $z_{\scriptsize{\textrm{abs}}} = 2.8750$, $z_{\scriptsize{\textrm{abs}}}$, $\beta$ = 0.0624):} The EW of the \mbox{C\,{\sc iv}} $\lambda$1548 ($\Delta$EW = 0.76 $\pm$ 0.16 $\textrm{\AA}$) line doubles between observations as it becomes significantly deeper and wider.  Additionally, pixel flux variations are apparent in the \mbox{C\,{\sc iv}} $\lambda$1551 line and an unidentified line at $\lambda_{obs}\approx 5840$ $\textrm{\AA}$.

\paragraph*{\it J161540.76+460451.0 (System 25, $z_{\scriptsize{\textrm{abs}}} = 1.3705$, $\beta$ = 0.0585):} The \mbox{Fe\,{\sc ii}} $\lambda$2344 ($\Delta$EW = 0.72 $\pm$ 0.19 $\textrm{\AA}$) and $\lambda$2587 ($\Delta$EW = 0.85 $\pm$ 0.19 $\textrm{\AA}$) lines with similar strengths both widen between observations.    A third significantly variable line, \mbox{Mg\,{\sc ii}} $\lambda$2804 ($\Delta$EW = --0.69 $\pm$ 0.20 $\textrm{\AA}$), has an opposite change of similar magnitude that could indicate density variations between elements in the absorbing gas cloud.

\paragraph*{\it J015733.87--004824.4 (System 26, $z_{\scriptsize{\textrm{abs}}} = 1.4156$, $\beta$ = 0.0546):} The \mbox{Al\,{\sc ii}} $\lambda$1671 ($\Delta$EW = 0.35 $\pm$ 0.11 $\textrm{\AA}$) and \mbox{Al\,{\sc iii}} $\lambda$1855 ($\Delta$EW = 0.24 $\pm$ 0.06 $\textrm{\AA}$) lines with similar strengths both become wider and deeper between observations.  Pixel flux variations are also visible in the \mbox{Mg\,{\sc ii}} $\lambda \lambda$2796, 2804 doublet and the \mbox{Fe\,{\sc ii}} $\lambda$1608 line.

\paragraph*{\it J114857.22+594153.9 (System 27, $z_{\scriptsize{\textrm{abs}}} = 1.9800$, $\beta$ = 0.0469):} The \mbox{C\,{\sc ii}} $\lambda$1335 ($\Delta$EW = 0.61 $\pm$ 0.14 $\textrm{\AA}$) line widens between observations.  The \mbox{Si\,{\sc iv}} $\lambda \lambda$1394, 1403 doublet and the \mbox{Fe\,{\sc ii}} $\lambda$1608 line all display pixel flux variations.

\paragraph*{\it J120142.99+004925.0 (System 28, $z_{\scriptsize{\textrm{abs}}} = 1.4181$, $\beta$ = 0.0438):} The \mbox{Fe\,{\sc ii}} $\lambda$2383 ($\Delta$EW = --0.87 $\pm$ 0.18 $\textrm{\AA}$) line shows clear line blending changes as the width narrows by more than a factor of two between observations.  Redshift variations occurred in the \mbox{Mg\,{\sc i}} $\lambda$2853 line while the remaining lines are mostly unchanged.

\paragraph*{\it J124708.42+500320.7 (System 29, $z_{\scriptsize{\textrm{abs}}} = 2.1327$, $\beta$ = 0.0432):} The \mbox{C\,{\sc i}} $\lambda$1277 ($\Delta$EW = 0.40 $\pm$ 0.08 $\textrm{\AA}$) and $\lambda$1657 ($\Delta$EW = 0.66 $\pm$ 0.13 $\textrm{\AA}$) lines both display significant line blending variations where at least two distinct components in the first observation appear to blend together in the second.  A third significantly variable line, \mbox{Fe\,{\sc ii}} $\lambda$2600 ($\Delta$EW = 0.79 $\pm$ 0.20 $\textrm{\AA}$), also exhibits this trend.  Similar pixel flux variations are observed in most of the remaining 16 detected lines although 11 are flagged for proximity to masked pixels, bad continuum fits, or both.

\paragraph*{\it J212812.33--081529.3 (System 30, $z_{\scriptsize{\textrm{abs}}} = 1.8138$, $\beta$ = 0.0235):} The \mbox{C\,{\sc iv}} $\lambda \lambda$1548, 1551 ($\Delta$EW = 0.25 $\pm$ 0.06 $\textrm{\AA}$, $\Delta$EW = --0.22 $\pm$ 0.05 $\textrm{\AA}$) doublet lines have opposite variations between observations which could be the result of density decreases leading to reduced line saturation (see Section \ref{secnvar}).  Pixel flux variations are observed in the \mbox{Si\,{\sc iv}} $\lambda \lambda$1394, 1403 doublet and the \mbox{Si\,{\sc ii}} $\lambda$1527 line.

\paragraph*{\it J084107.24+333921.7 (System 31, $z_{\scriptsize{\textrm{abs}}} = 2.9777$, $\beta$ = 0.0228):} This system has the highest redshift of all the variable NAL systems in Table 3.  The \mbox{C\,{\sc iv}} $\lambda $1548 ($\Delta$EW = --0.98 $\pm$ 0.24 $\textrm{\AA}$) line becomes narrower and shallower between observations.  Similar variations are evident in the red line ($\lambda$1551) of this doublet.  Overall, both spectra are quite noisy which prevents the detection of possible variability in other absorption lines.

\paragraph*{\it J170428.65+242918.0 (System 32, $z_{\scriptsize{\textrm{abs}}} = 1.7462$, $\beta$ = 0.0190):} The \mbox{C\,{\sc iv}} $\lambda$1551 ($\Delta$EW = --0.18 $\pm$ 0.04 $\textrm{\AA}$) line narrows between observations.  This line has numerous alternate identifications including a \mbox{C\,{\sc iv}} $\lambda$1548 line in a grade C system at $z_{\scriptsize{\textrm{abs}}}$ = 1.7513.  The $\lambda$1551 line in this alternate system displays comparable and consistent variations with the first line.  This common variability, combined with the lack of observed variability in the primary system lines, makes it probable that the alternate identification is correct even though there are not many other absorption lines detected in this system.   

\paragraph*{\it J231055.32+004817.1 (System 33, $z_{\scriptsize{\textrm{abs}}} = 2.9528$, $\beta$ = 0.0127):} The \mbox{C\,{\sc iv}} $\lambda$1551 ($\Delta$EW = 0.48 $\pm$ 0.08 $\textrm{\AA}$) line widens between observations.  Similar variations are visible in the blue line ($\lambda$1548) of this doublet.

\clearpage
\begin{figure*}
\begin{center}
\includegraphics[width=84mm]{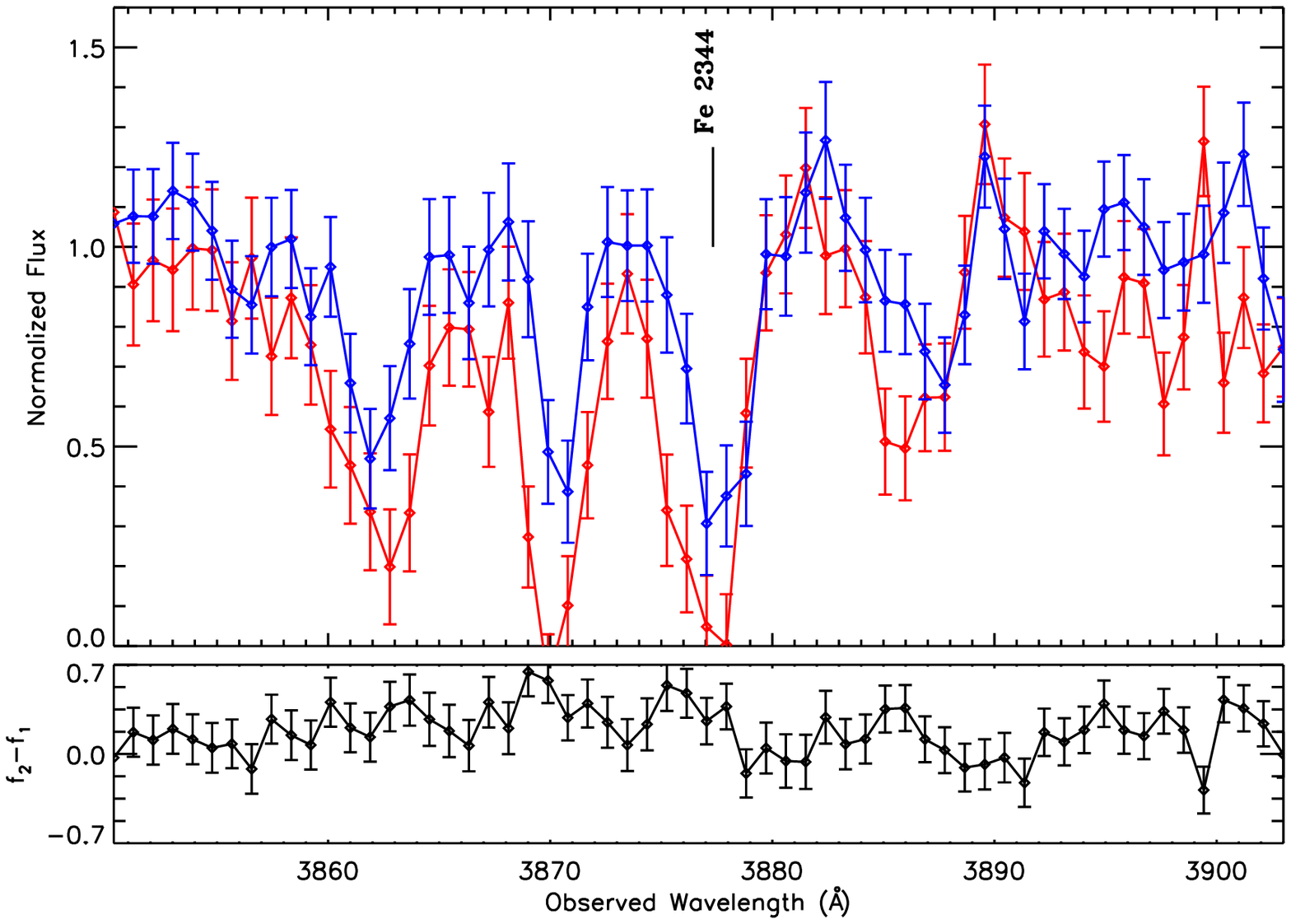}
\includegraphics[width=84mm]{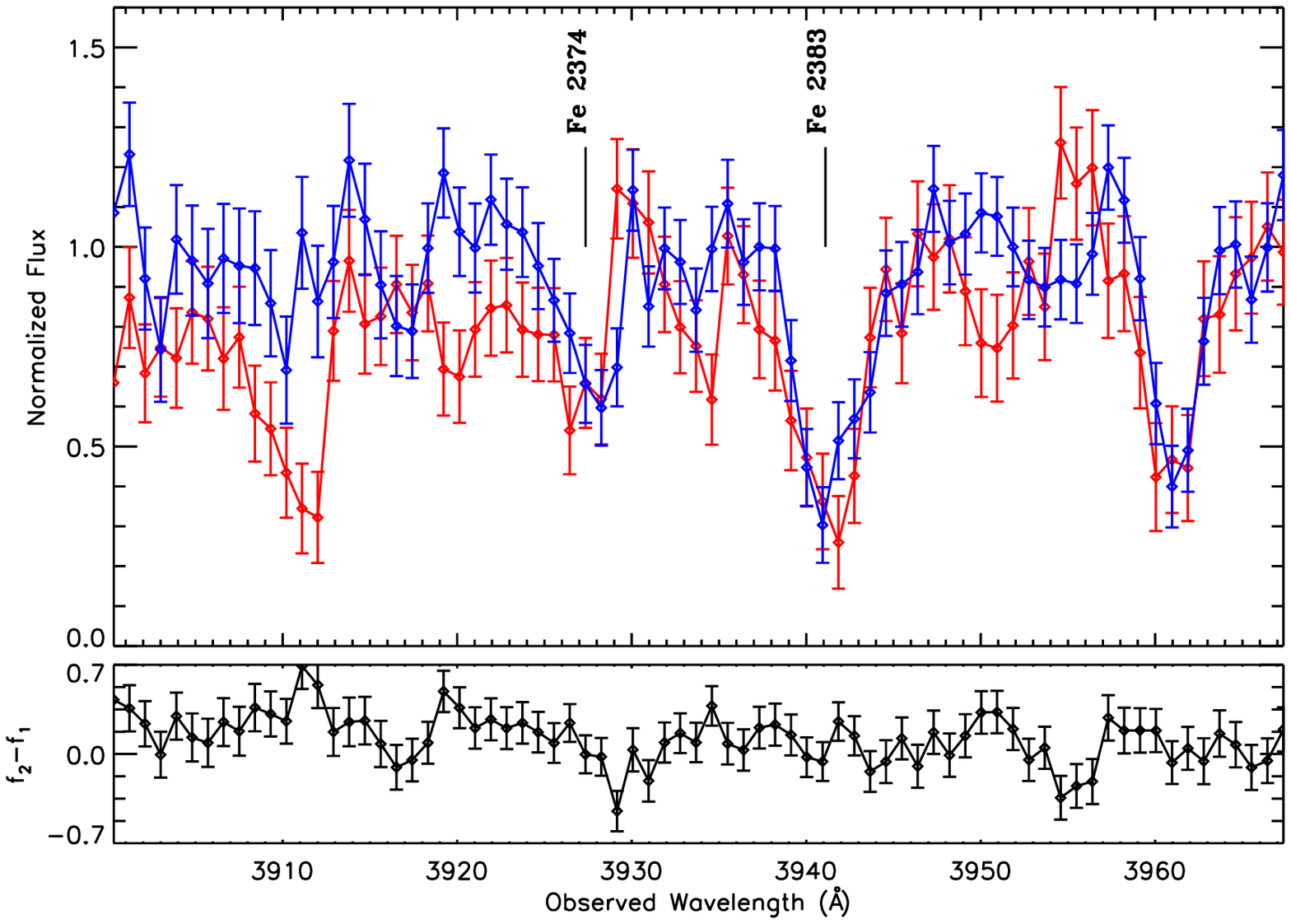}
\includegraphics[width=84mm]{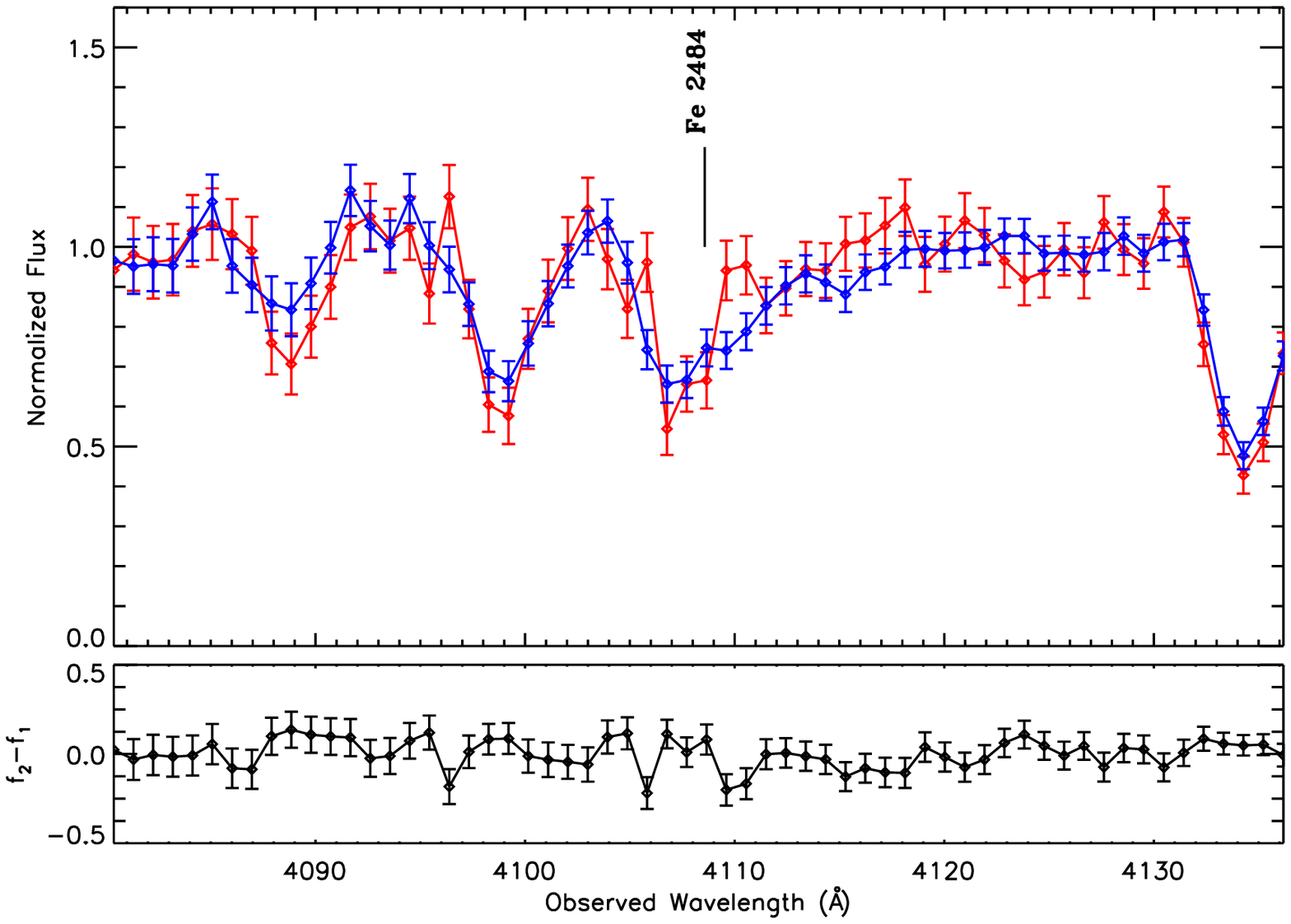}
\includegraphics[width=84mm]{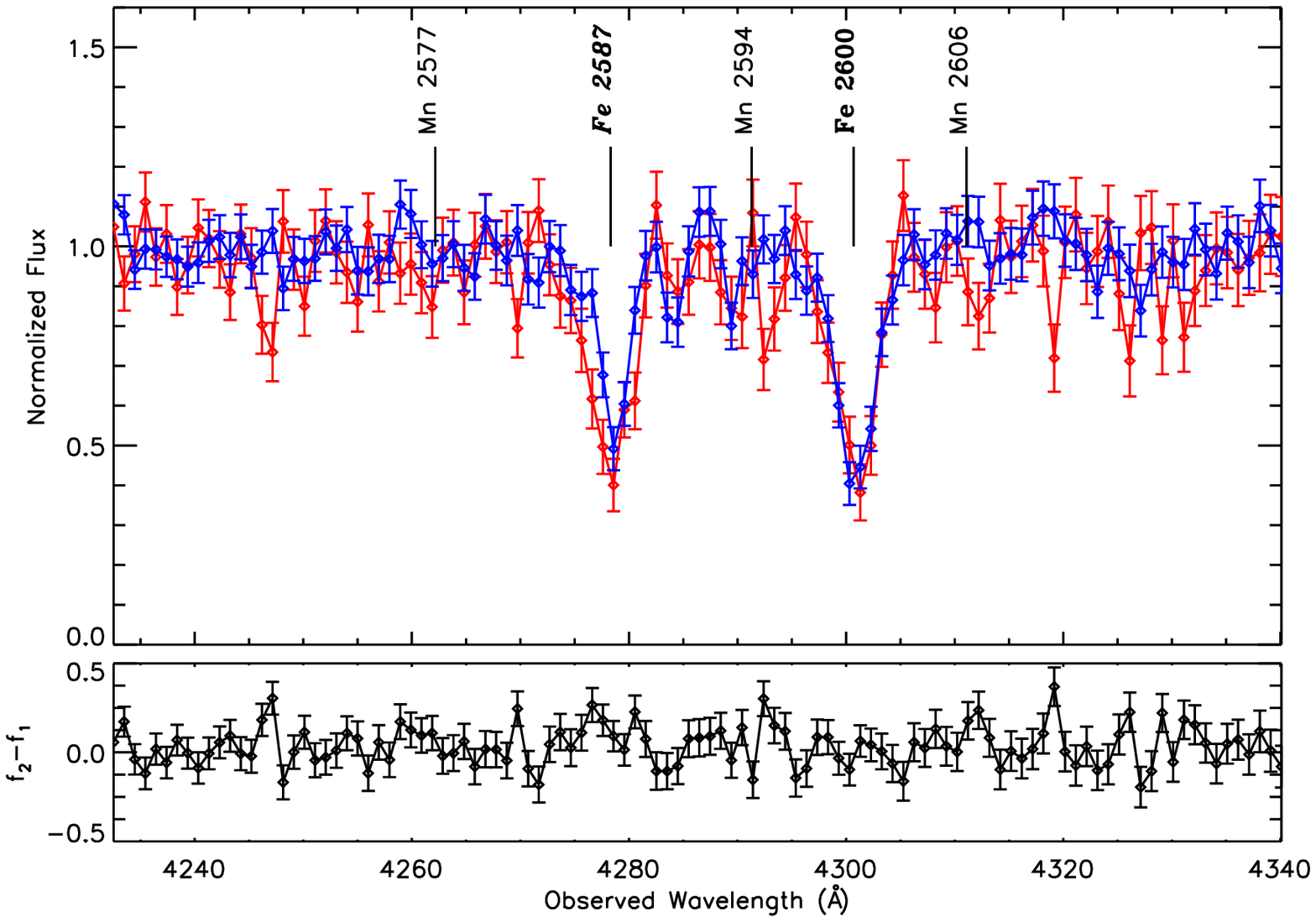}
\includegraphics[width=84mm]{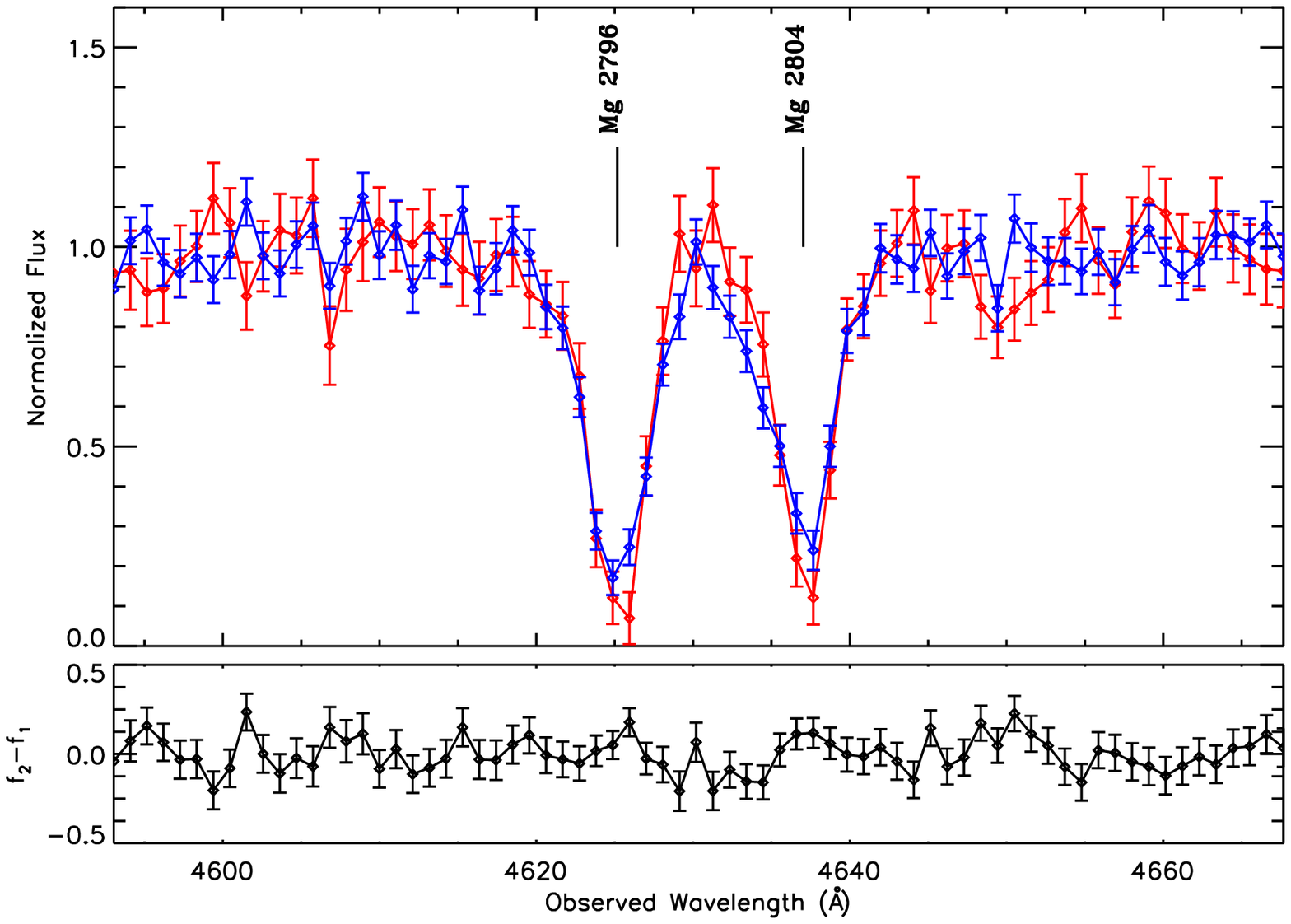}
\includegraphics[width=84mm]{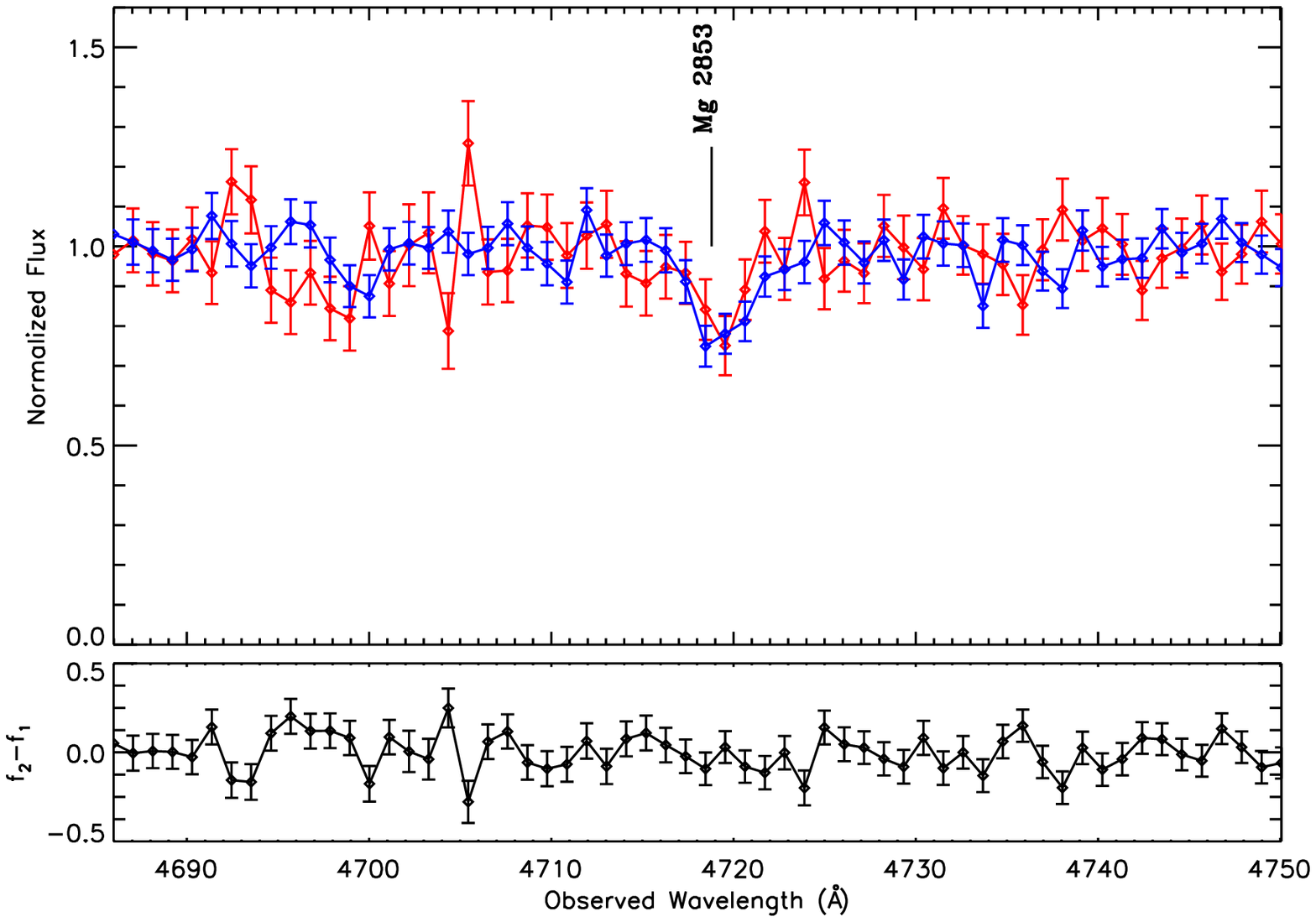}
\caption[Two-epoch normalized spectra of SDSS J082033.97+432751.8]{Two-epoch normalized spectra of the variable NAL system at $\beta$ = 0.6190 in SDSS J082033.97+432751.8.  The top panel shows the normalized pixel flux values with 1$\sigma$ error bars (first observations are red and second are blue), the bottom panel plots the difference spectrum of the two observation epochs, and shaded backgrounds identify masked pixels not included in our search for absorption line variability.  Line identifications for significantly variable absorption lines are italicised, lines detected in both observation epochs are in bold font, and undetected lines are in regular font (see Table A.1 for ion labels). \label{figvs1}}
\end{center}
\end{figure*}

\clearpage
\begin{figure*}
\begin{center}
\includegraphics[width=84mm]{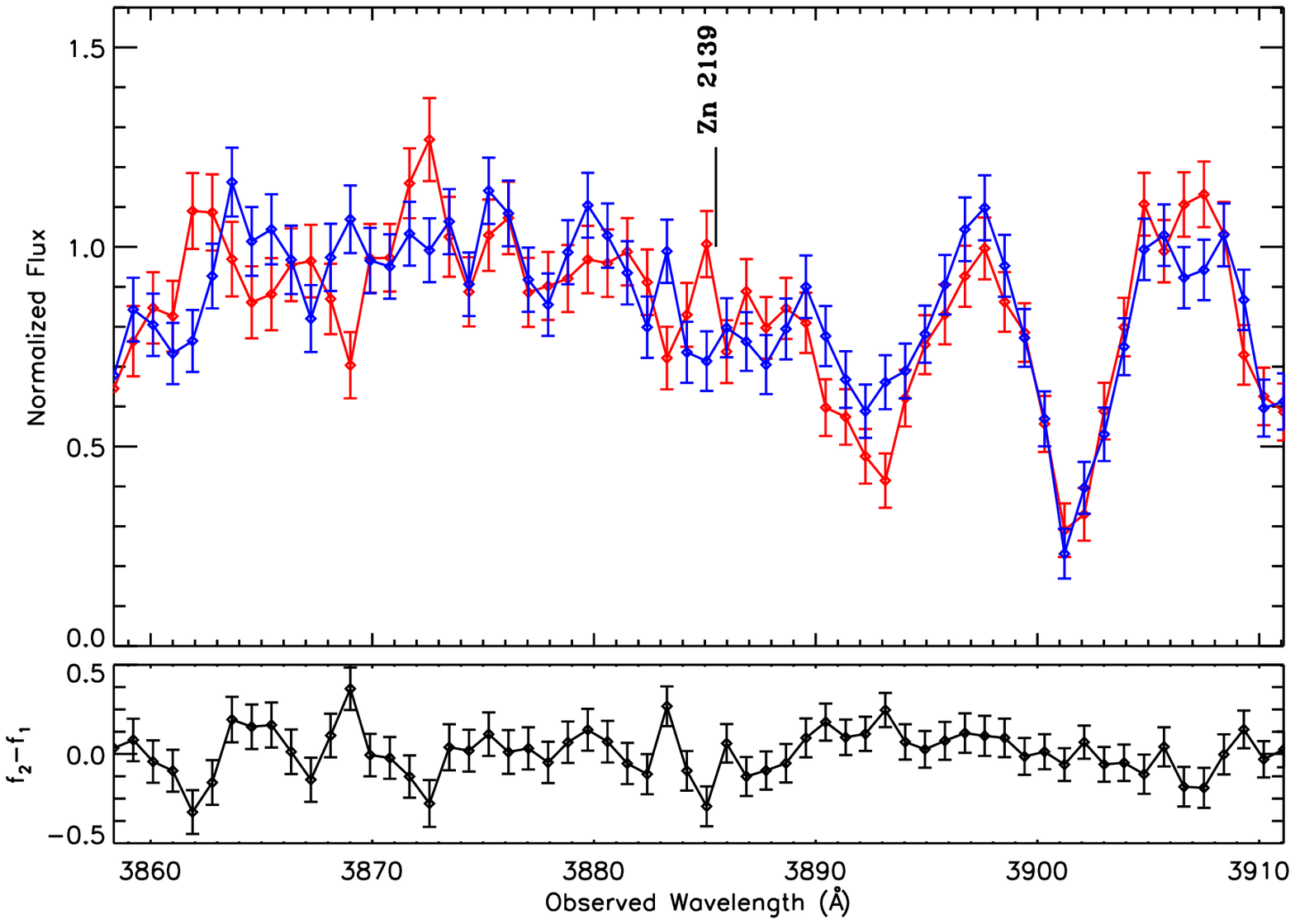}
\includegraphics[width=84mm]{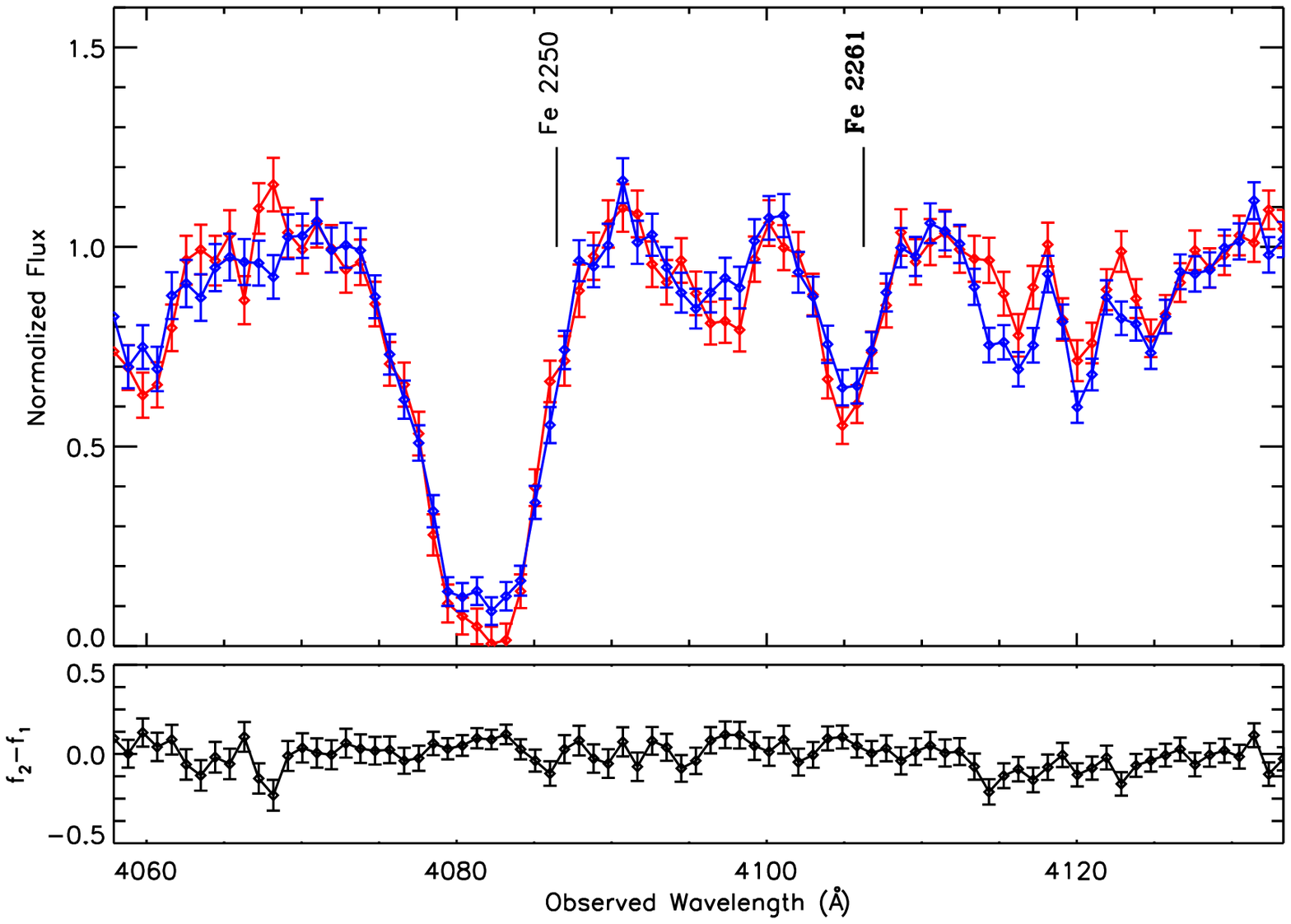}
\includegraphics[width=84mm]{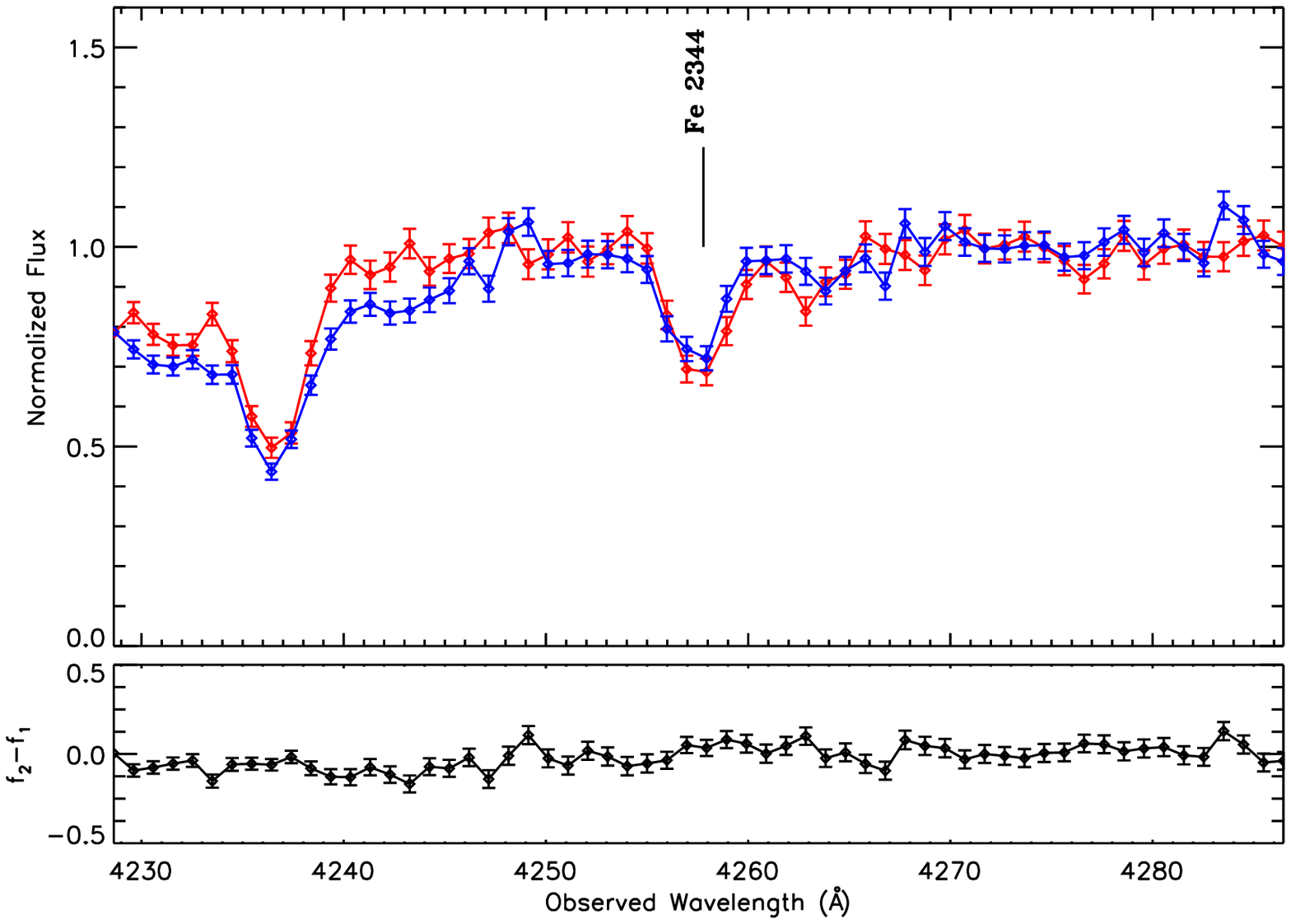}
\includegraphics[width=84mm]{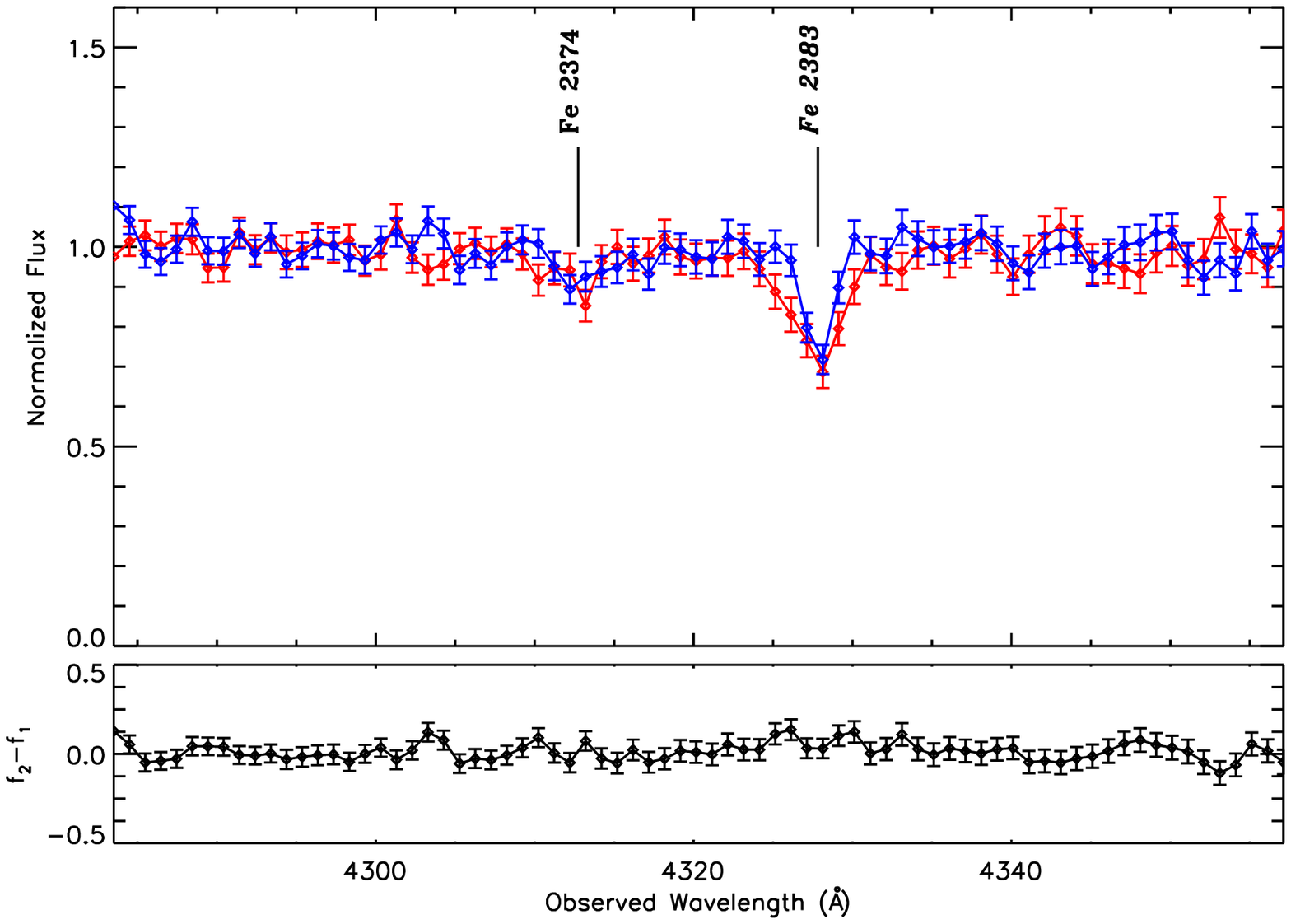}
\includegraphics[width=84mm]{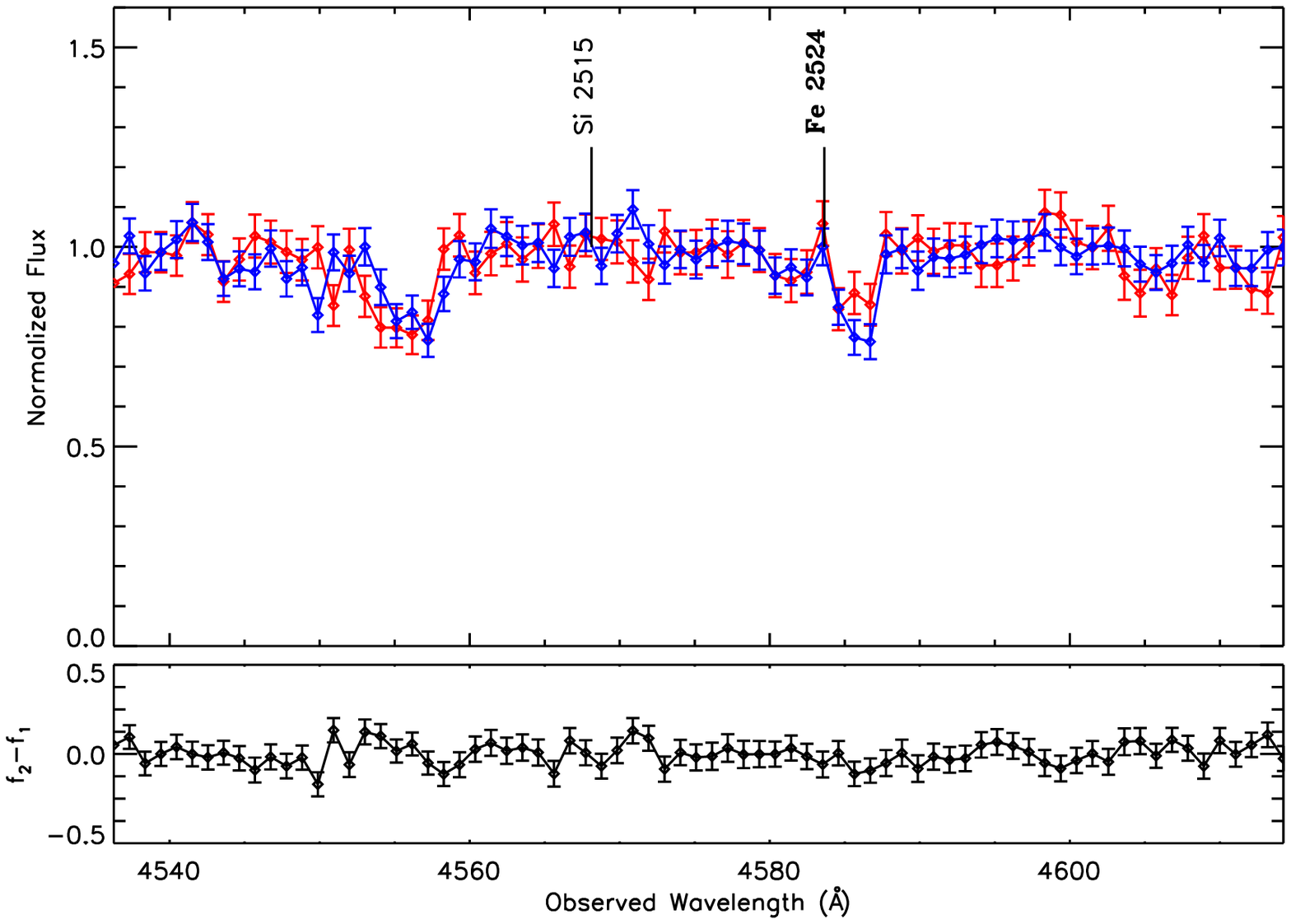}
\includegraphics[width=84mm]{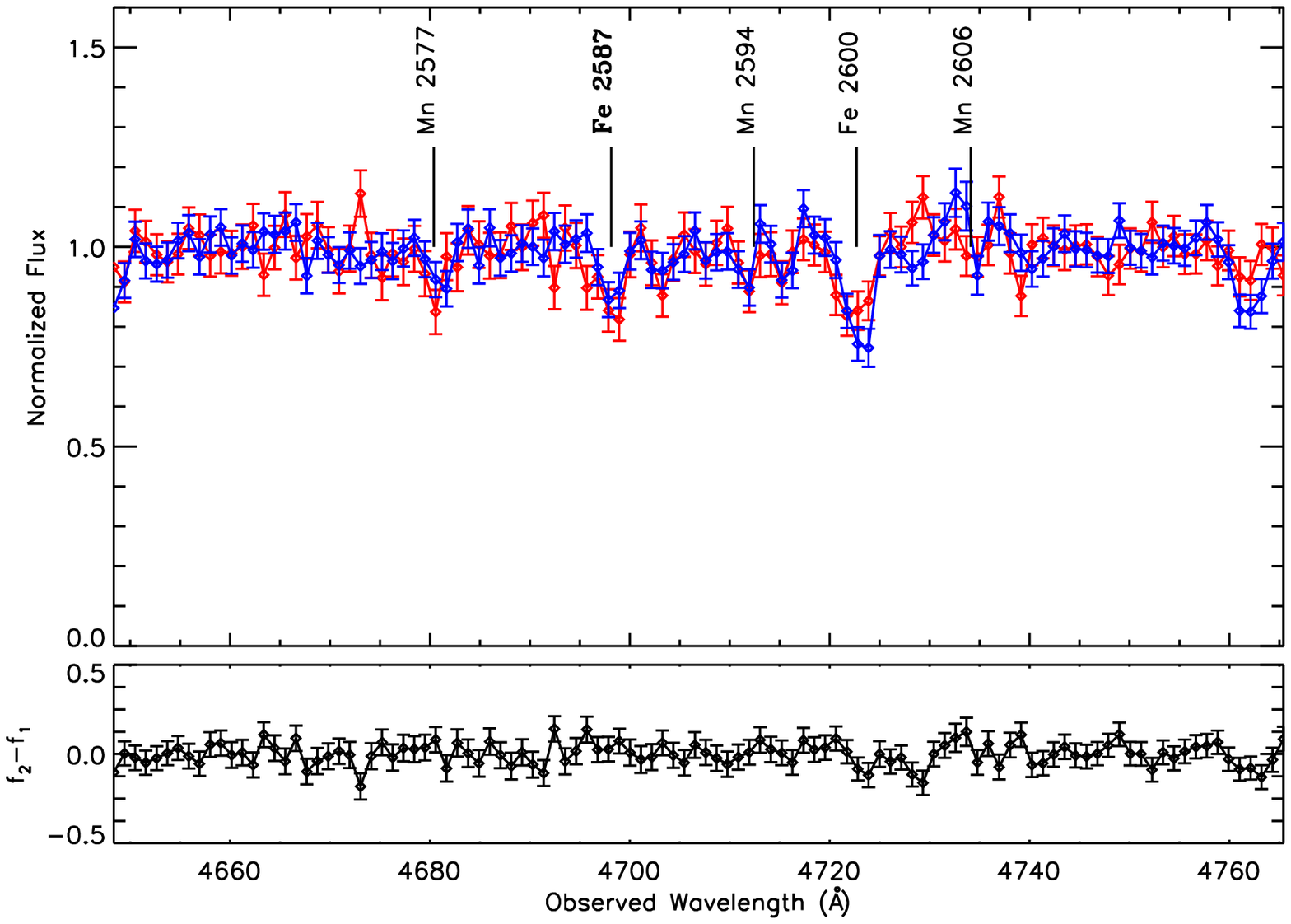}
\caption[Two-epoch normalized spectra of SDSS J140323.39-000606.9]{Two-epoch normalized spectra of the variable NAL system at $\beta$ = 0.5701 in SDSS J140323.39-000606.9.  The top panel shows the normalized pixel flux values with 1$\sigma$ error bars (first observations are red and second are blue), the bottom panel plots the difference spectrum of the two observation epochs, and shaded backgrounds identify masked pixels not included in our search for absorption line variability.  Line identifications for significantly variable absorption lines are italicised, lines detected in both observation epochs are in bold font, and undetected lines are in regular font (see Table A.1 for ion labels).  Continued in next figure. \label{figvs2}}
\end{center}
\end{figure*}

\begin{figure*}
\ContinuedFloat
\begin{center}
\includegraphics[width=84mm]{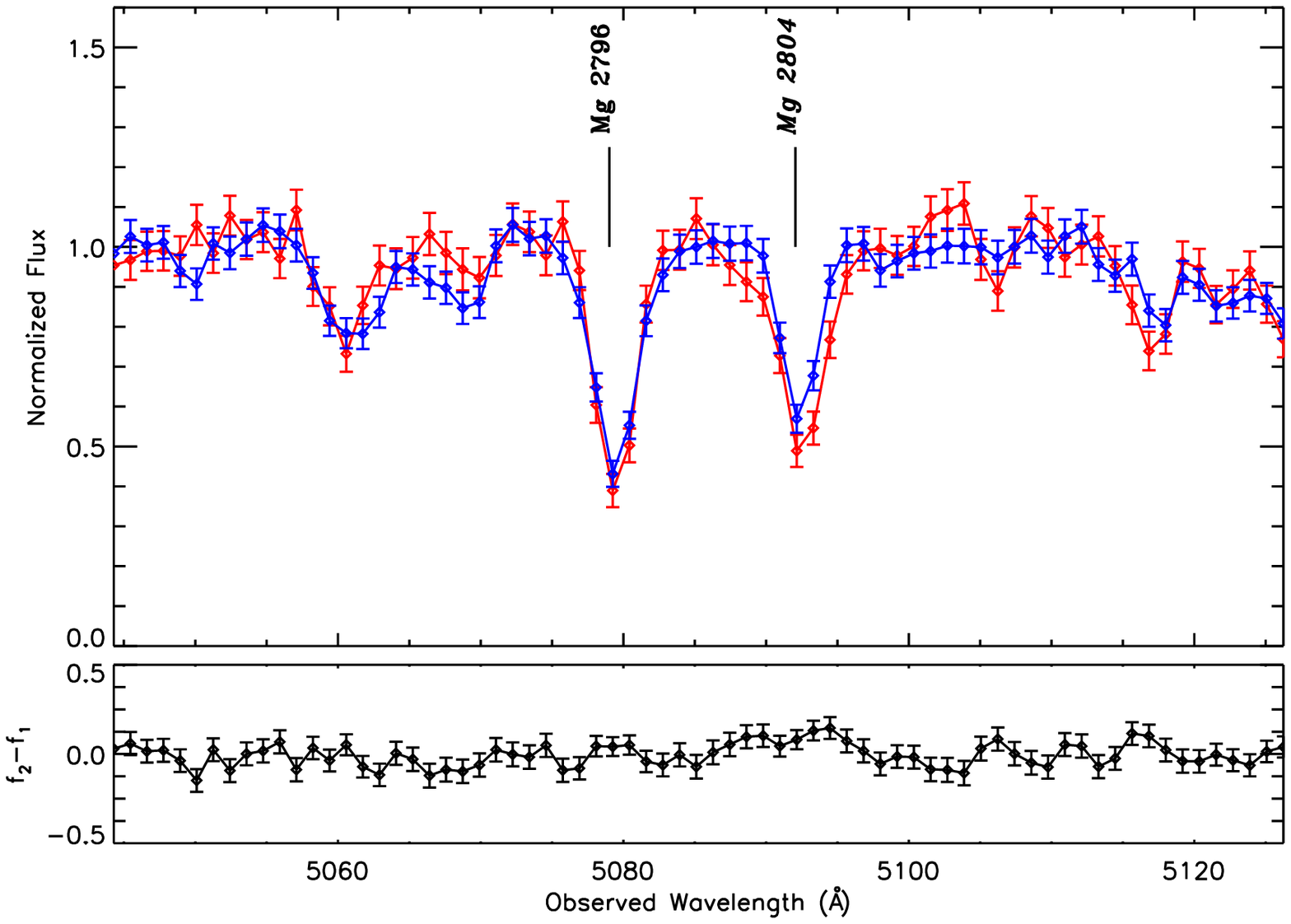}
\caption[]{Two-epoch normalized spectra of the variable NAL system at $\beta$ = 0.5701 in SDSS J140323.39-000606.9.  The top panel shows the normalized pixel flux values with 1$\sigma$ error bars (first observations are red and second are blue), the bottom panel plots the difference spectrum of the two observation epochs, and shaded backgrounds identify masked pixels not included in our search for absorption line variability.  Line identifications for significantly variable absorption lines are italicised, lines detected in both observation epochs are in bold font, and undetected lines are in regular font (see Table A.1 for ion labels).  Continued from previous figure.}
\vspace{3.5cm}
\end{center}
\end{figure*}

\clearpage
\begin{figure*}
\begin{center}
\includegraphics[width=84mm]{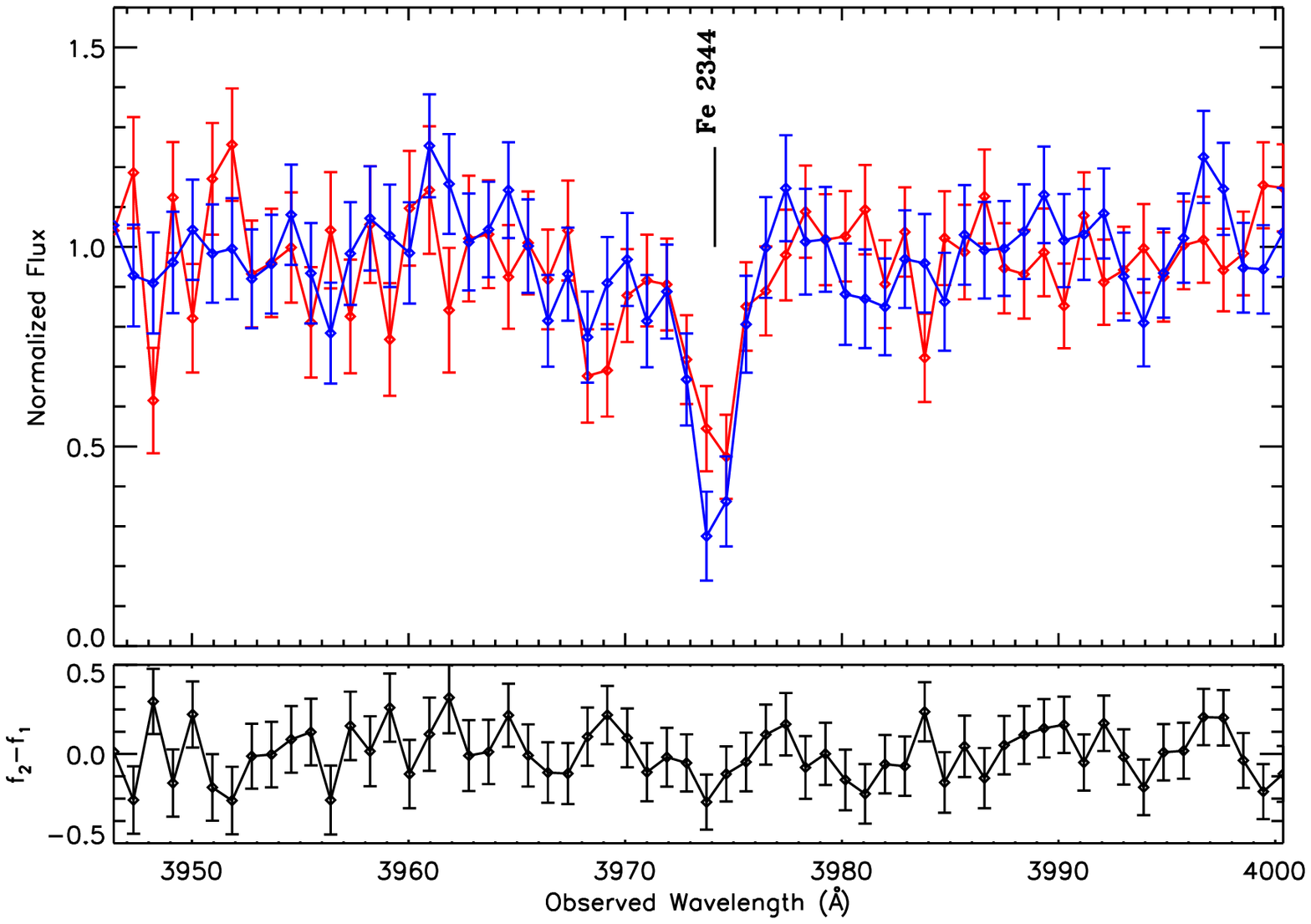}
\includegraphics[width=84mm]{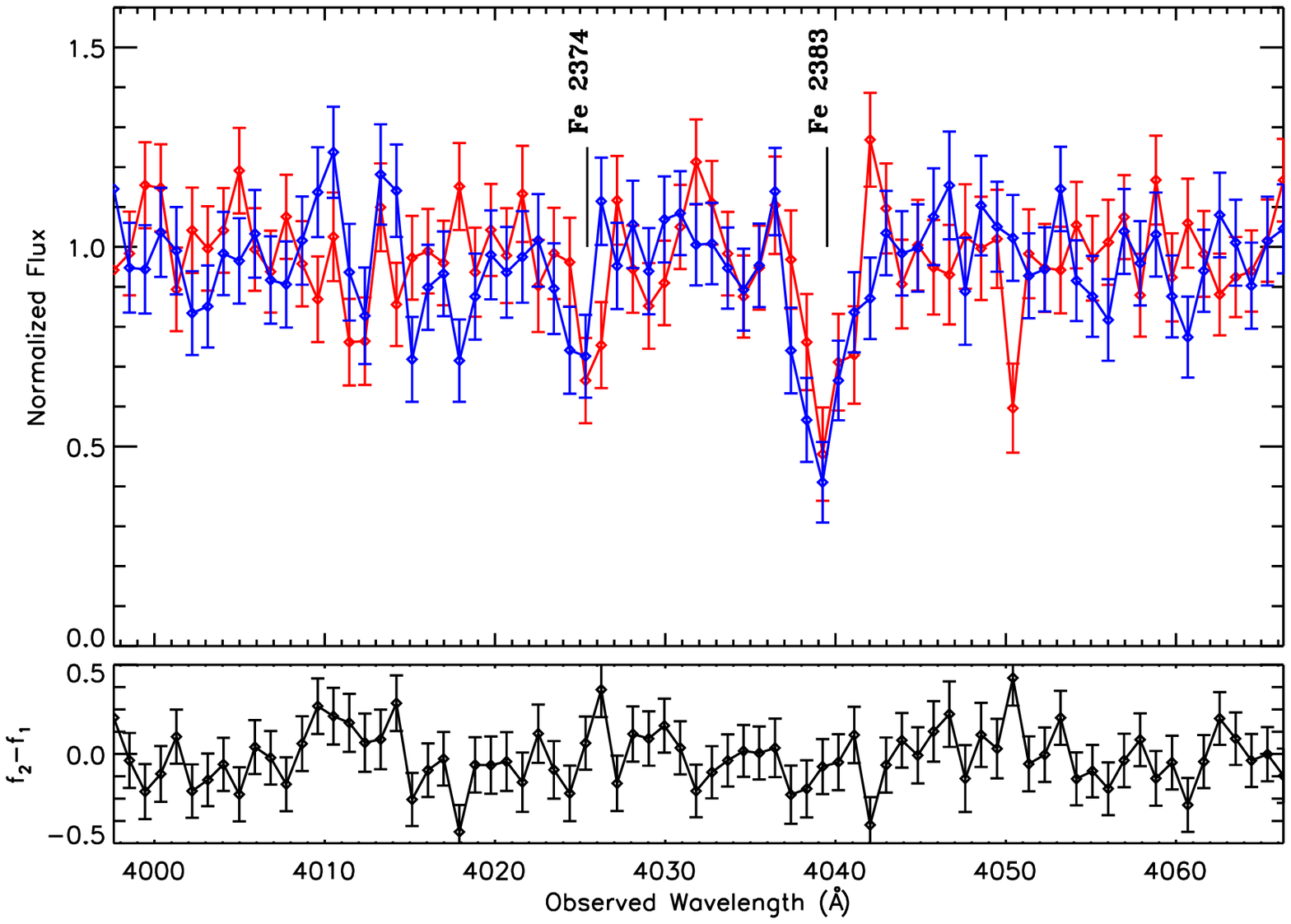}
\includegraphics[width=84mm]{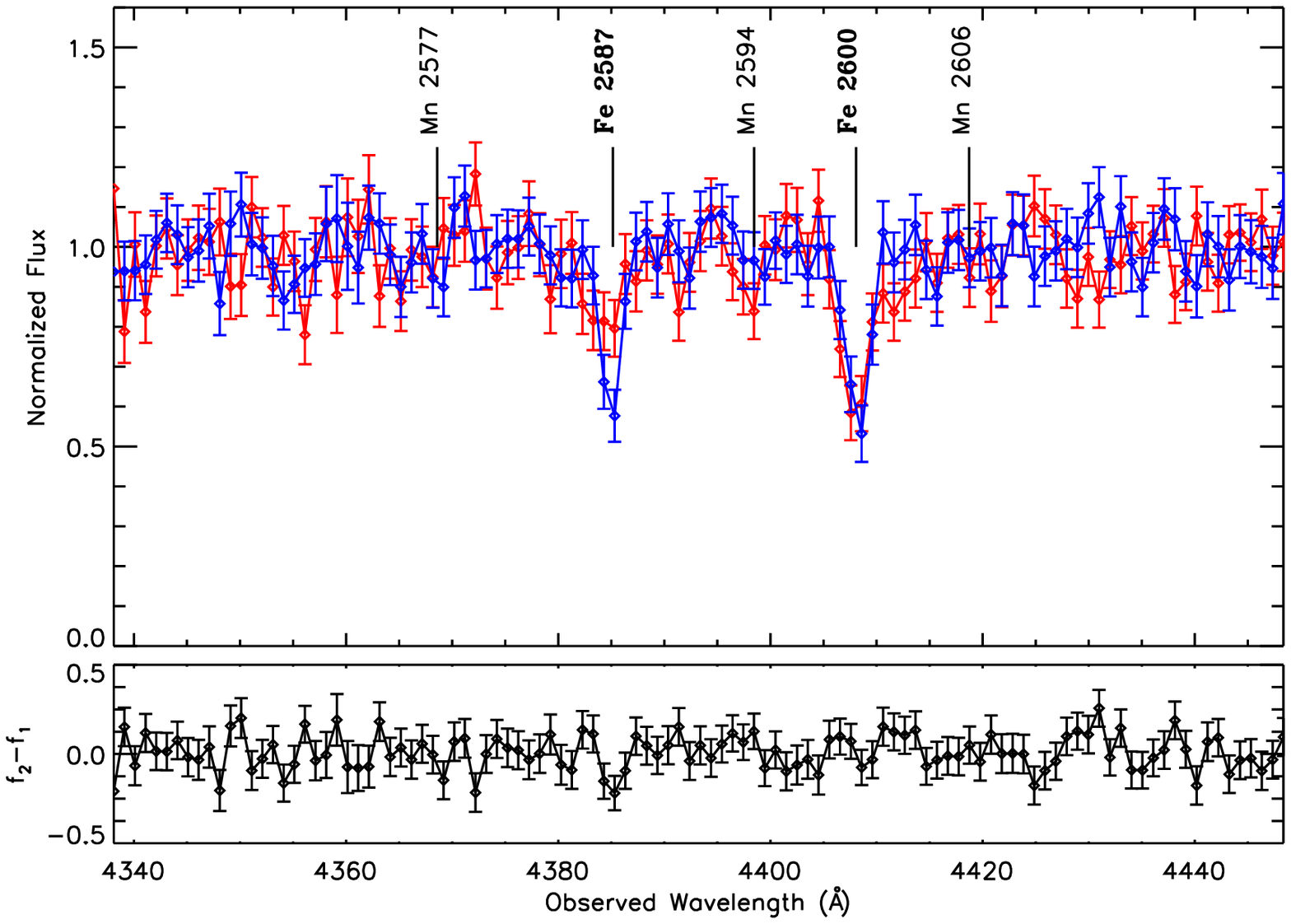}
\includegraphics[width=84mm]{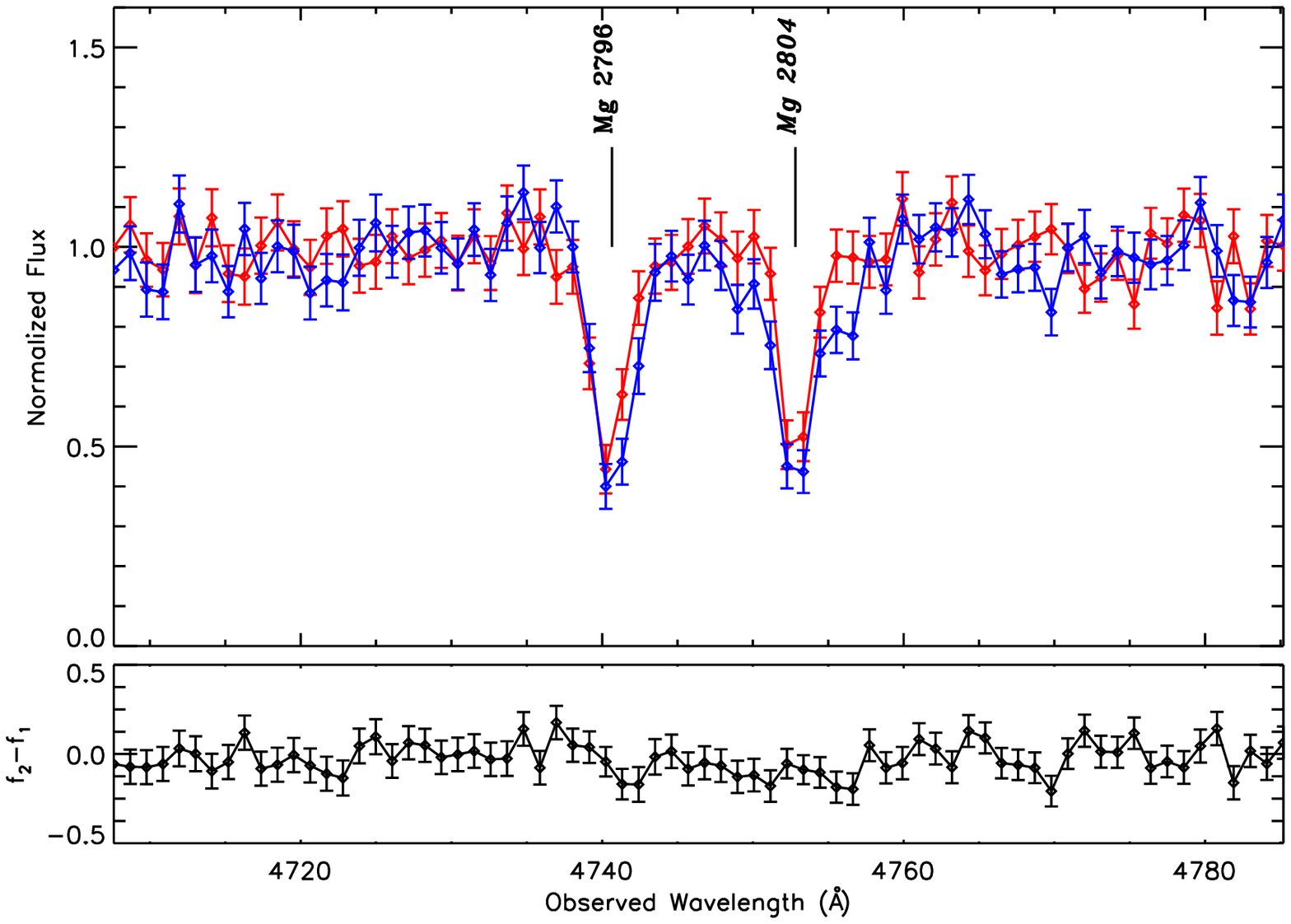}
\caption[Two-epoch normalized spectra of SDSS J151652.69-005834.8]{Two-epoch normalized spectra of the variable NAL system at $\beta$ = 0.4426 in SDSS J151652.69-005834.8.  The top panel shows the normalized pixel flux values with 1$\sigma$ error bars (first observations are red and second are blue), the bottom panel plots the difference spectrum of the two observation epochs, and shaded backgrounds identify masked pixels not included in our search for absorption line variability.  Line identifications for significantly variable absorption lines are italicised, lines detected in both observation epochs are in bold font, and undetected lines are in regular font (see Table A.1 for ion labels). \label{figvs3}}
\end{center}
\end{figure*}

\clearpage
\begin{figure*}
\begin{center}
\includegraphics[width=84mm]{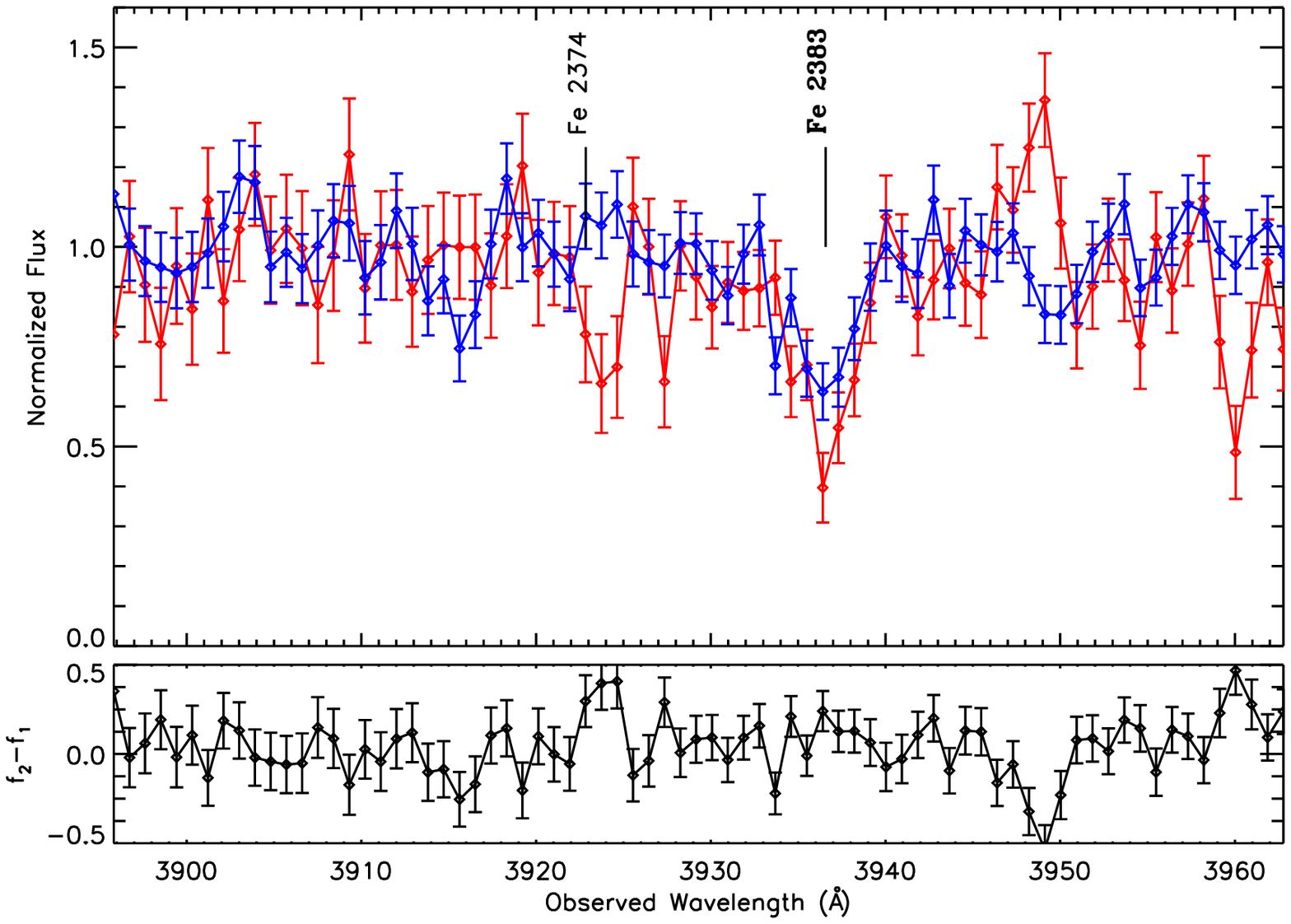}
\includegraphics[width=84mm]{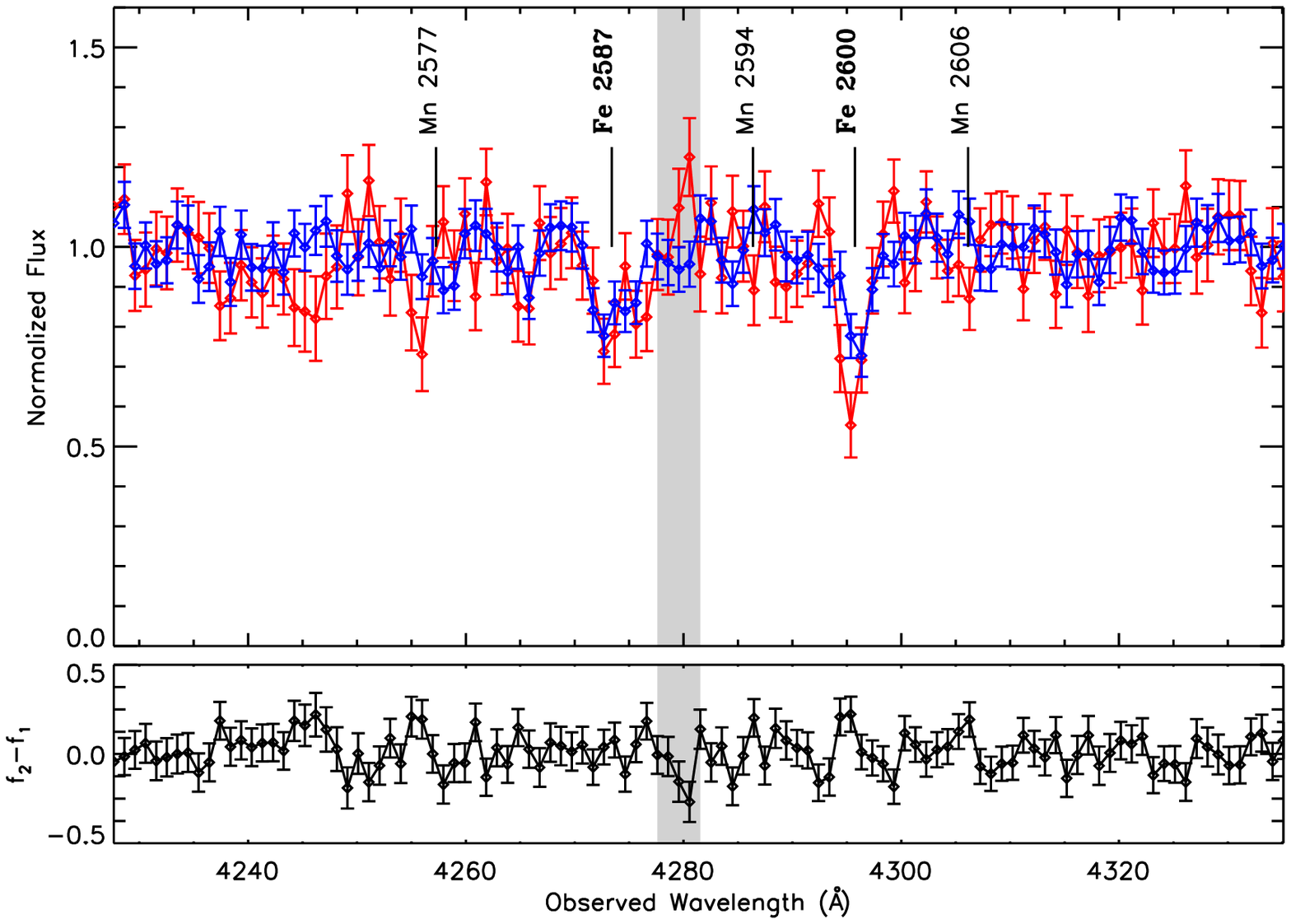}
\includegraphics[width=84mm]{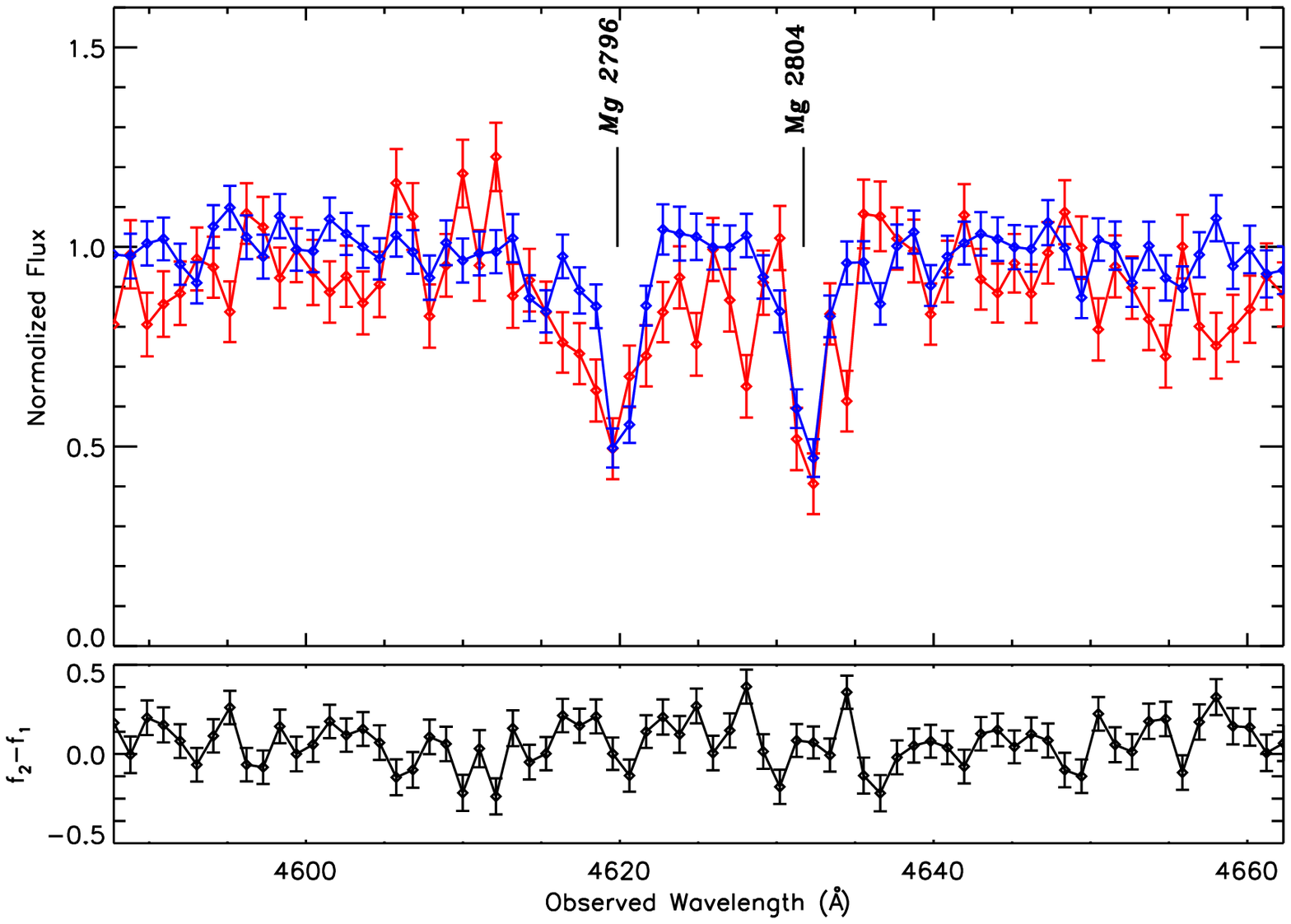}
\includegraphics[width=84mm]{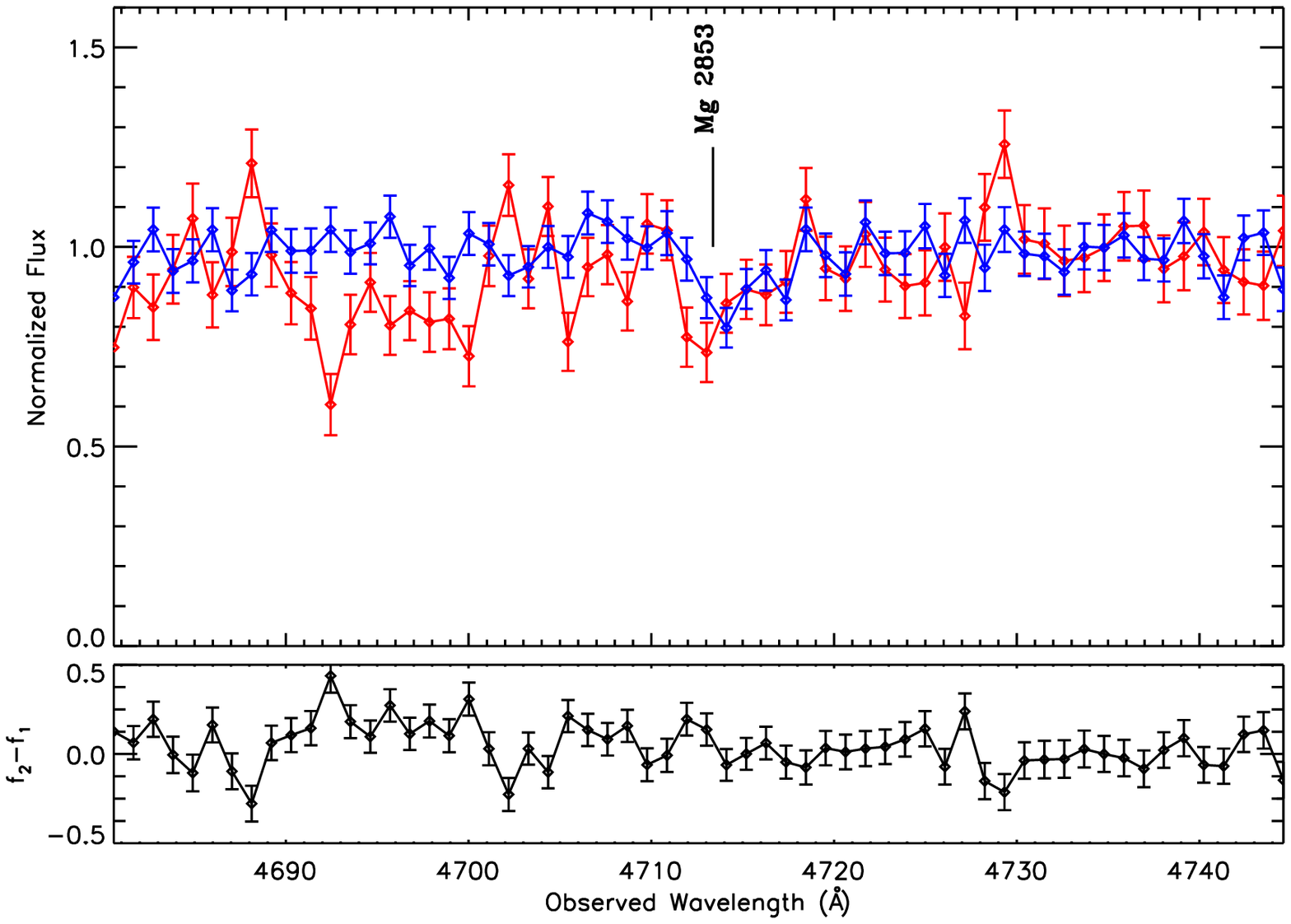}
\caption[Two-epoch normalized spectra of SDSS J215421.13-074430.0]{Two-epoch normalized spectra of the variable NAL system at $\beta$ = 0.4359 in SDSS J215421.13-074430.0.  The top panel shows the normalized pixel flux values with 1$\sigma$ error bars (first observations are red and second are blue), the bottom panel plots the difference spectrum of the two observation epochs, and shaded backgrounds identify masked pixels not included in our search for absorption line variability.  Line identifications for significantly variable absorption lines are italicised, lines detected in both observation epochs are in bold font, and undetected lines are in regular font (see Table A.1 for ion labels).  \label{figvs4}}
\end{center}
\end{figure*}

\clearpage
\begin{figure*}
\begin{center}
\includegraphics[width=84mm]{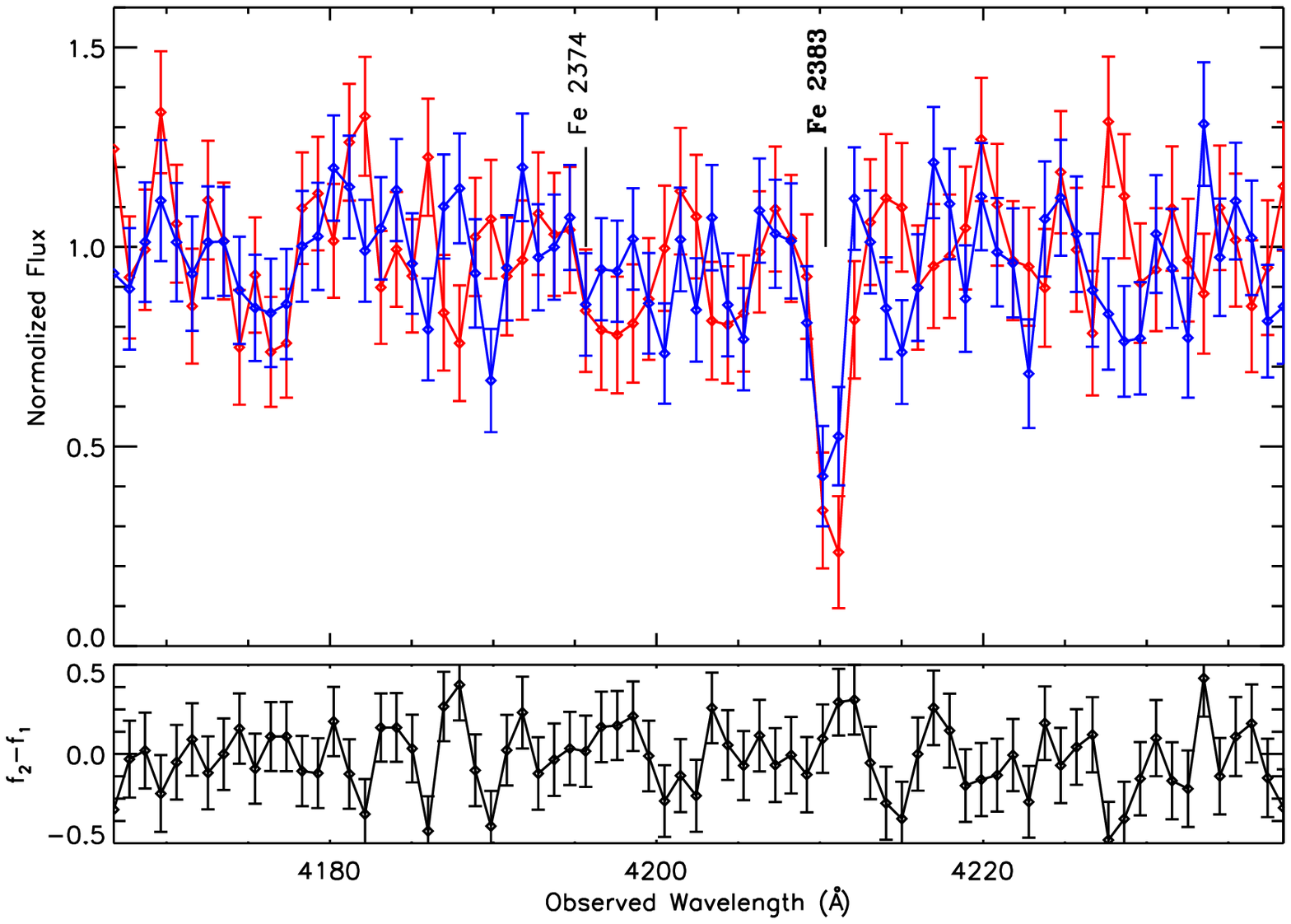}
\includegraphics[width=84mm]{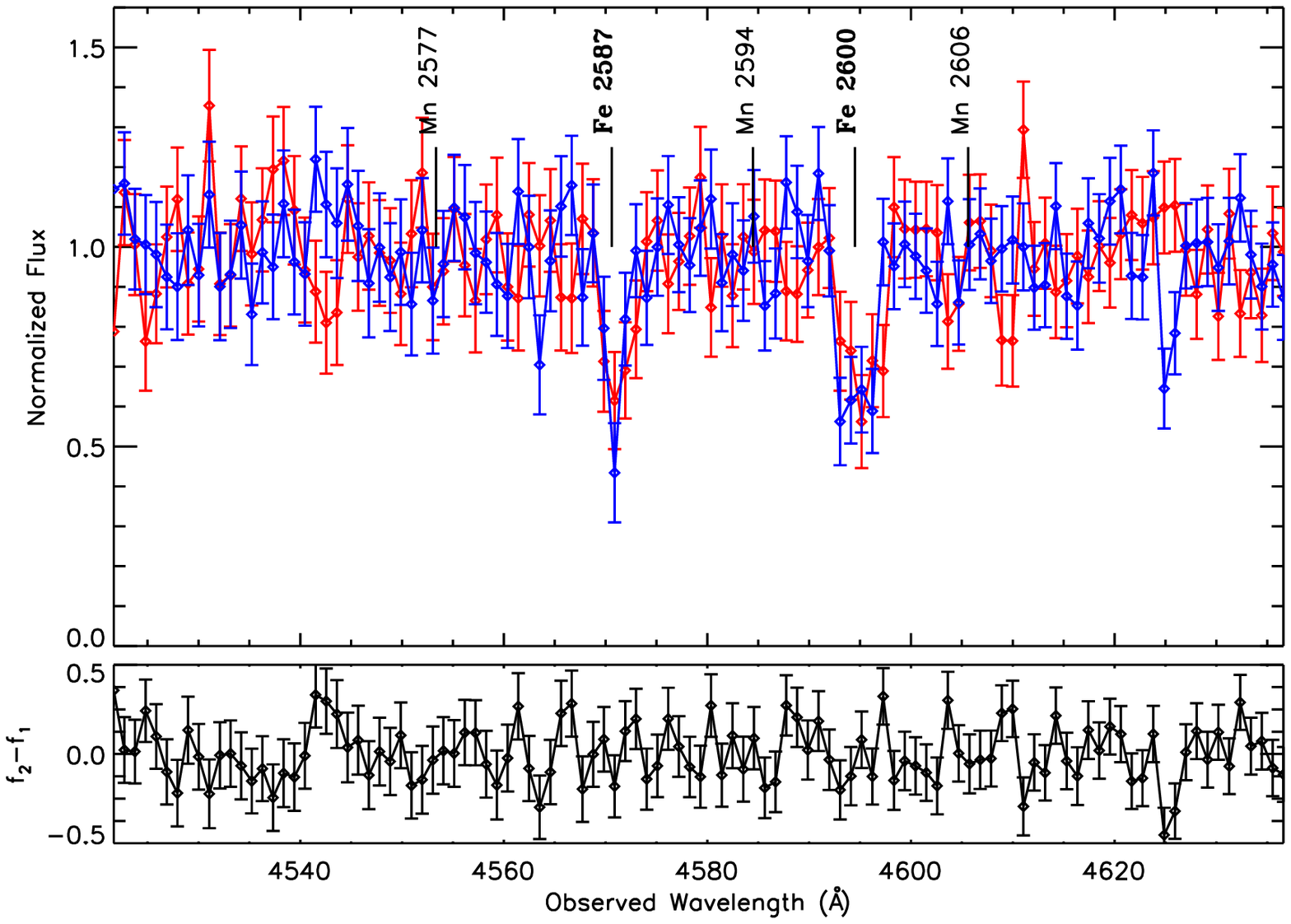}
\includegraphics[width=84mm]{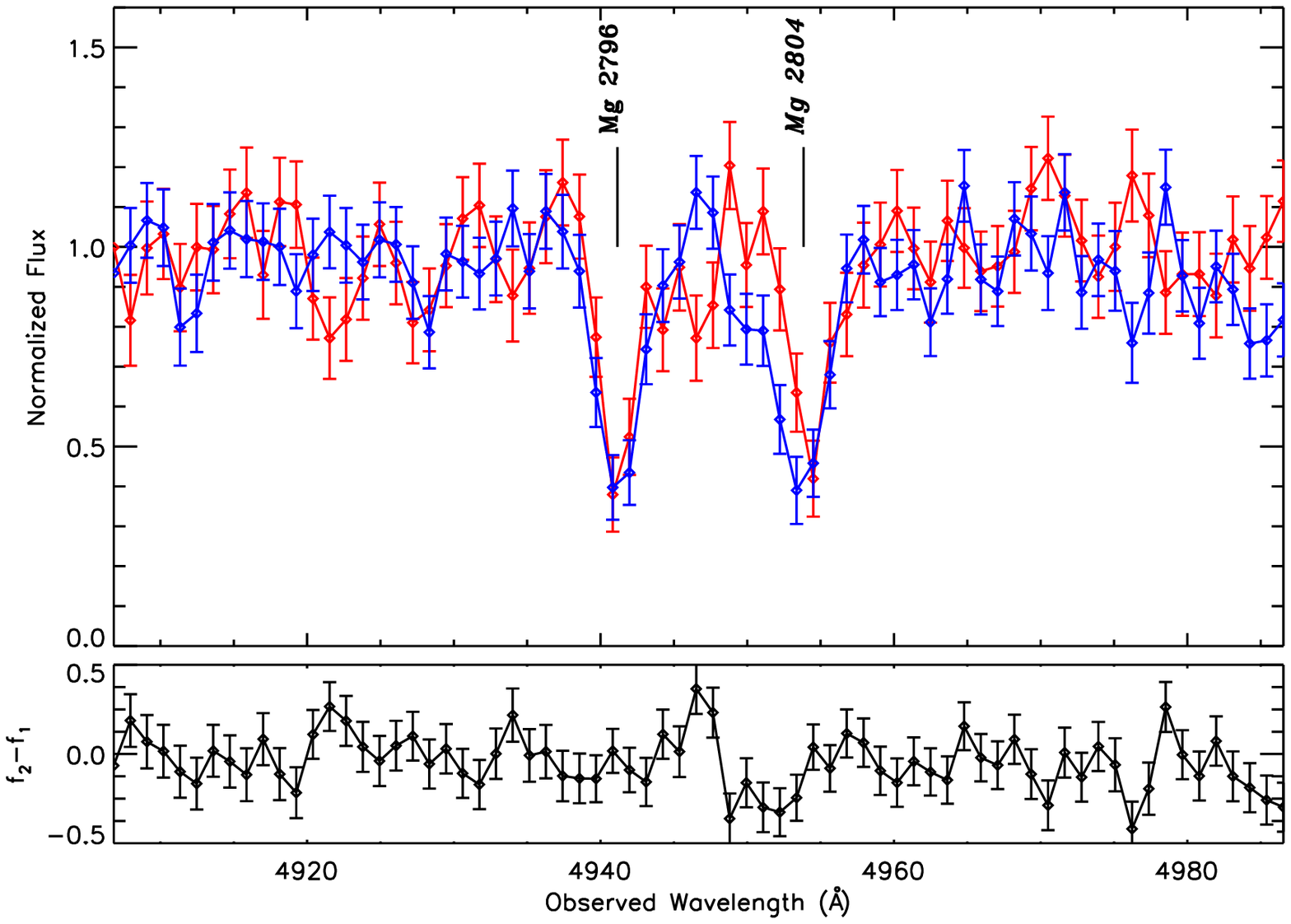}
\includegraphics[width=84mm]{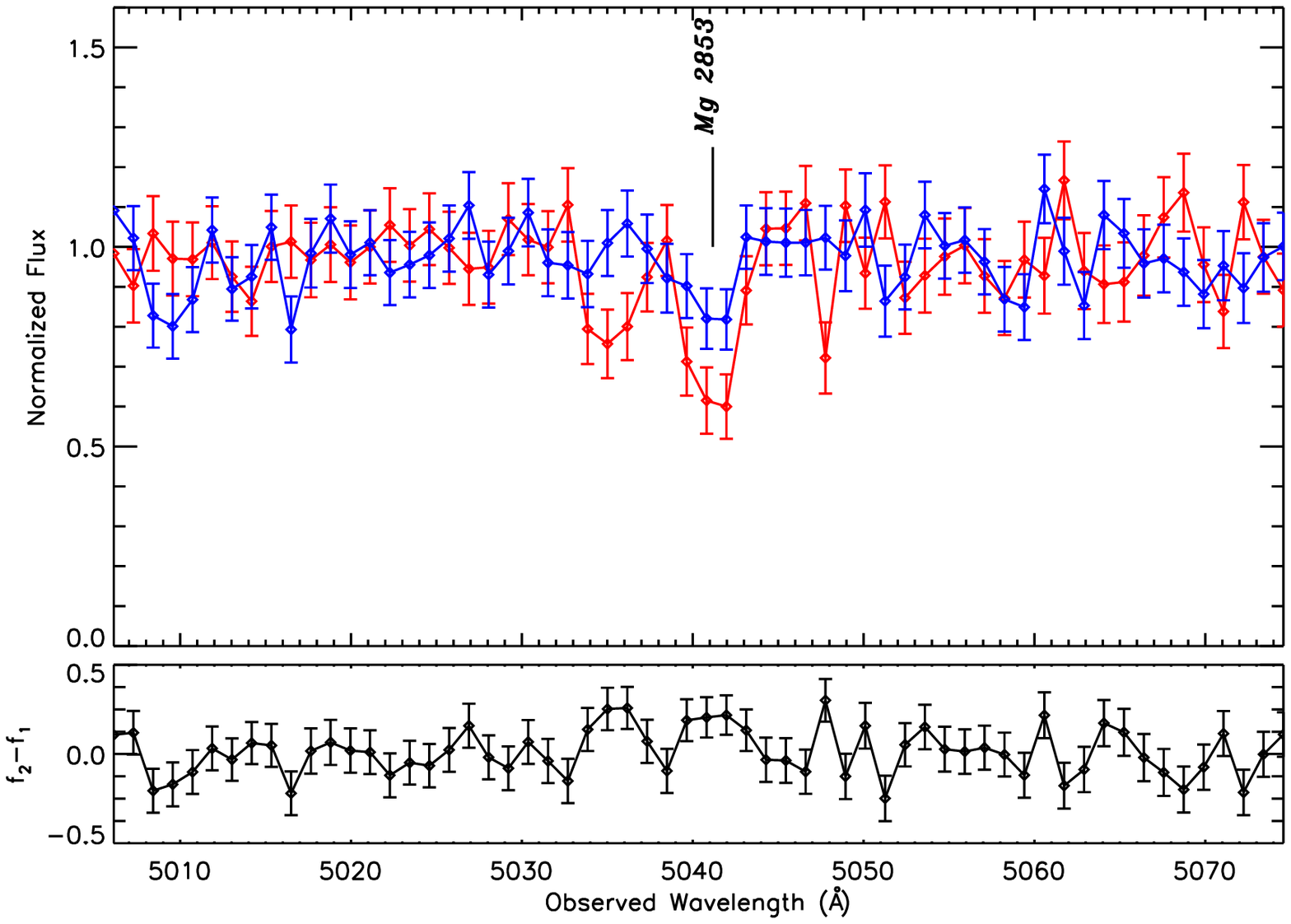}
\caption[Two-epoch normalized spectra of SDSS J025743.72+011144.5]{Two-epoch normalized spectra of the variable NAL system at $\beta$ = 0.4037 in SDSS J025743.72+011144.5.  The top panel shows the normalized pixel flux values with 1$\sigma$ error bars (first observations are red and second are blue), the bottom panel plots the difference spectrum of the two observation epochs, and shaded backgrounds identify masked pixels not included in our search for absorption line variability.  Line identifications for significantly variable absorption lines are italicised, lines detected in both observation epochs are in bold font, and undetected lines are in regular font (see Table A.1 for ion labels).  \label{figvs5}}
\end{center}
\end{figure*}

\clearpage
\begin{figure*}
\begin{center}
\includegraphics[width=84mm]{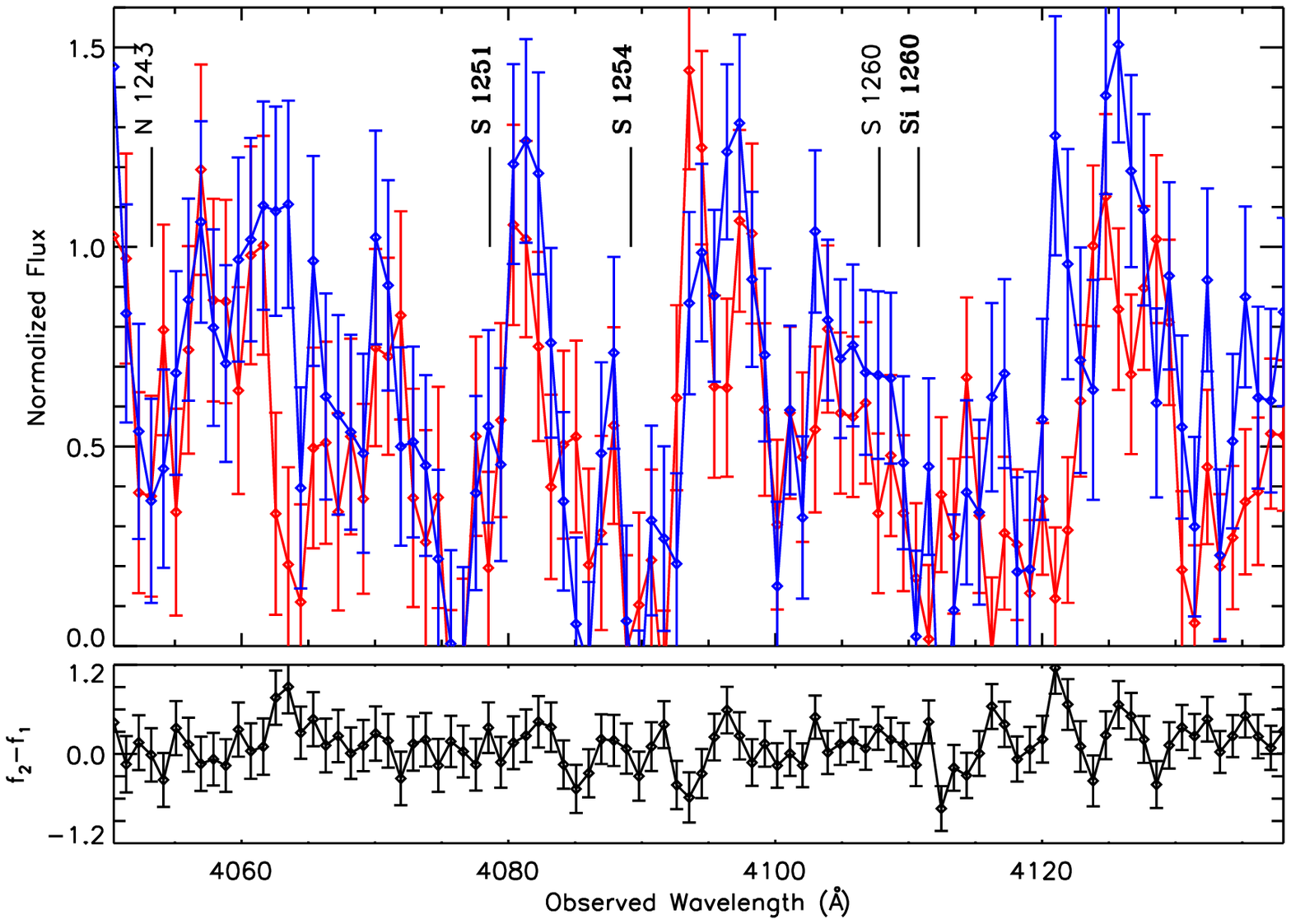}
\includegraphics[width=84mm]{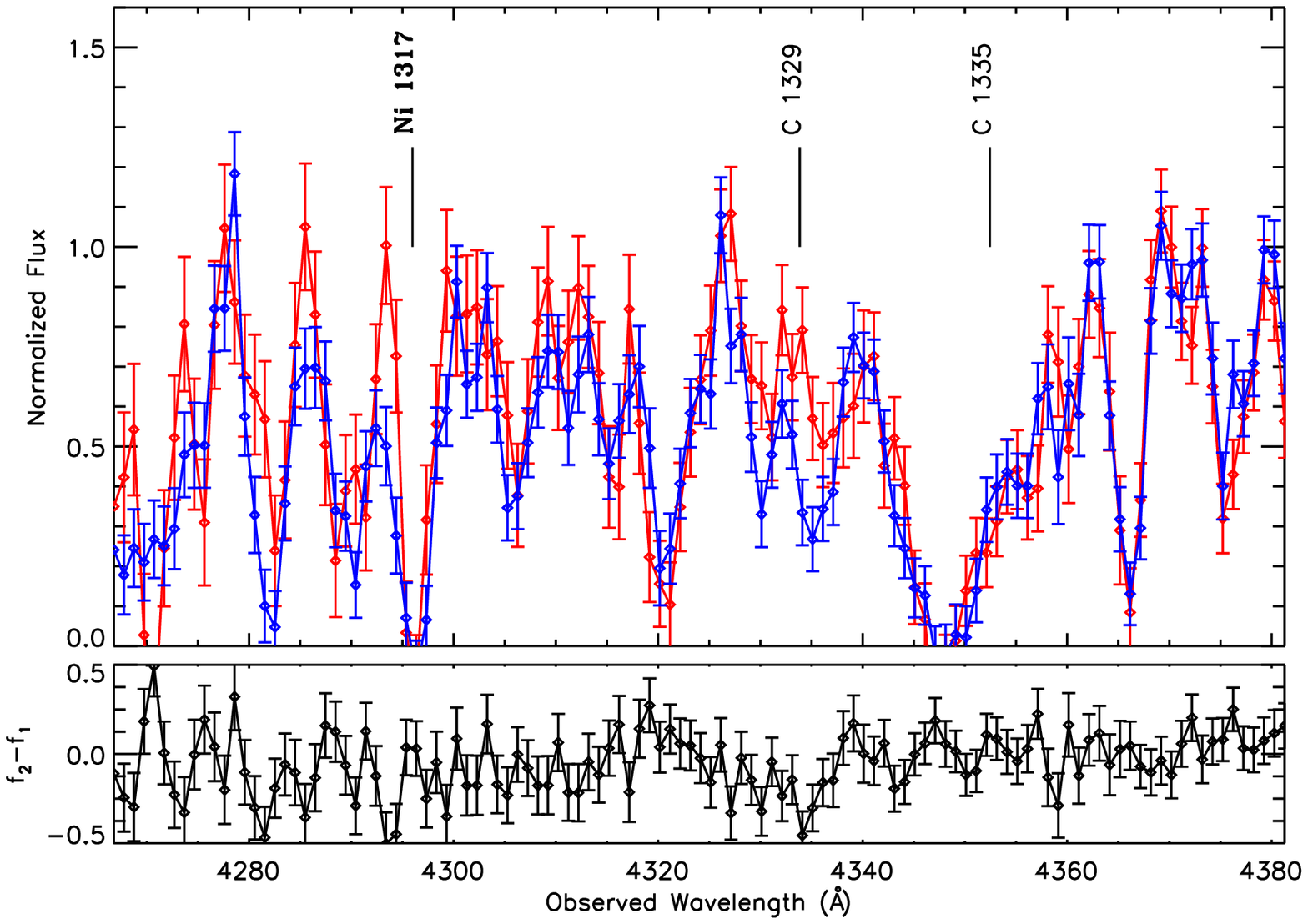}
\includegraphics[width=84mm]{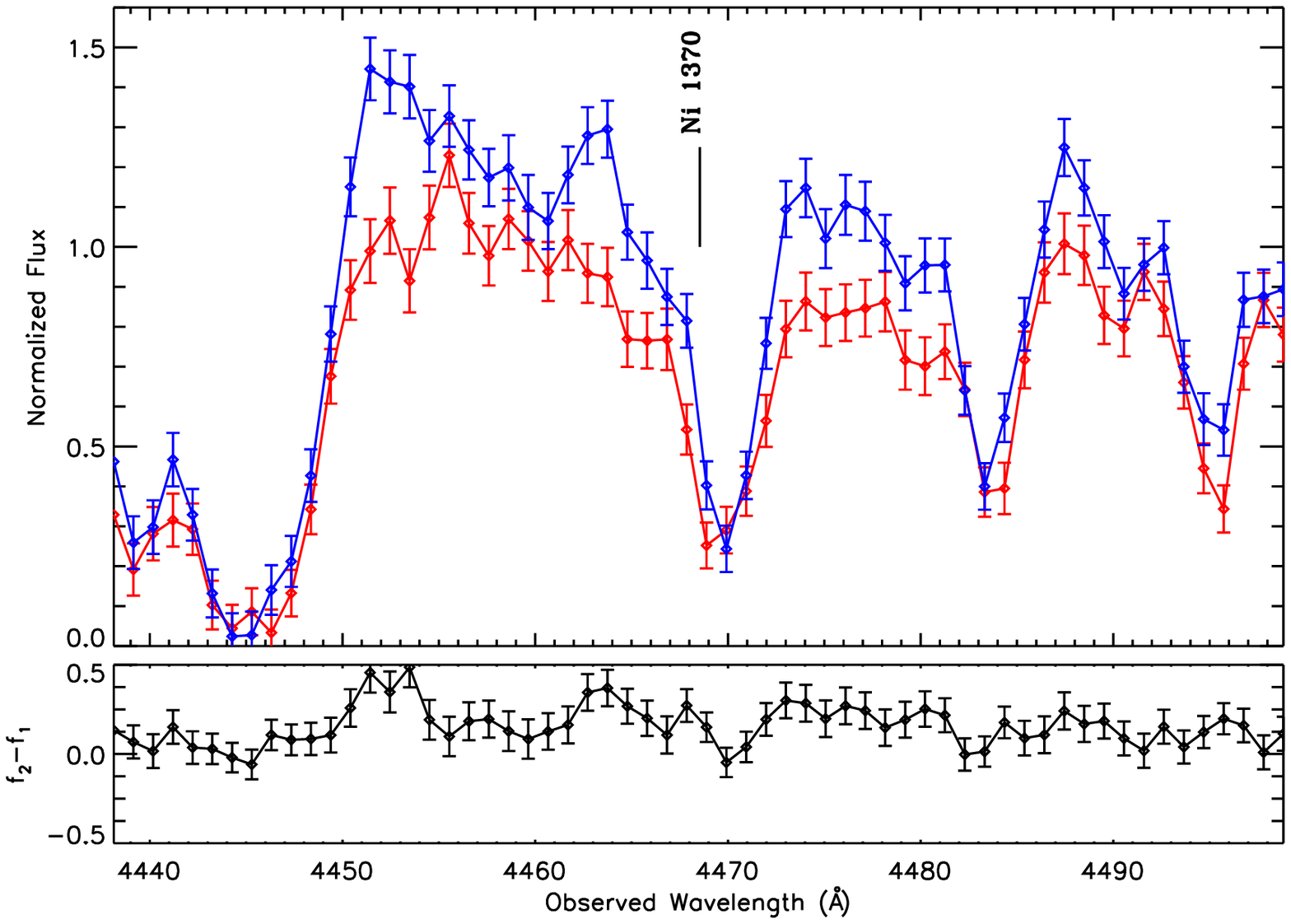}
\includegraphics[width=84mm]{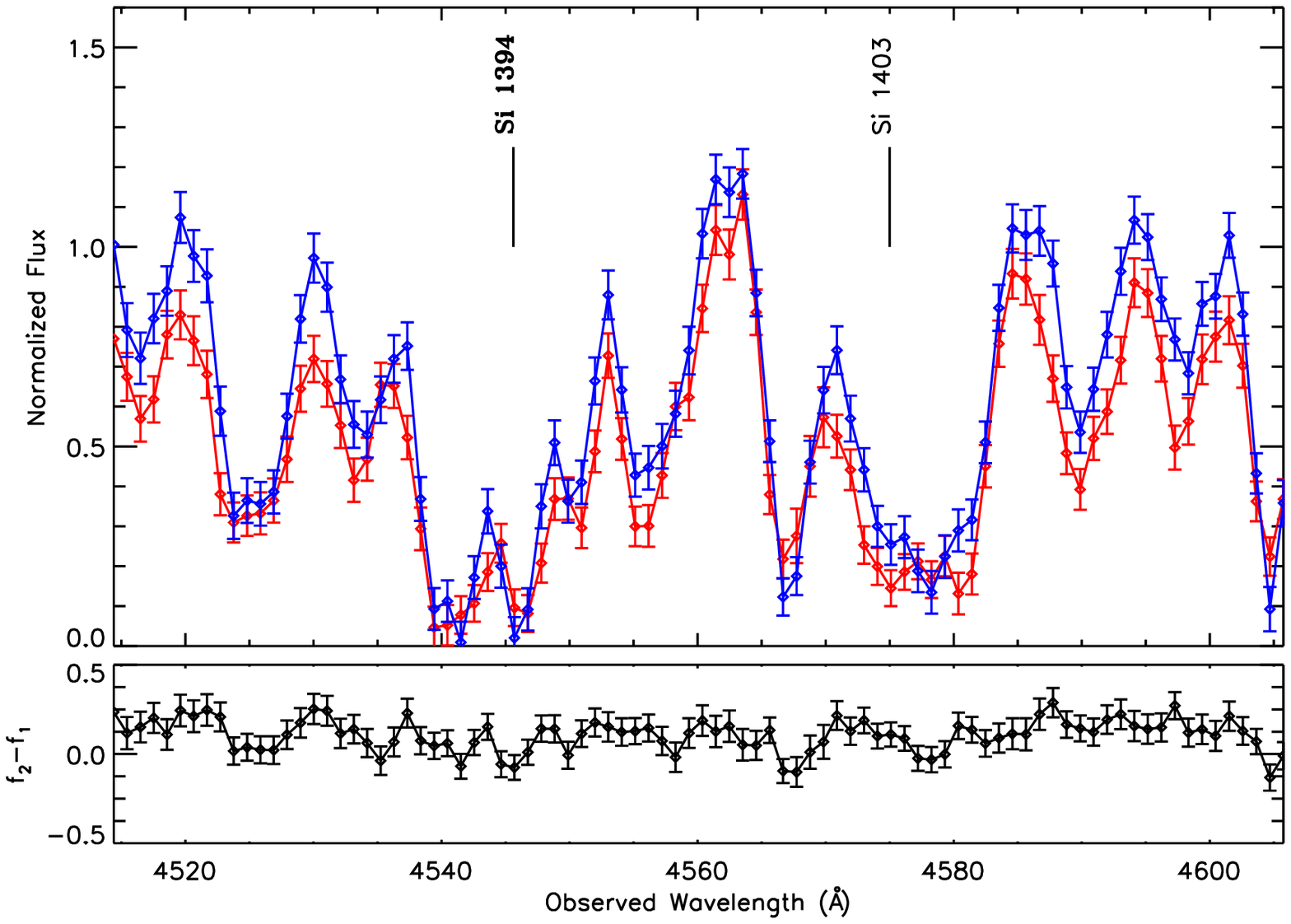}
\includegraphics[width=84mm]{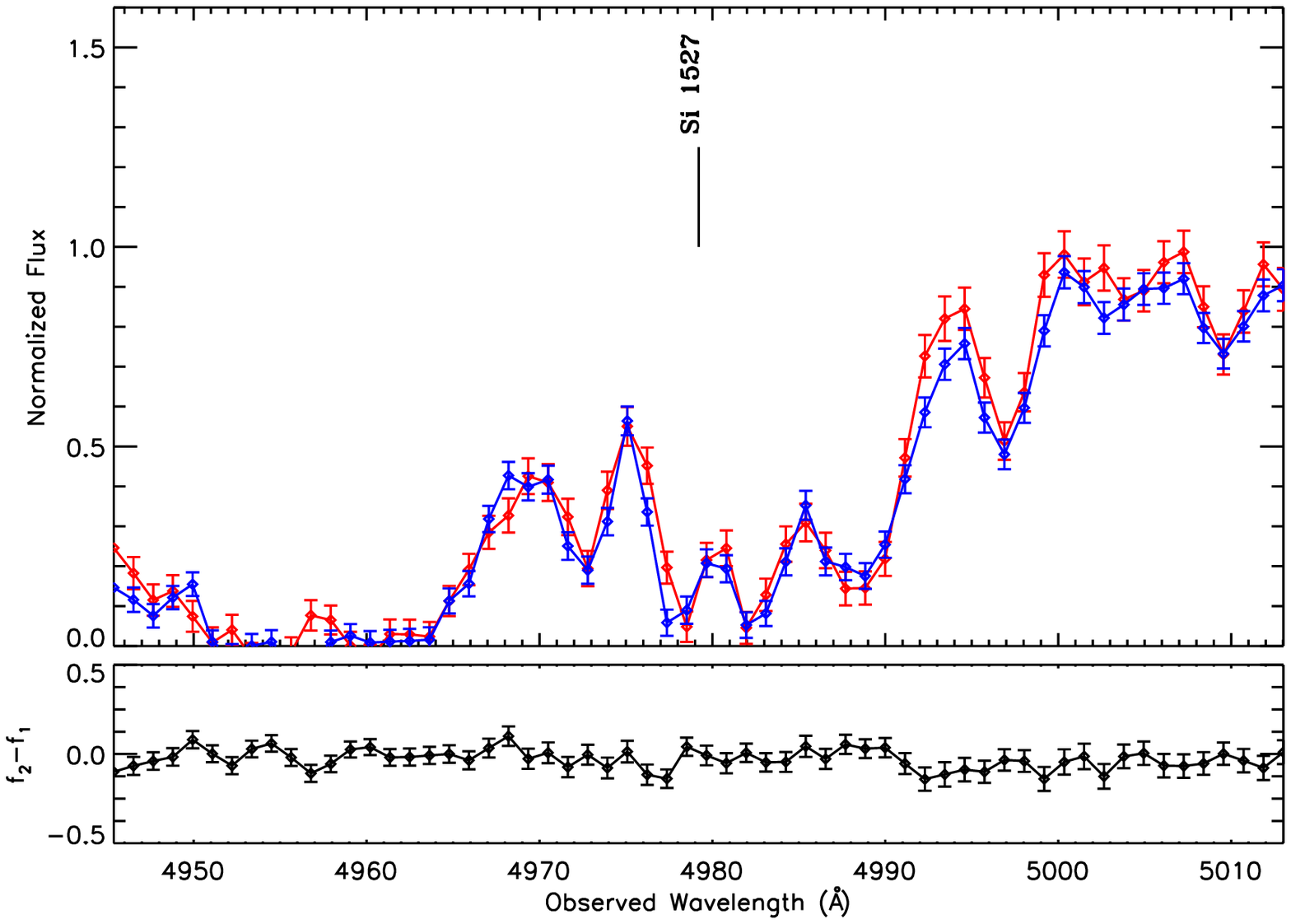}
\includegraphics[width=84mm]{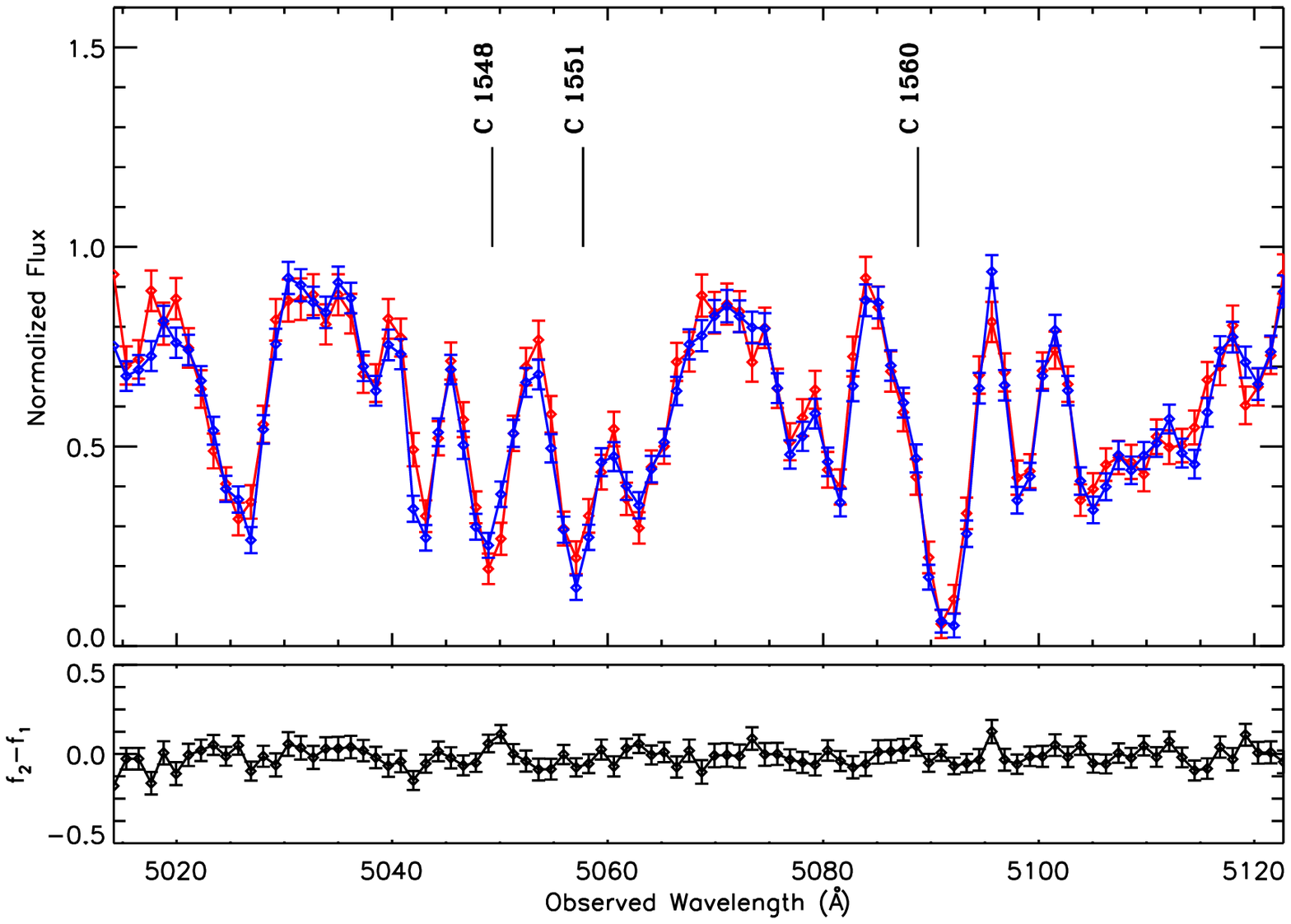}
\caption[Two-epoch normalized spectra of SDSS J012403.77+004432.6]{Two-epoch normalized spectra of the variable NAL system at $\beta$ = 0.3722 in SDSS J012403.77+004432.6.  The top panel shows the normalized pixel flux values with 1$\sigma$ error bars (first observations are red and second are blue), the bottom panel plots the difference spectrum of the two observation epochs, and shaded backgrounds identify masked pixels not included in our search for absorption line variability.  Line identifications for significantly variable absorption lines are italicised, lines detected in both observation epochs are in bold font, and undetected lines are in regular font (see Table A.1 for ion labels).  Continued in next figure.  \label{figvs6}}
\end{center}
\end{figure*}

\begin{figure*}
\ContinuedFloat
\begin{center}
\includegraphics[width=84mm]{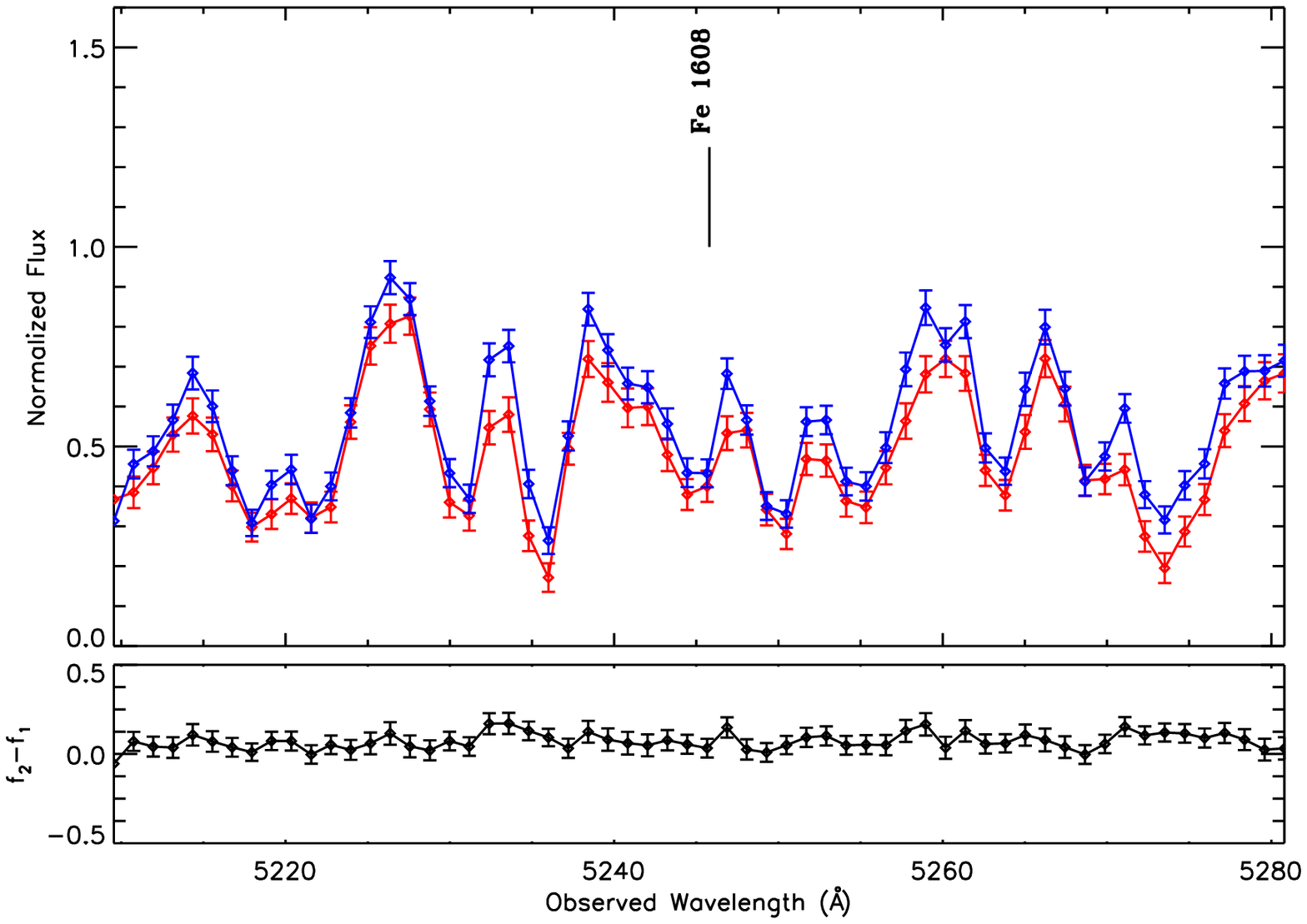}
\includegraphics[width=84mm]{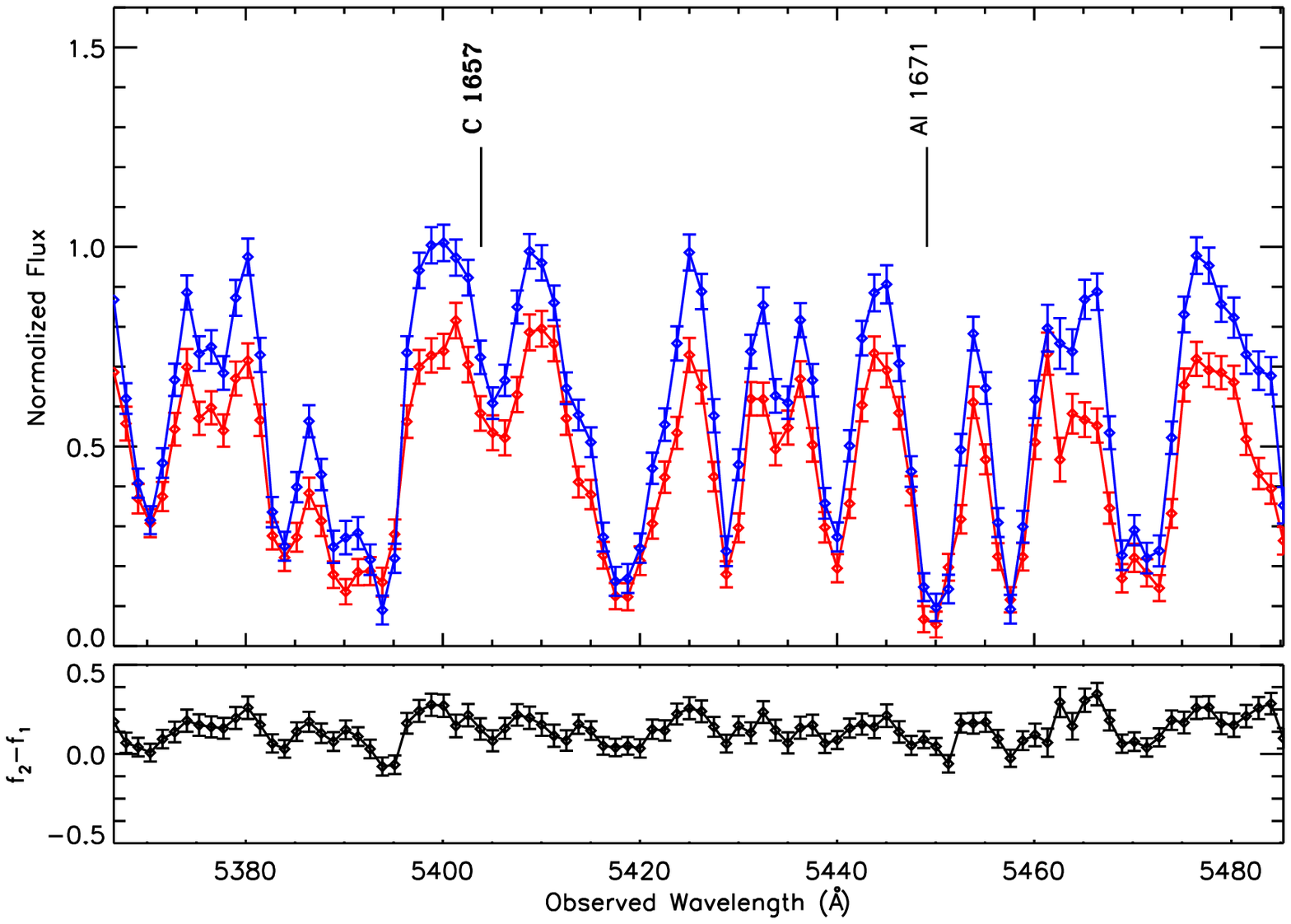}
\includegraphics[width=84mm]{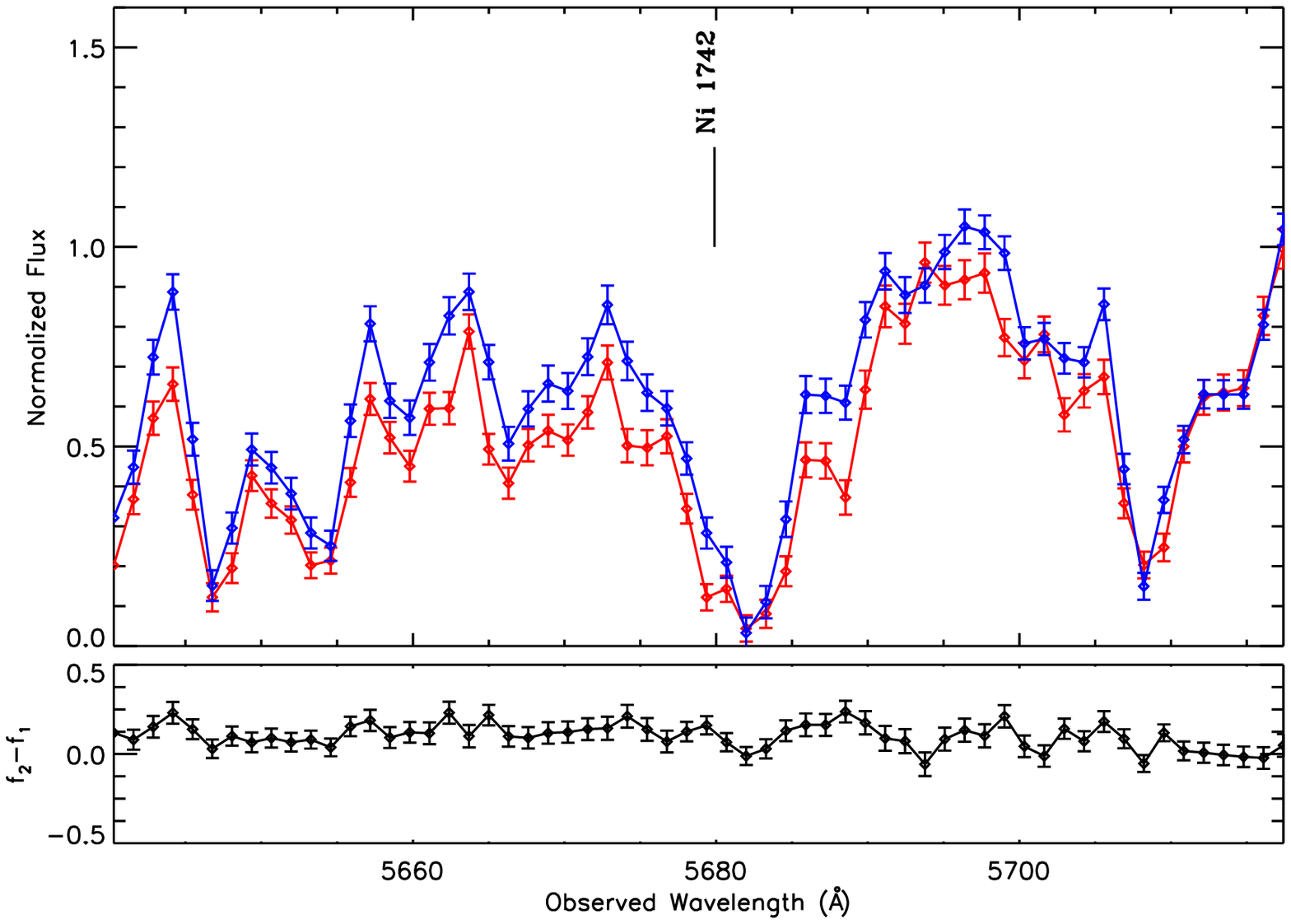}
\includegraphics[width=84mm]{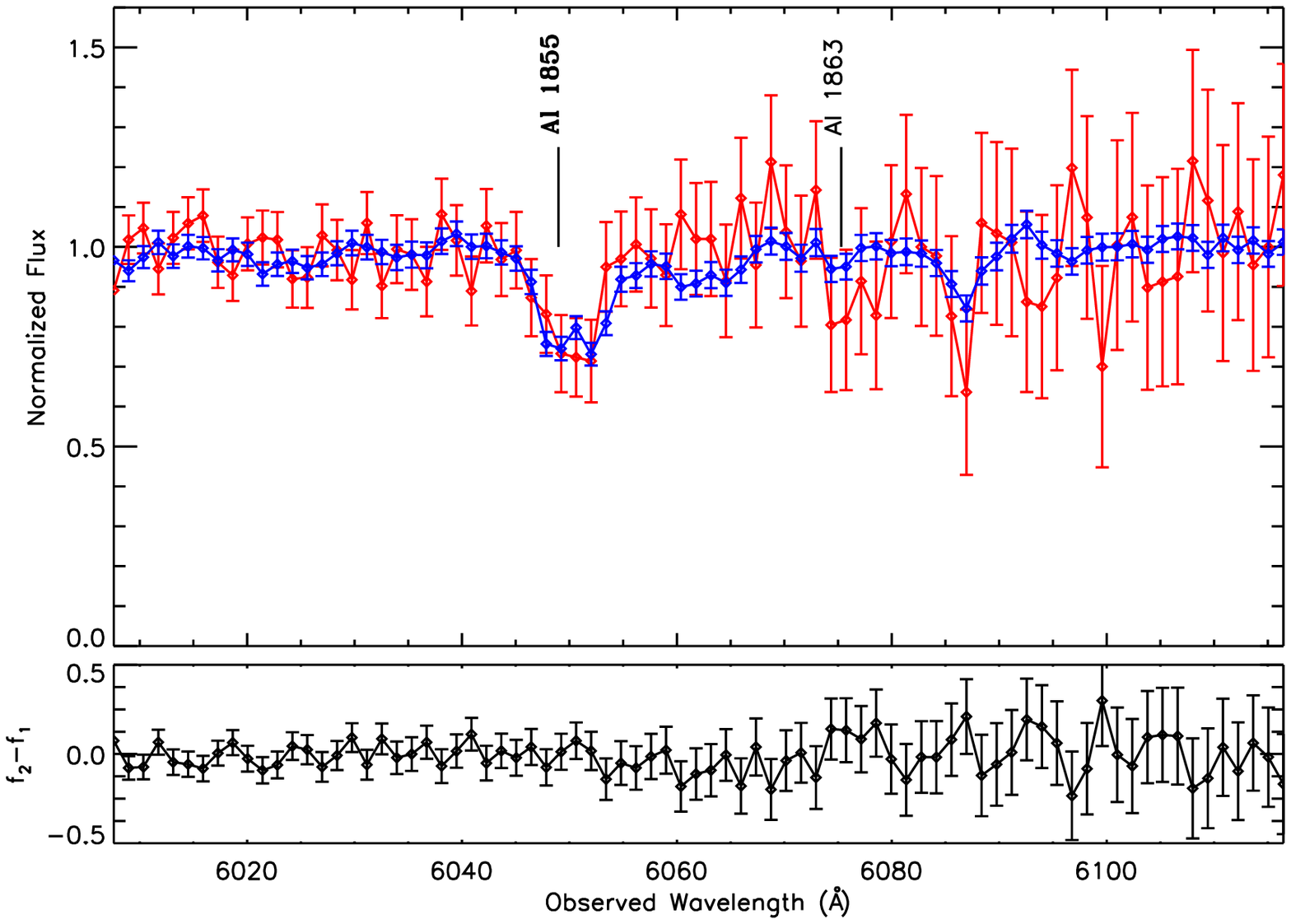}
\includegraphics[width=84mm]{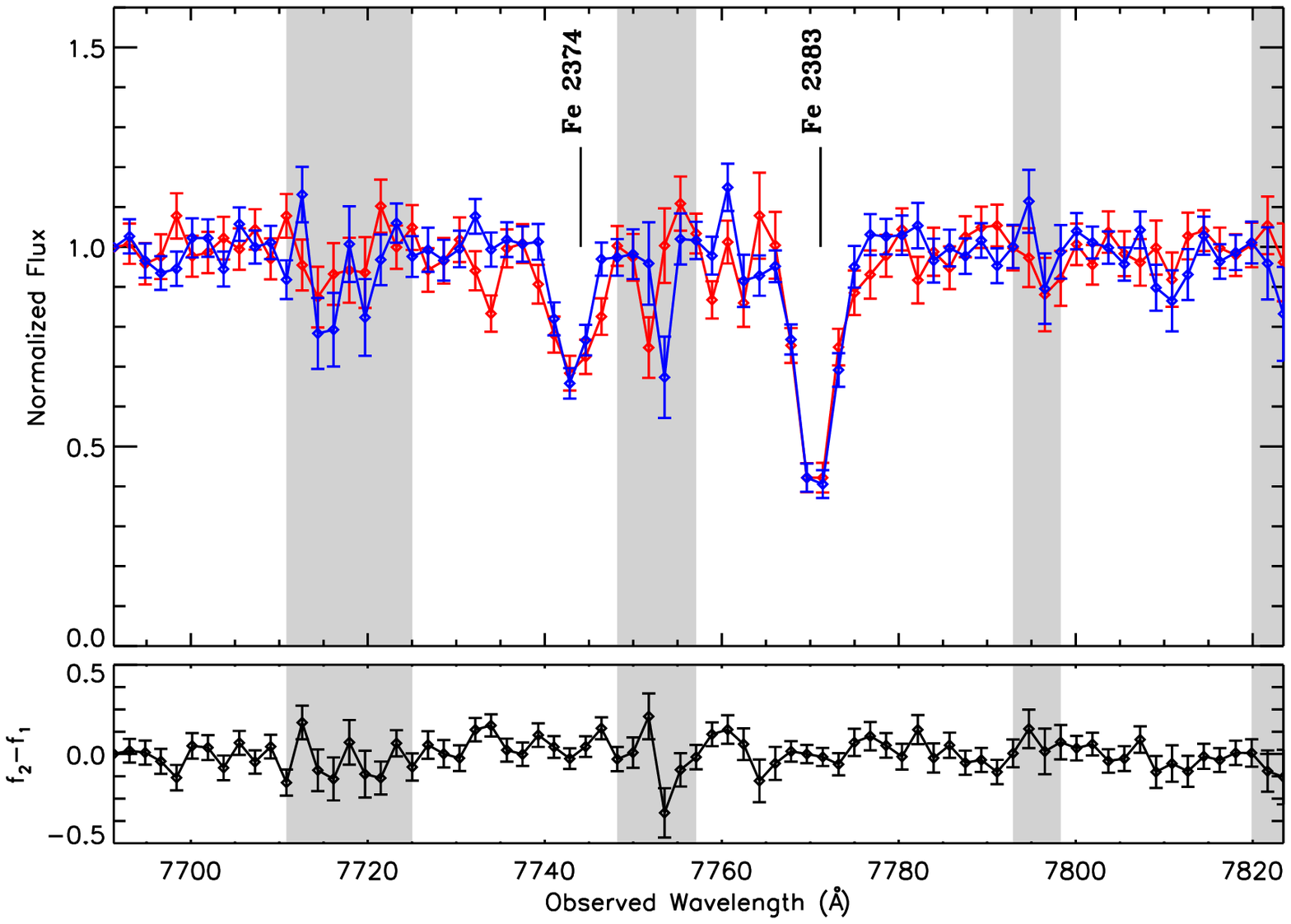}
\includegraphics[width=84mm]{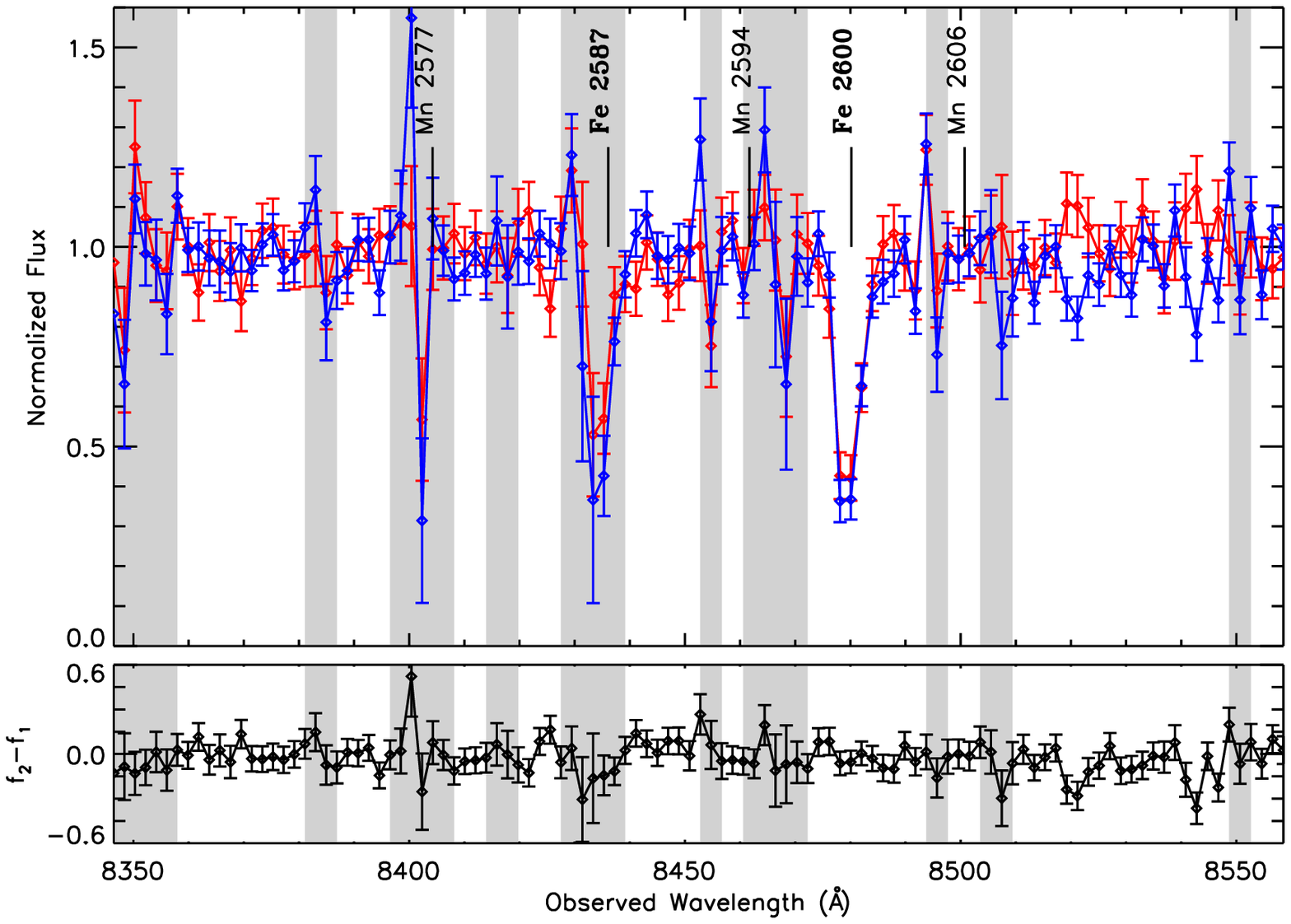}
\caption[]{Two-epoch normalized spectra of the variable NAL system at $\beta$ = 0.3722 in SDSS J012403.77+004432.6.  The top panel shows the normalized pixel flux values with 1$\sigma$ error bars (first observations are red and second are blue), the bottom panel plots the difference spectrum of the two observation epochs, and shaded backgrounds identify masked pixels not included in our search for absorption line variability.  Line identifications for significantly variable absorption lines are italicised, lines detected in both observation epochs are in bold font, and undetected lines are in regular font (see Table A.1 for ion labels).  Continued from previous figure.}
\end{center}
\end{figure*}

\begin{figure*}
\ContinuedFloat
\begin{center}
\includegraphics[width=84mm]{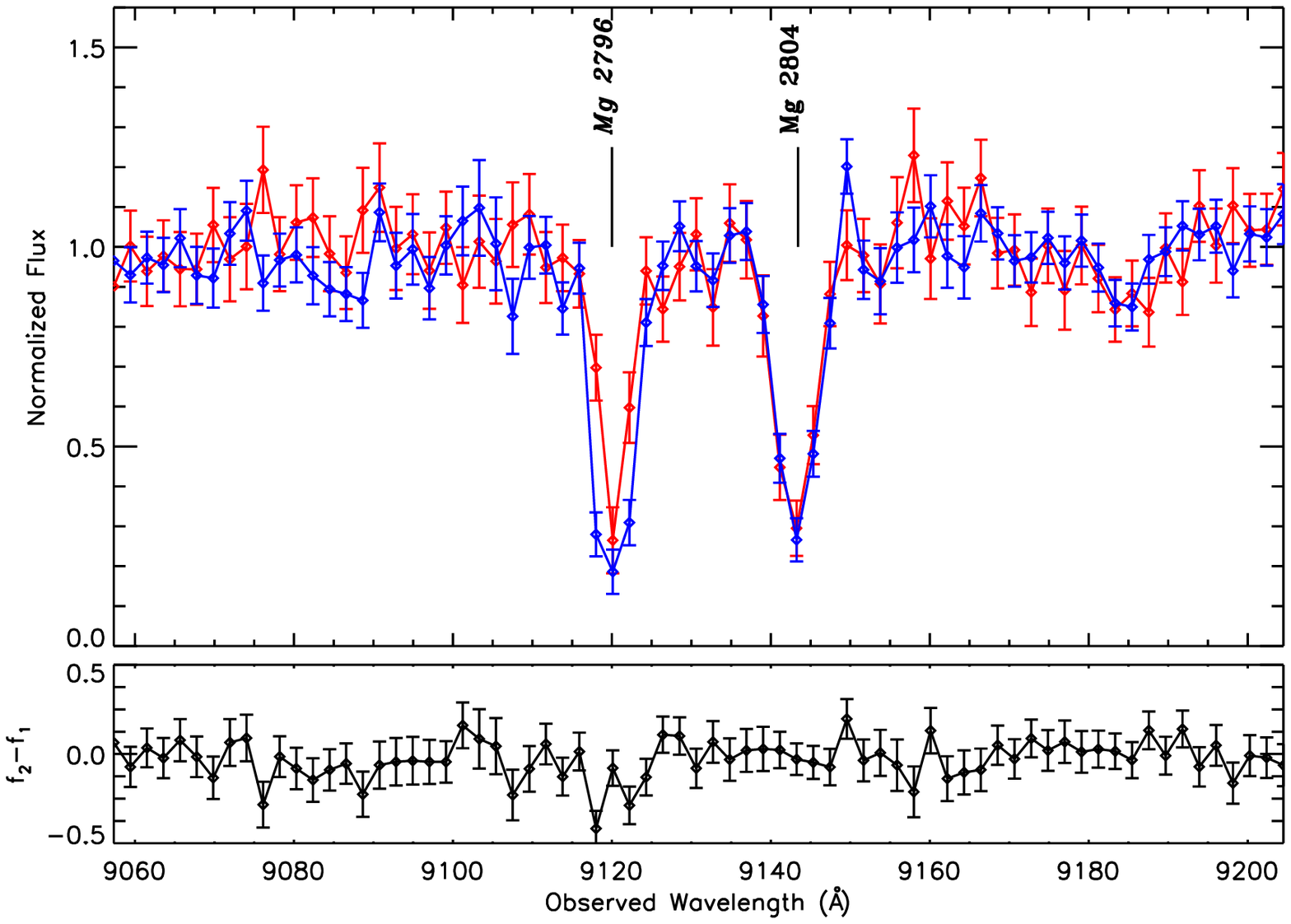}
\caption[]{Two-epoch normalized spectra of the variable NAL system at $\beta$ = 0.3722 in SDSS J012403.77+004432.6.  The top panel shows the normalized pixel flux values with 1$\sigma$ error bars (first observations are red and second are blue), the bottom panel plots the difference spectrum of the two observation epochs, and shaded backgrounds identify masked pixels not included in our search for absorption line variability.  Line identifications for significantly variable absorption lines are italicised, lines detected in both observation epochs are in bold font, and undetected lines are in regular font (see Table A.1 for ion labels).  Continued from previous figure.}
\vspace{3.5cm}
\end{center}
\end{figure*}

\clearpage
\begin{figure*}
\begin{center}
\includegraphics[width=84mm]{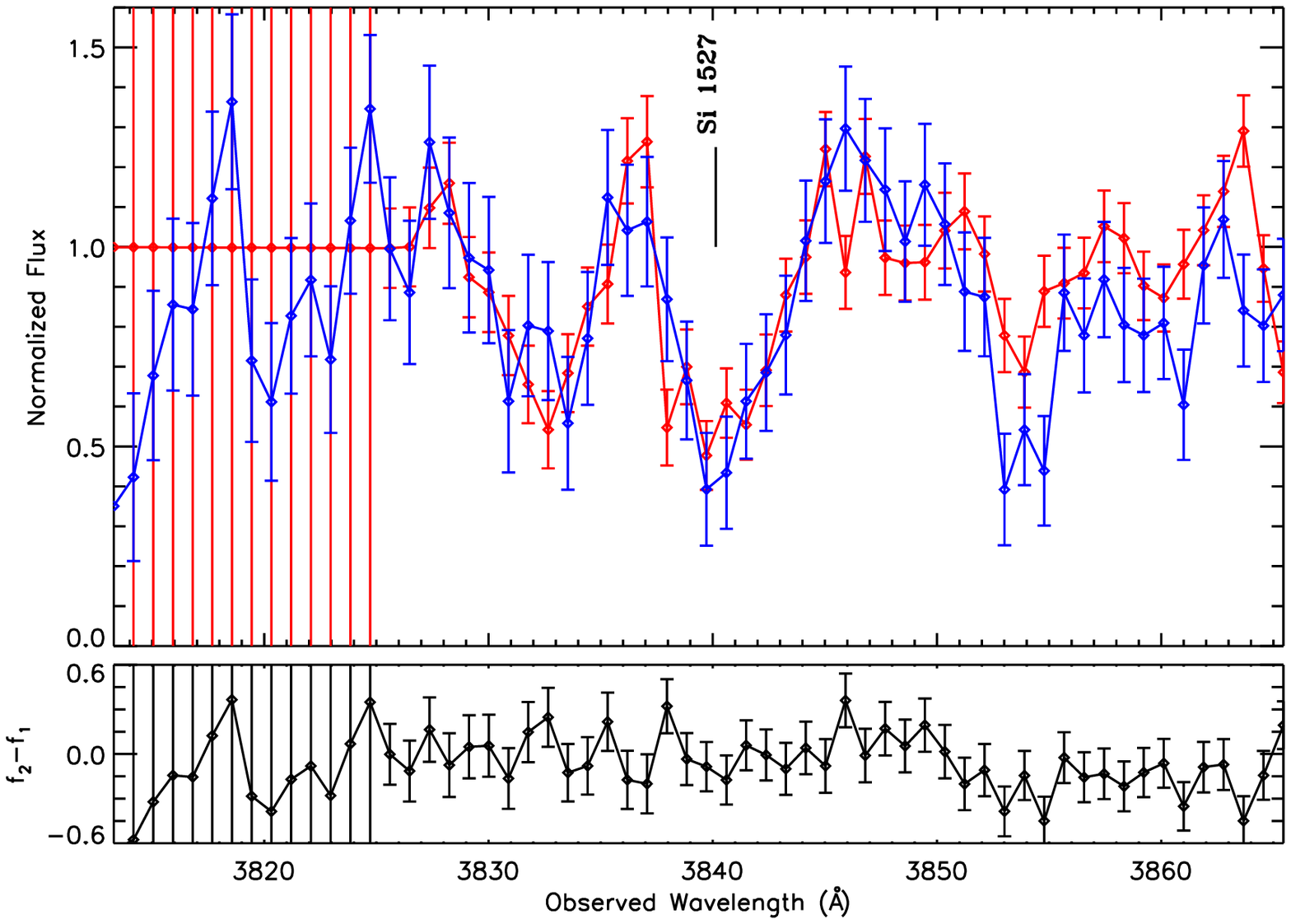}
\includegraphics[width=84mm]{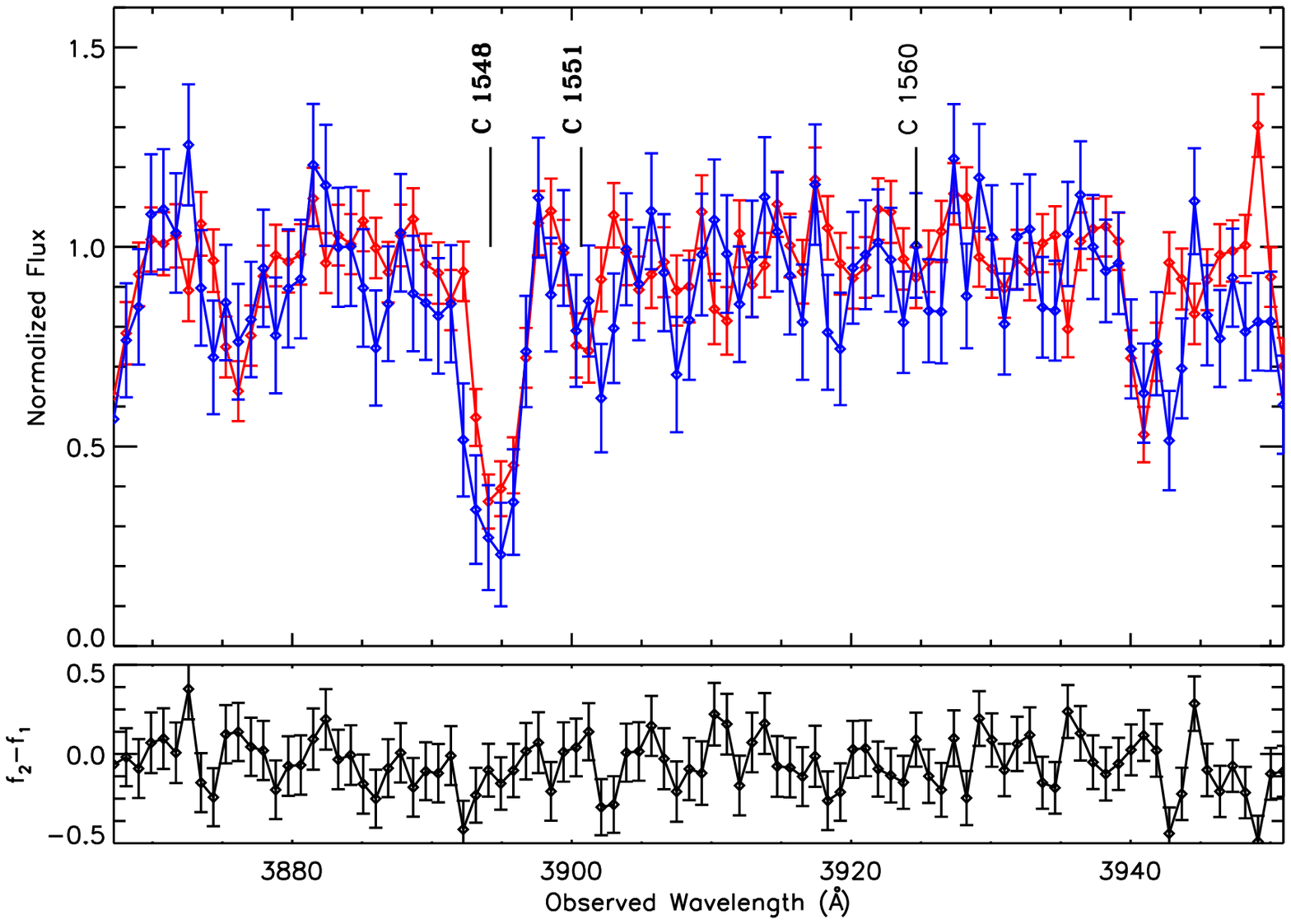}
\includegraphics[width=84mm]{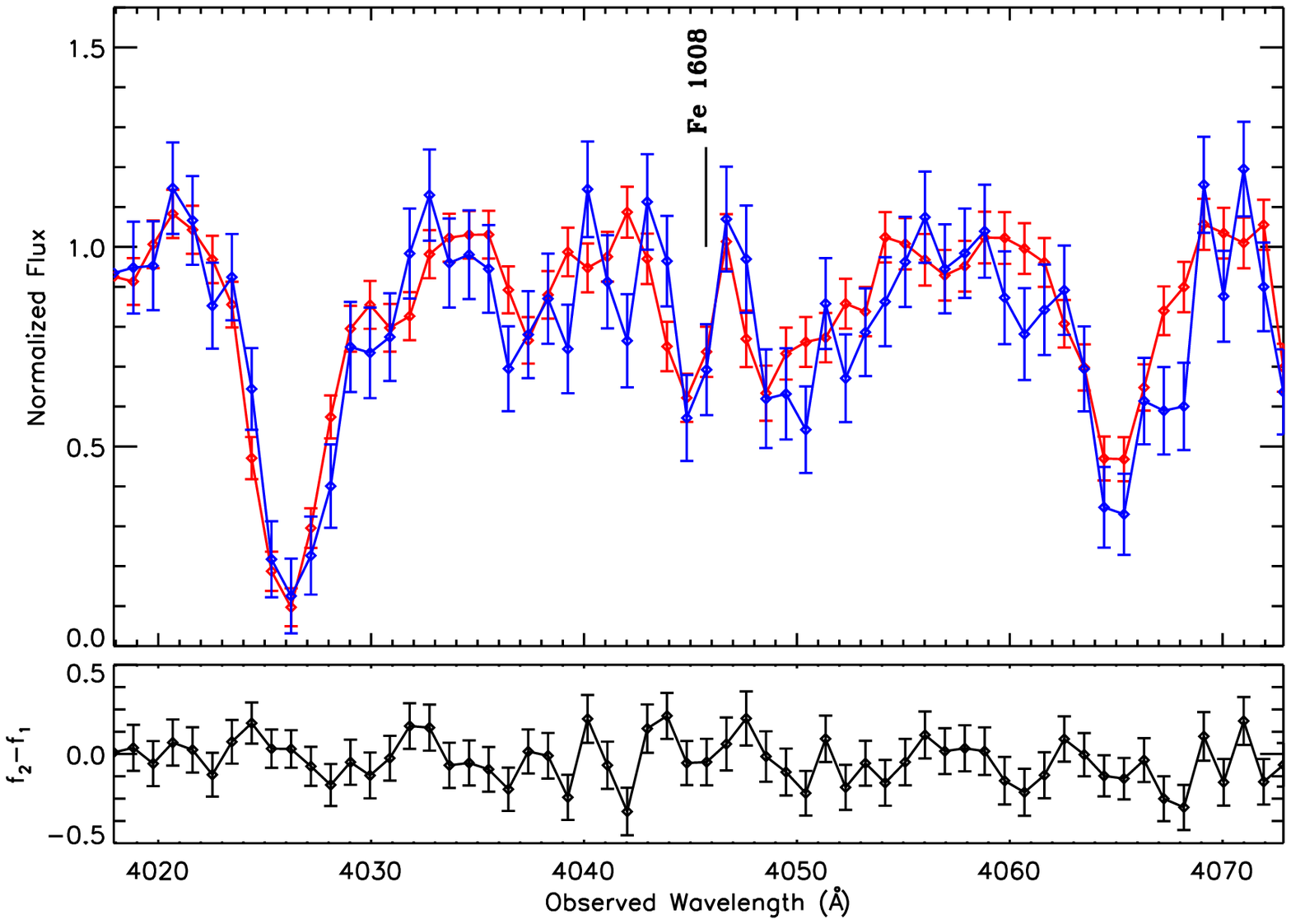}
\includegraphics[width=84mm]{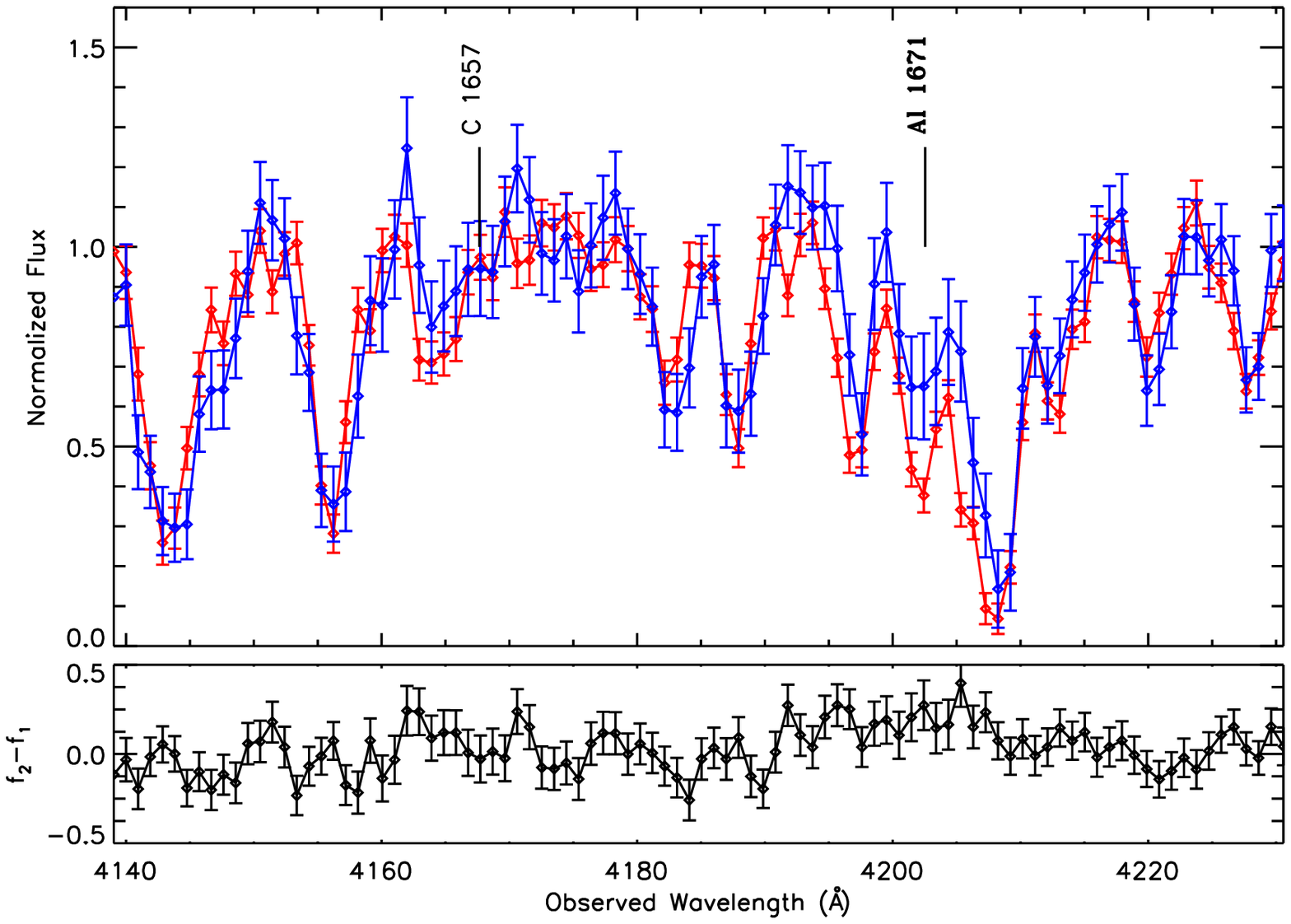}
\includegraphics[width=84mm]{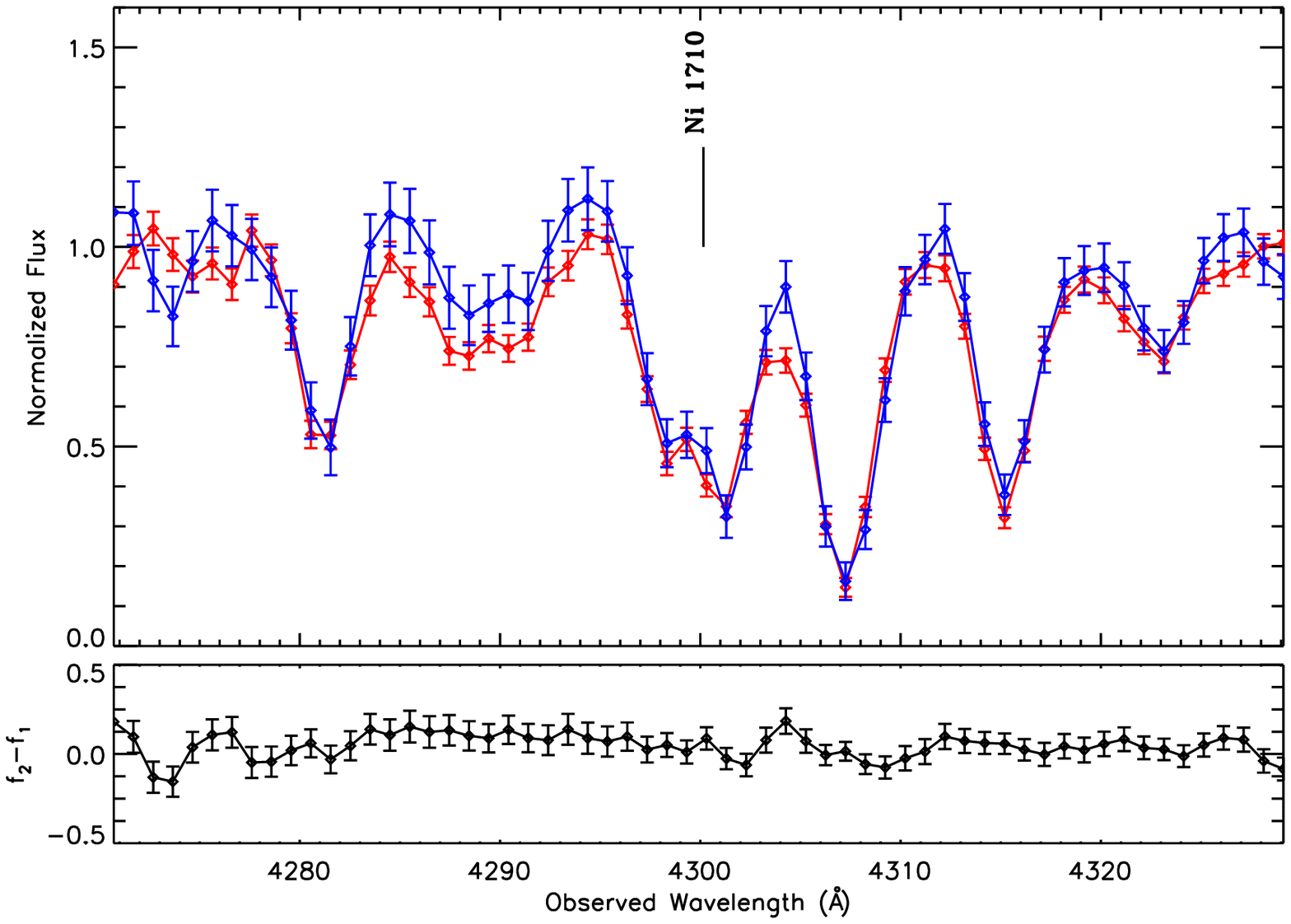}
\includegraphics[width=84mm]{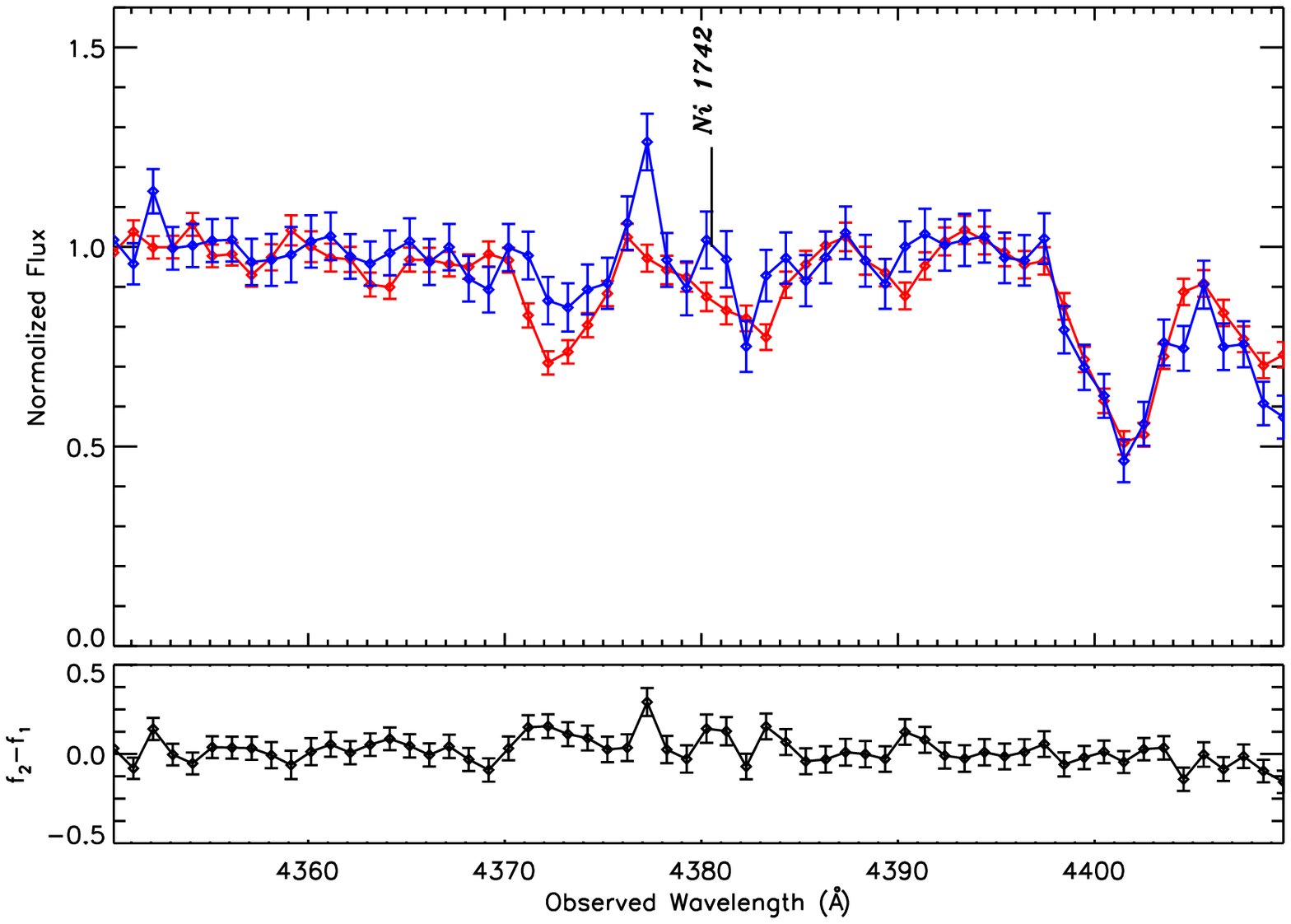}
\caption[Two-epoch normalized spectra of SDSS J092655.98+254830.5]{Two-epoch normalized spectra of the variable NAL system at $\beta$ = 0.3383 in SDSS J092655.98+254830.5.  The top panel shows the normalized pixel flux values with 1$\sigma$ error bars (first observations are red and second are blue), the bottom panel plots the difference spectrum of the two observation epochs, and shaded backgrounds identify masked pixels not included in our search for absorption line variability.  Line identifications for significantly variable absorption lines are italicised, lines detected in both observation epochs are in bold font, and undetected lines are in regular font (see Table A.1 for ion labels).  Continued in next figure.  \label{figvs7}}
\end{center}
\end{figure*}

\begin{figure*}
\ContinuedFloat
\begin{center}
\includegraphics[width=84mm]{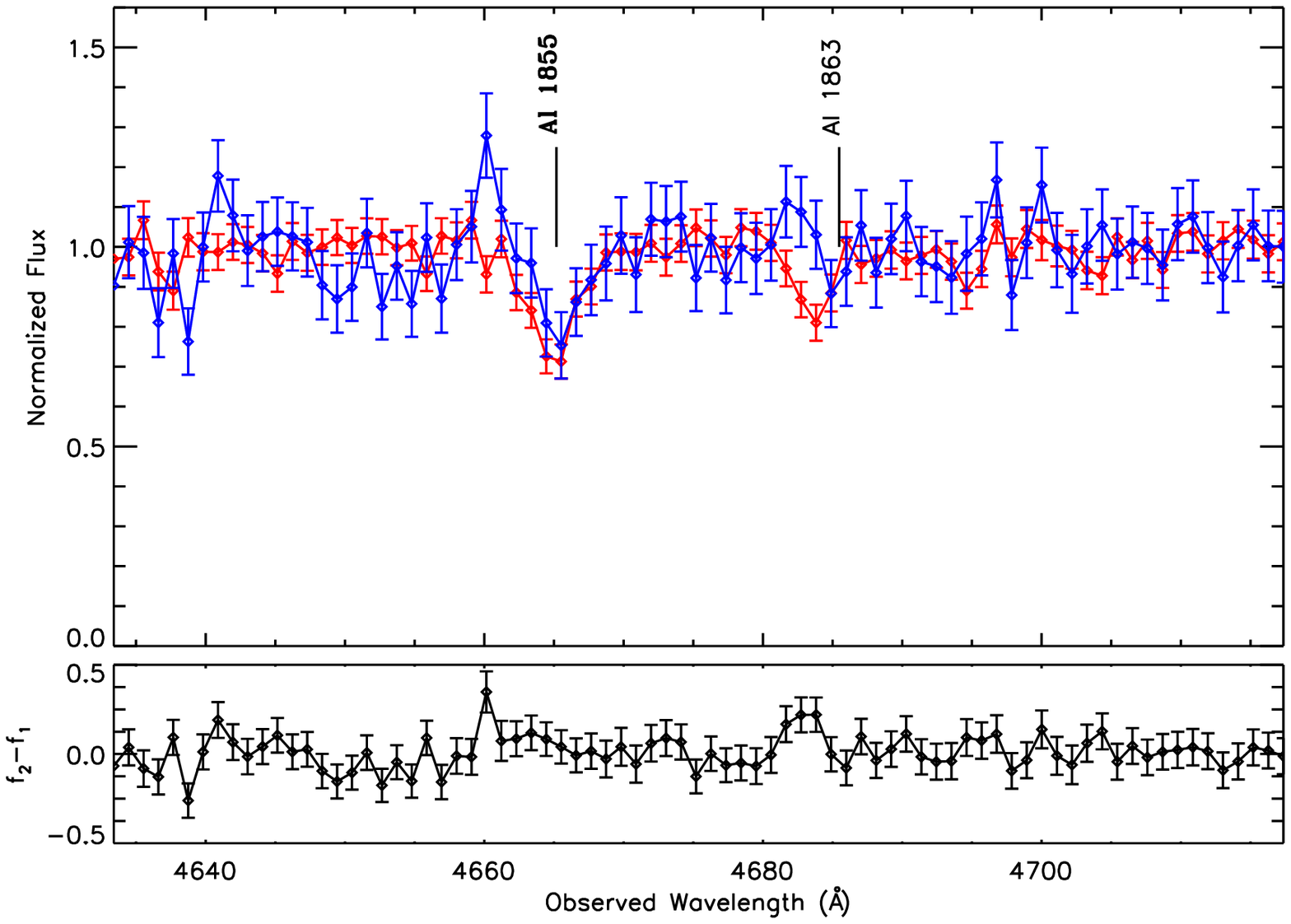}
\includegraphics[width=84mm]{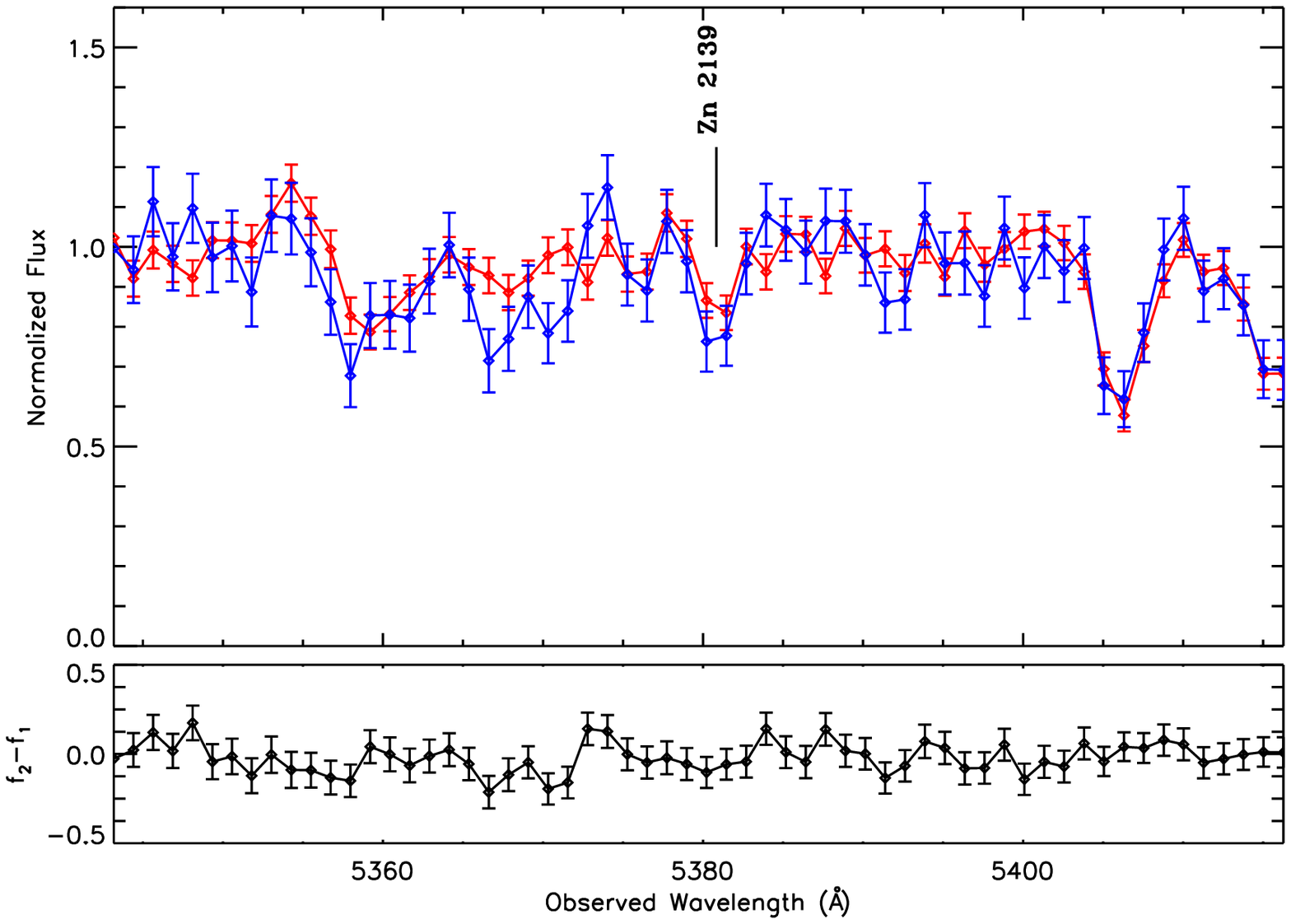}
\includegraphics[width=84mm]{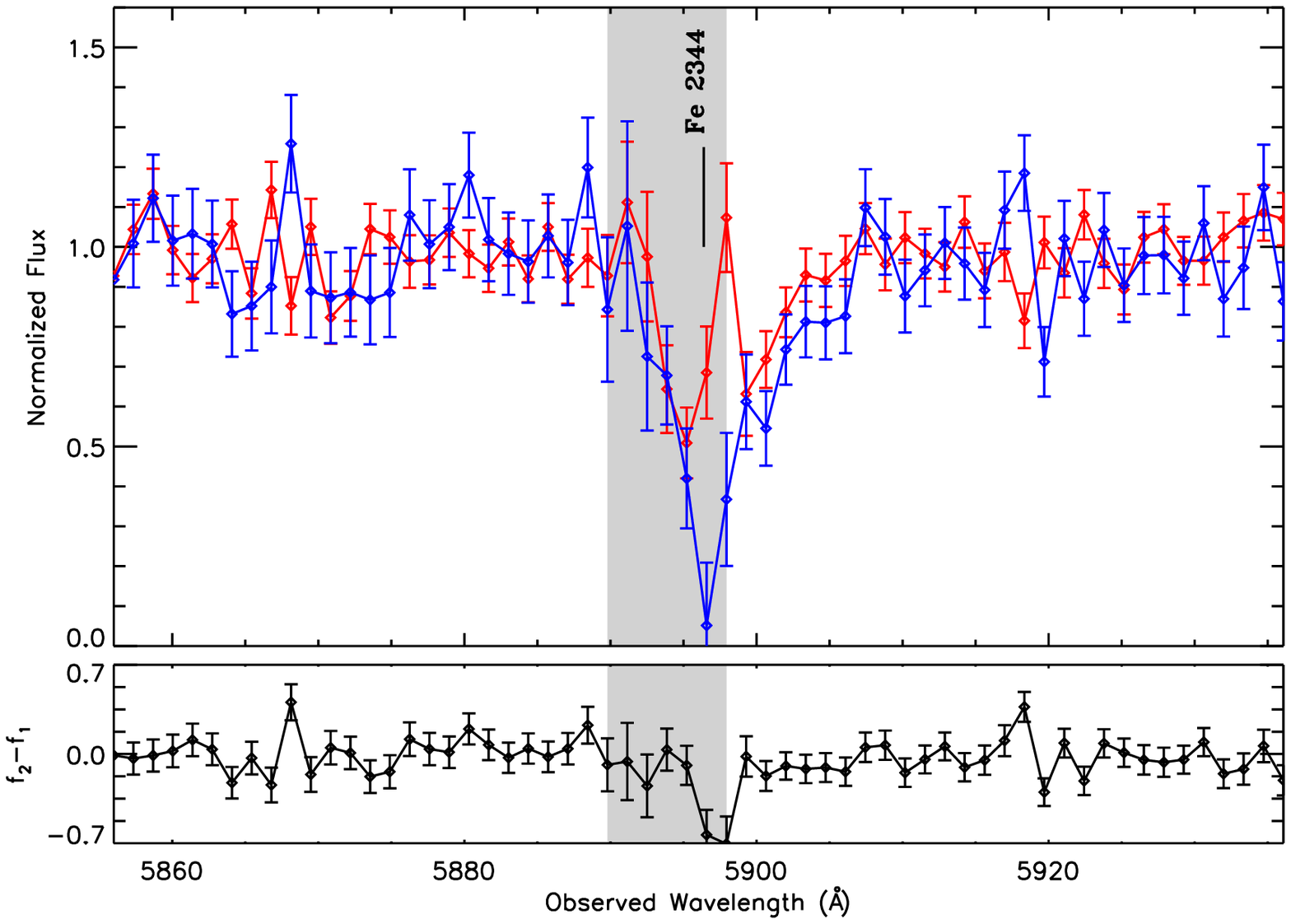}
\includegraphics[width=84mm]{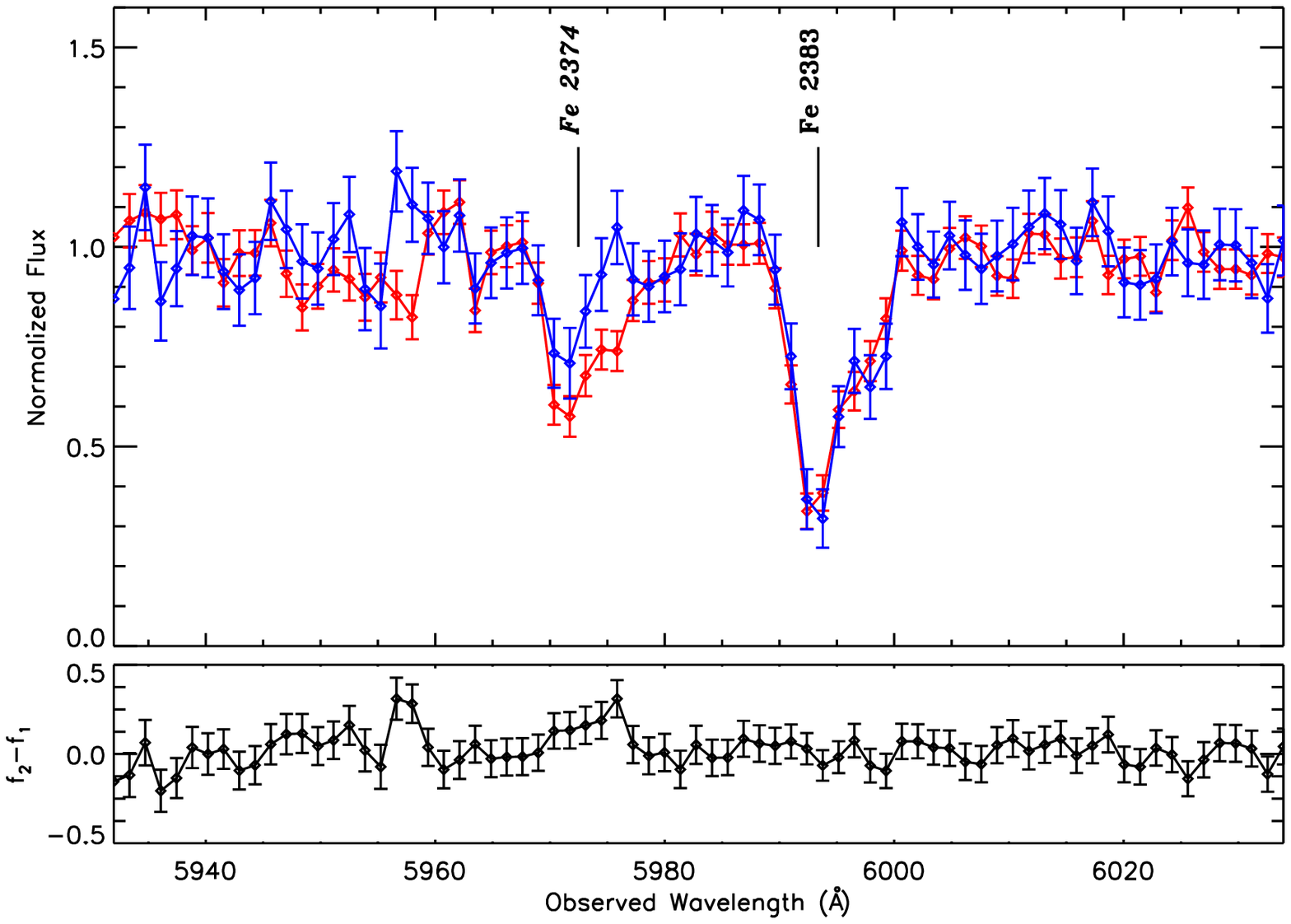}
\includegraphics[width=84mm]{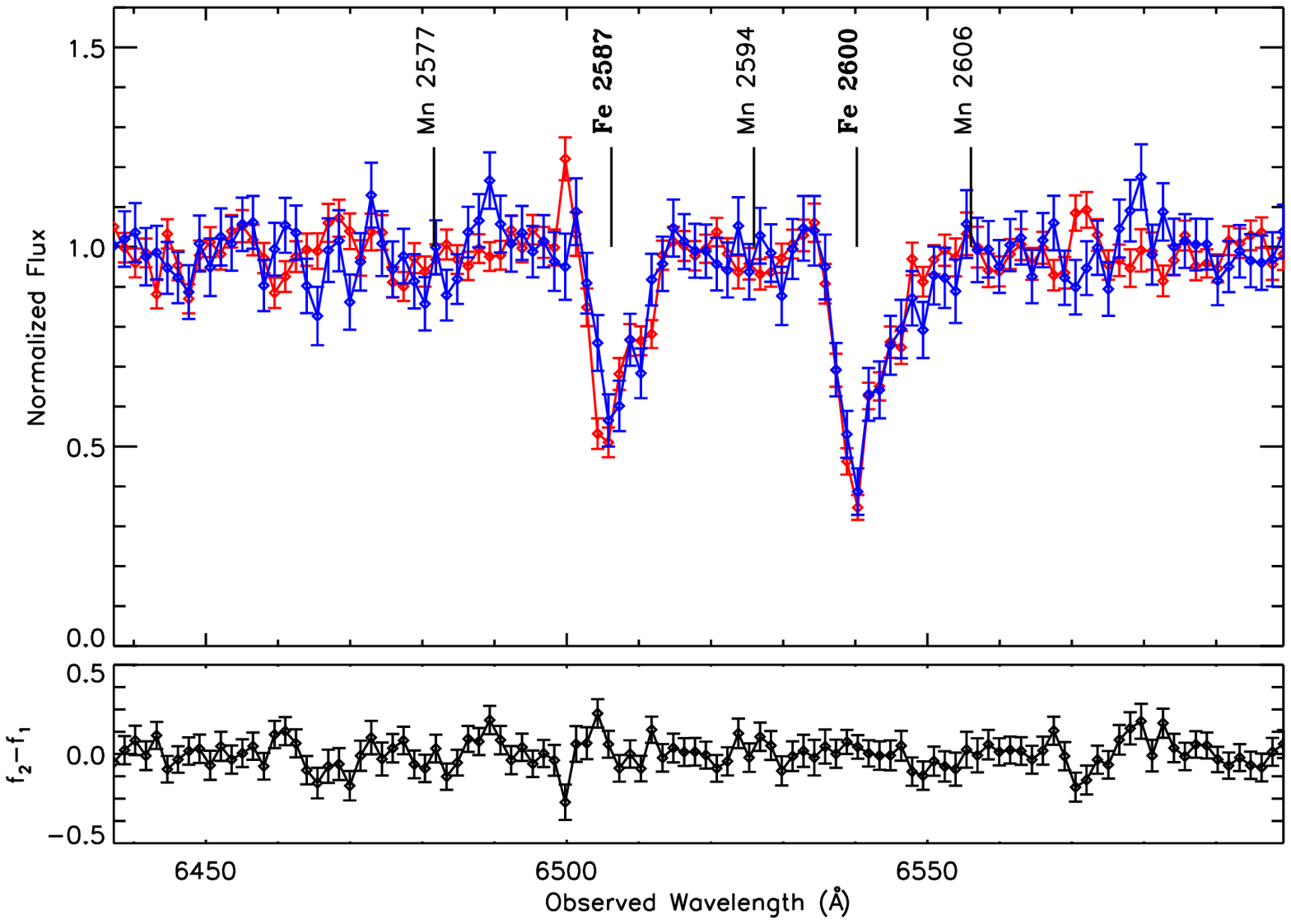}
\includegraphics[width=84mm]{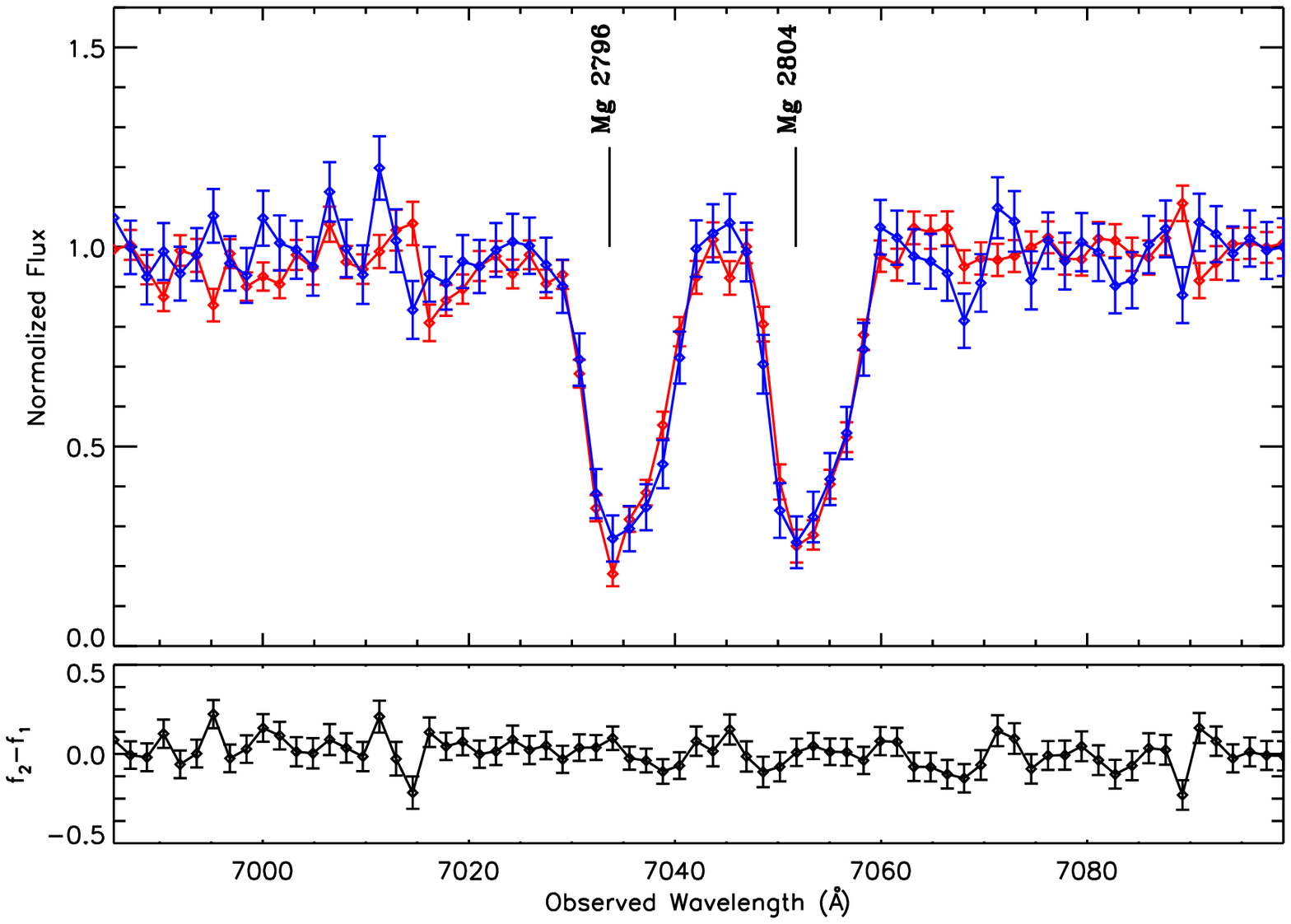}
\caption[]{Two-epoch normalized spectra of the variable NAL system at $\beta$ = 0.3383 in SDSS J092655.98+254830.5.  The top panel shows the normalized pixel flux values with 1$\sigma$ error bars (first observations are red and second are blue), the bottom panel plots the difference spectrum of the two observation epochs, and shaded backgrounds identify masked pixels not included in our search for absorption line variability.  Line identifications for significantly variable absorption lines are italicised, lines detected in both observation epochs are in bold font, and undetected lines are in regular font (see Table A.1 for ion labels).  Continued from previous figure.}
\end{center}
\end{figure*}

\clearpage
\begin{figure*}
\begin{center}
\includegraphics[width=84mm]{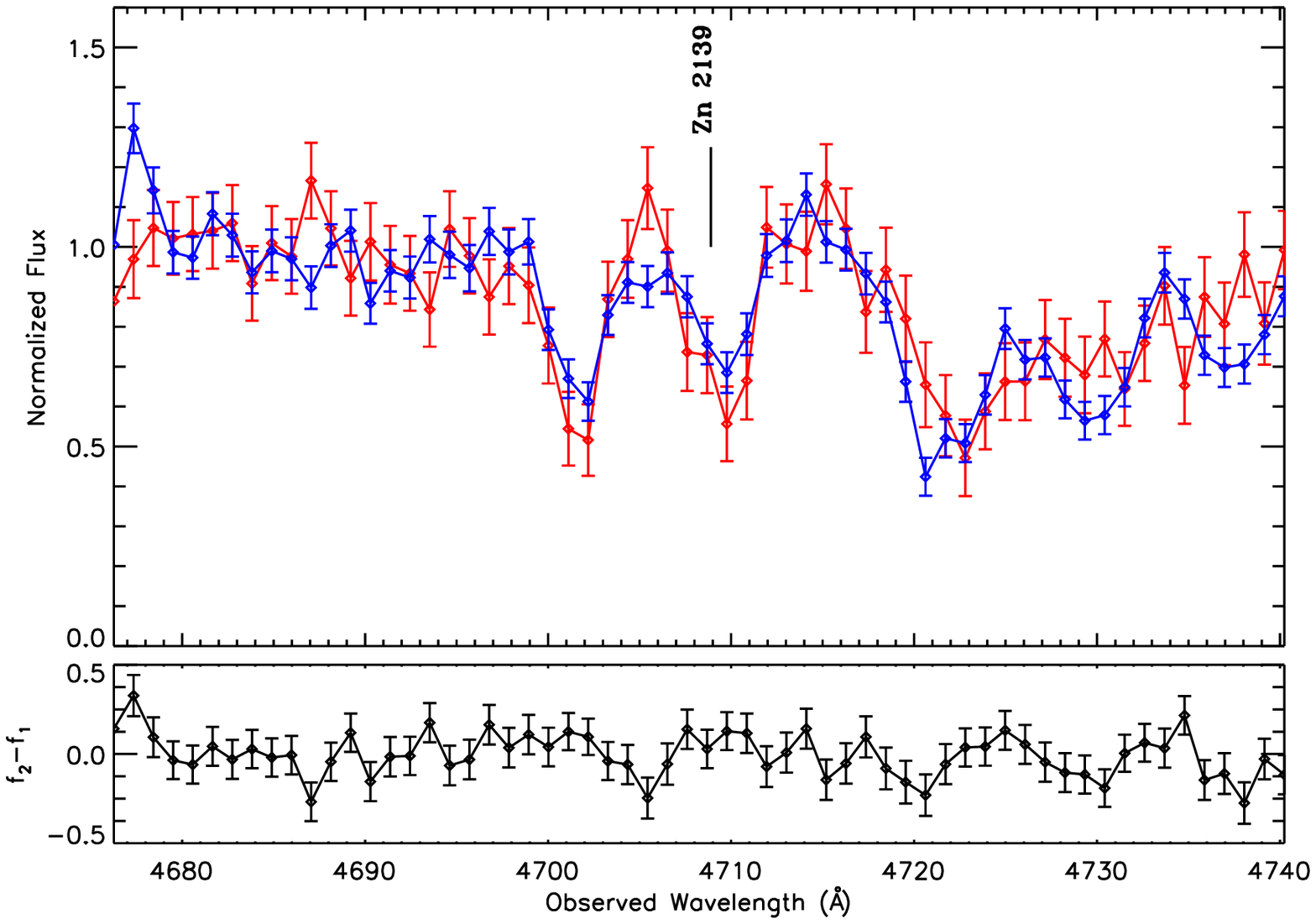}
\includegraphics[width=84mm]{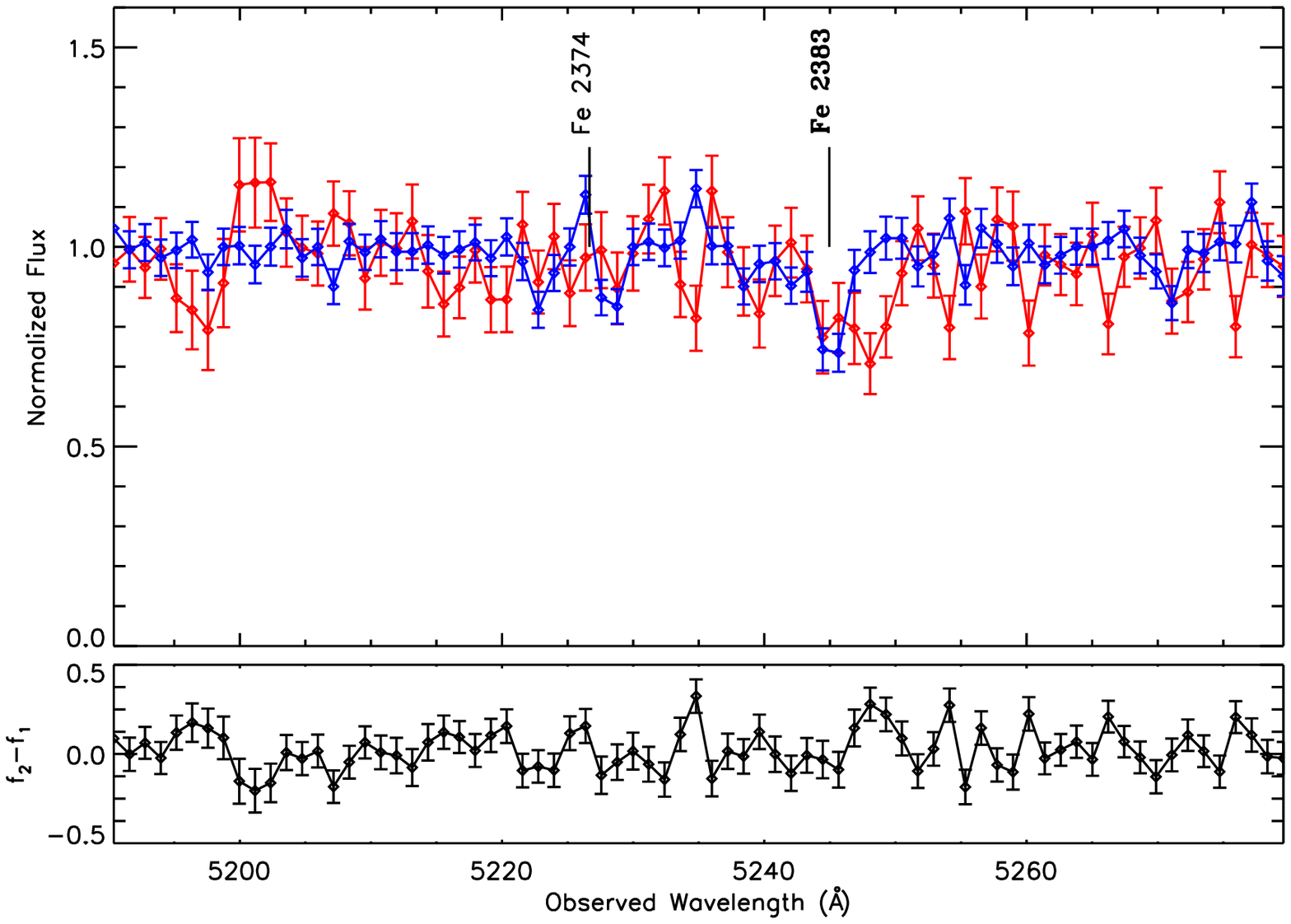}
\includegraphics[width=84mm]{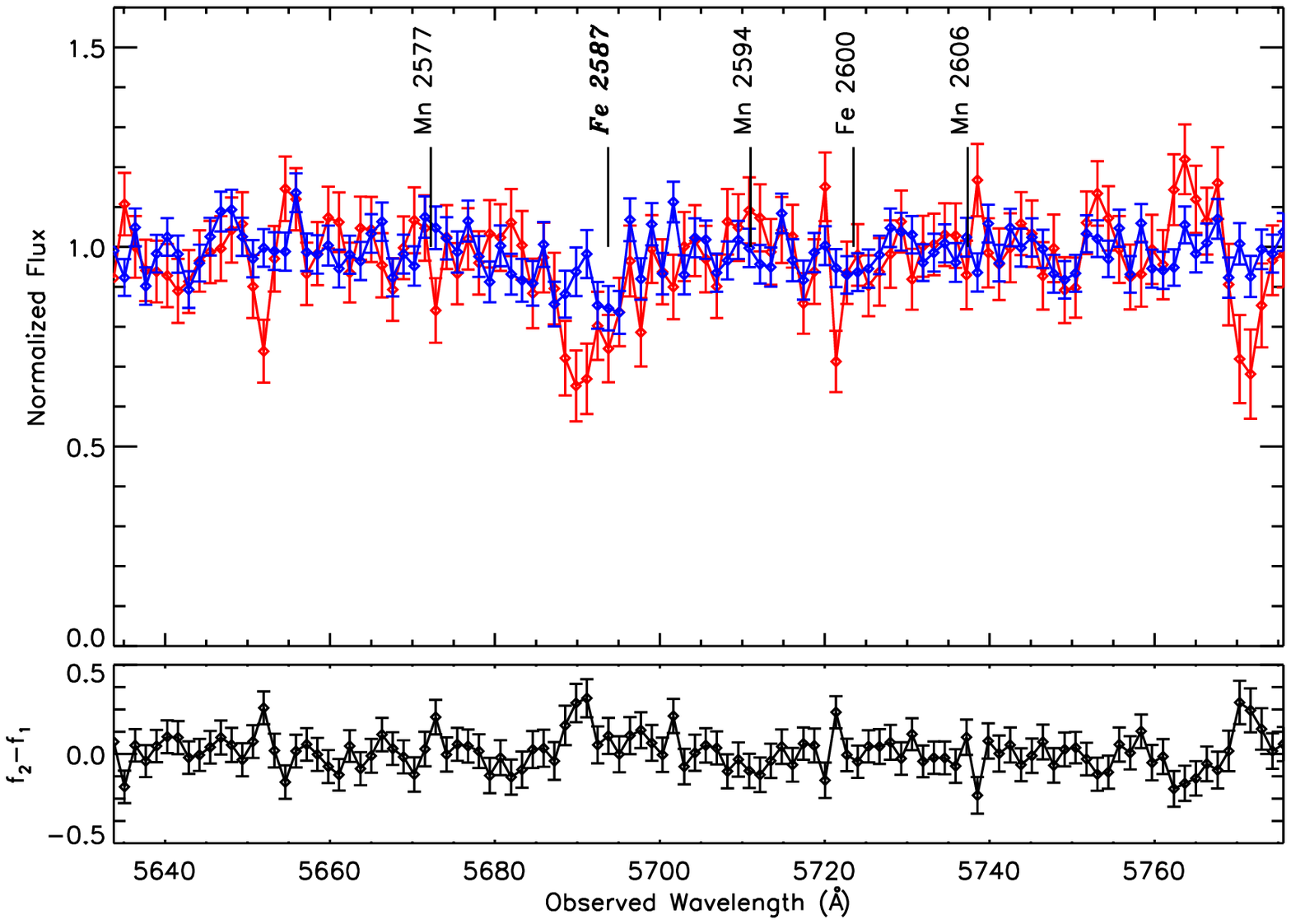}
\includegraphics[width=84mm]{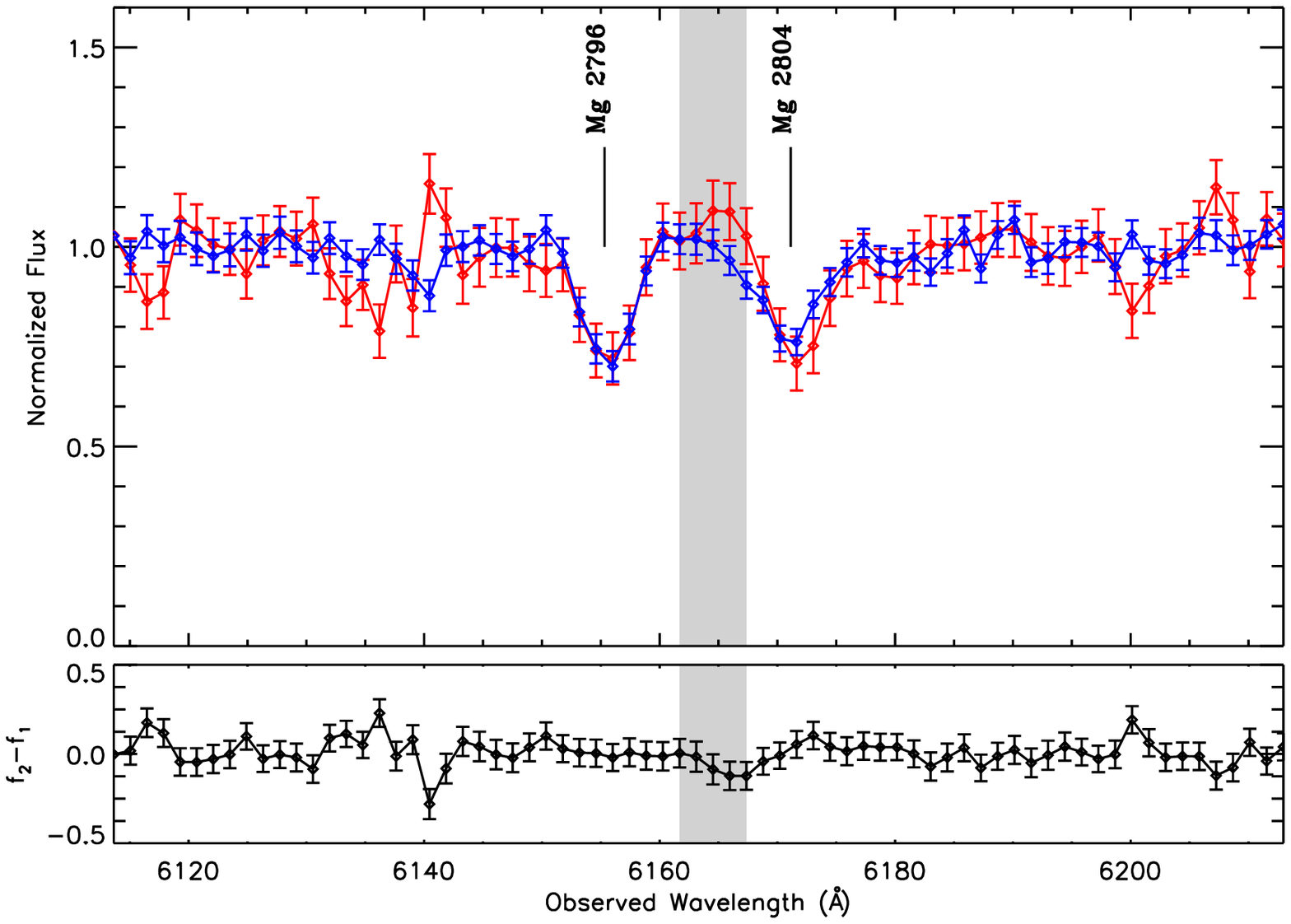}
\caption[Two-epoch normalized spectra of SDSS J075105.17+272116.8]{Two-epoch normalized spectra of the variable NAL system at $\beta$ = 0.3340 in SDSS J075105.17+272116.8.  The top panel shows the normalized pixel flux values with 1$\sigma$ error bars (first observations are red and second are blue), the bottom panel plots the difference spectrum of the two observation epochs, and shaded backgrounds identify masked pixels not included in our search for absorption line variability.  Line identifications for significantly variable absorption lines are italicised, lines detected in both observation epochs are in bold font, and undetected lines are in regular font (see Table A.1 for ion labels).  \label{figvs8}}
\end{center}
\end{figure*}

\clearpage
\begin{figure*}
\begin{center}
\includegraphics[width=84mm]{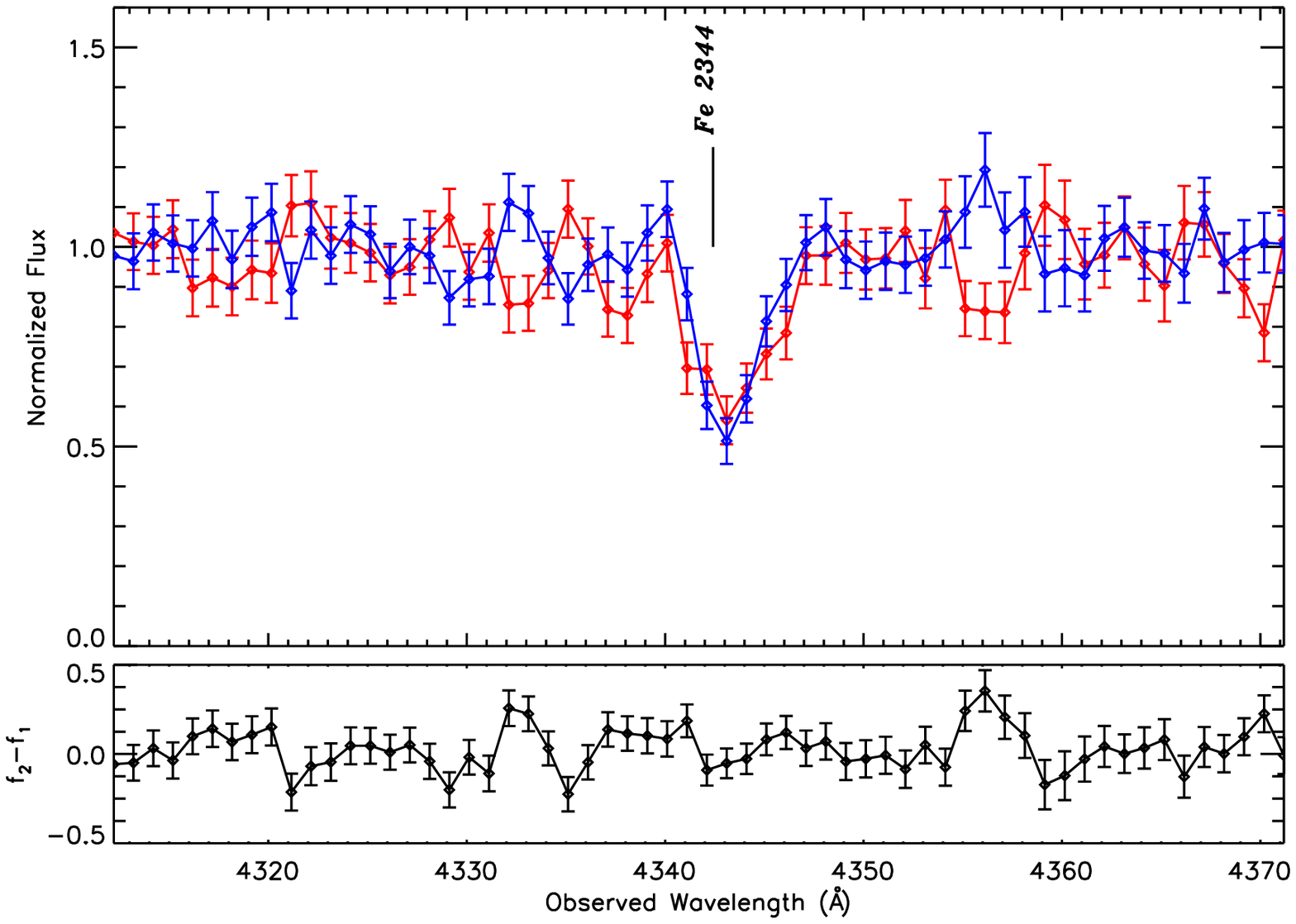}
\includegraphics[width=84mm]{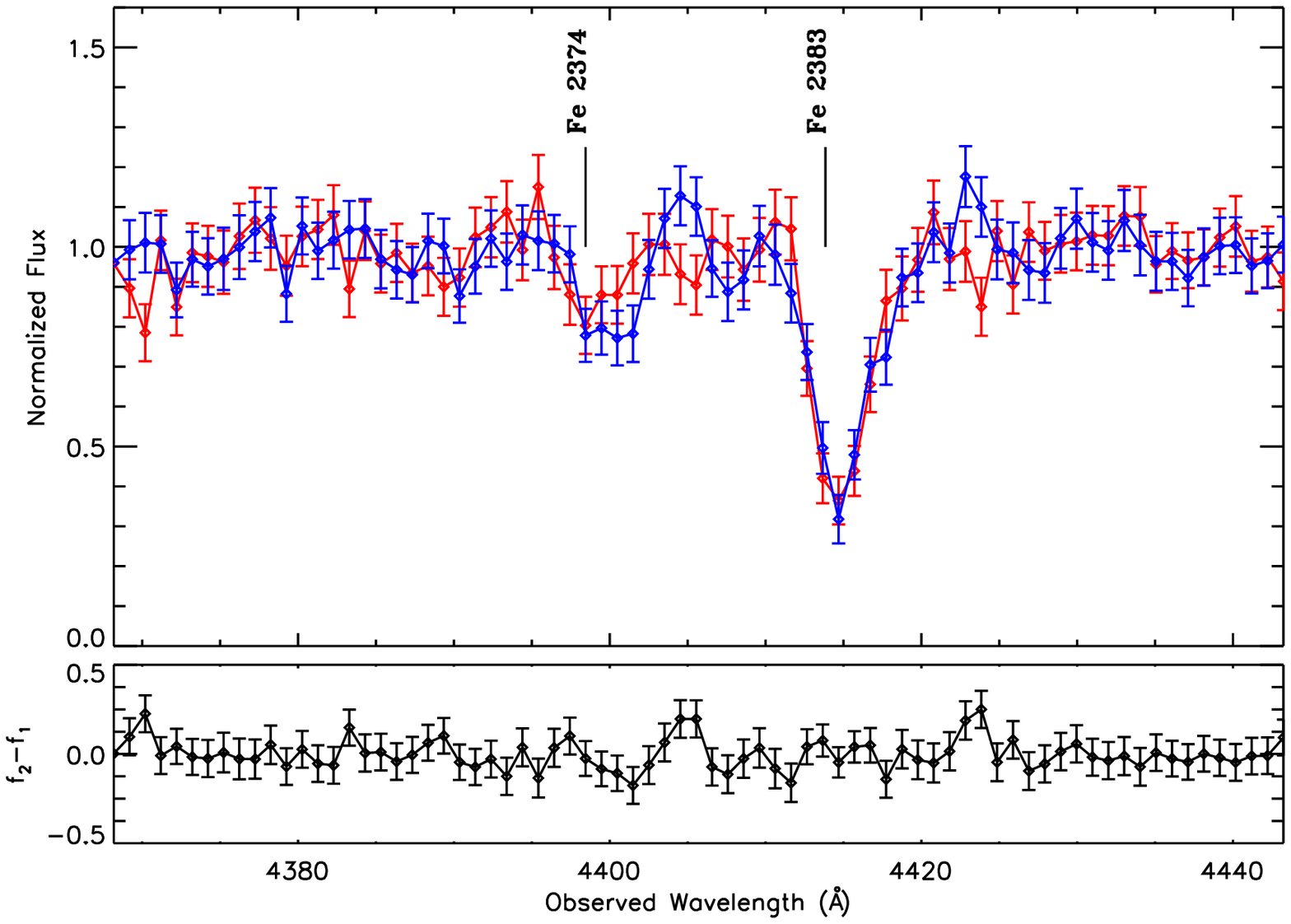}
\includegraphics[width=84mm]{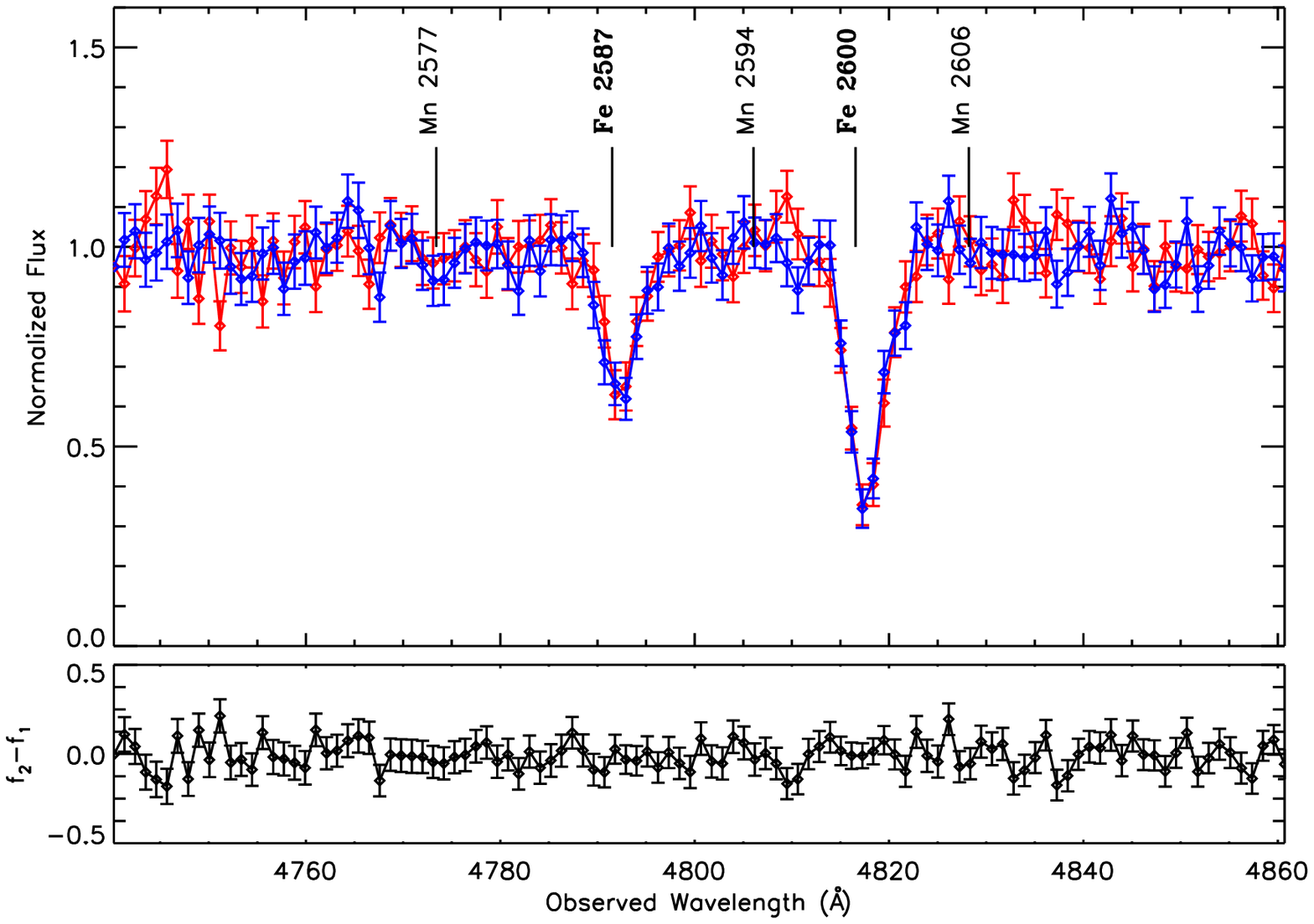}
\includegraphics[width=84mm]{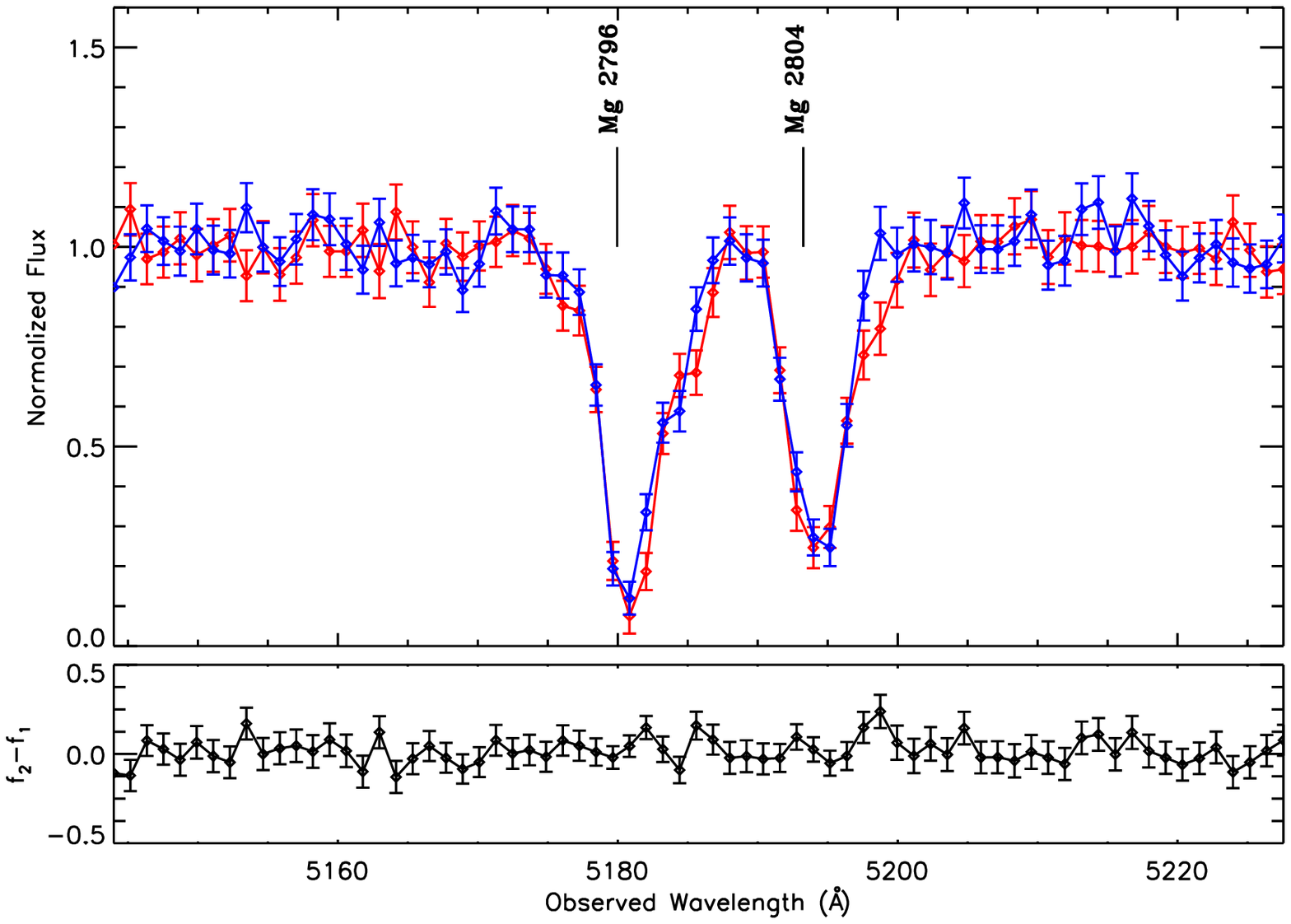}
\includegraphics[width=84mm]{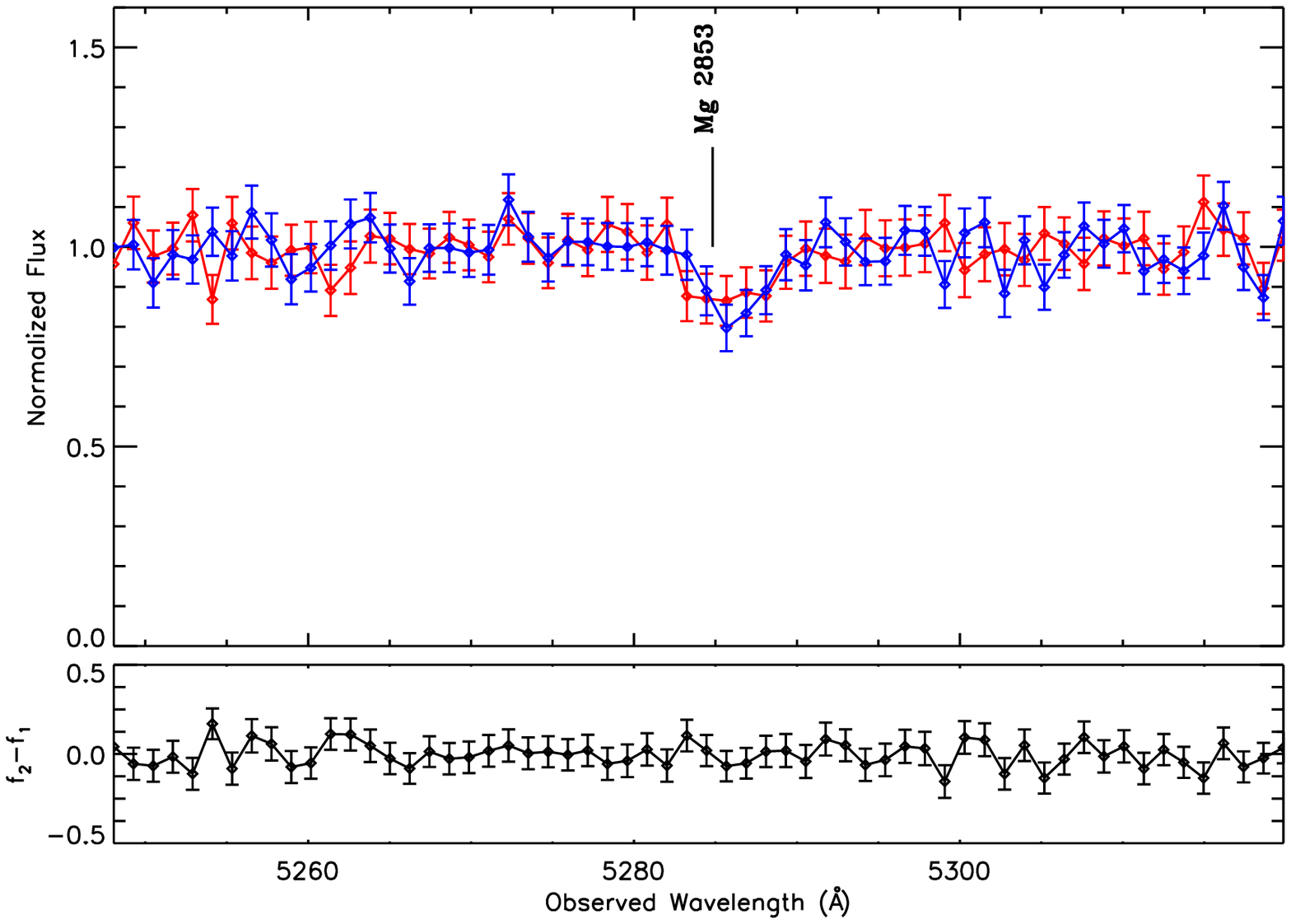}
\caption[Two-epoch normalized spectra of SDSS J024603.68-003211.7]{Two-epoch normalized spectra of the variable NAL system at $\beta$ = 0.3274 in SDSS J024603.68-003211.7.  The top panel shows the normalized pixel flux values with 1$\sigma$ error bars (first observations are red and second are blue), the bottom panel plots the difference spectrum of the two observation epochs, and shaded backgrounds identify masked pixels not included in our search for absorption line variability.  Line identifications for significantly variable absorption lines are italicised, lines detected in both observation epochs are in bold font, and undetected lines are in regular font (see Table A.1 for ion labels).  \label{figvs9}}
\end{center}
\end{figure*}

\clearpage
\begin{figure*}
\begin{center}
\includegraphics[width=84mm]{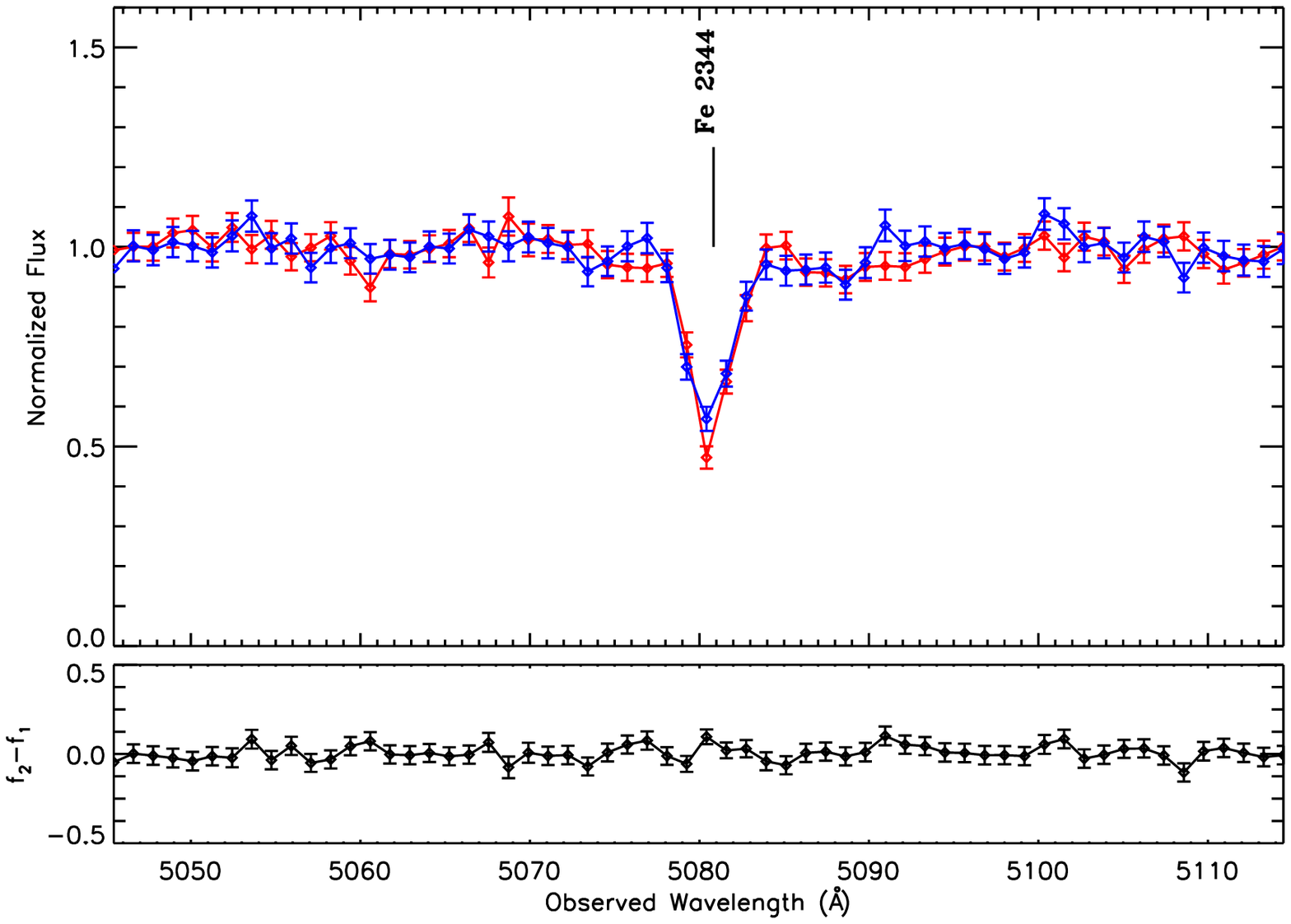}
\includegraphics[width=84mm]{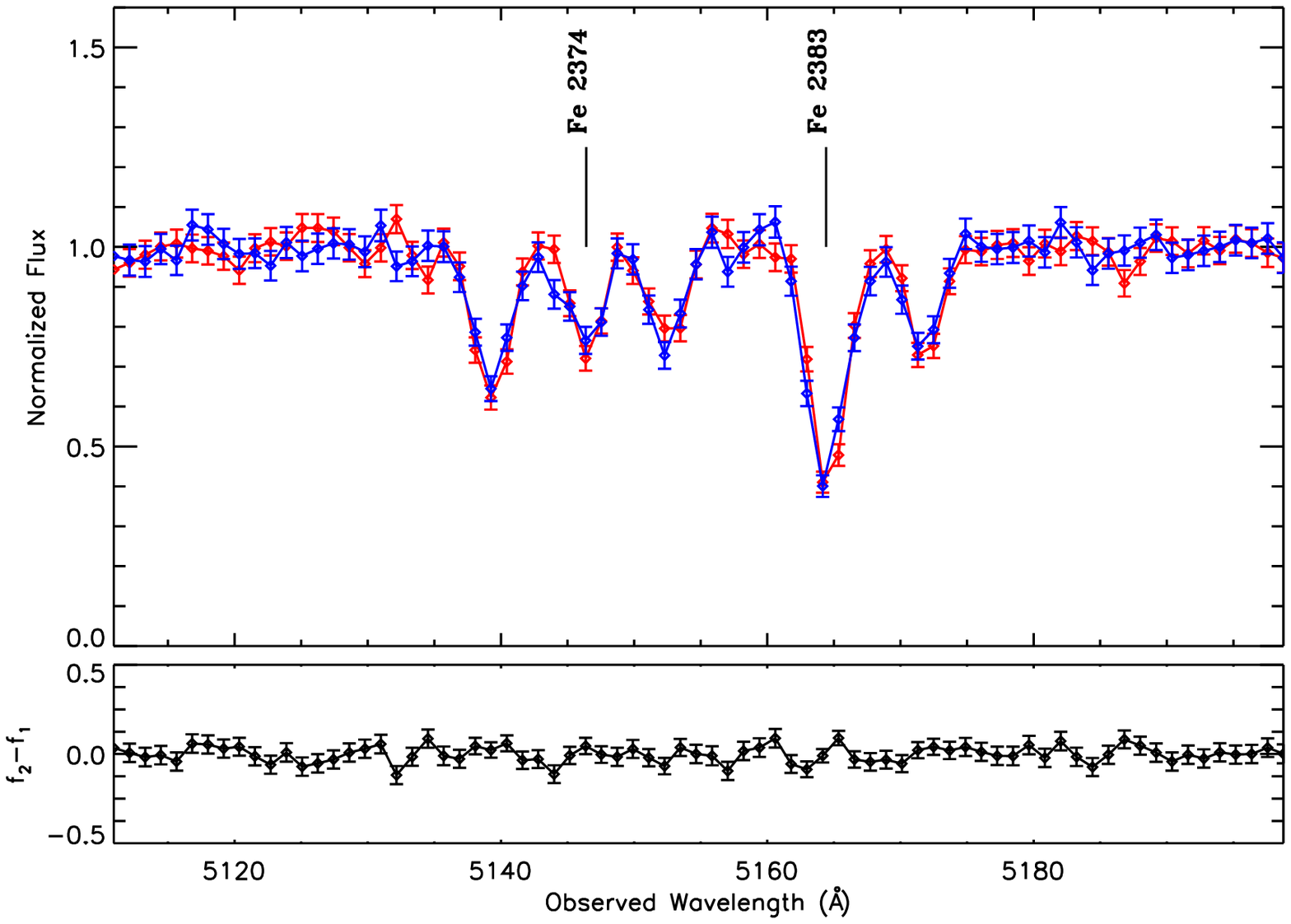}
\includegraphics[width=84mm]{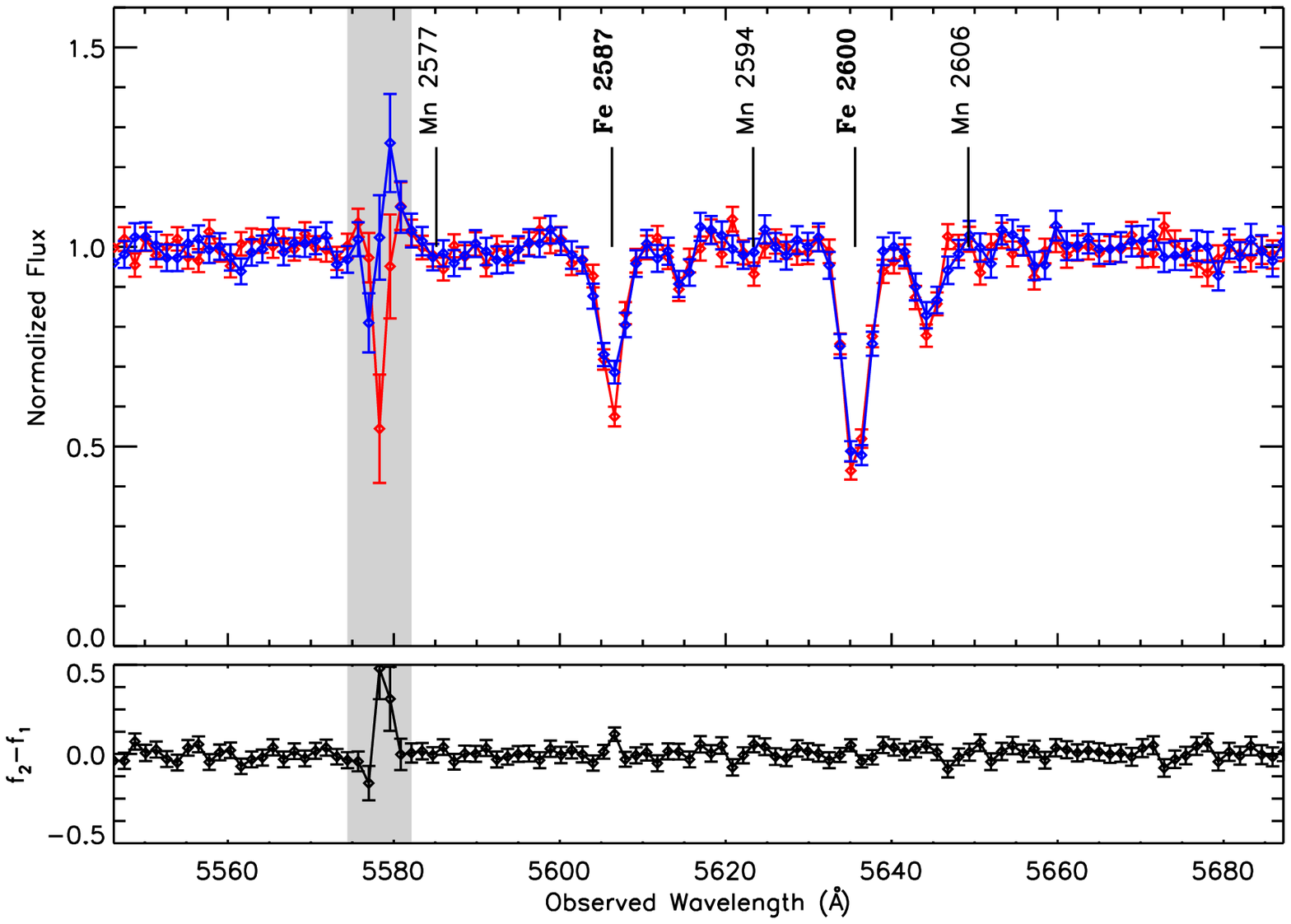}
\includegraphics[width=84mm]{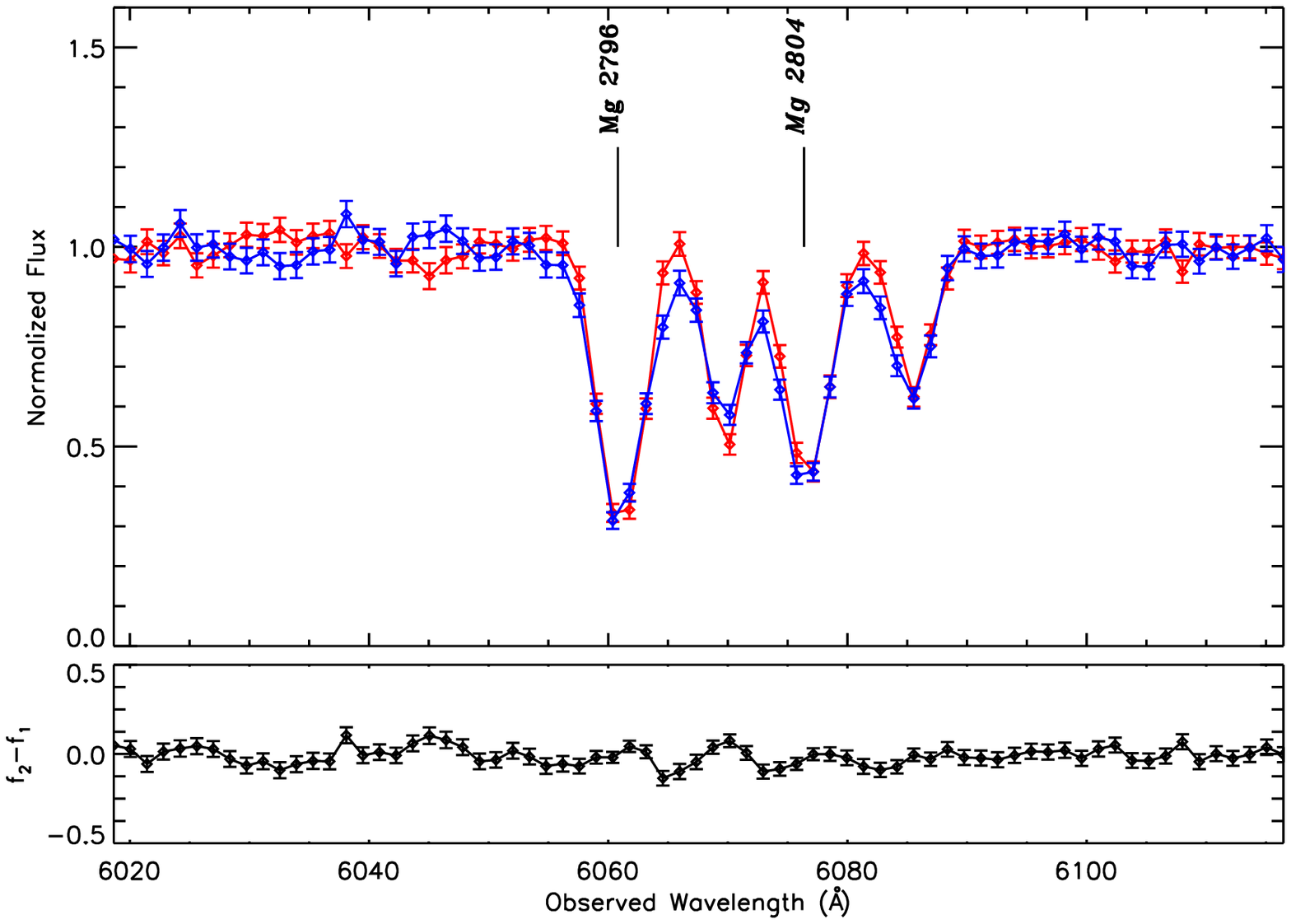}
\includegraphics[width=84mm]{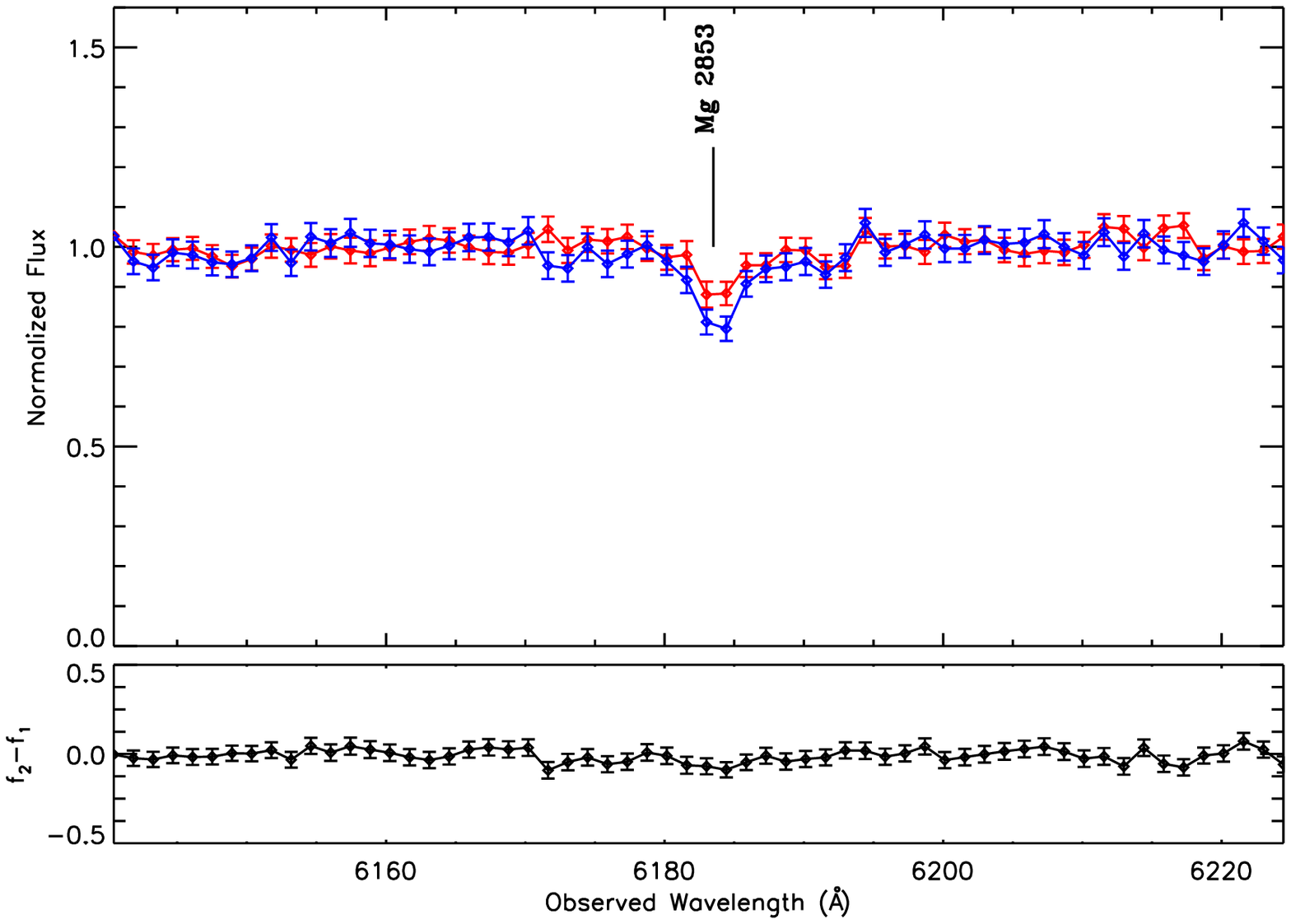}
\caption[Two-epoch normalized spectra of SDSS J171748.76+275532.5]{Two-epoch normalized spectra of the variable NAL system at $\beta$ = 0.2957 in SDSS J171748.76+275532.5.  The top panel shows the normalized pixel flux values with 1$\sigma$ error bars (first observations are red and second are blue), the bottom panel plots the difference spectrum of the two observation epochs, and shaded backgrounds identify masked pixels not included in our search for absorption line variability.  Line identifications for significantly variable absorption lines are italicised, lines detected in both observation epochs are in bold font, and undetected lines are in regular font (see Table A.1 for ion labels).  \label{figvs10}}
\end{center}
\end{figure*}

\clearpage
\begin{figure*}
\begin{center}
\includegraphics[width=84mm]{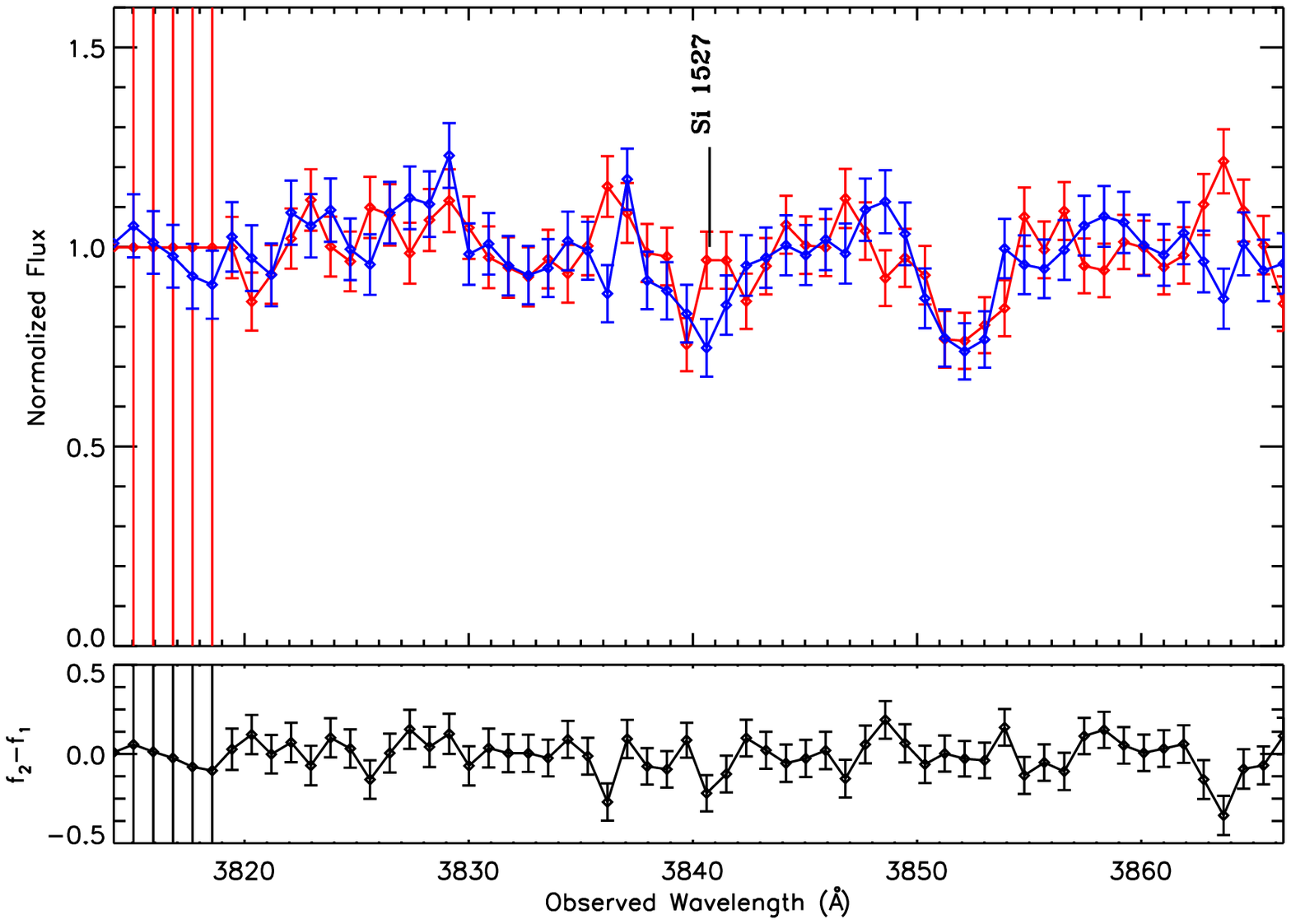}
\includegraphics[width=84mm]{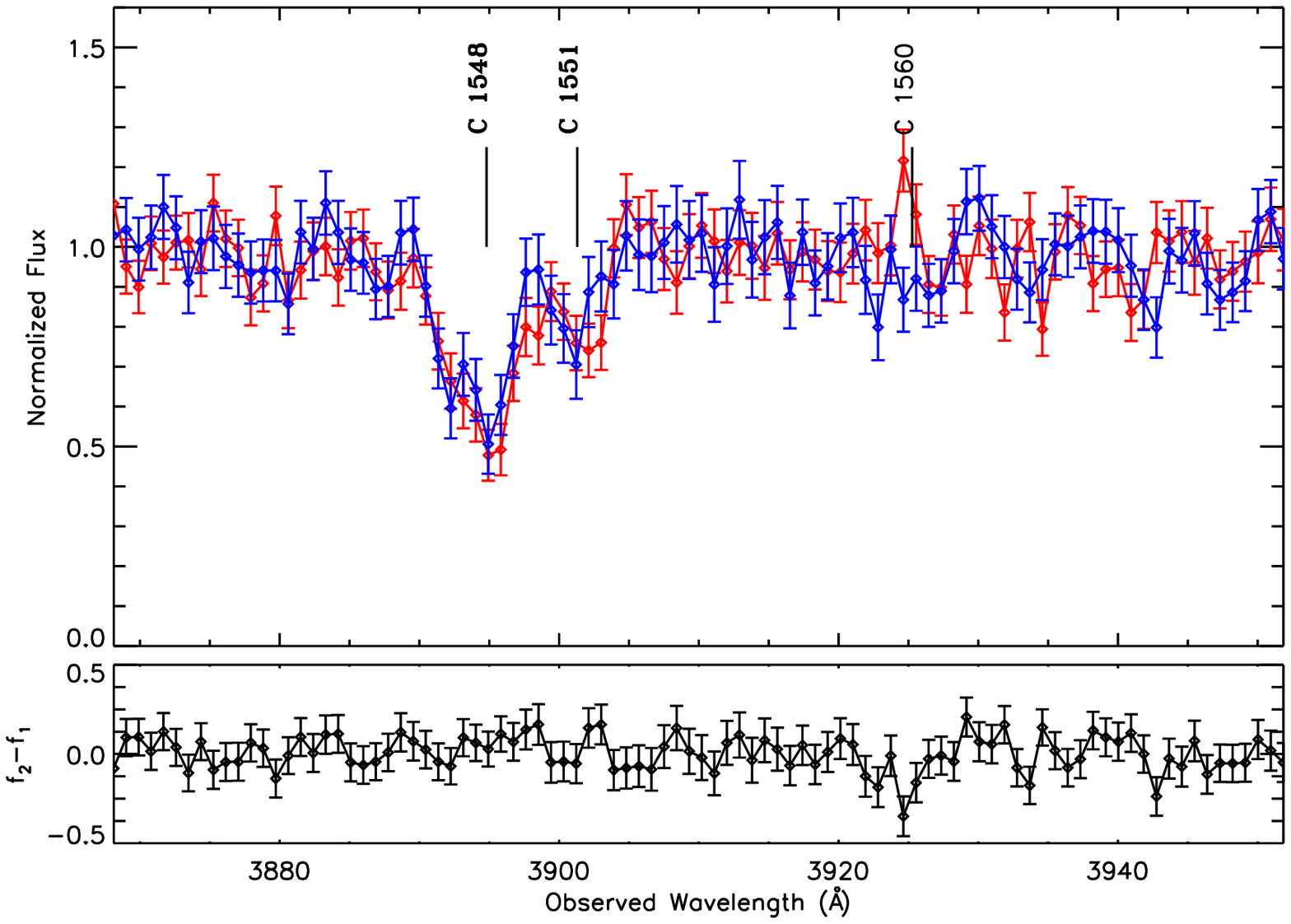}
\includegraphics[width=84mm]{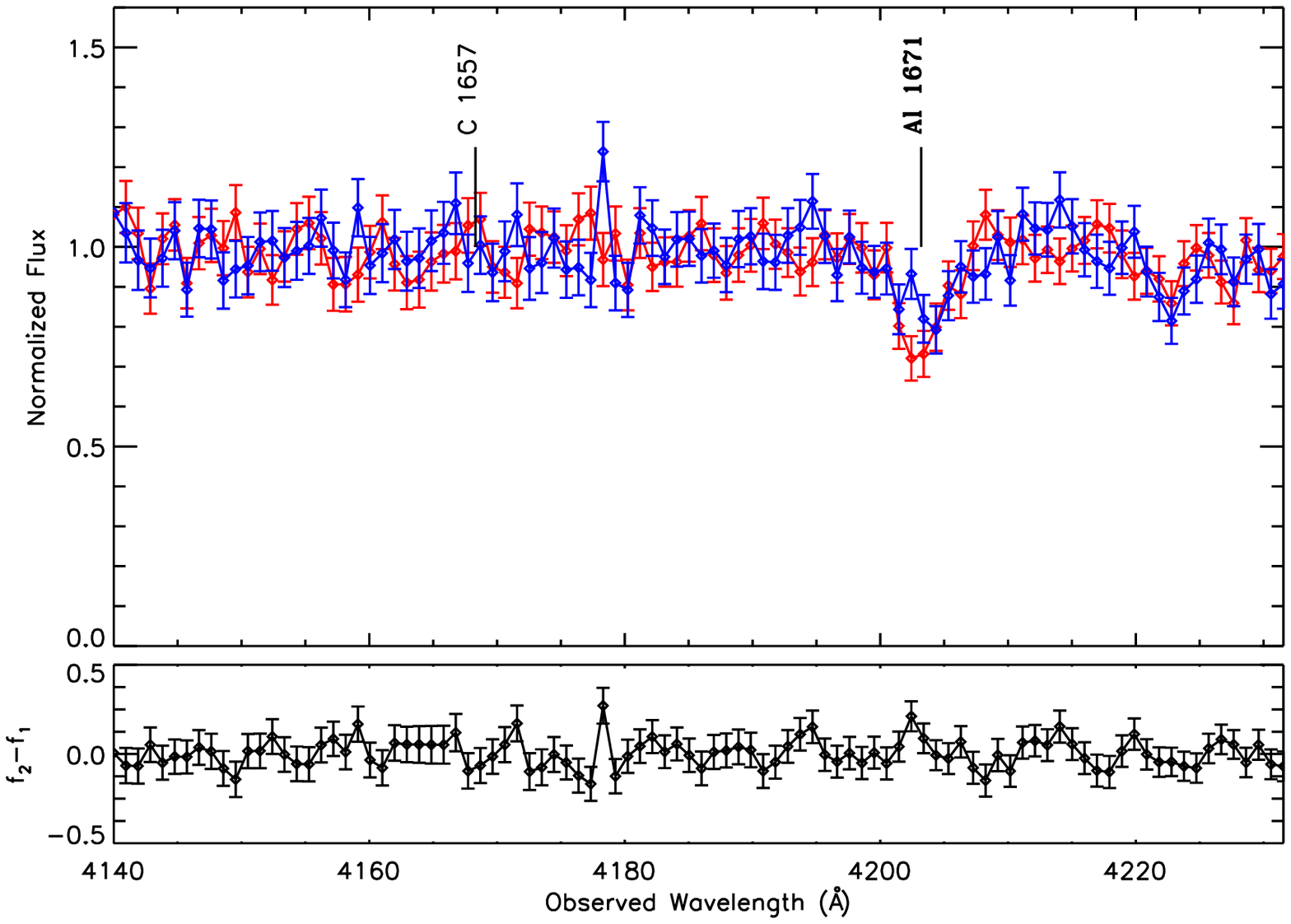}
\includegraphics[width=84mm]{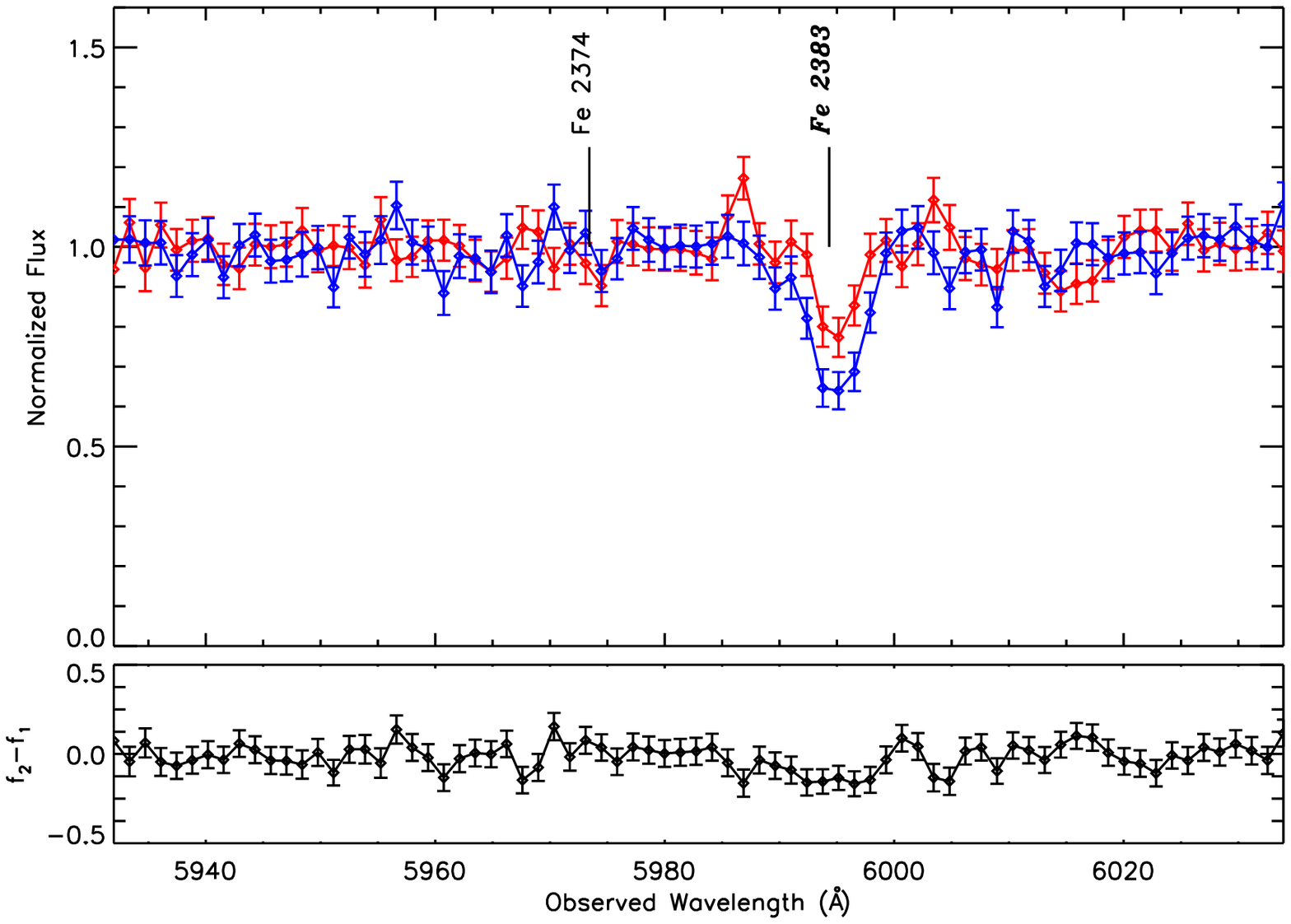}
\includegraphics[width=84mm]{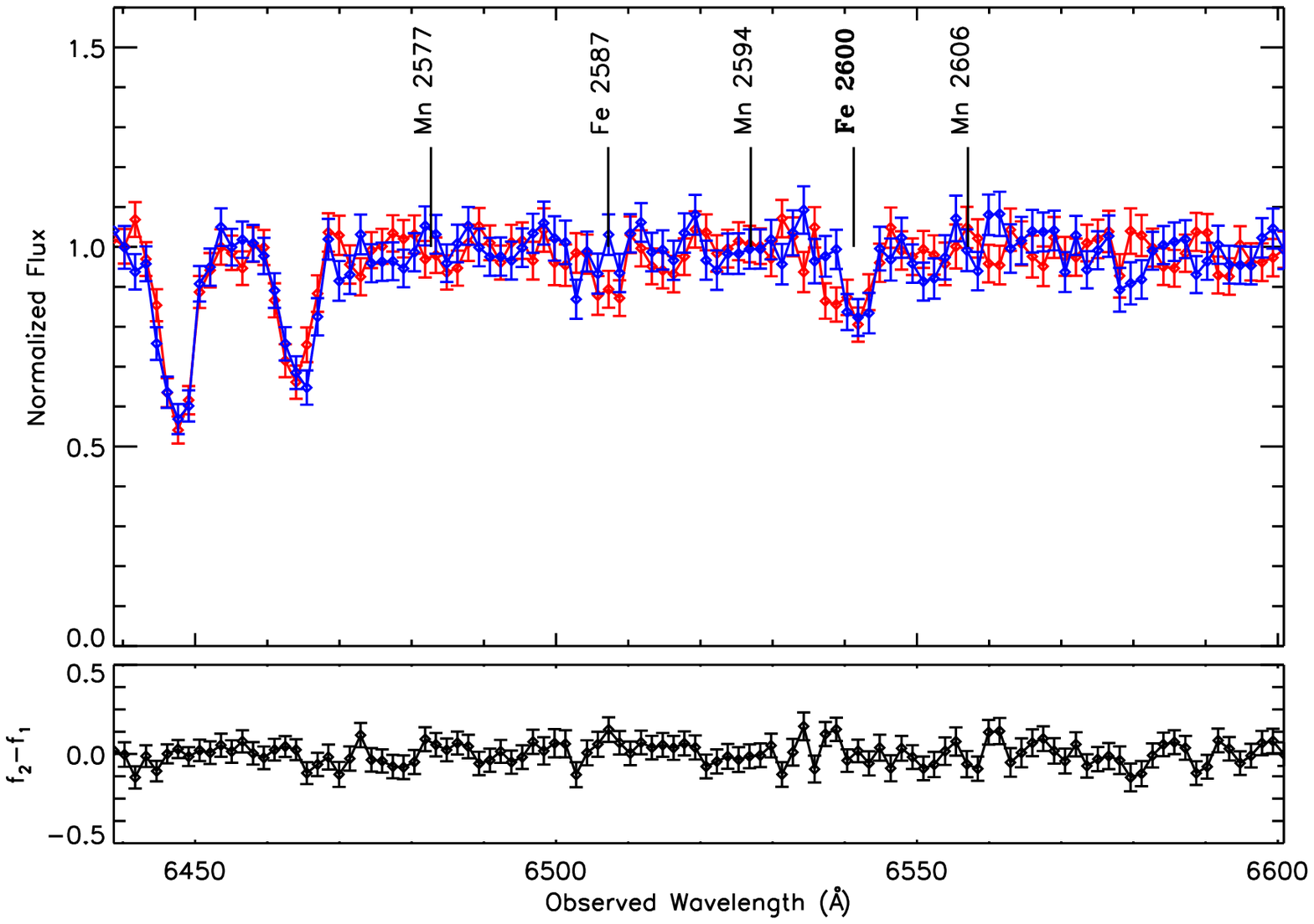}
\includegraphics[width=84mm]{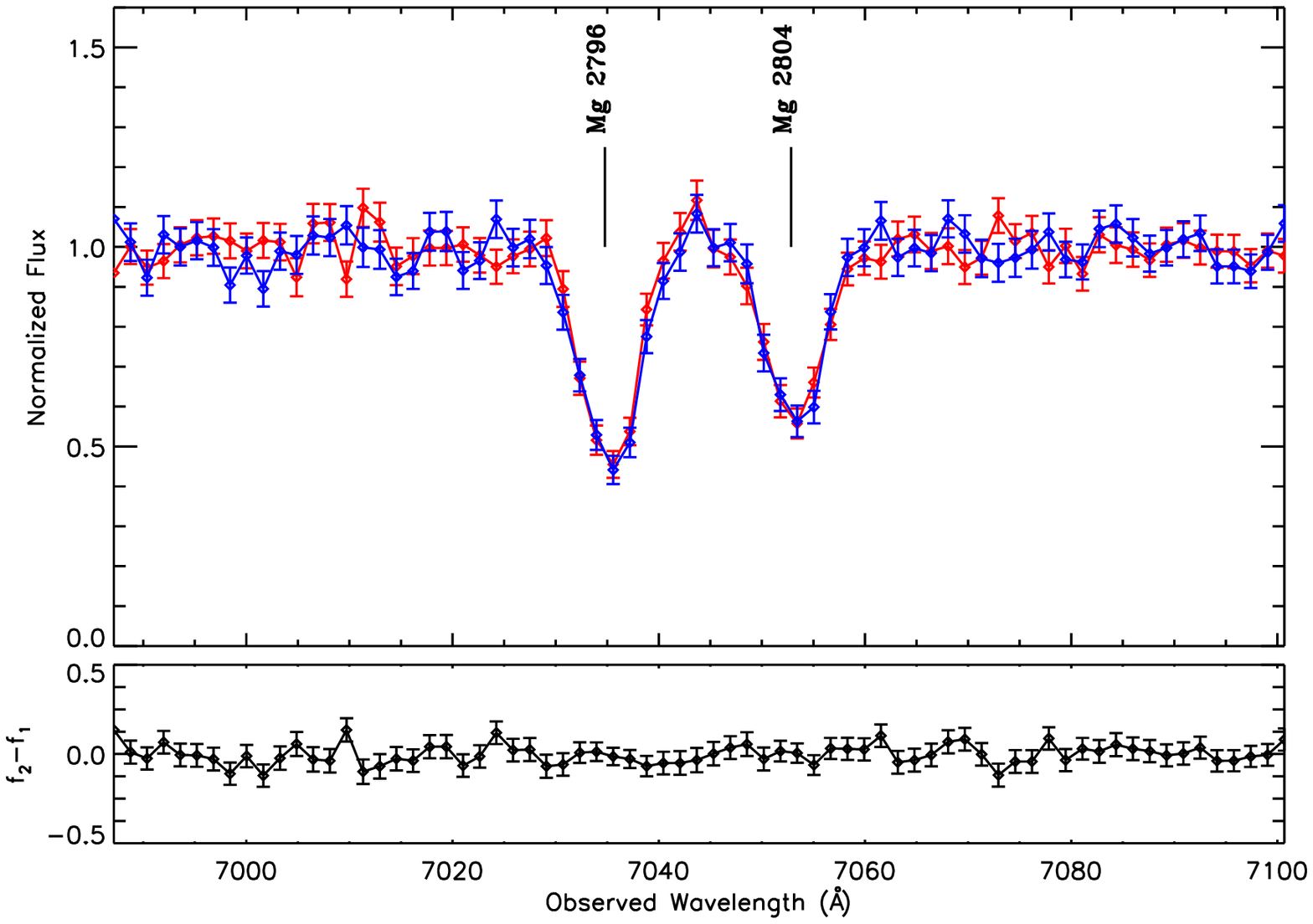}
\caption[Two-epoch normalized spectra of SDSS J131347.68+294201.3]{Two-epoch normalized spectra of the variable NAL system at $\beta$ = 0.1938 in SDSS J131347.68+294201.3.  The top panel shows the normalized pixel flux values with 1$\sigma$ error bars (first observations are red and second are blue), the bottom panel plots the difference spectrum of the two observation epochs, and shaded backgrounds identify masked pixels not included in our search for absorption line variability.  Line identifications for significantly variable absorption lines are italicised, lines detected in both observation epochs are in bold font, and undetected lines are in regular font (see Table A.1 for ion labels). \label{figvs11}}
\end{center}
\end{figure*}

\clearpage
\begin{figure*}
\begin{center}
\includegraphics[width=84mm]{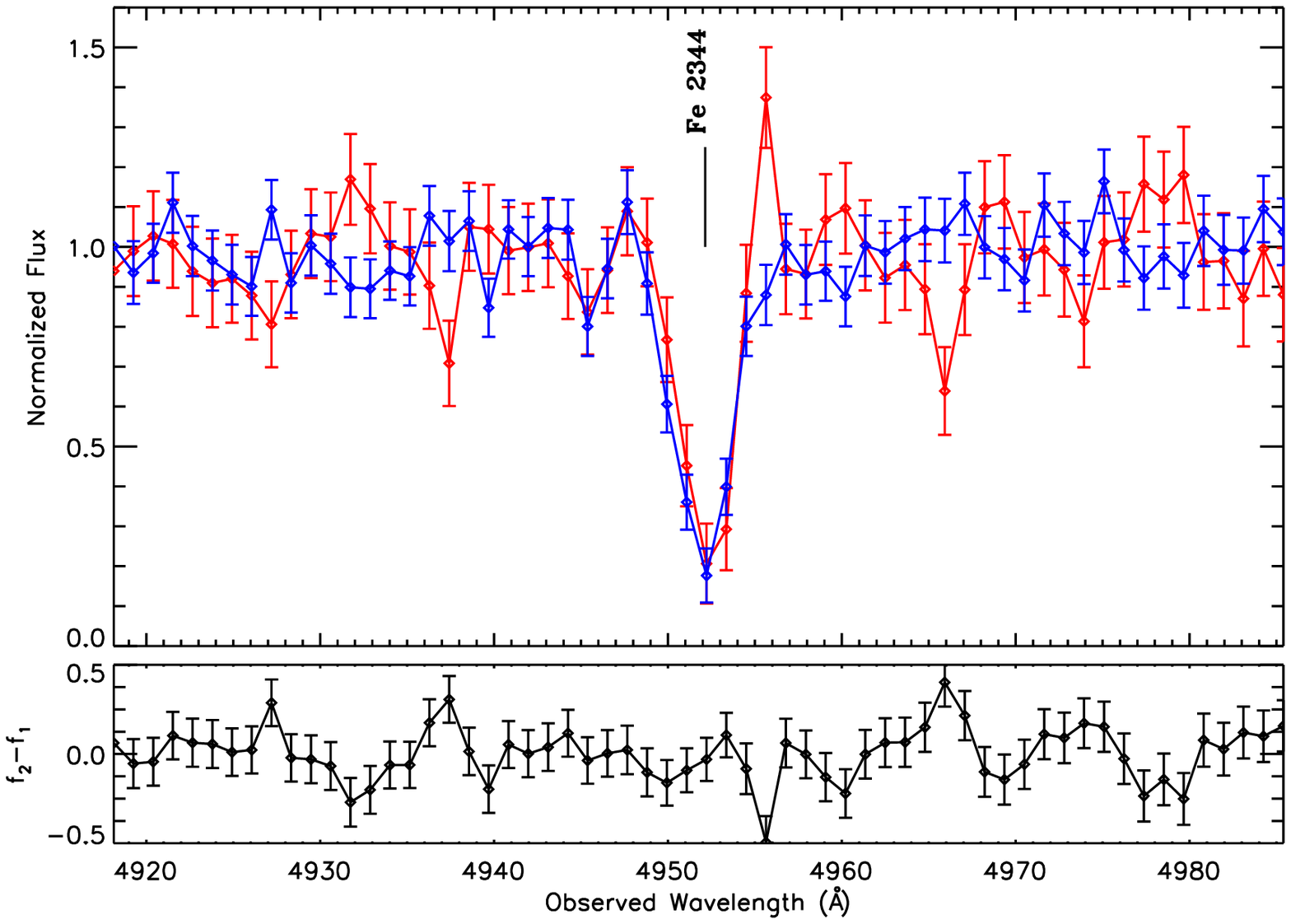}
\includegraphics[width=84mm]{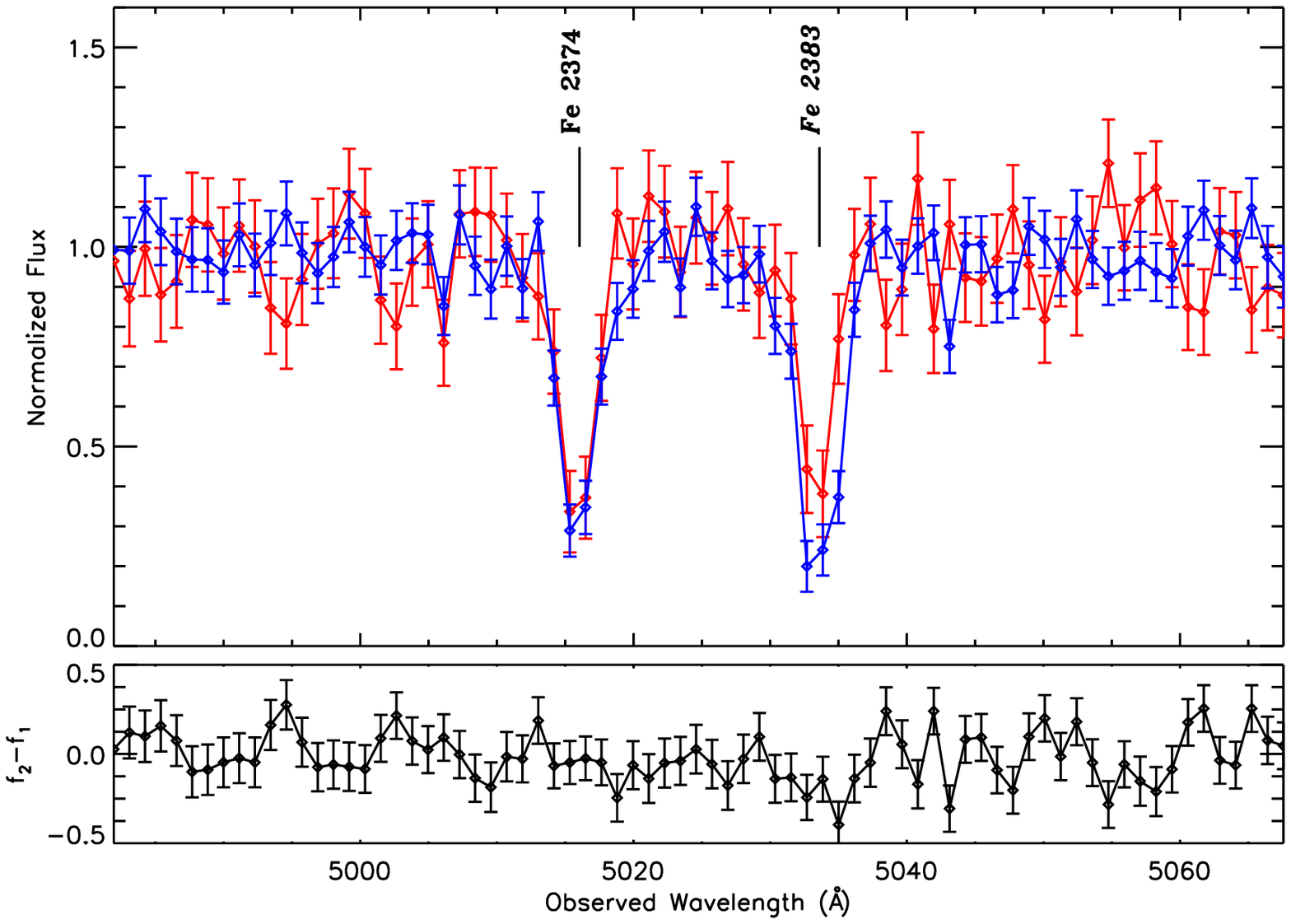}
\includegraphics[width=84mm]{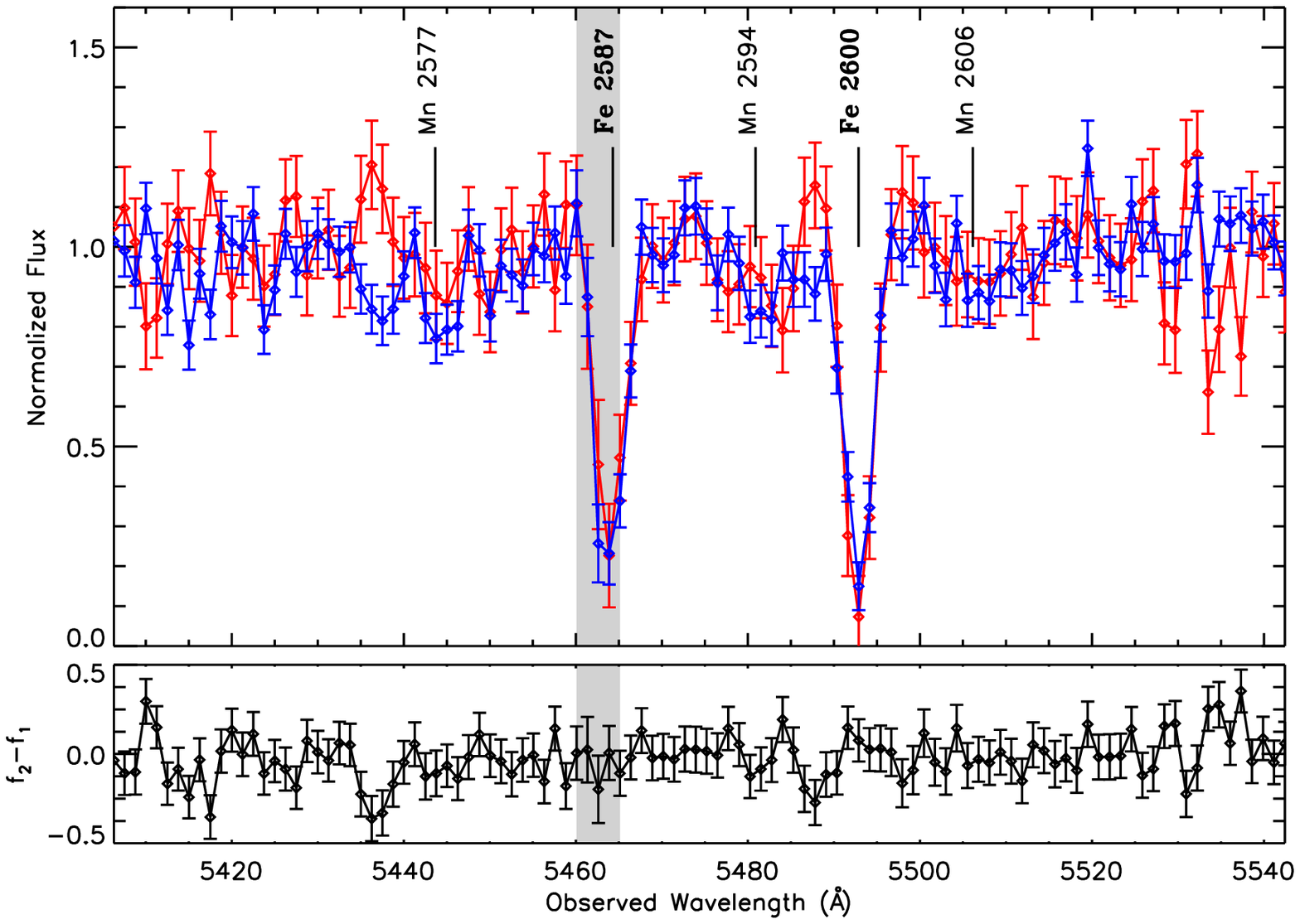}
\includegraphics[width=84mm]{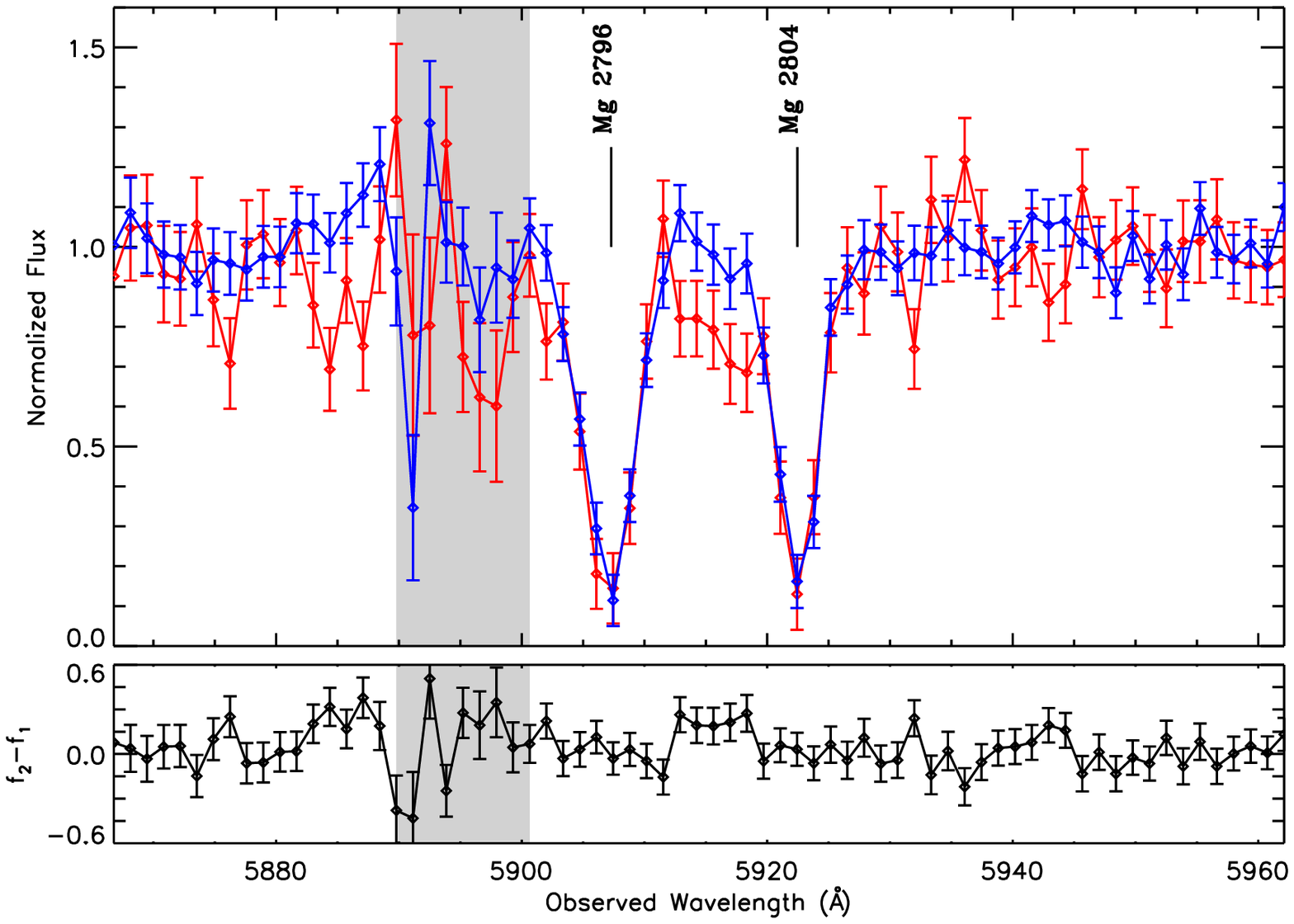}
\includegraphics[width=84mm]{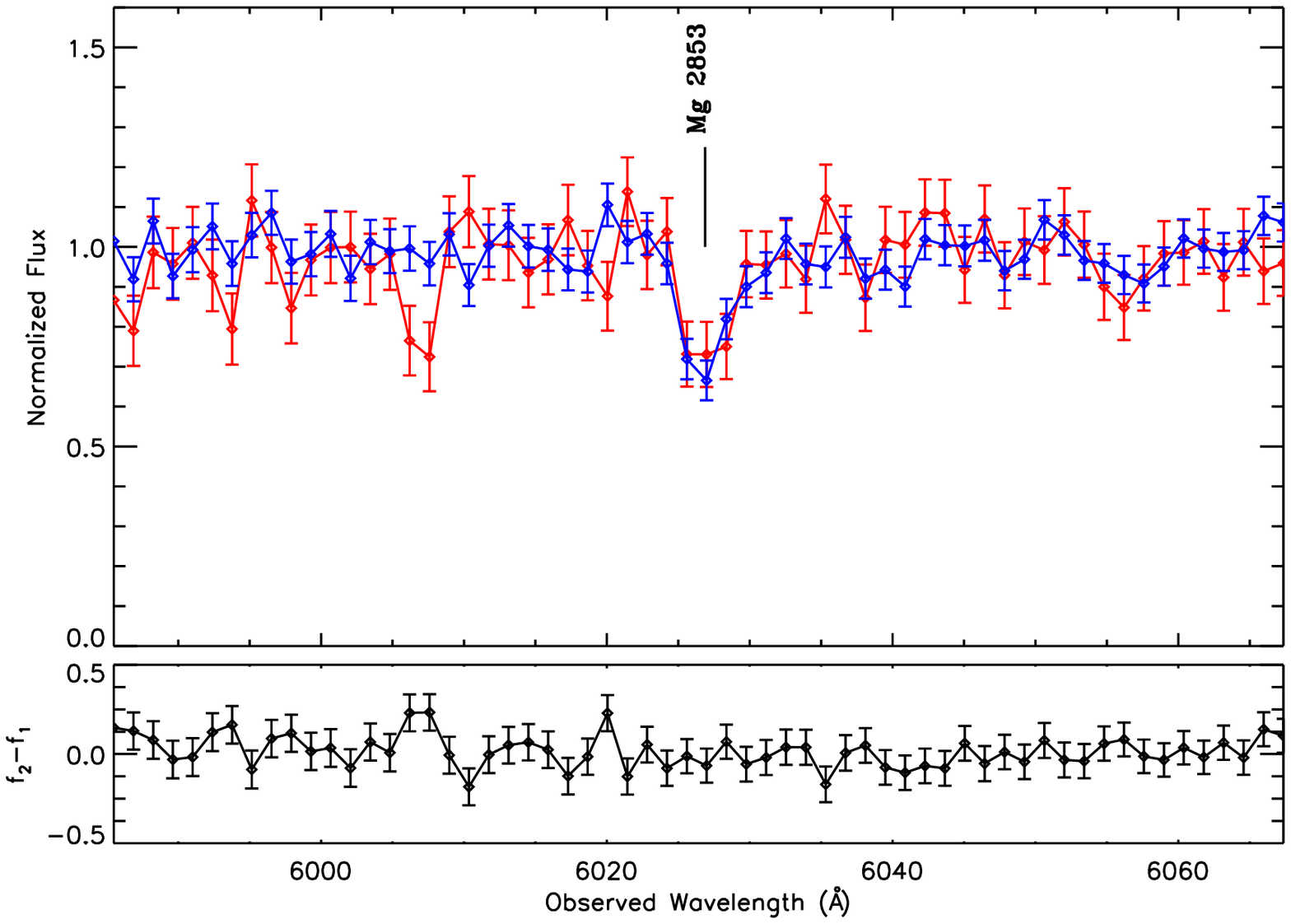}
\includegraphics[width=84mm]{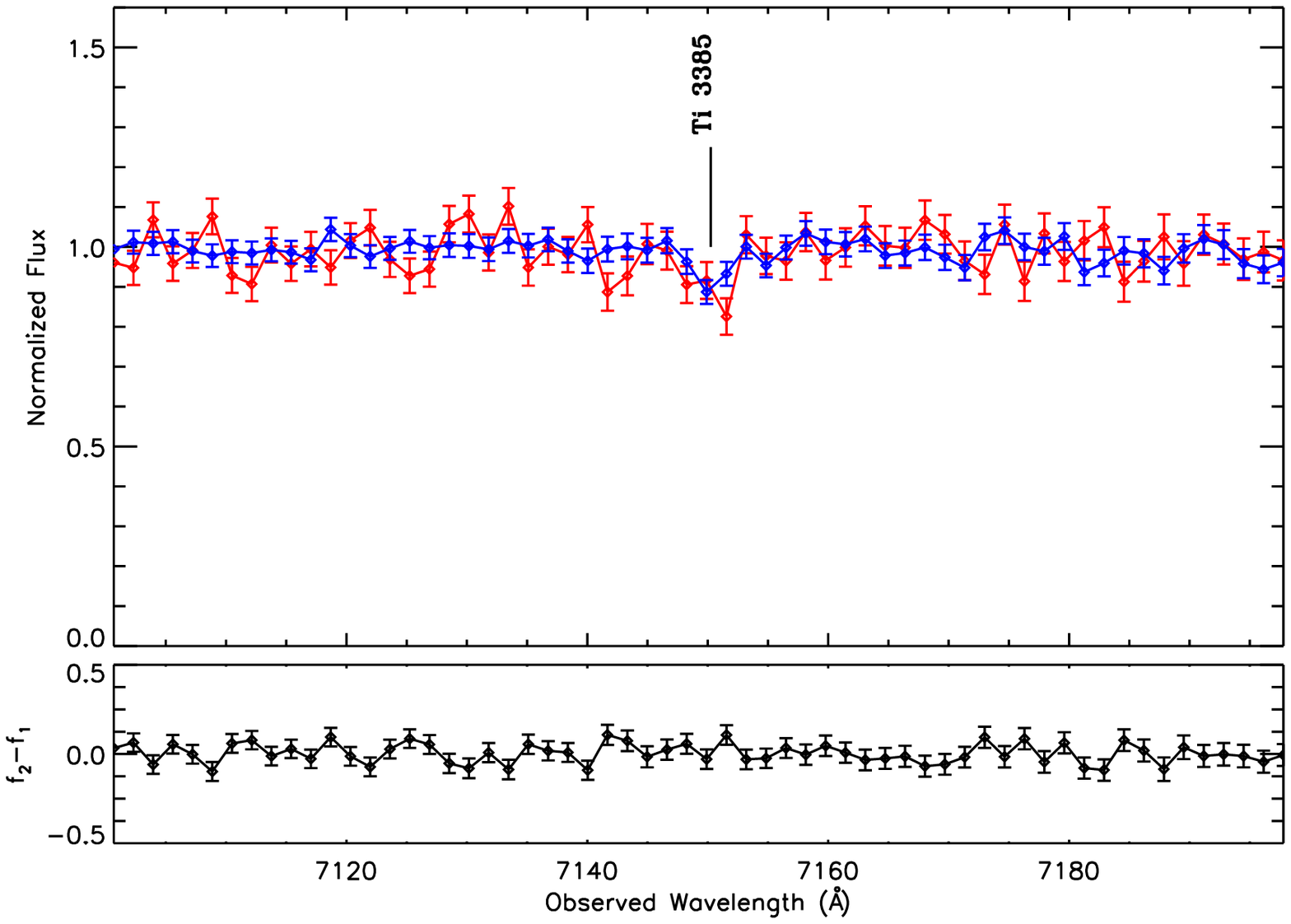}
\caption[Two-epoch normalized spectra of SDSS J024534.07+010813.7]{Two-epoch normalized spectra of the variable NAL system at $\beta$ = 0.1780 in SDSS J024534.07+010813.7.  The top panel shows the normalized pixel flux values with 1$\sigma$ error bars (first observations are red and second are blue), the bottom panel plots the difference spectrum of the two observation epochs, and shaded backgrounds identify masked pixels not included in our search for absorption line variability.  Line identifications for significantly variable absorption lines are italicised, lines detected in both observation epochs are in bold font, and undetected lines are in regular font (see Table A.1 for ion labels).  Continued in next figure.  \label{figvs12}}
\end{center}
\end{figure*}

\begin{figure*}
\ContinuedFloat
\begin{center}
\includegraphics[width=84mm]{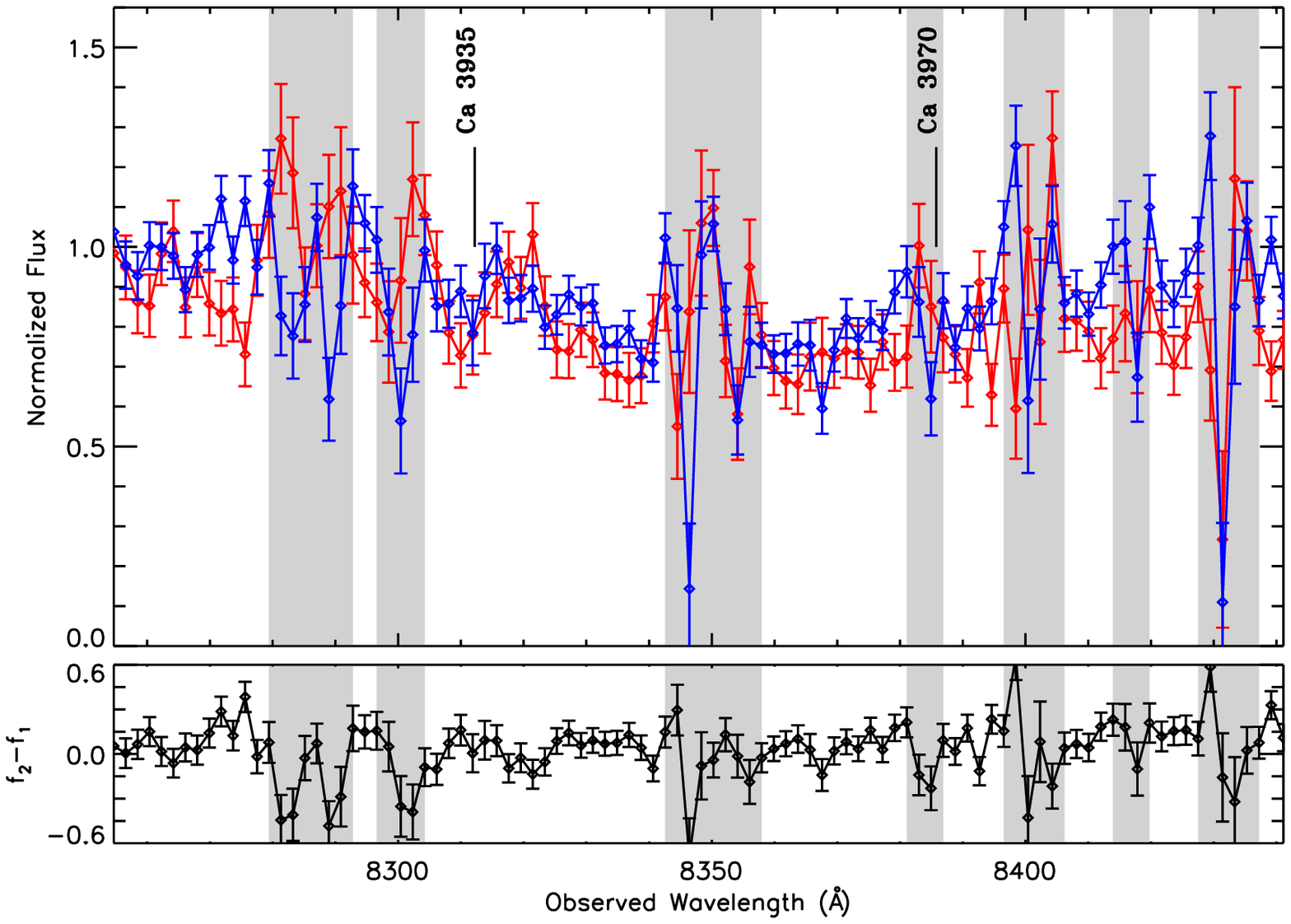}
\caption[]{Two-epoch normalized spectra of the variable NAL system at $\beta$ = 0.1780 in SDSS J024534.07+010813.7.  The top panel shows the normalized pixel flux values with 1$\sigma$ error bars (first observations are red and second are blue), the bottom panel plots the difference spectrum of the two observation epochs, and shaded backgrounds identify masked pixels not included in our search for absorption line variability.  Line identifications for significantly variable absorption lines are italicised, lines detected in both observation epochs are in bold font, and undetected lines are in regular font (see Table A.1 for ion labels).  Continued from previous figure.}
\vspace{3.5cm}
\end{center}
\end{figure*}

\clearpage
\begin{figure*}
\begin{center}
\includegraphics[width=84mm]{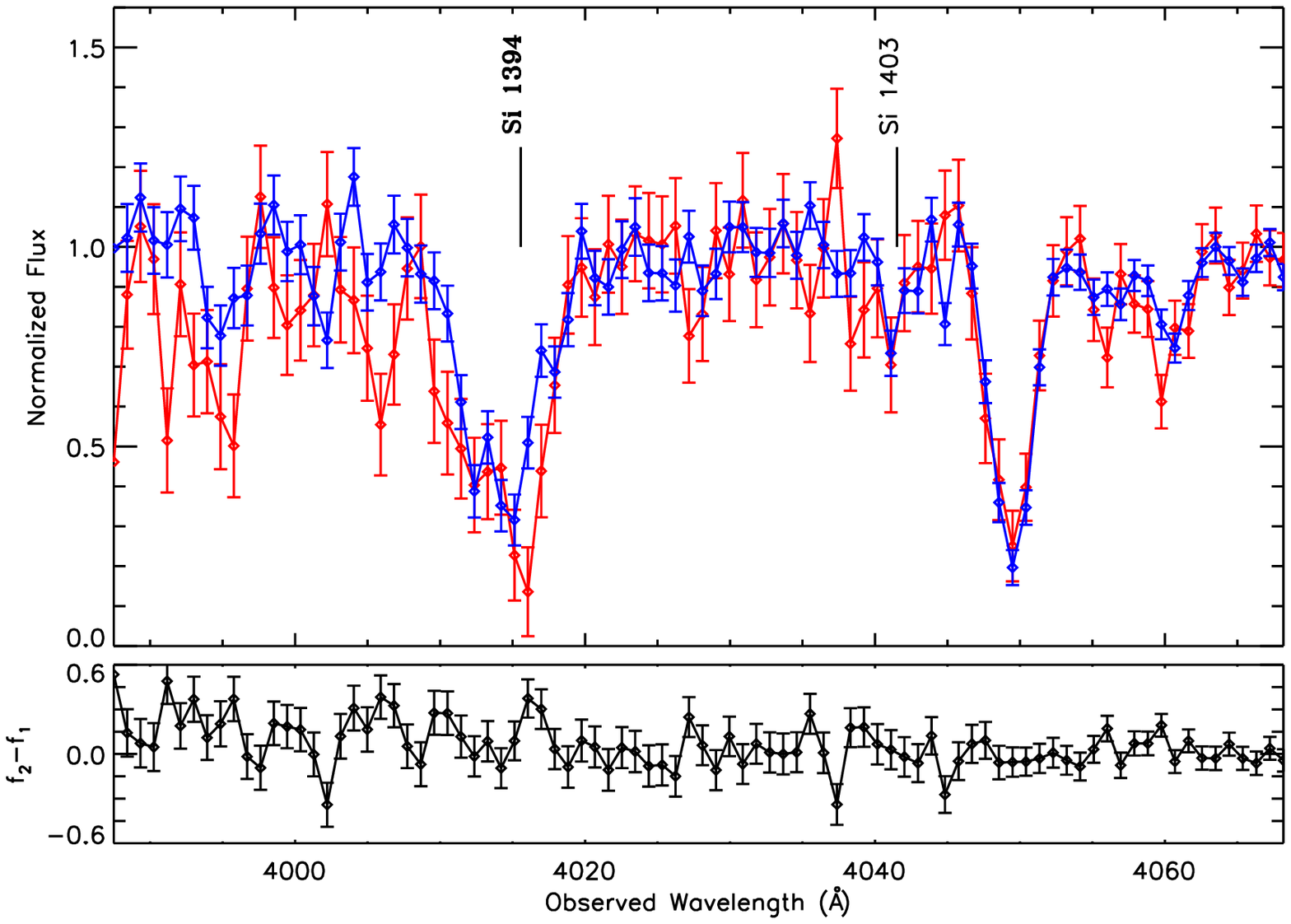}
\includegraphics[width=84mm]{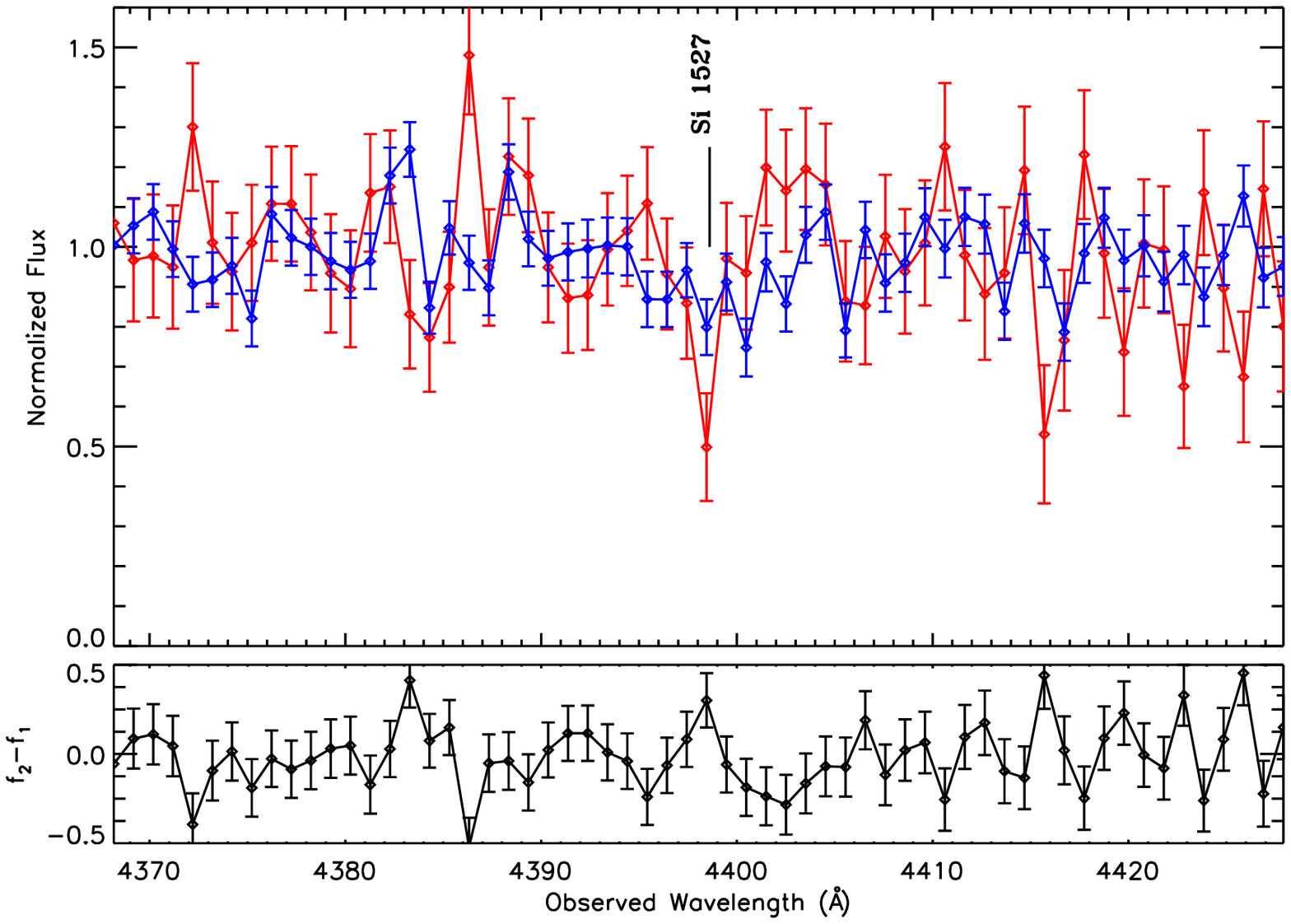}
\includegraphics[width=84mm]{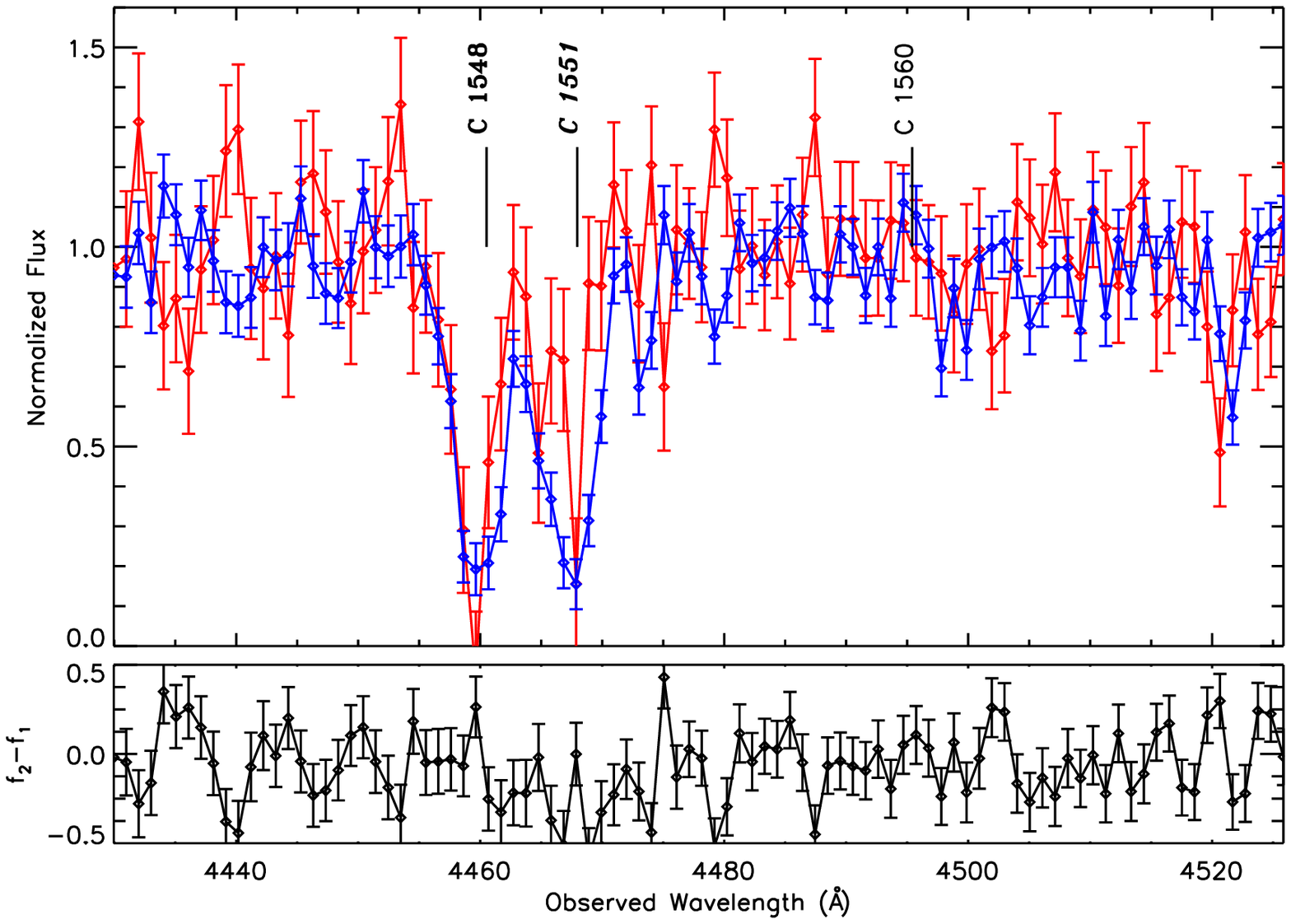}
\includegraphics[width=84mm]{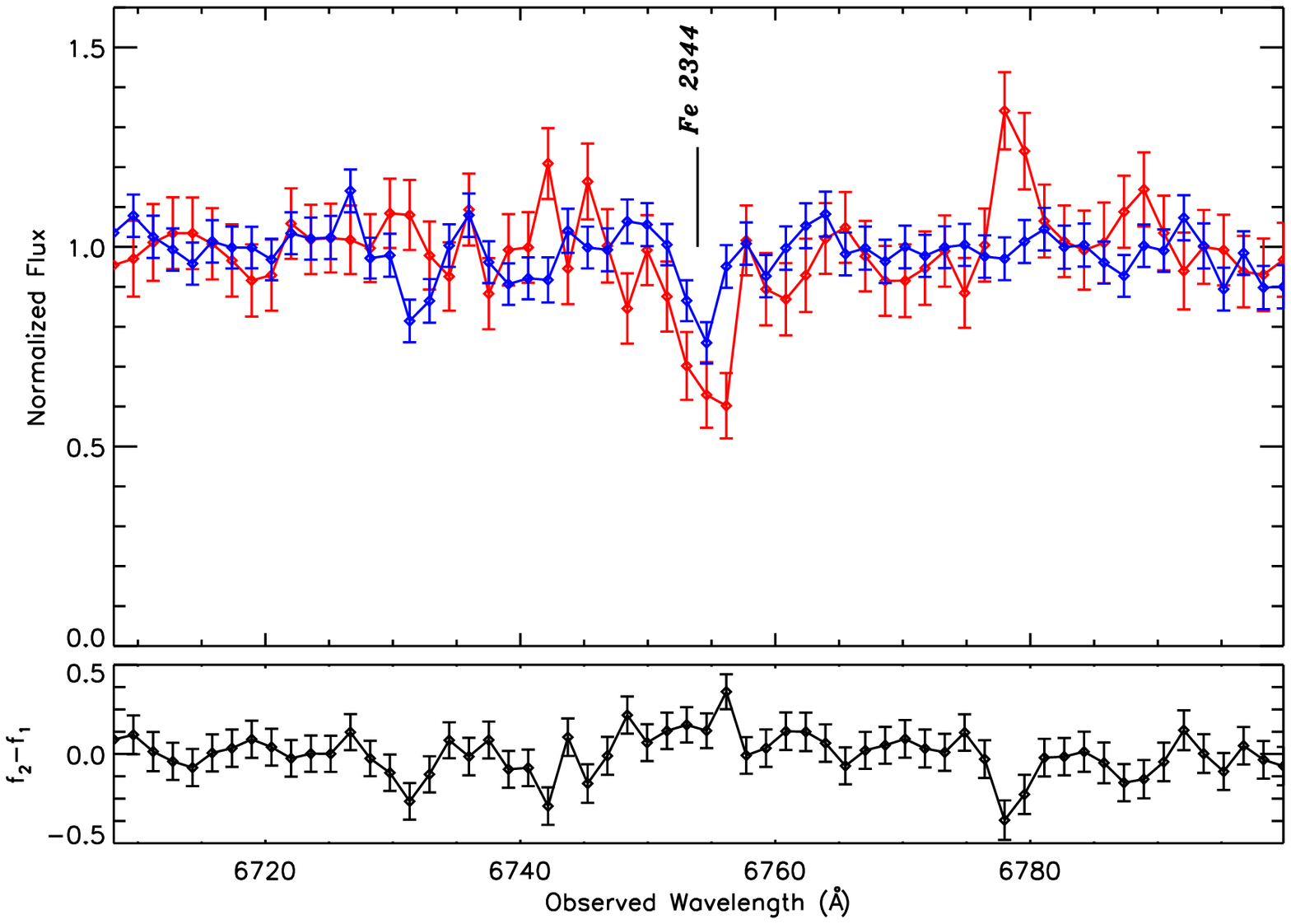}
\includegraphics[width=84mm]{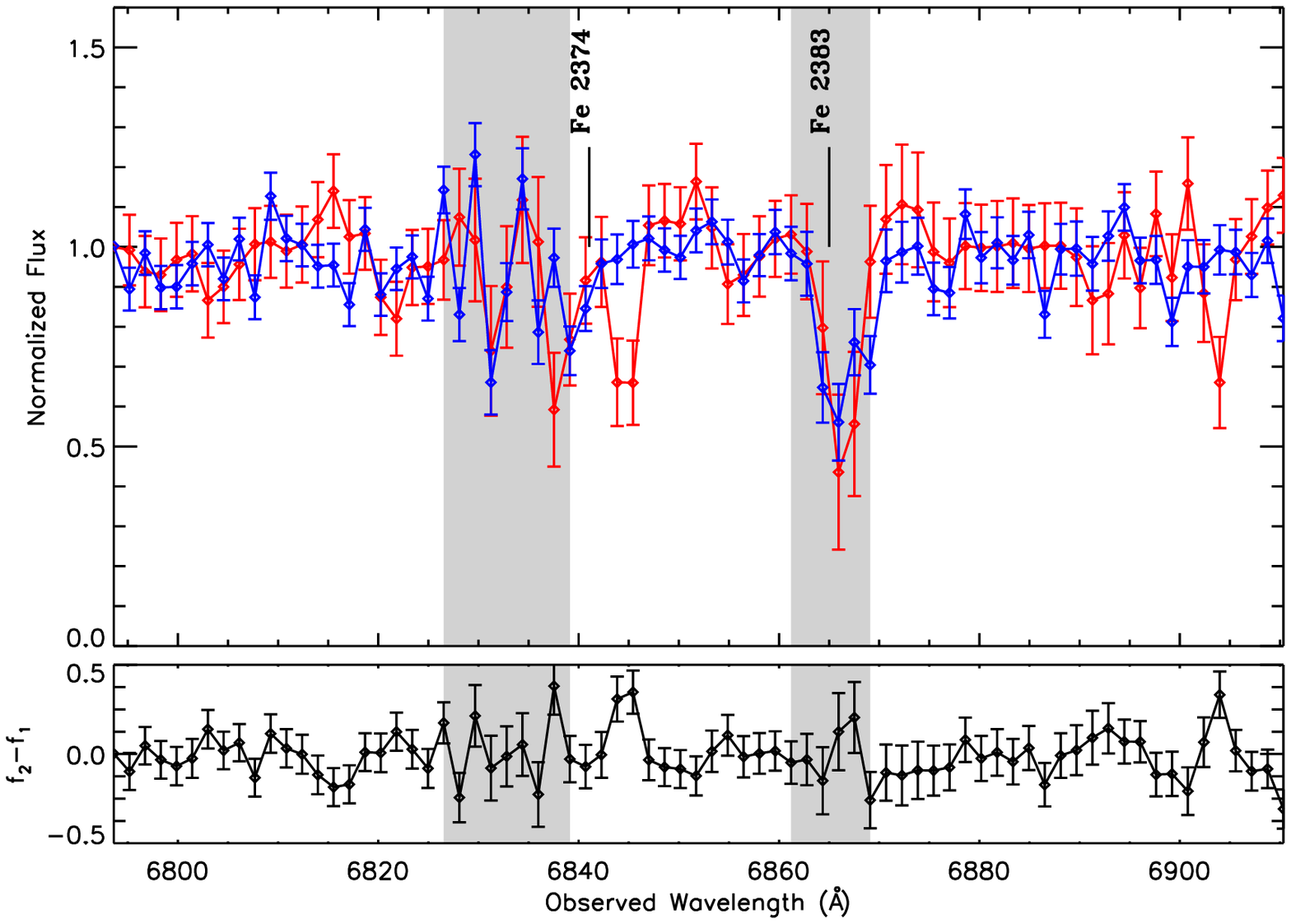}
\includegraphics[width=84mm]{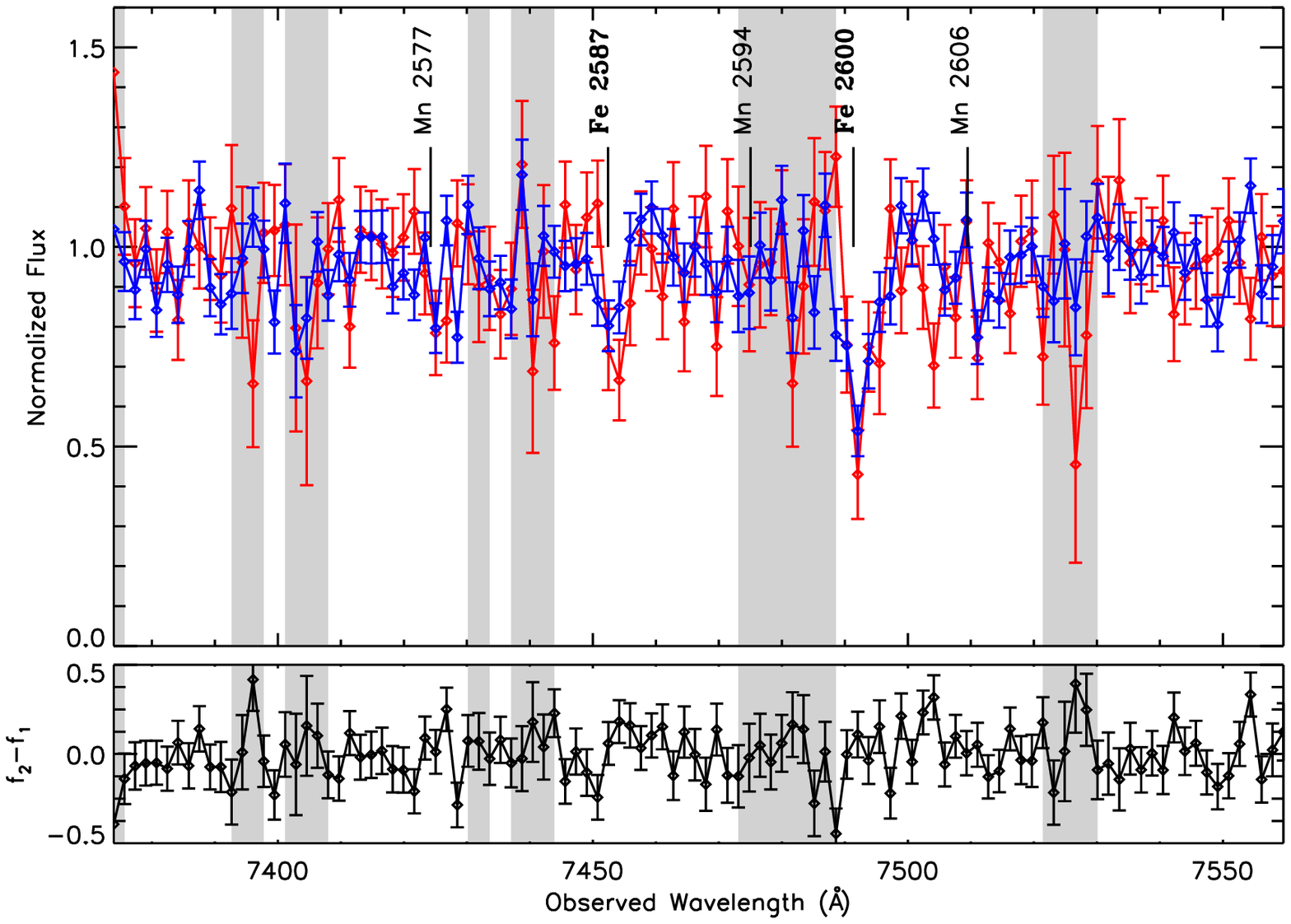}
\caption[Two-epoch normalized spectra of SDSS J004806.05+004623.6]{Two-epoch normalized spectra of the variable NAL system at $\beta$ = 0.1536 in SDSS J004806.05+004623.6.  The top panel shows the normalized pixel flux values with 1$\sigma$ error bars (first observations are red and second are blue), the bottom panel plots the difference spectrum of the two observation epochs, and shaded backgrounds identify masked pixels not included in our search for absorption line variability.  Line identifications for significantly variable absorption lines are italicised, lines detected in both observation epochs are in bold font, and undetected lines are in regular font (see Table A.1 for ion labels).  Continued in next figure. \label{figvs13}}
\end{center}
\end{figure*}

\begin{figure*}
\ContinuedFloat
\begin{center}
\includegraphics[width=84mm]{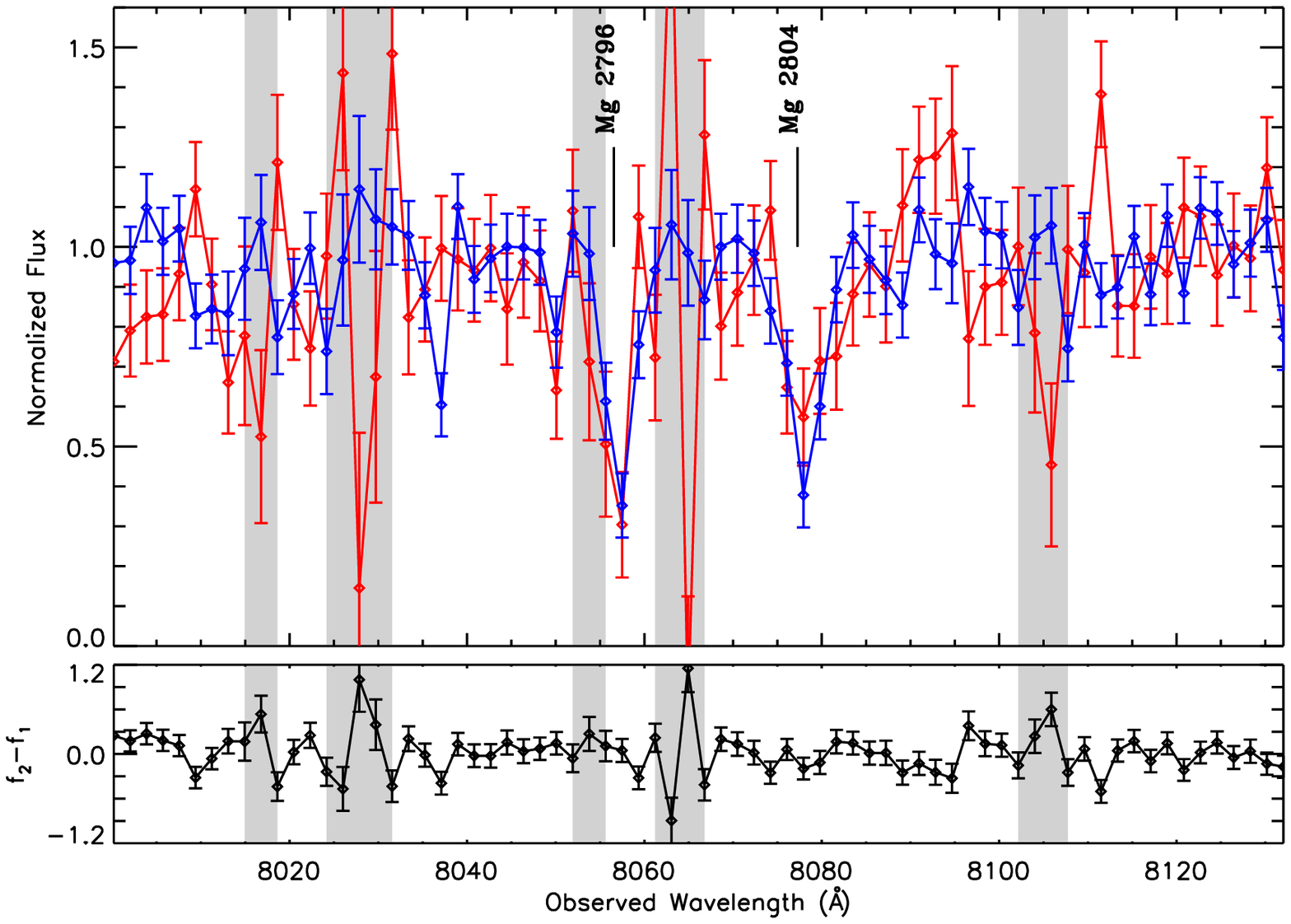}
\caption[]{Two-epoch normalized spectra of the variable NAL system at $\beta$ = 0.1536 in SDSS J004806.05+004623.6.  The top panel shows the normalized pixel flux values with 1$\sigma$ error bars (first observations are red and second are blue), the bottom panel plots the difference spectrum of the two observation epochs, and shaded backgrounds identify masked pixels not included in our search for absorption line variability.  Line identifications for significantly variable absorption lines are italicised, lines detected in both observation epochs are in bold font, and undetected lines are in regular font (see Table A.1 for ion labels).  Continued from previous figure.}
\vspace{3.5cm}
\end{center}
\end{figure*}

\begin{figure*}
\begin{center}
\includegraphics[width=84mm]{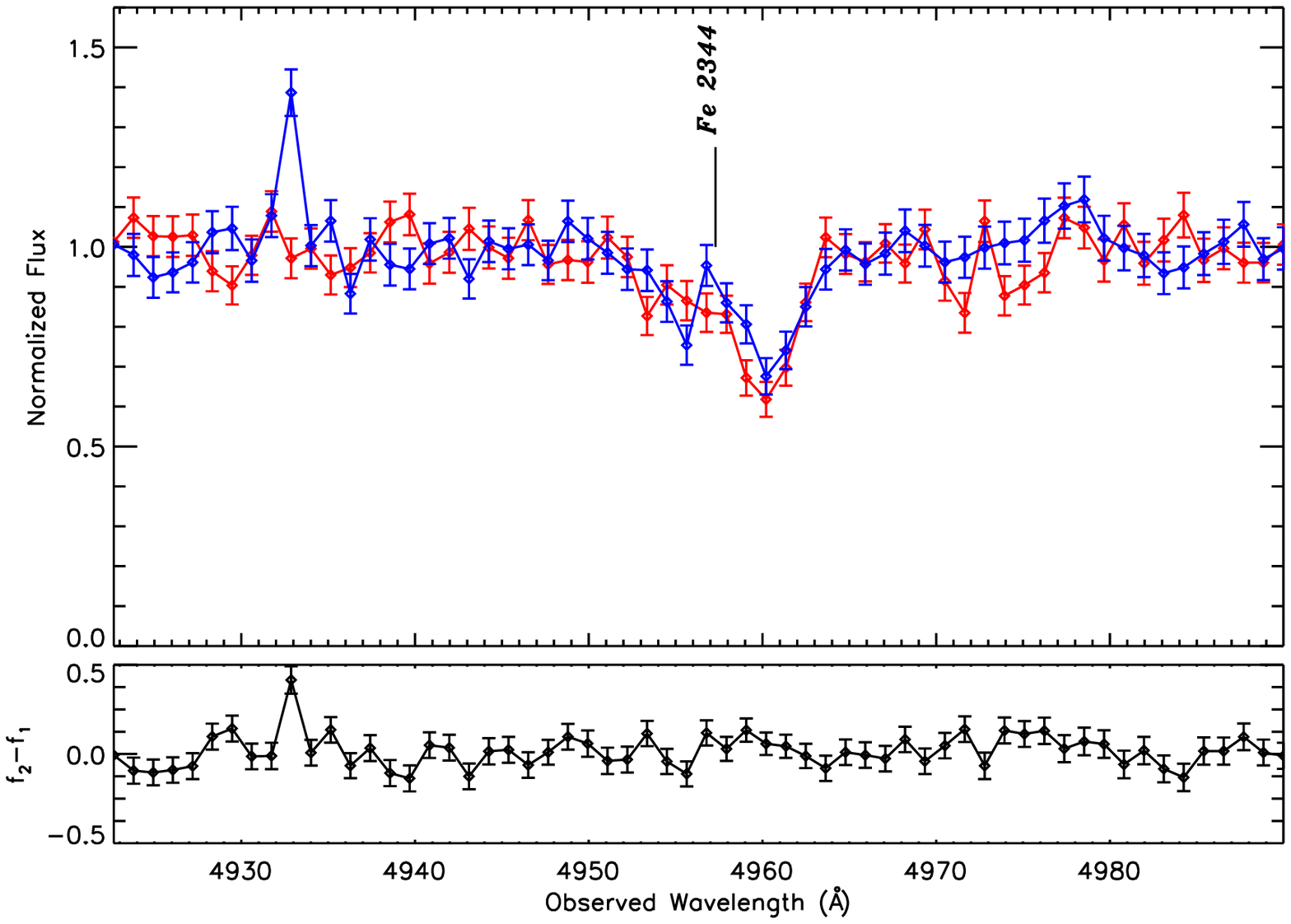}
\includegraphics[width=84mm]{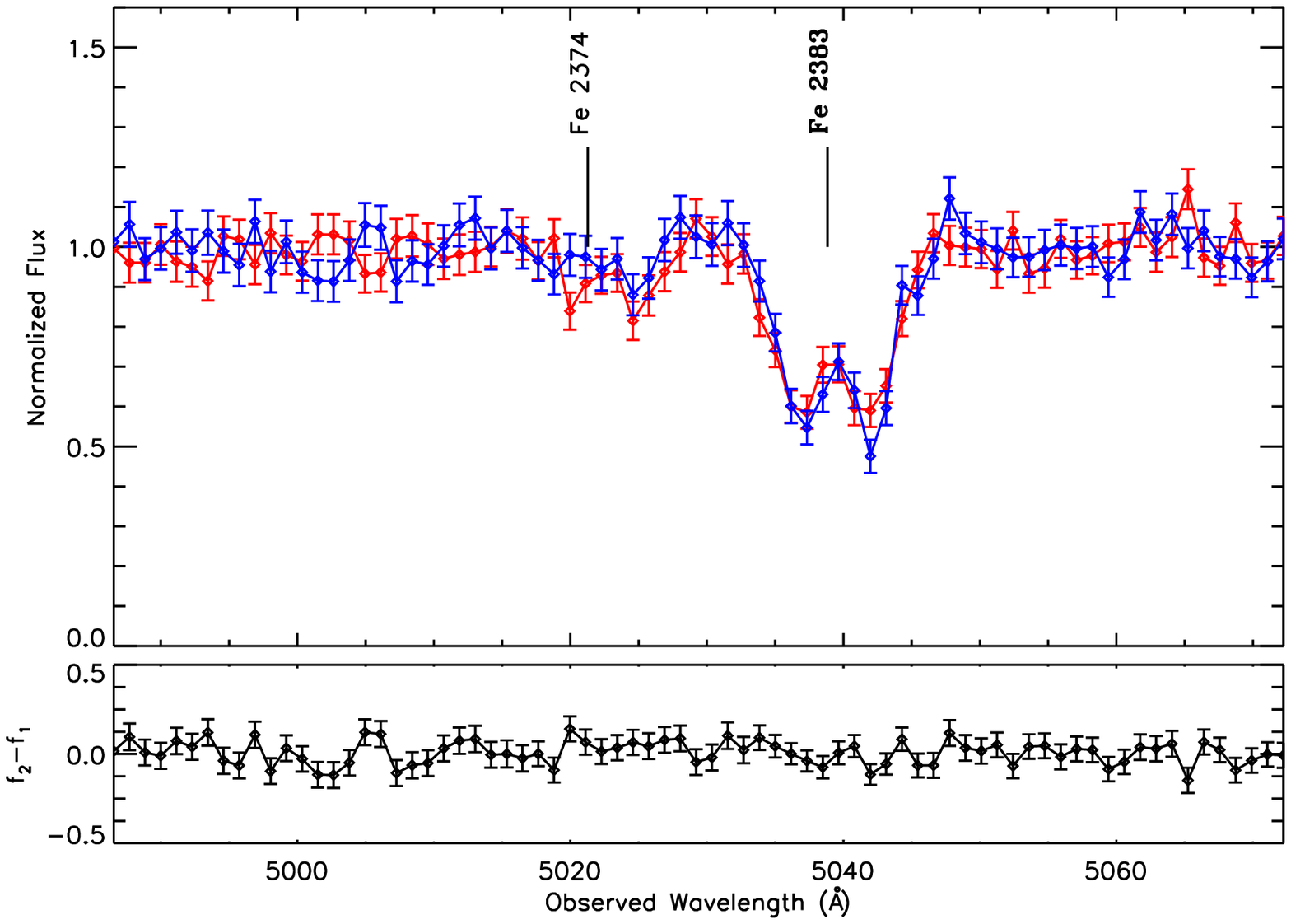}
\includegraphics[width=84mm]{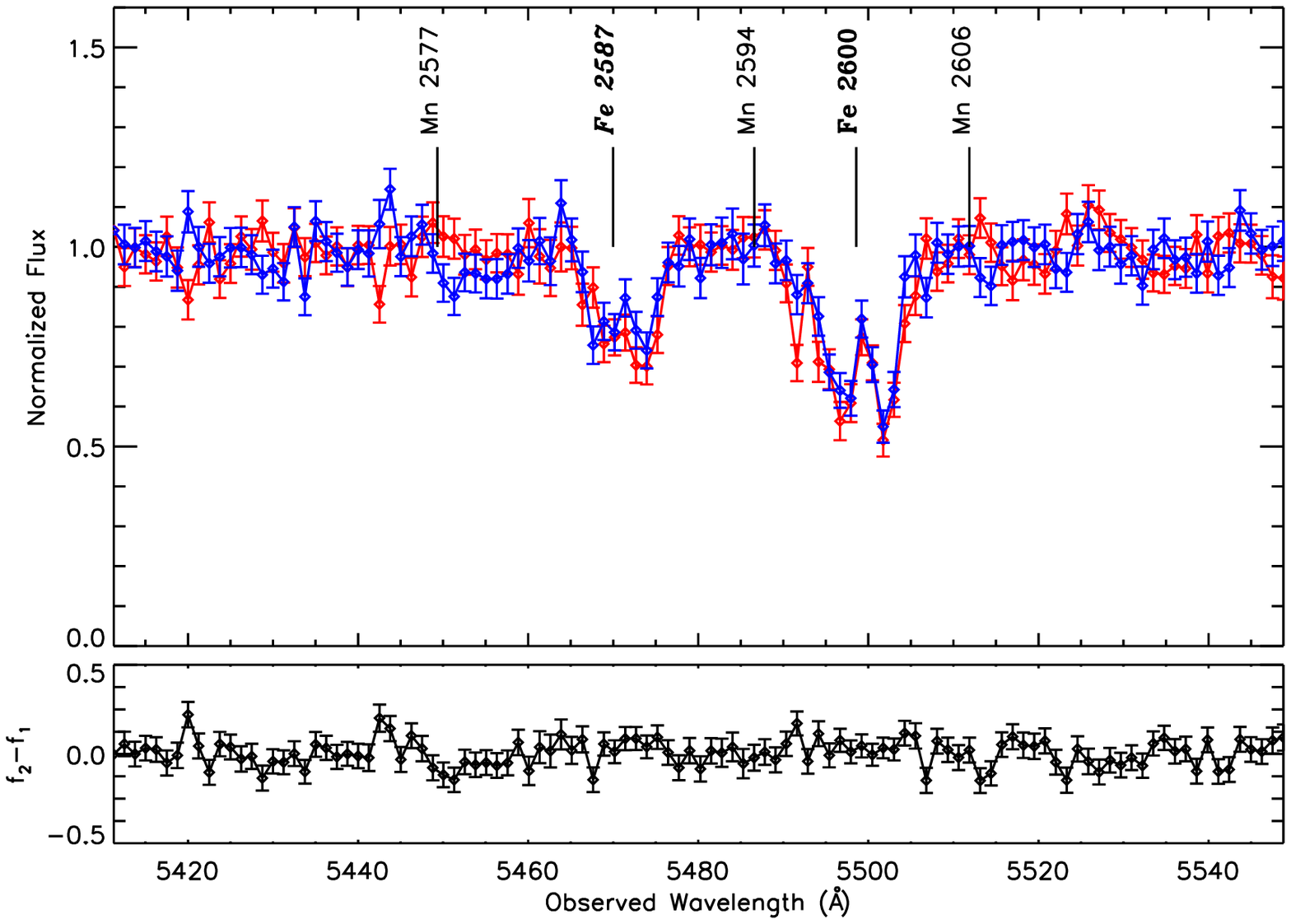}
\includegraphics[width=84mm]{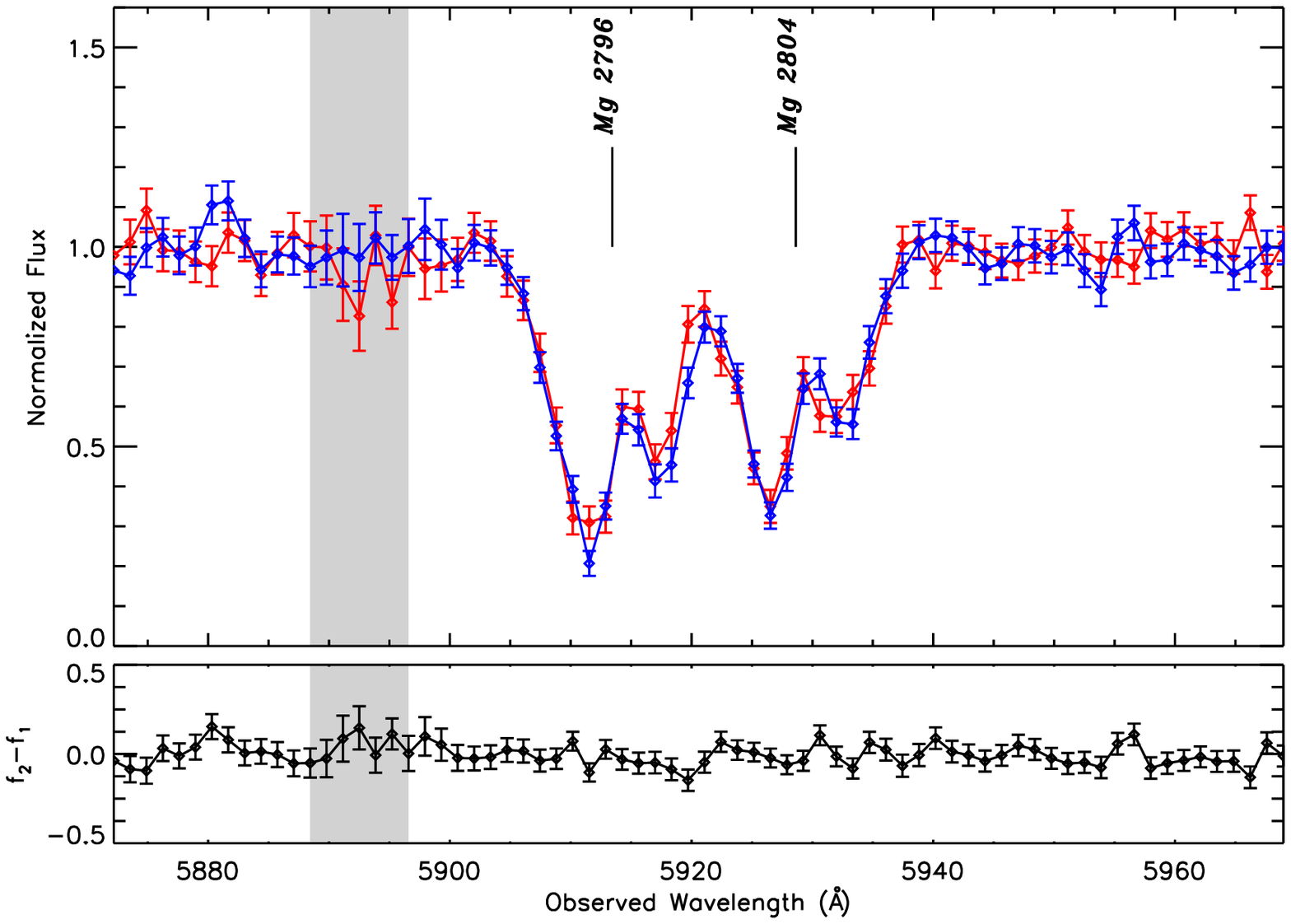}
\includegraphics[width=84mm]{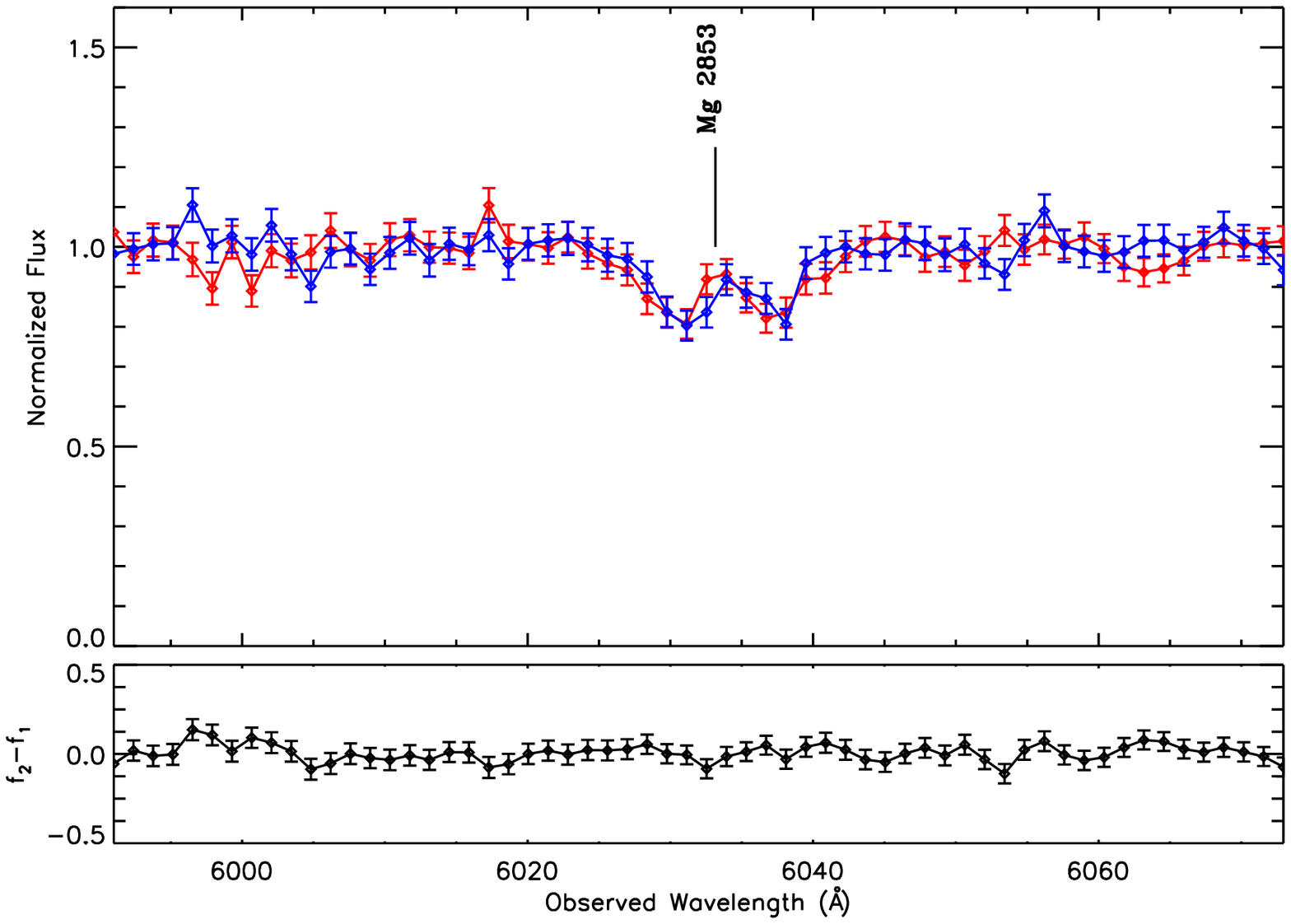}
\caption[Two-epoch normalized spectra of SDSS J152555.81+010835.4]{Two-epoch normalized spectra of the variable NAL system at $\beta$ = 0.1513 in SDSS J152555.81+010835.4.  The top panel shows the normalized pixel flux values with 1$\sigma$ error bars (first observations are red and second are blue), the bottom panel plots the difference spectrum of the two observation epochs, and shaded backgrounds identify masked pixels not included in our search for absorption line variability.  Line identifications for significantly variable absorption lines are italicised, lines detected in both observation epochs are in bold font, and undetected lines are in regular font (see Table A.1 for ion labels).  \label{figvs14}}
\end{center}
\end{figure*}

\begin{figure*}
\begin{center}
\includegraphics[width=84mm]{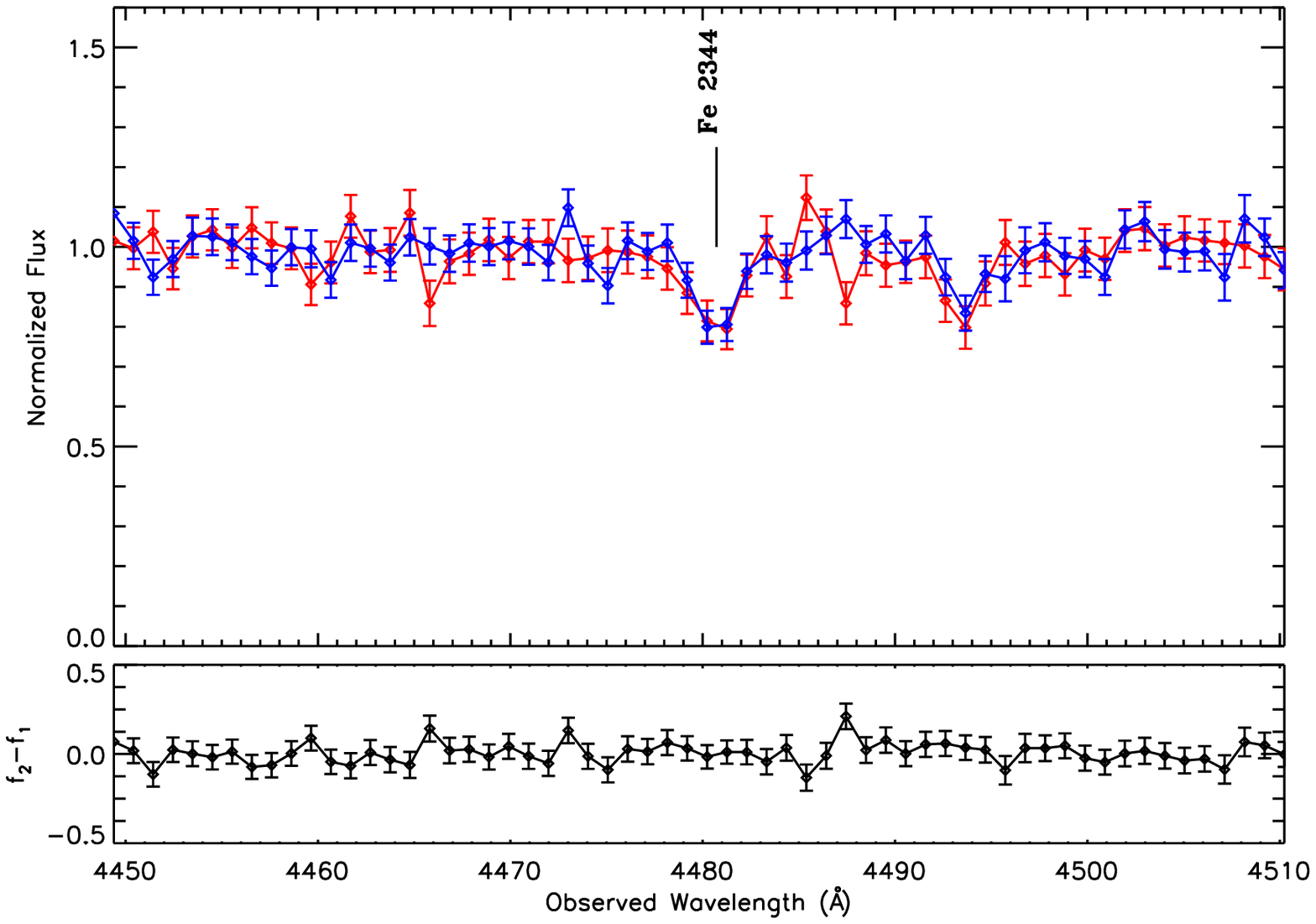}
\includegraphics[width=84mm]{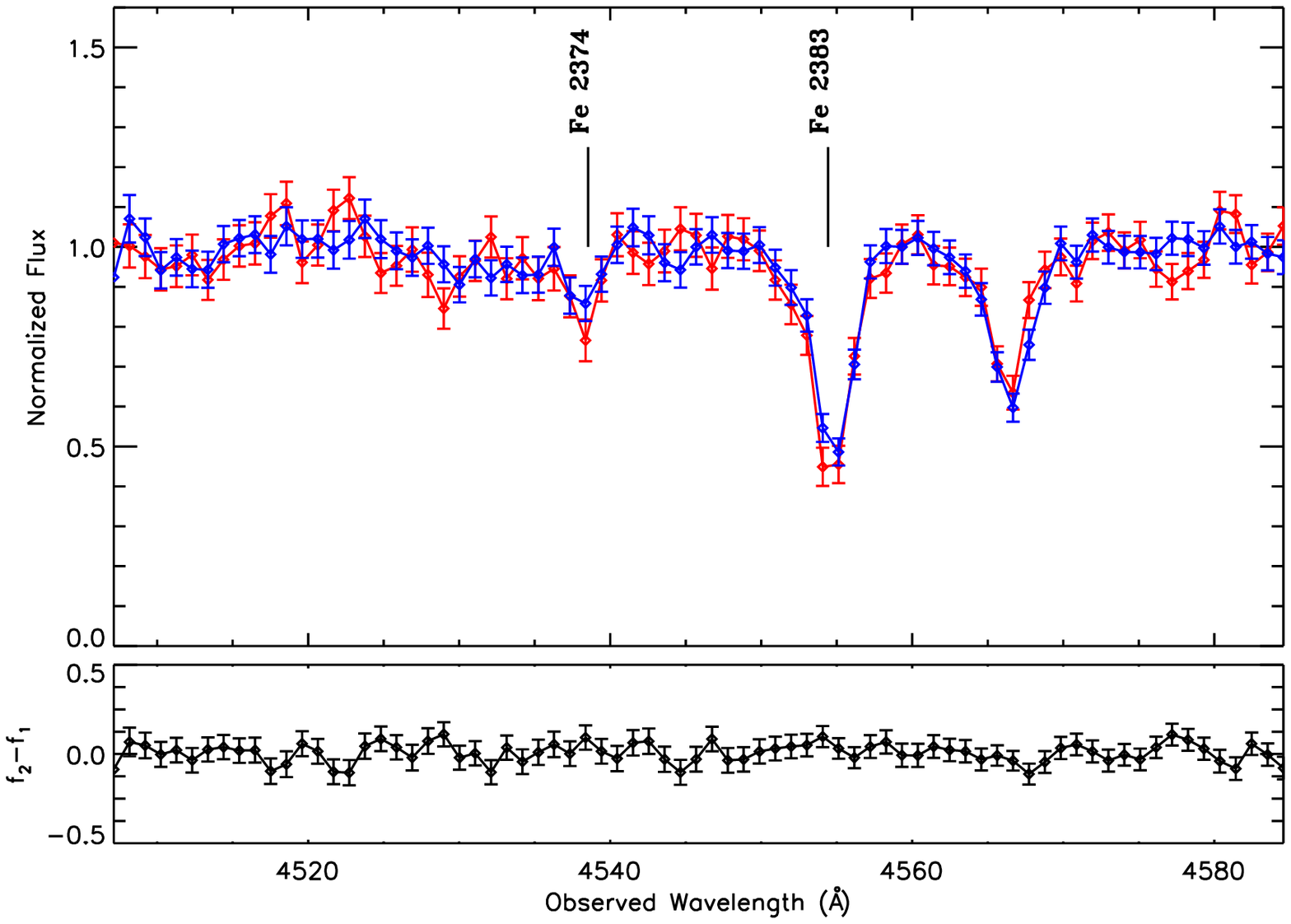}
\includegraphics[width=84mm]{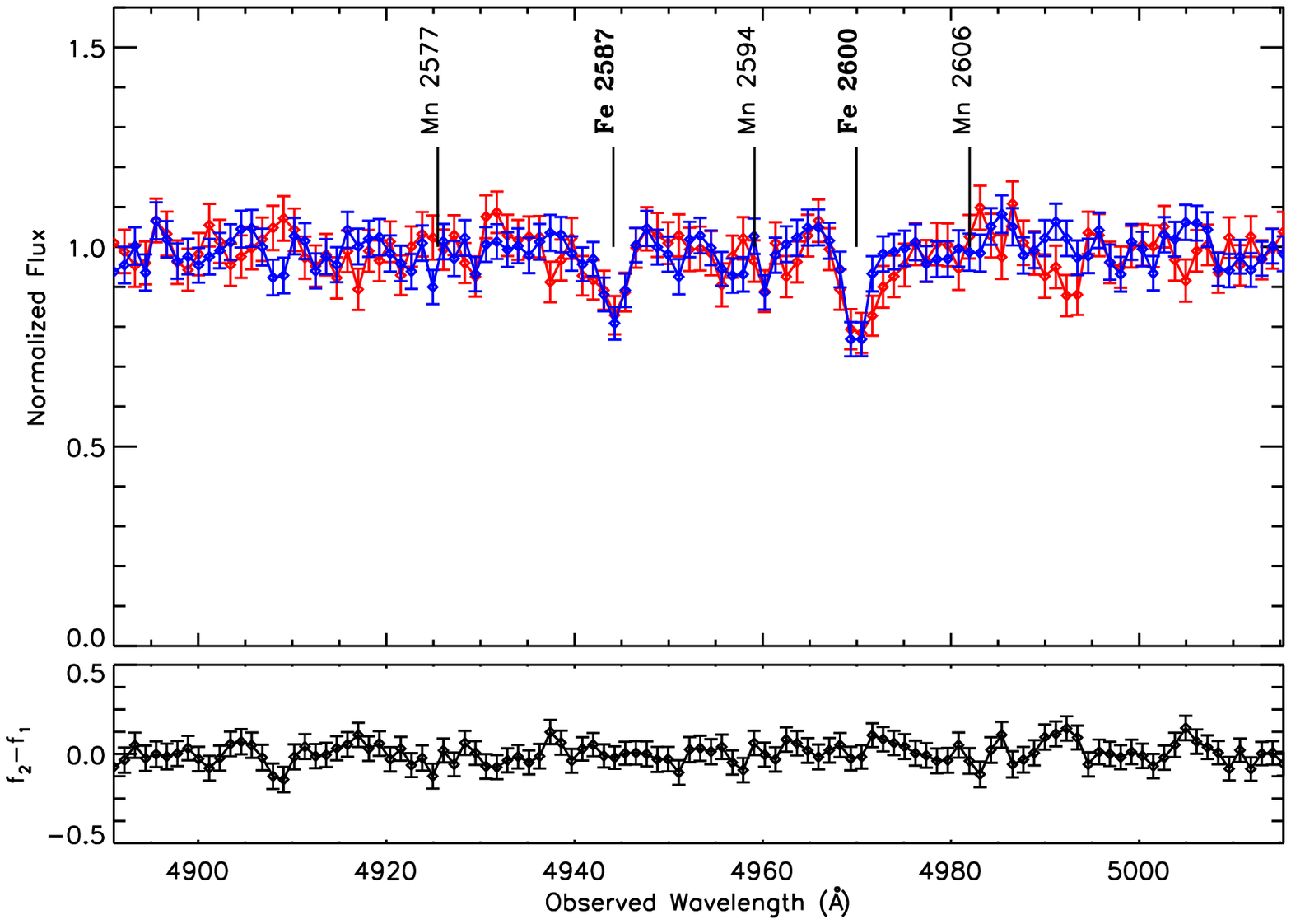}
\includegraphics[width=84mm]{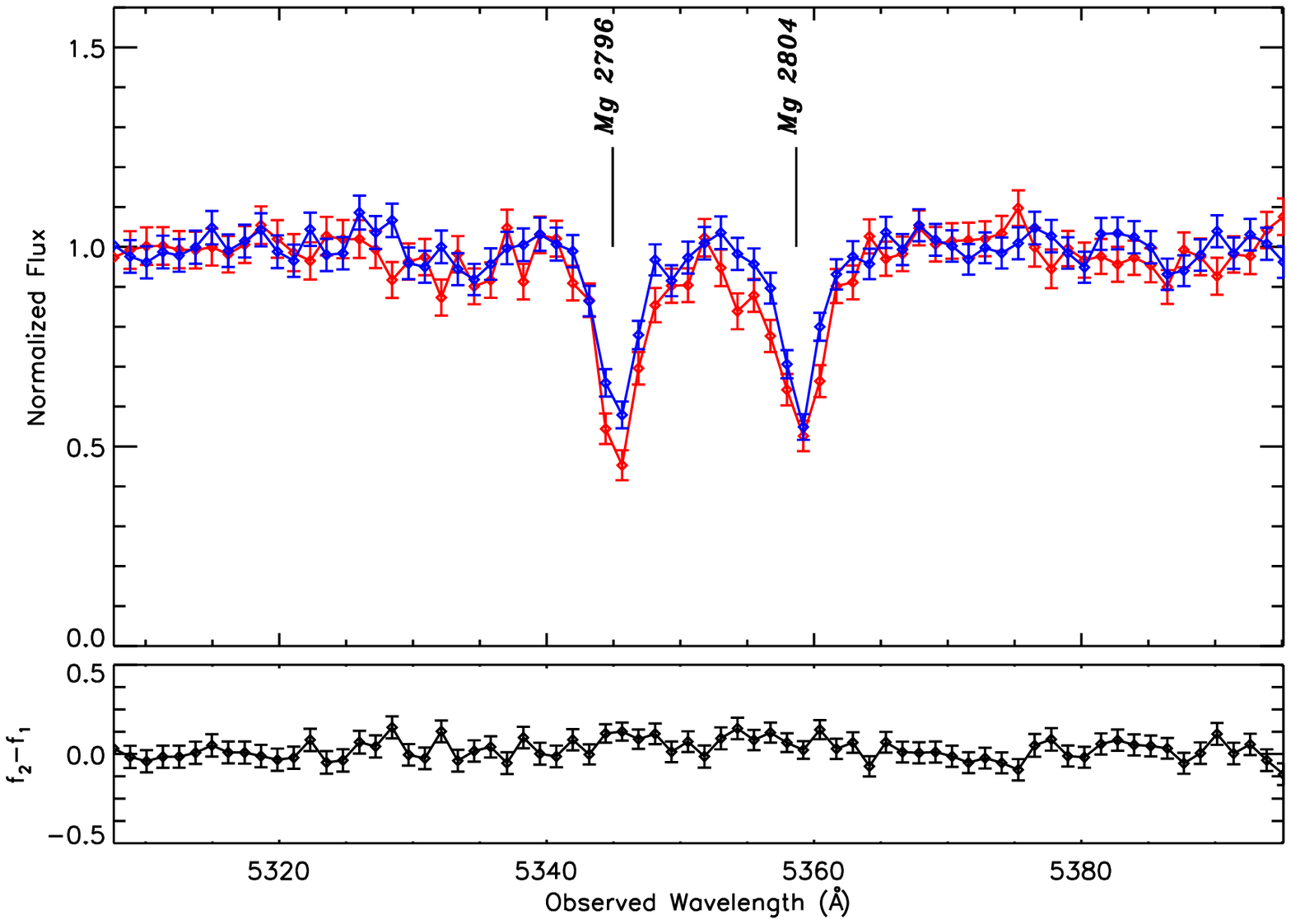}
\caption[Two-epoch normalized spectra of SDSS J143826.73+642859.8]{Two-epoch normalized spectra of the variable NAL system at $\beta$ = 0.1492 in SDSS J143826.73+642859.8.  The top panel shows the normalized pixel flux values with 1$\sigma$ error bars (first observations are red and second are blue), the bottom panel plots the difference spectrum of the two observation epochs, and shaded backgrounds identify masked pixels not included in our search for absorption line variability.  Line identifications for significantly variable absorption lines are italicised, lines detected in both observation epochs are in bold font, and undetected lines are in regular font (see Table A.1 for ion labels).  \label{figvs15}}
\end{center}
\end{figure*}

\begin{figure*}
\begin{center}
\includegraphics[width=84mm]{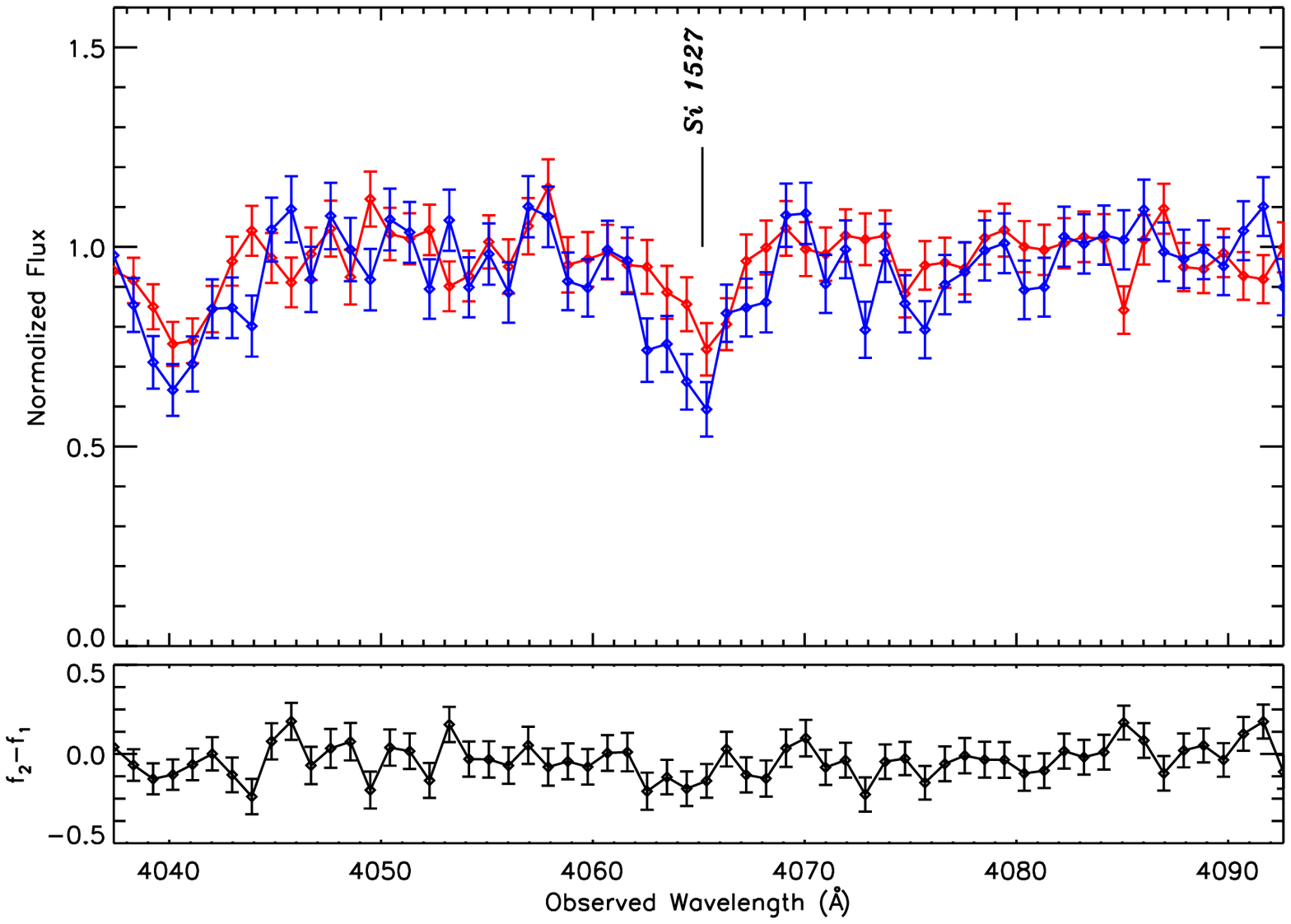}
\includegraphics[width=84mm]{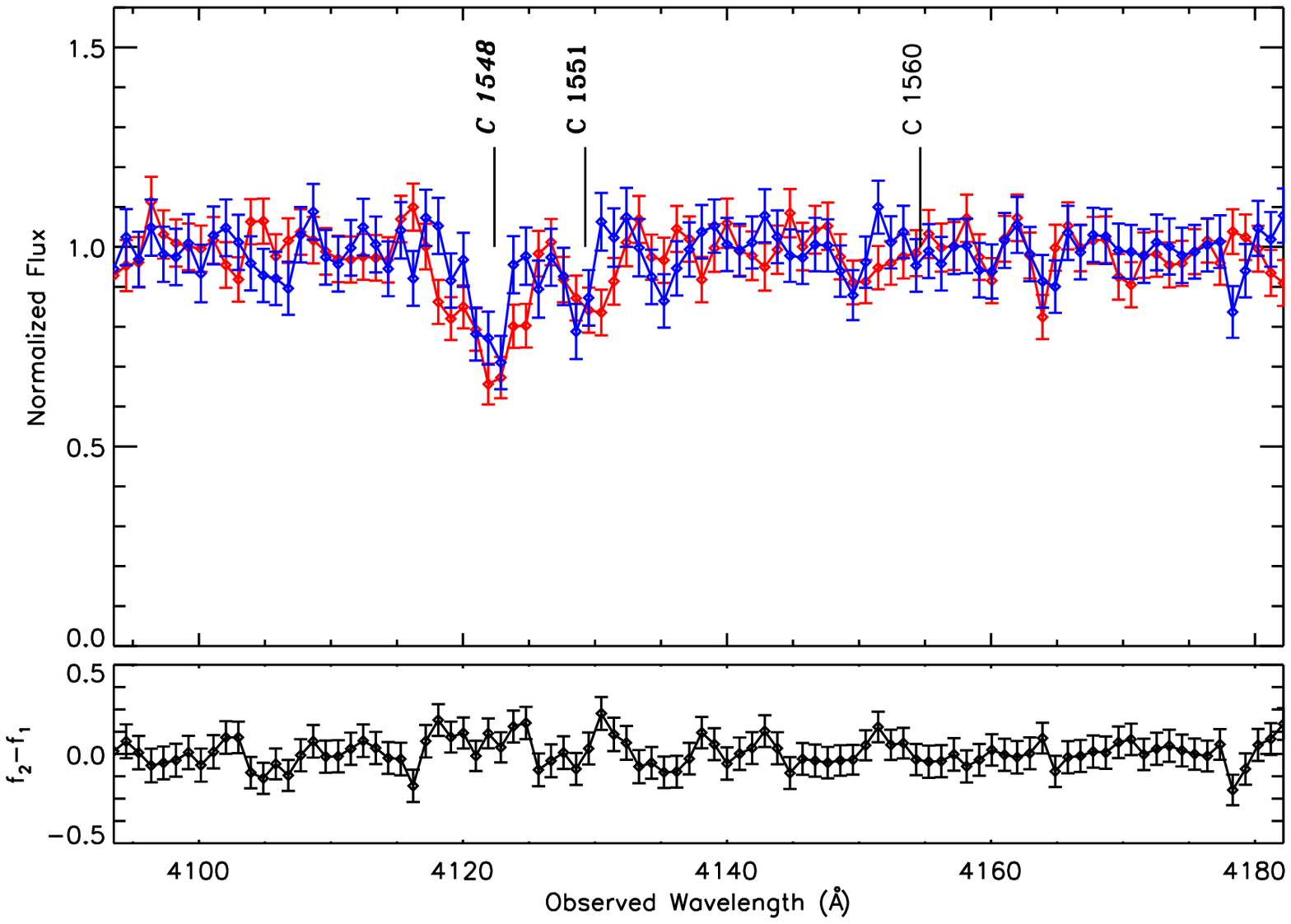}
\includegraphics[width=84mm]{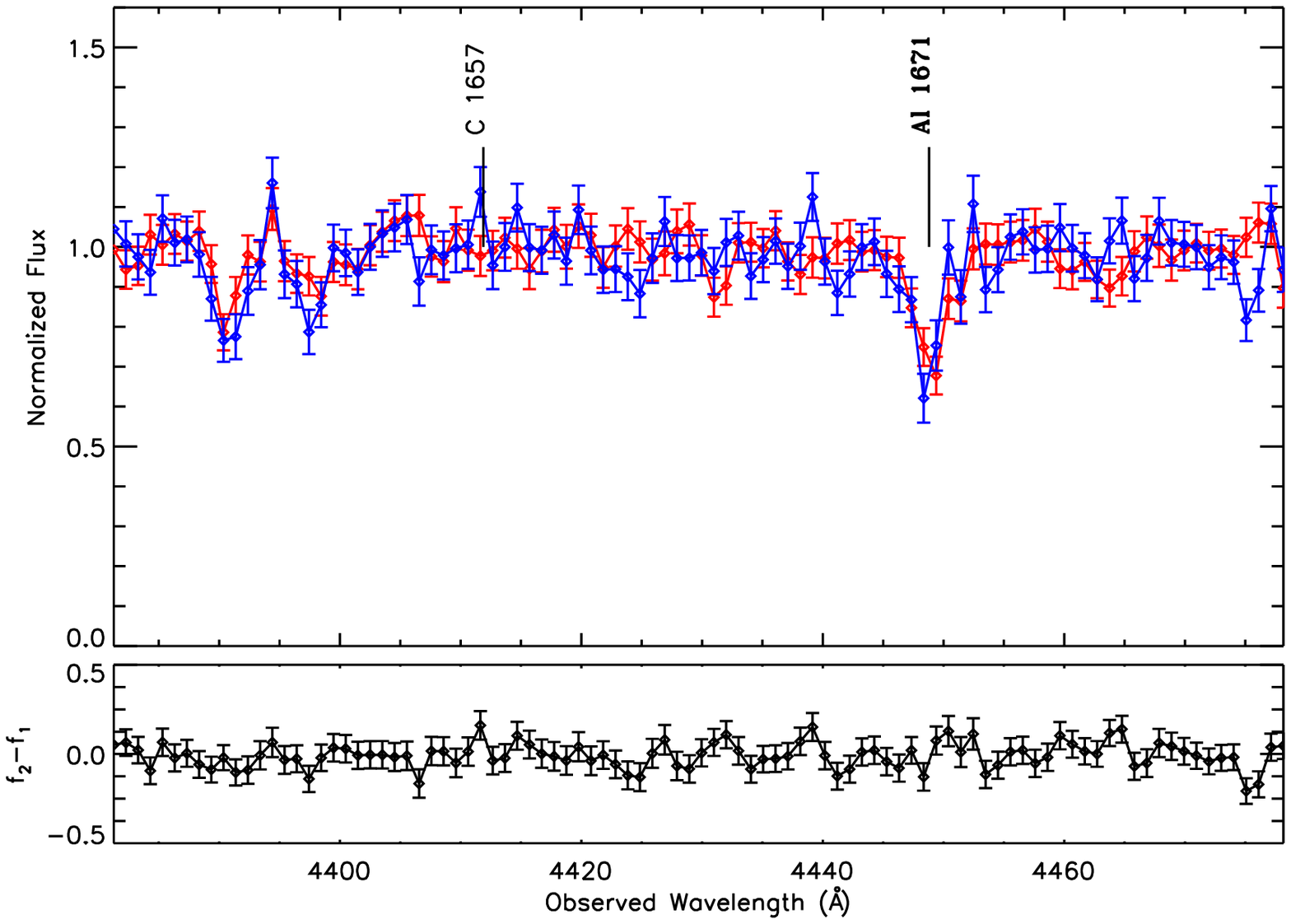}
\includegraphics[width=84mm]{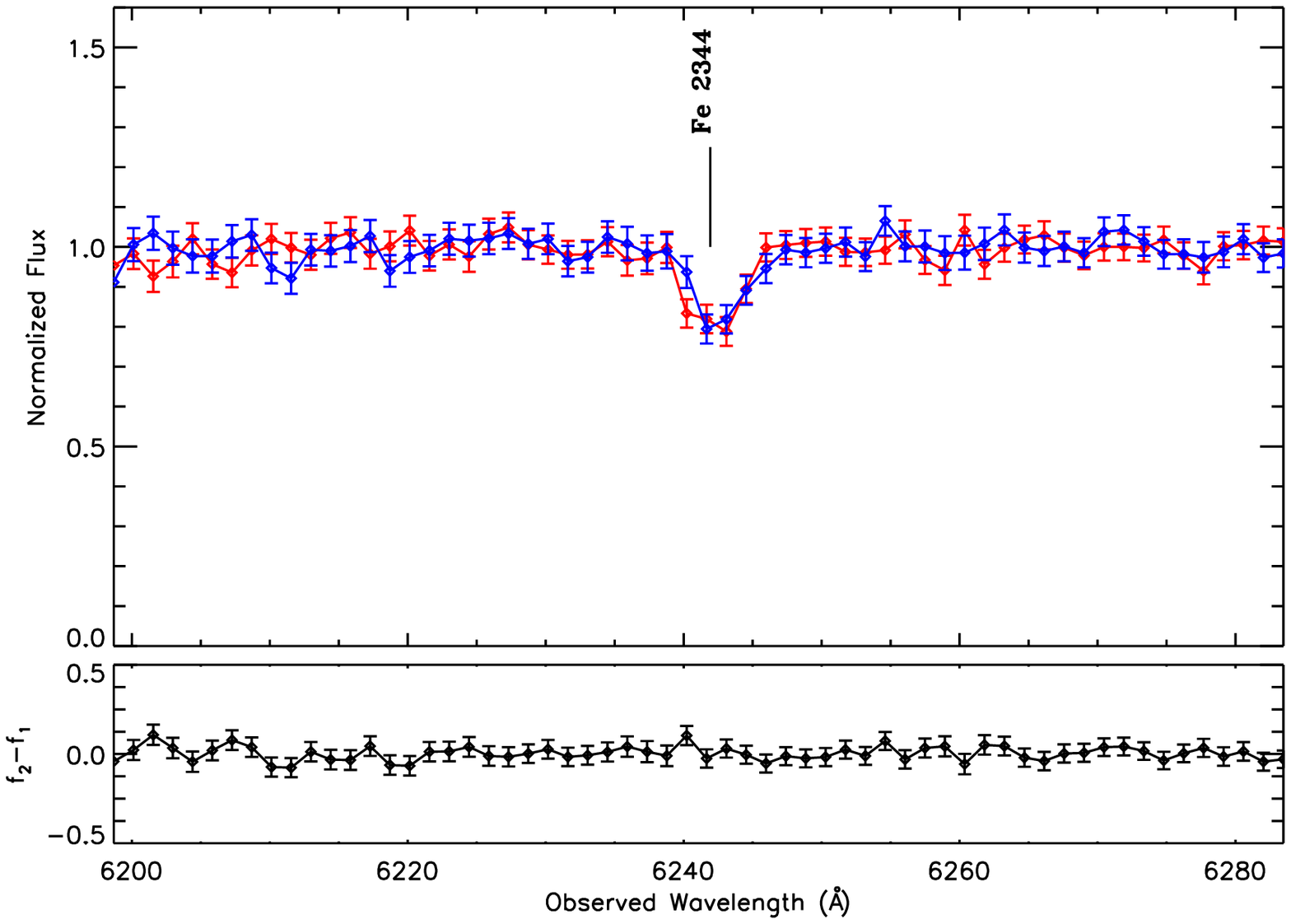}
\includegraphics[width=84mm]{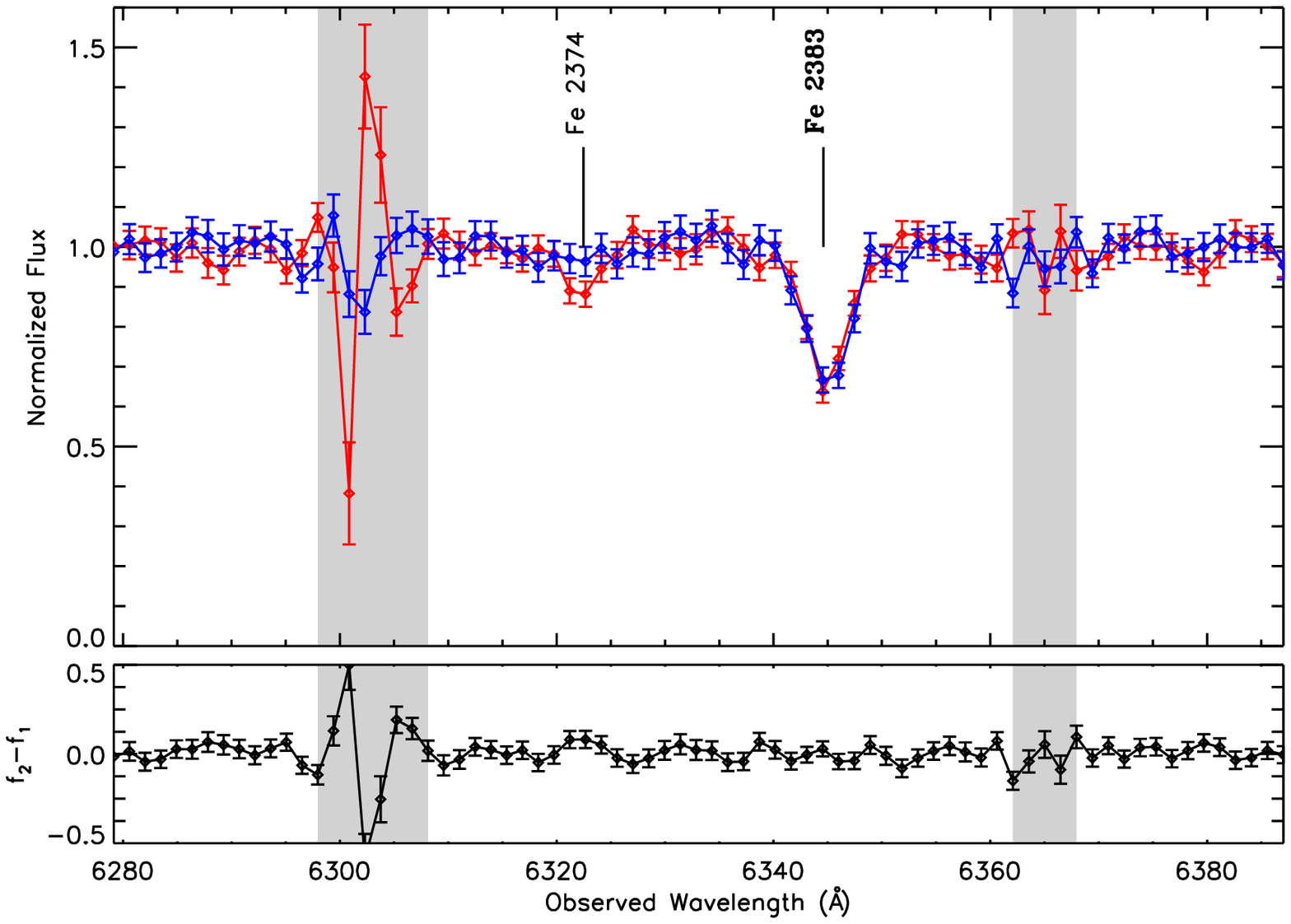}
\includegraphics[width=84mm]{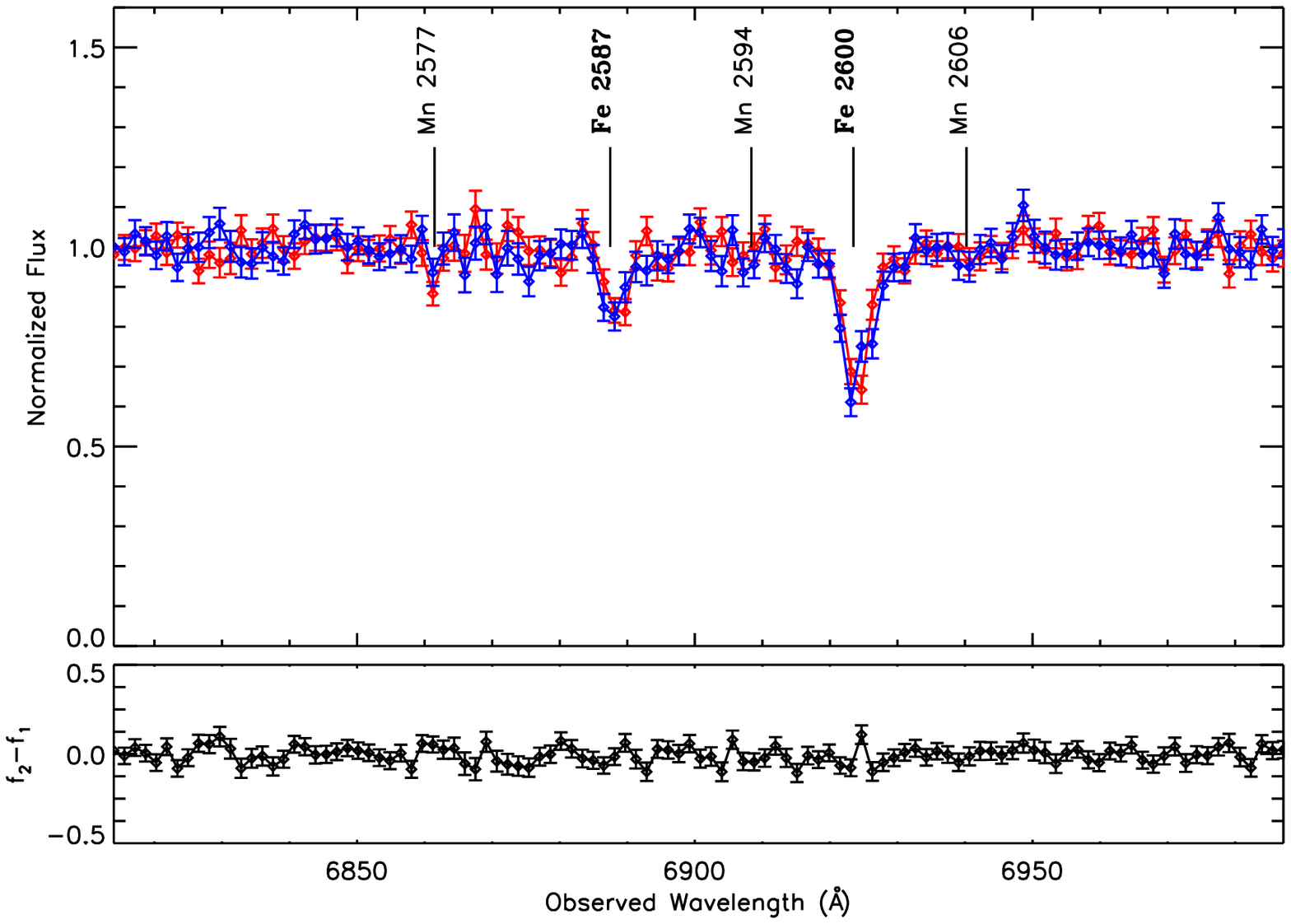}
\caption[Two-epoch normalized spectra of SDSS J143229.24-010616.0]{Two-epoch normalized spectra of the variable NAL system at $\beta$ = 0.1468 in SDSS J143229.24-010616.0.  The top panel shows the normalized pixel flux values with 1$\sigma$ error bars (first observations are red and second are blue), the bottom panel plots the difference spectrum of the two observation epochs, and shaded backgrounds identify masked pixels not included in our search for absorption line variability.  Line identifications for significantly variable absorption lines are italicised, lines detected in both observation epochs are in bold font, and undetected lines are in regular font (see Table A.1 for ion labels).  Continued in next figure.  \label{figvs16}}
\end{center}
\end{figure*}

\begin{figure*}
\ContinuedFloat
\begin{center}
\includegraphics[width=84mm]{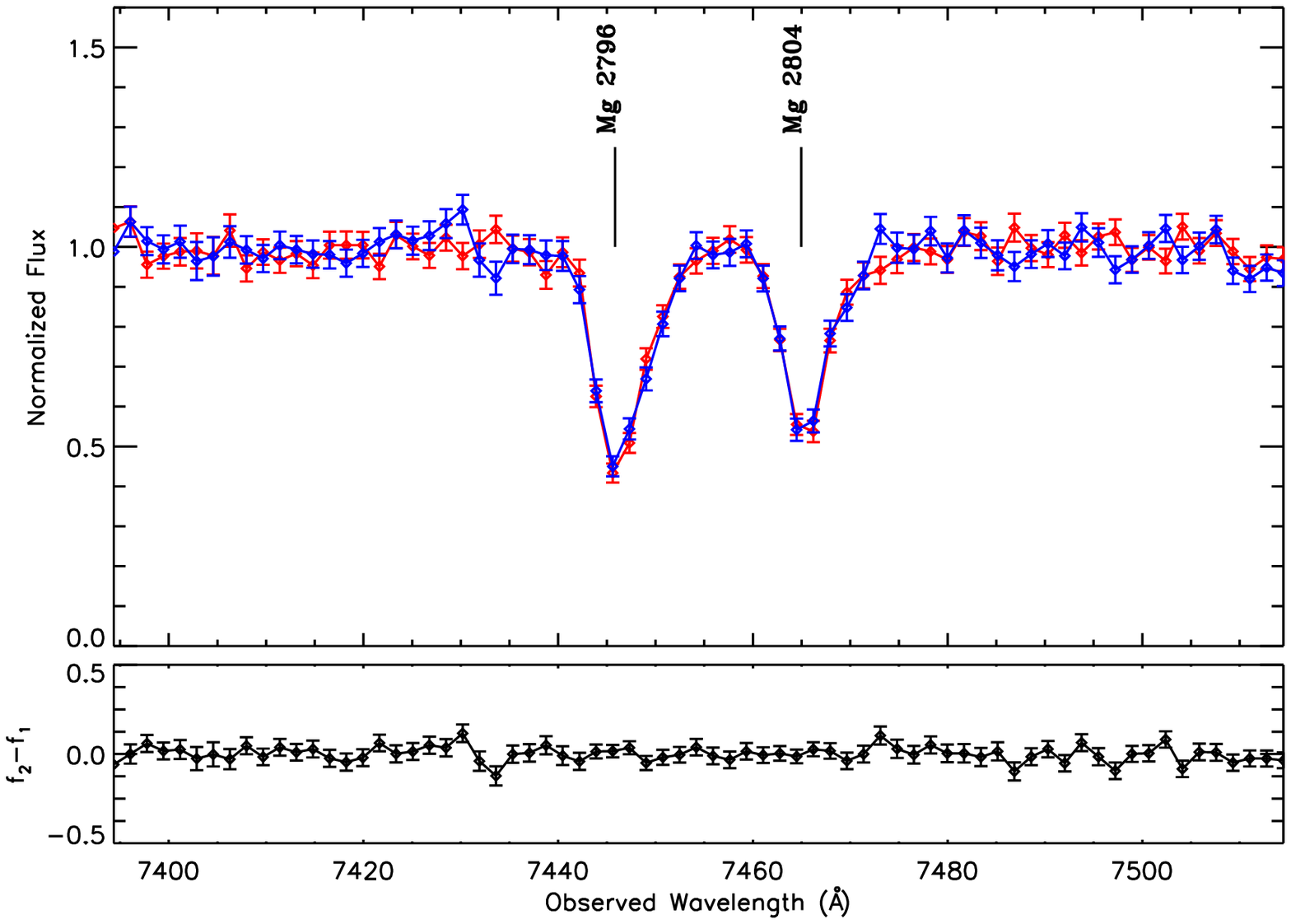}
\includegraphics[width=84mm]{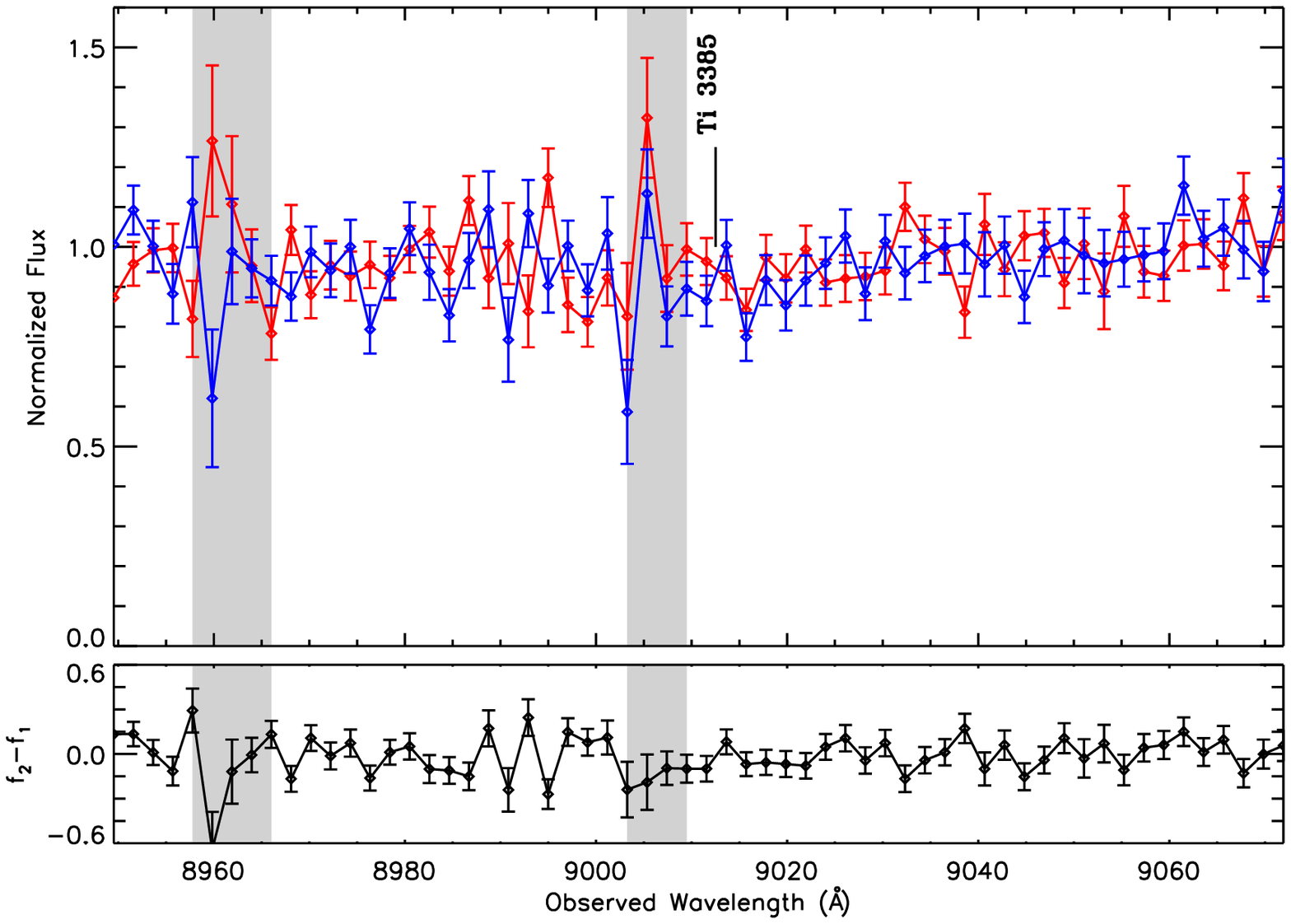}
\caption[]{Two-epoch normalized spectra of the variable NAL system at $\beta$ = 0.1468 in SDSS J143229.24-010616.0.  The top panel shows the normalized pixel flux values with 1$\sigma$ error bars (first observations are red and second are blue), the bottom panel plots the difference spectrum of the two observation epochs, and shaded backgrounds identify masked pixels not included in our search for absorption line variability.  Line identifications for significantly variable absorption lines are italicised, lines detected in both observation epochs are in bold font, and undetected lines are in regular font (see Table A.1 for ion labels).  Continued from previous figure.}
\vspace{3.5cm}
\end{center}
\end{figure*}

\begin{figure*}
\begin{center}
\includegraphics[width=84mm]{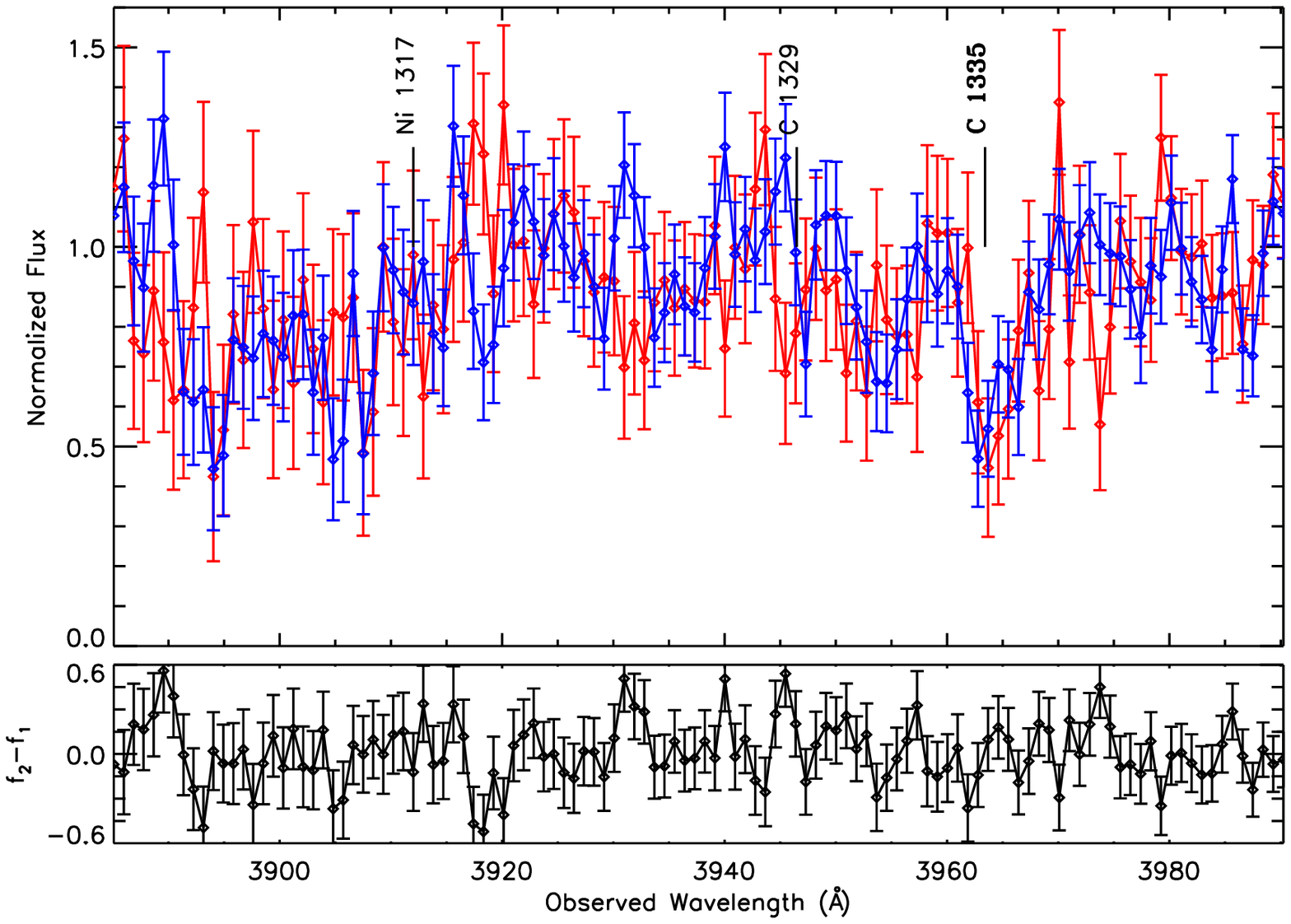}
\includegraphics[width=84mm]{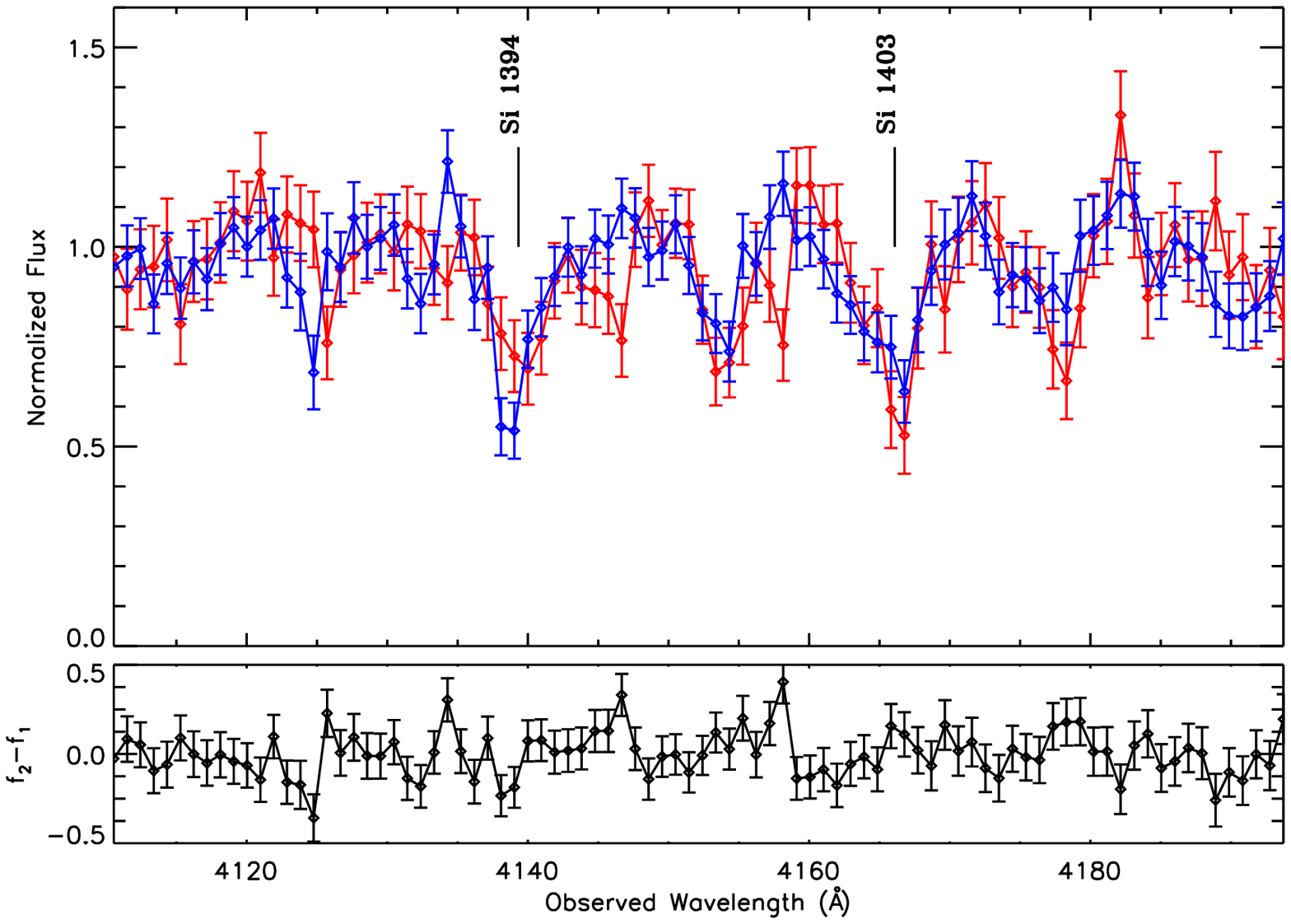}
\includegraphics[width=84mm]{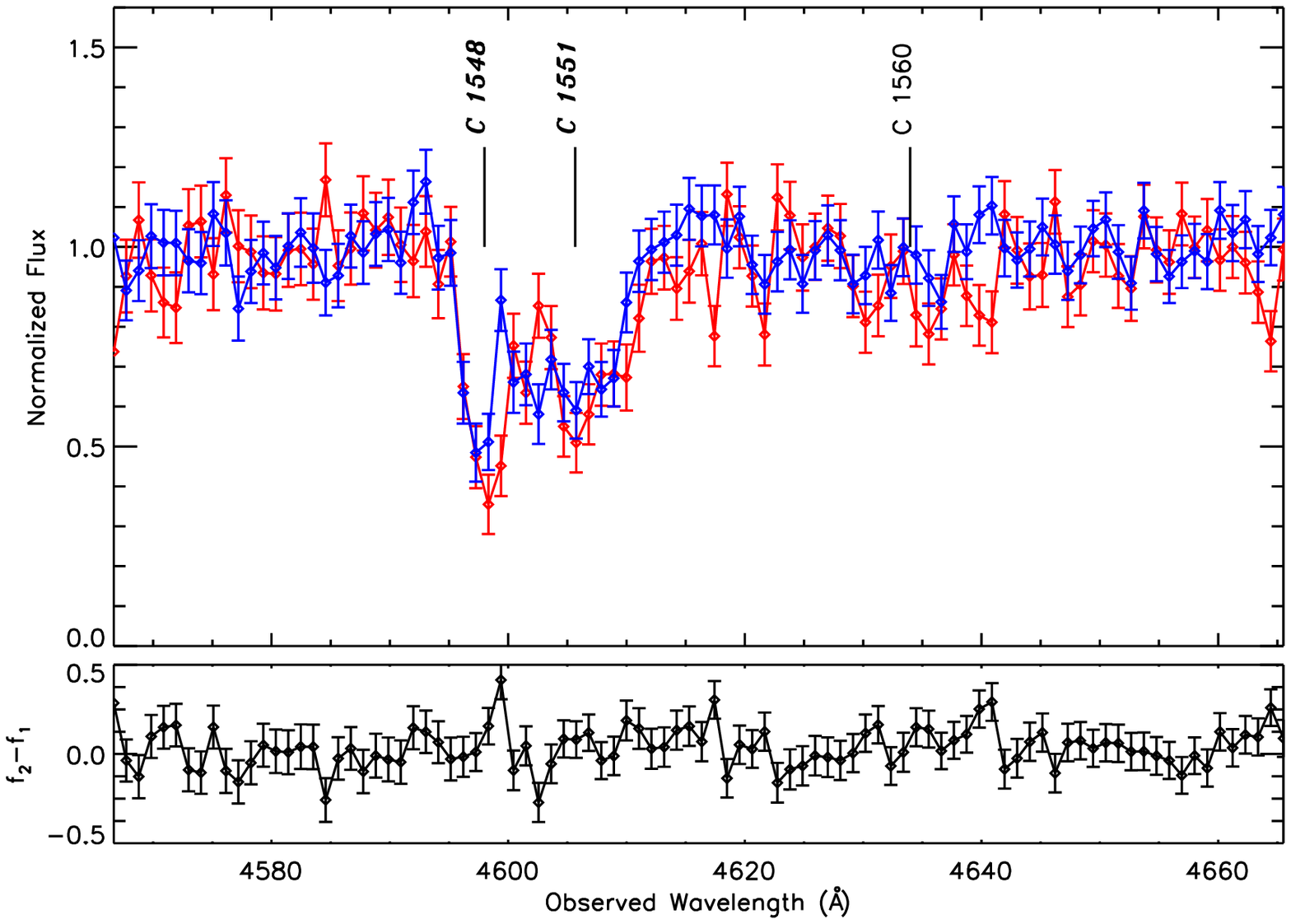}
\includegraphics[width=84mm]{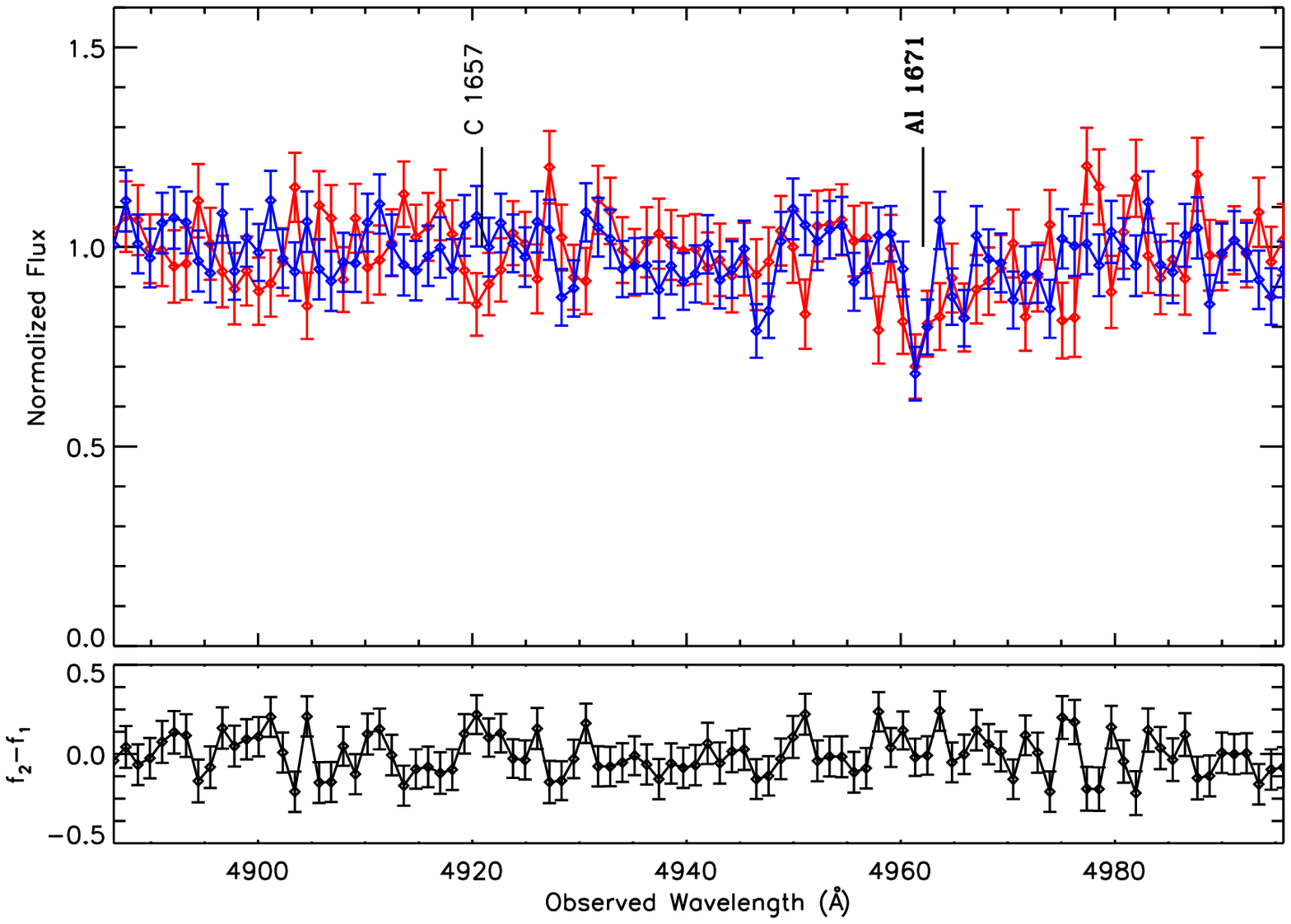}
\includegraphics[width=84mm]{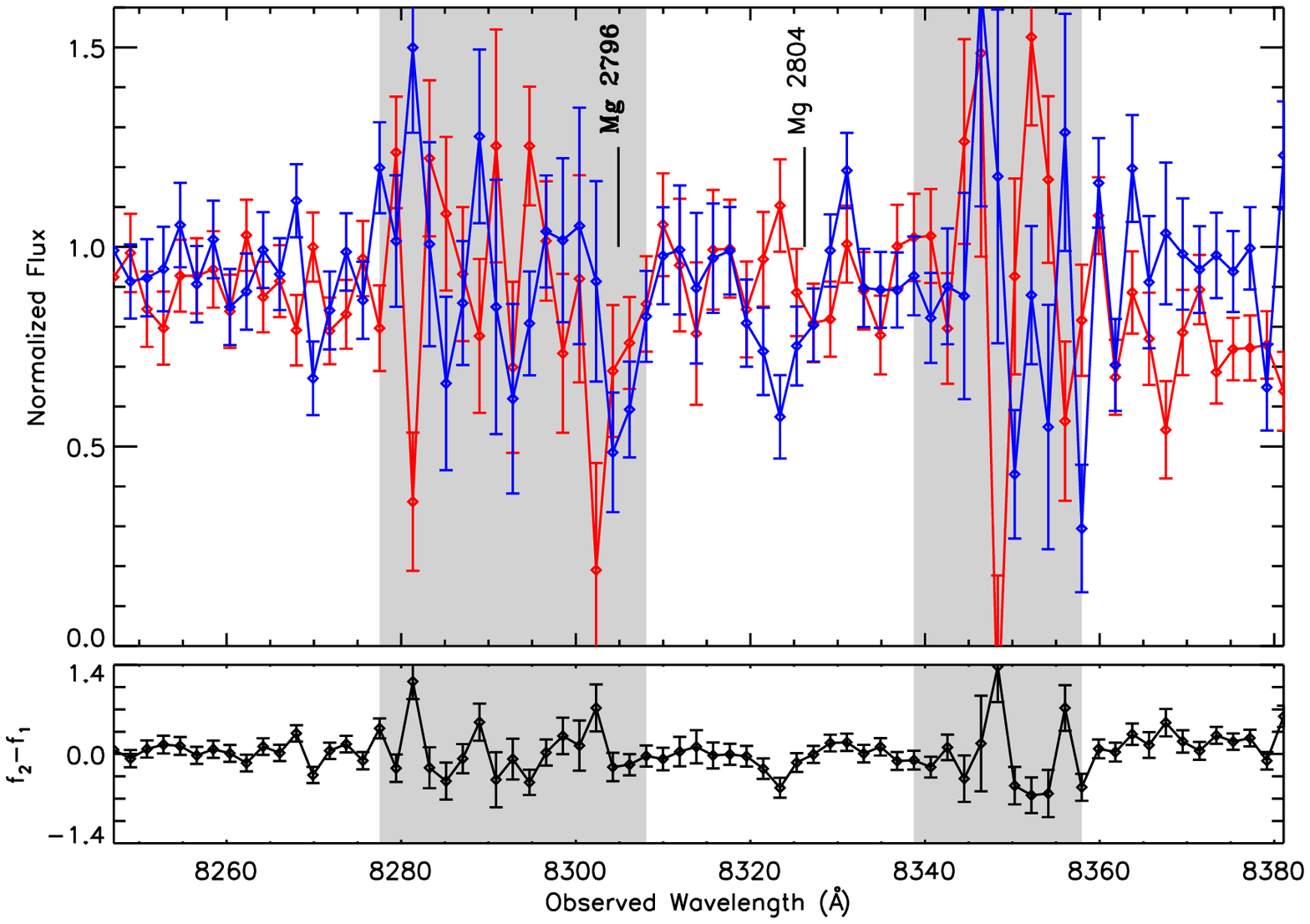}
\caption[Two-epoch normalized spectra of SDSS J022632.51-003841.4]{Two-epoch normalized spectra of the variable NAL system at $\beta$ = 0.1208 in SDSS J022632.51-003841.4.  The top panel shows the normalized pixel flux values with 1$\sigma$ error bars (first observations are red and second are blue), the bottom panel plots the difference spectrum of the two observation epochs, and shaded backgrounds identify masked pixels not included in our search for absorption line variability.  Line identifications for significantly variable absorption lines are italicised, lines detected in both observation epochs are in bold font, and undetected lines are in regular font (see Table A.1 for ion labels).  \label{figvs17}}
\end{center}
\end{figure*}

\begin{figure*}
\begin{center}
\includegraphics[width=84mm]{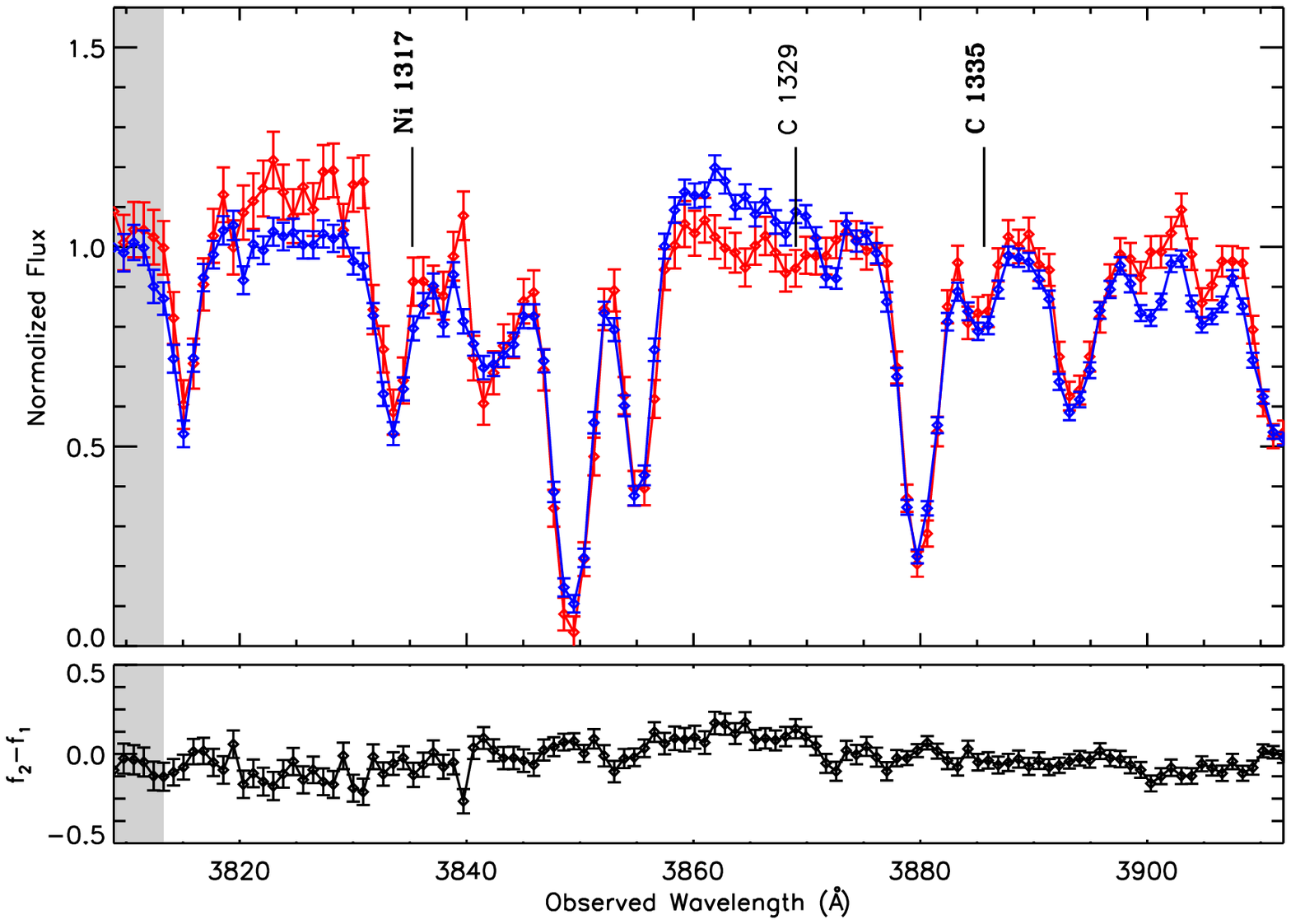}
\includegraphics[width=84mm]{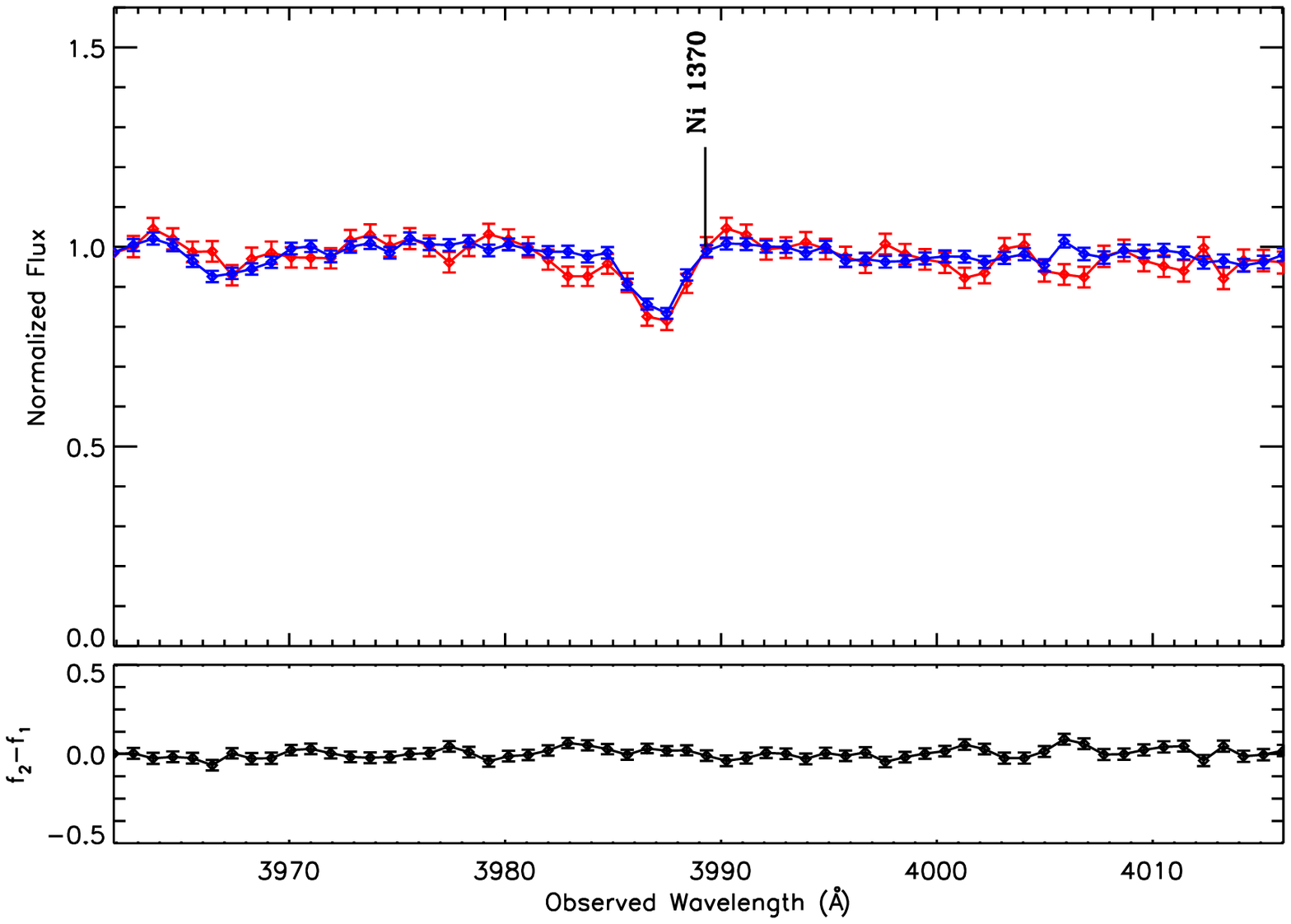}
\includegraphics[width=84mm]{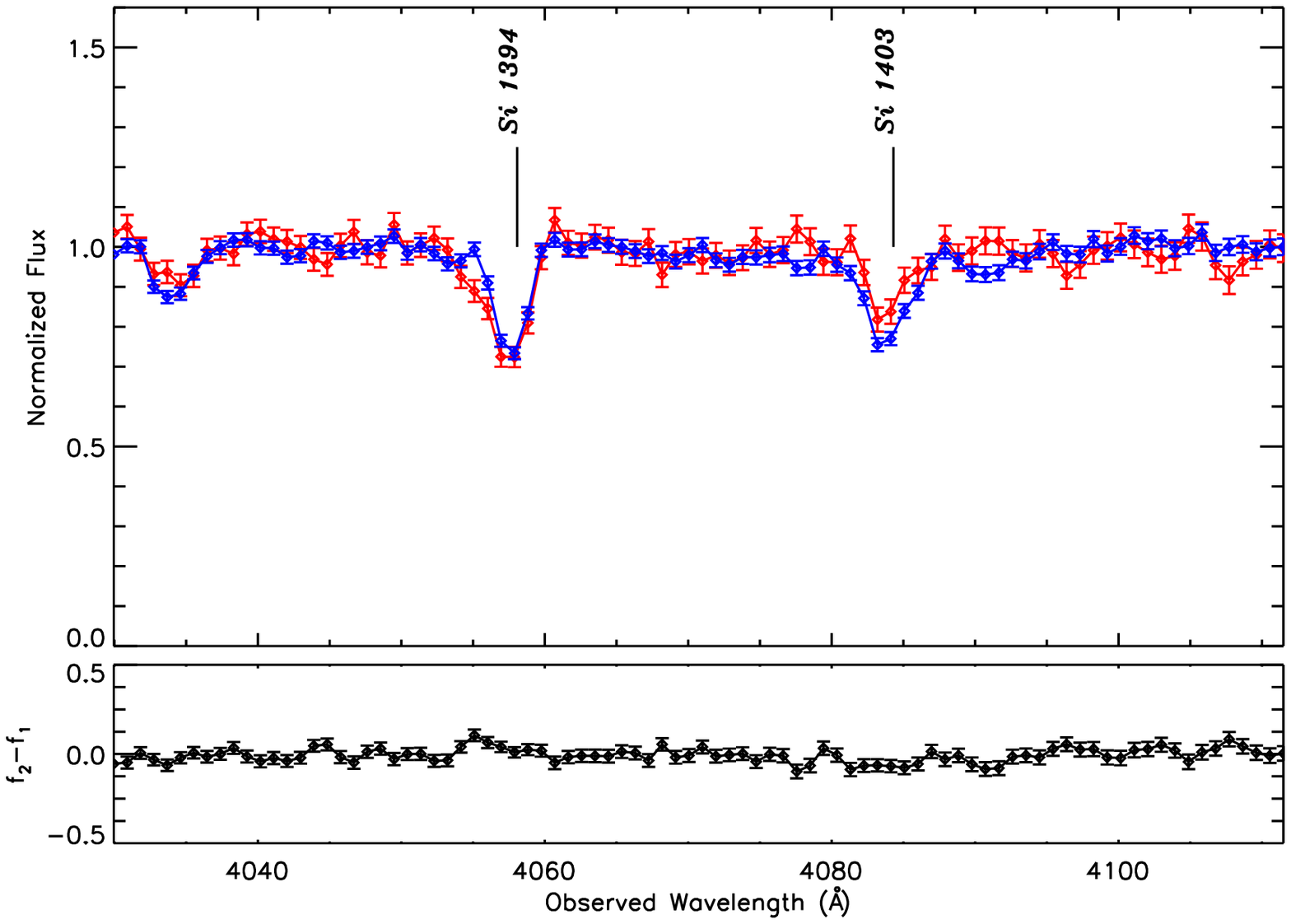}
\includegraphics[width=84mm]{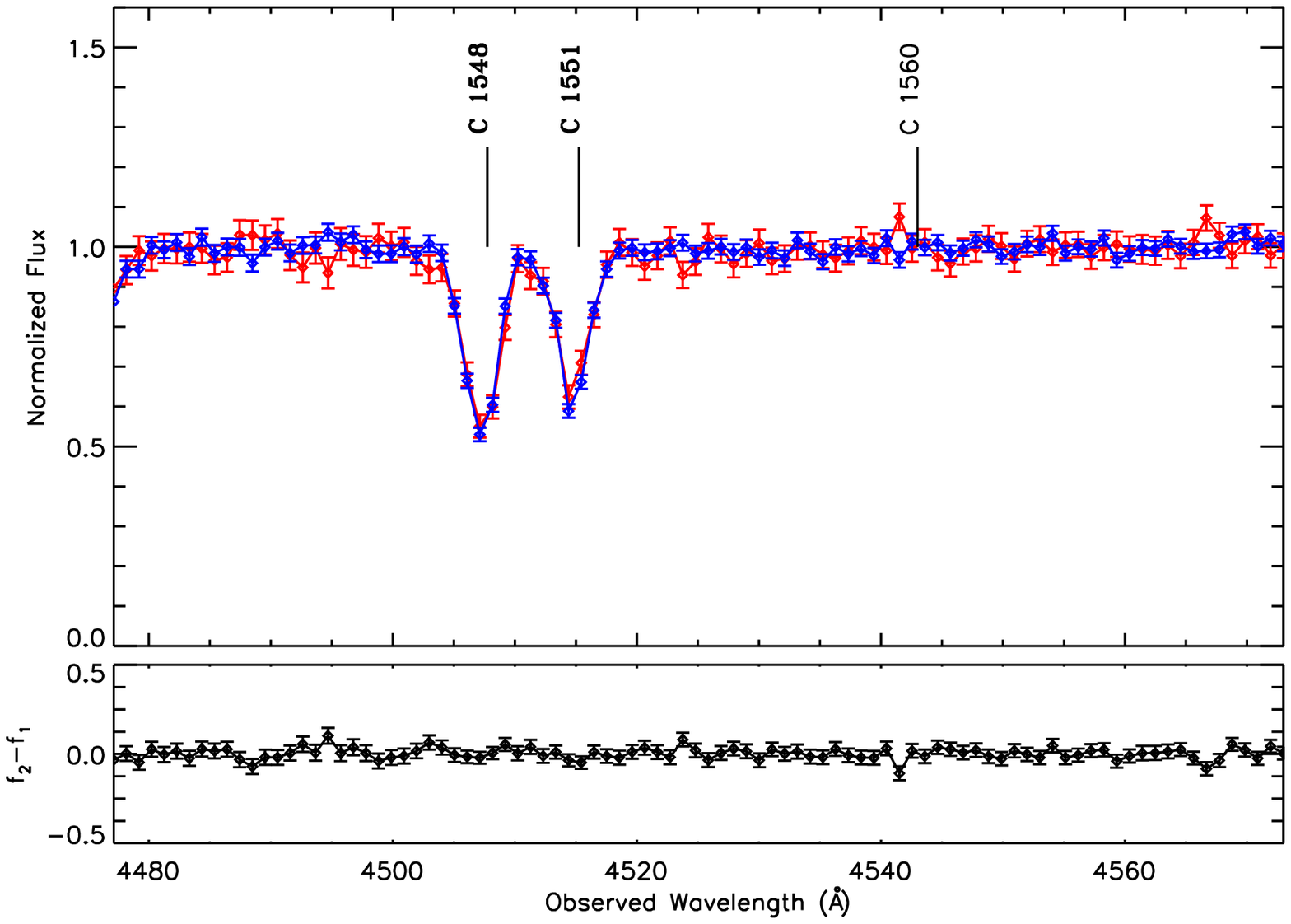}
\includegraphics[width=84mm]{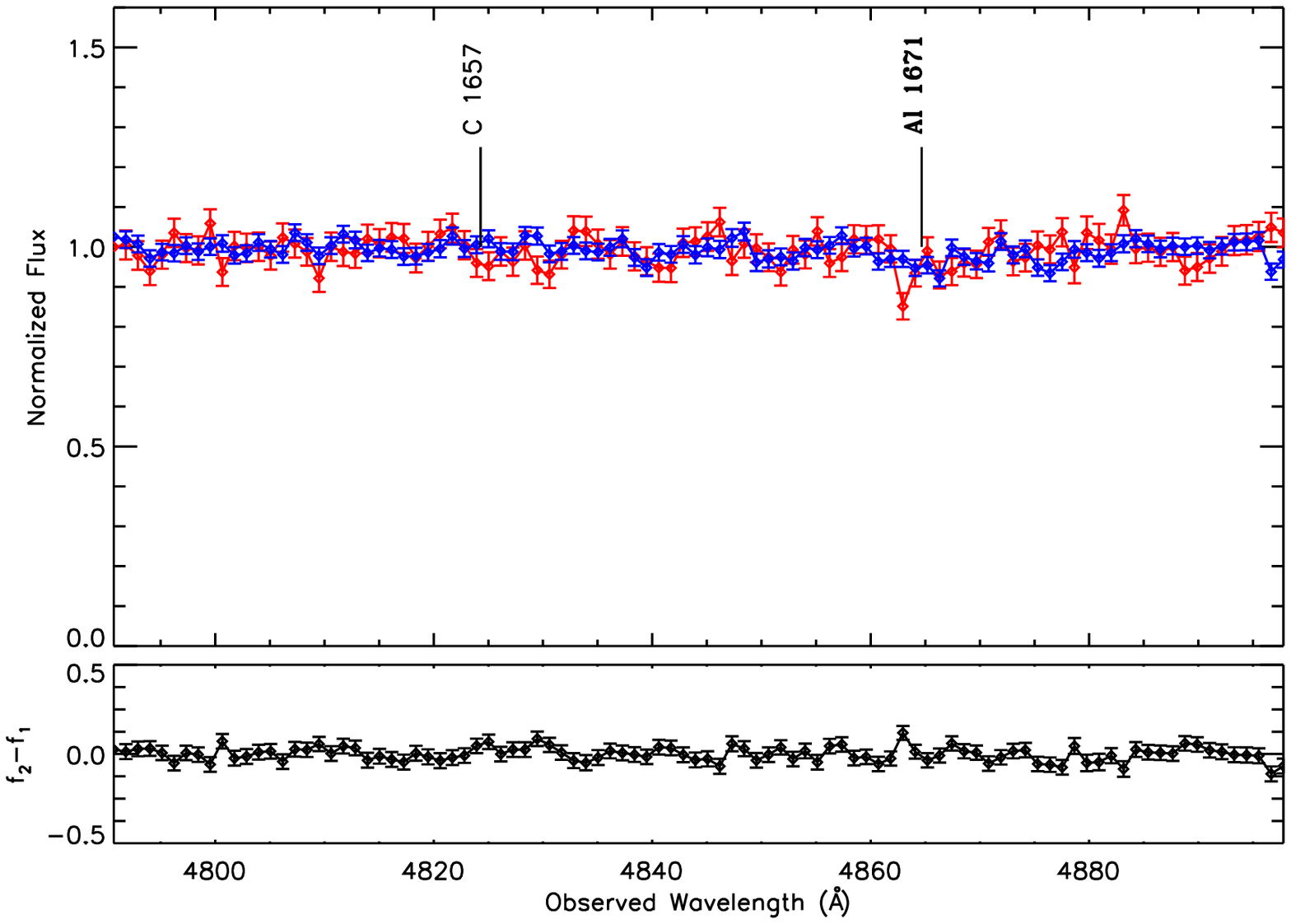}
\includegraphics[width=84mm]{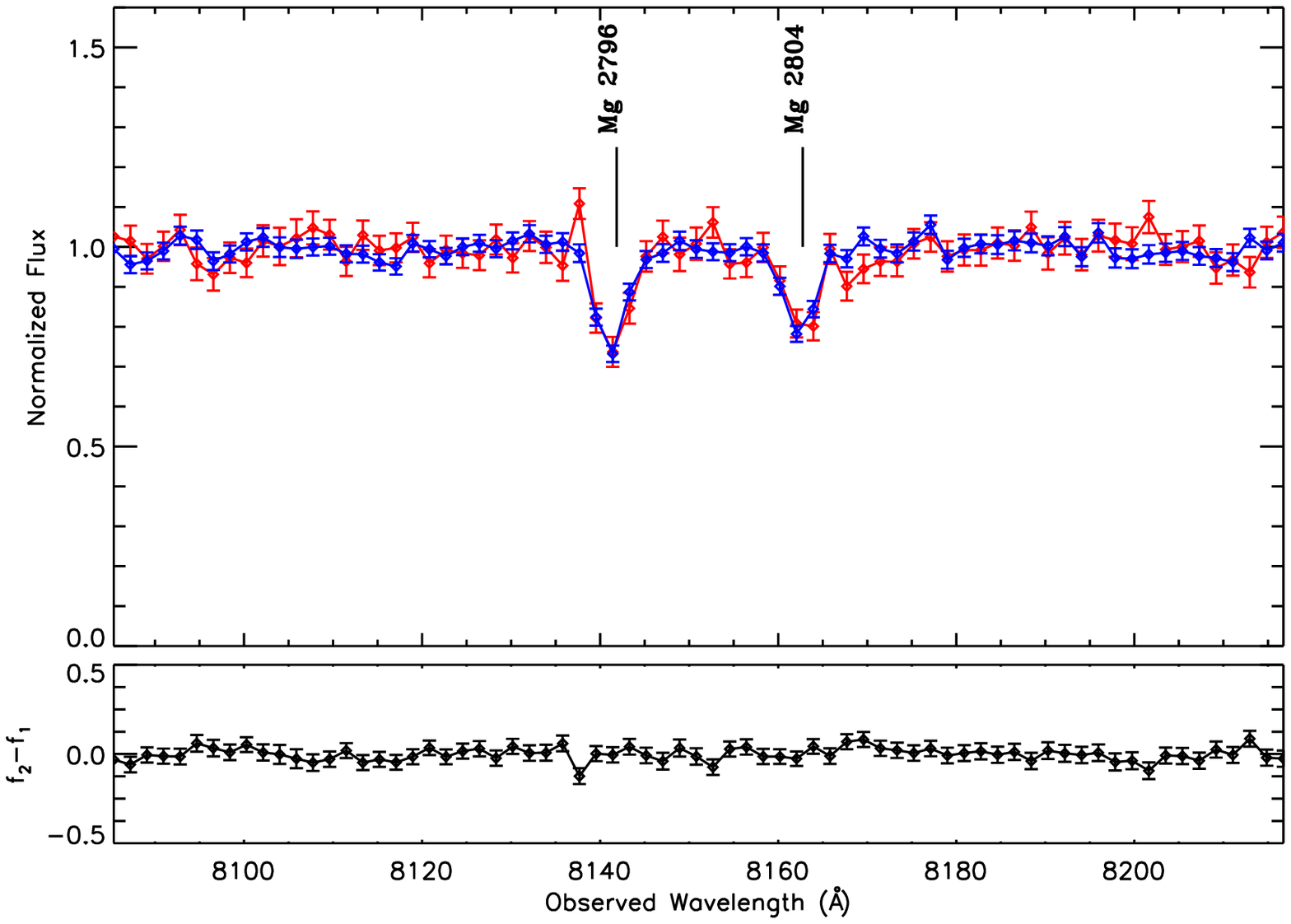}
\caption[Two-epoch normalized spectra of SDSS J005202.40+010129.2]{Two-epoch normalized spectra of the variable NAL system at $\beta$ = 0.1179 in SDSS J005202.40+010129.2.  The top panel shows the normalized pixel flux values with 1$\sigma$ error bars (first observations are red and second are blue), the bottom panel plots the difference spectrum of the two observation epochs, and shaded backgrounds identify masked pixels not included in our search for absorption line variability.  Line identifications for significantly variable absorption lines are italicised, lines detected in both observation epochs are in bold font, and undetected lines are in regular font (see Table A.1 for ion labels).  \label{figvs18}}
\end{center}
\end{figure*}

\begin{figure*}
\begin{center}
\includegraphics[width=84mm]{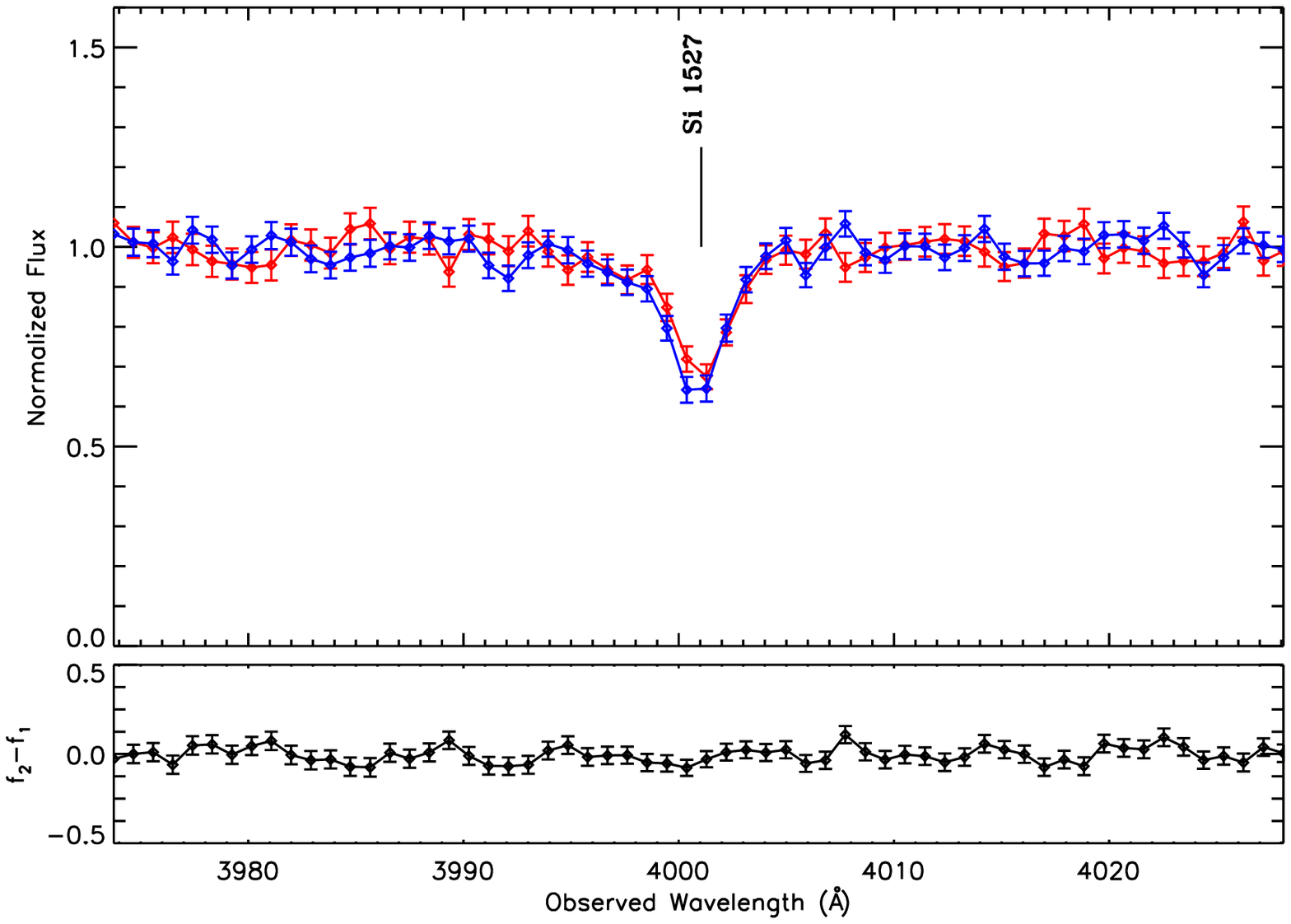}
\includegraphics[width=84mm]{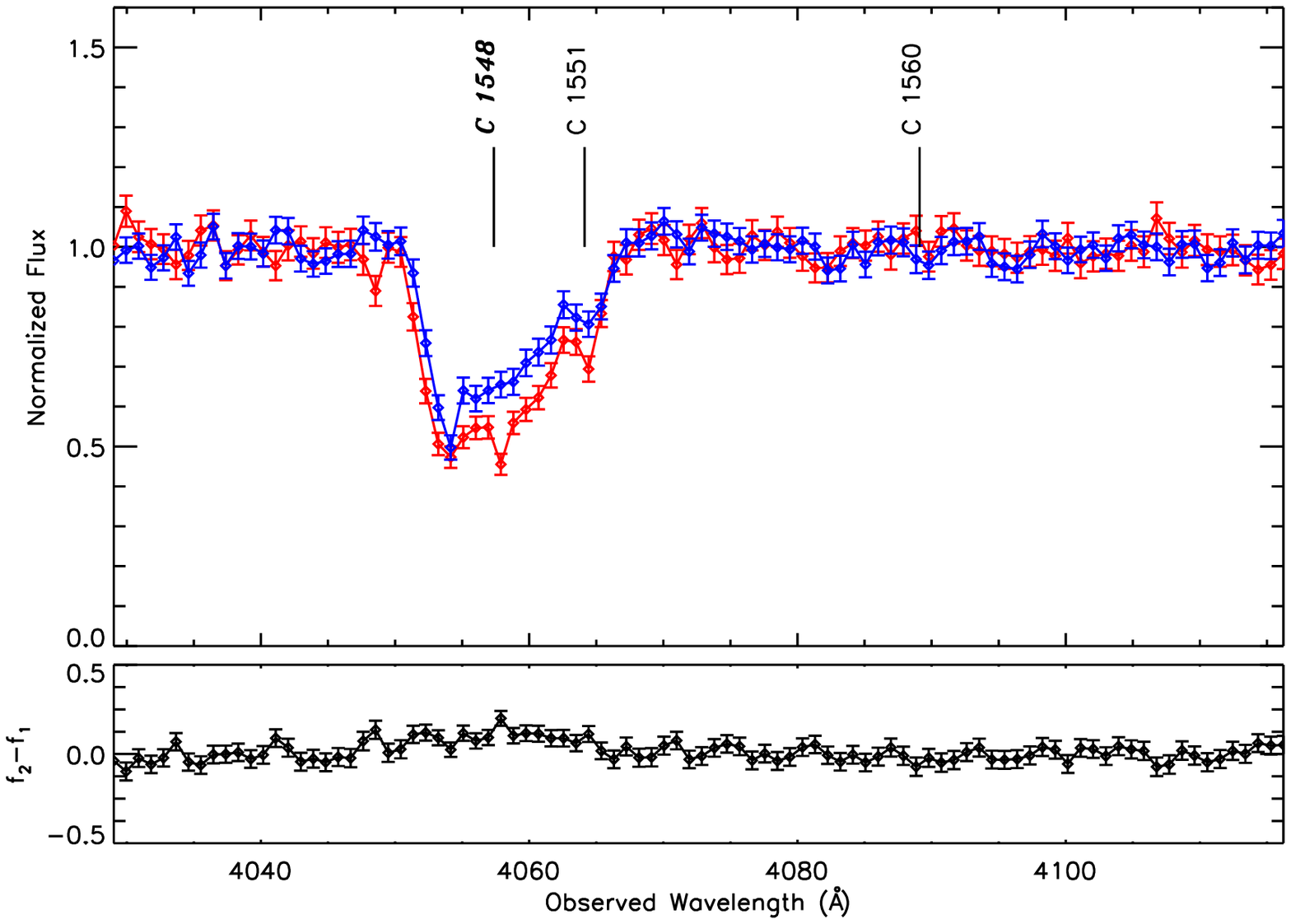}
\includegraphics[width=84mm]{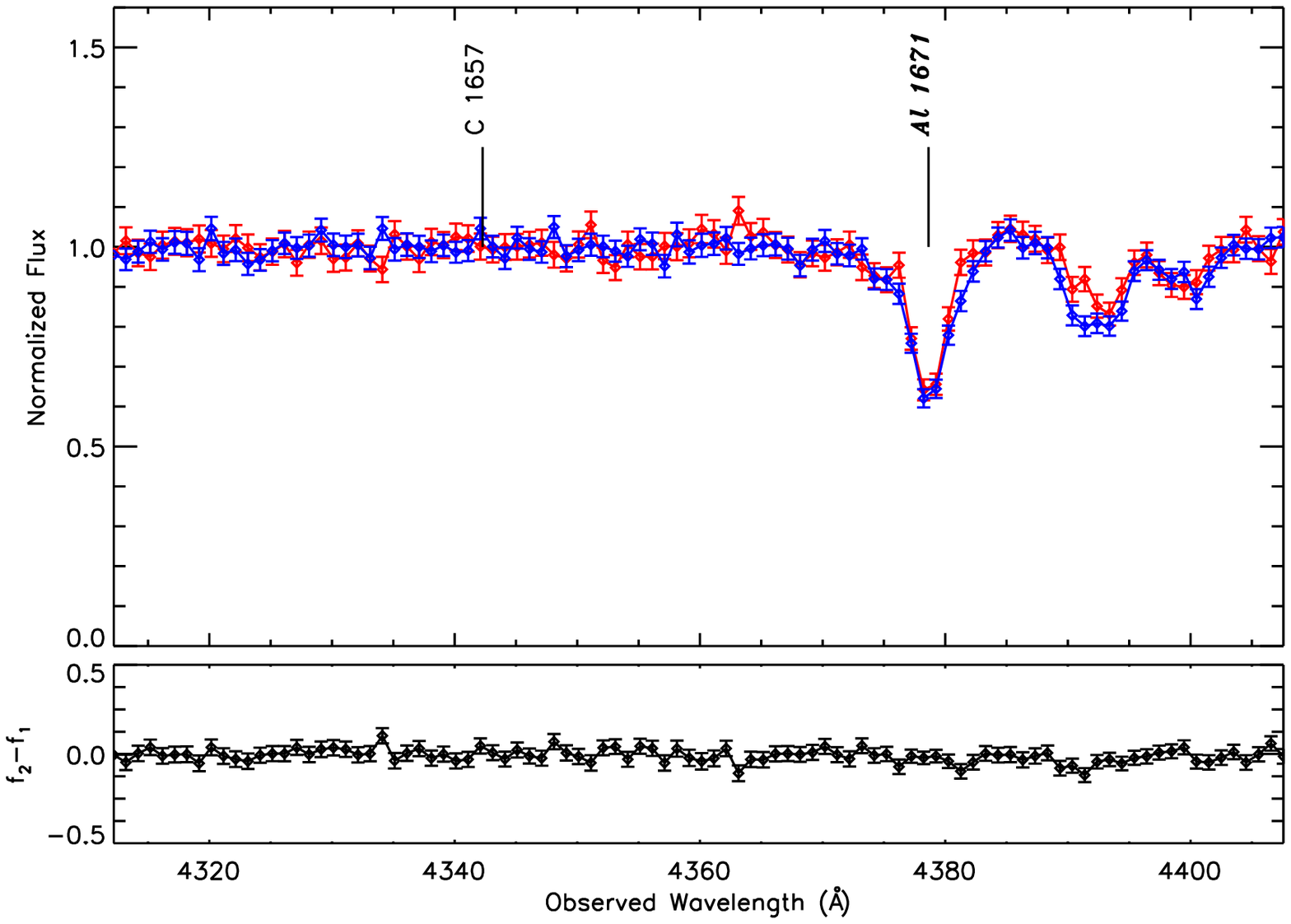}
\includegraphics[width=84mm]{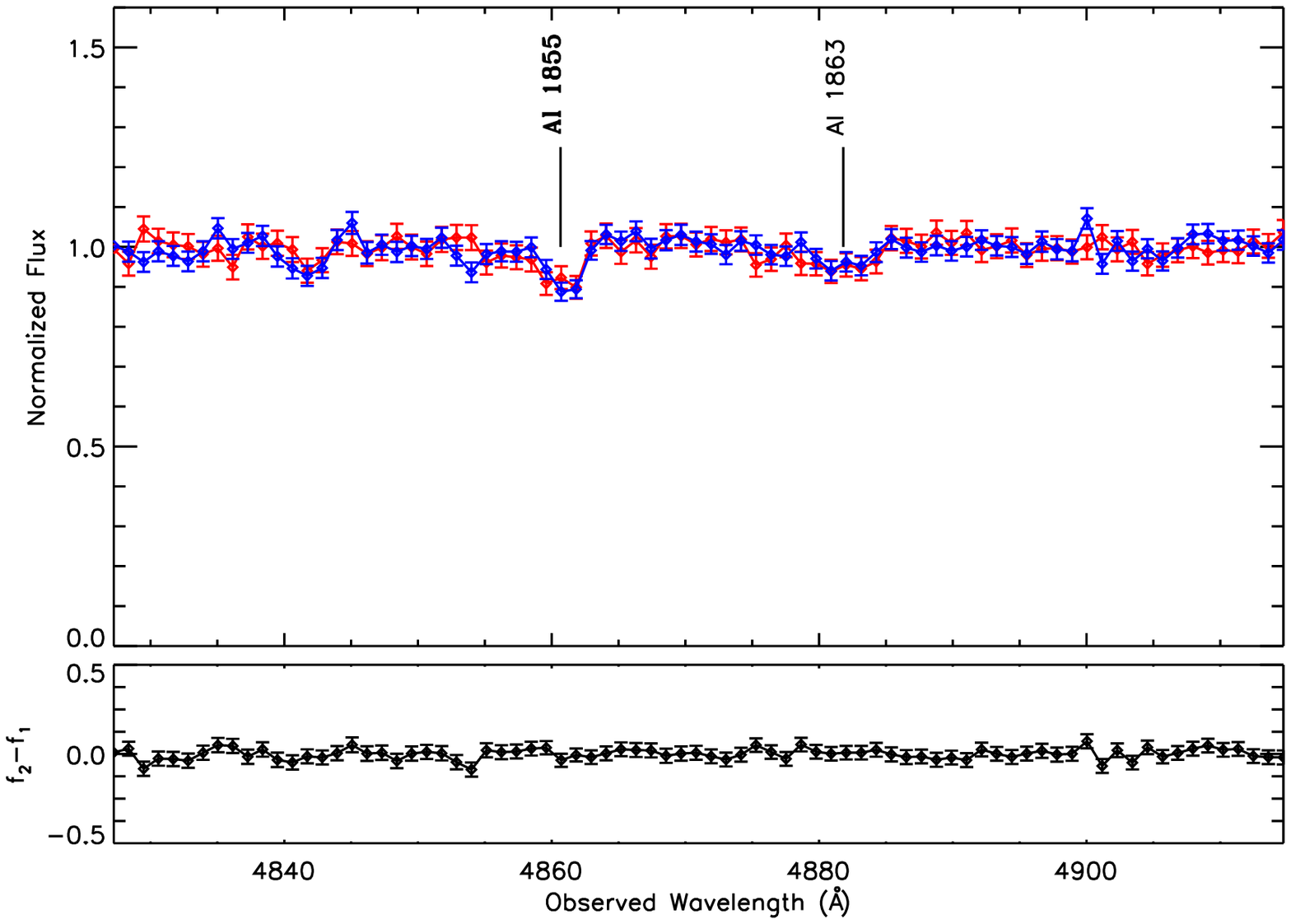}
\includegraphics[width=84mm]{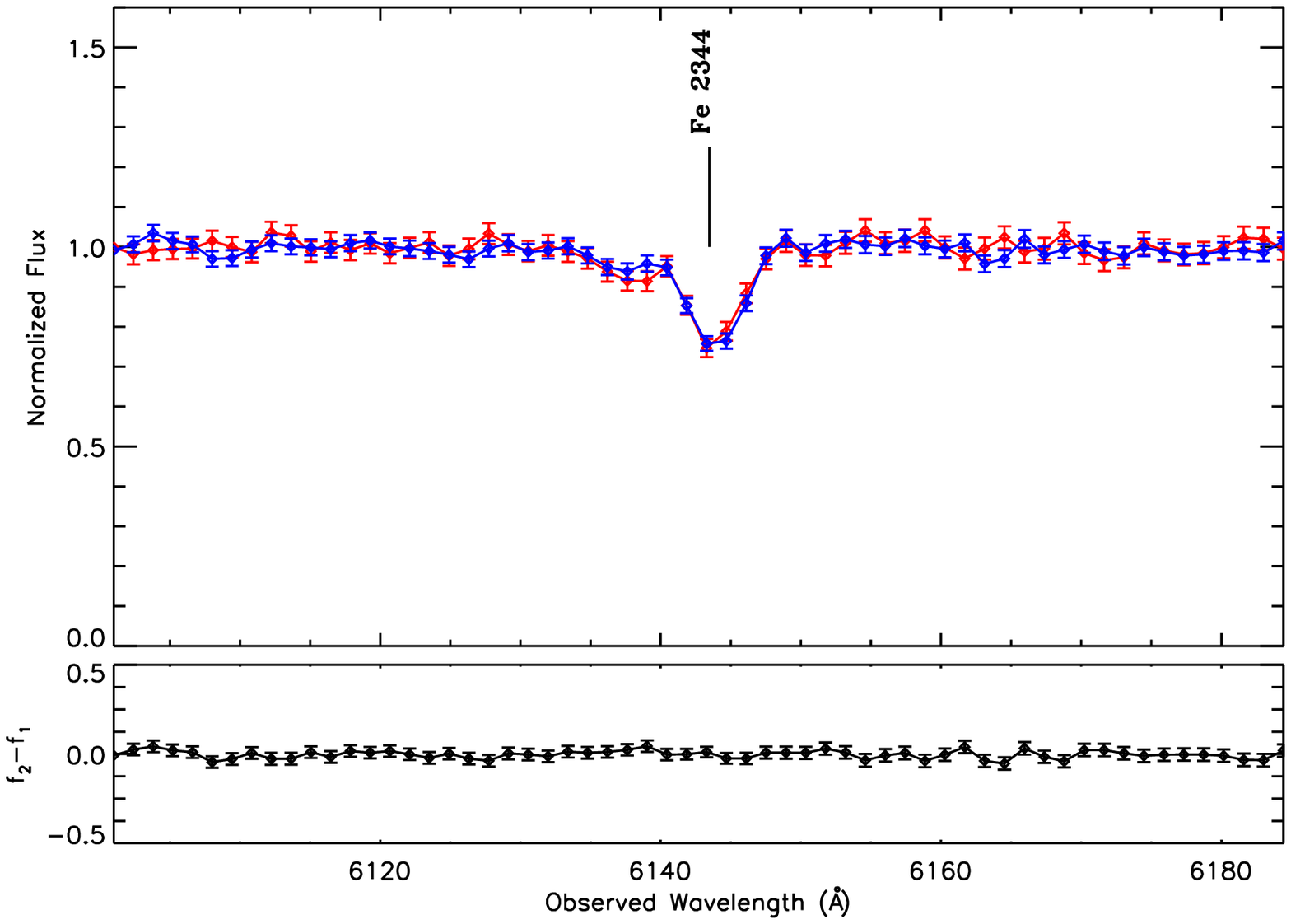}
\includegraphics[width=84mm]{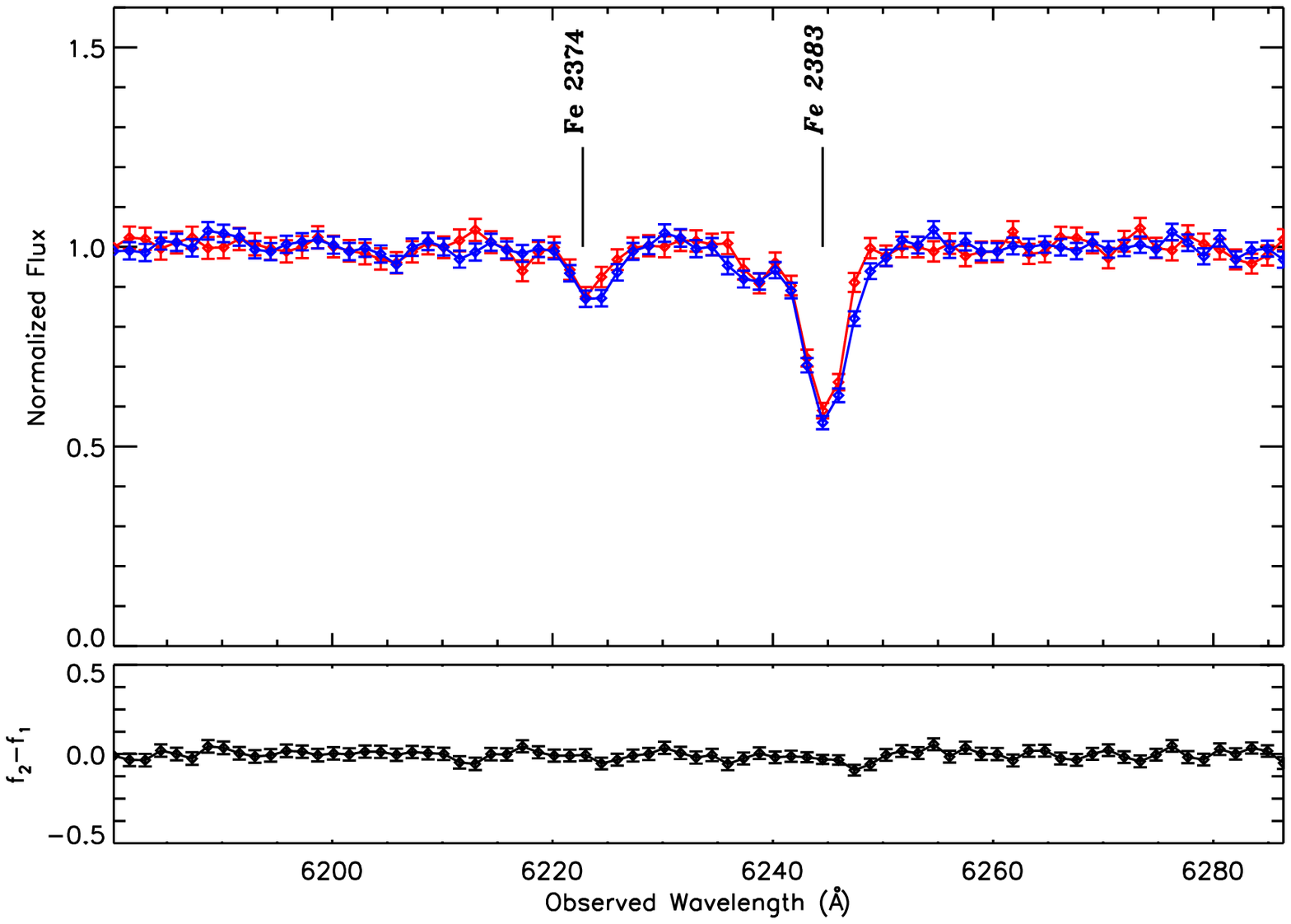}
\caption[Two-epoch normalized spectra of SDSS J004023.76+140807.3]{Two-epoch normalized spectra of the variable NAL system at $\beta$ = 0.1027 in SDSS J004023.76+140807.3.  The top panel shows the normalized pixel flux values with 1$\sigma$ error bars (first observations are red and second are blue), the bottom panel plots the difference spectrum of the two observation epochs, and shaded backgrounds identify masked pixels not included in our search for absorption line variability.  Line identifications for significantly variable absorption lines are italicised, lines detected in both observation epochs are in bold font, and undetected lines are in regular font (see Table A.1 for ion labels).  Continued in next figure.  \label{figvs19}}
\end{center}
\end{figure*}

\begin{figure*}
\ContinuedFloat
\begin{center}
\includegraphics[width=84mm]{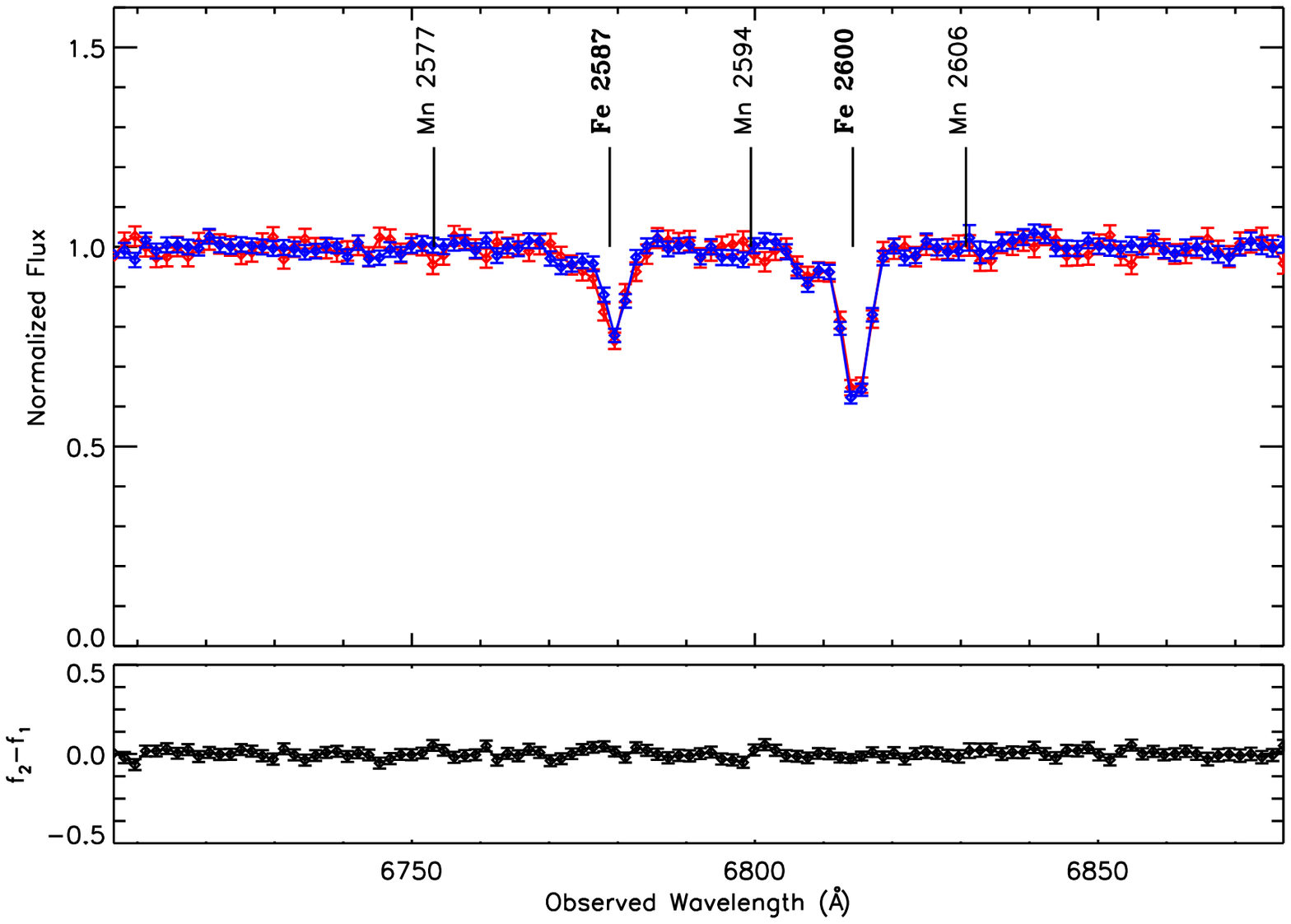}
\includegraphics[width=84mm]{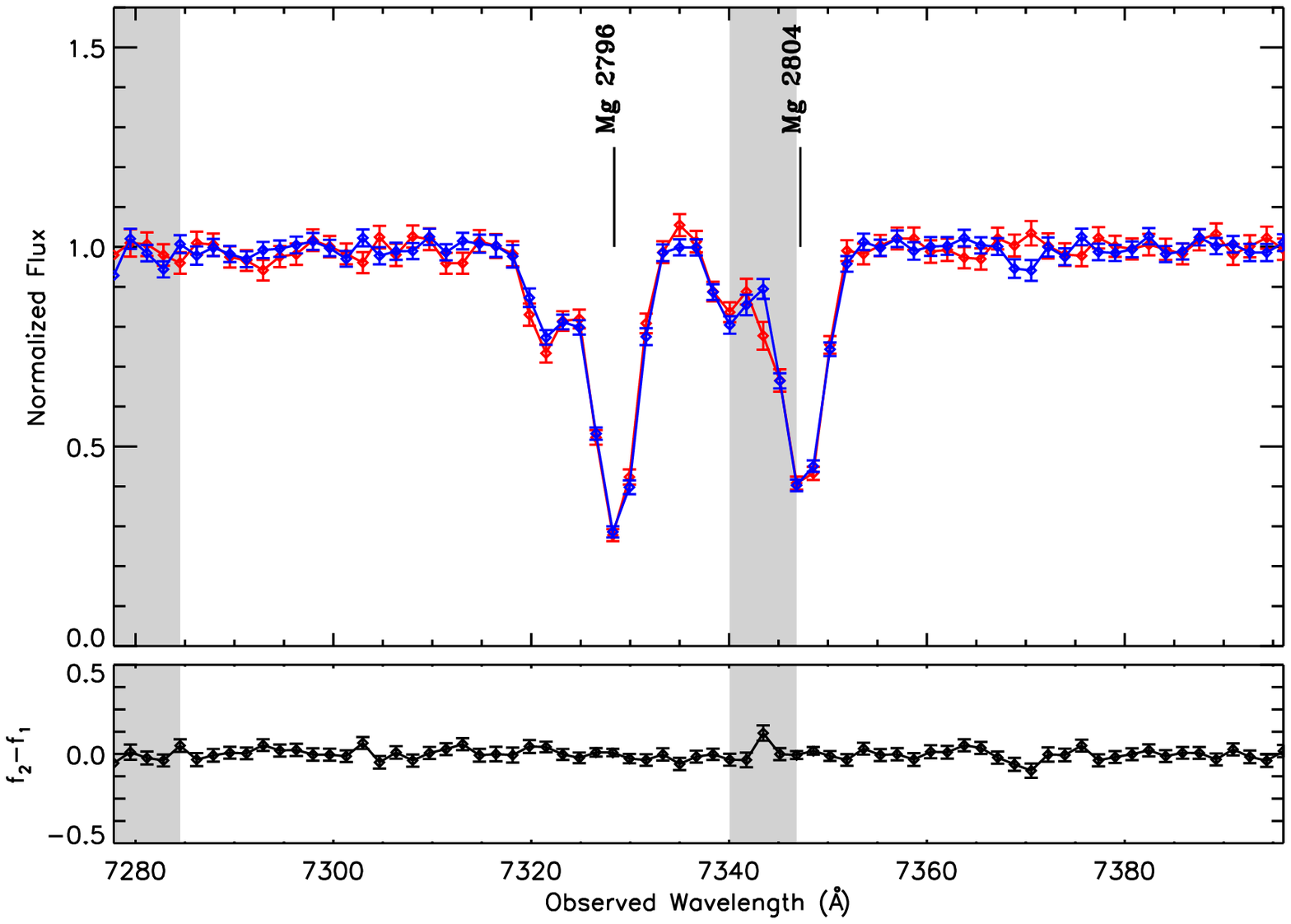}
\includegraphics[width=84mm]{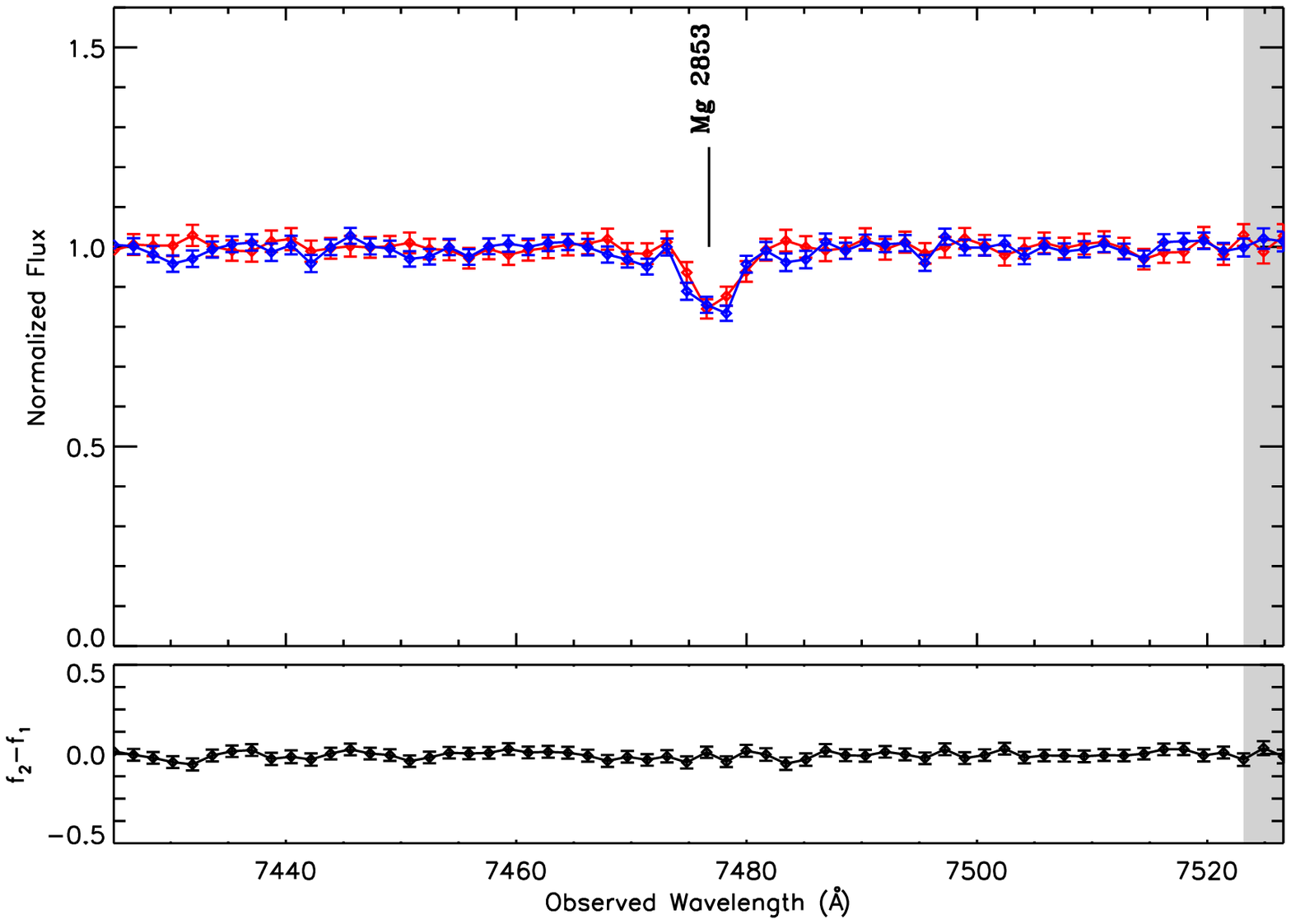}
\caption[]{Two-epoch normalized spectra of the variable NAL system at $\beta$ = 0.1027 in SDSS J004023.76+140807.3.  The top panel shows the normalized pixel flux values with 1$\sigma$ error bars (first observations are red and second are blue), the bottom panel plots the difference spectrum of the two observation epochs, and shaded backgrounds identify masked pixels not included in our search for absorption line variability.  Line identifications for significantly variable absorption lines are italicised, lines detected in both observation epochs are in bold font, and undetected lines are in regular font (see Table A.1 for ion labels).  Continued from previous figure.}
\end{center}
\end{figure*}

\begin{figure*}
\begin{center}
\includegraphics[width=84mm]{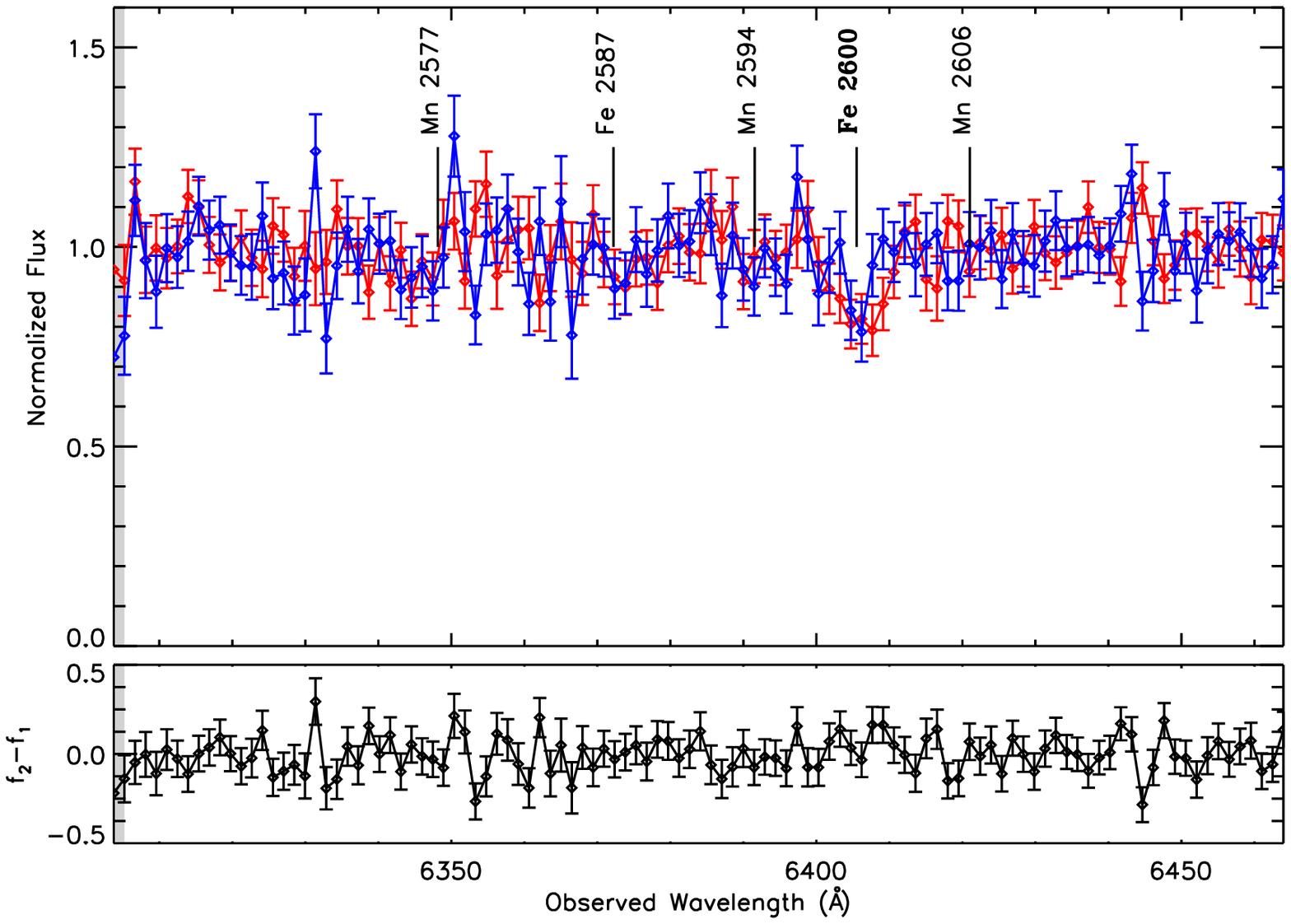}
\includegraphics[width=84mm]{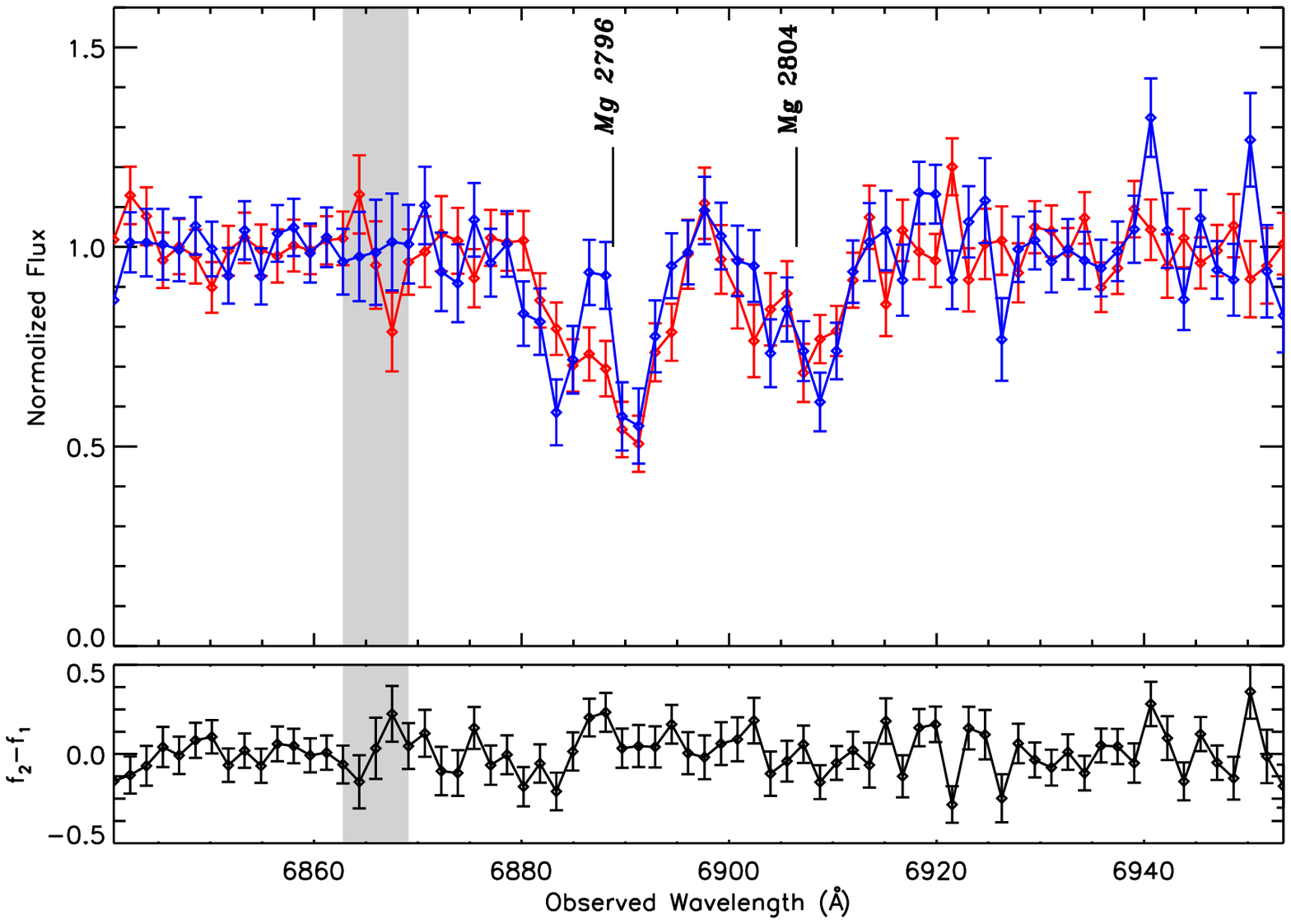}
\includegraphics[width=84mm]{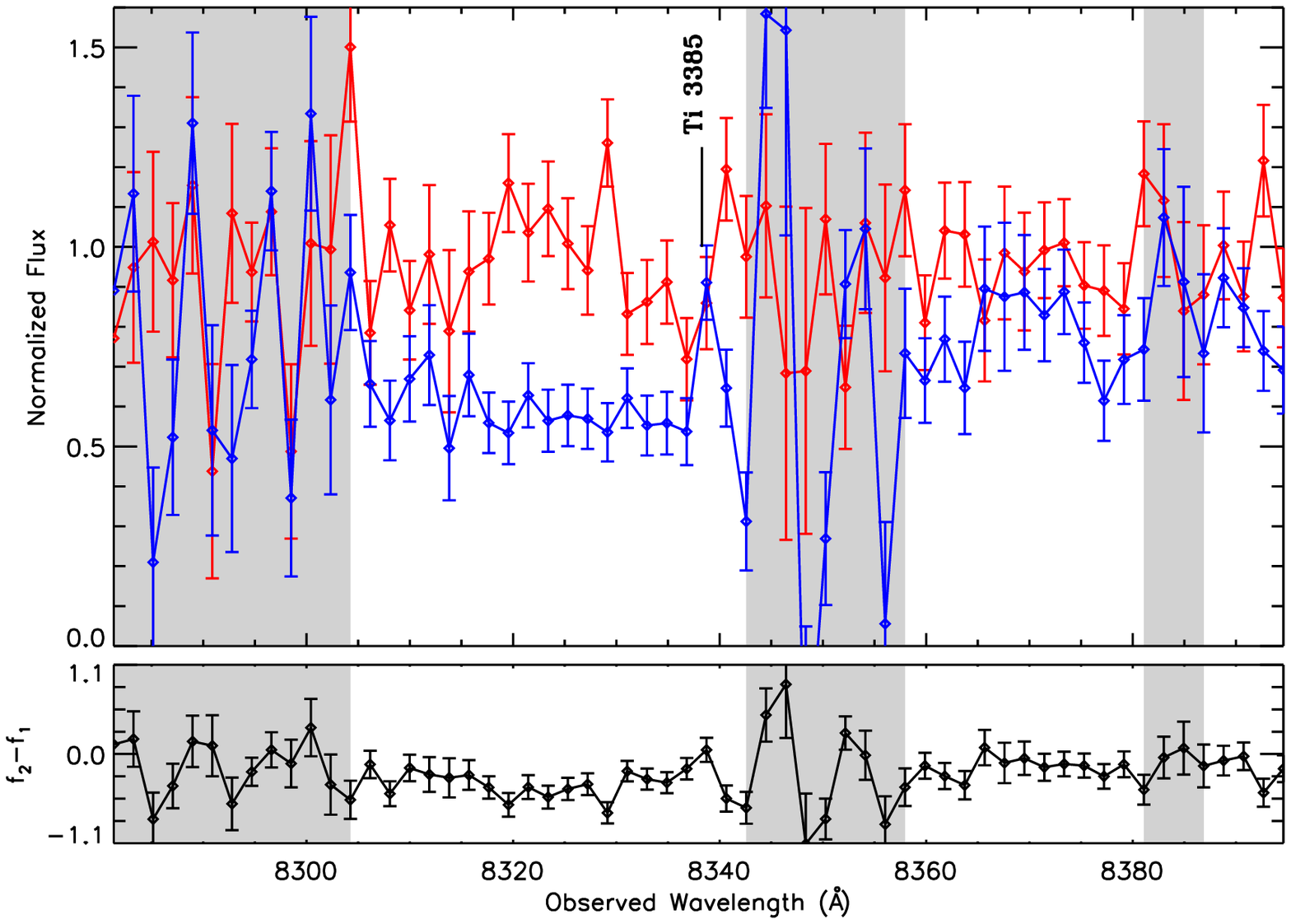}
\caption[Two-epoch normalized spectra of SDSS J023620.79-003342.2]{Two-epoch normalized spectra of the variable NAL system at $\beta$ = 0.0808 in SDSS J023620.79-003342.2.  The top panel shows the normalized pixel flux values with 1$\sigma$ error bars (first observations are red and second are blue), the bottom panel plots the difference spectrum of the two observation epochs, and shaded backgrounds identify masked pixels not included in our search for absorption line variability.  Line identifications for significantly variable absorption lines are italicised, lines detected in both observation epochs are in bold font, and undetected lines are in regular font (see Table A.1 for ion labels).  \label{figvs20}}
\end{center}
\end{figure*}

\begin{figure*}
\begin{center}
\includegraphics[width=84mm]{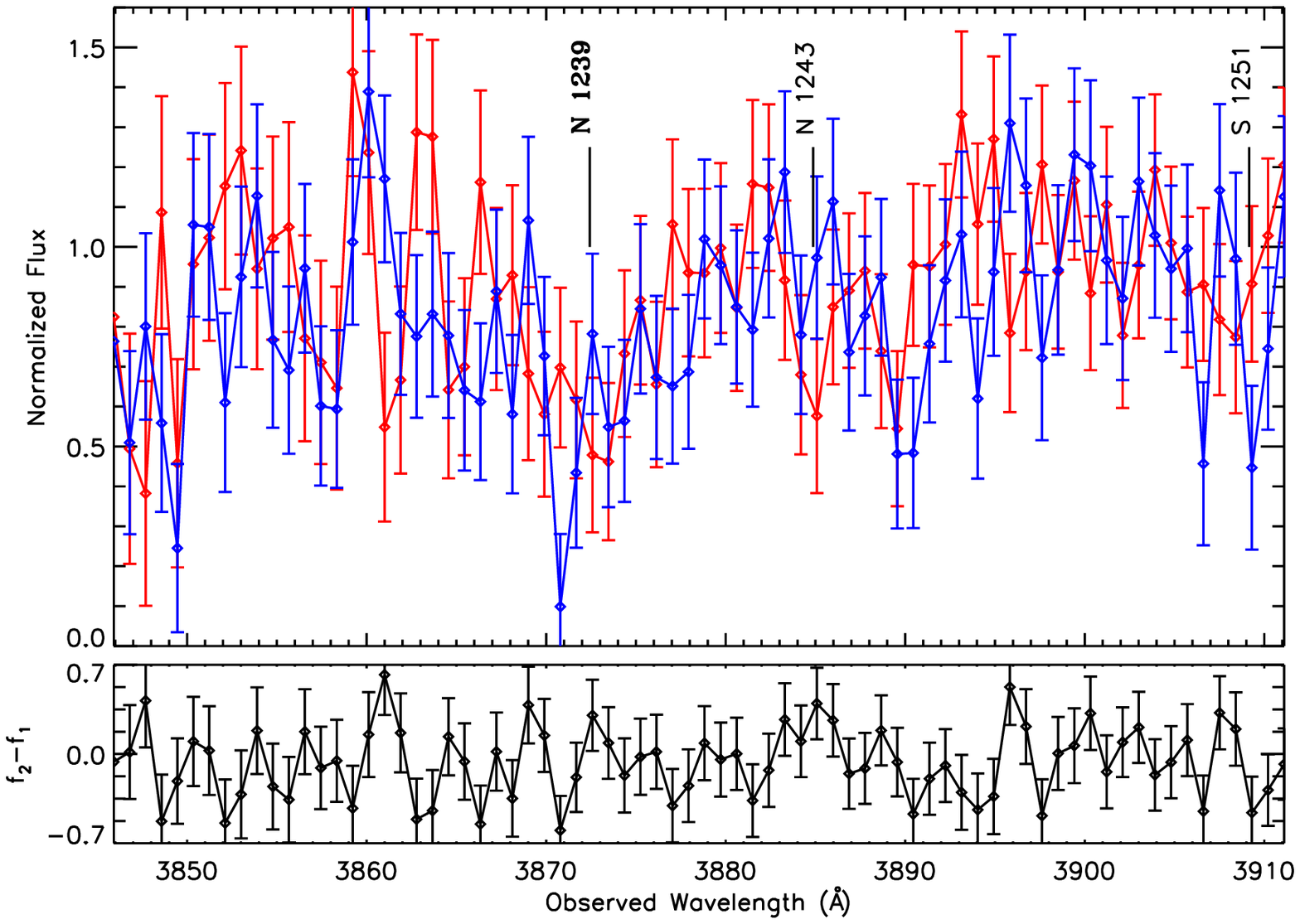}
\includegraphics[width=84mm]{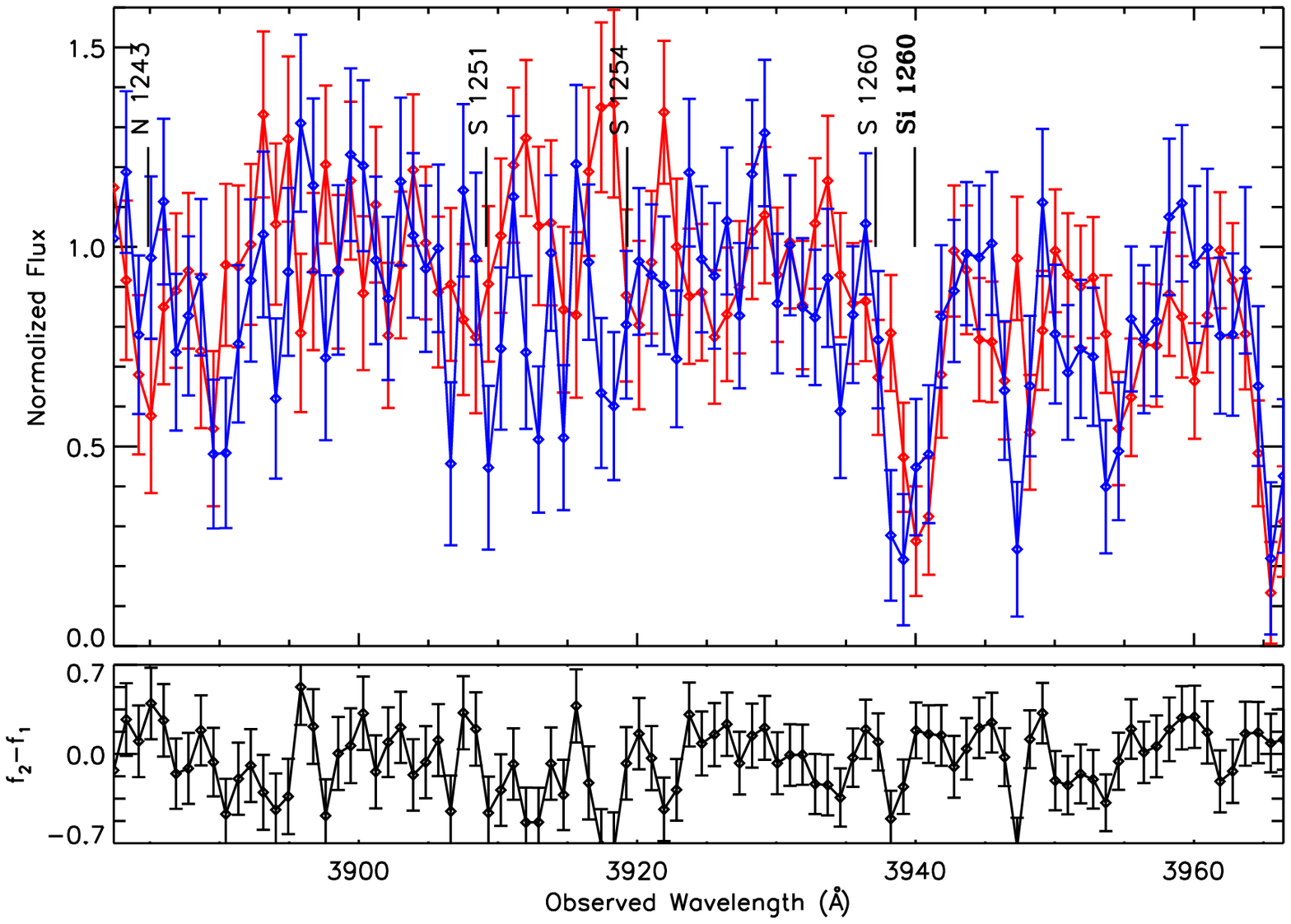}
\includegraphics[width=84mm]{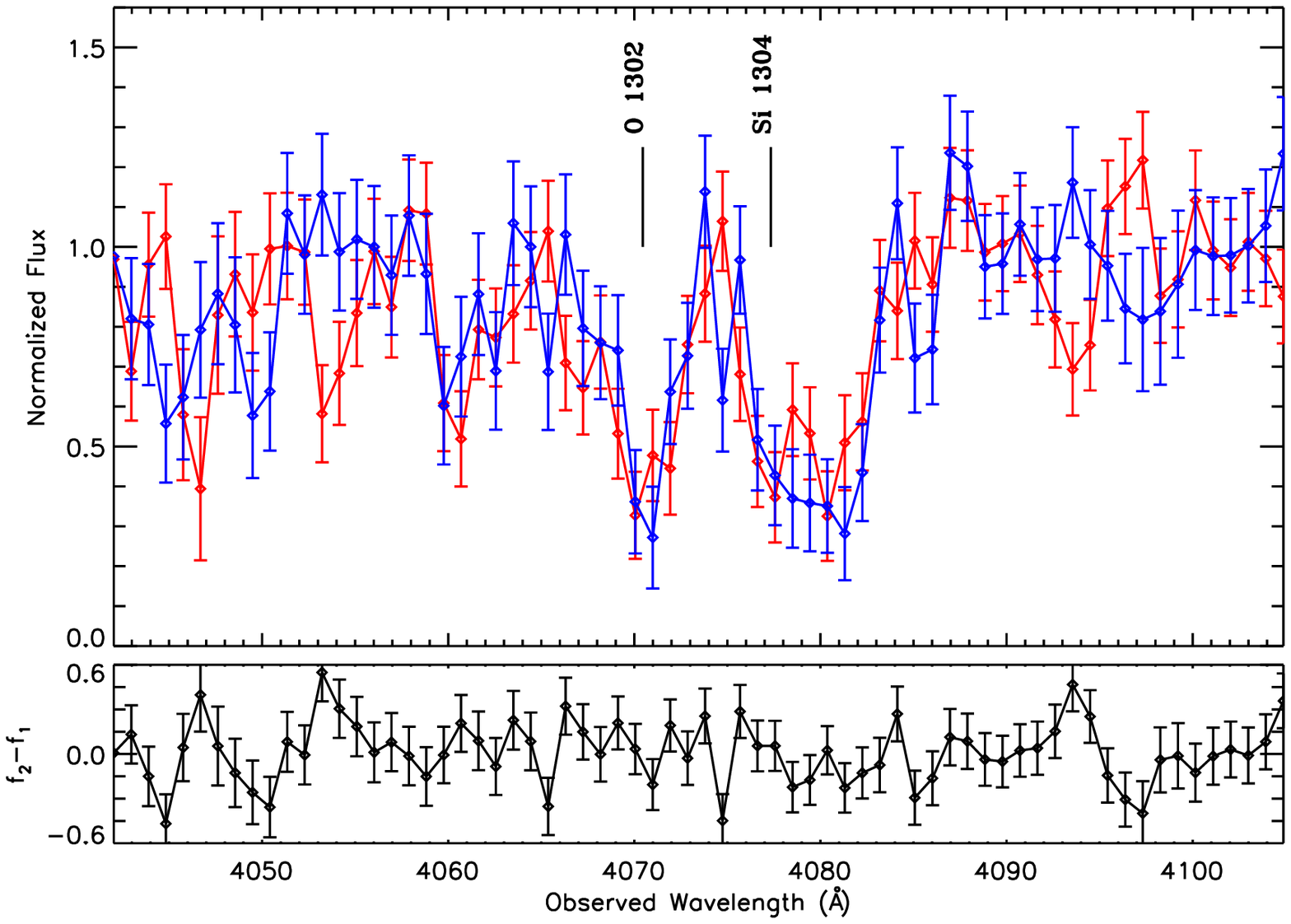}
\includegraphics[width=84mm]{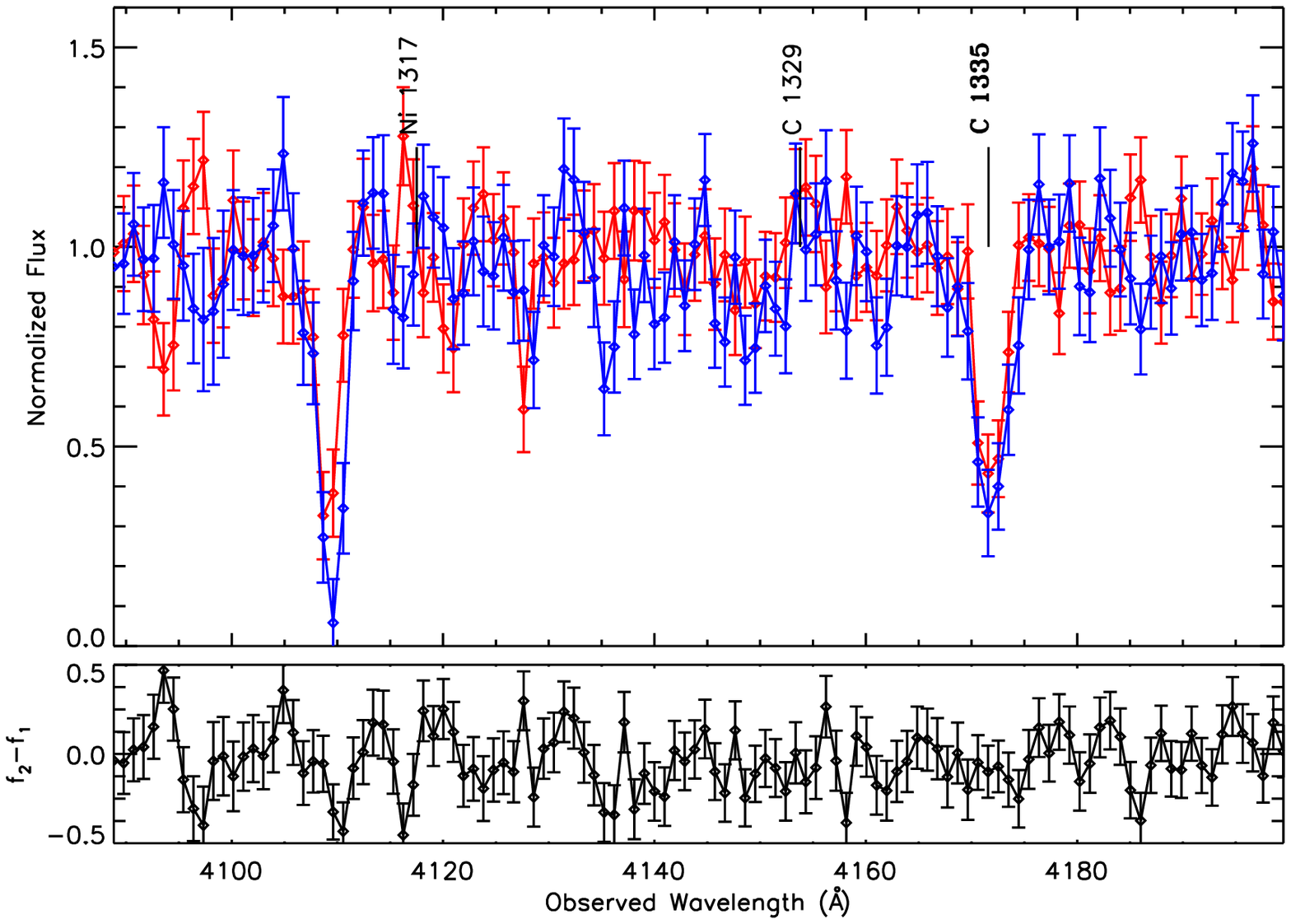}
\includegraphics[width=84mm]{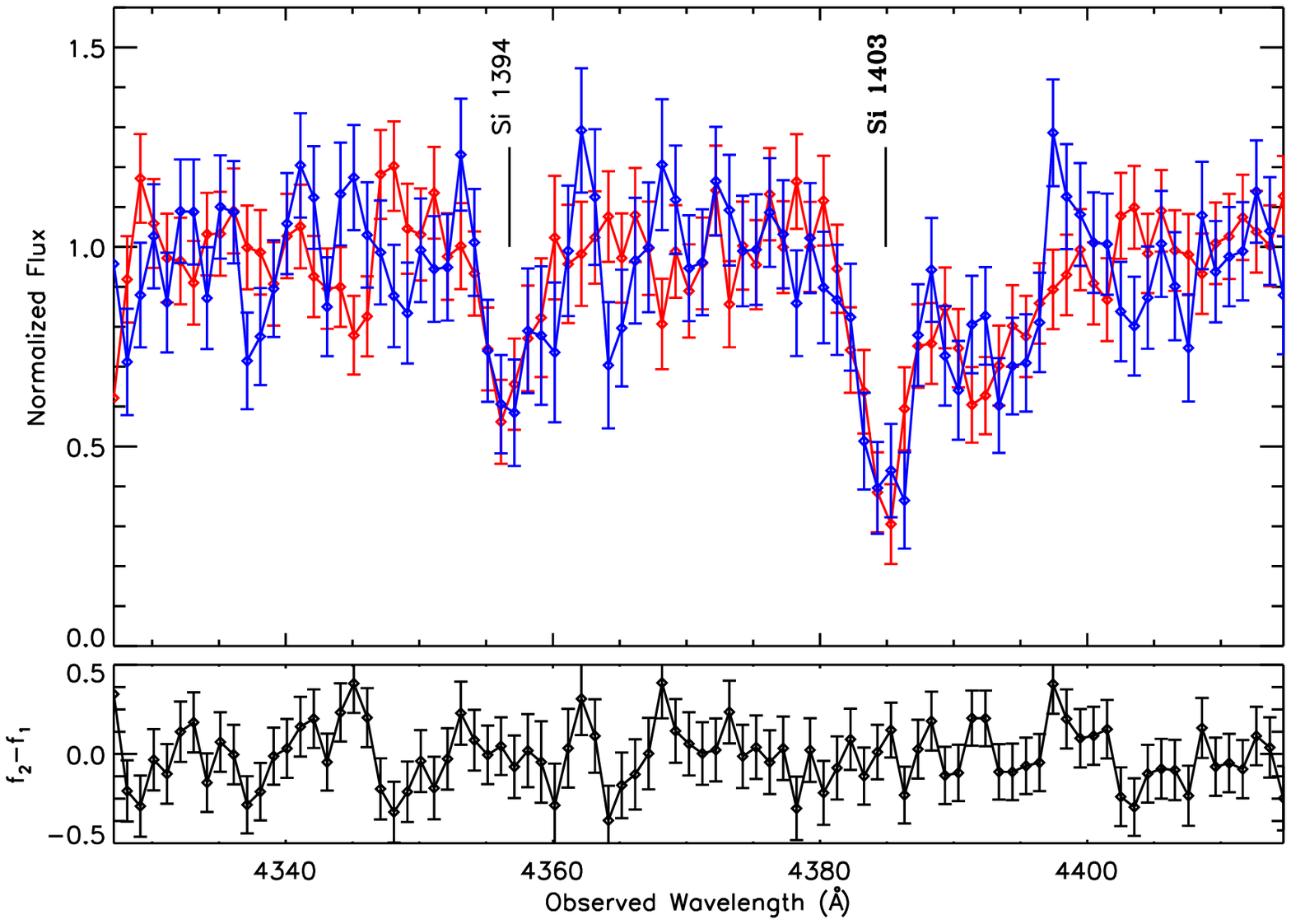}
\includegraphics[width=84mm]{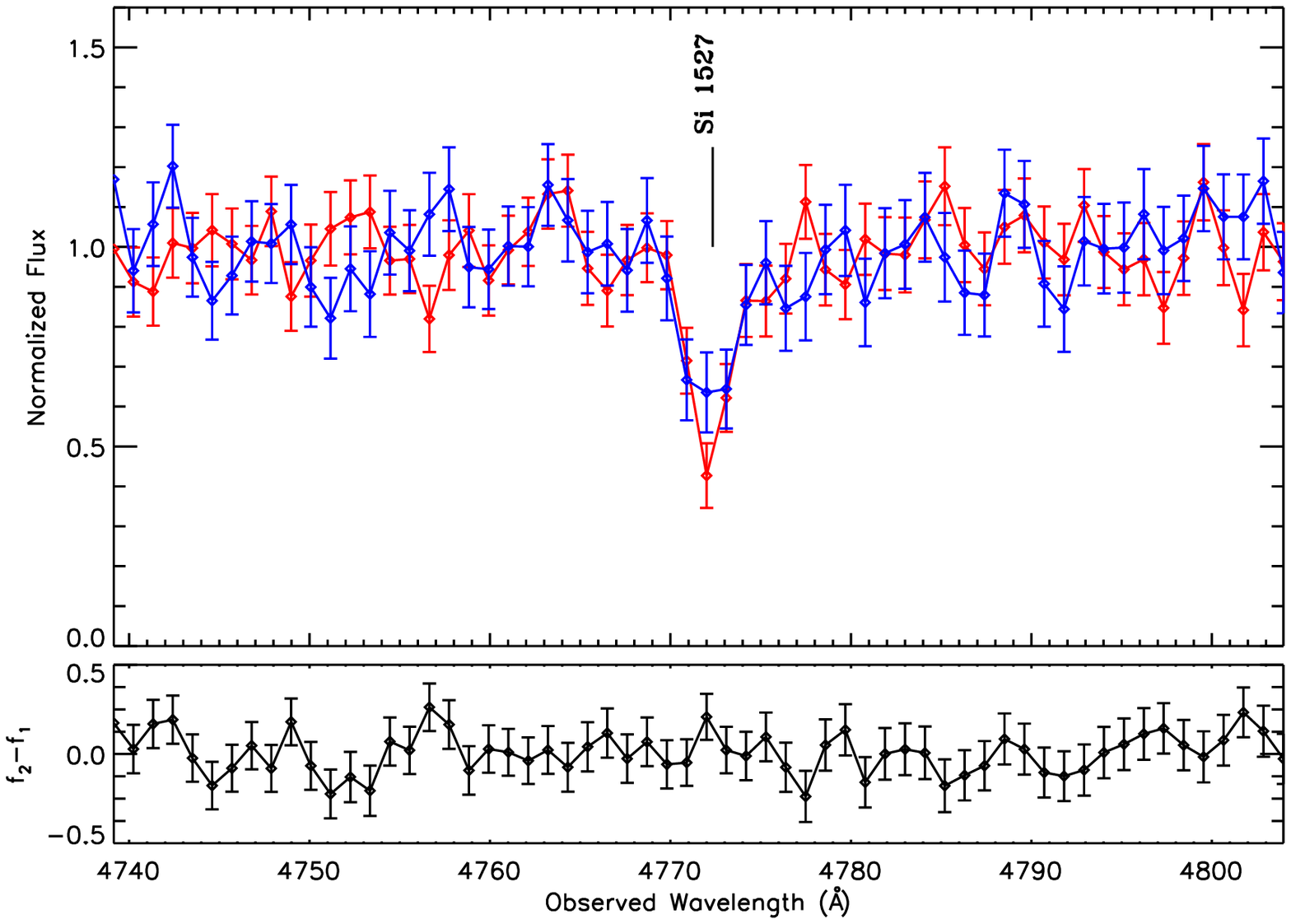}
\caption[Two-epoch normalized spectra of SDSS J024154.42-004757.6]{Two-epoch normalized spectra of the variable NAL system at $\beta$ = 0.0794 in SDSS J024154.42-004757.6.  The top panel shows the normalized pixel flux values with 1$\sigma$ error bars (first observations are red and second are blue), the bottom panel plots the difference spectrum of the two observation epochs, and shaded backgrounds identify masked pixels not included in our search for absorption line variability.  Line identifications for significantly variable absorption lines are italicised, lines detected in both observation epochs are in bold font, and undetected lines are in regular font (see Table A.1 for ion labels).  Continued in next figure.  \label{figvs21}}
\end{center}
\end{figure*}

\begin{figure*}
\ContinuedFloat
\begin{center}
\includegraphics[width=84mm]{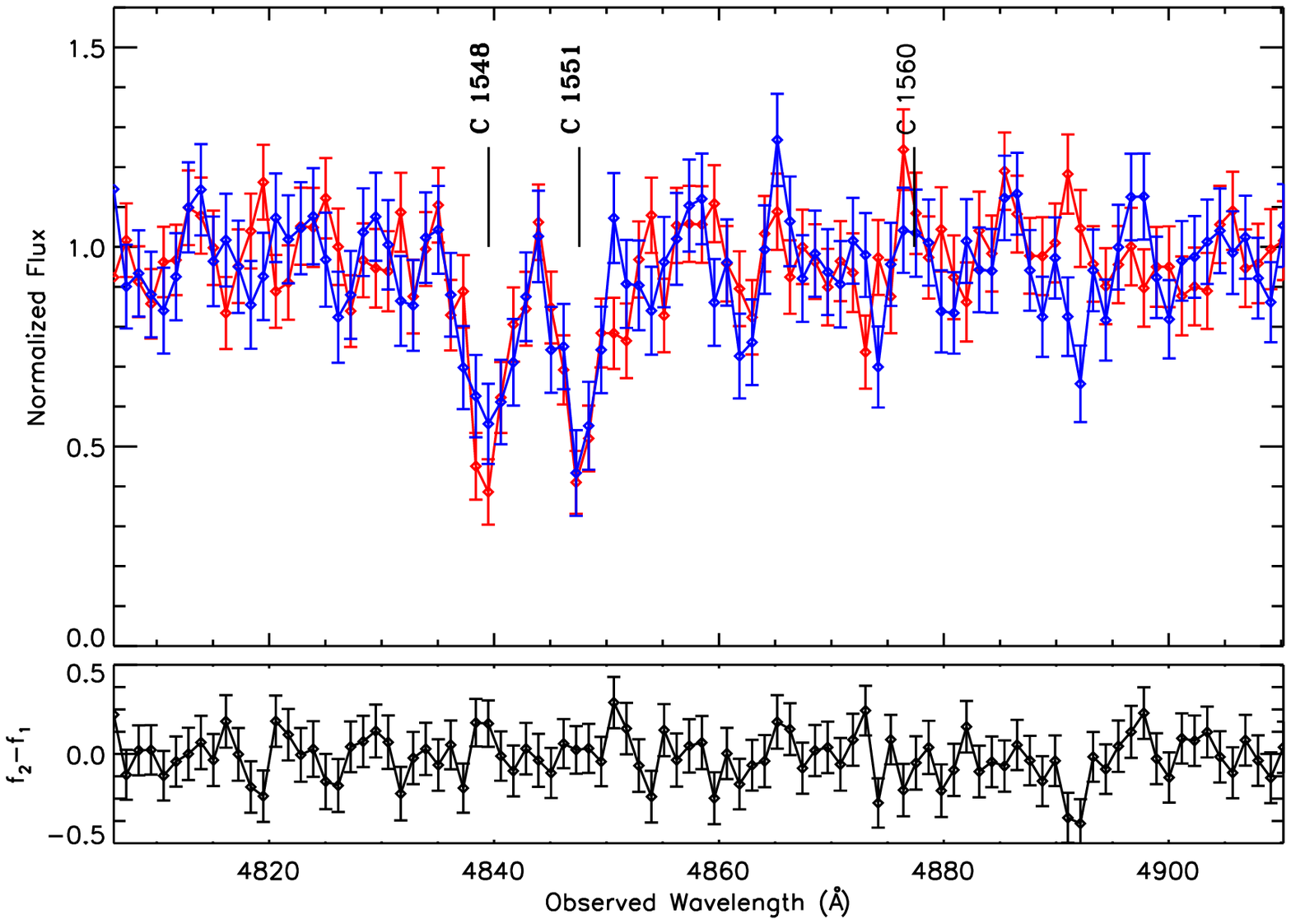}
\includegraphics[width=84mm]{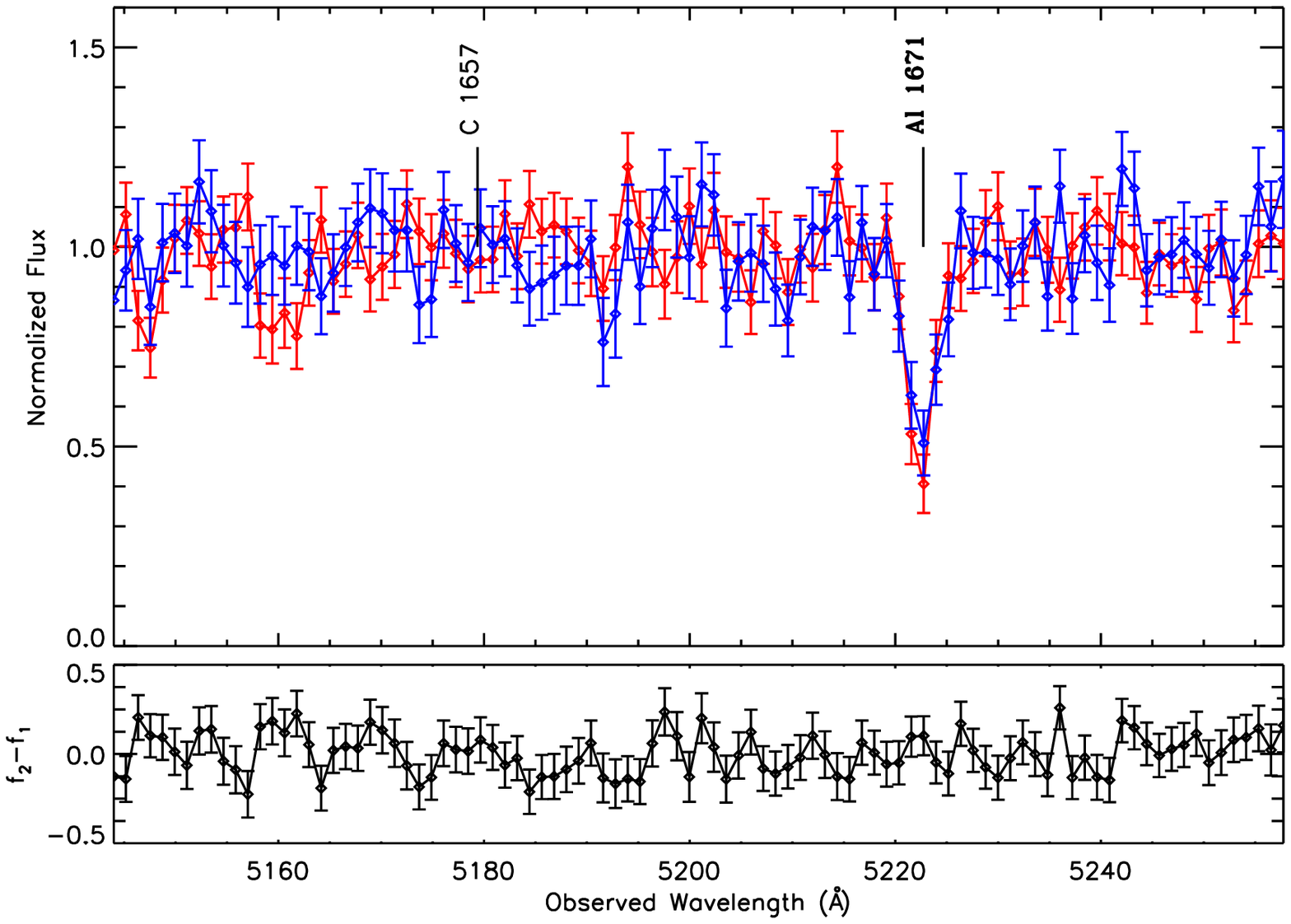}
\includegraphics[width=84mm]{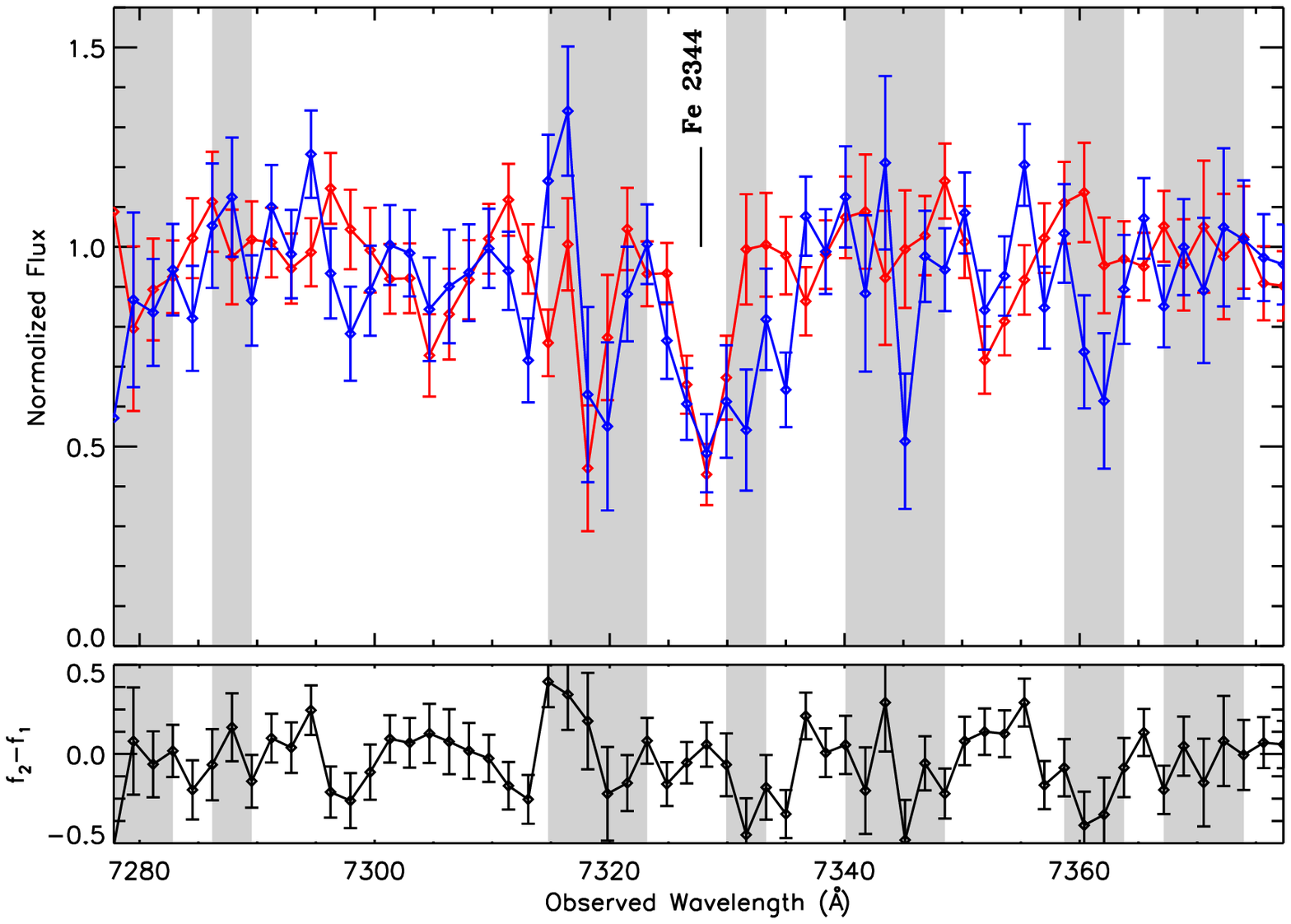}
\includegraphics[width=84mm]{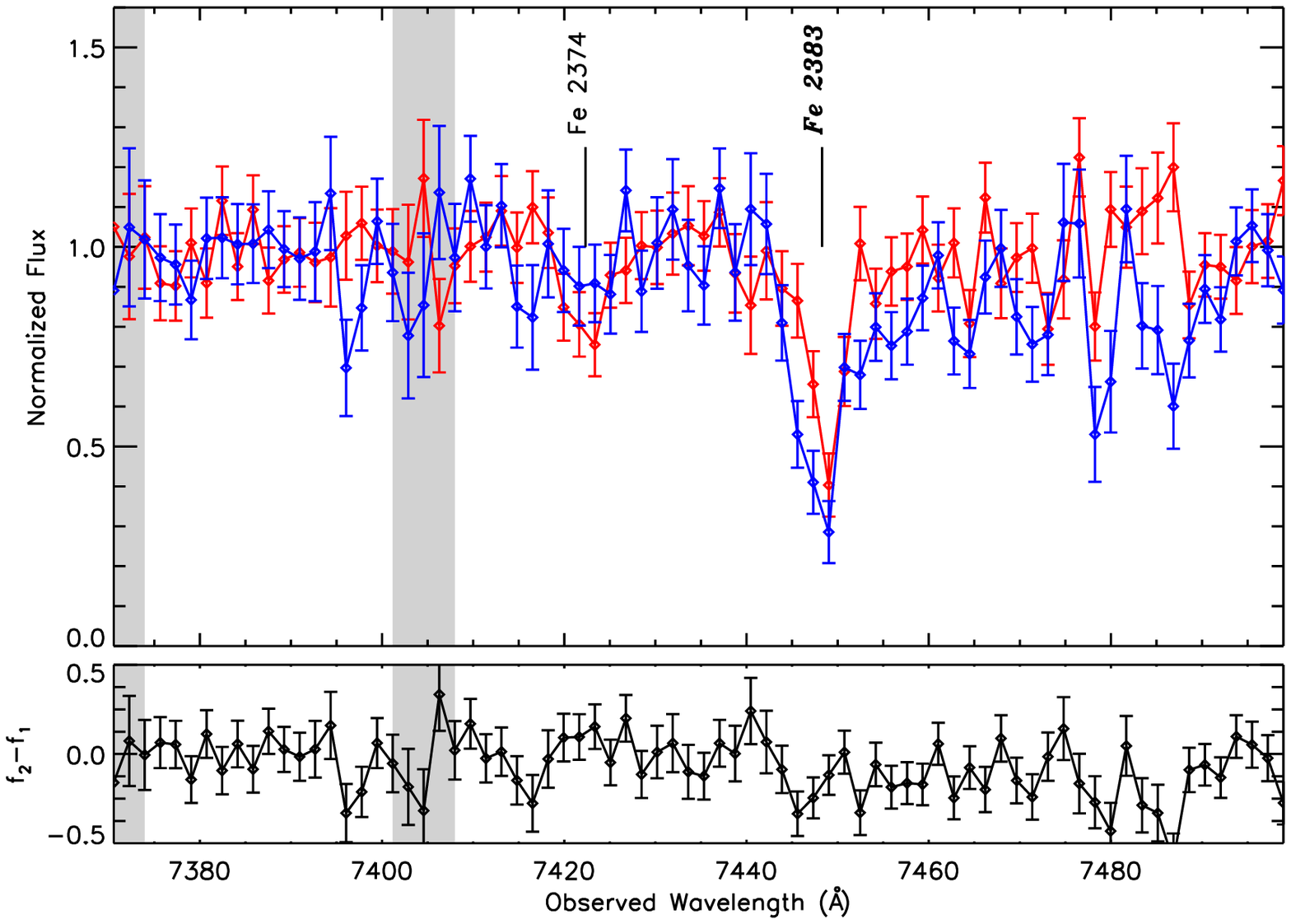}
\includegraphics[width=84mm]{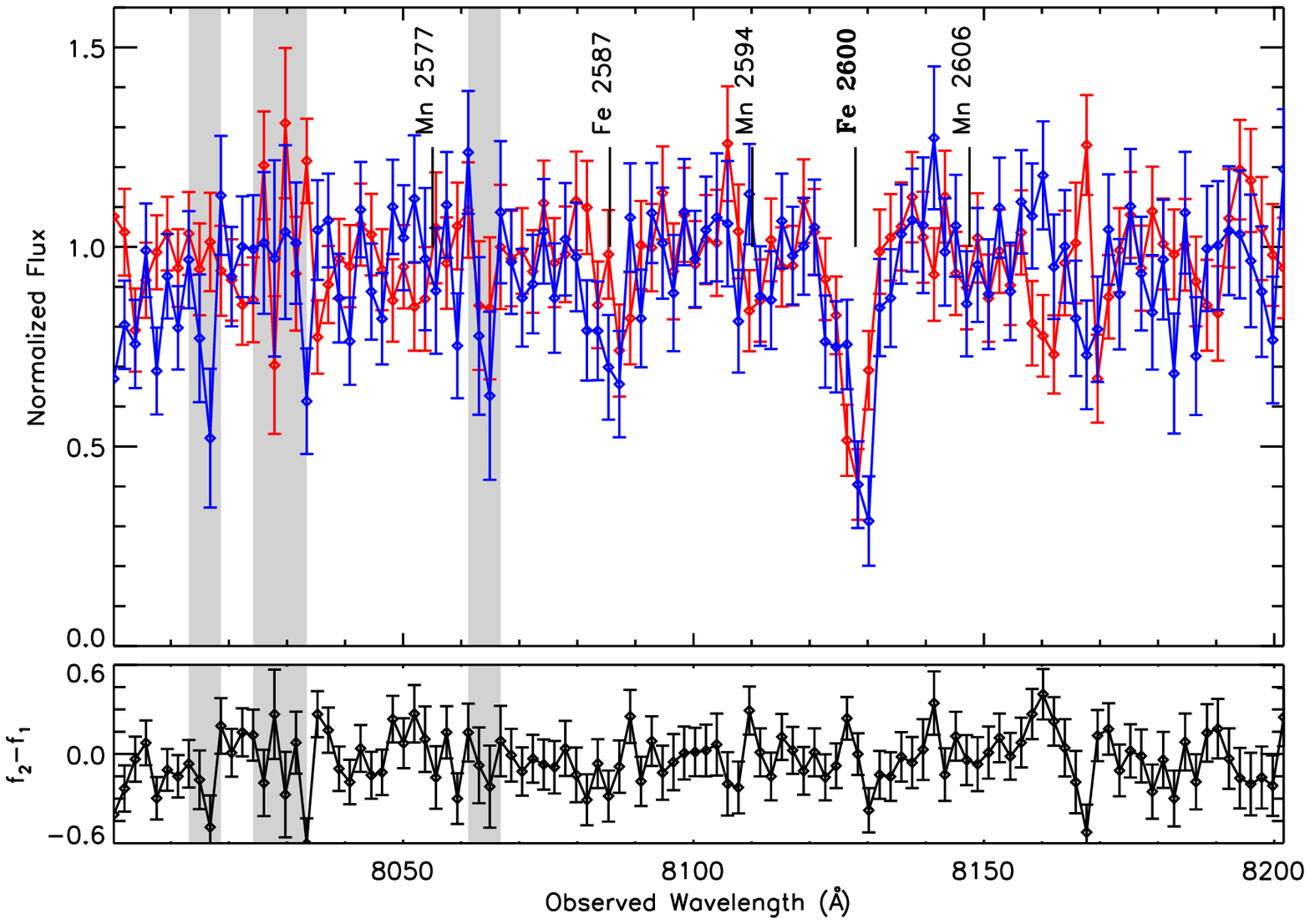}
\includegraphics[width=84mm]{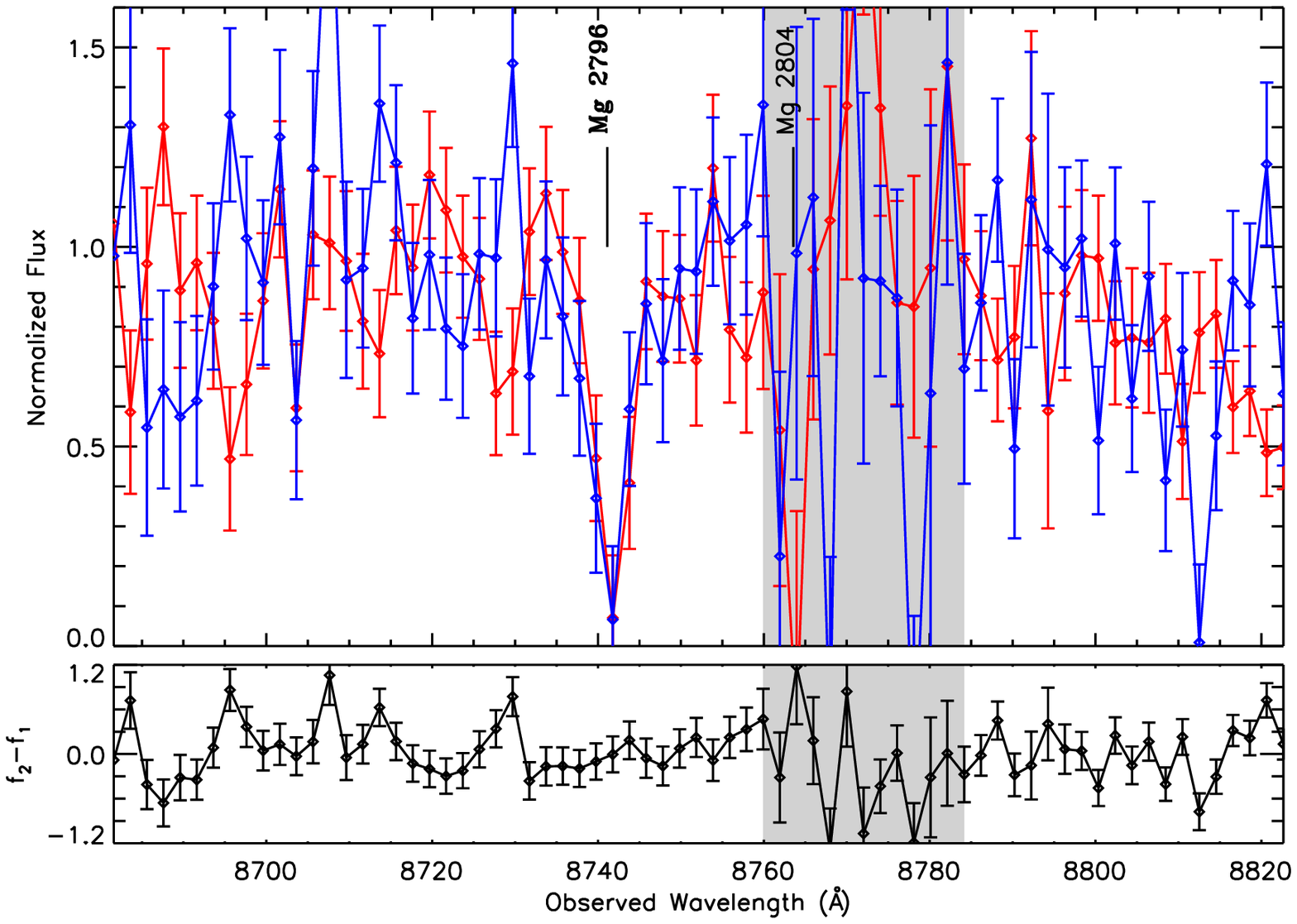}
\caption[]{Two-epoch normalized spectra of the variable NAL system at $\beta$ = 0.0794 in SDSS J024154.42-004757.6.  The top panel shows the normalized pixel flux values with 1$\sigma$ error bars (first observations are red and second are blue), the bottom panel plots the difference spectrum of the two observation epochs, and shaded backgrounds identify masked pixels not included in our search for absorption line variability.  Line identifications for significantly variable absorption lines are italicised, lines detected in both observation epochs are in bold font, and undetected lines are in regular font (see Table A.1 for ion labels).  Continued from previous figure.}
\end{center}
\end{figure*}

\begin{figure*}
\begin{center}
\includegraphics[width=84mm]{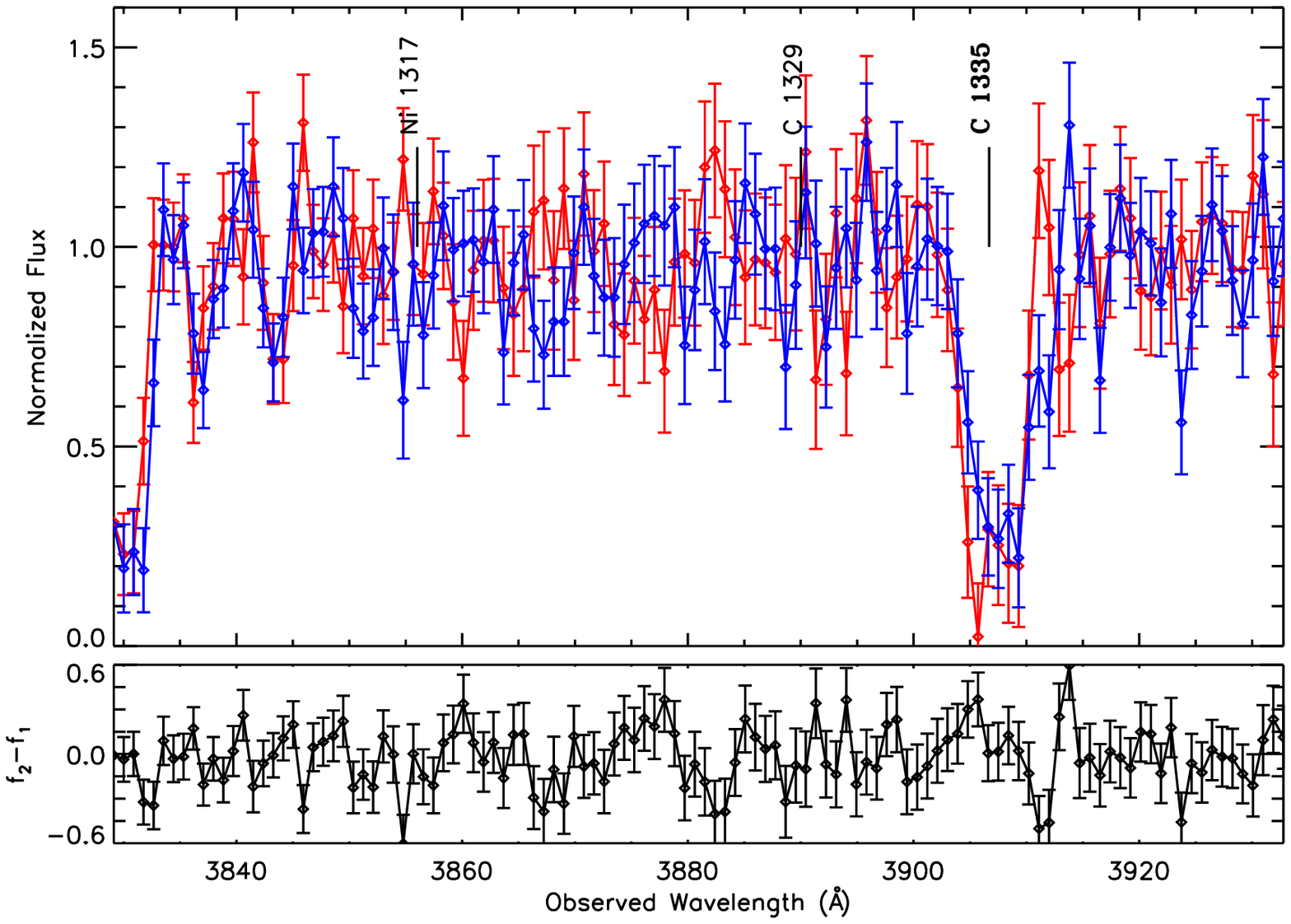}
\includegraphics[width=84mm]{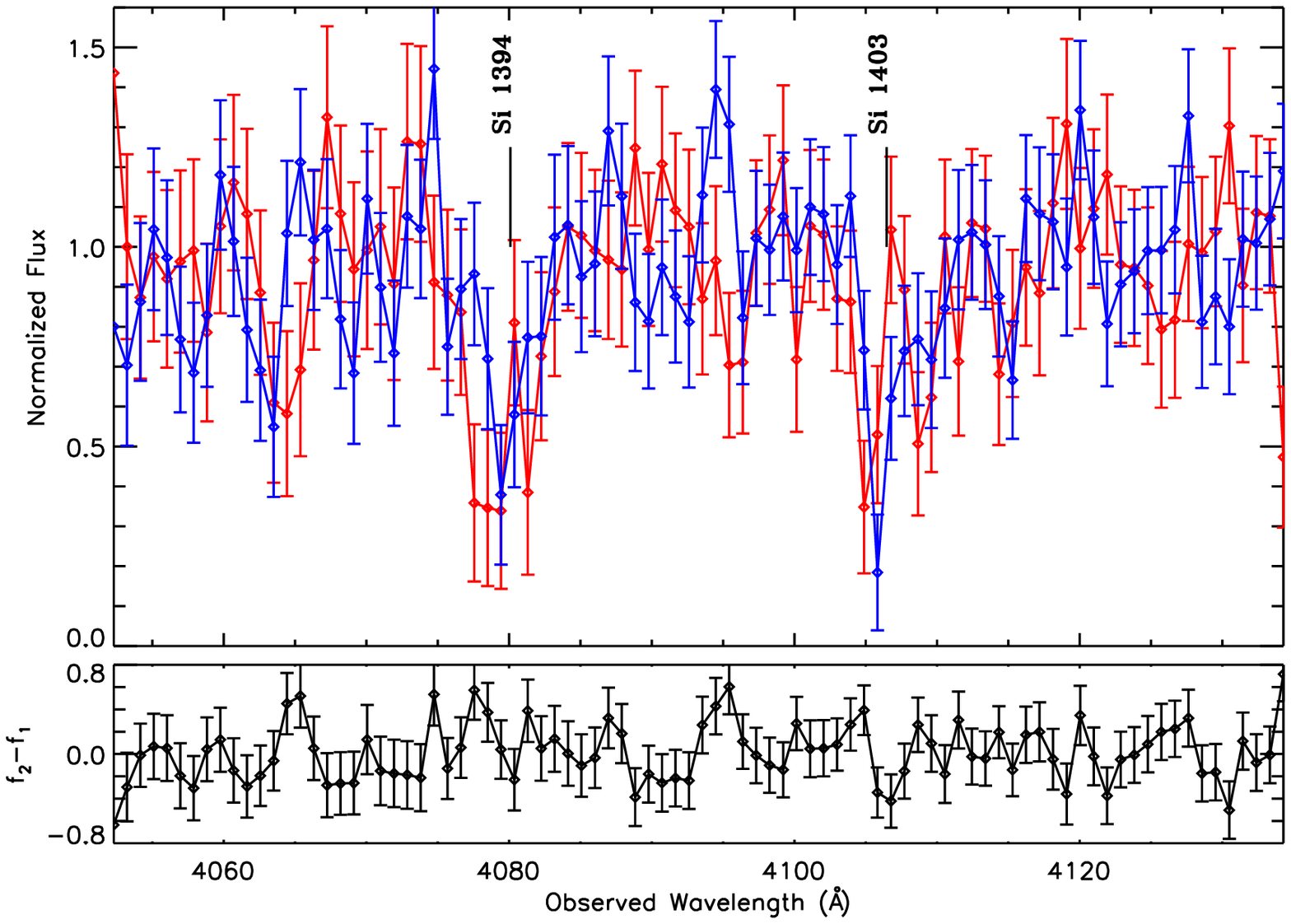}
\includegraphics[width=84mm]{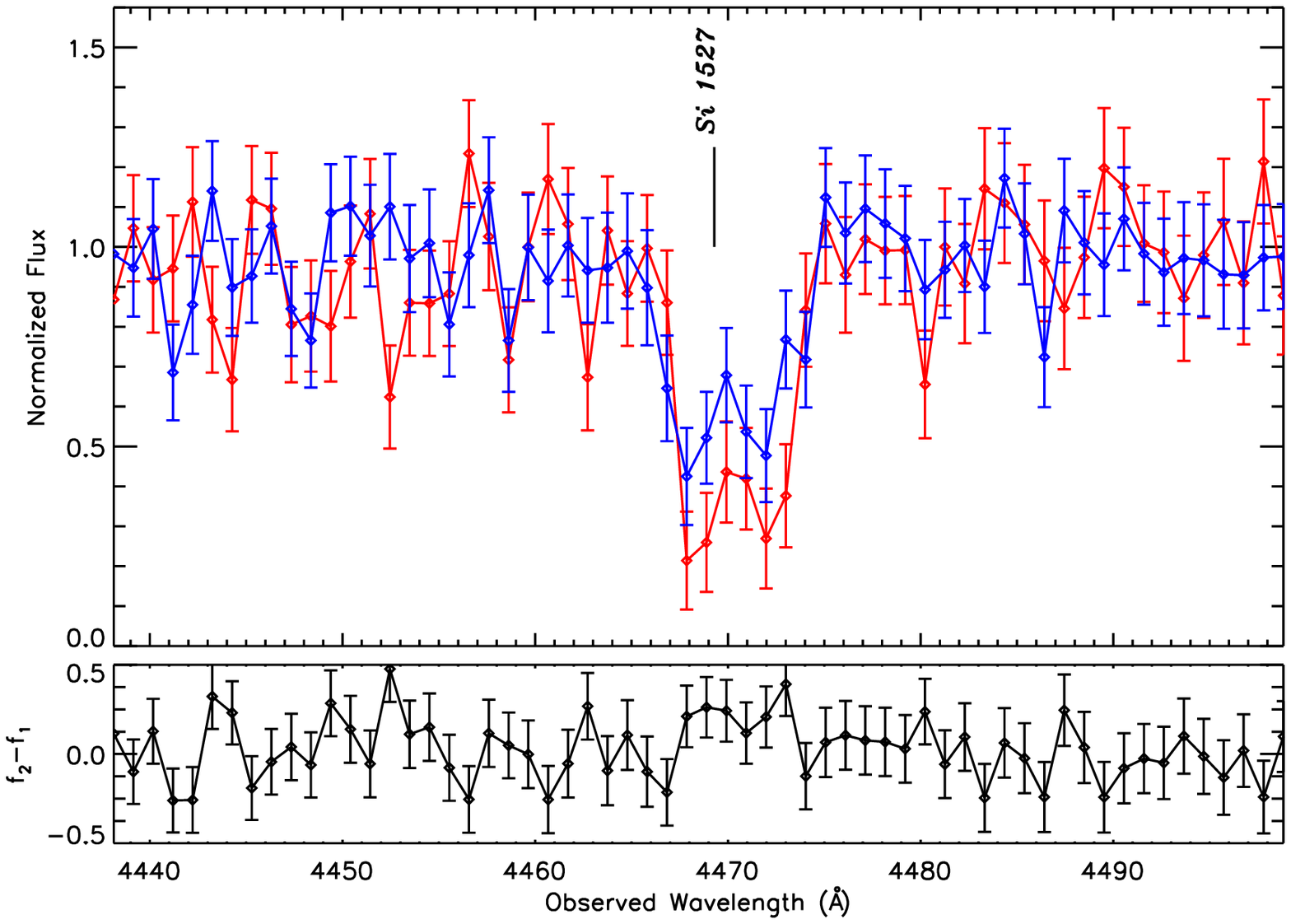}
\includegraphics[width=84mm]{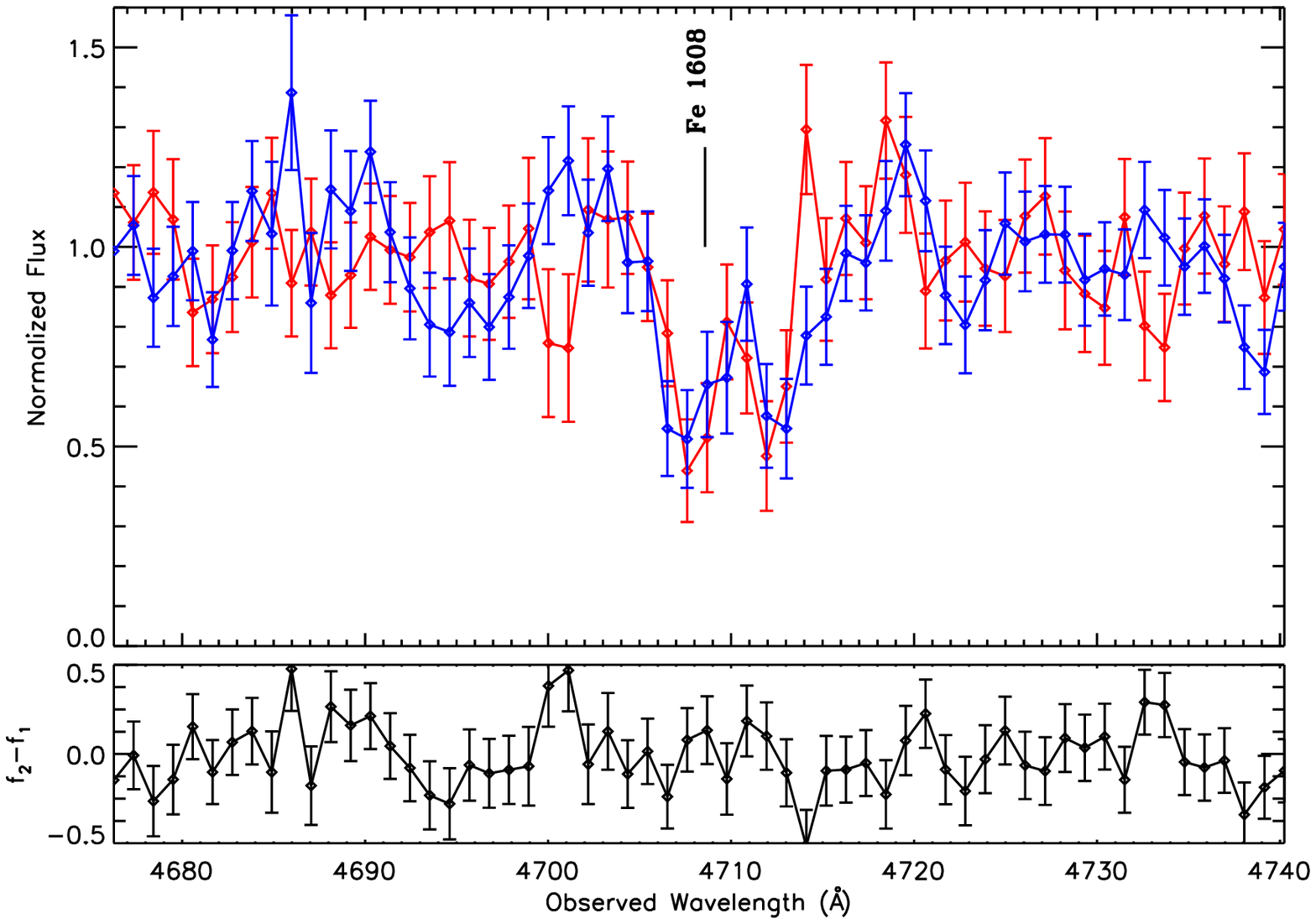}
\includegraphics[width=84mm]{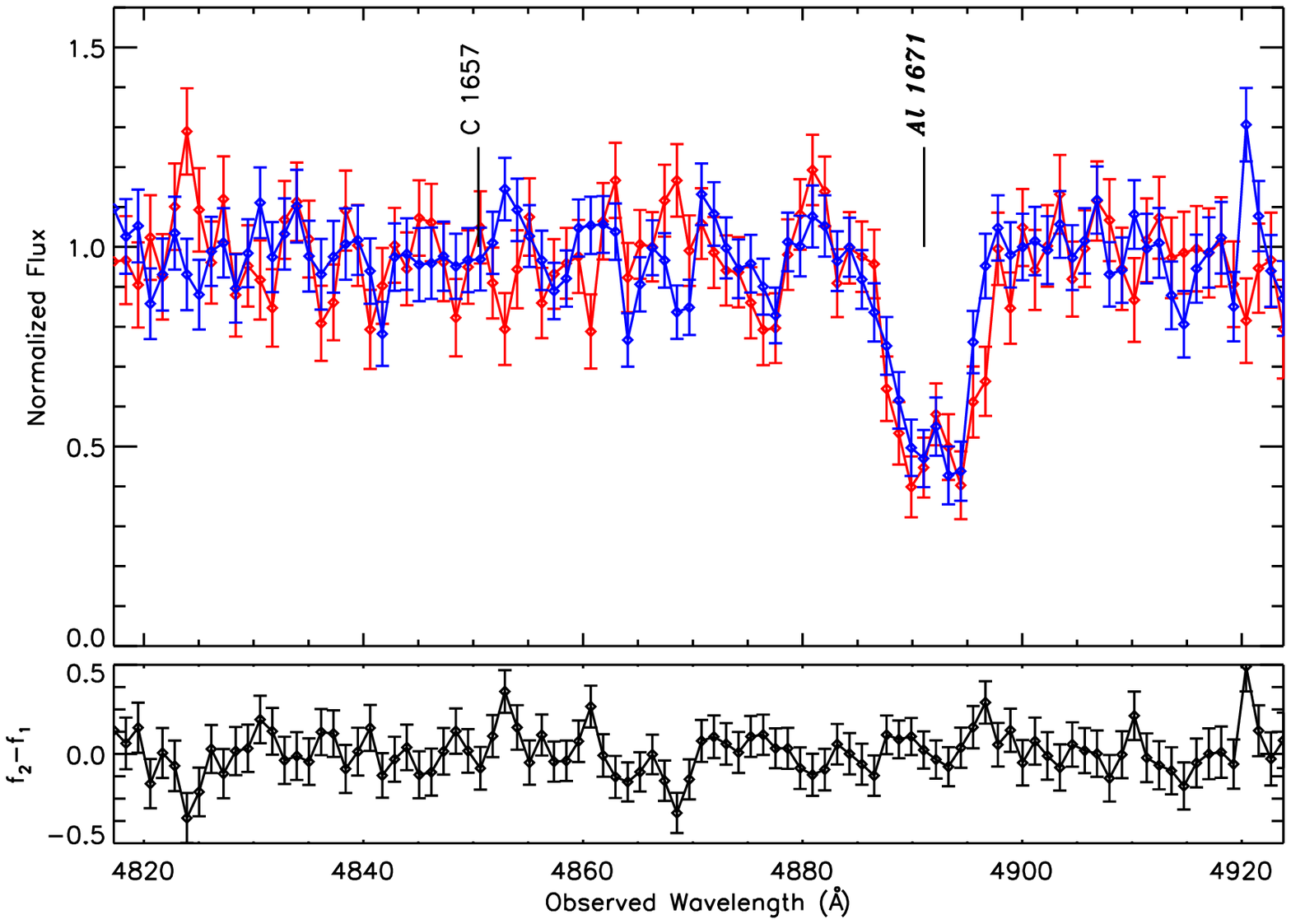}
\includegraphics[width=84mm]{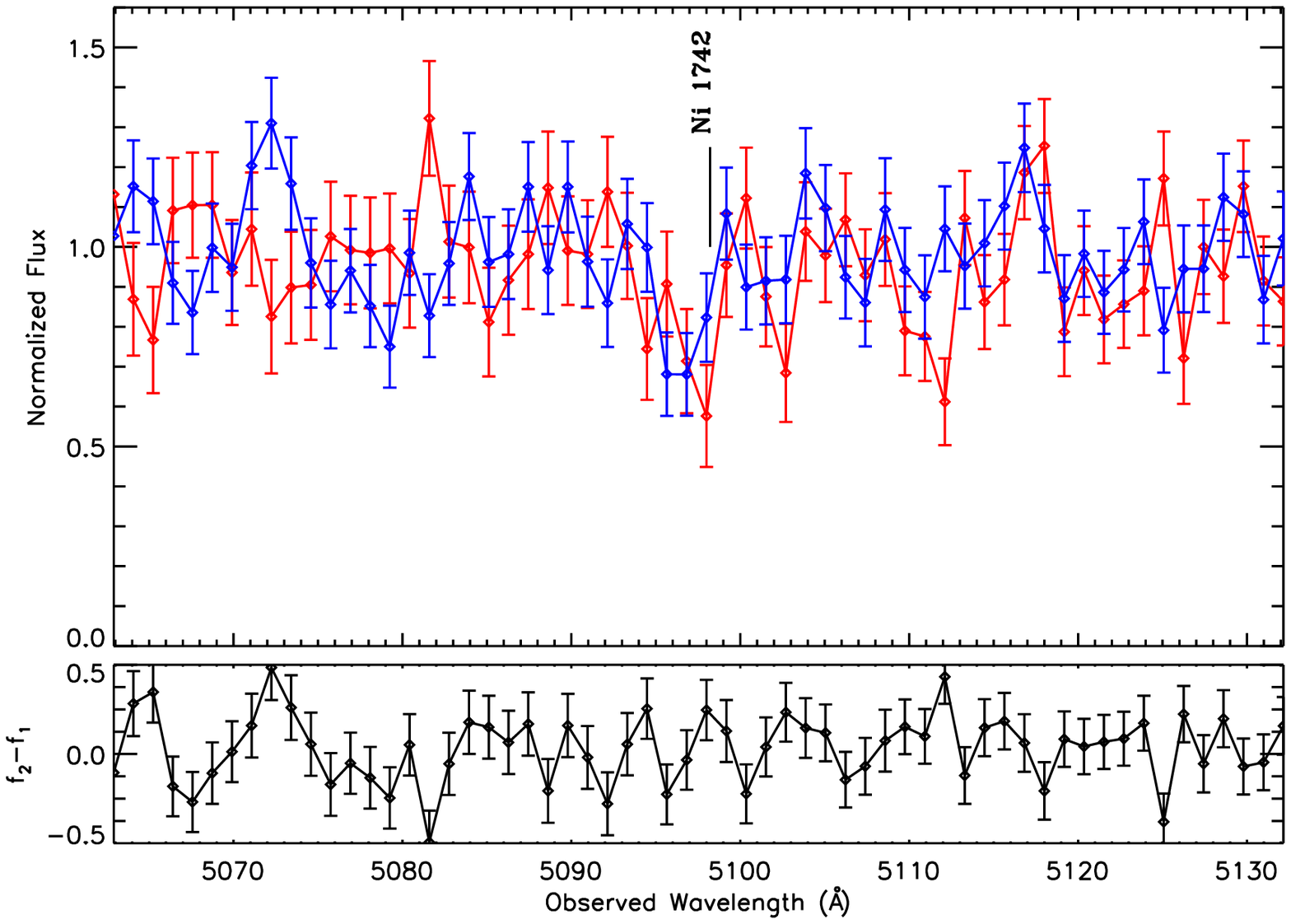}
\caption[Two-epoch normalized spectra of SDSS J131712.01+020224.9]{Two-epoch normalized spectra of the variable NAL system at $\beta$ = 0.0751 in SDSS J131712.01+020224.9.  The top panel shows the normalized pixel flux values with 1$\sigma$ error bars (first observations are red and second are blue), the bottom panel plots the difference spectrum of the two observation epochs, and shaded backgrounds identify masked pixels not included in our search for absorption line variability.  Line identifications for significantly variable absorption lines are italicised, lines detected in both observation epochs are in bold font, and undetected lines are in regular font (see Table A.1 for ion labels).  Continued in next figure.  \label{figvs22}}
\end{center}
\end{figure*}

\begin{figure*}
\ContinuedFloat
\begin{center}
\includegraphics[width=84mm]{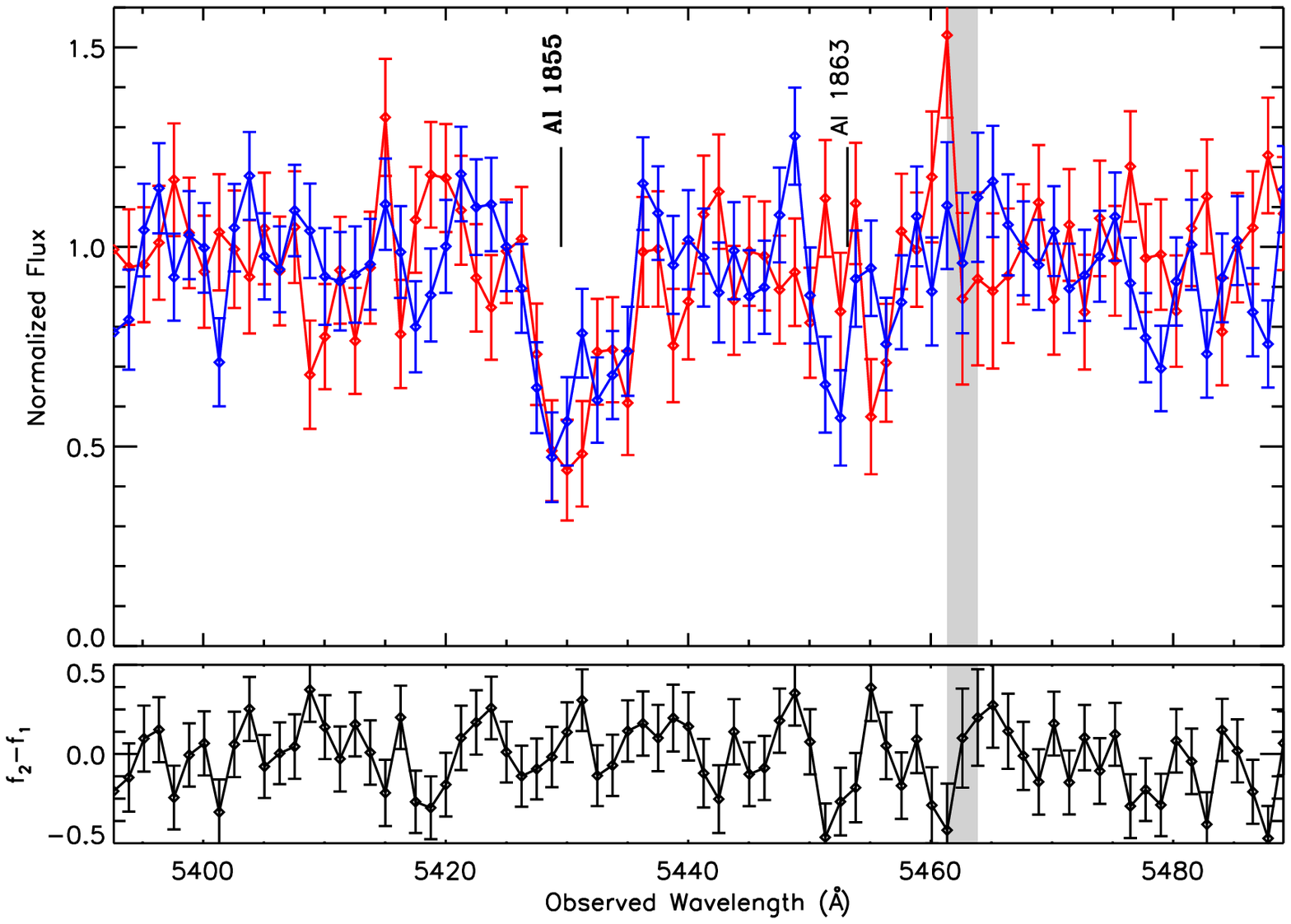}
\includegraphics[width=84mm]{nz1742_0525-52295-026-0526-52312-271_1_9274.eps}
\includegraphics[width=84mm]{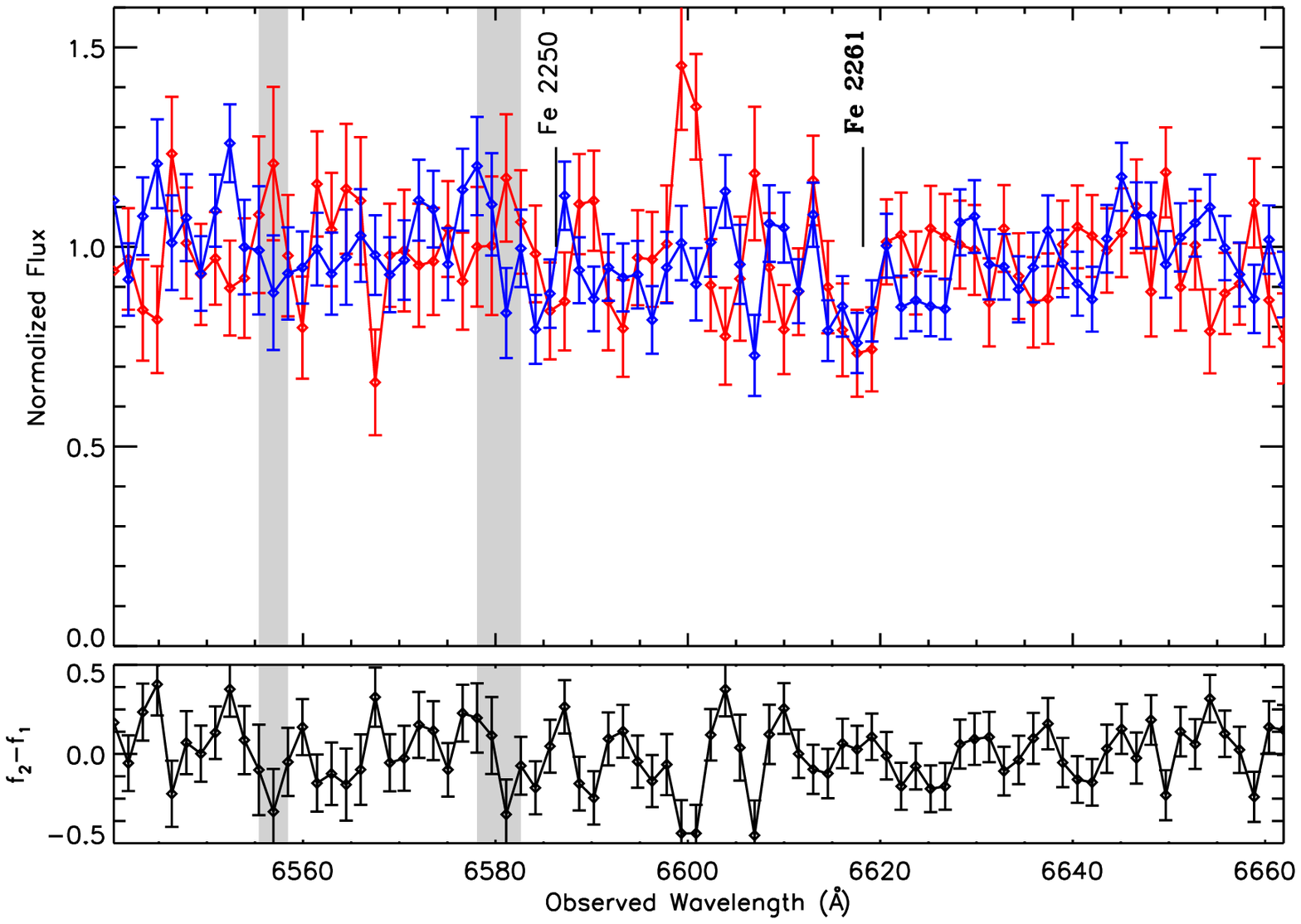}
\includegraphics[width=84mm]{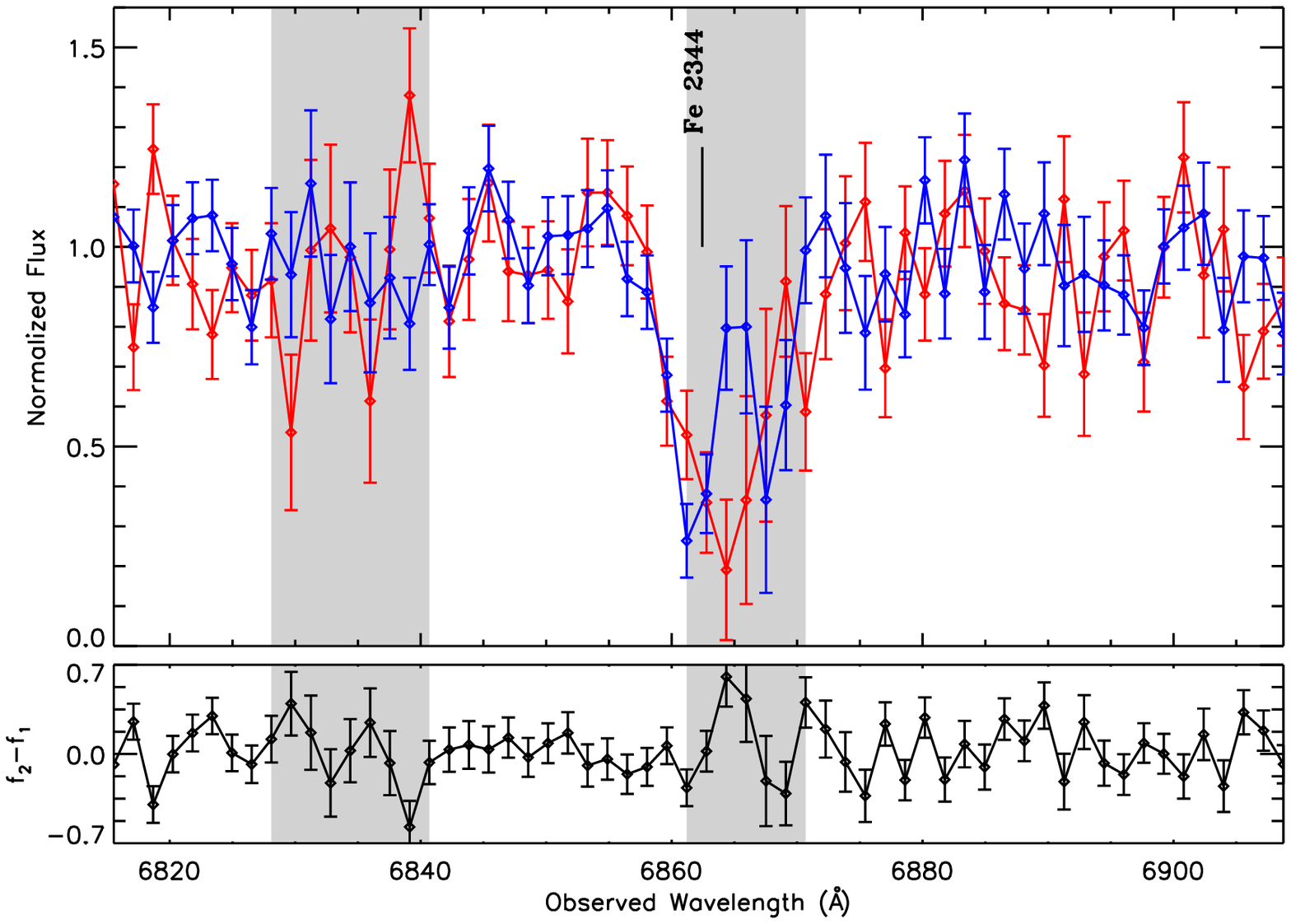}
\includegraphics[width=84mm]{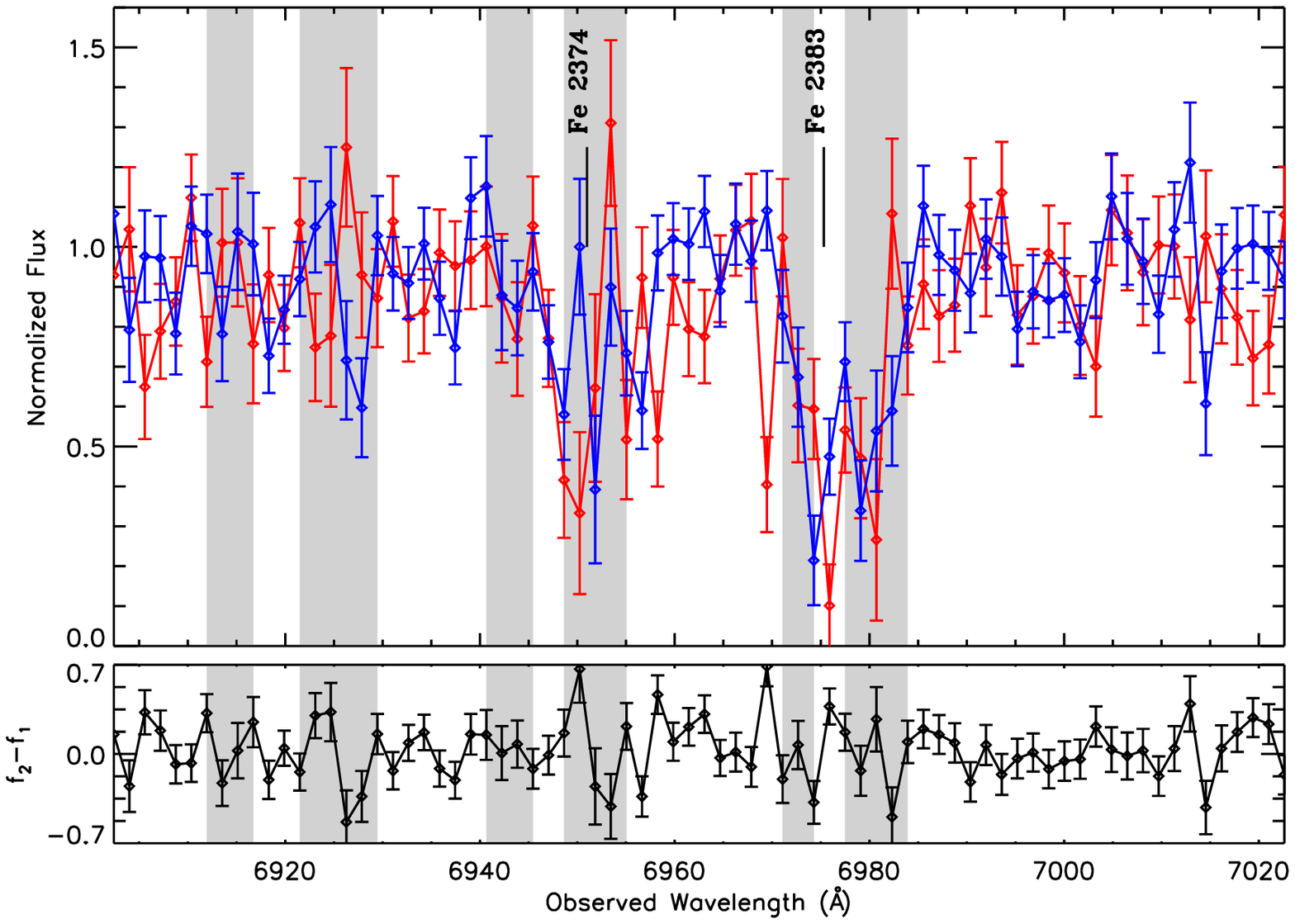}
\includegraphics[width=84mm]{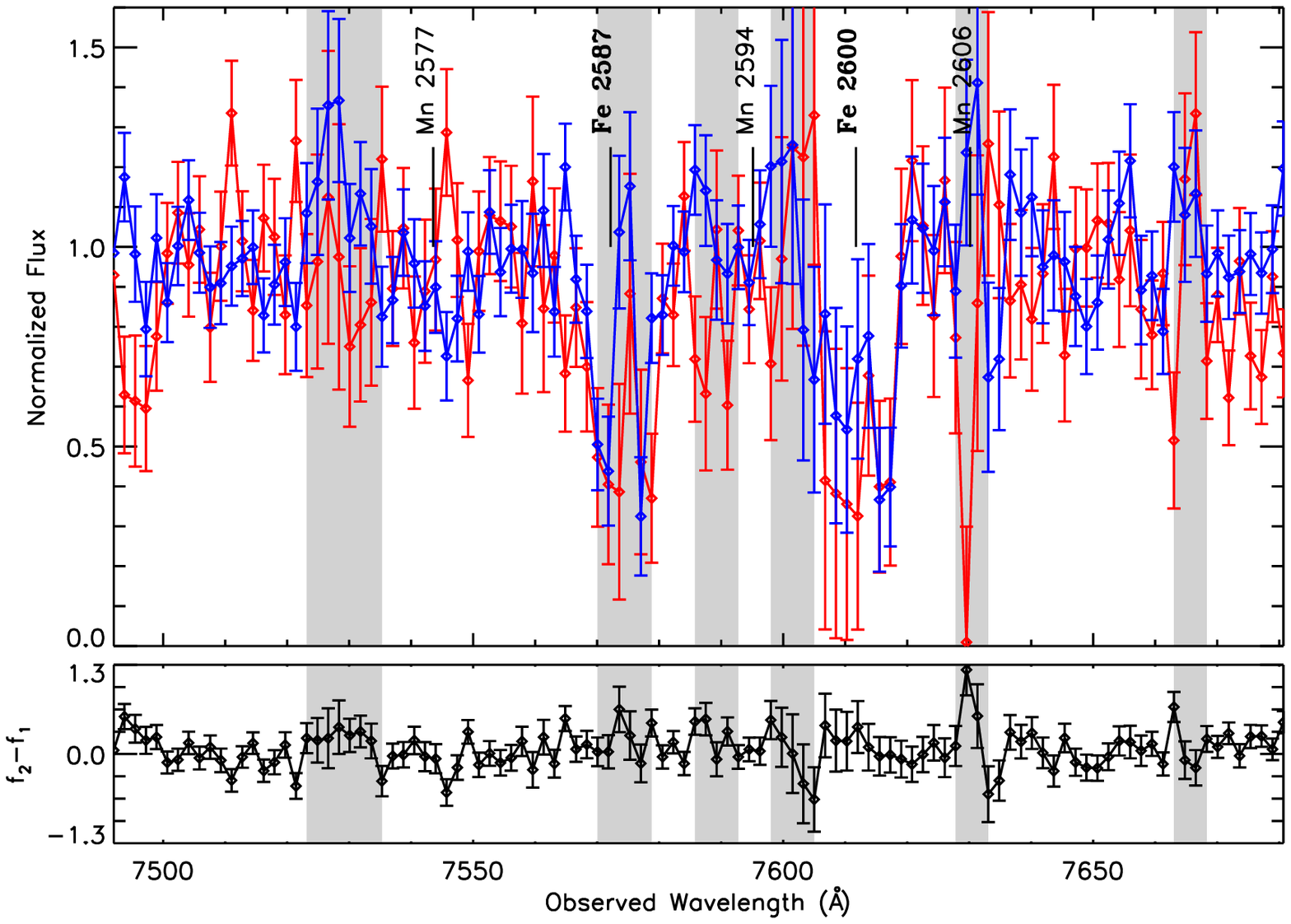}
\caption[]{Two-epoch normalized spectra of the variable NAL system at $\beta$ = 0.0751 in SDSS J131712.01+020224.9.  The top panel shows the normalized pixel flux values with 1$\sigma$ error bars (first observations are red and second are blue), the bottom panel plots the difference spectrum of the two observation epochs, and shaded backgrounds identify masked pixels not included in our search for absorption line variability.  Line identifications for significantly variable absorption lines are italicised, lines detected in both observation epochs are in bold font, and undetected lines are in regular font (see Table A.1 for ion labels).  Continued from previous figure.}
\end{center}
\end{figure*}

\begin{figure*}
\ContinuedFloat
\begin{center}
\includegraphics[width=84mm]{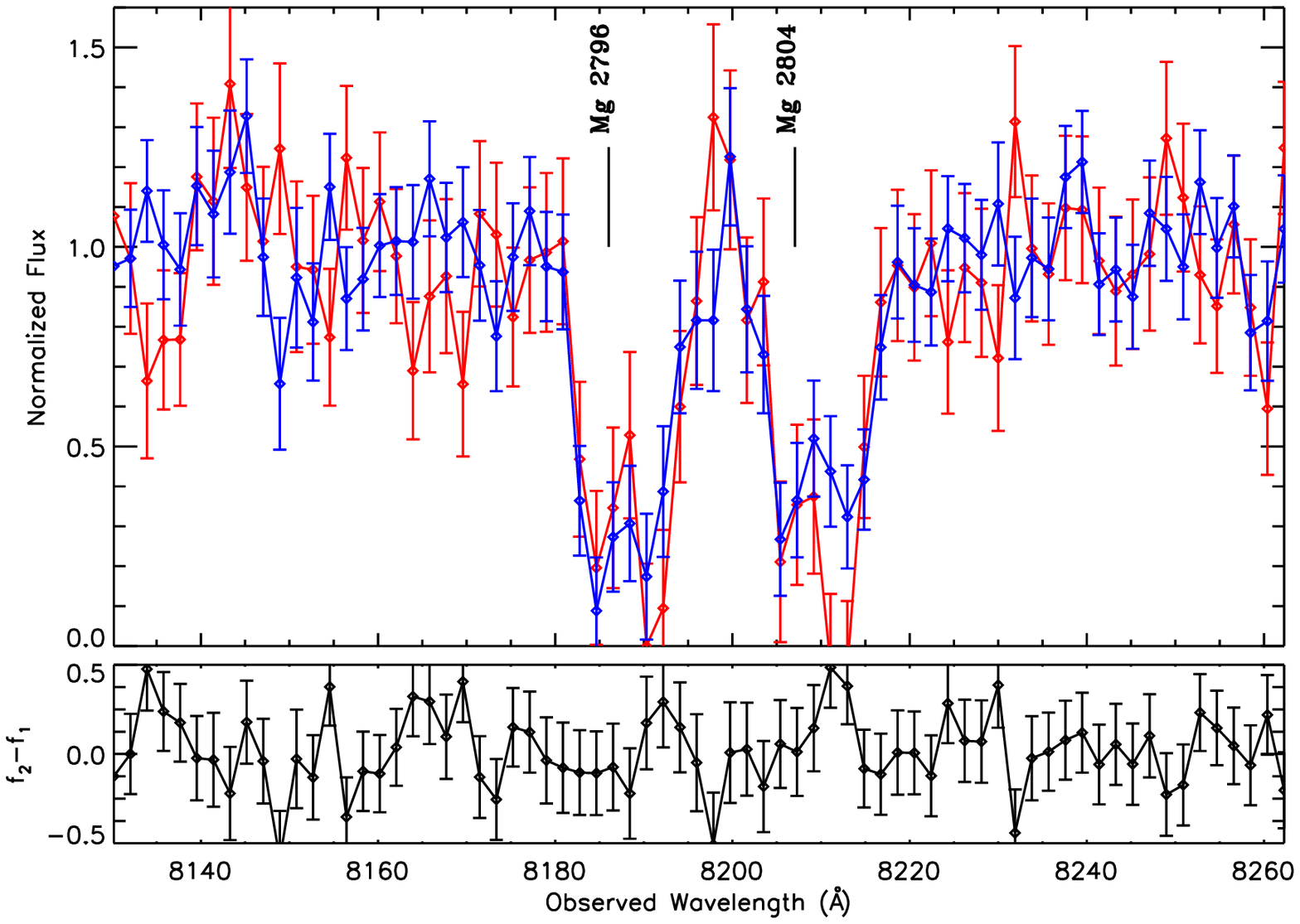}
\caption[]{Two-epoch normalized spectra of the variable NAL system at $\beta$ = 0.0751 in SDSS J131712.01+020224.9.  The top panel shows the normalized pixel flux values with 1$\sigma$ error bars (first observations are red and second are blue), the bottom panel plots the difference spectrum of the two observation epochs, and shaded backgrounds identify masked pixels not included in our search for absorption line variability.  Line identifications for significantly variable absorption lines are italicised, lines detected in both observation epochs are in bold font, and undetected lines are in regular font (see Table A.1 for ion labels).  Continued from previous figure.}
\vspace{3.5cm}
\end{center}
\end{figure*}

\begin{figure*}
\begin{center}
\includegraphics[width=84mm]{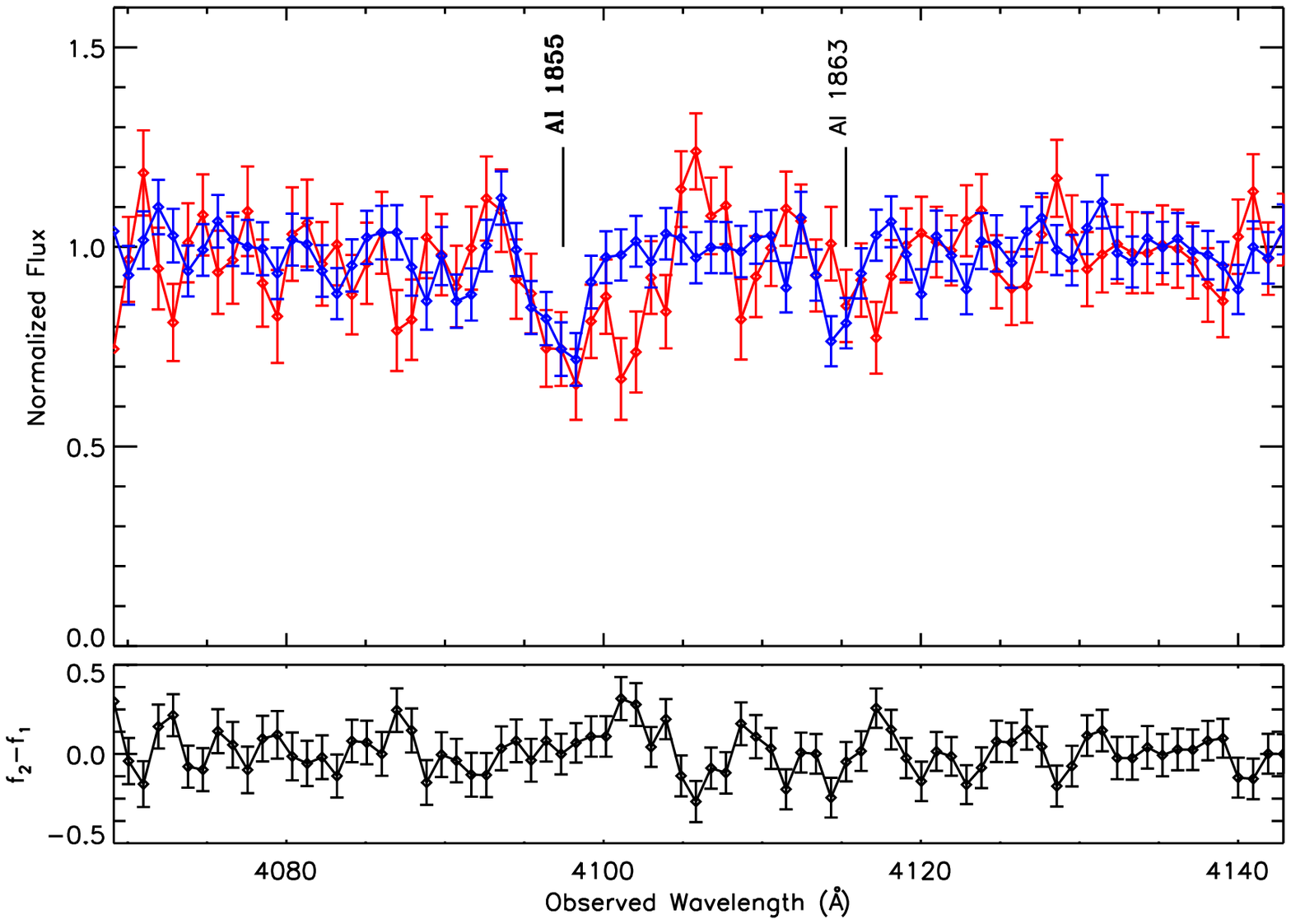}
\includegraphics[width=84mm]{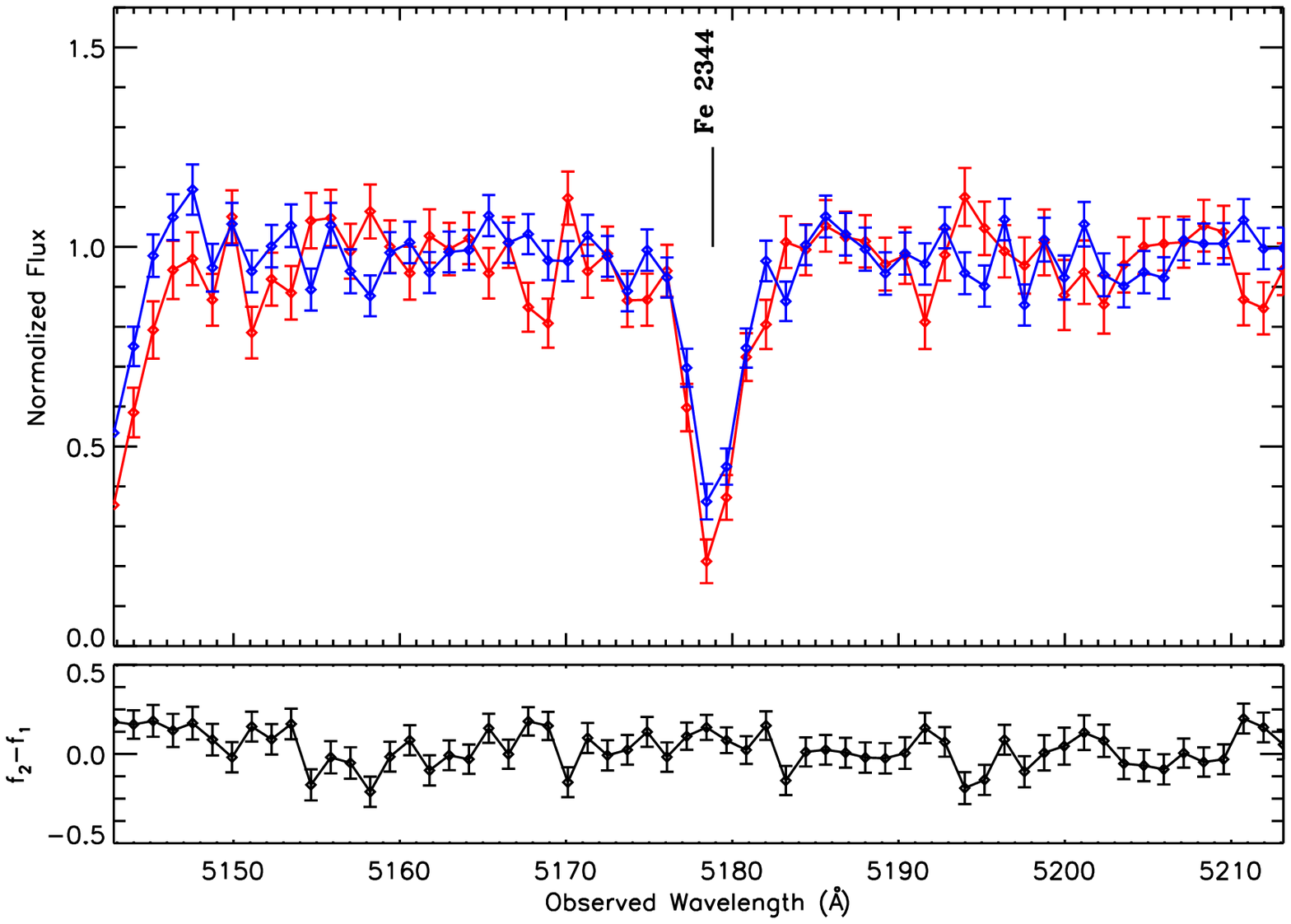}
\includegraphics[width=84mm]{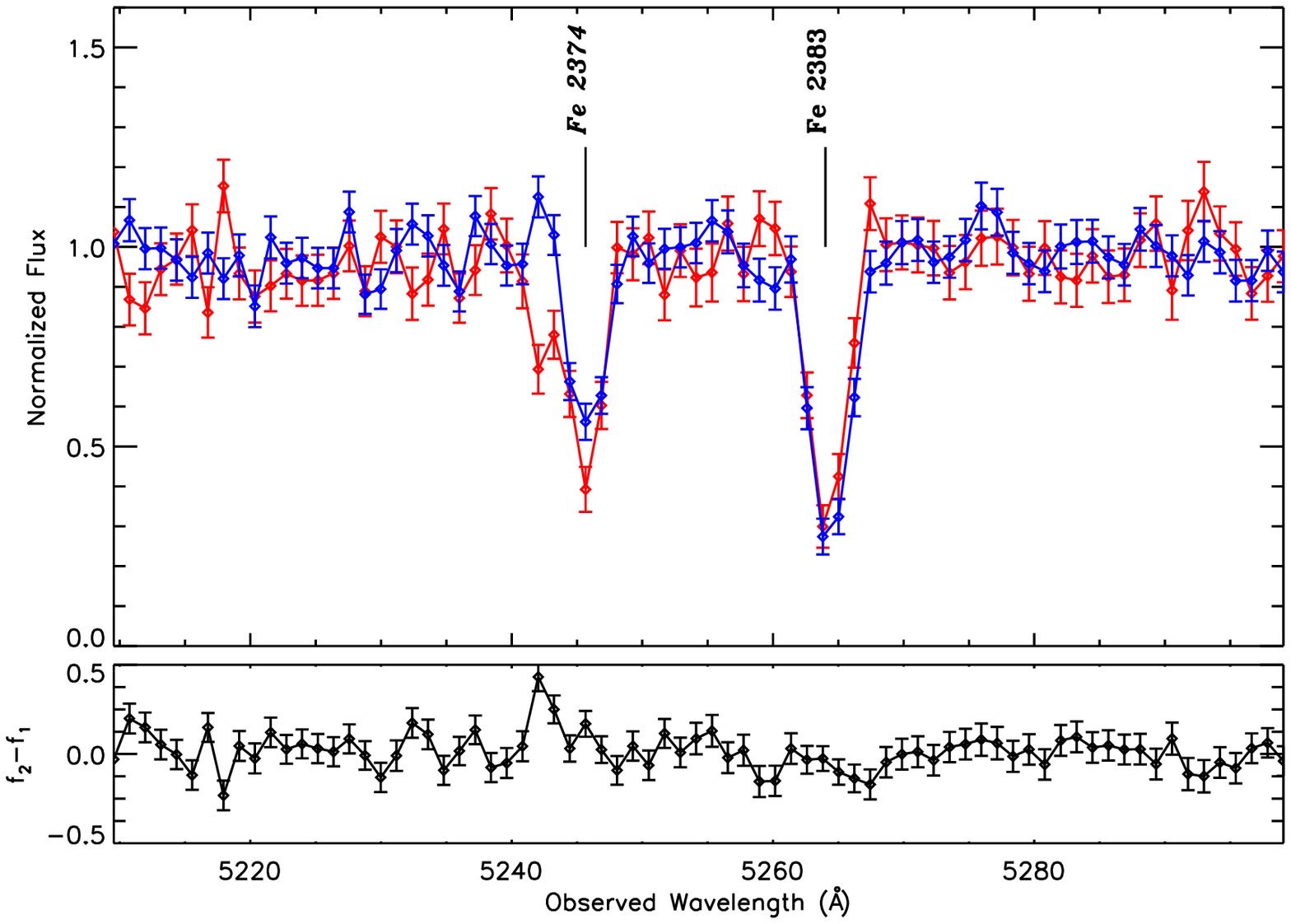}
\includegraphics[width=84mm]{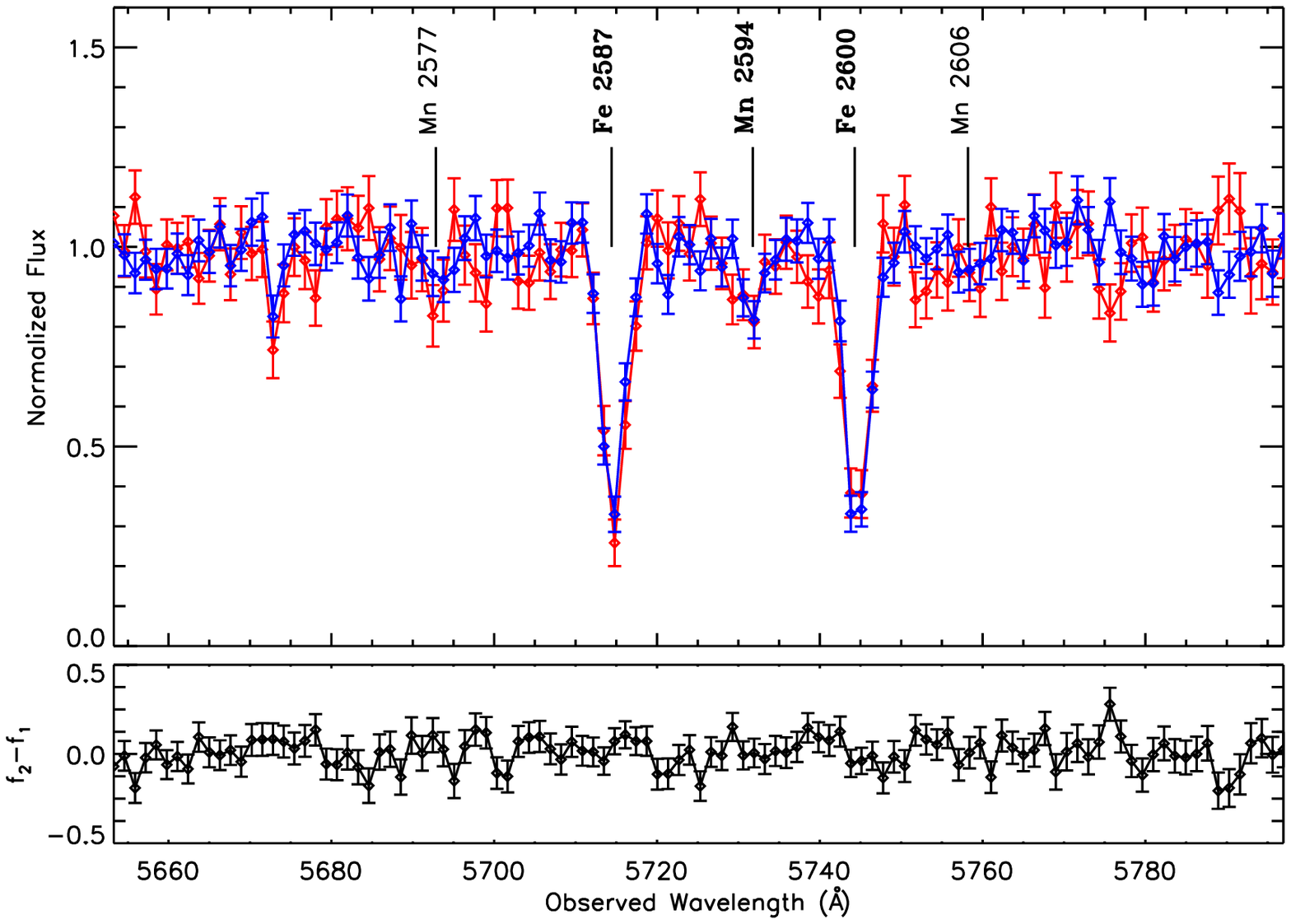}
\includegraphics[width=84mm]{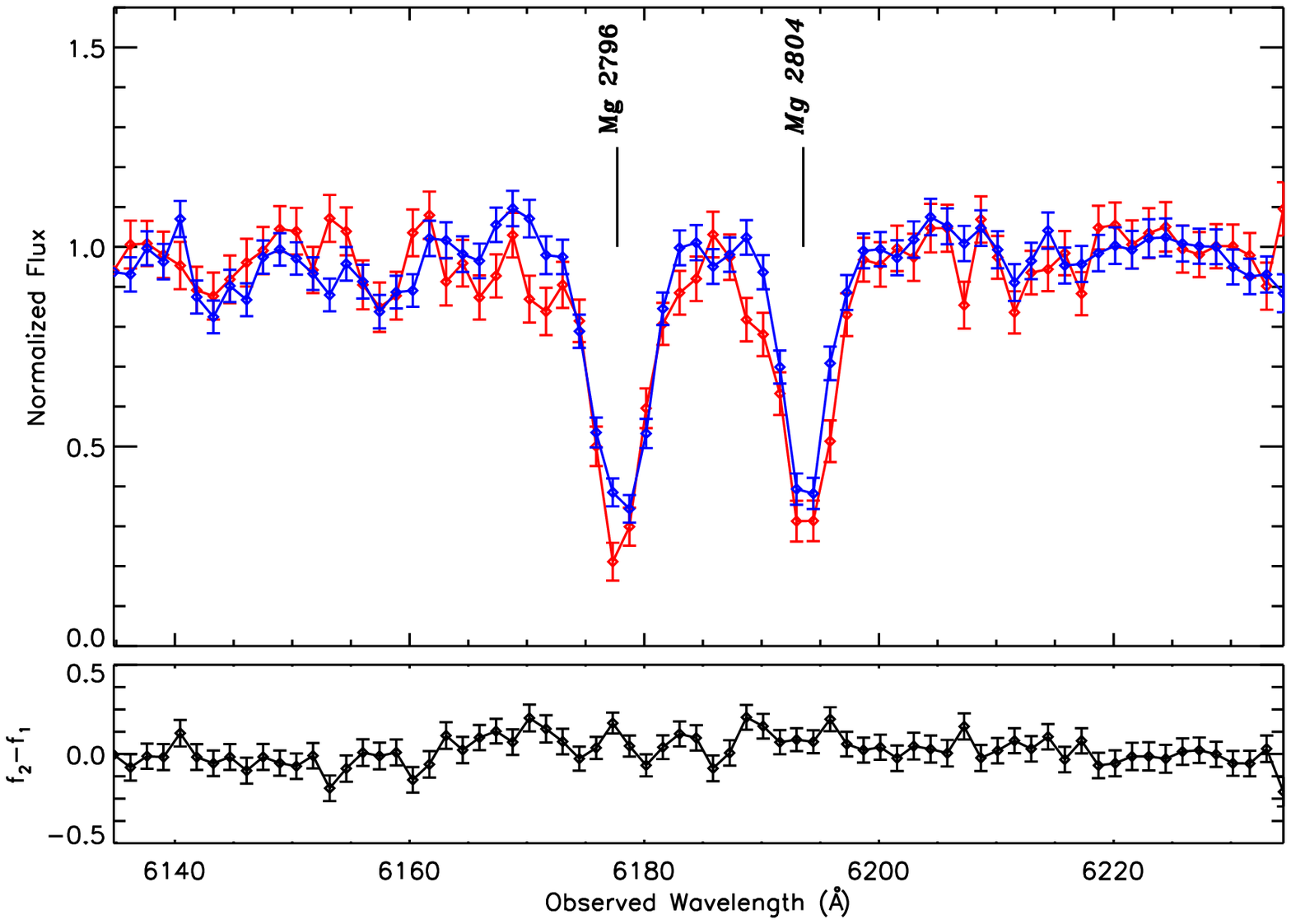}
\includegraphics[width=84mm]{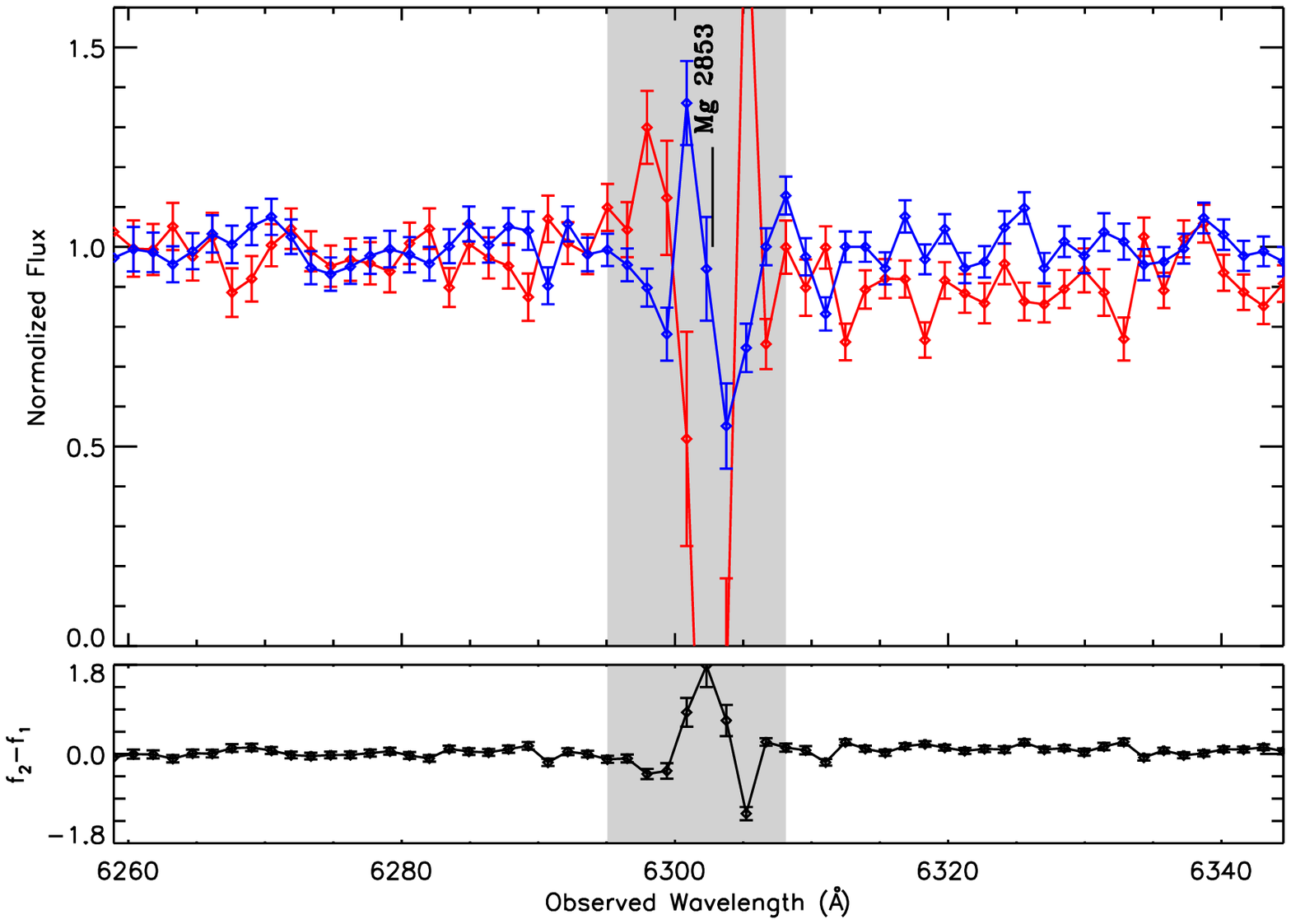}
\caption[Two-epoch normalized spectra of SDSS J171244.11+555949.7]{Two-epoch normalized spectra of the variable NAL system at $\beta$ = 0.0660 in SDSS J171244.11+555949.7.  The top panel shows the normalized pixel flux values with 1$\sigma$ error bars (first observations are red and second are blue), the bottom panel plots the difference spectrum of the two observation epochs, and shaded backgrounds identify masked pixels not included in our search for absorption line variability.  Line identifications for significantly variable absorption lines are italicised, lines detected in both observation epochs are in bold font, and undetected lines are in regular font (see Table A.1 for ion labels).  Continued in next figure.  \label{figvs23}}
\end{center}
\end{figure*}

\begin{figure*}
\ContinuedFloat
\begin{center}
\includegraphics[width=84mm]{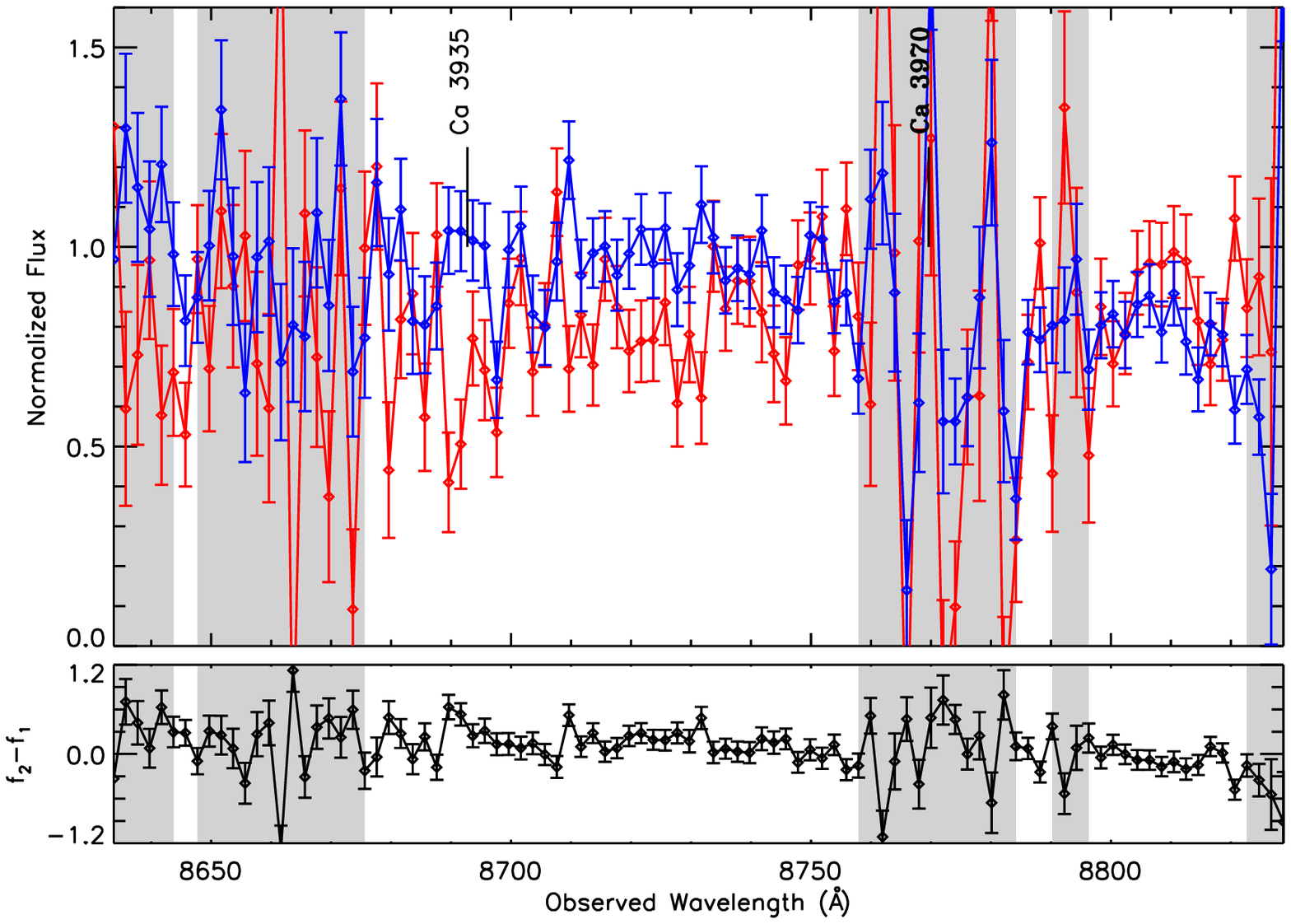}
\caption[]{Two-epoch normalized spectra of the variable NAL system at $\beta$ = 0.0660 in SDSS J171244.11+555949.7.  The top panel shows the normalized pixel flux values with 1$\sigma$ error bars (first observations are red and second are blue), the bottom panel plots the difference spectrum of the two observation epochs, and shaded backgrounds identify masked pixels not included in our search for absorption line variability.  Line identifications for significantly variable absorption lines are italicised, lines detected in both observation epochs are in bold font, and undetected lines are in regular font (see Table A.1 for ion labels).  Continued from previous figure.}
\vspace{3.5cm}
\end{center}
\end{figure*}

\begin{figure*}
\begin{center}
\includegraphics[width=84mm]{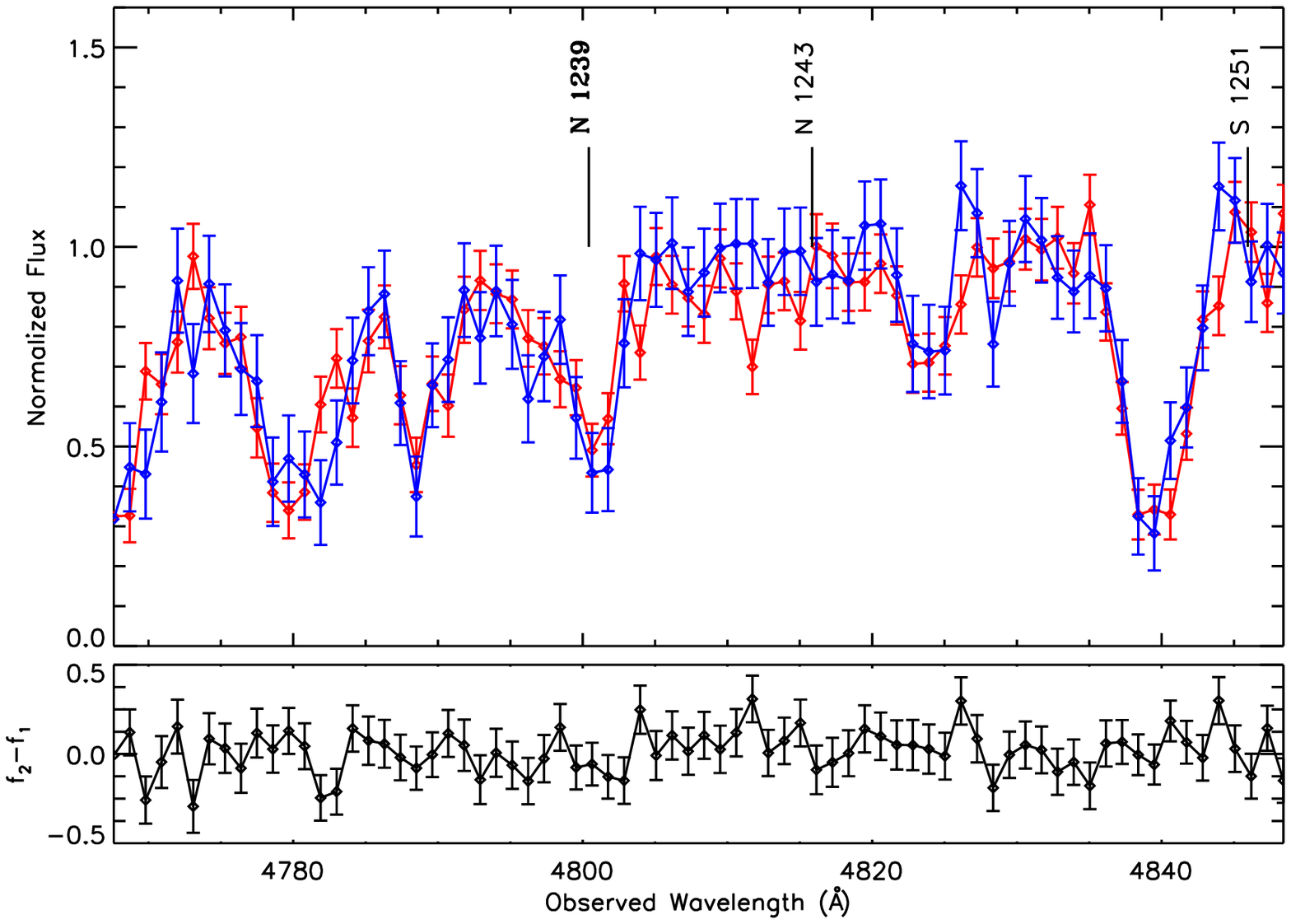}
\includegraphics[width=84mm]{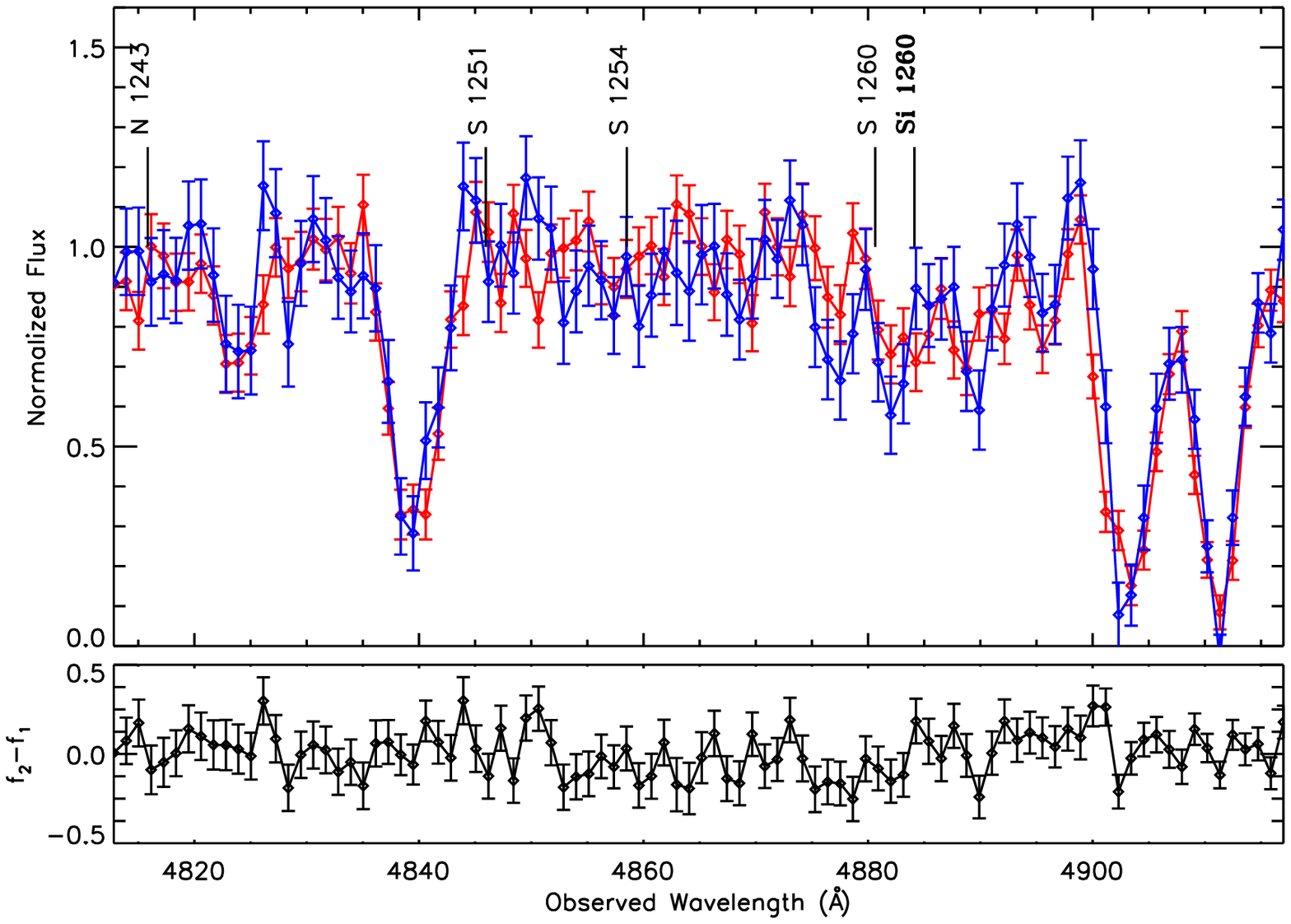}
\includegraphics[width=84mm]{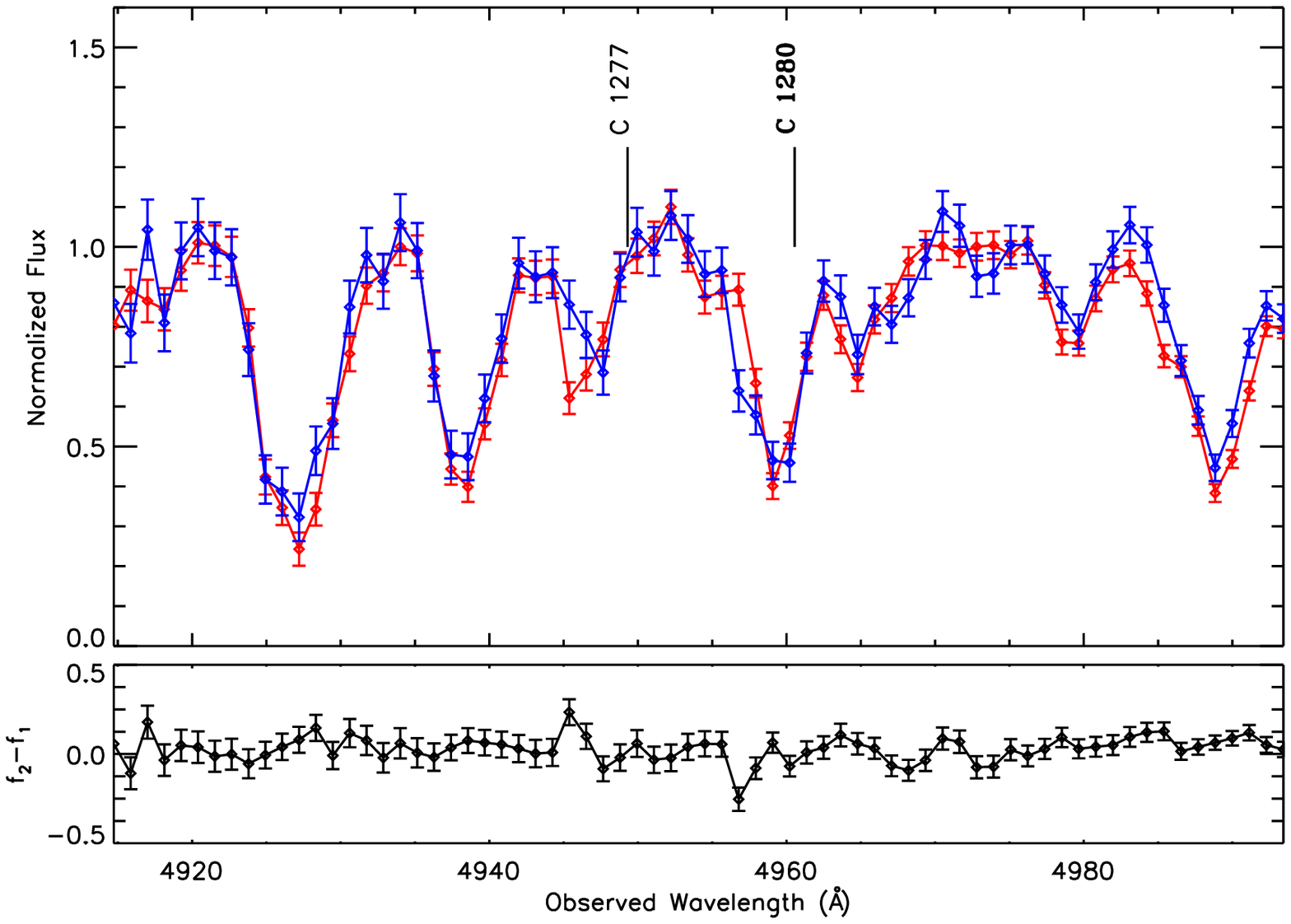}
\includegraphics[width=84mm]{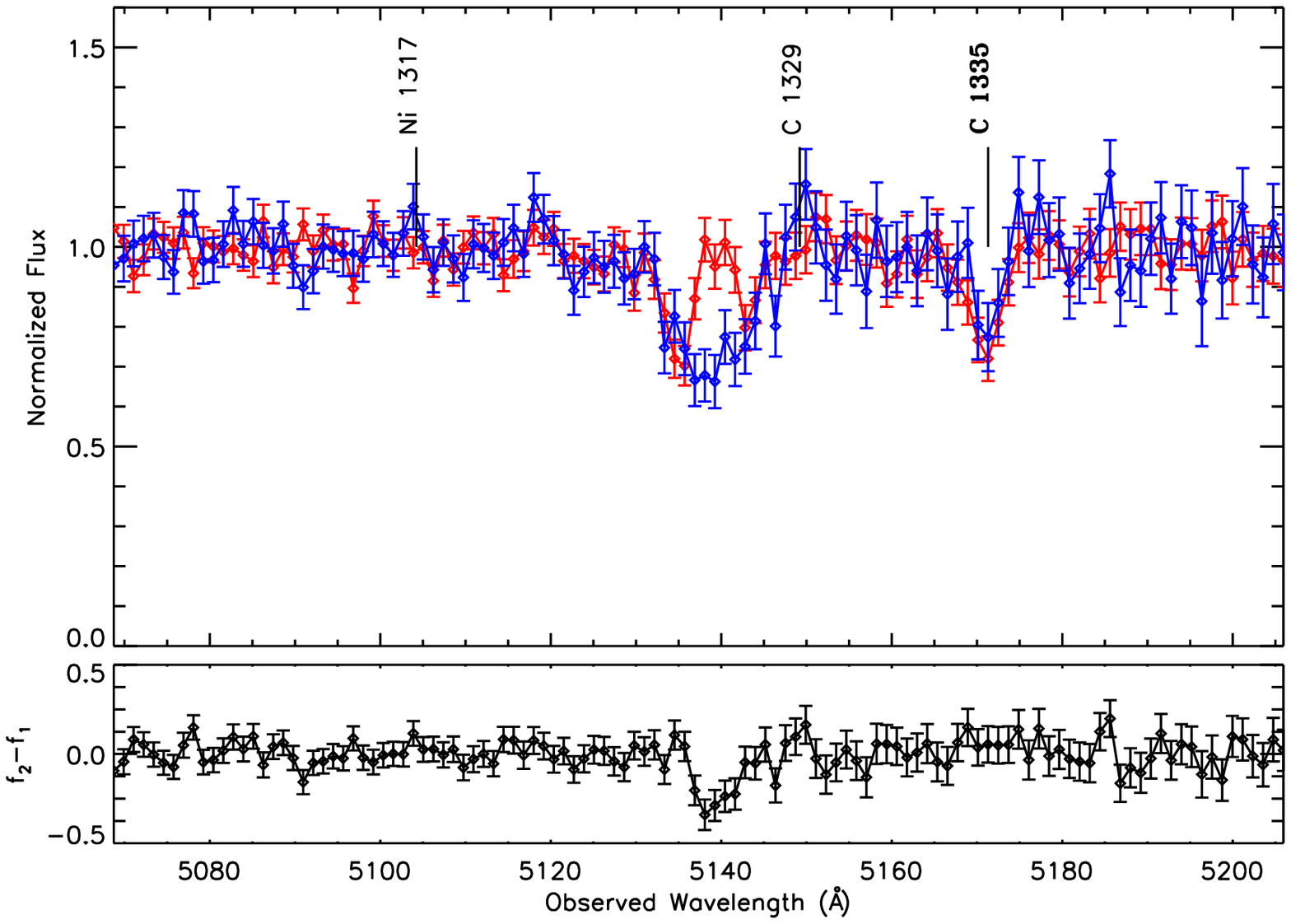}
\includegraphics[width=84mm]{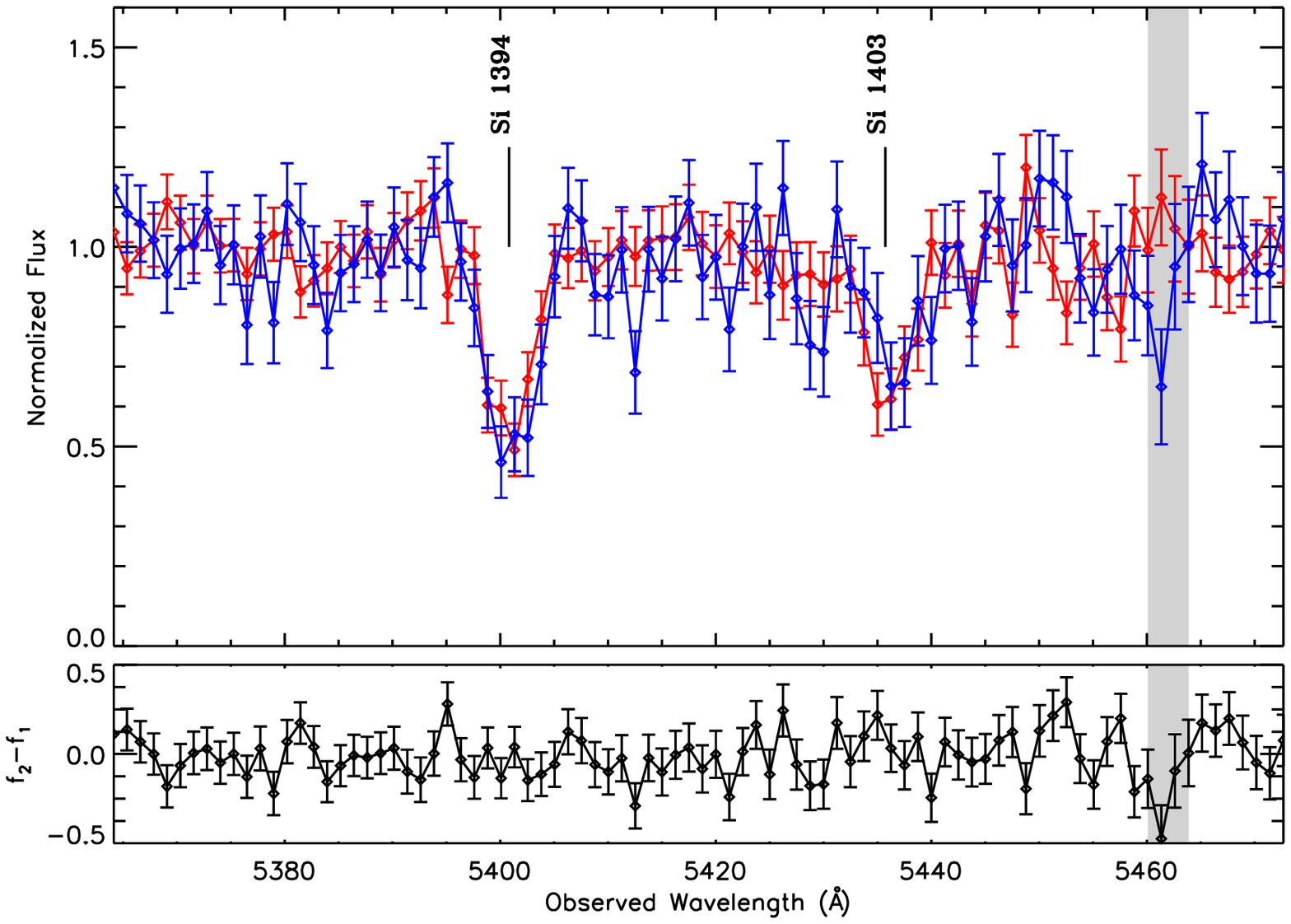}
\includegraphics[width=84mm]{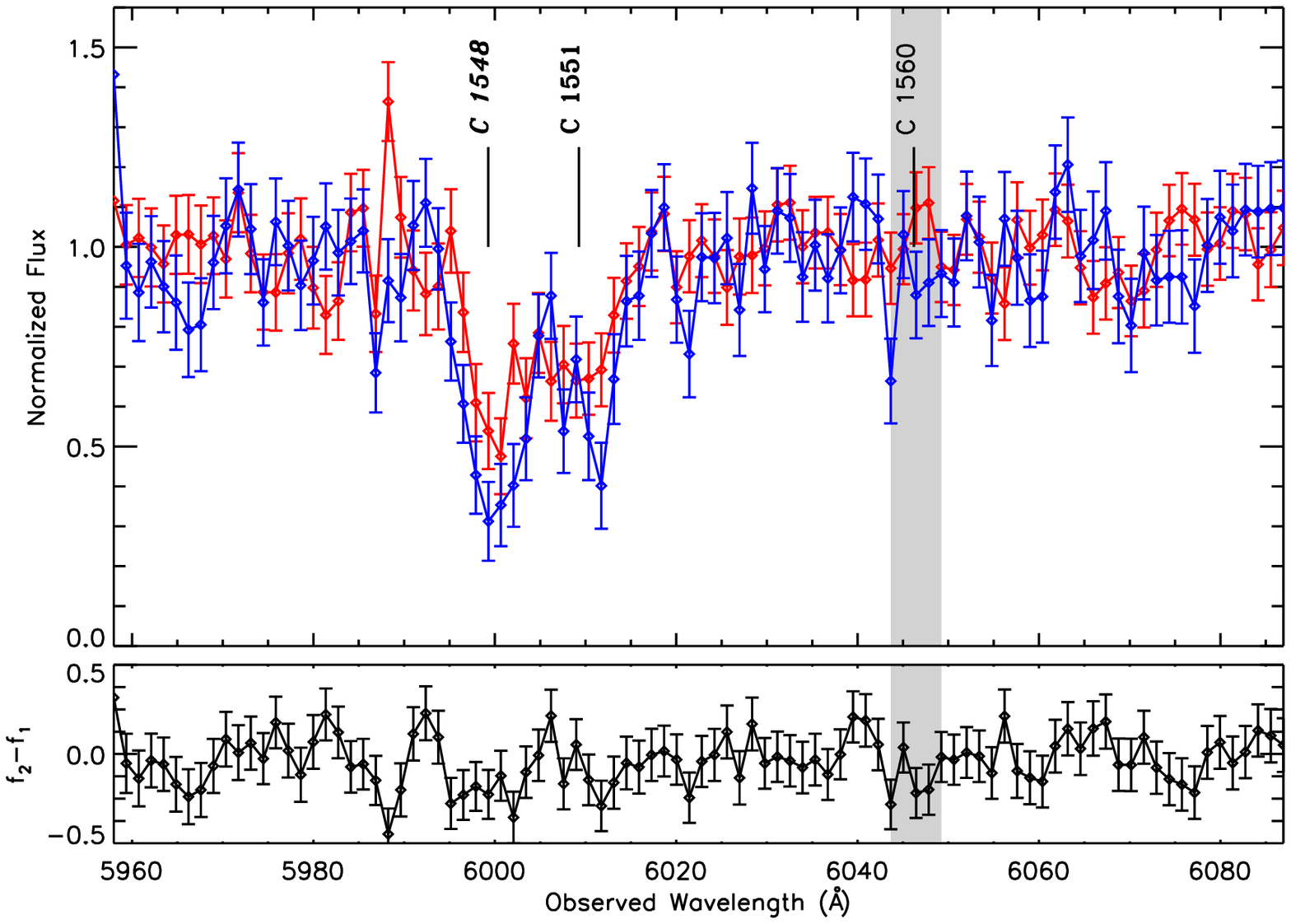}
\caption[Two-epoch normalized spectra of SDSS J102046.62+282707.1]{Two-epoch normalized spectra of the variable NAL system at $\beta$ = 0.0624 in SDSS J102046.62+282707.1.  The top panel shows the normalized pixel flux values with 1$\sigma$ error bars (first observations are red and second are blue), the bottom panel plots the difference spectrum of the two observation epochs, and shaded backgrounds identify masked pixels not included in our search for absorption line variability.  Line identifications for significantly variable absorption lines are italicised, lines detected in both observation epochs are in bold font, and undetected lines are in regular font (see Table A.1 for ion labels).  \label{figvs24}}
\end{center}
\end{figure*}

\begin{figure*}
\begin{center}
\includegraphics[width=84mm]{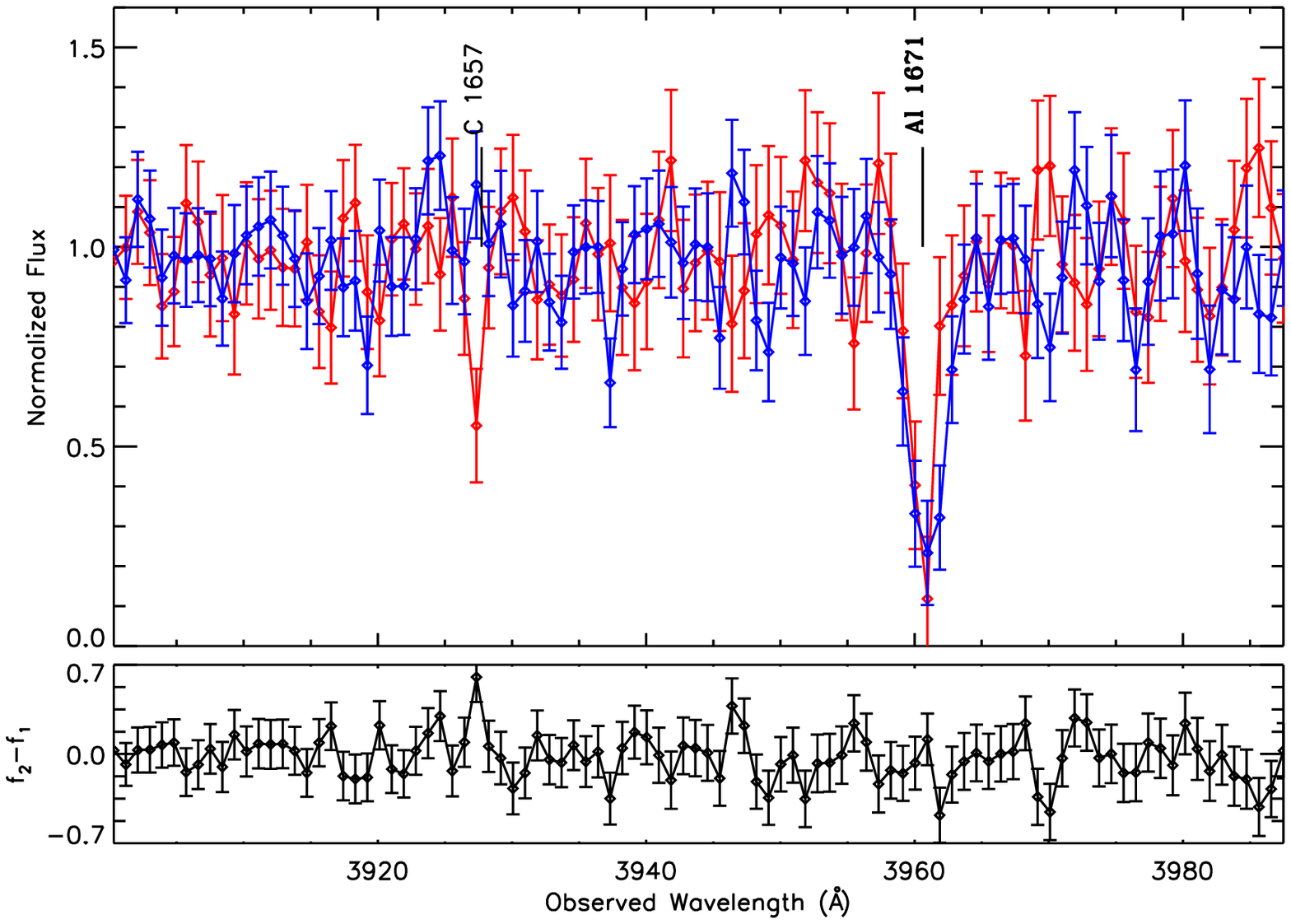}
\includegraphics[width=84mm]{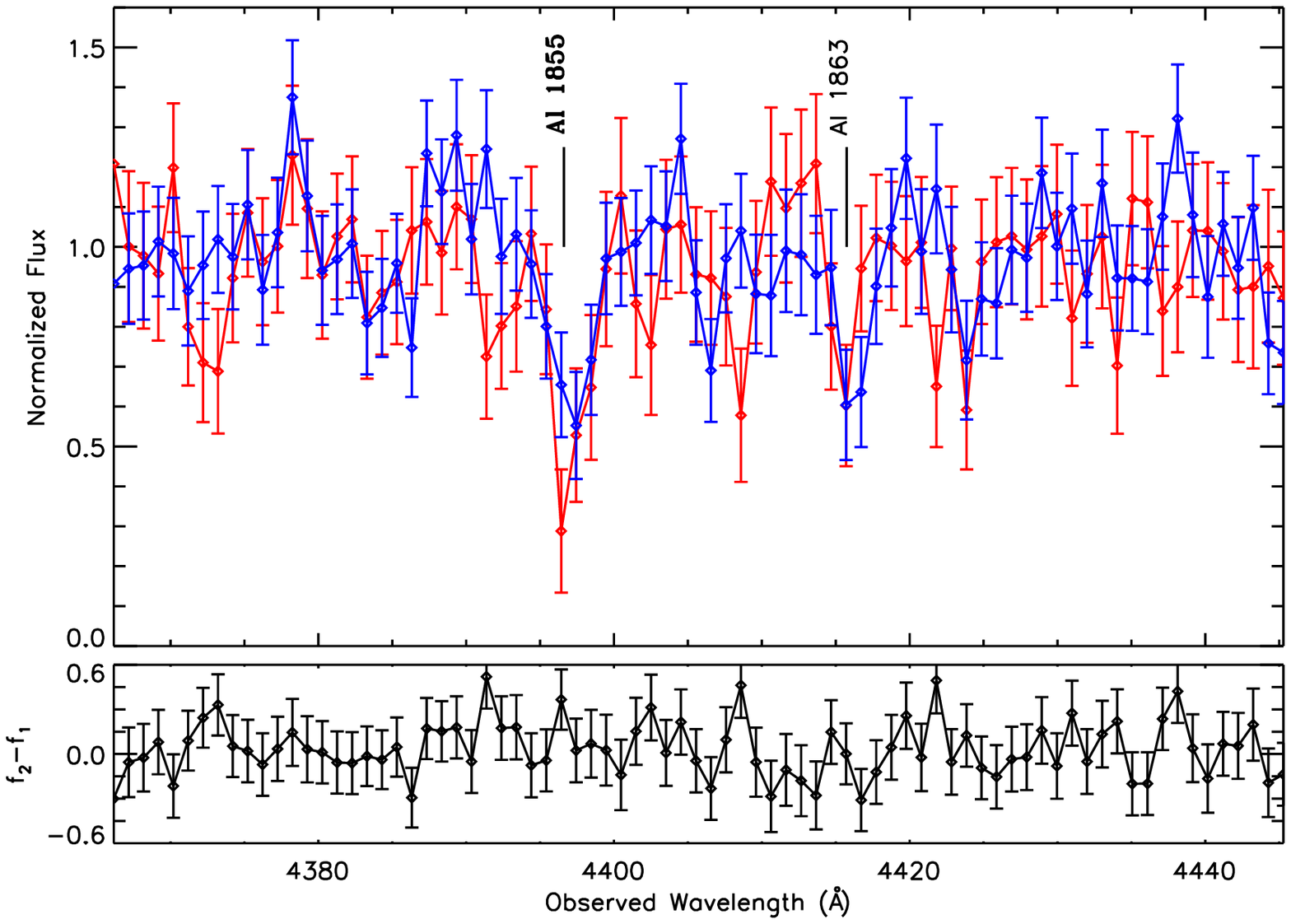}
\includegraphics[width=84mm]{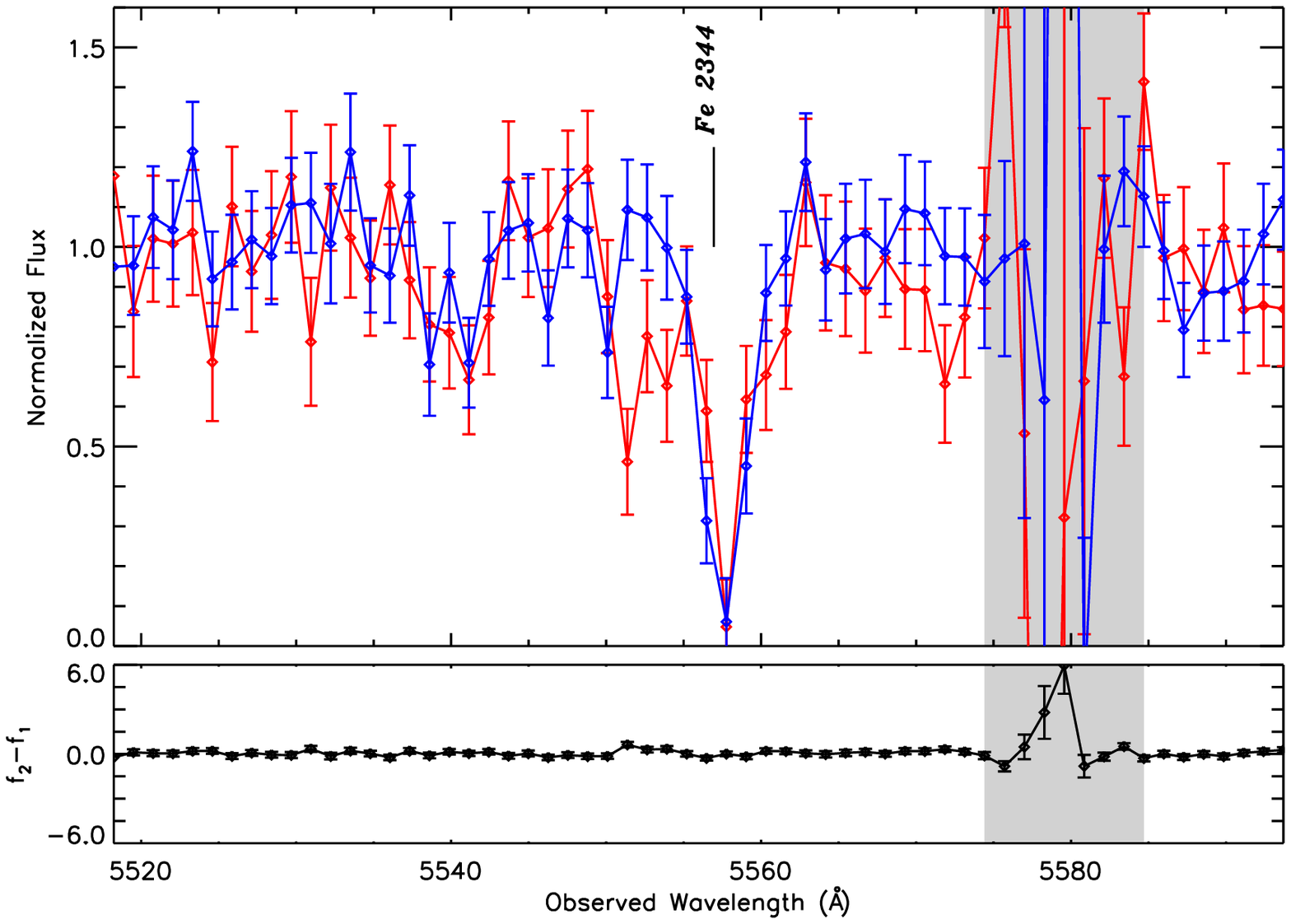}
\includegraphics[width=84mm]{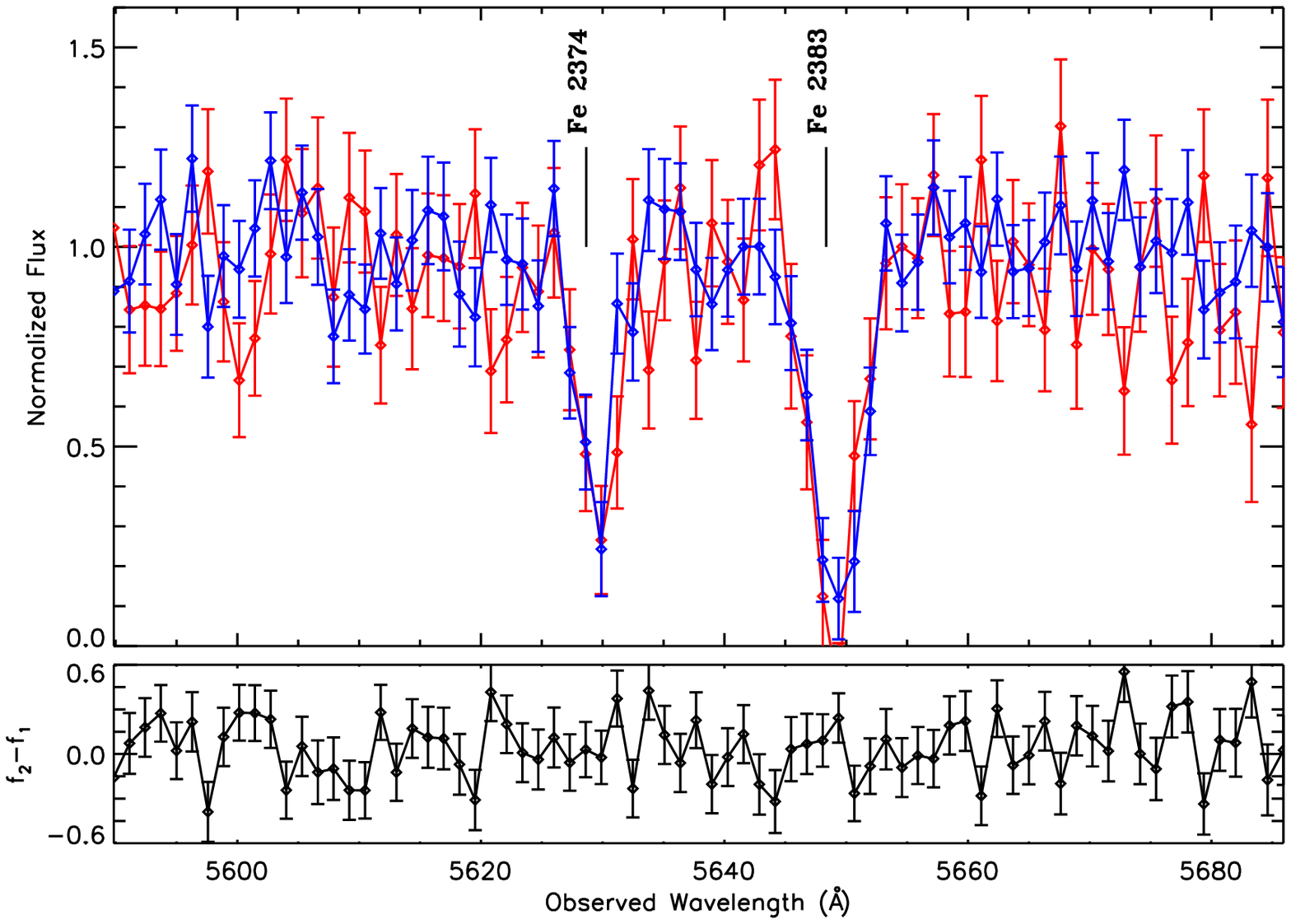}
\includegraphics[width=84mm]{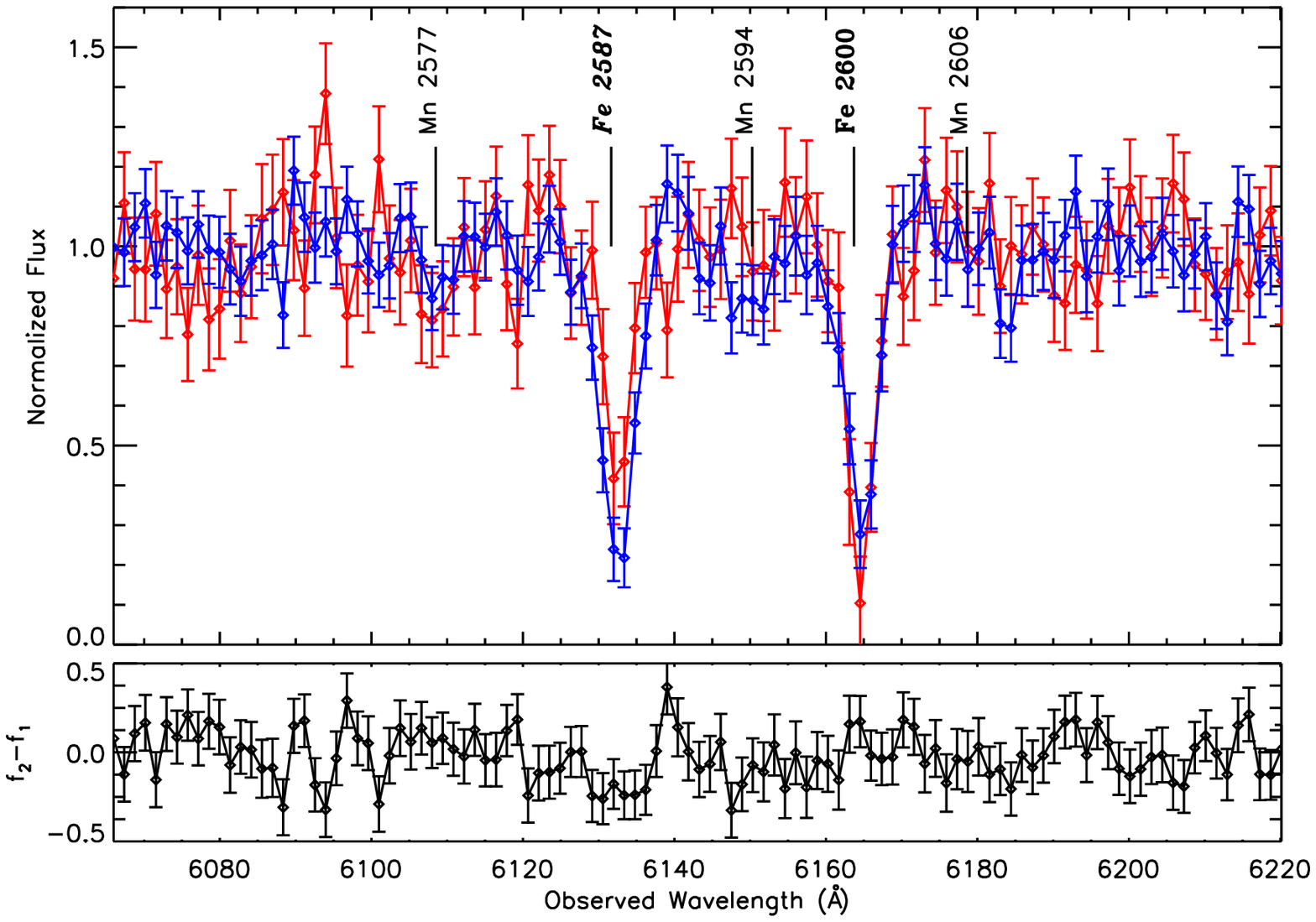}
\includegraphics[width=84mm]{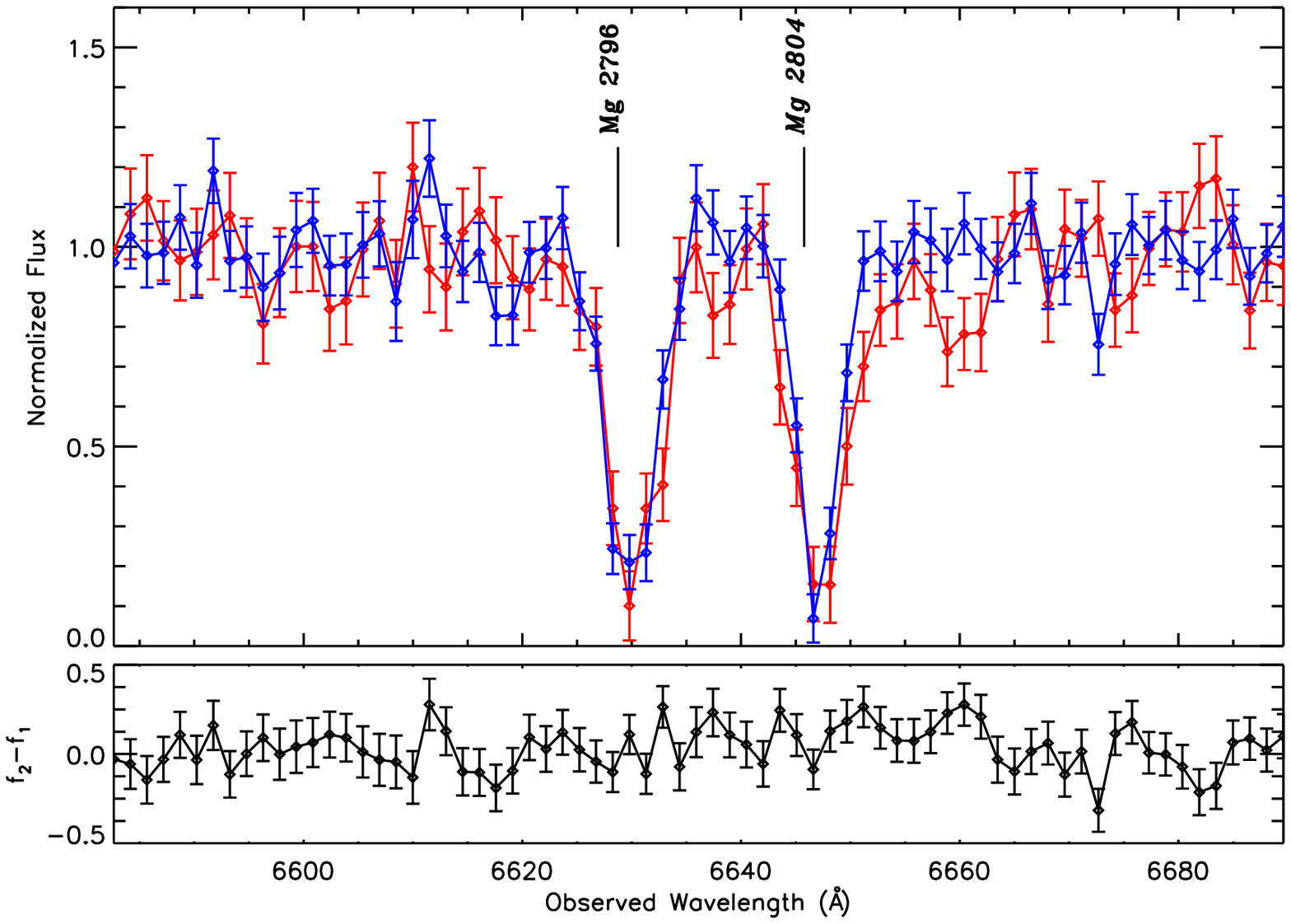}
\caption[Two-epoch normalized spectra of SDSS J161540.76+460451.0]{Two-epoch normalized spectra of the variable NAL system at $\beta$ = 0.0585 in SDSS J161540.76+460451.0.  The top panel shows the normalized pixel flux values with 1$\sigma$ error bars (first observations are red and second are blue), the bottom panel plots the difference spectrum of the two observation epochs, and shaded backgrounds identify masked pixels not included in our search for absorption line variability.  Line identifications for significantly variable absorption lines are italicised, lines detected in both observation epochs are in bold font, and undetected lines are in regular font (see Table A.1 for ion labels).  Continued in next figure.  \label{figvs25}}
\end{center}
\end{figure*}

\begin{figure*}
\ContinuedFloat
\begin{center}
\includegraphics[width=84mm]{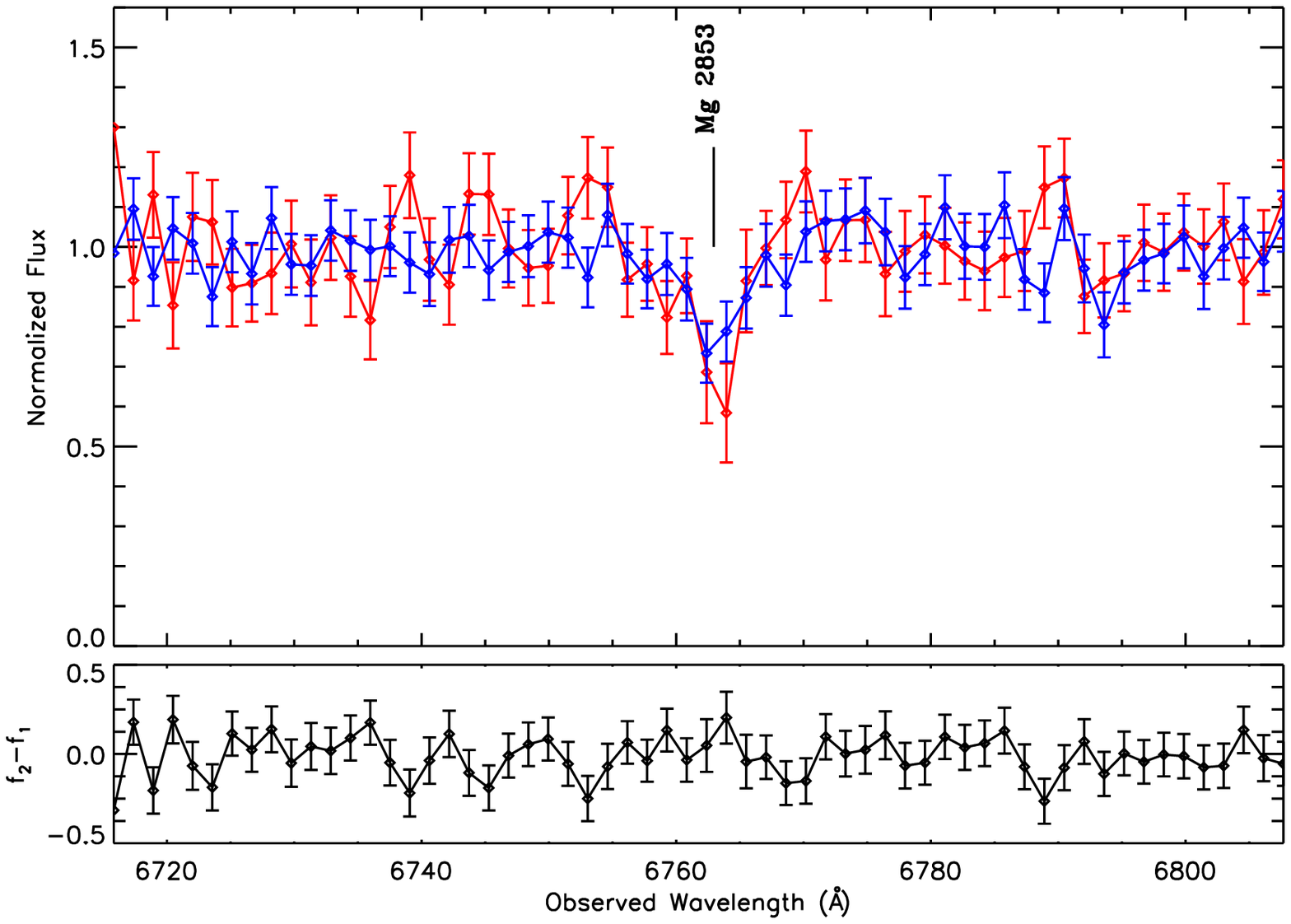}
\caption[]{Two-epoch normalized spectra of the variable NAL system at $\beta$ = 0.0585 in SDSS J161540.76+460451.0.  The top panel shows the normalized pixel flux values with 1$\sigma$ error bars (first observations are red and second are blue), the bottom panel plots the difference spectrum of the two observation epochs, and shaded backgrounds identify masked pixels not included in our search for absorption line variability.  Line identifications for significantly variable absorption lines are italicised, lines detected in both observation epochs are in bold font, and undetected lines are in regular font (see Table A.1 for ion labels).  Continued from previous figure.}
\vspace{3.5cm}
\end{center}
\end{figure*}

\begin{figure*}
\begin{center}
\includegraphics[width=84mm]{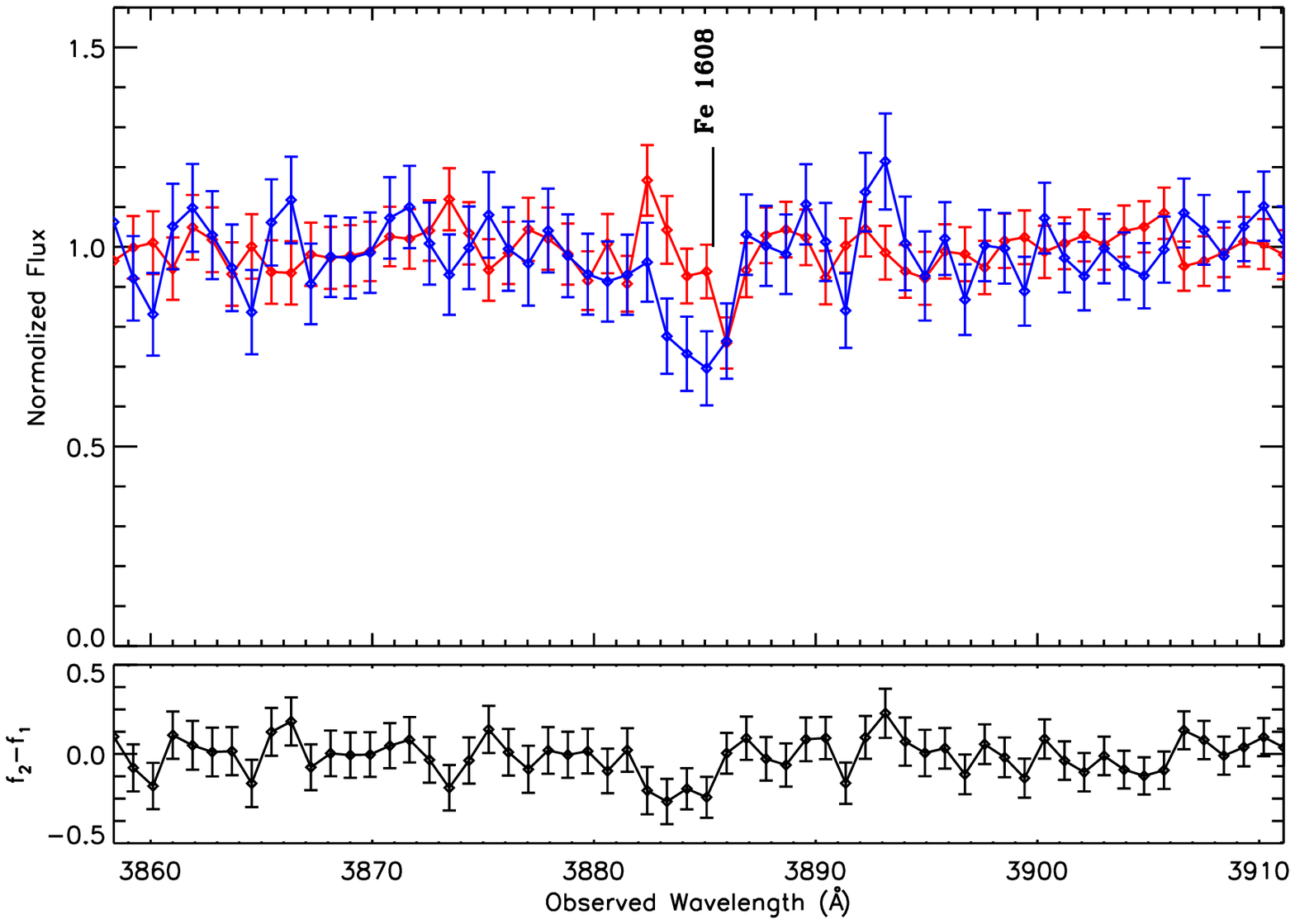}
\includegraphics[width=84mm]{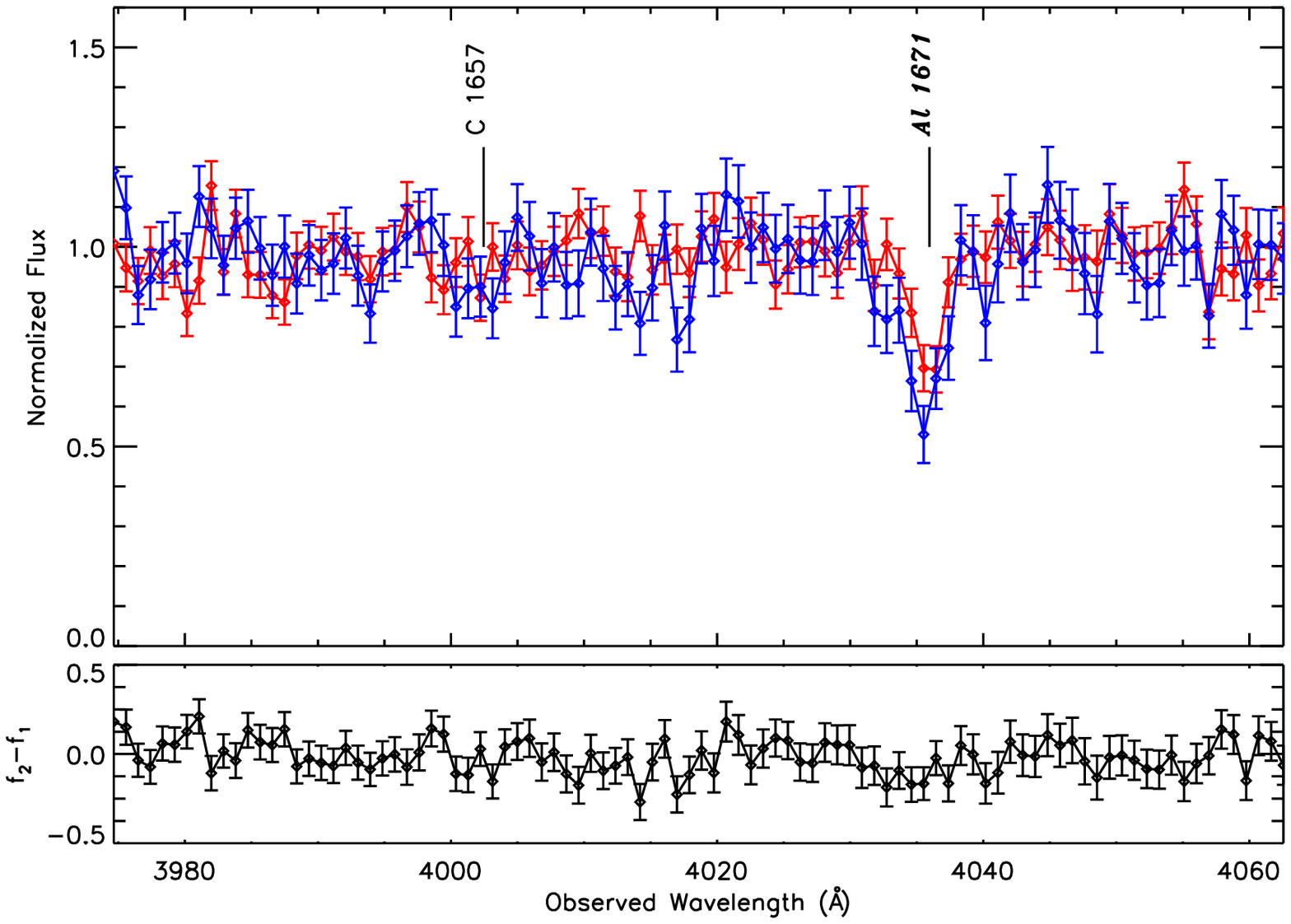}
\includegraphics[width=84mm]{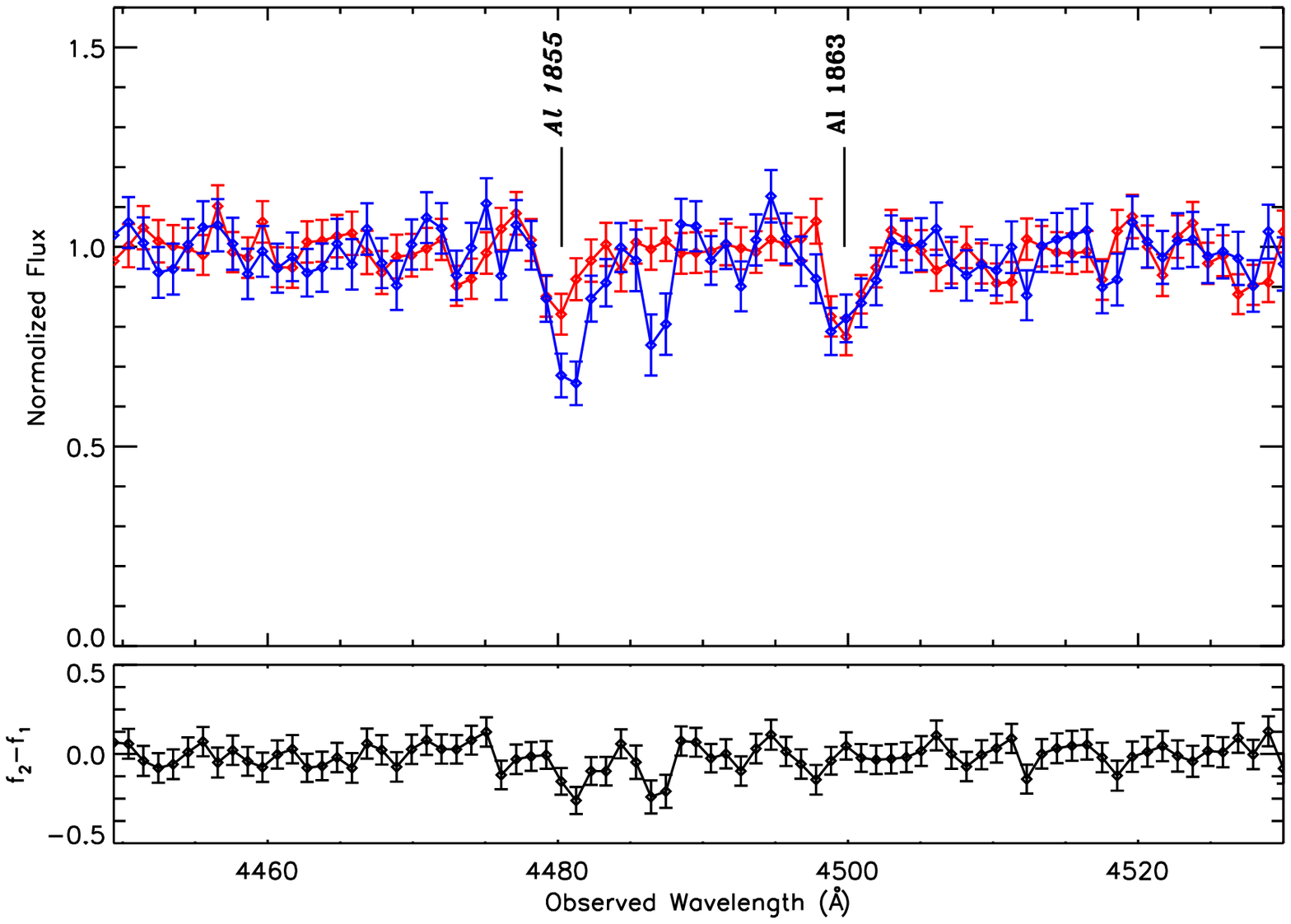}
\includegraphics[width=84mm]{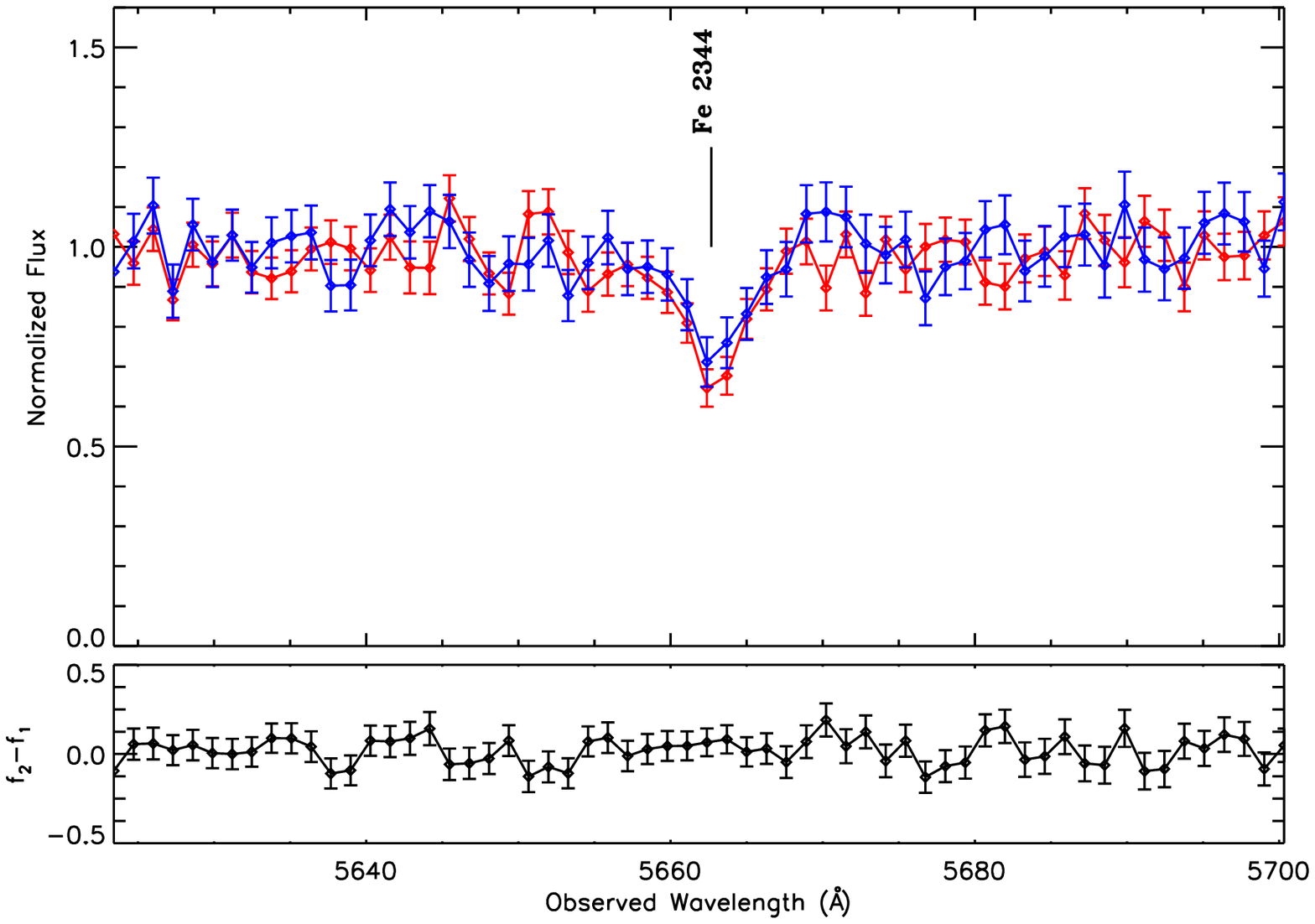}
\includegraphics[width=84mm]{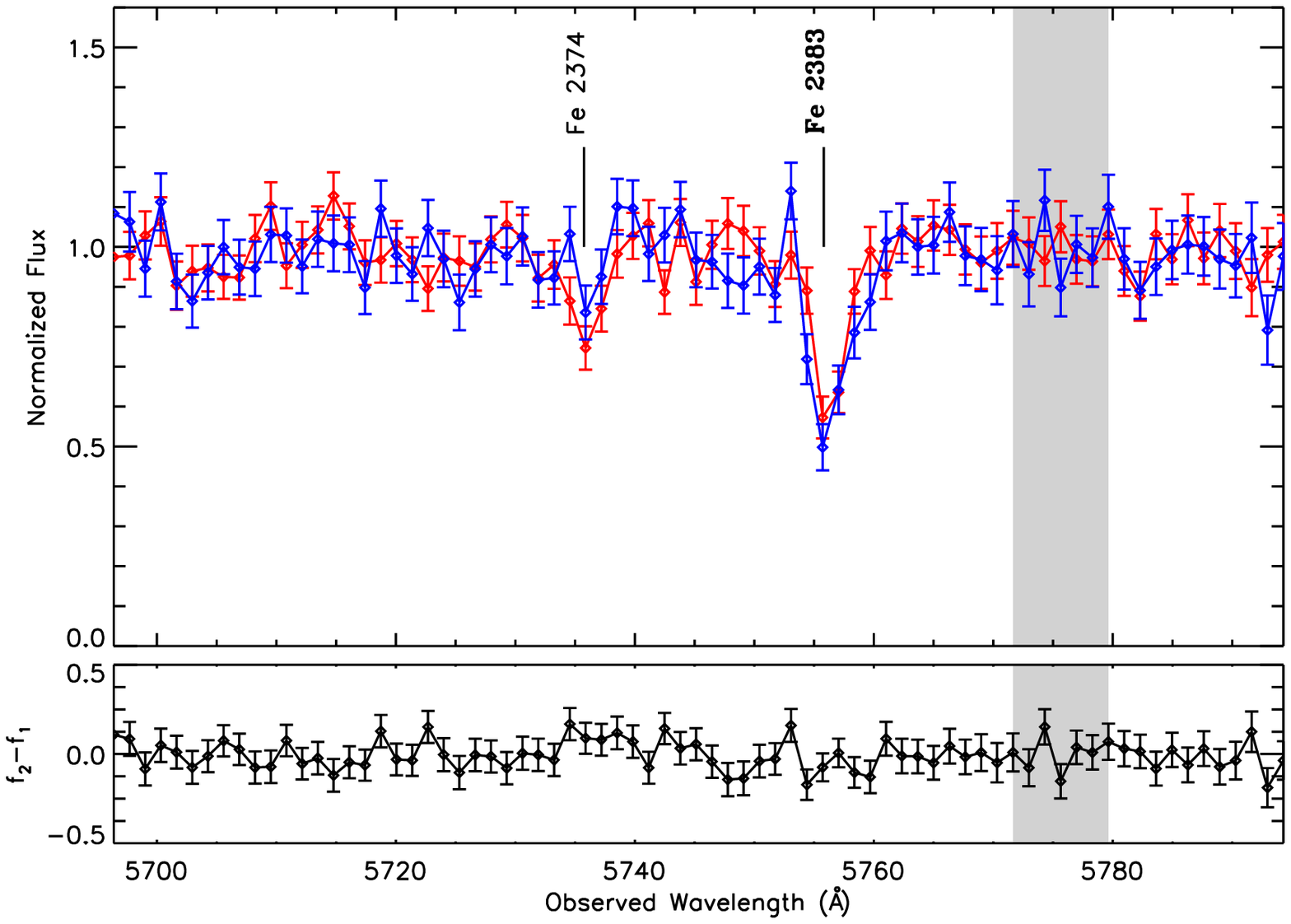}
\includegraphics[width=84mm]{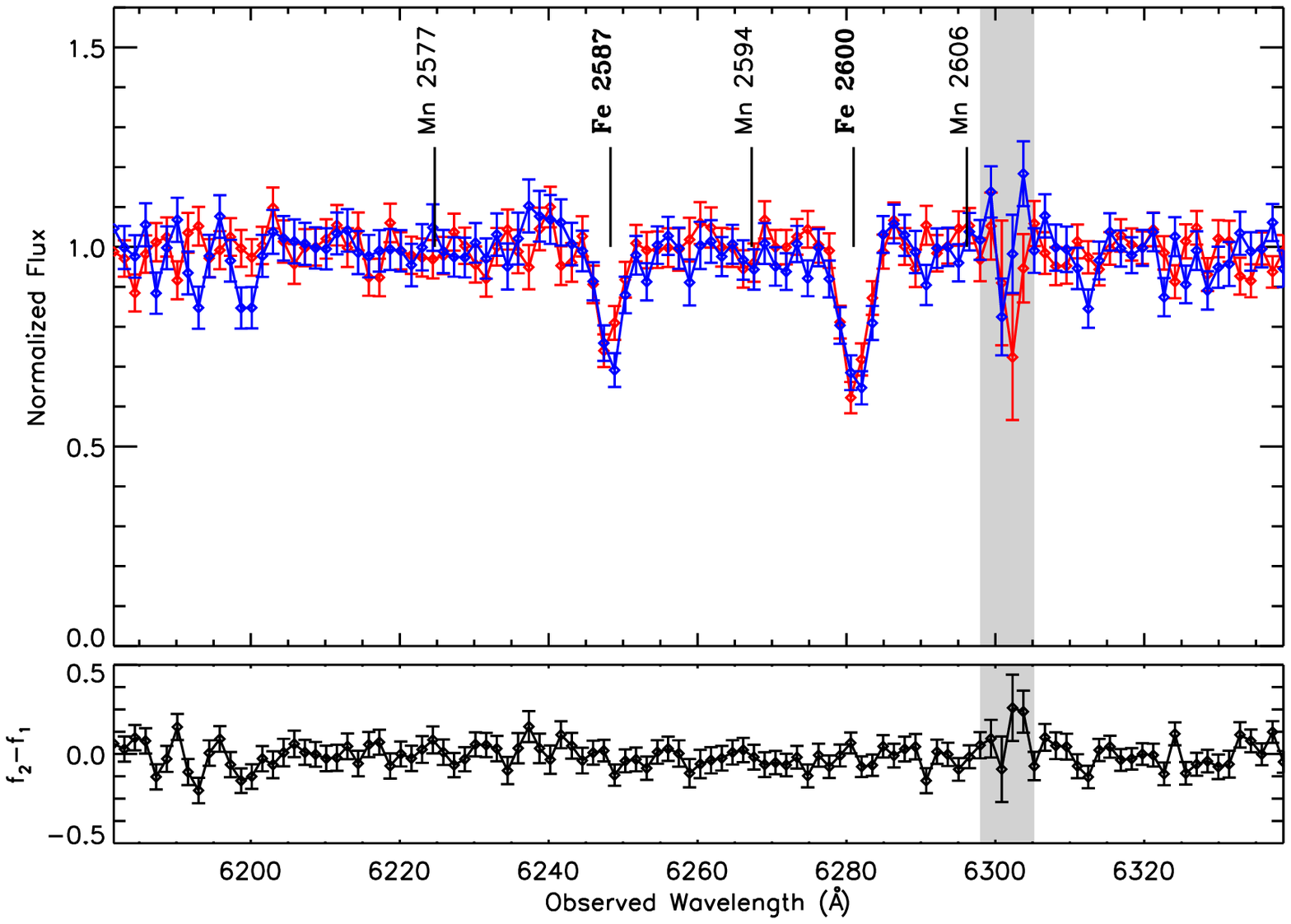}
\caption[Two-epoch normalized spectra of SDSS J015733.87-004824.4]{Two-epoch normalized spectra of the variable NAL system at $\beta$ = 0.0546 in SDSS J015733.87-004824.4.  The top panel shows the normalized pixel flux values with 1$\sigma$ error bars (first observations are red and second are blue), the bottom panel plots the difference spectrum of the two observation epochs, and shaded backgrounds identify masked pixels not included in our search for absorption line variability.  Line identifications for significantly variable absorption lines are italicised, lines detected in both observation epochs are in bold font, and undetected lines are in regular font (see Table A.1 for ion labels).  Continued in next figure.  \label{figvs26}}
\end{center}
\end{figure*}

\begin{figure*}
\ContinuedFloat
\begin{center}
\includegraphics[width=84mm]{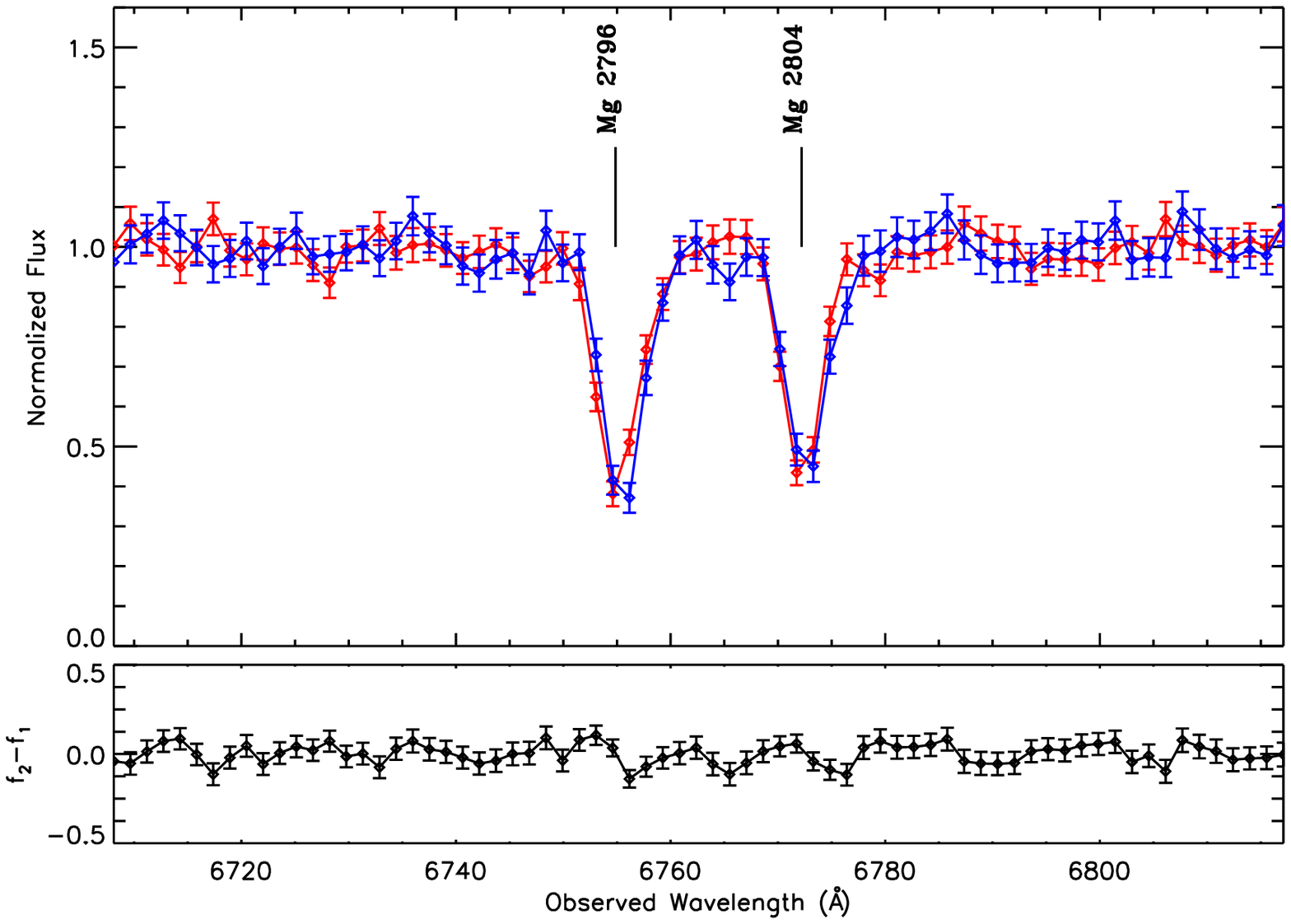}
\caption[]{Two-epoch normalized spectra of the variable NAL system at $\beta$ = 0.0546 in SDSS J015733.87-004824.4.  The top panel shows the normalized pixel flux values with 1$\sigma$ error bars (first observations are red and second are blue), the bottom panel plots the difference spectrum of the two observation epochs, and shaded backgrounds identify masked pixels not included in our search for absorption line variability.  Line identifications for significantly variable absorption lines are italicised, lines detected in both observation epochs are in bold font, and undetected lines are in regular font (see Table A.1 for ion labels).  Continued from previous figure.}
\vspace{3.5cm}
\end{center}
\end{figure*}

\begin{figure*}
\begin{center}
\includegraphics[width=84mm]{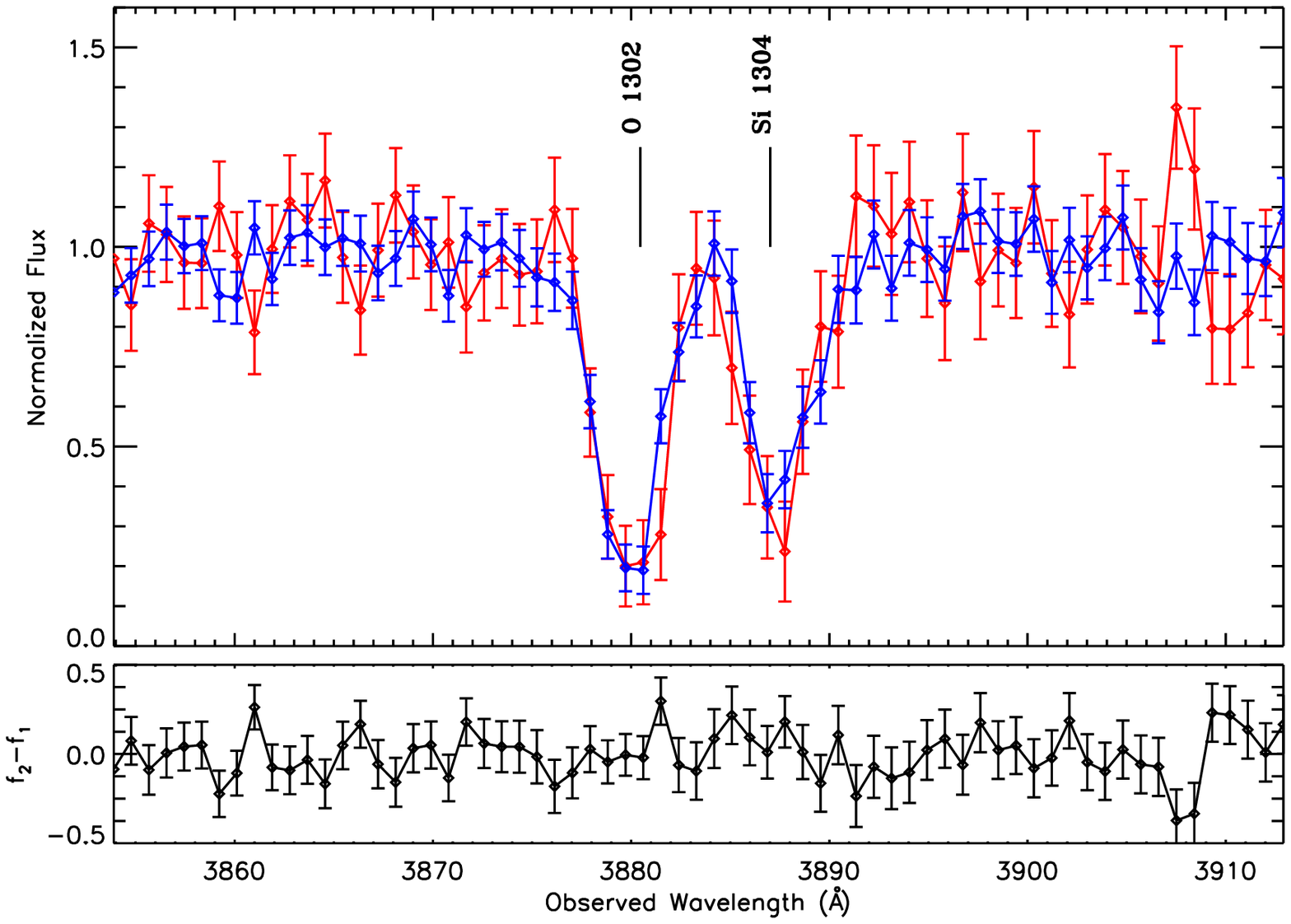}
\includegraphics[width=84mm]{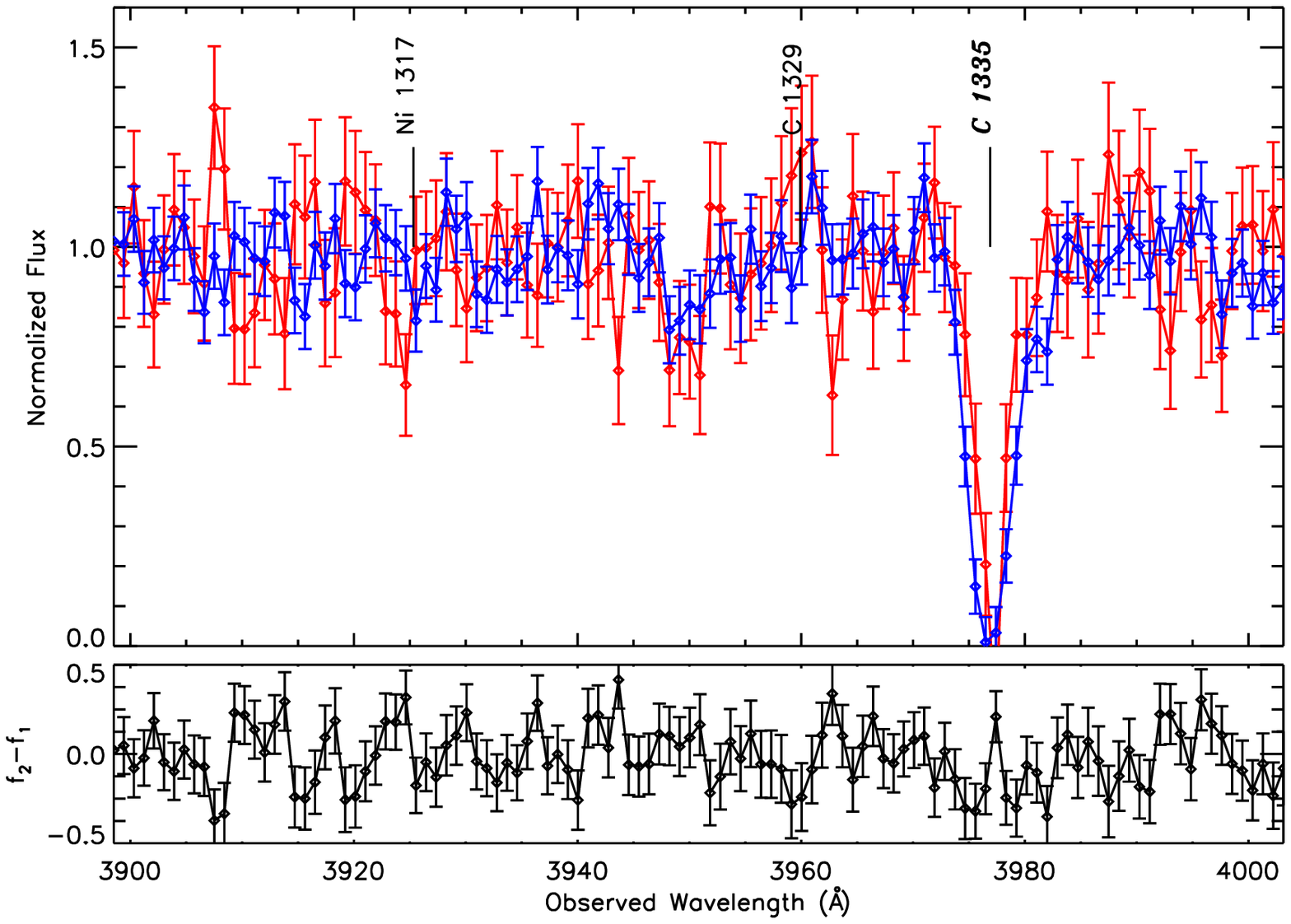}
\includegraphics[width=84mm]{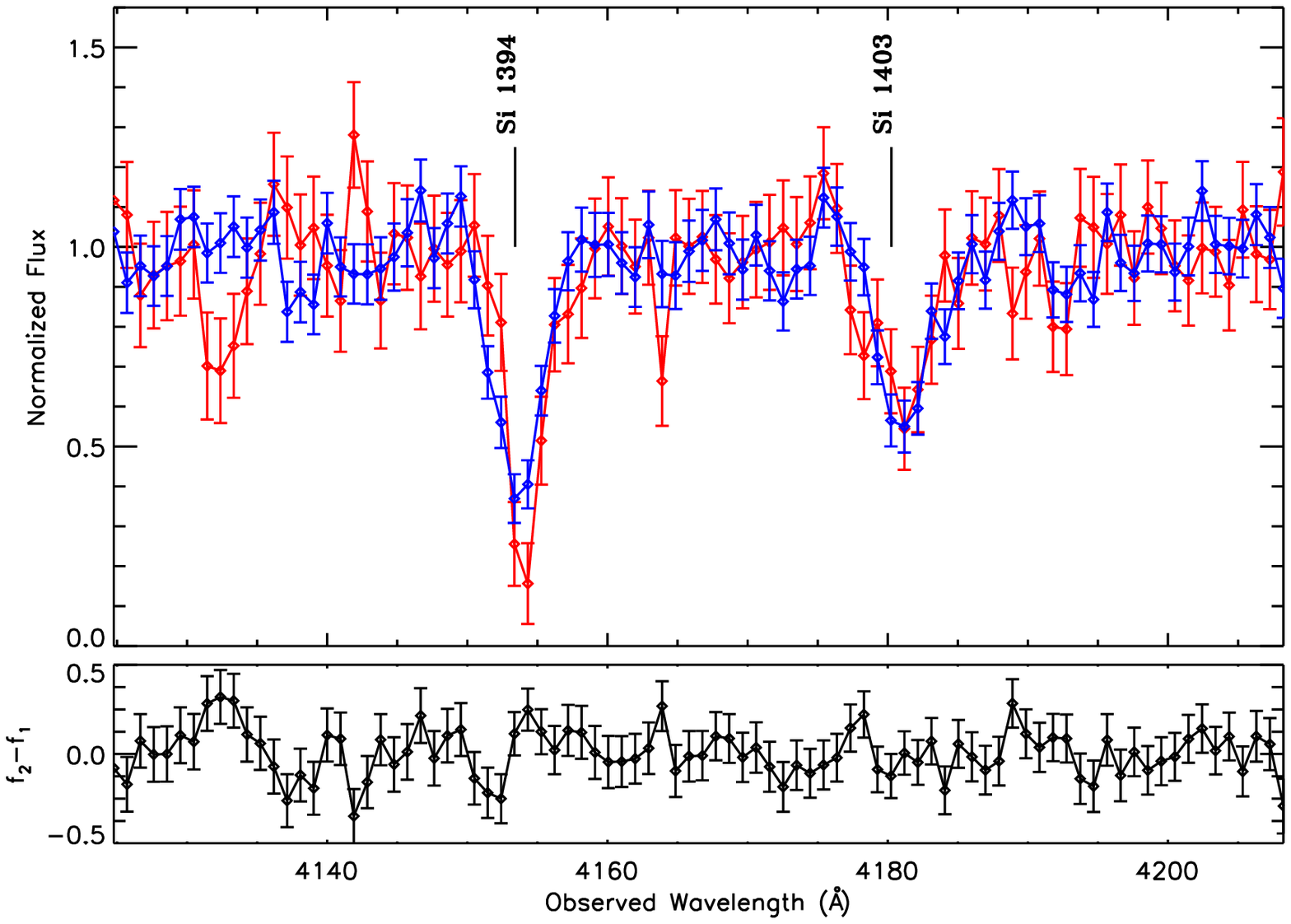}
\includegraphics[width=84mm]{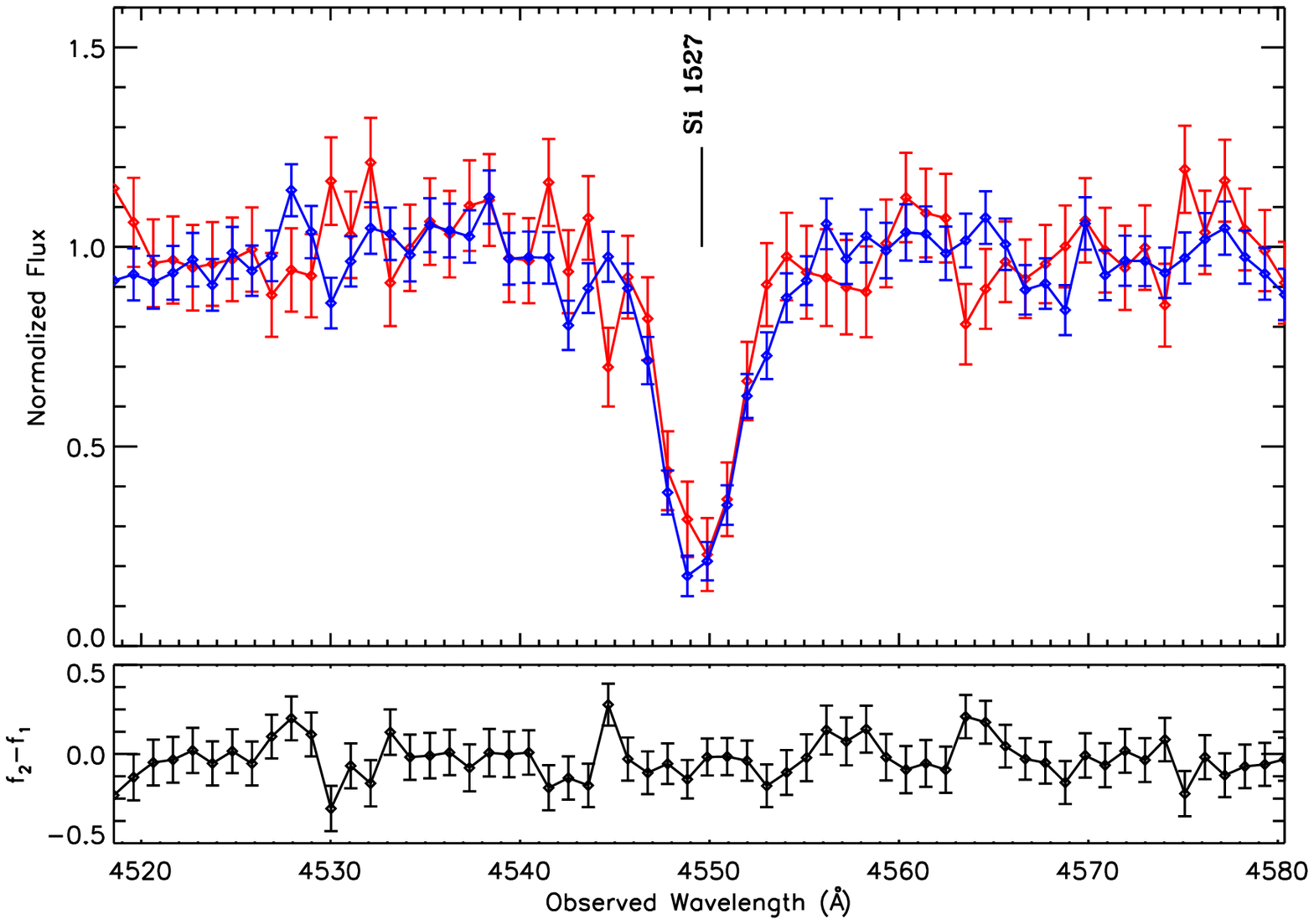}
\includegraphics[width=84mm]{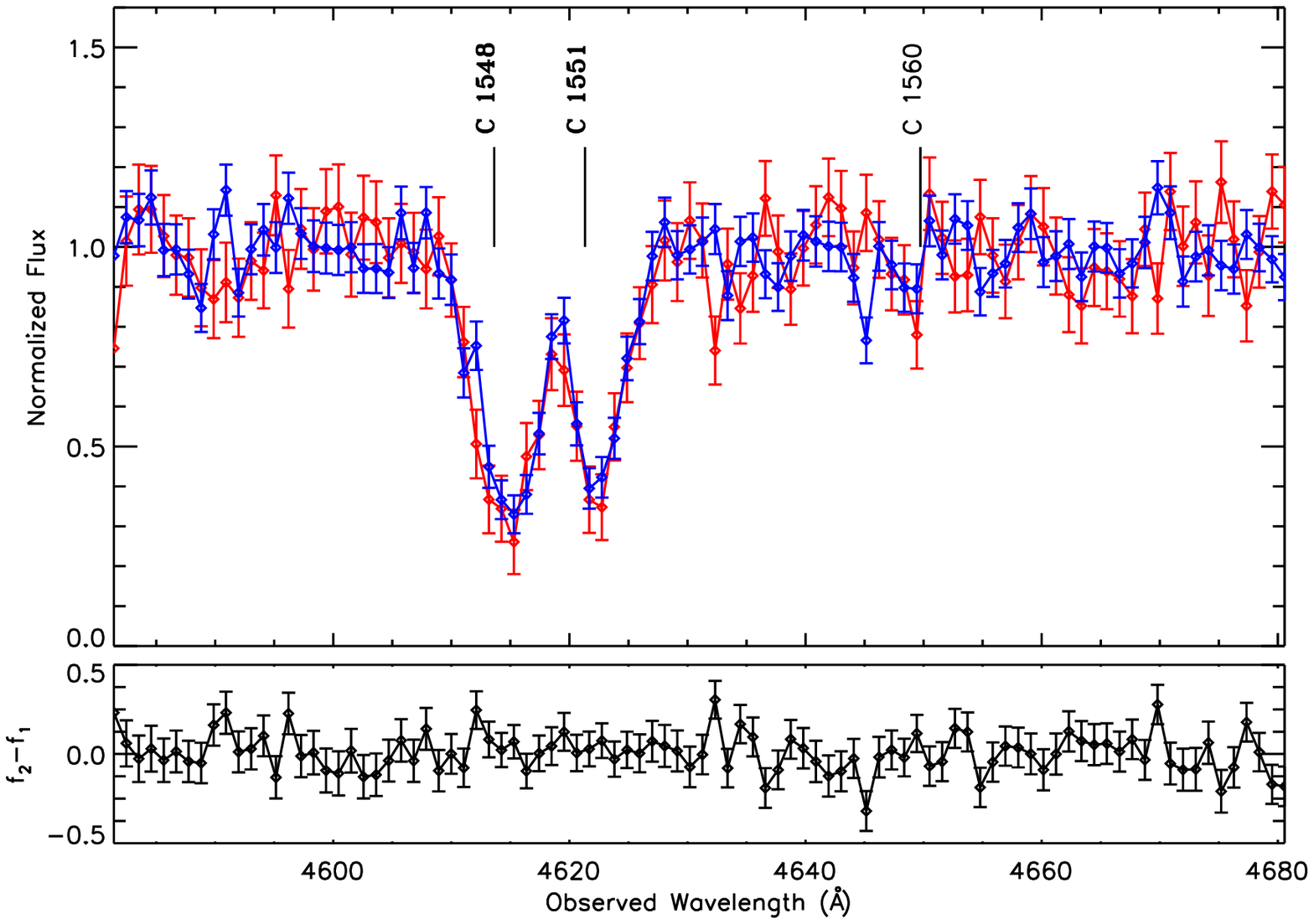}
\includegraphics[width=84mm]{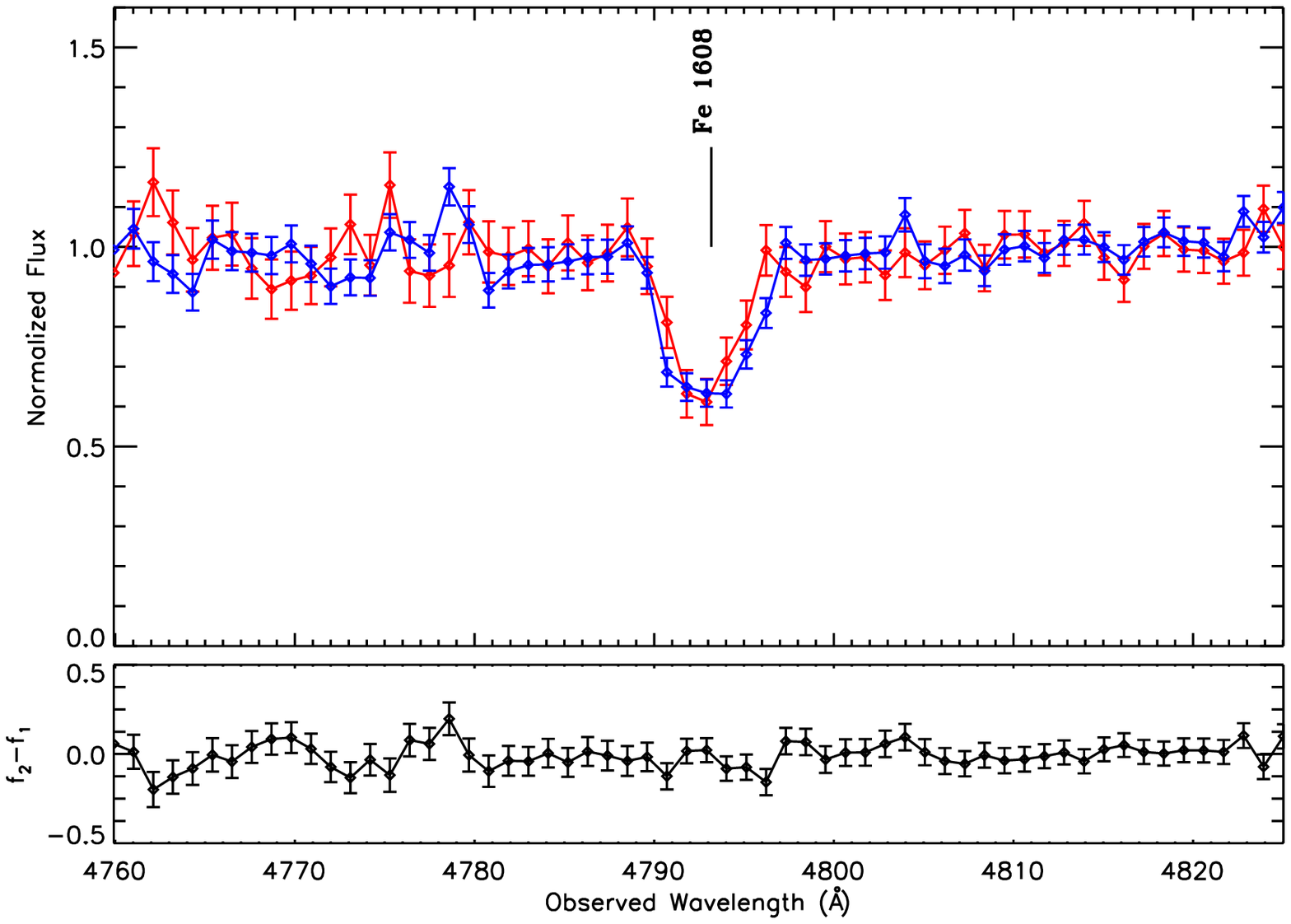}
\caption[Two-epoch normalized spectra of SDSS J114857.22+594153.9]{Two-epoch normalized spectra of the variable NAL system at $\beta$ = 0.0469 in SDSS J114857.22+594153.9.  The top panel shows the normalized pixel flux values with 1$\sigma$ error bars (first observations are red and second are blue), the bottom panel plots the difference spectrum of the two observation epochs, and shaded backgrounds identify masked pixels not included in our search for absorption line variability.  Line identifications for significantly variable absorption lines are italicised, lines detected in both observation epochs are in bold font, and undetected lines are in regular font (see Table A.1 for ion labels).  Continued in next figure.  \label{figvs27}}
\end{center}
\end{figure*}

\begin{figure*}
\ContinuedFloat
\begin{center}
\includegraphics[width=84mm]{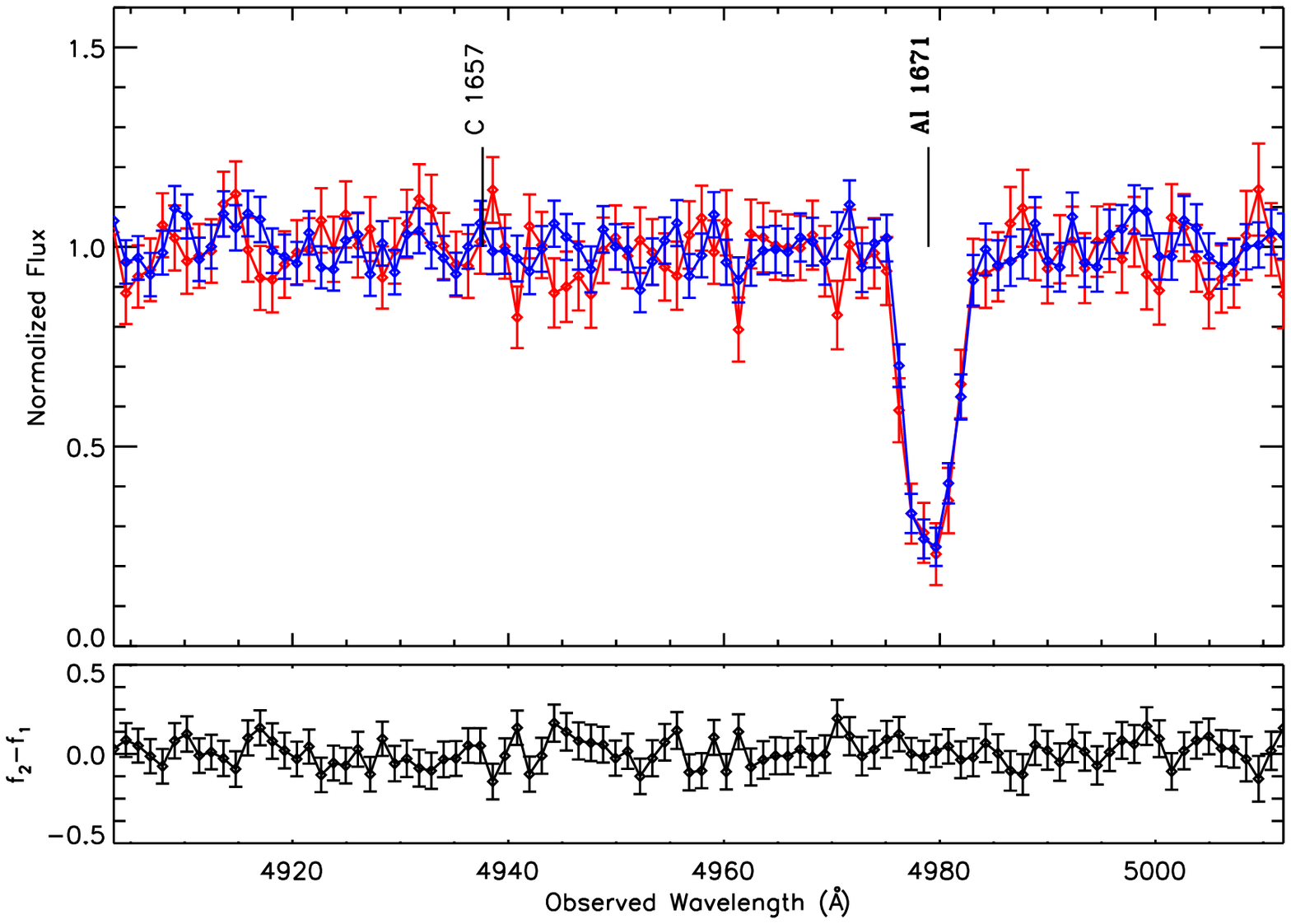}
\includegraphics[width=84mm]{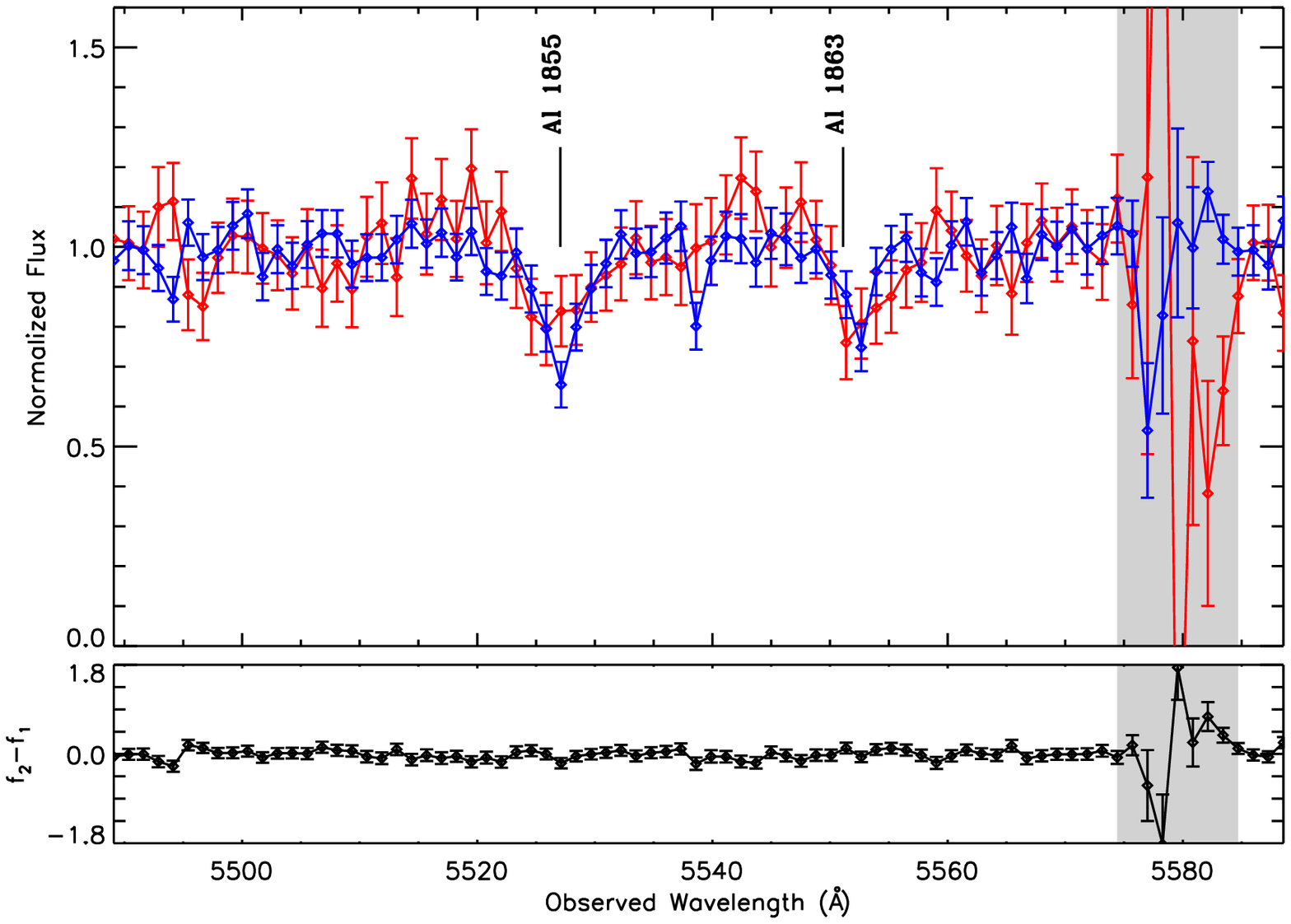}
\includegraphics[width=84mm]{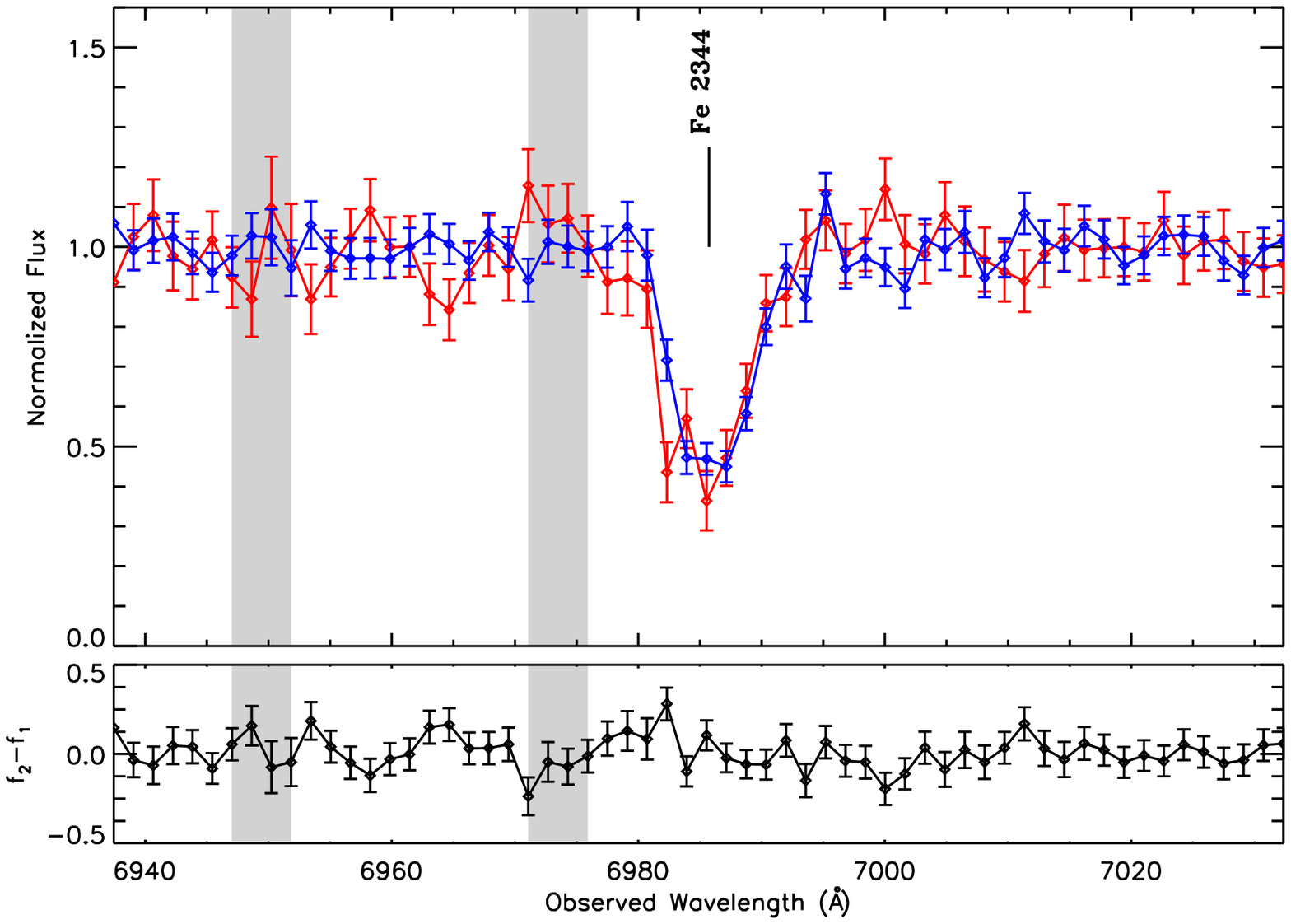}
\includegraphics[width=84mm]{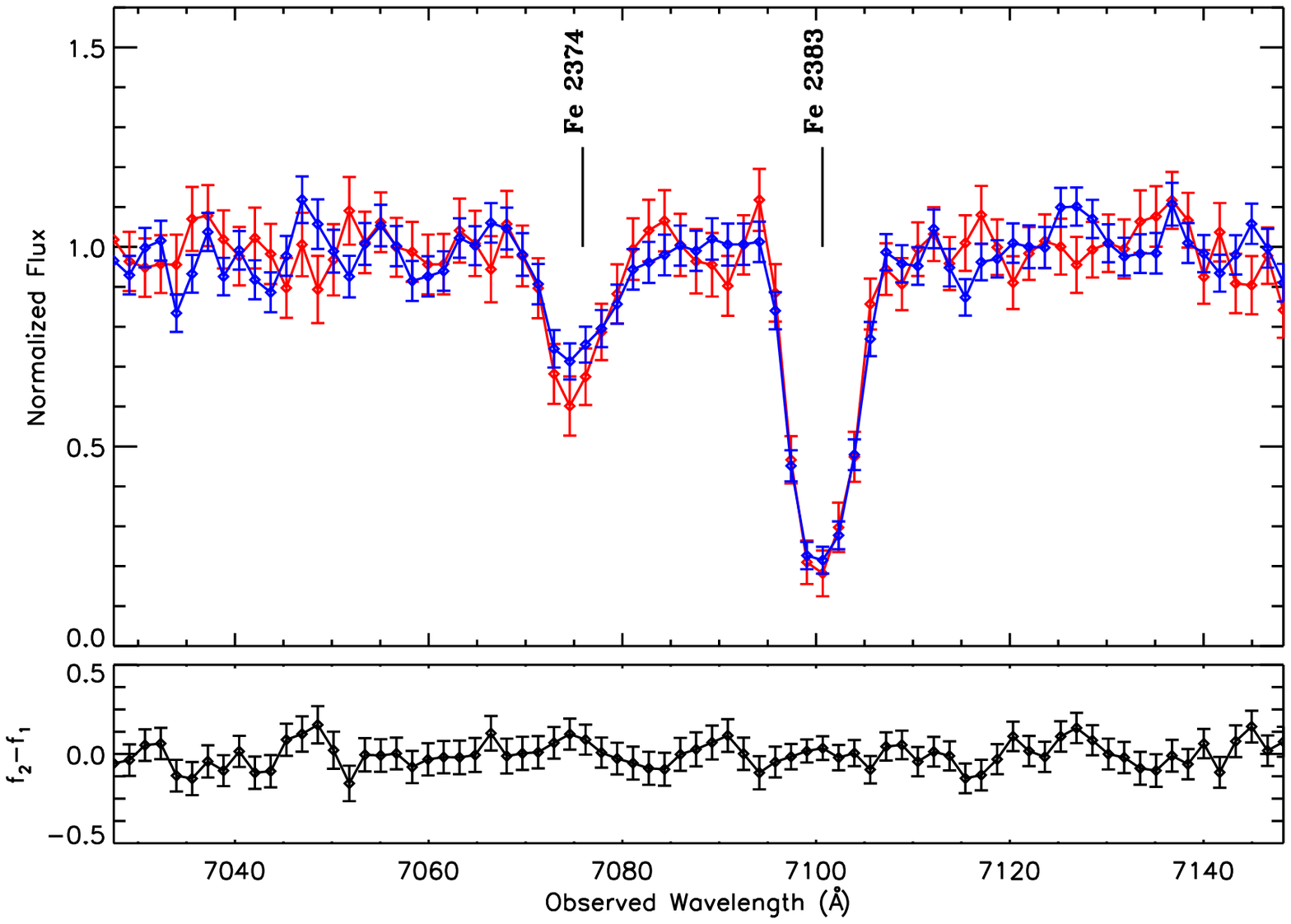}
\includegraphics[width=84mm]{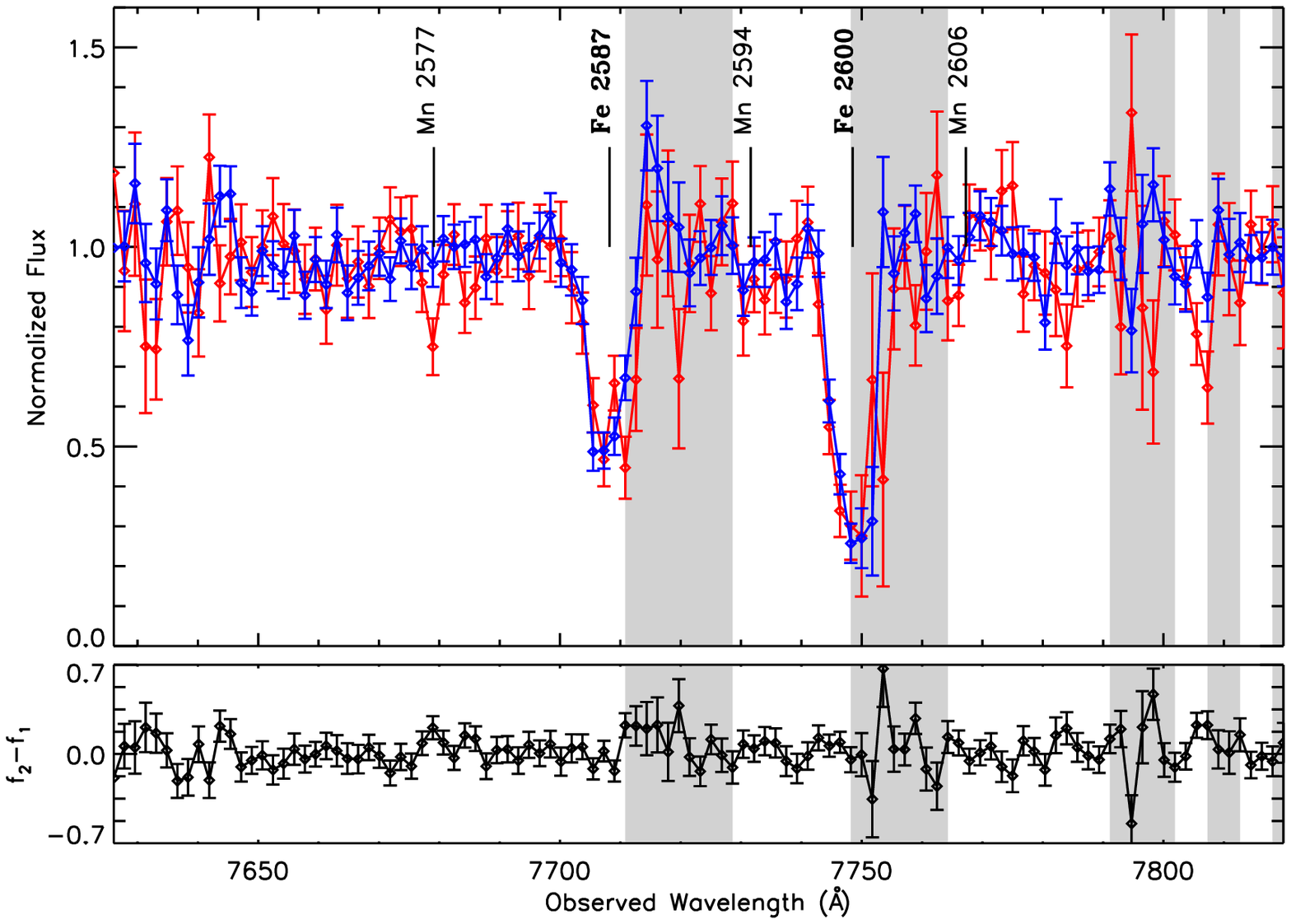}
\includegraphics[width=84mm]{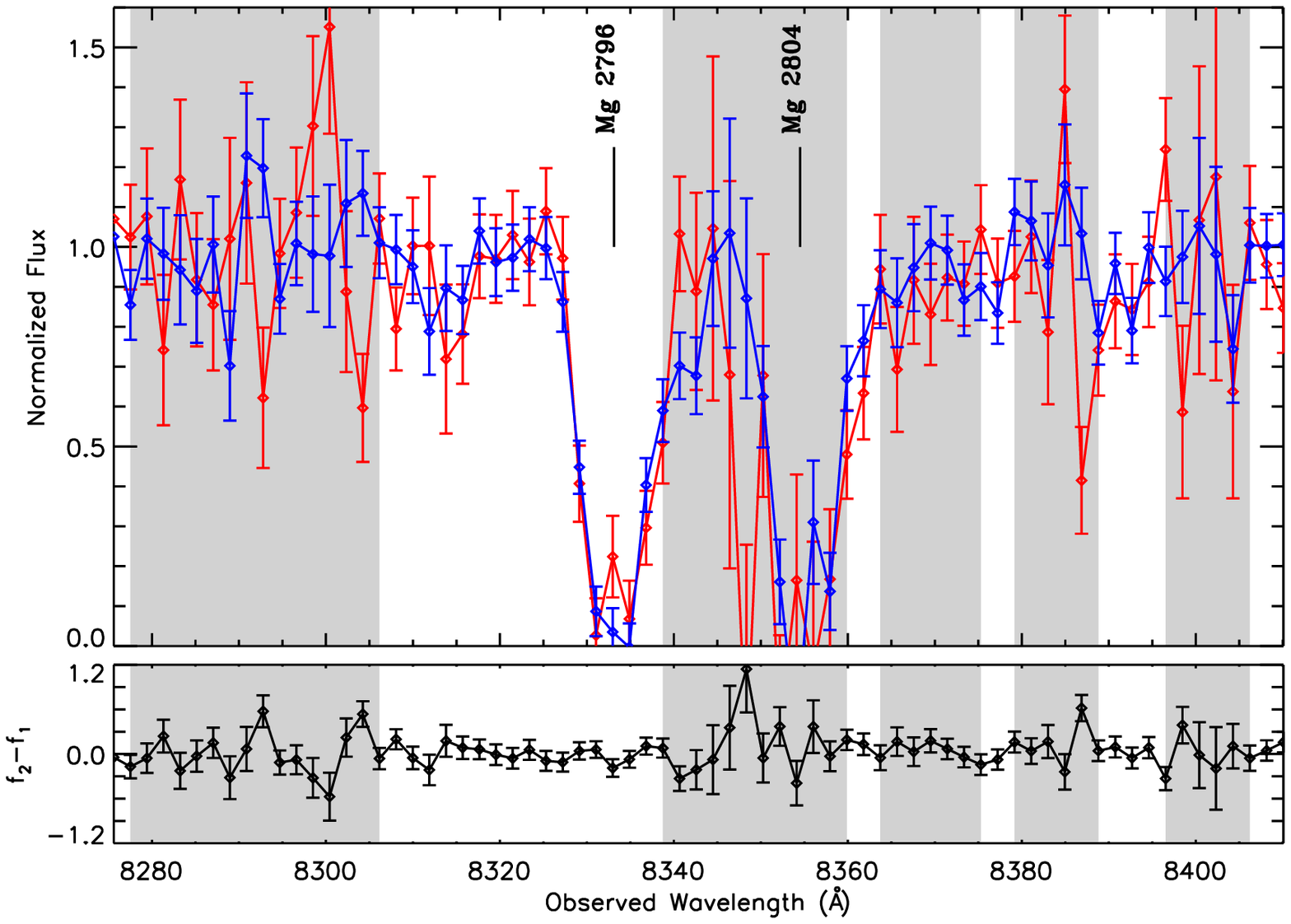}
\caption[]{Two-epoch normalized spectra of the variable NAL system at $\beta$ = 0.0469 in SDSS J114857.22+594153.9.  The top panel shows the normalized pixel flux values with 1$\sigma$ error bars (first observations are red and second are blue), the bottom panel plots the difference spectrum of the two observation epochs, and shaded backgrounds identify masked pixels not included in our search for absorption line variability.  Line identifications for significantly variable absorption lines are italicised, lines detected in both observation epochs are in bold font, and undetected lines are in regular font (see Table A.1 for ion labels).  Continued from previous figure.}
\end{center}
\end{figure*}

\begin{figure*}
\begin{center}
\includegraphics[width=84mm]{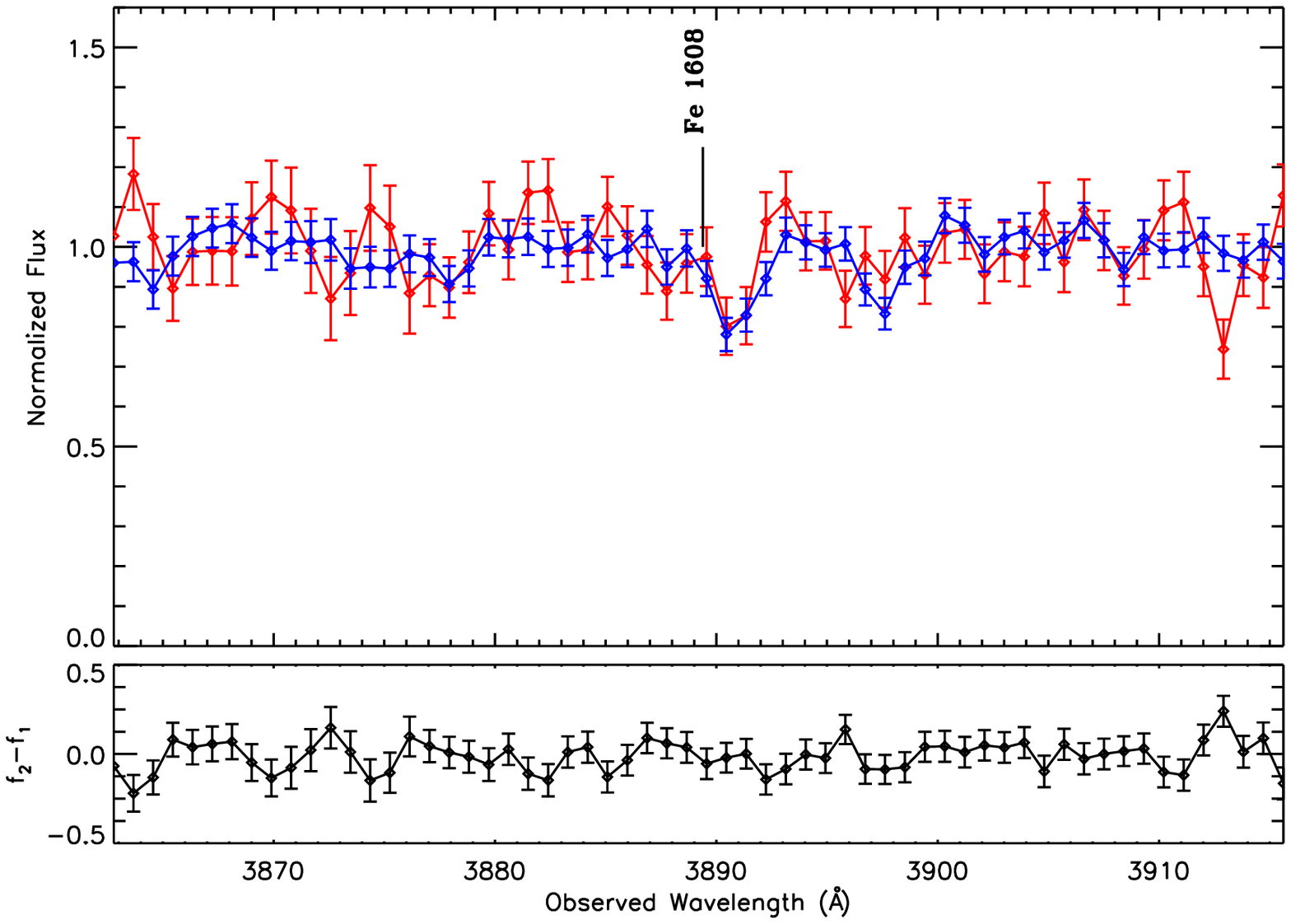}
\includegraphics[width=84mm]{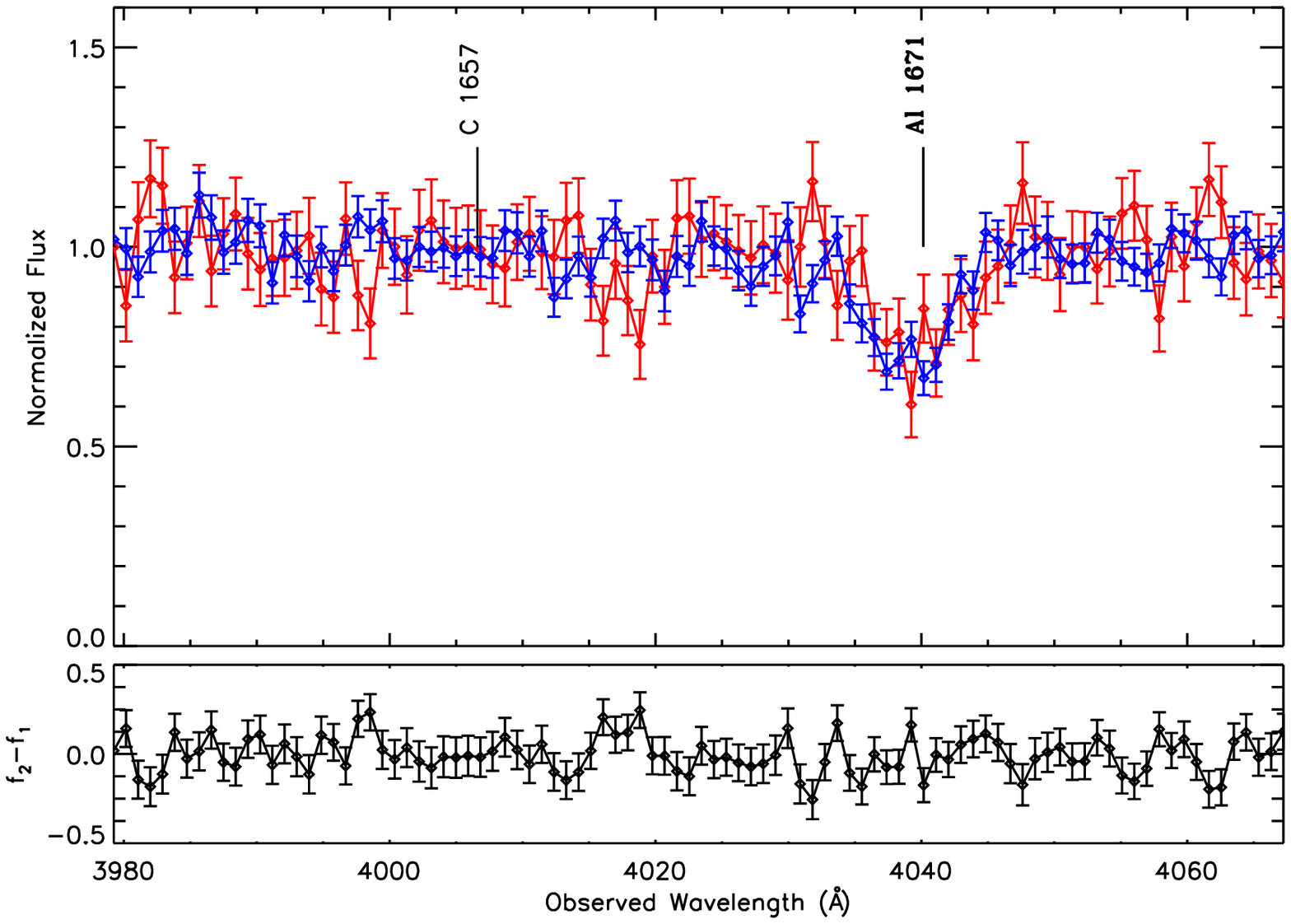}
\includegraphics[width=84mm]{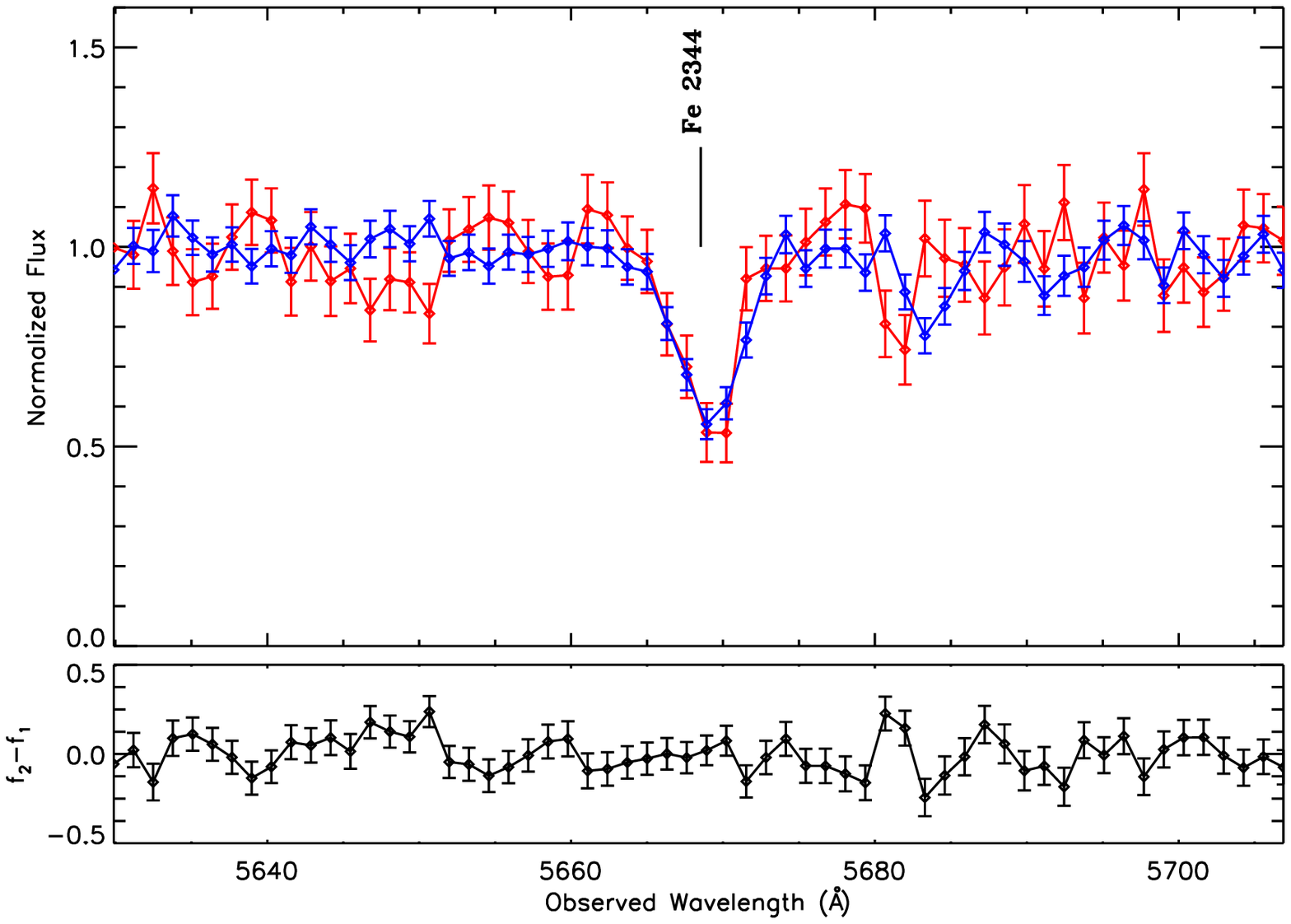}
\includegraphics[width=84mm]{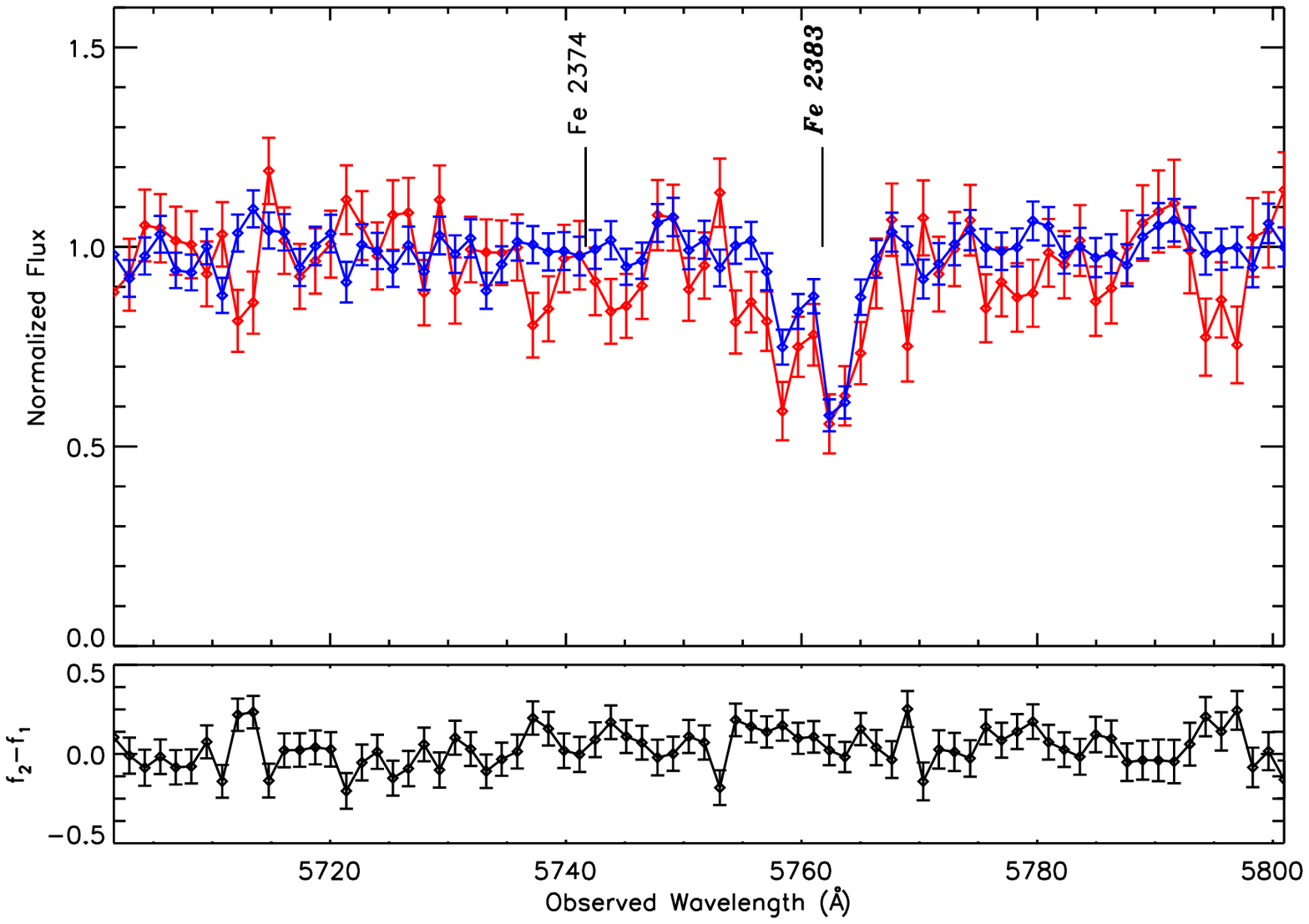}
\includegraphics[width=84mm]{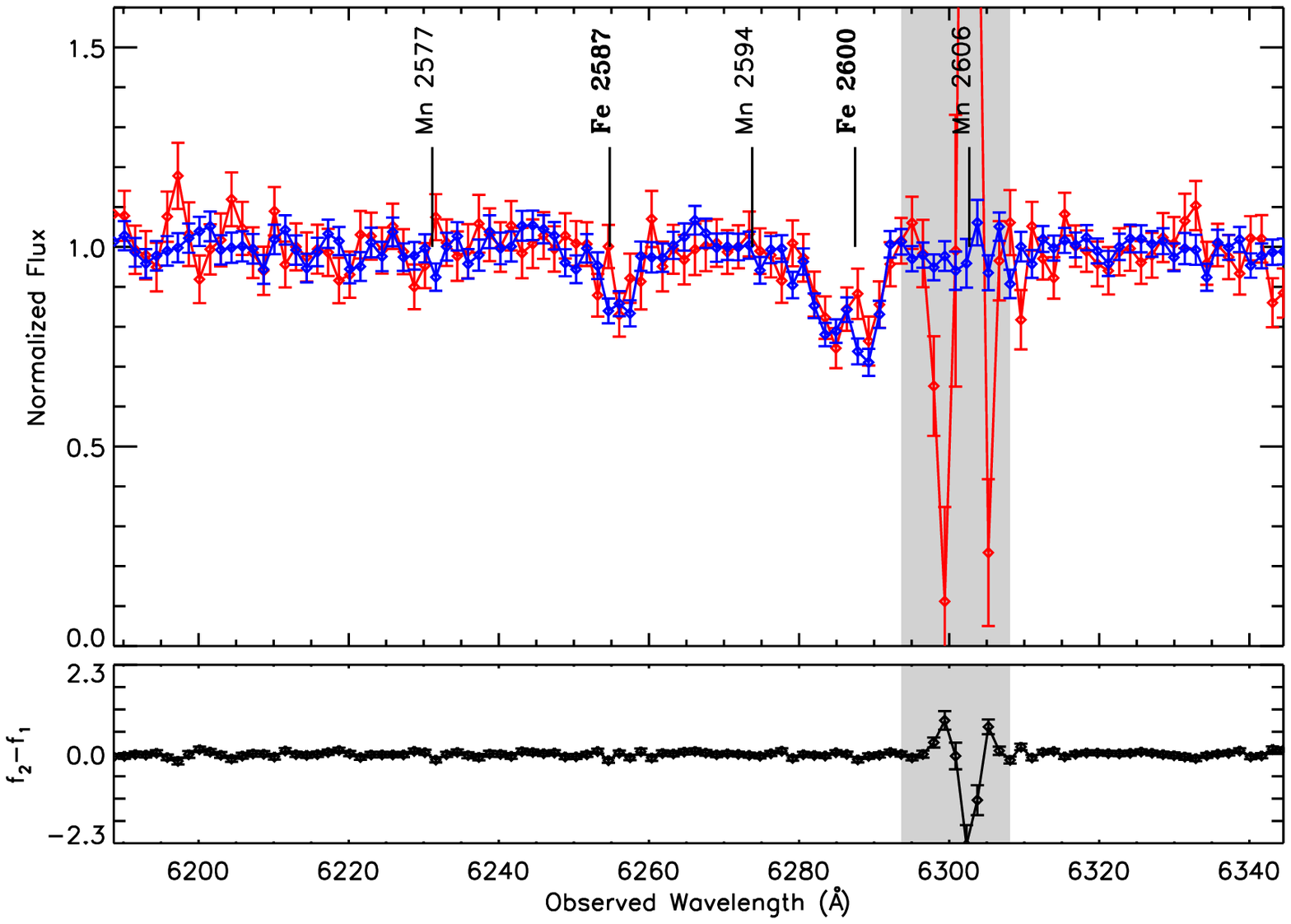}
\includegraphics[width=84mm]{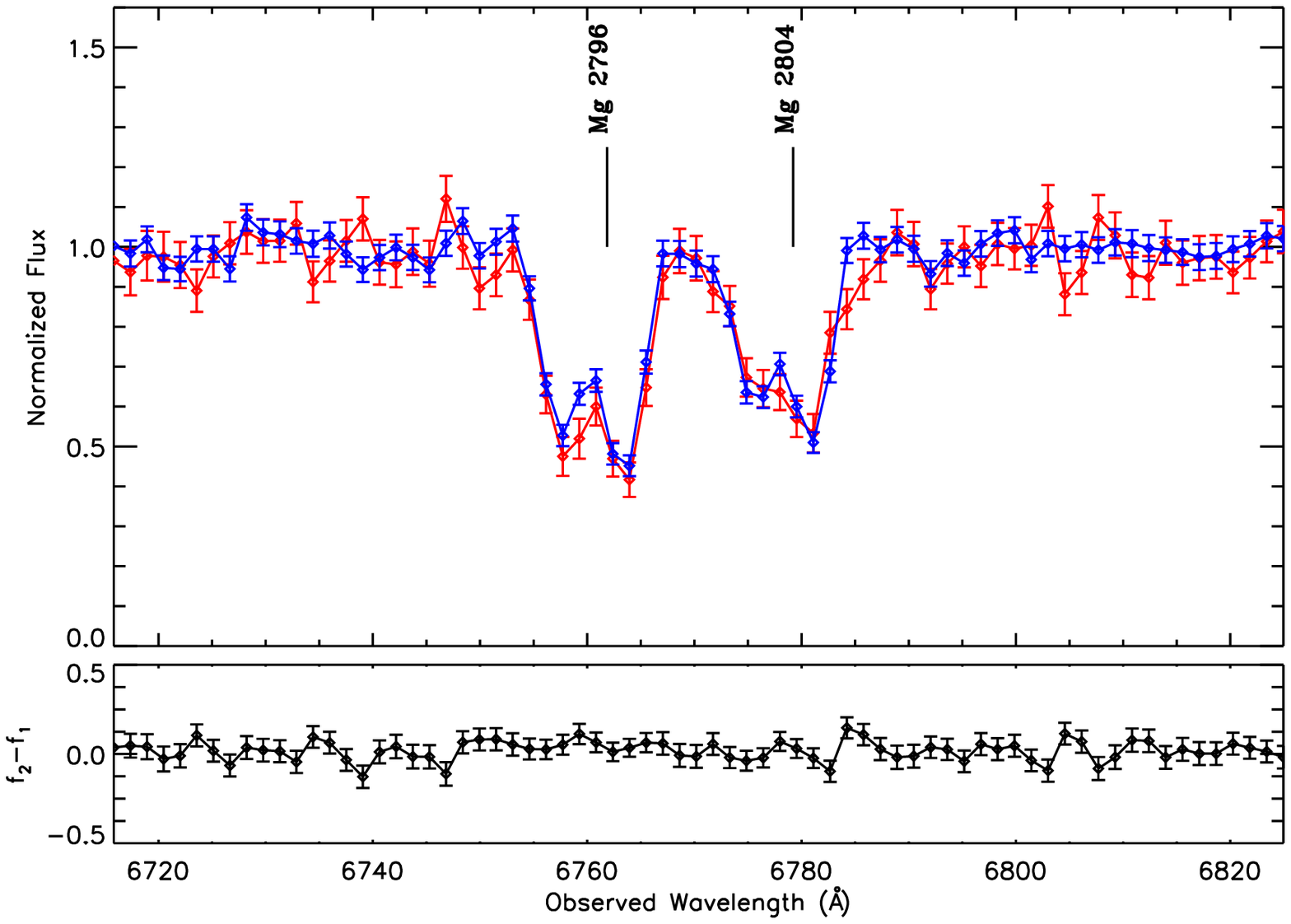}
\caption[Two-epoch normalized spectra of SDSS J120142.99+004925.0]{Two-epoch normalized spectra of the variable NAL system at $\beta$ = 0.0438 in SDSS J120142.99+004925.0.  The top panel shows the normalized pixel flux values with 1$\sigma$ error bars (first observations are red and second are blue), the bottom panel plots the difference spectrum of the two observation epochs, and shaded backgrounds identify masked pixels not included in our search for absorption line variability.  Line identifications for significantly variable absorption lines are italicised, lines detected in both observation epochs are in bold font, and undetected lines are in regular font (see Table A.1 for ion labels).  Continued in next figure.  \label{figvs28}}
\end{center}
\end{figure*}

\begin{figure*}
\ContinuedFloat
\begin{center}
\includegraphics[width=84mm]{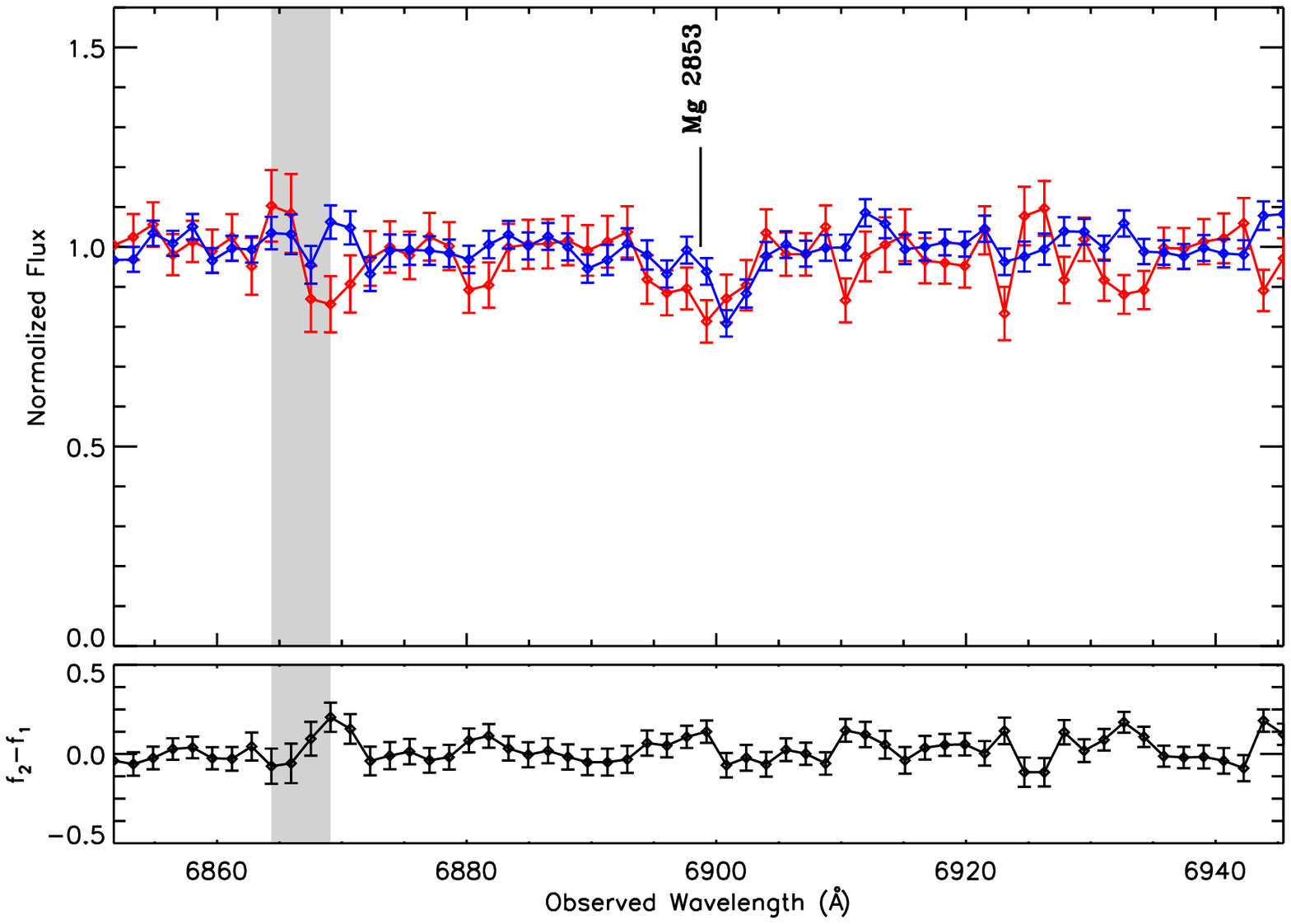}
\caption[]{Two-epoch normalized spectra of the variable NAL system at $\beta$ = 0.0438 in SDSS J120142.99+004925.0.  The top panel shows the normalized pixel flux values with 1$\sigma$ error bars (first observations are red and second are blue), the bottom panel plots the difference spectrum of the two observation epochs, and shaded backgrounds identify masked pixels not included in our search for absorption line variability.  Line identifications for significantly variable absorption lines are italicised, lines detected in both observation epochs are in bold font, and undetected lines are in regular font (see Table A.1 for ion labels).  Continued from previous figure.}
\vspace{3.5cm}
\end{center}
\end{figure*}

\begin{figure*}
\begin{center}
\includegraphics[width=84mm]{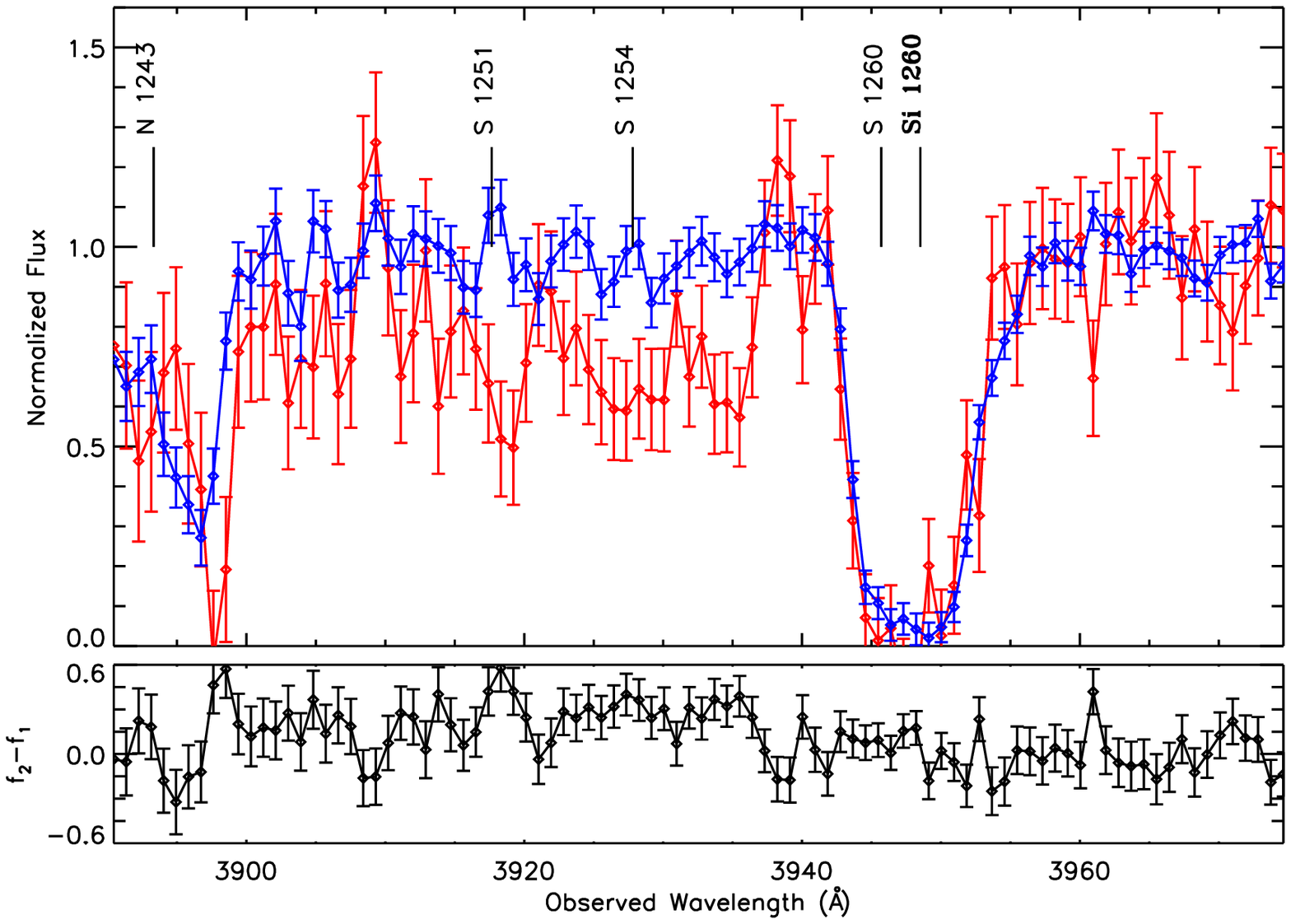}
\includegraphics[width=84mm]{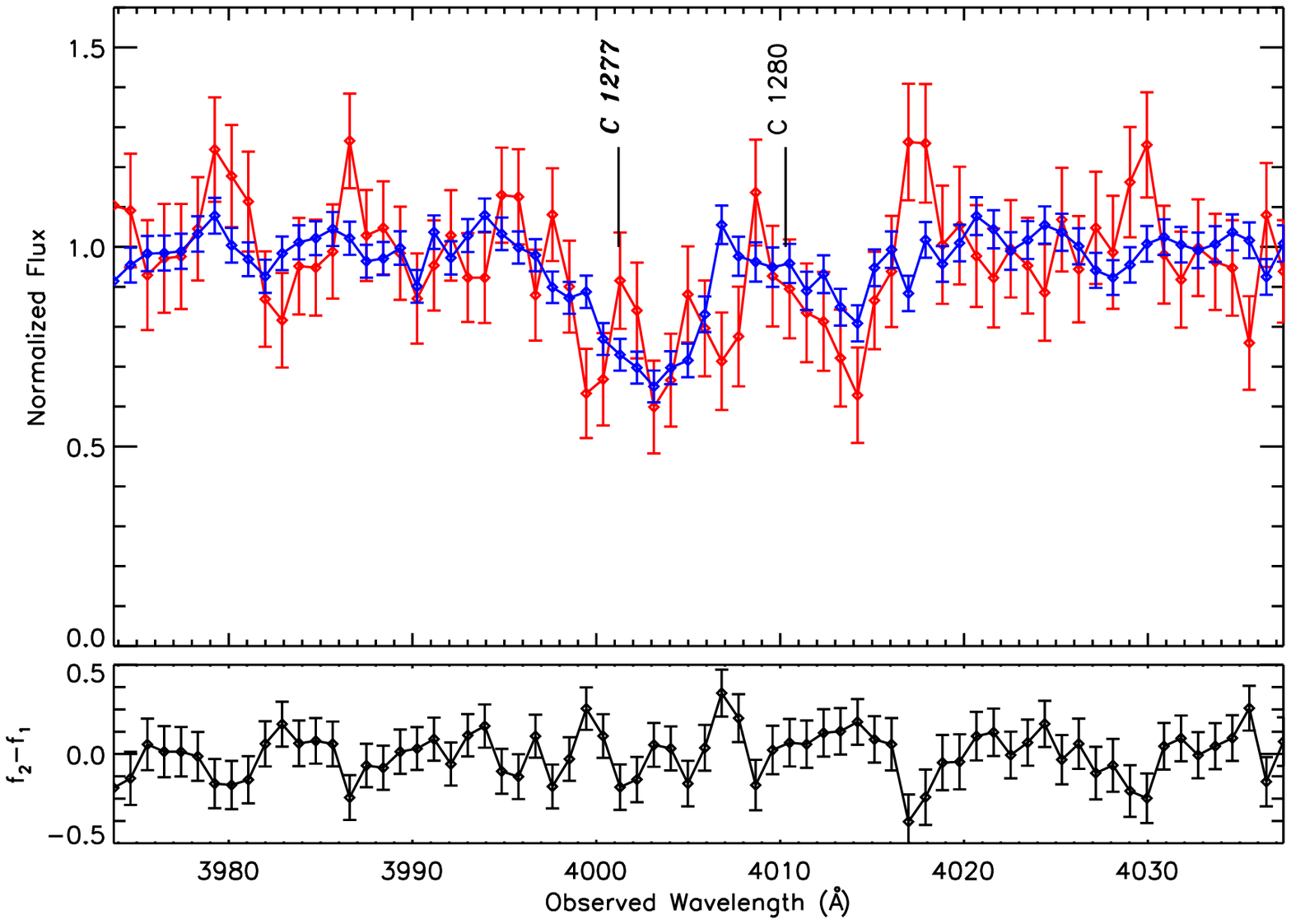}
\includegraphics[width=84mm]{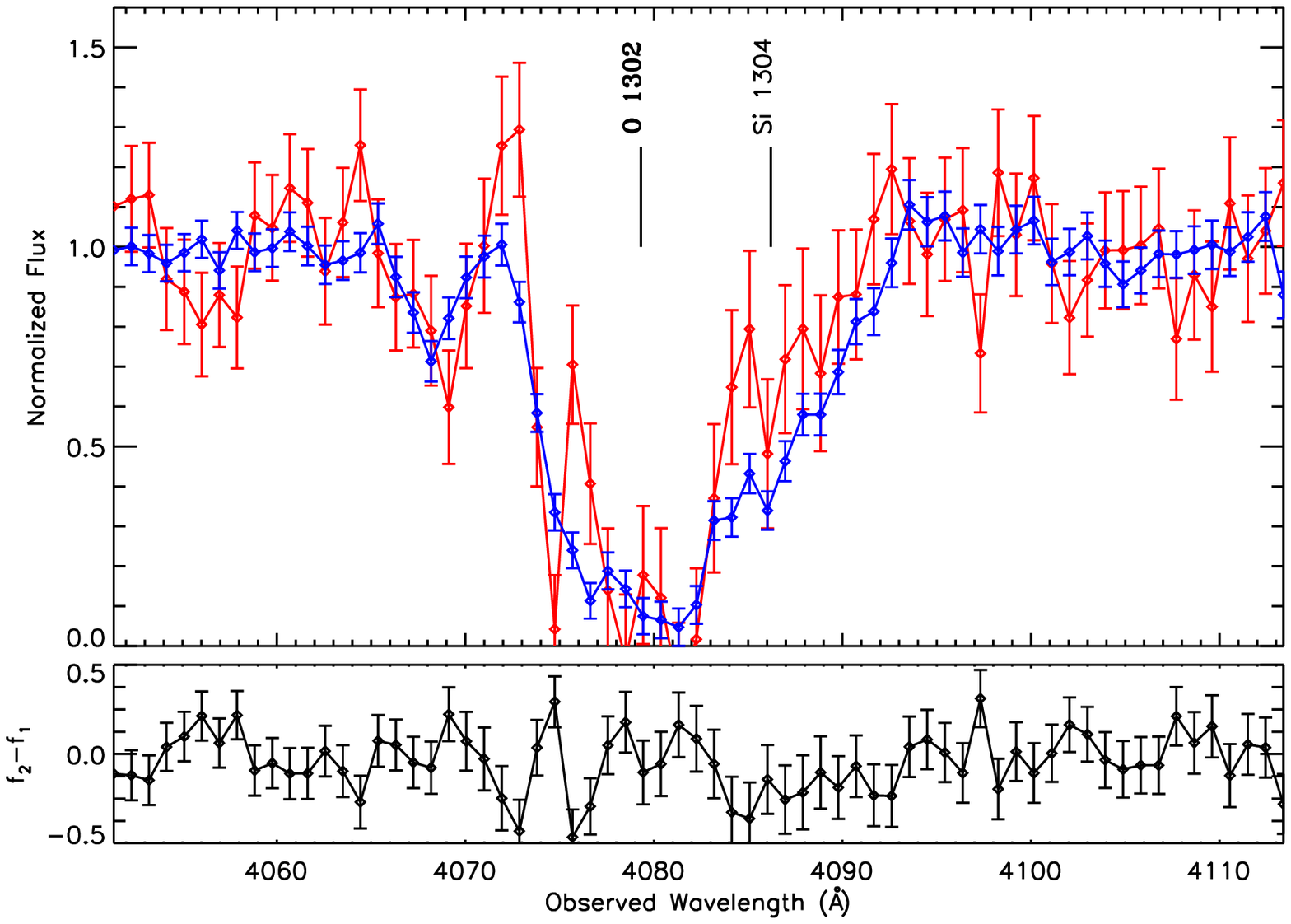}
\includegraphics[width=84mm]{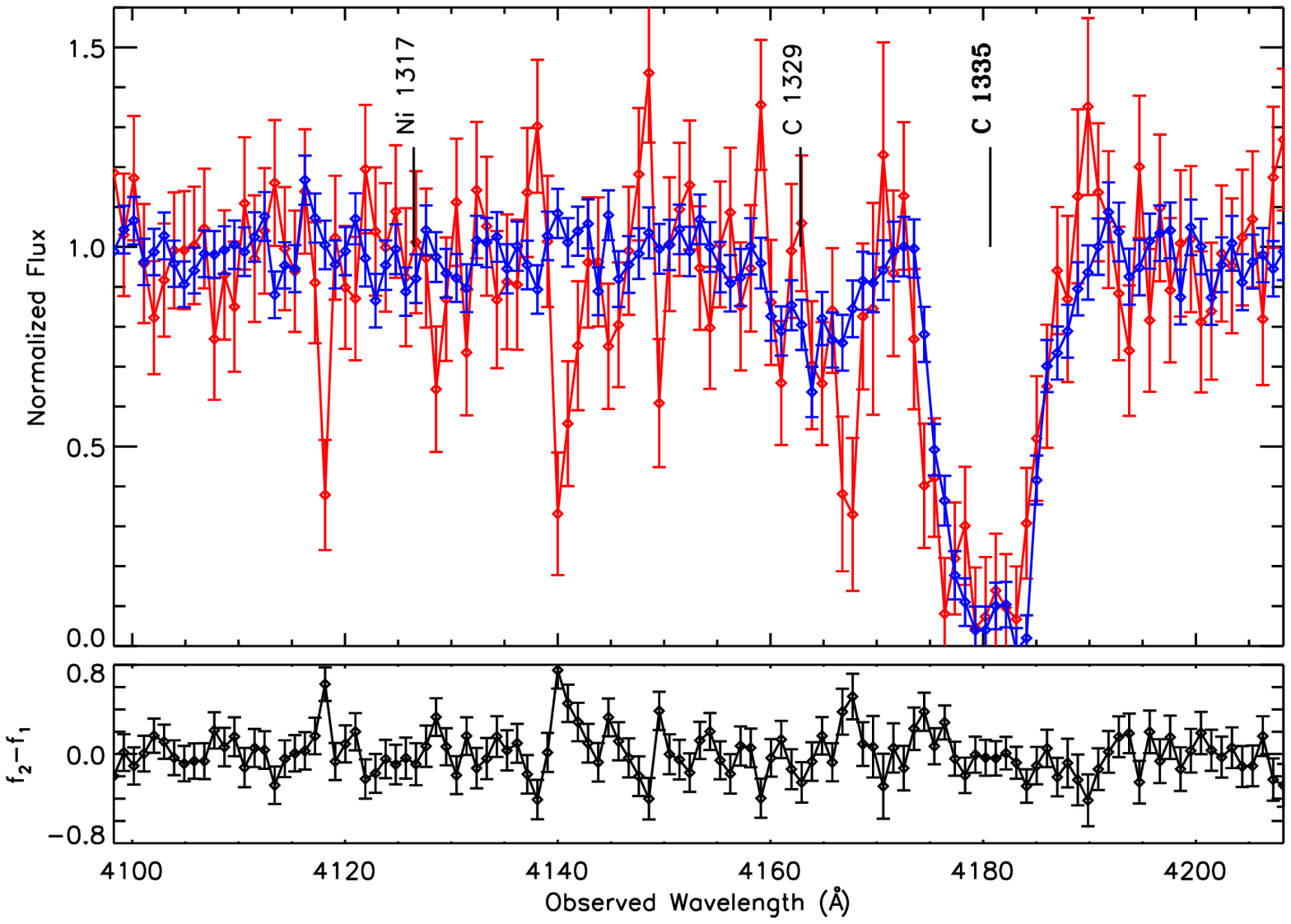}
\includegraphics[width=84mm]{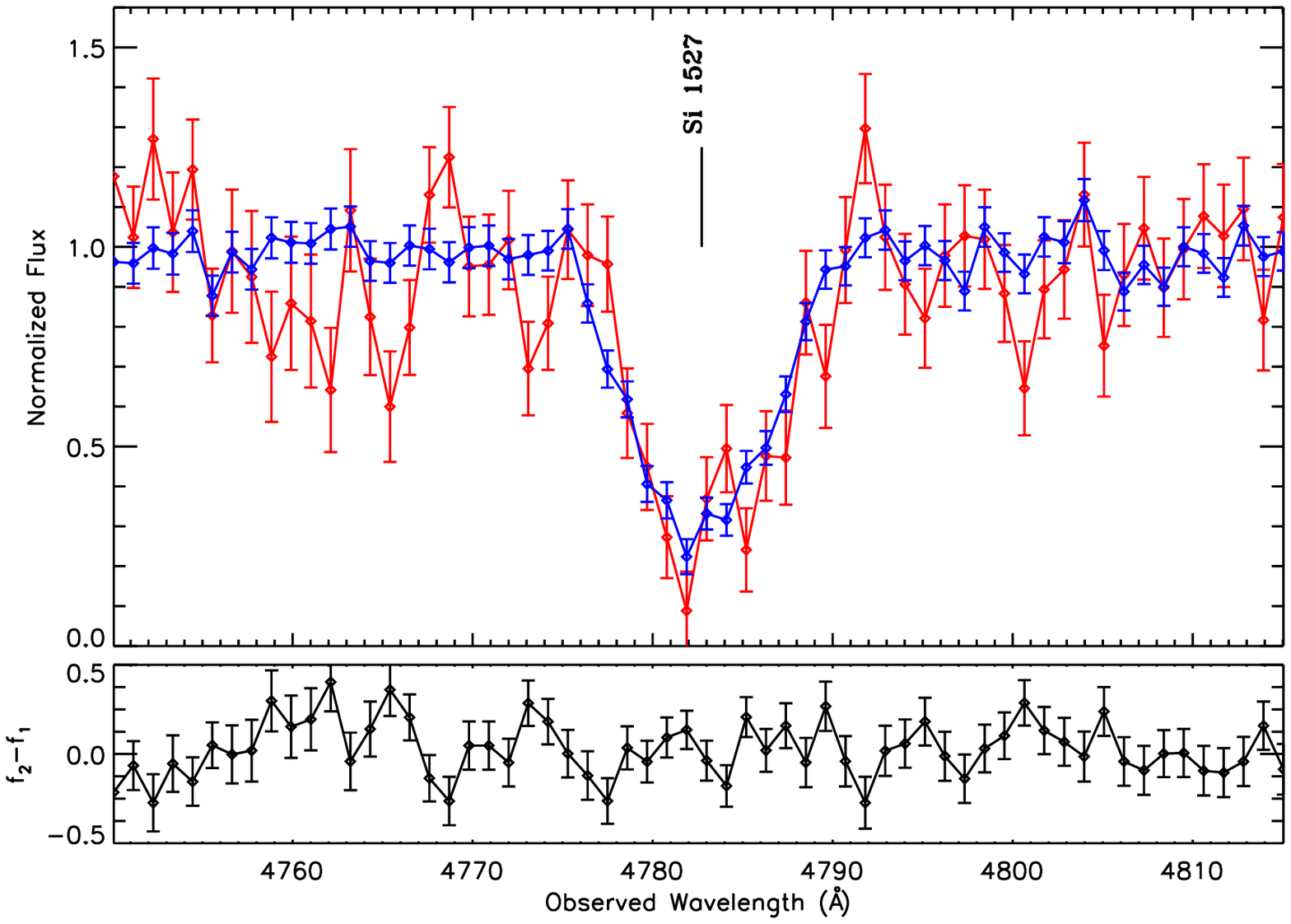}
\includegraphics[width=84mm]{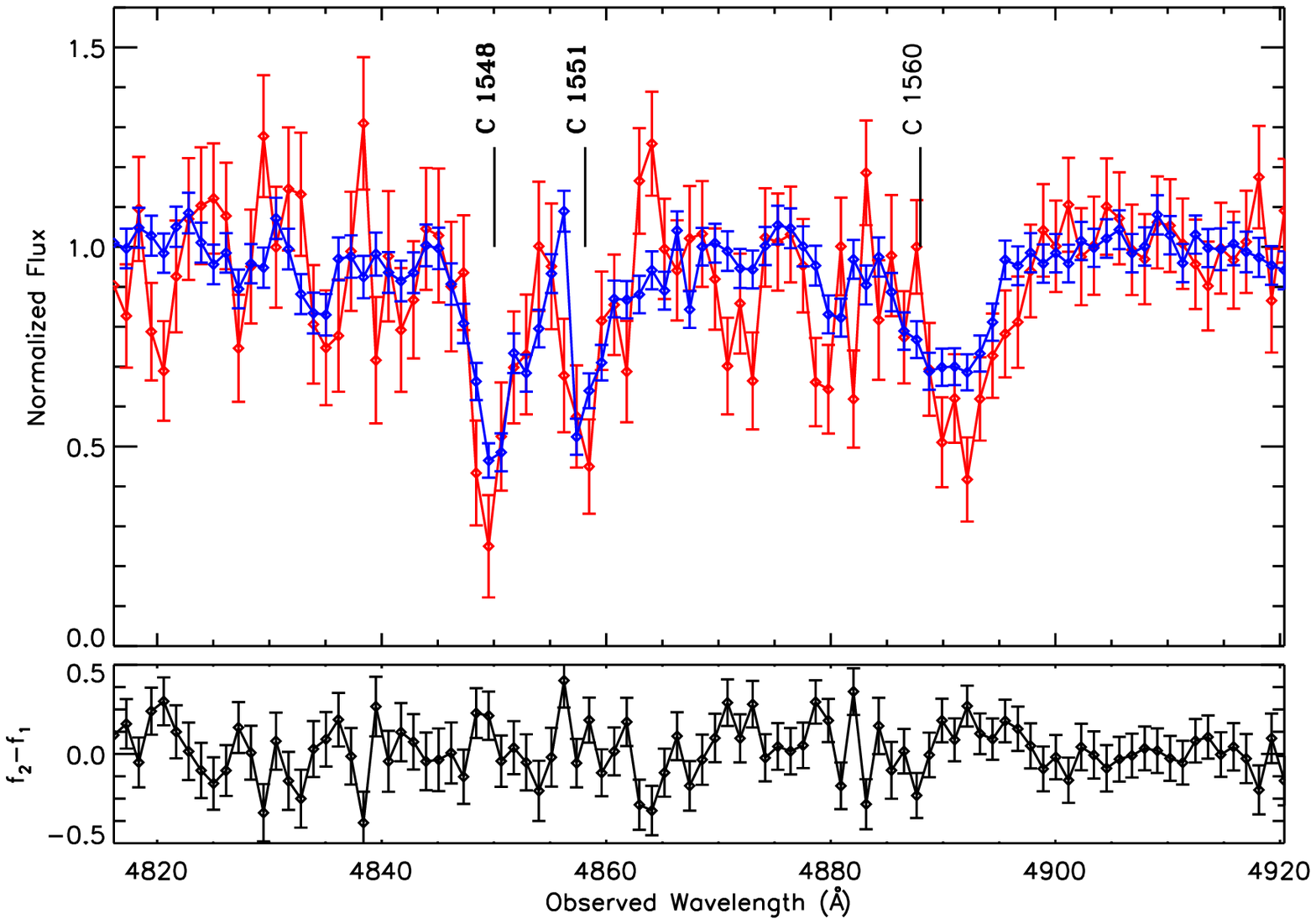}
\caption[Two-epoch normalized spectra of SDSS J124708.42+500320.7]{Two-epoch normalized spectra of the variable NAL system at $\beta$ = 0.0432 in SDSS J124708.42+500320.7.  The top panel shows the normalized pixel flux values with 1$\sigma$ error bars (first observations are red and second are blue), the bottom panel plots the difference spectrum of the two observation epochs, and shaded backgrounds identify masked pixels not included in our search for absorption line variability.  Line identifications for significantly variable absorption lines are italicised, lines detected in both observation epochs are in bold font, and undetected lines are in regular font (see Table A.1 for ion labels).  Continued in next figure.  \label{figvs29}}
\end{center}
\end{figure*}

\begin{figure*}
\ContinuedFloat
\begin{center}
\includegraphics[width=84mm]{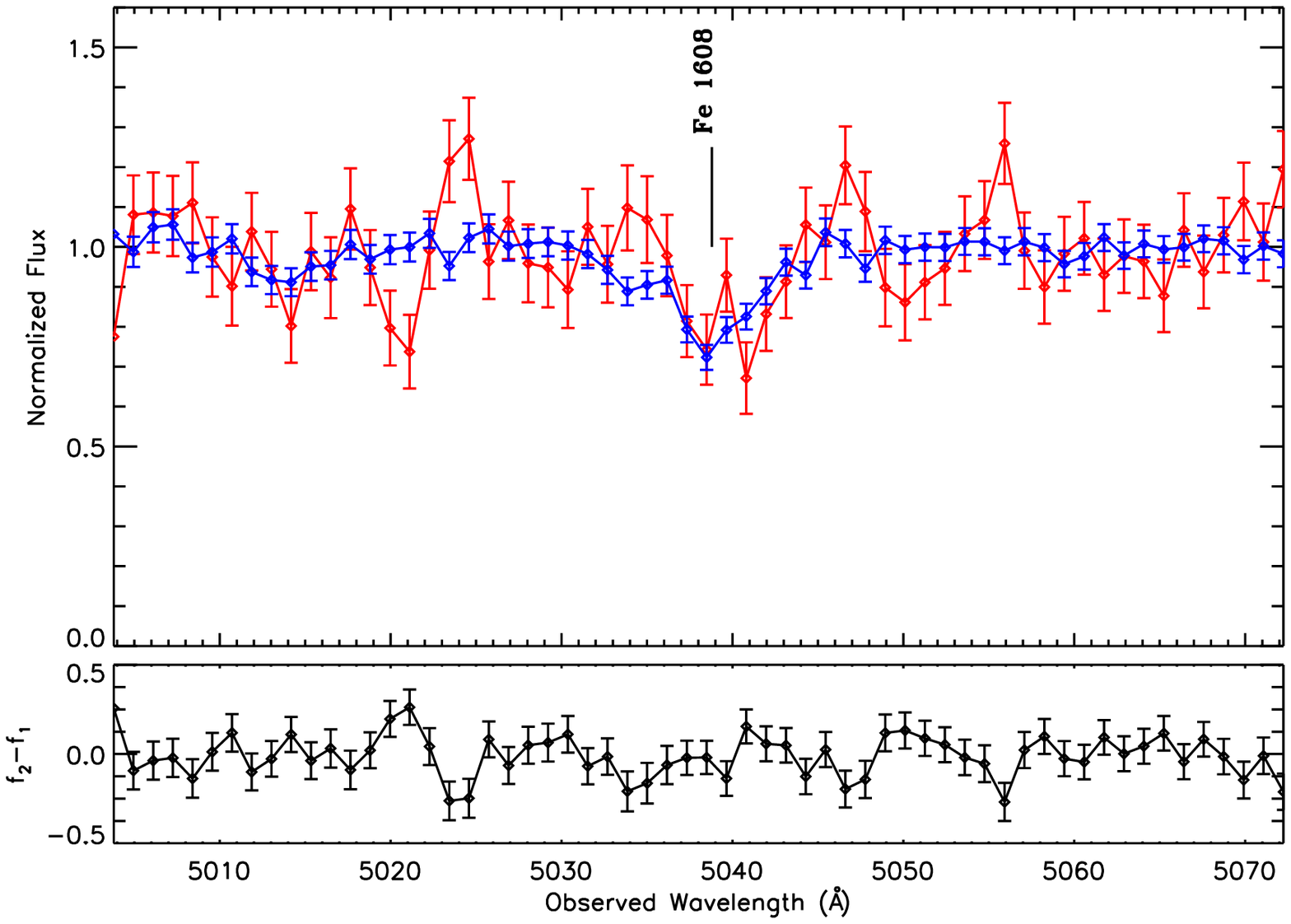}
\includegraphics[width=84mm]{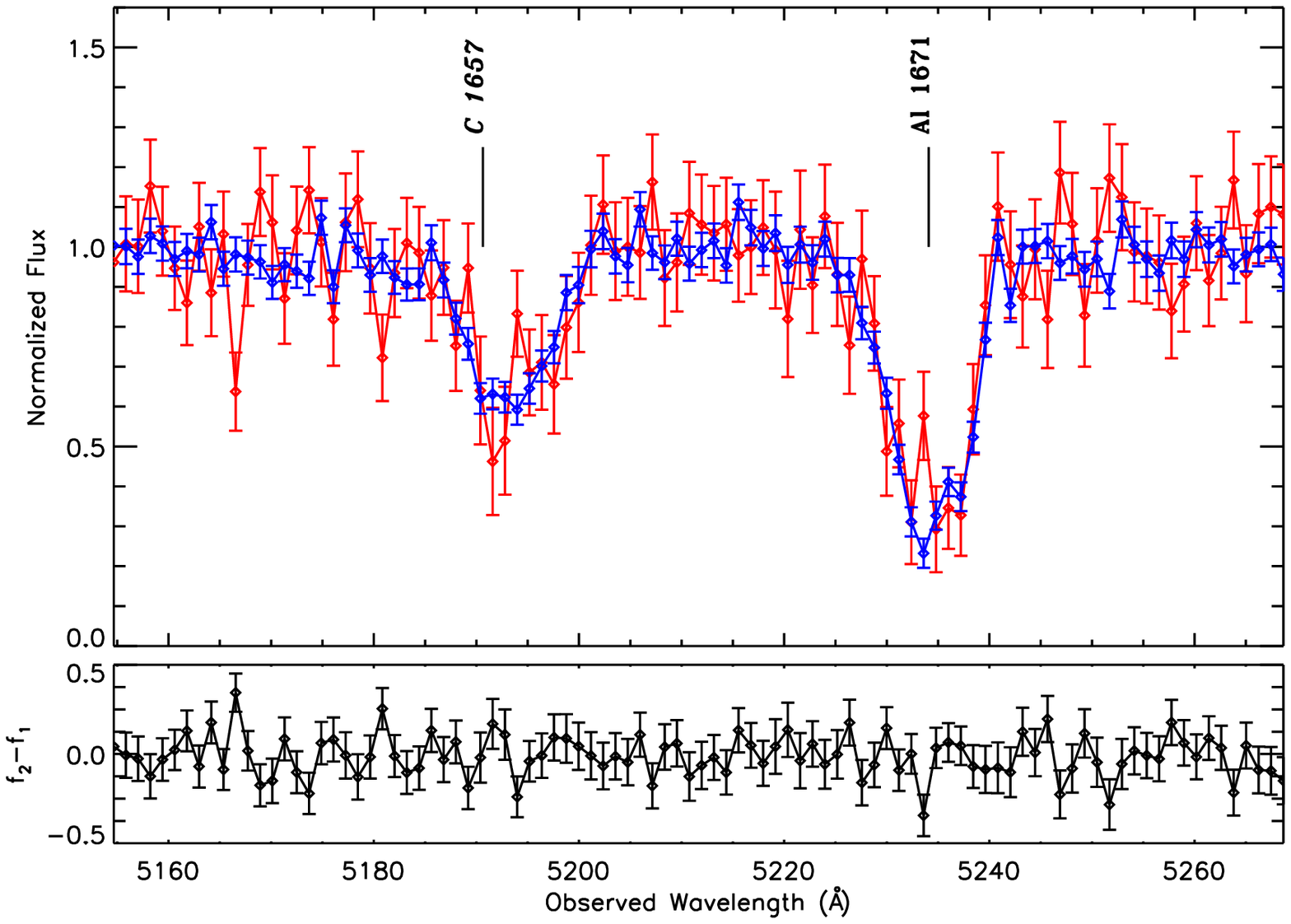}
\includegraphics[width=84mm]{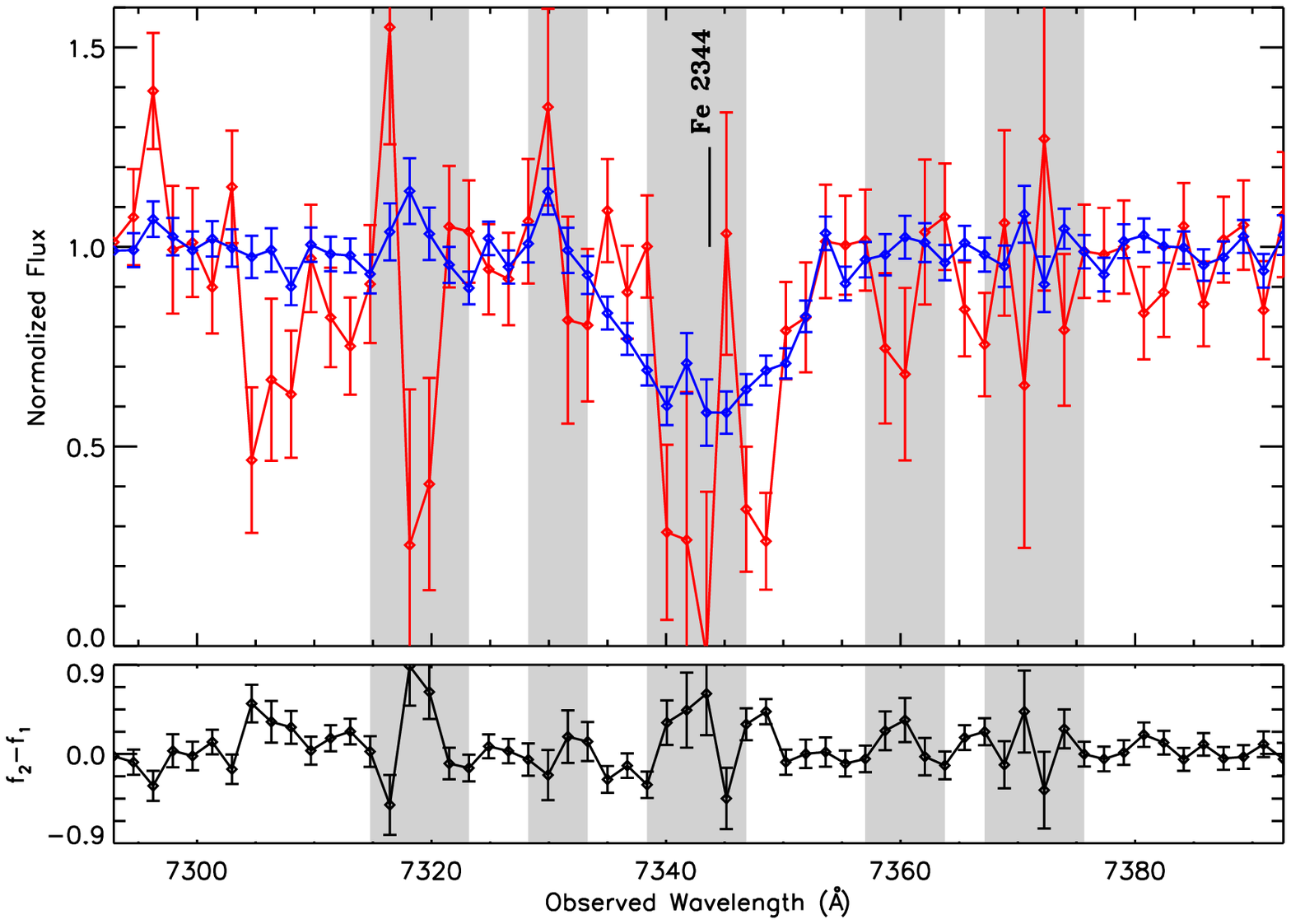}
\includegraphics[width=84mm]{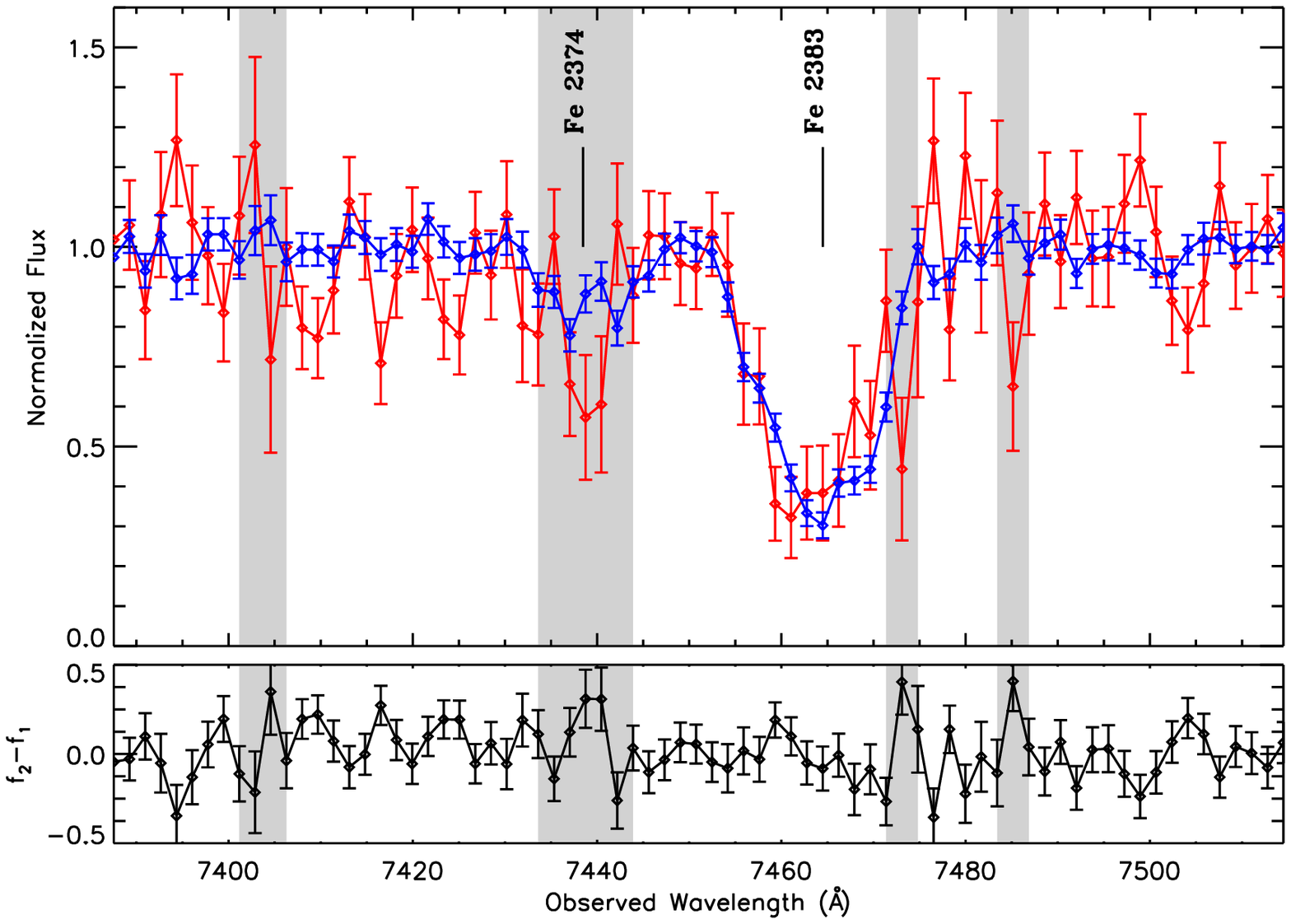}
\includegraphics[width=84mm]{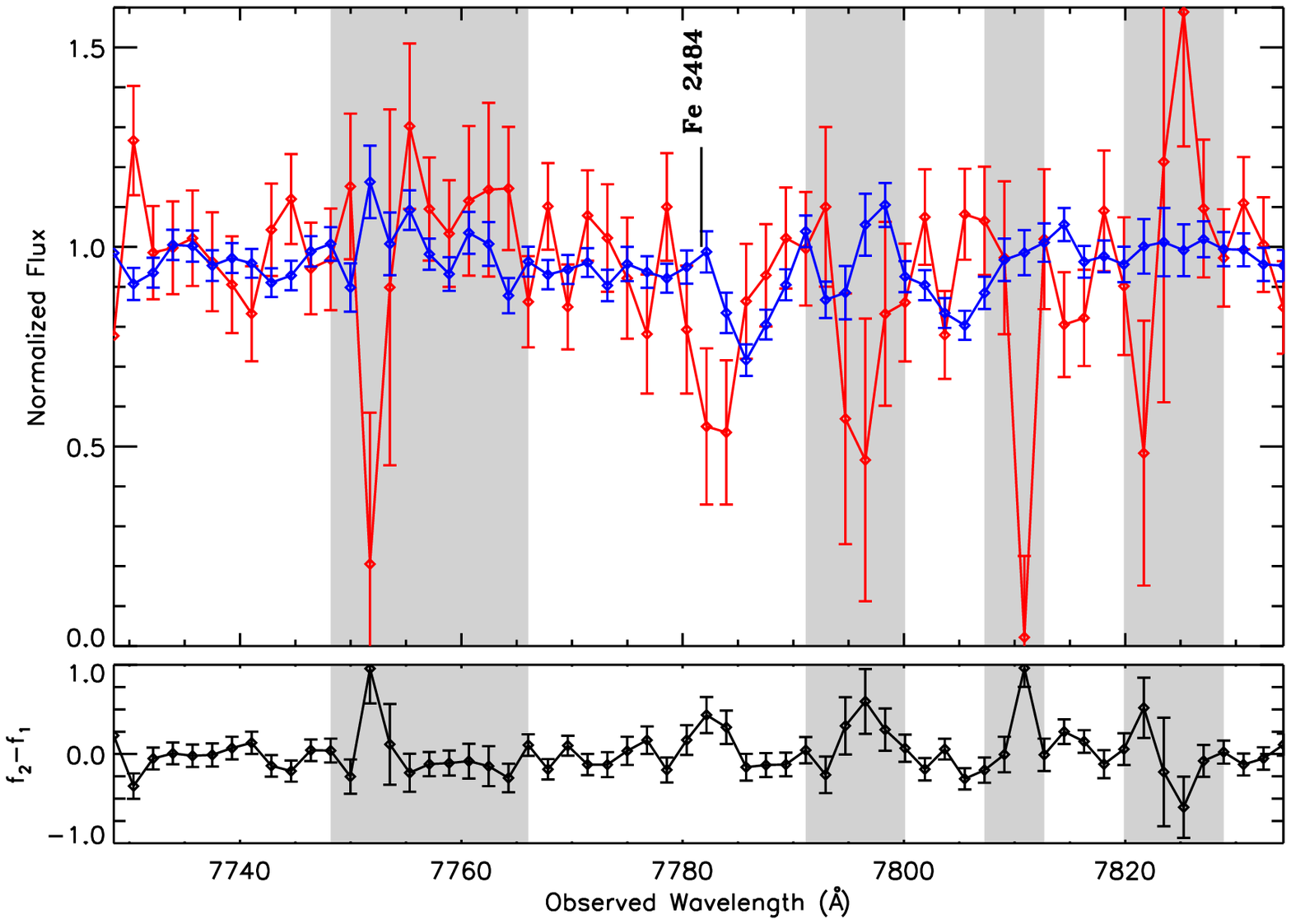}
\includegraphics[width=84mm]{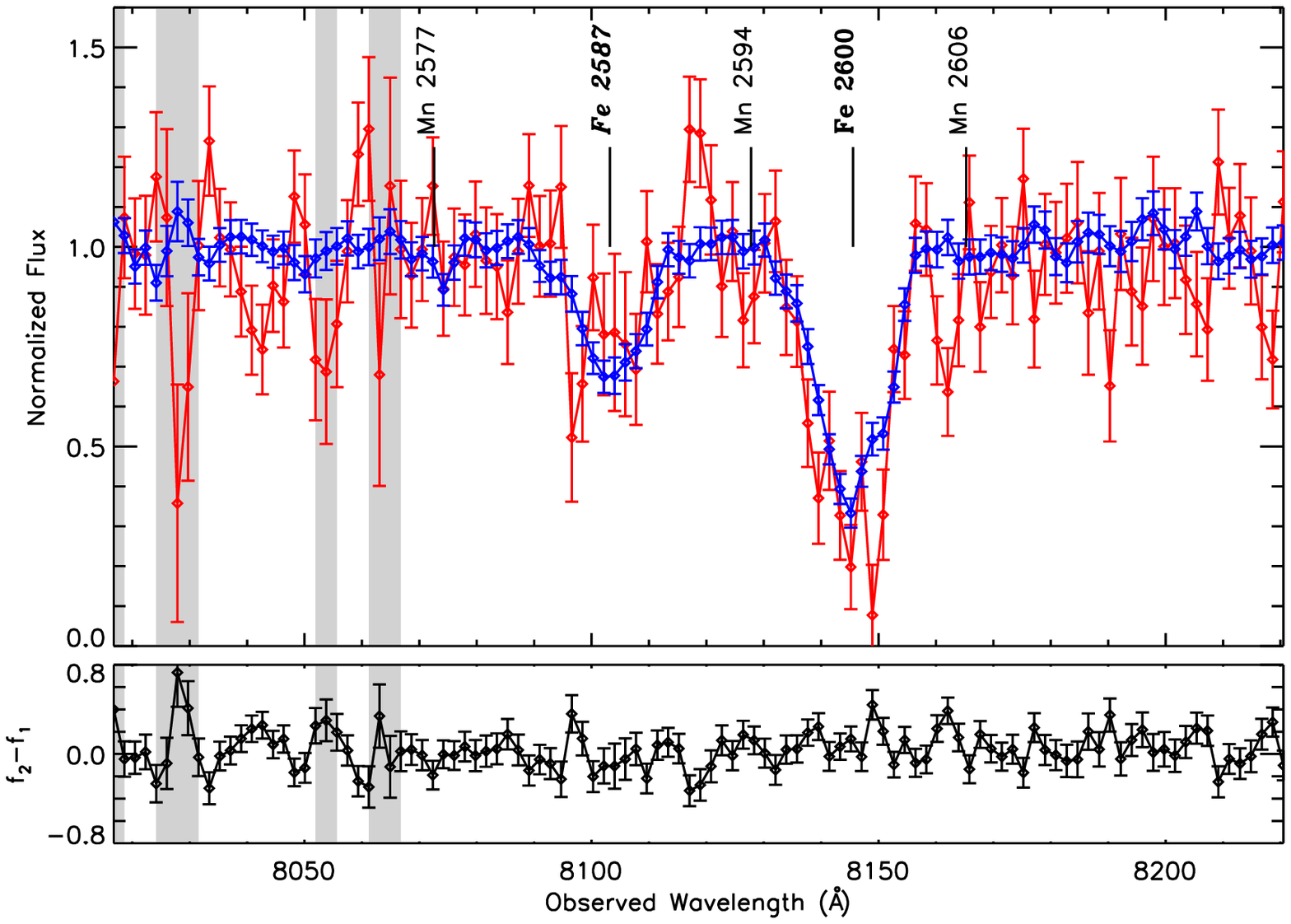}
\caption[]{Two-epoch normalized spectra of the variable NAL system at $\beta$ = 0.0432 in SDSS J124708.42+500320.7.  The top panel shows the normalized pixel flux values with 1$\sigma$ error bars (first observations are red and second are blue), the bottom panel plots the difference spectrum of the two observation epochs, and shaded backgrounds identify masked pixels not included in our search for absorption line variability.  Line identifications for significantly variable absorption lines are italicised, lines detected in both observation epochs are in bold font, and undetected lines are in regular font (see Table A.1 for ion labels).  Continued from previous figure.}
\end{center}
\end{figure*}

\begin{figure*}
\ContinuedFloat
\begin{center}
\includegraphics[width=84mm]{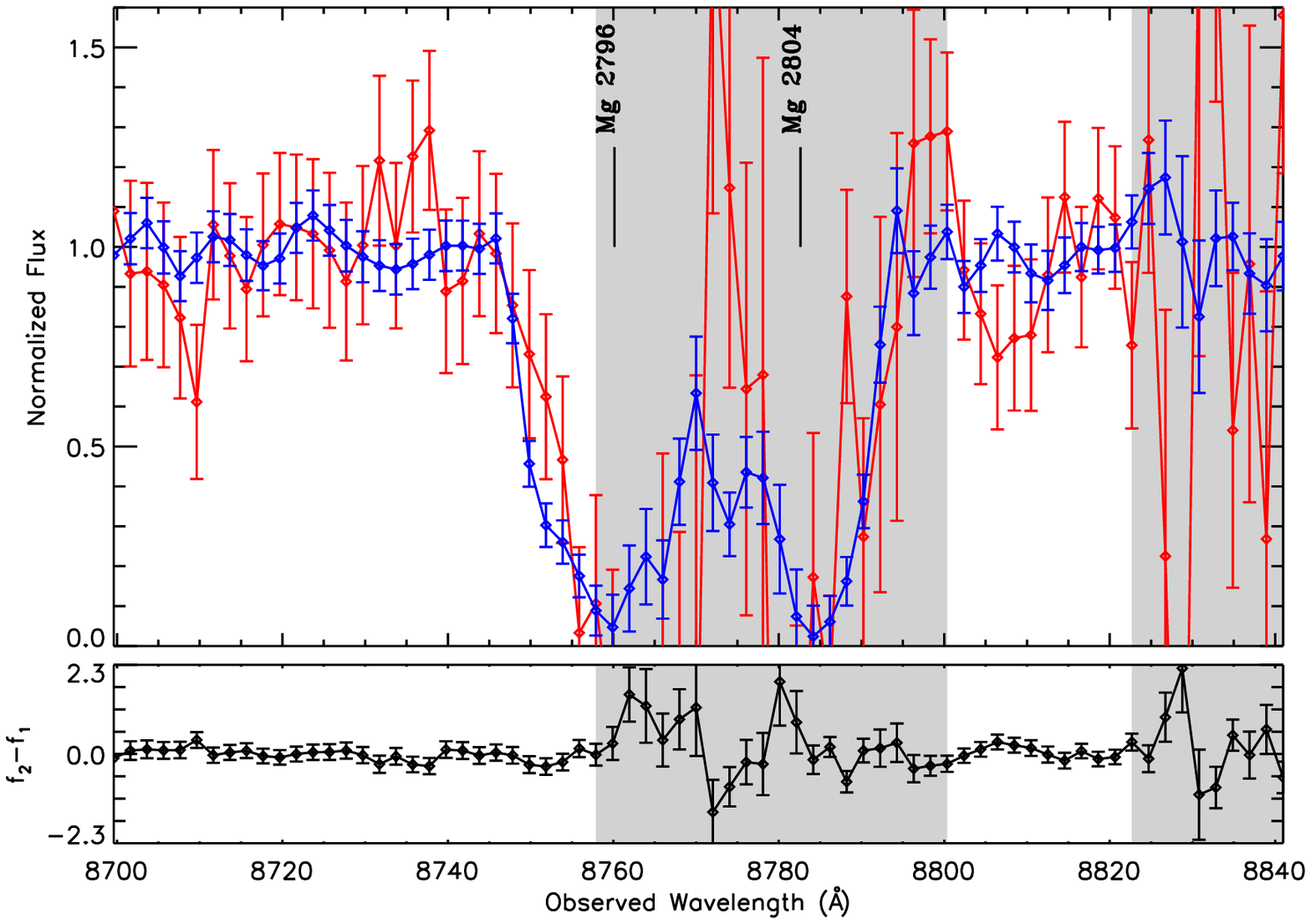}
\includegraphics[width=84mm]{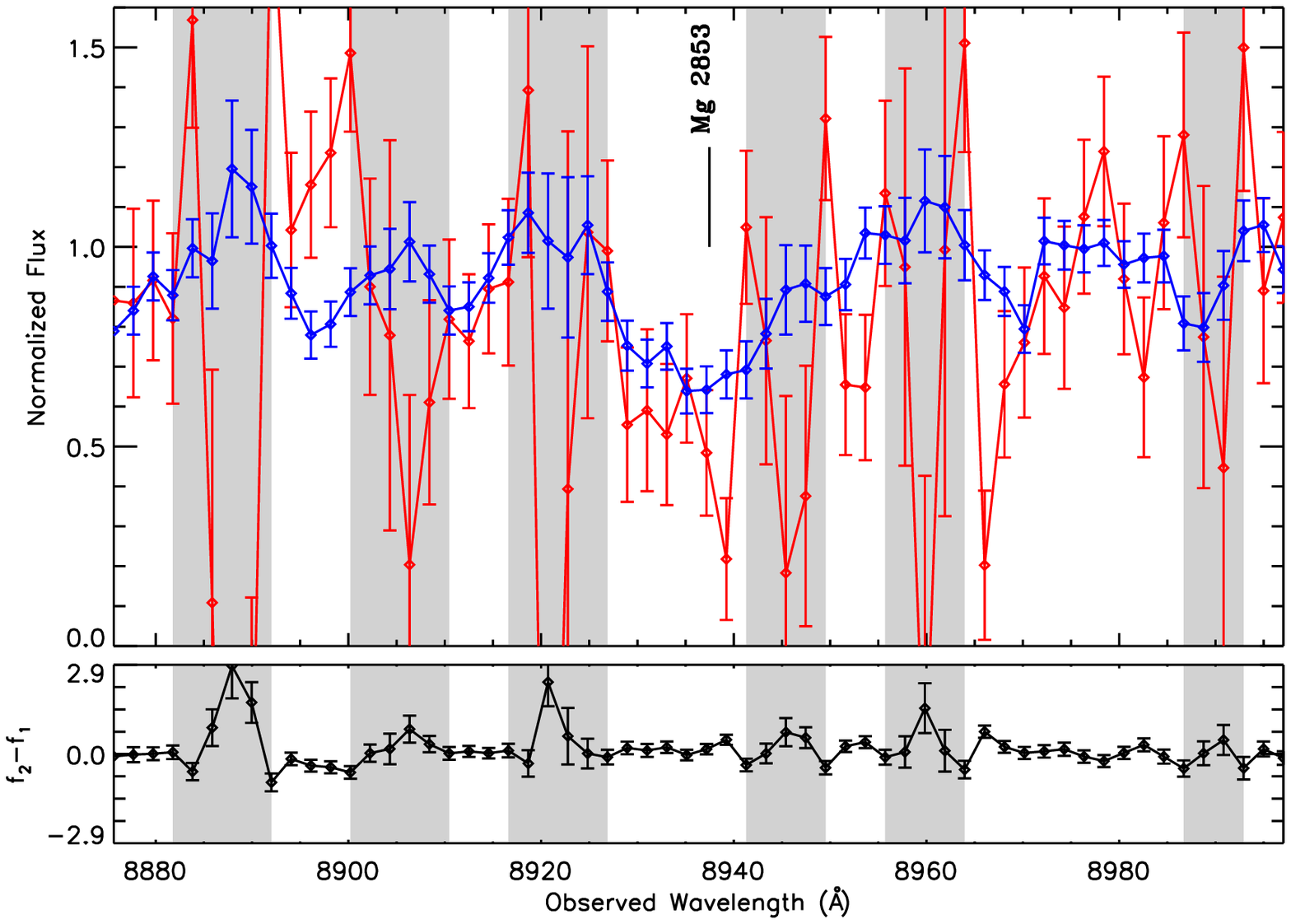}
\caption[]{Two-epoch normalized spectra of the variable NAL system at $\beta$ = 0.0432 in SDSS J124708.42+500320.7.  The top panel shows the normalized pixel flux values with 1$\sigma$ error bars (first observations are red and second are blue), the bottom panel plots the difference spectrum of the two observation epochs, and shaded backgrounds identify masked pixels not included in our search for absorption line variability.  Line identifications for significantly variable absorption lines are italicised, lines detected in both observation epochs are in bold font, and undetected lines are in regular font (see Table A.1 for ion labels).  Continued from previous figure.}
\vspace{3.5cm}
\end{center}
\end{figure*}

\begin{figure*}
\begin{center}
\includegraphics[width=84mm]{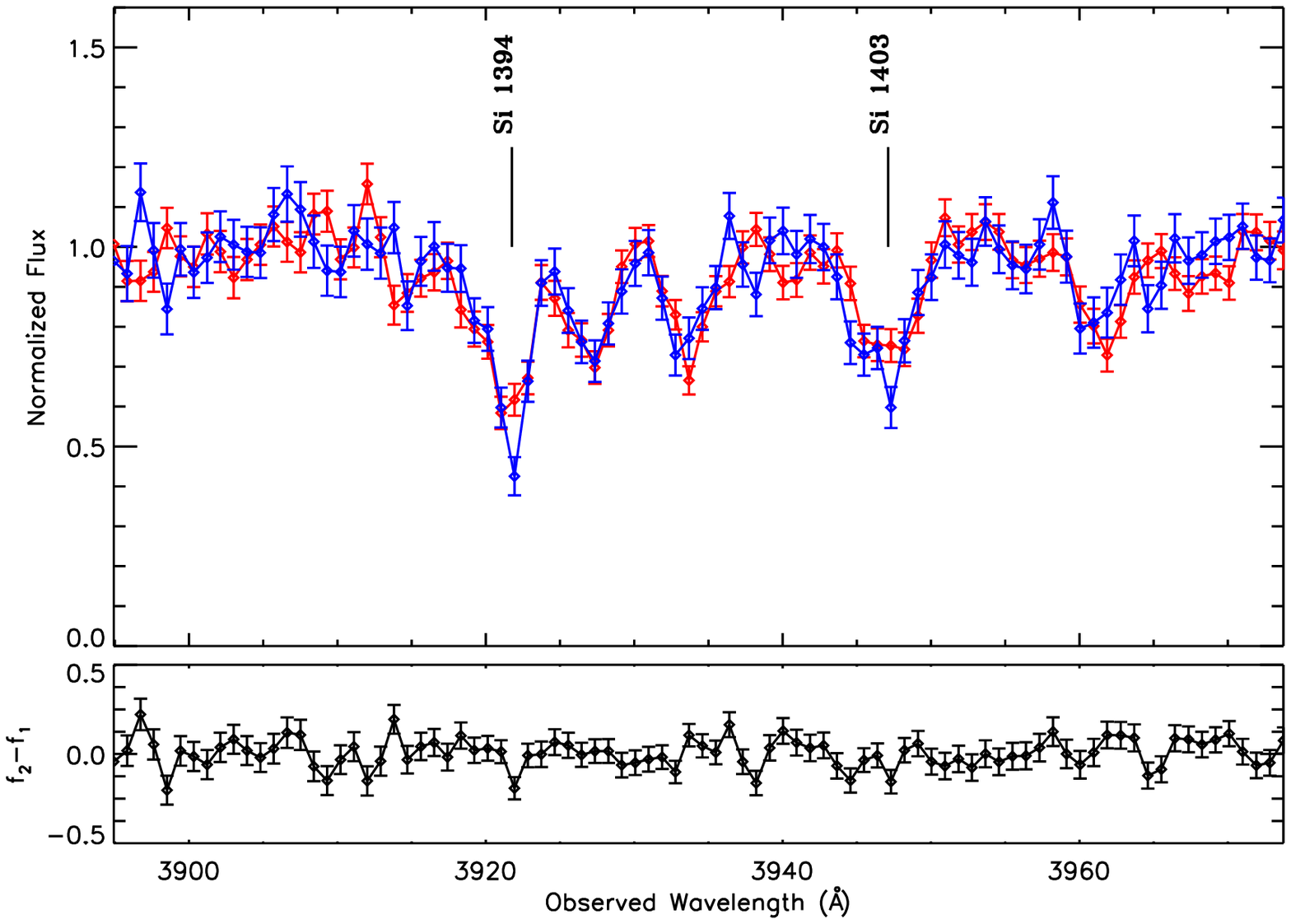}
\includegraphics[width=84mm]{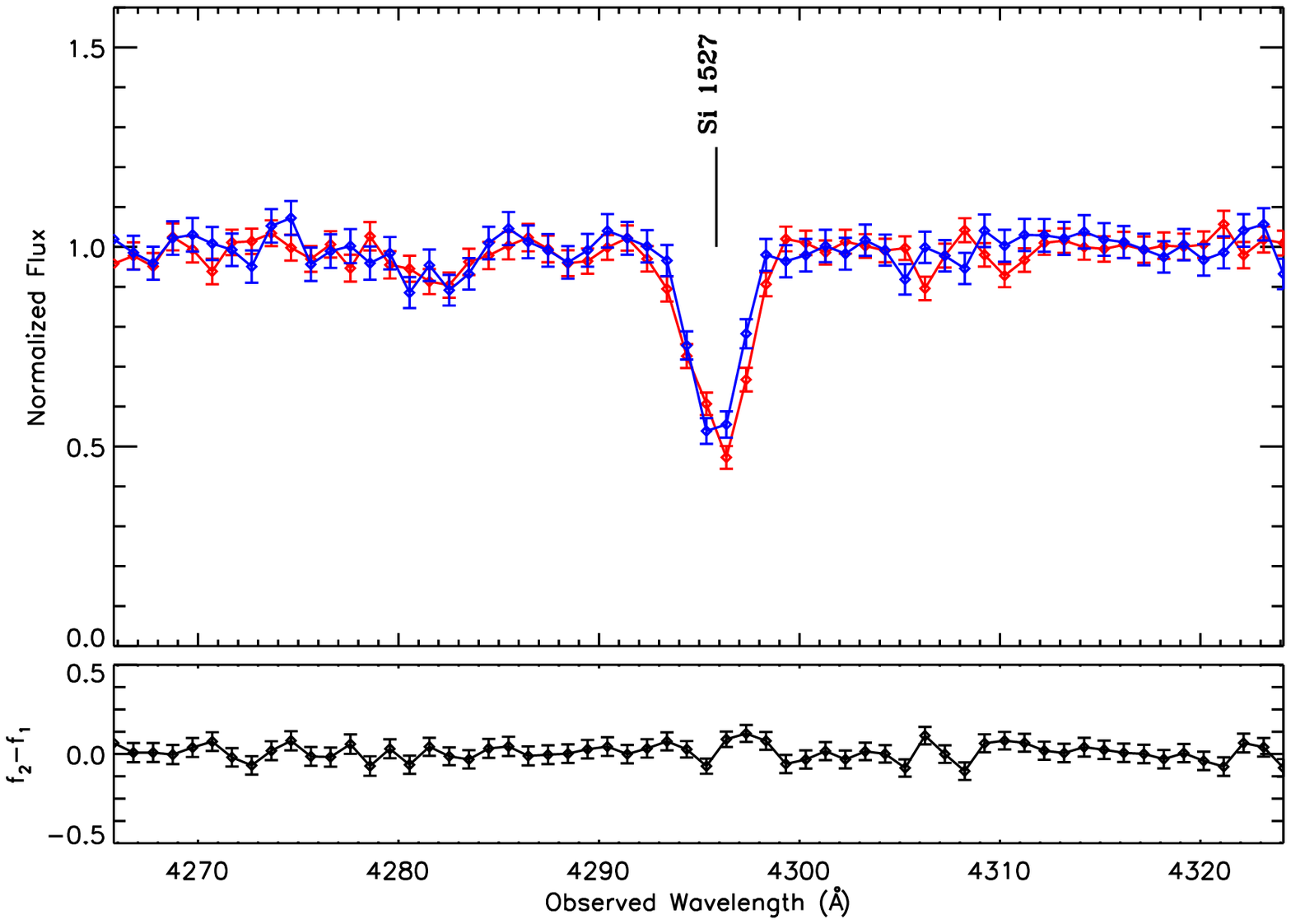}
\includegraphics[width=84mm]{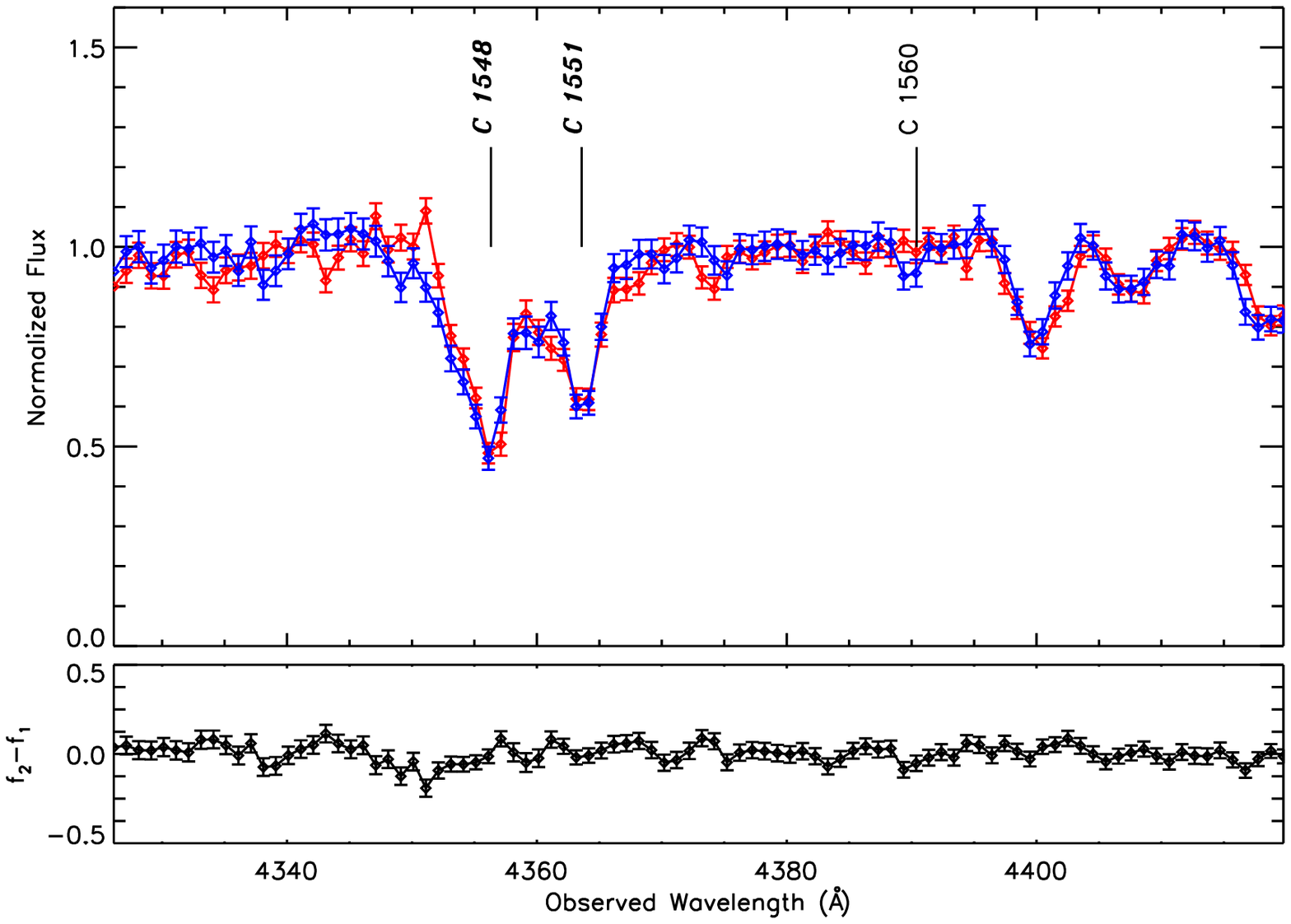}
\includegraphics[width=84mm]{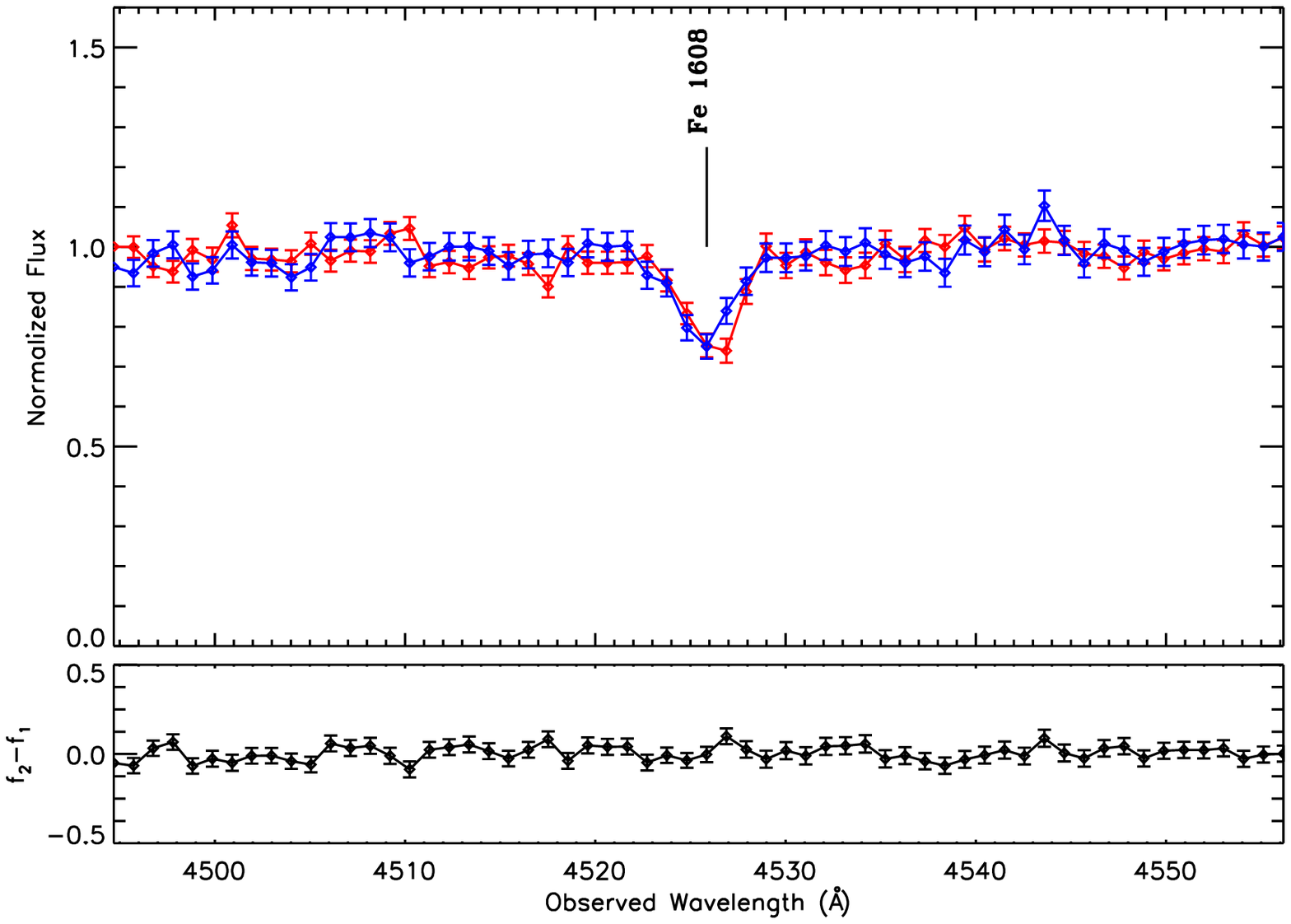}
\includegraphics[width=84mm]{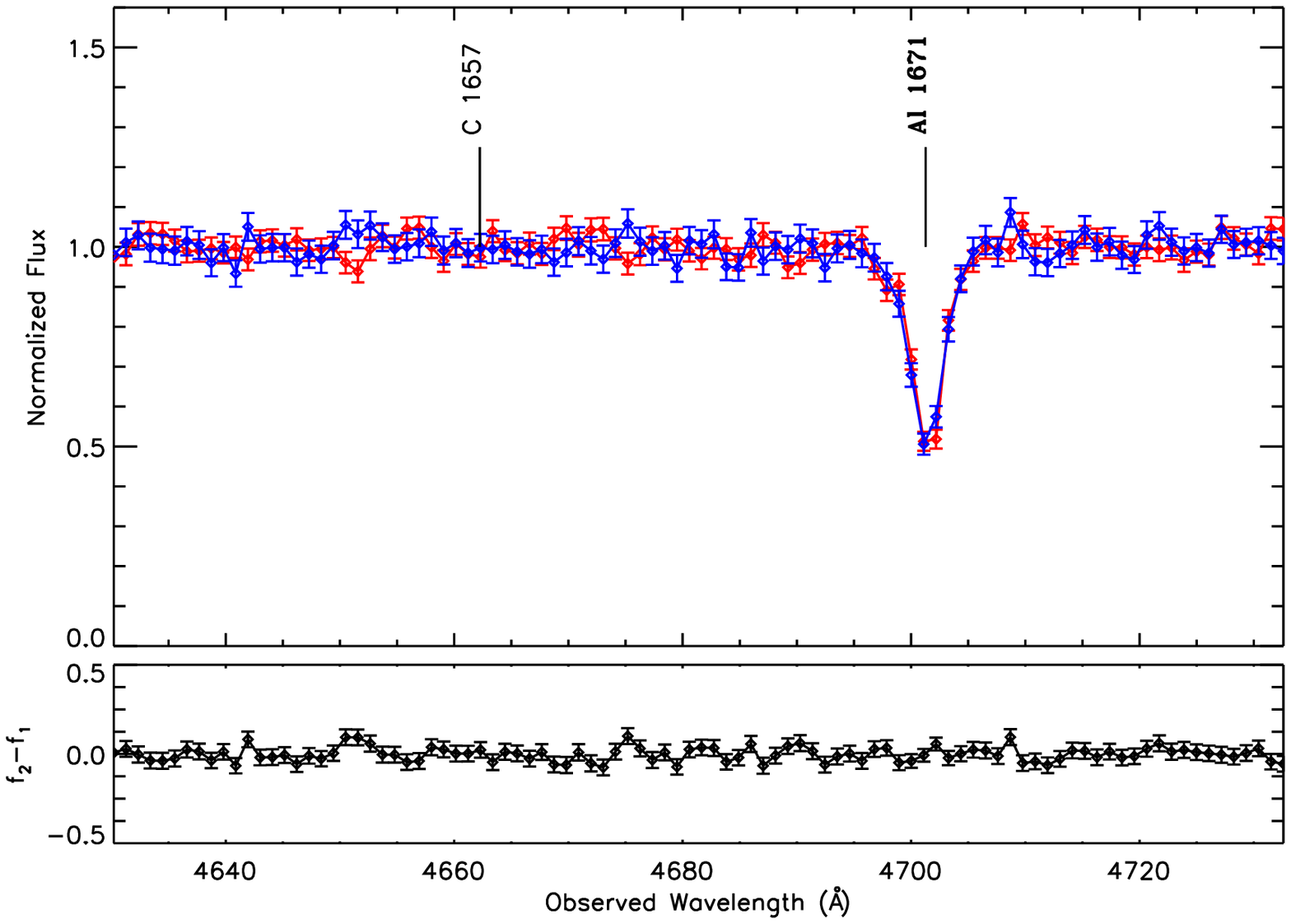}
\includegraphics[width=84mm]{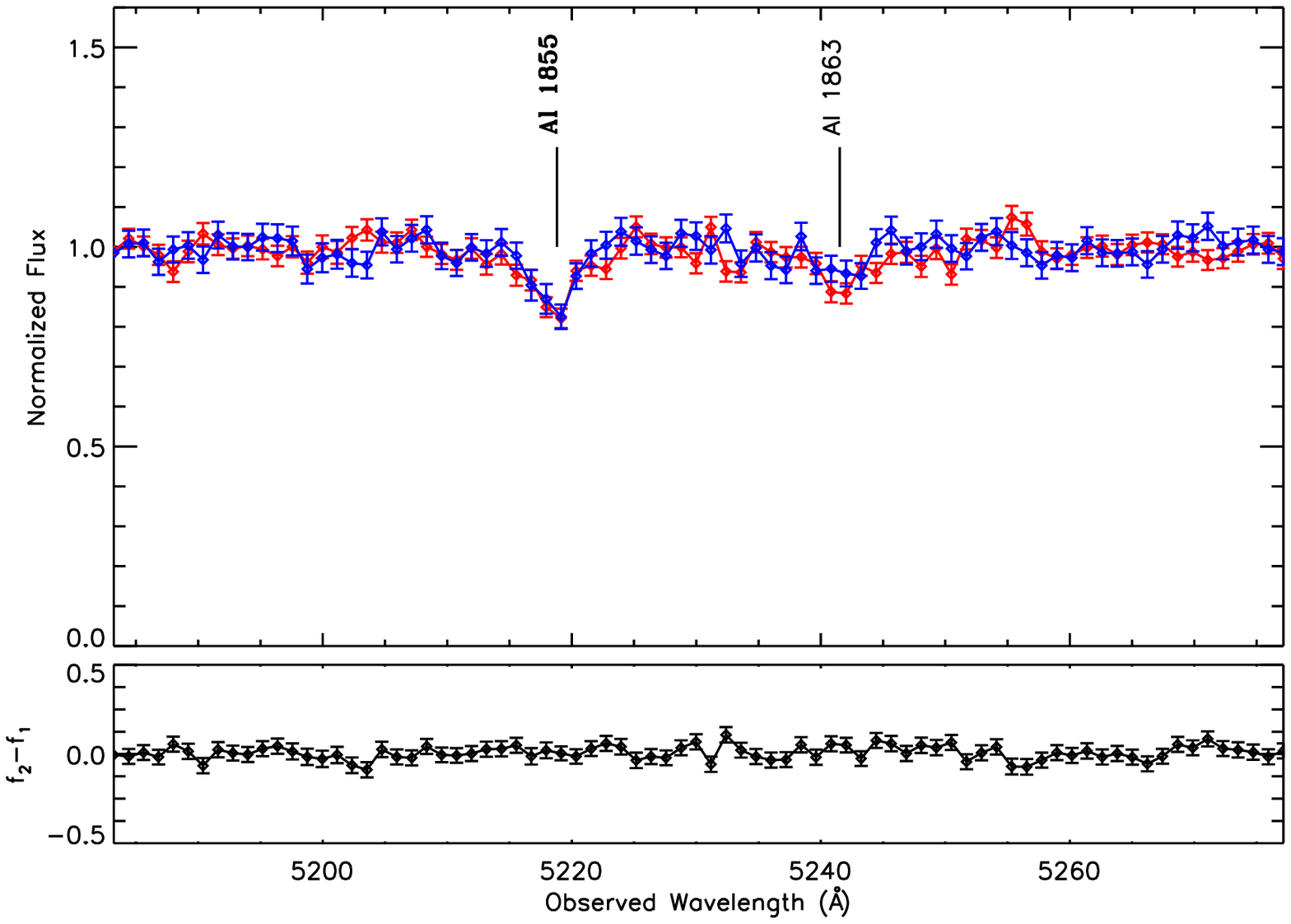}
\caption[Two-epoch normalized spectra of SDSS J212812.33-081529.3]{Two-epoch normalized spectra of the variable NAL system at $\beta$ = 0.0235 in SDSS J212812.33-081529.3.  The top panel shows the normalized pixel flux values with 1$\sigma$ error bars (first observations are red and second are blue), the bottom panel plots the difference spectrum of the two observation epochs, and shaded backgrounds identify masked pixels not included in our search for absorption line variability.  Line identifications for significantly variable absorption lines are italicised, lines detected in both observation epochs are in bold font, and undetected lines are in regular font (see Table A.1 for ion labels).  Continued in next figure.  \label{figvs30}}
\end{center}
\end{figure*}

\begin{figure*}
\ContinuedFloat
\begin{center}
\includegraphics[width=84mm]{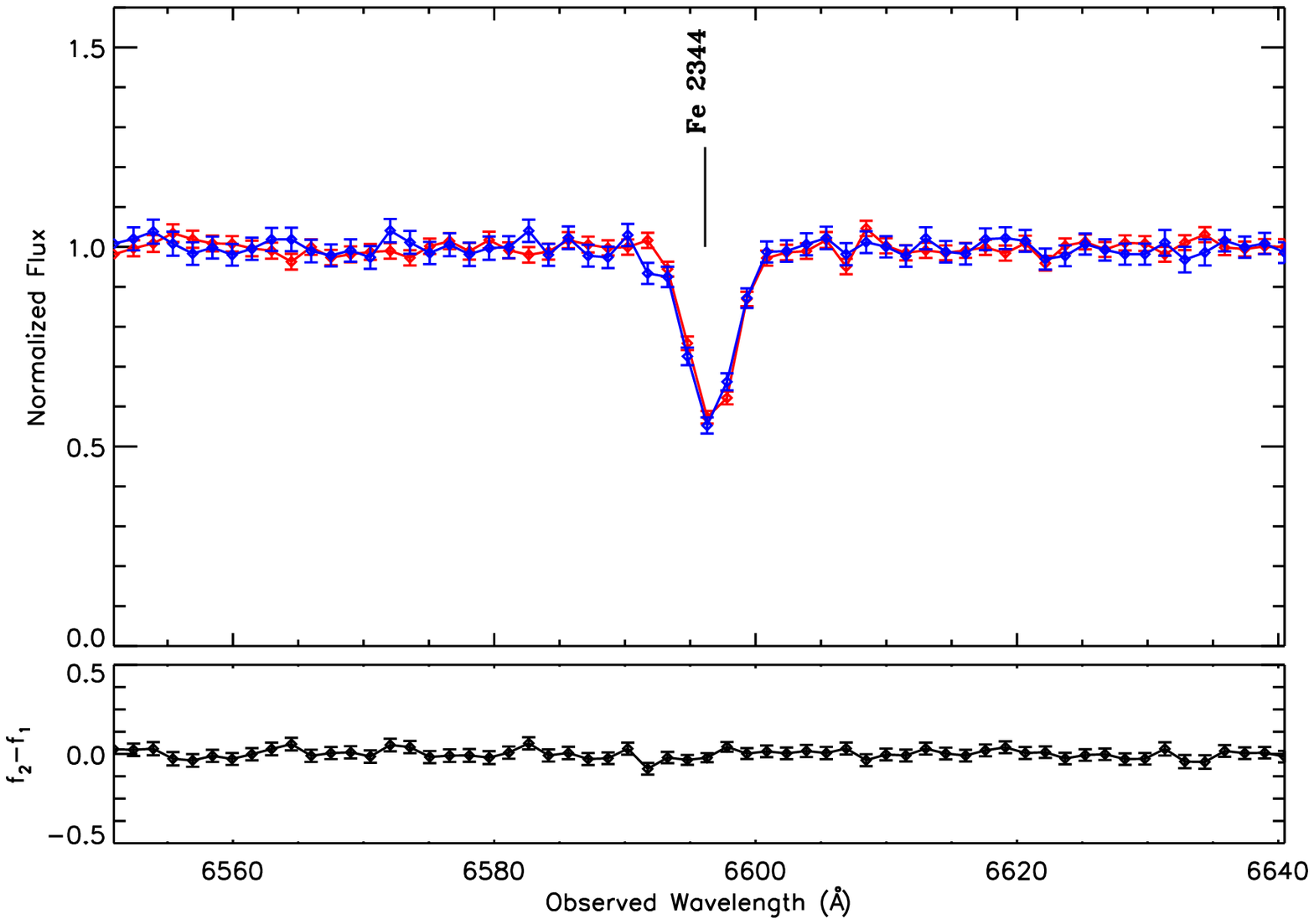}
\includegraphics[width=84mm]{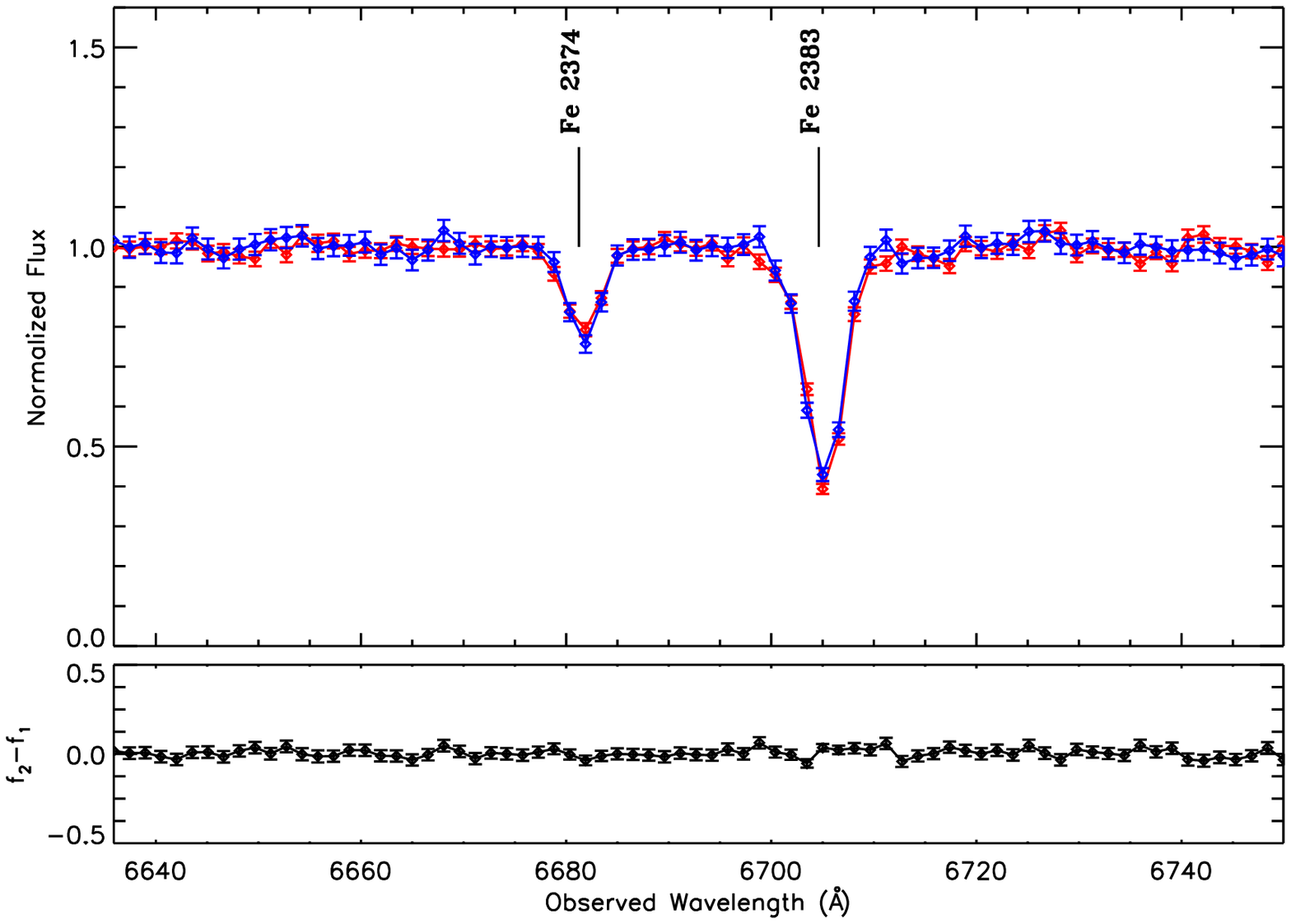}
\includegraphics[width=84mm]{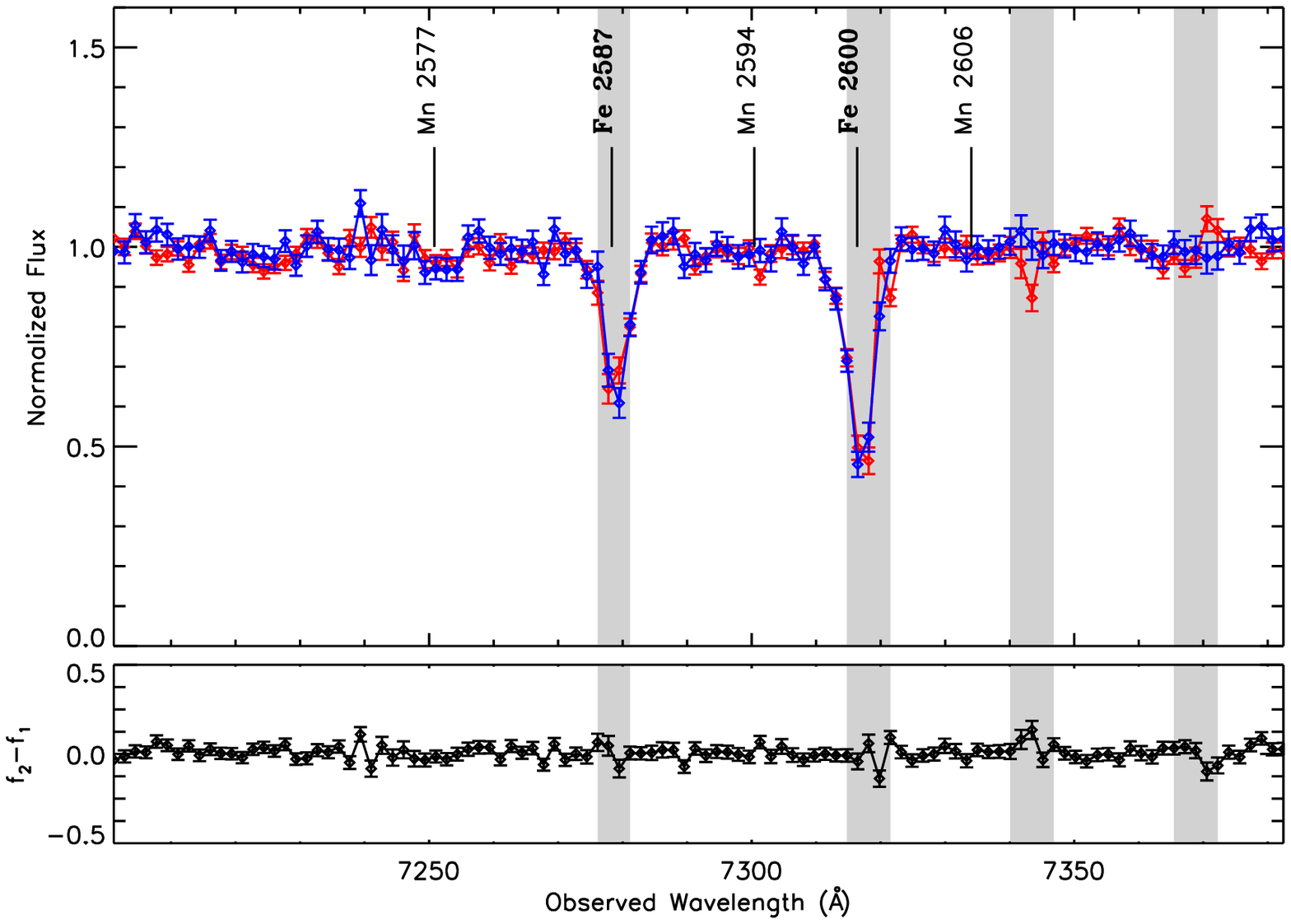}
\includegraphics[width=84mm]{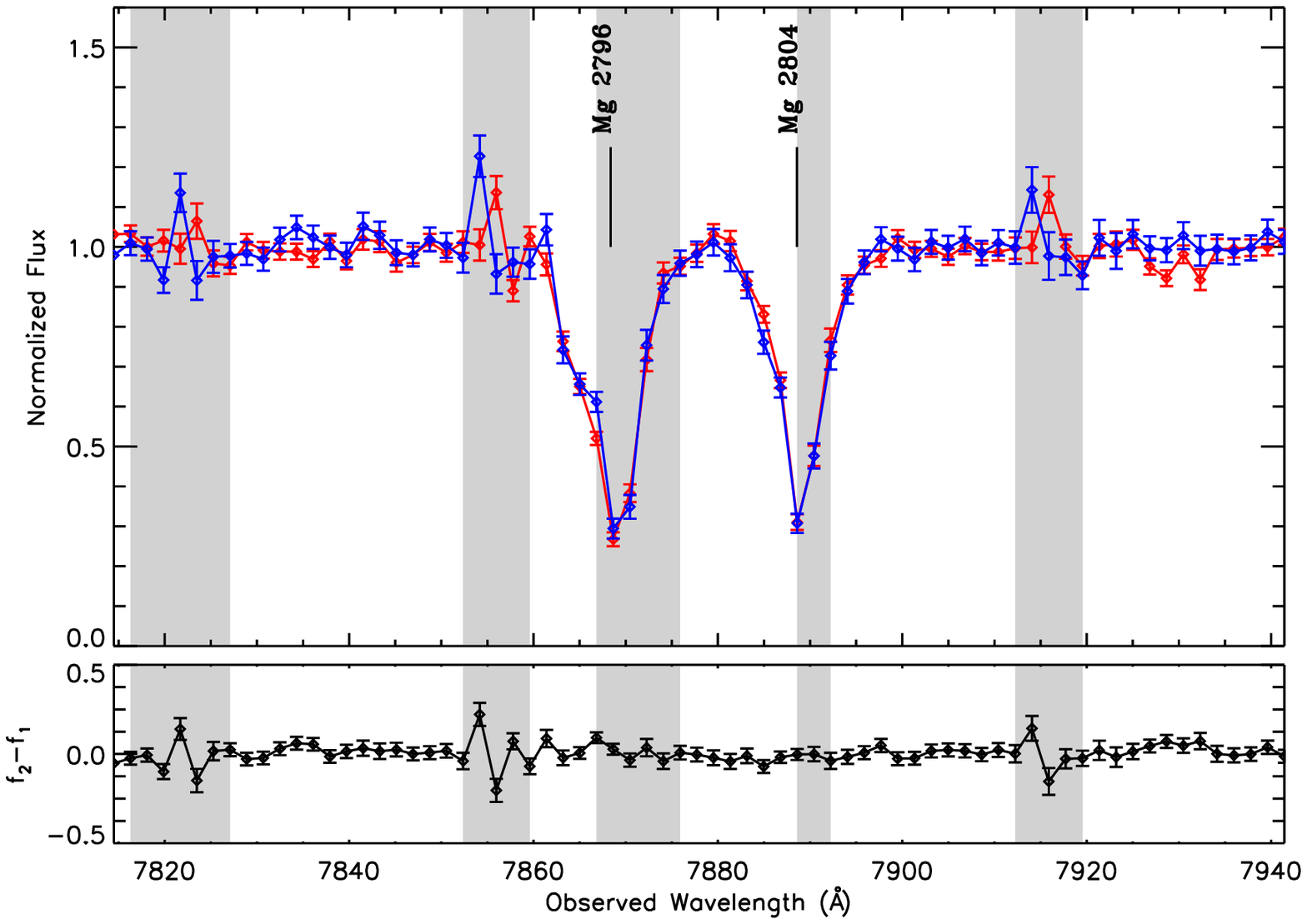}
\includegraphics[width=84mm]{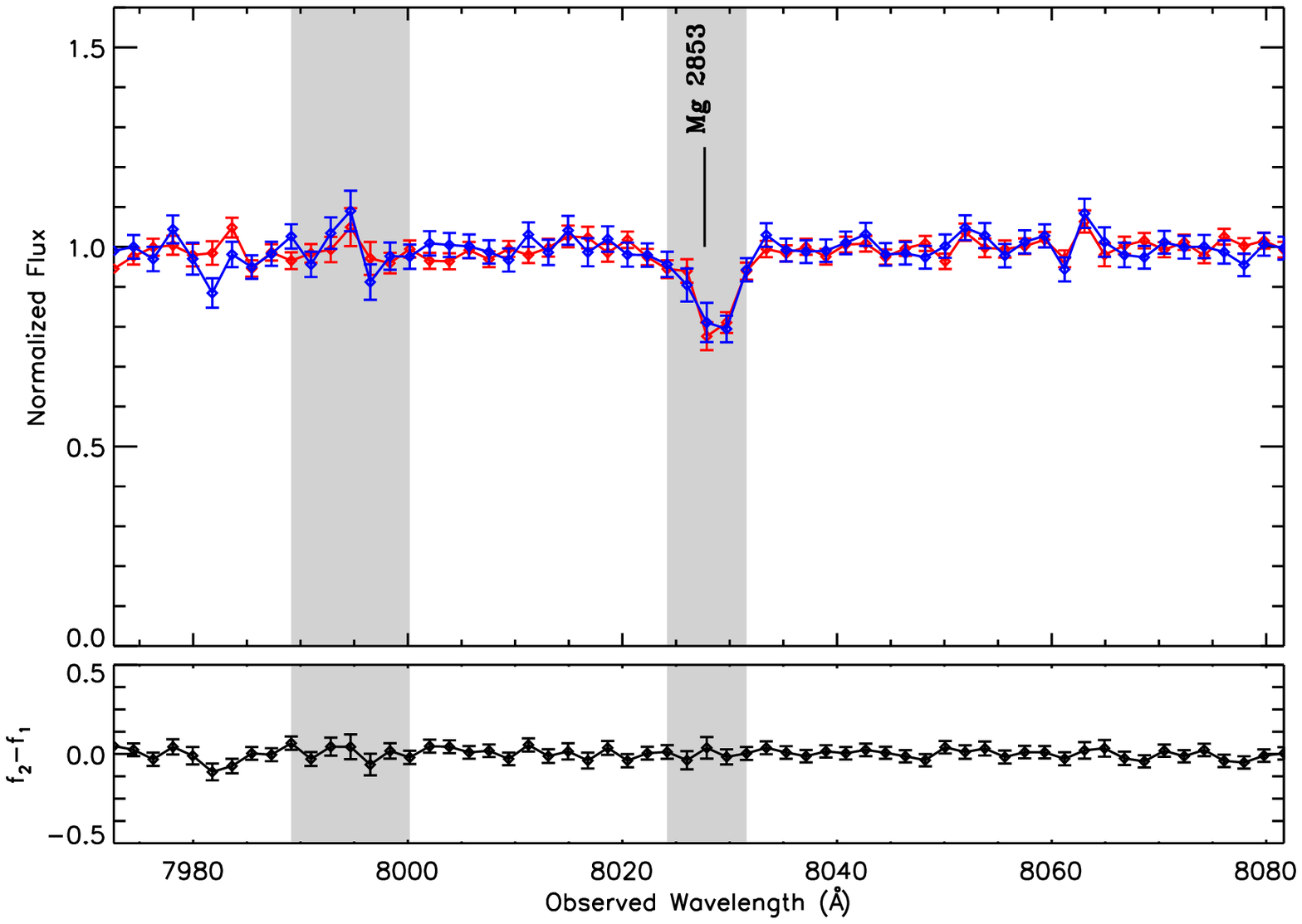}
\caption[]{Two-epoch normalized spectra of the variable NAL system at $\beta$ = 0.0235 in SDSS J212812.33-081529.3.  The top panel shows the normalized pixel flux values with 1$\sigma$ error bars (first observations are red and second are blue), the bottom panel plots the difference spectrum of the two observation epochs, and shaded backgrounds identify masked pixels not included in our search for absorption line variability.  Line identifications for significantly variable absorption lines are italicised, lines detected in both observation epochs are in bold font, and undetected lines are in regular font (see Table A.1 for ion labels).  Continued from previous figure.}
\end{center}
\end{figure*}

\begin{figure*}
\begin{center}
\includegraphics[width=84mm]{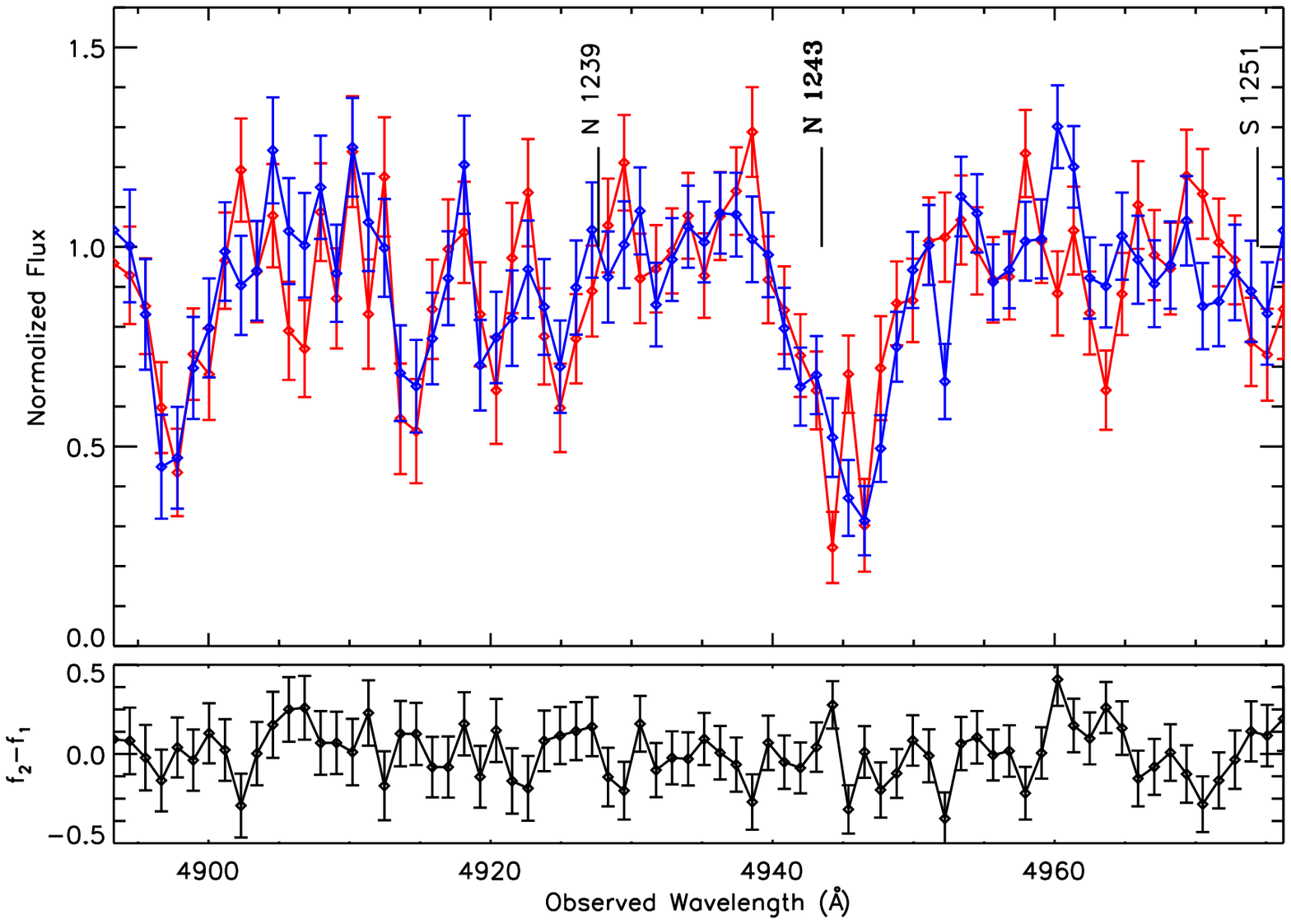}
\includegraphics[width=84mm]{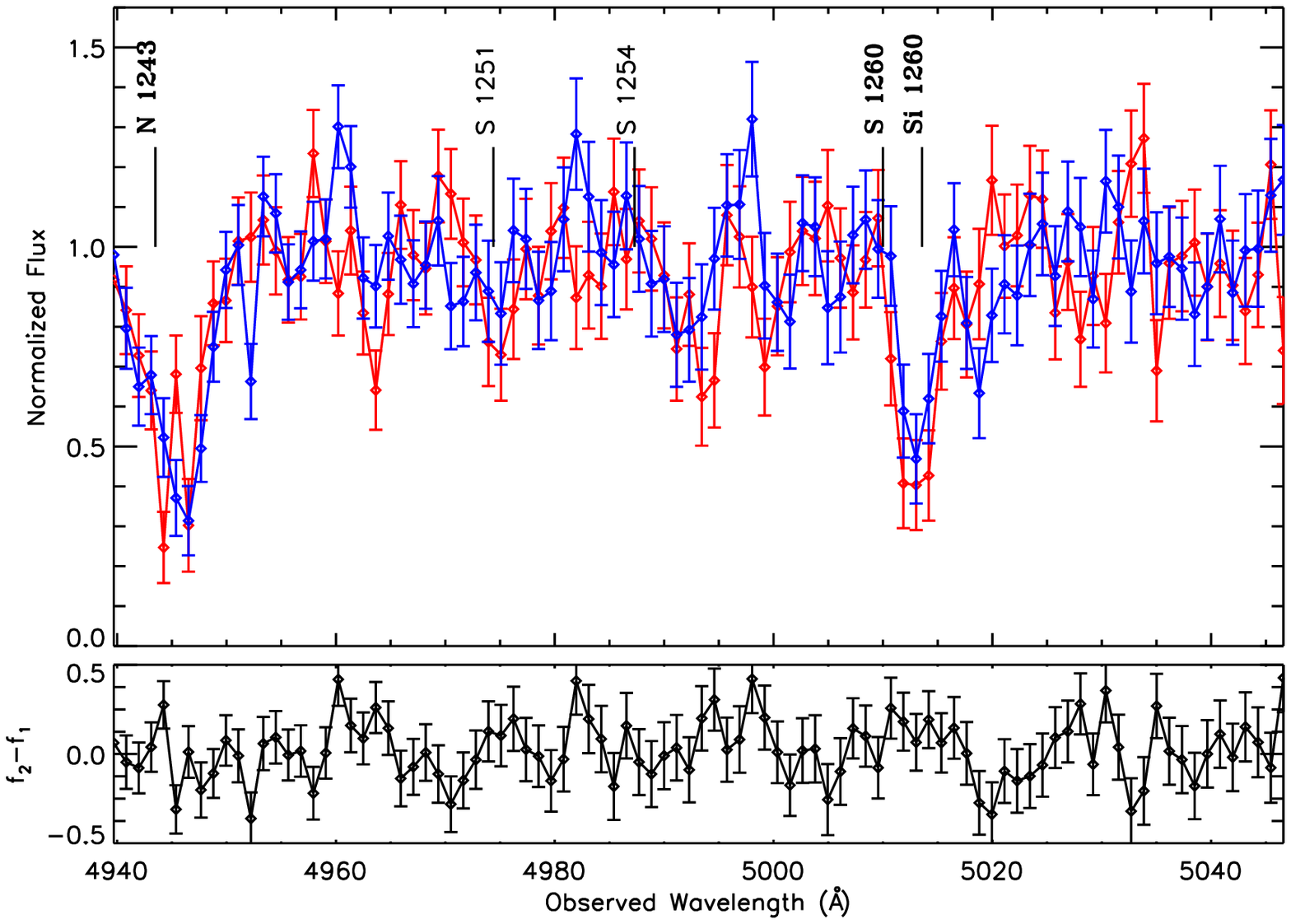}
\includegraphics[width=84mm]{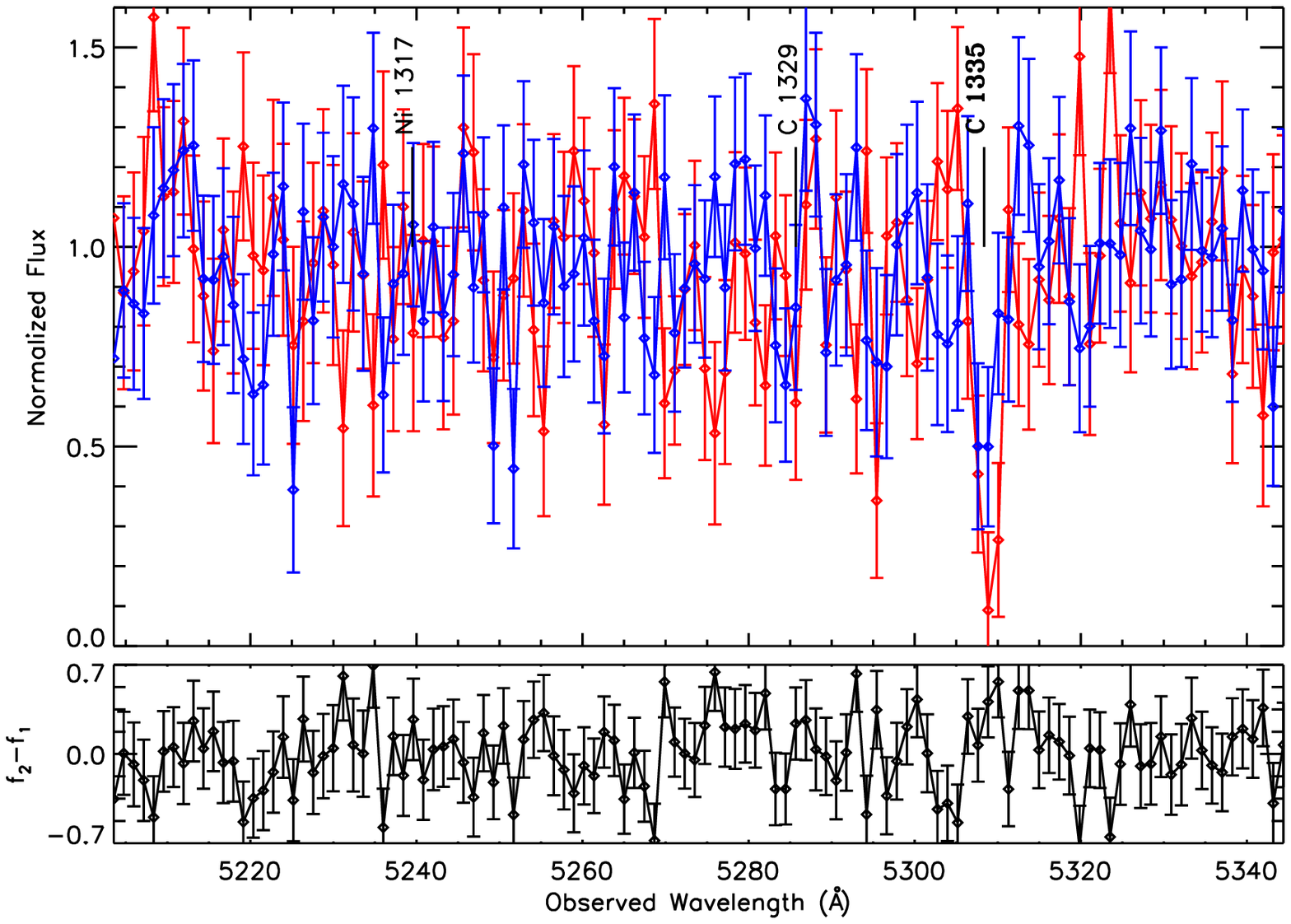}
\includegraphics[width=84mm]{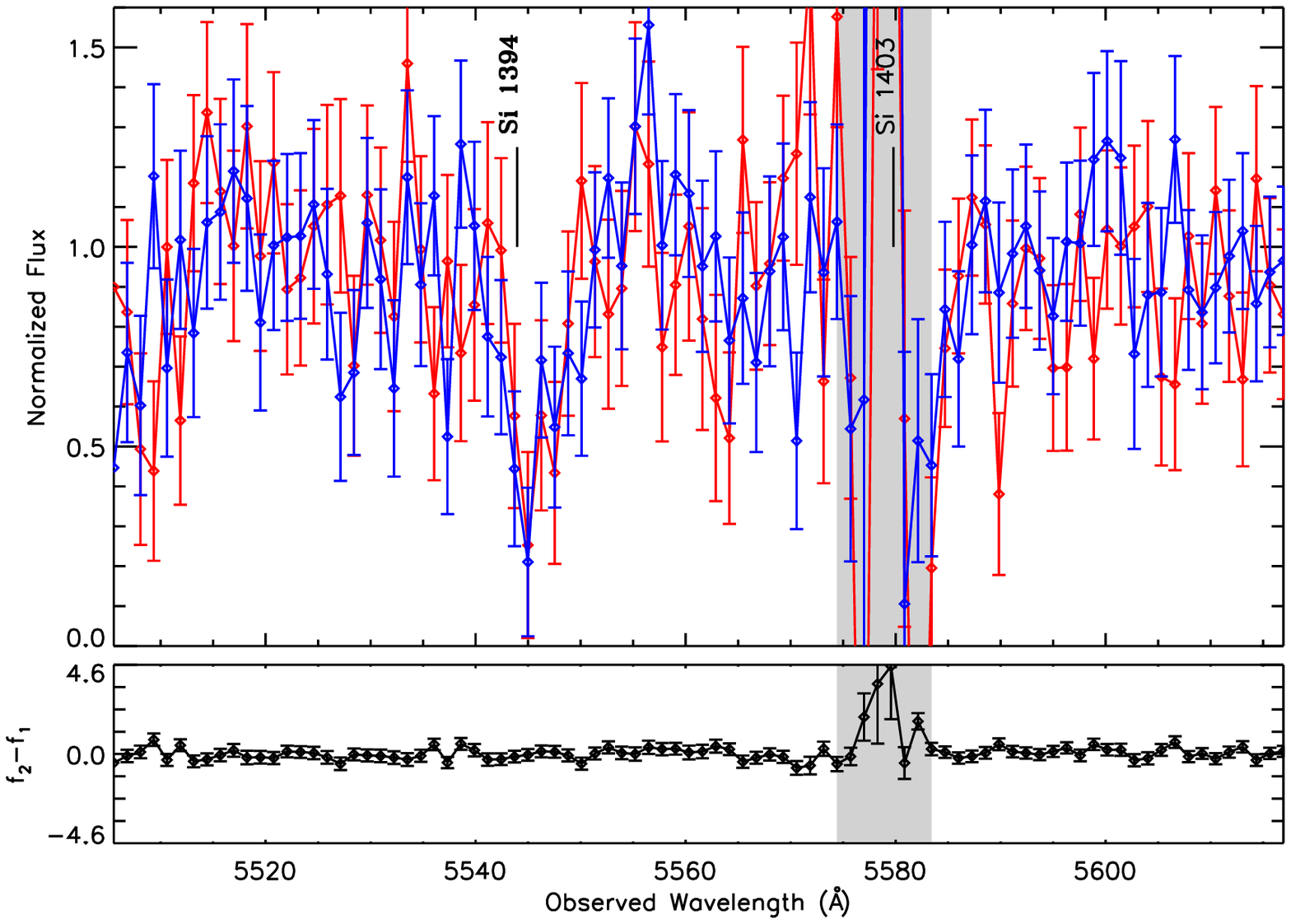}
\includegraphics[width=84mm]{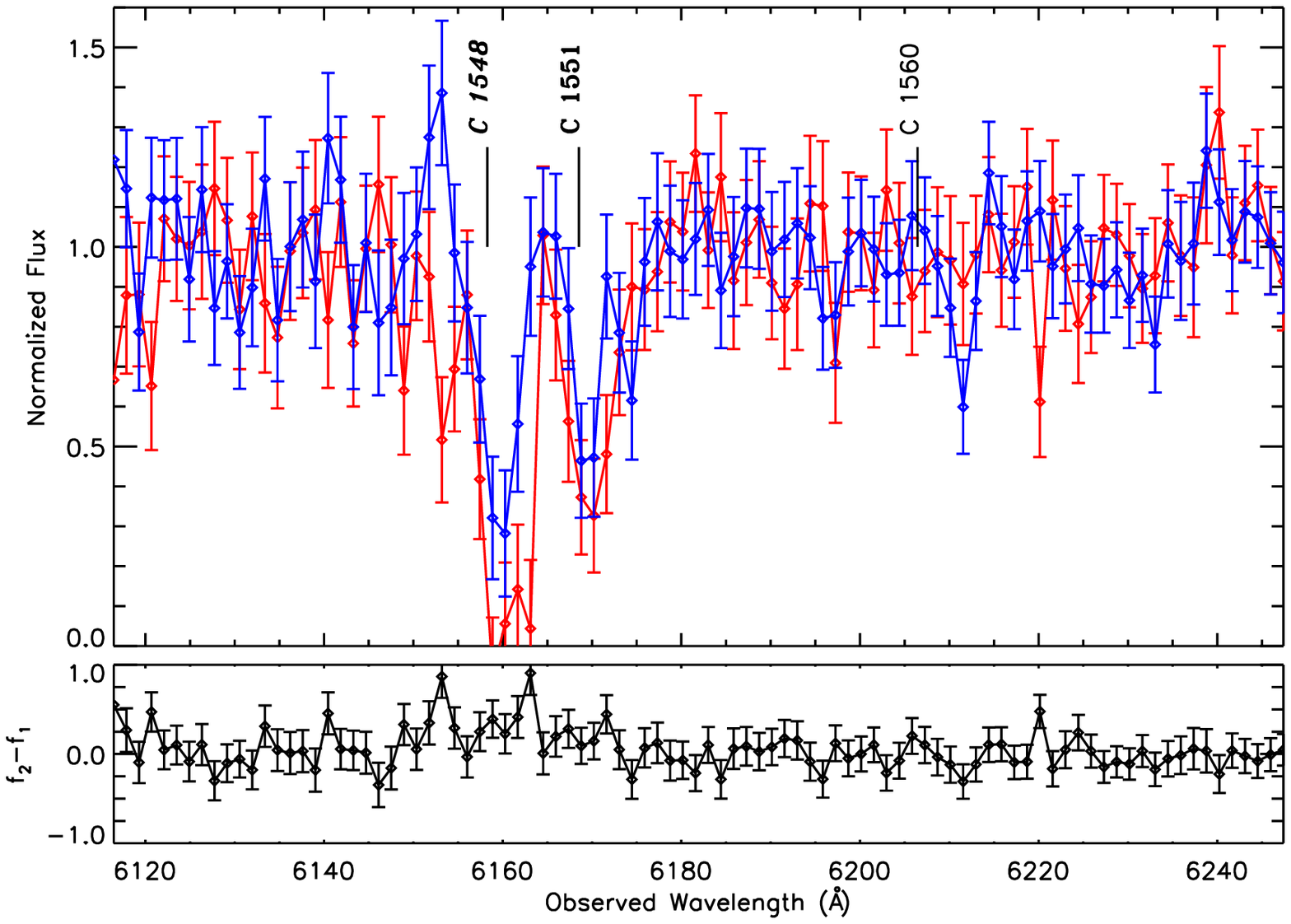}
\includegraphics[width=84mm]{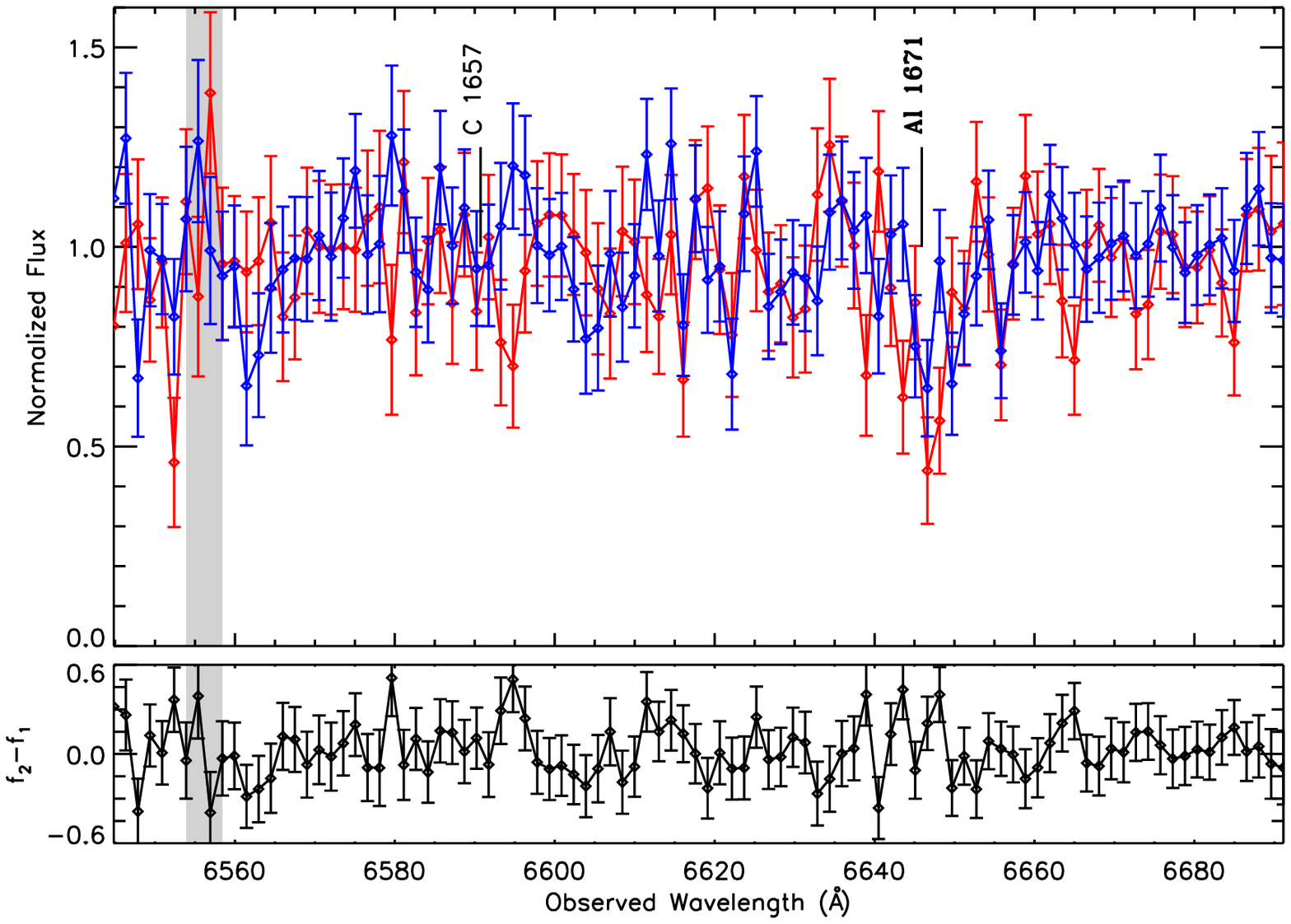}
\caption[Two-epoch normalized spectra of SDSS J084107.24+333921.7]{Two-epoch normalized spectra of the variable NAL system at $\beta$ = 0.0228 in SDSS J084107.24+333921.7.  The top panel shows the normalized pixel flux values with 1$\sigma$ error bars (first observations are red and second are blue), the bottom panel plots the difference spectrum of the two observation epochs, and shaded backgrounds identify masked pixels not included in our search for absorption line variability.  Line identifications for significantly variable absorption lines are italicised, lines detected in both observation epochs are in bold font, and undetected lines are in regular font (see Table A.1 for ion labels).  \label{figvs31}}
\end{center}
\end{figure*}

\begin{figure*}
\begin{center}
\includegraphics[width=84mm]{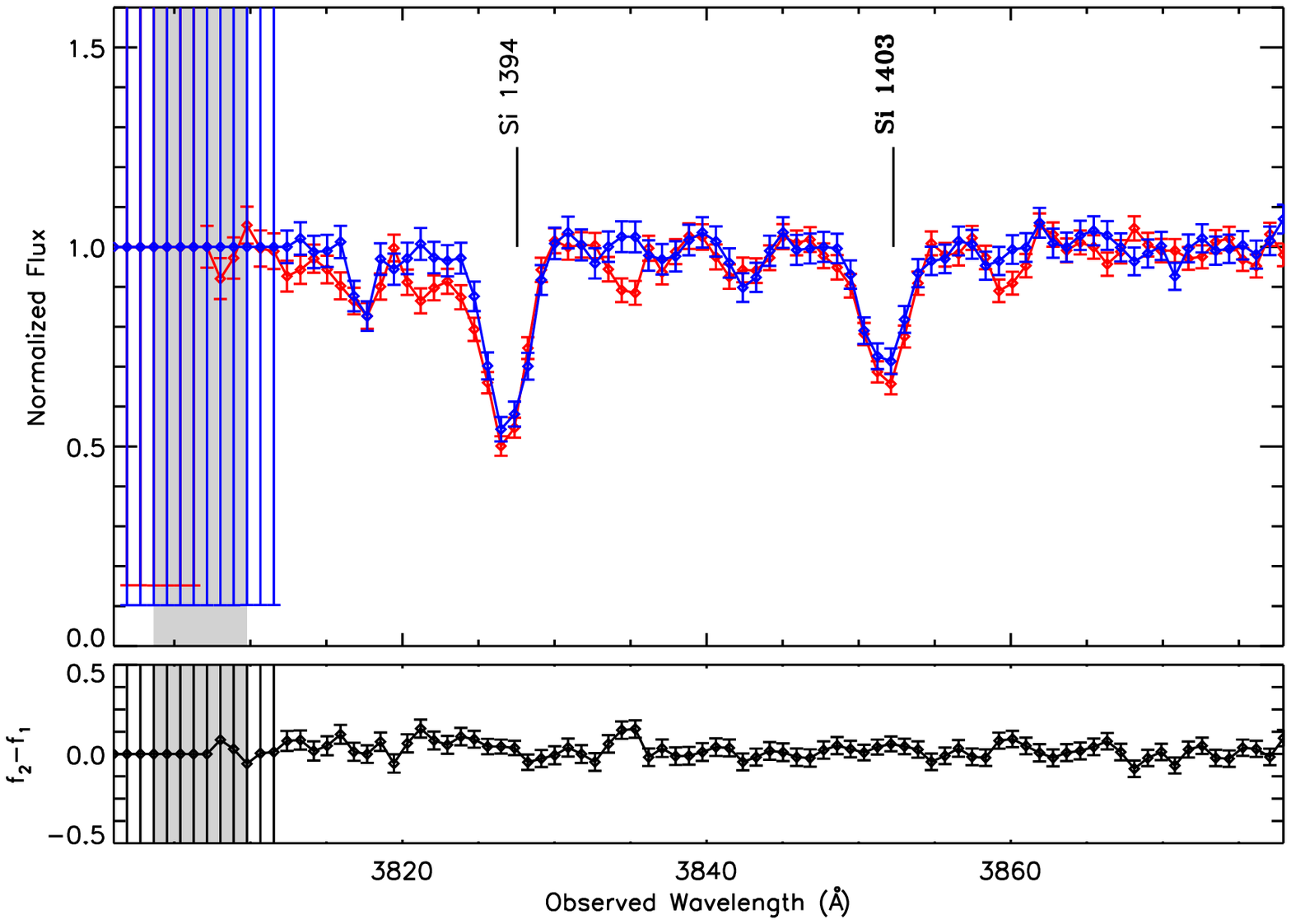}
\includegraphics[width=84mm]{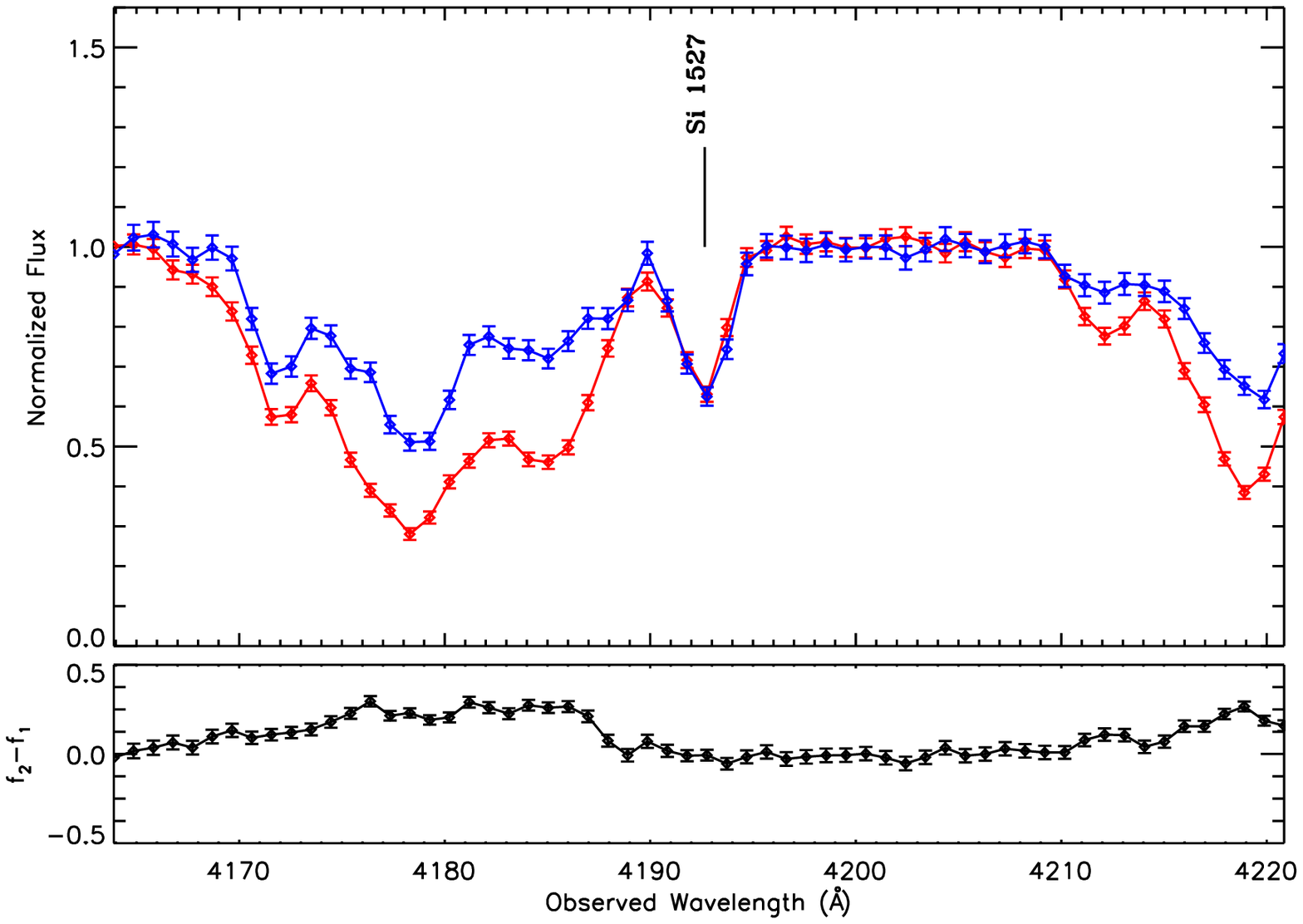}
\includegraphics[width=84mm]{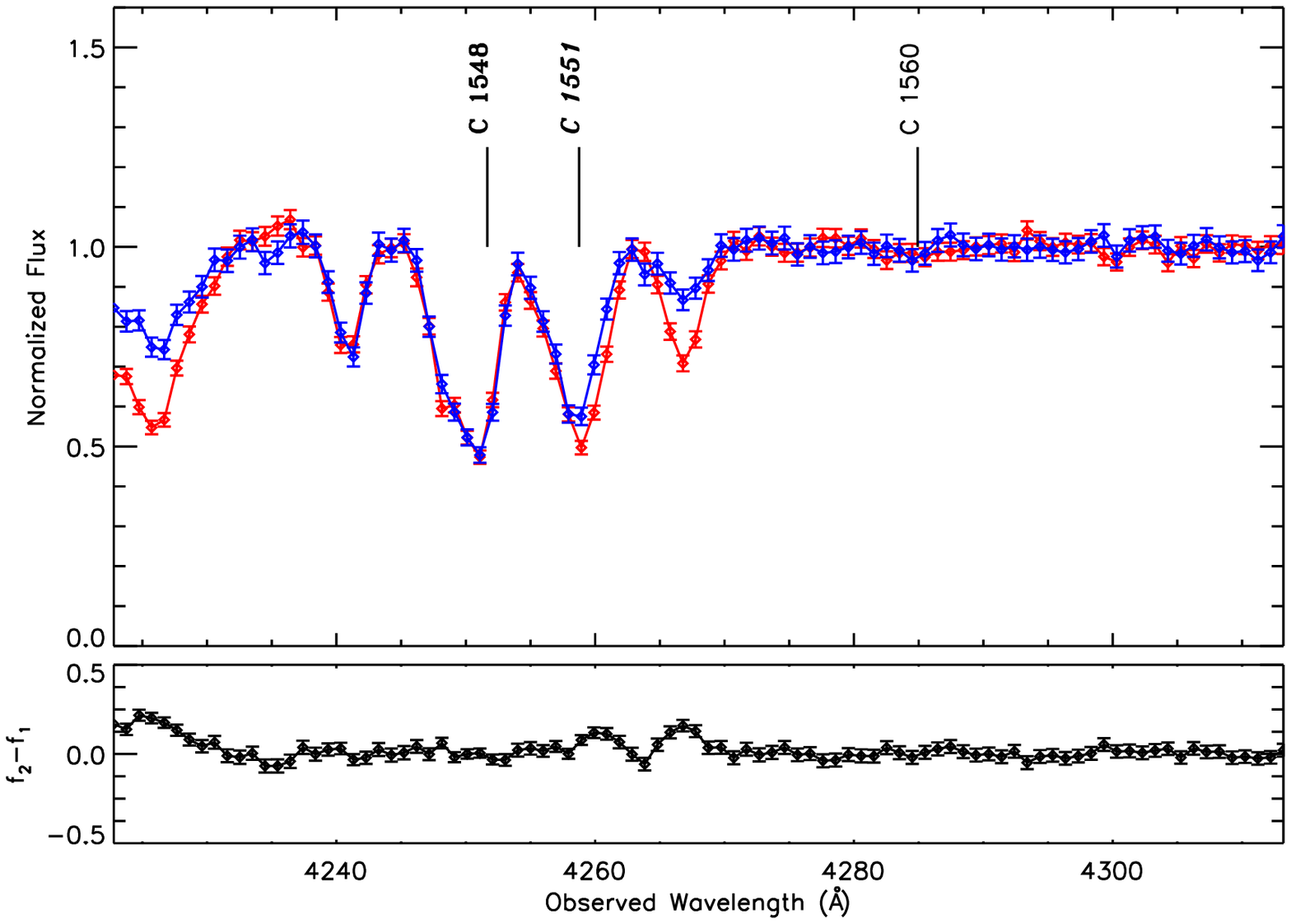}
\includegraphics[width=84mm]{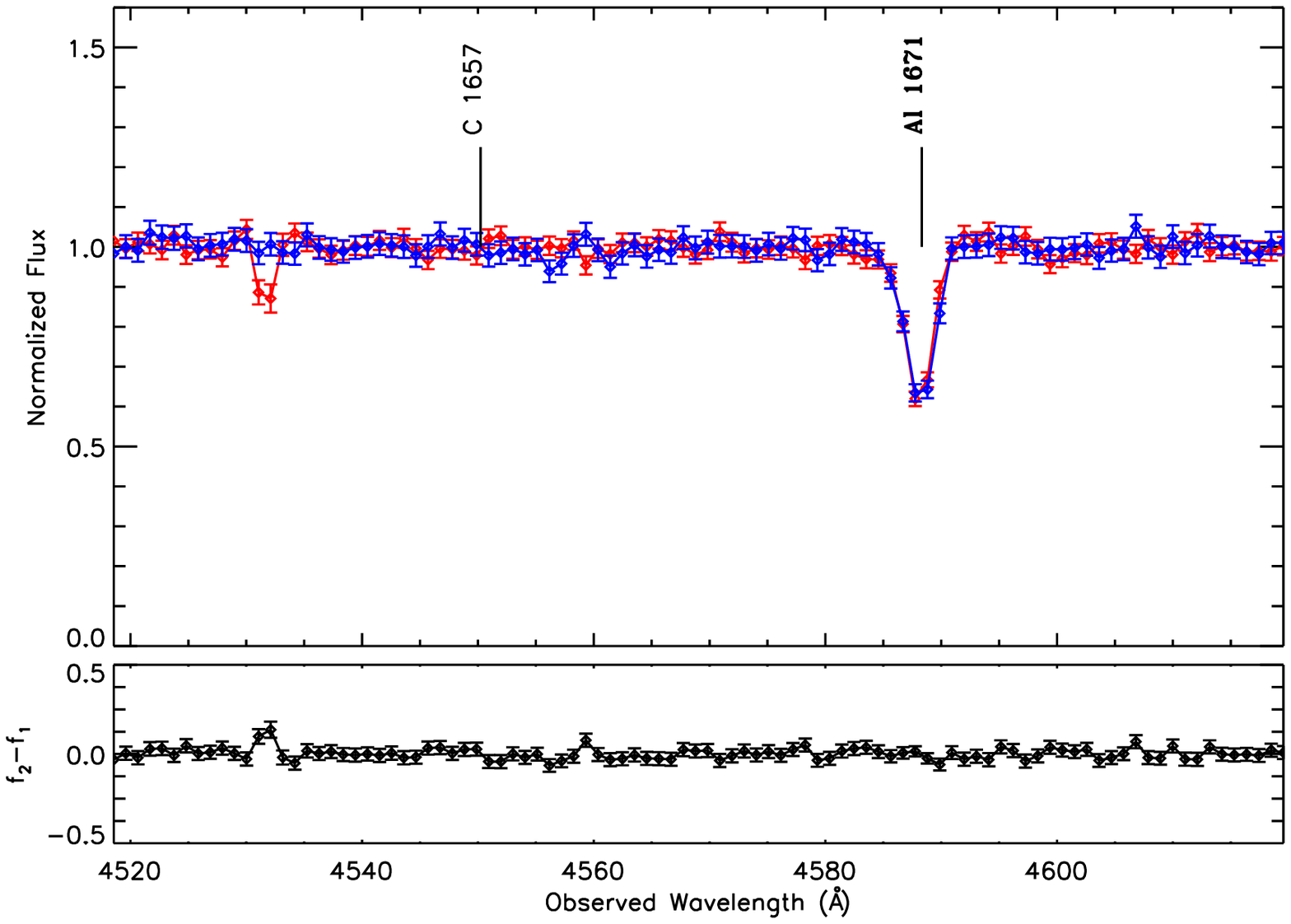}
\includegraphics[width=84mm]{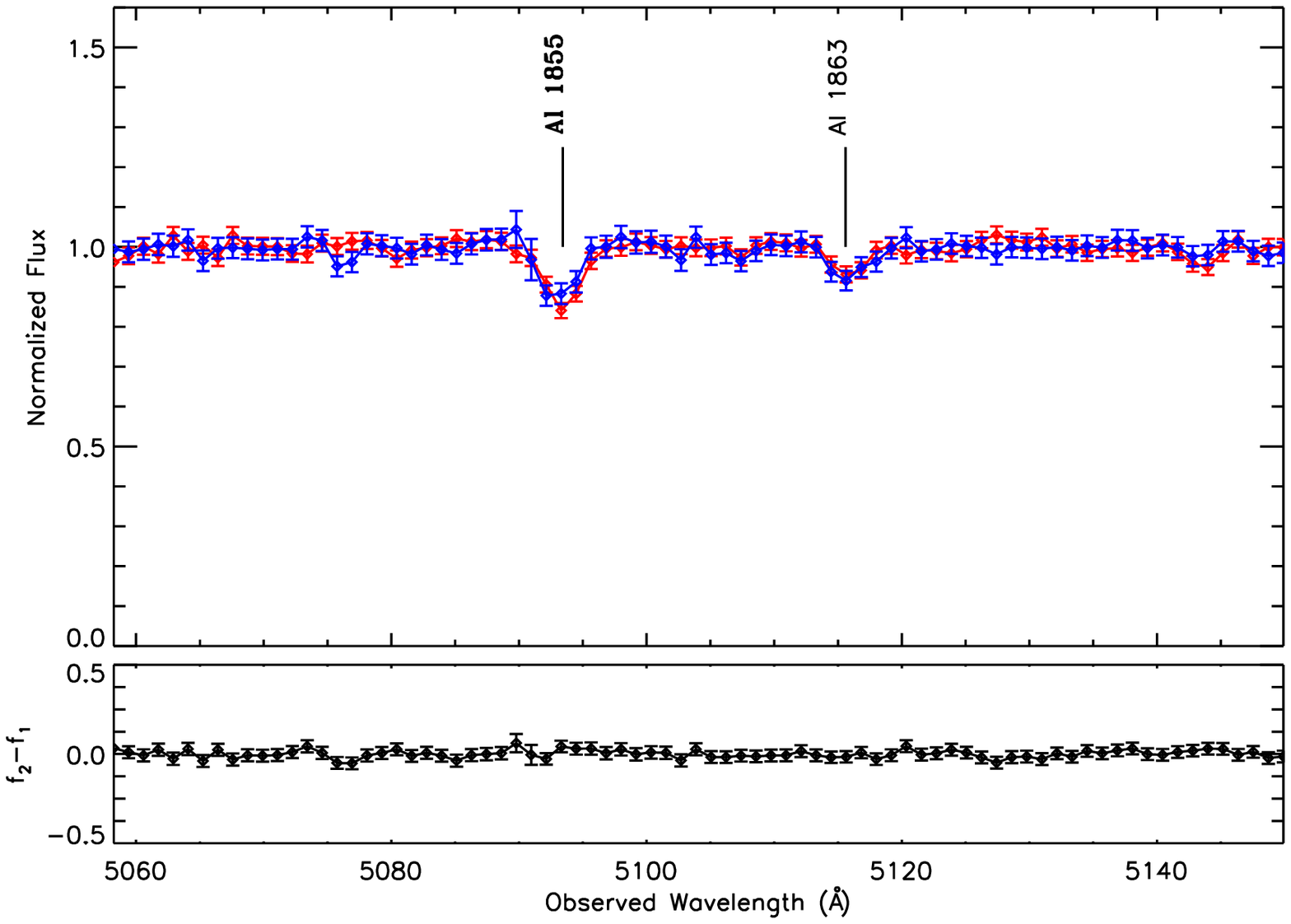}
\includegraphics[width=84mm]{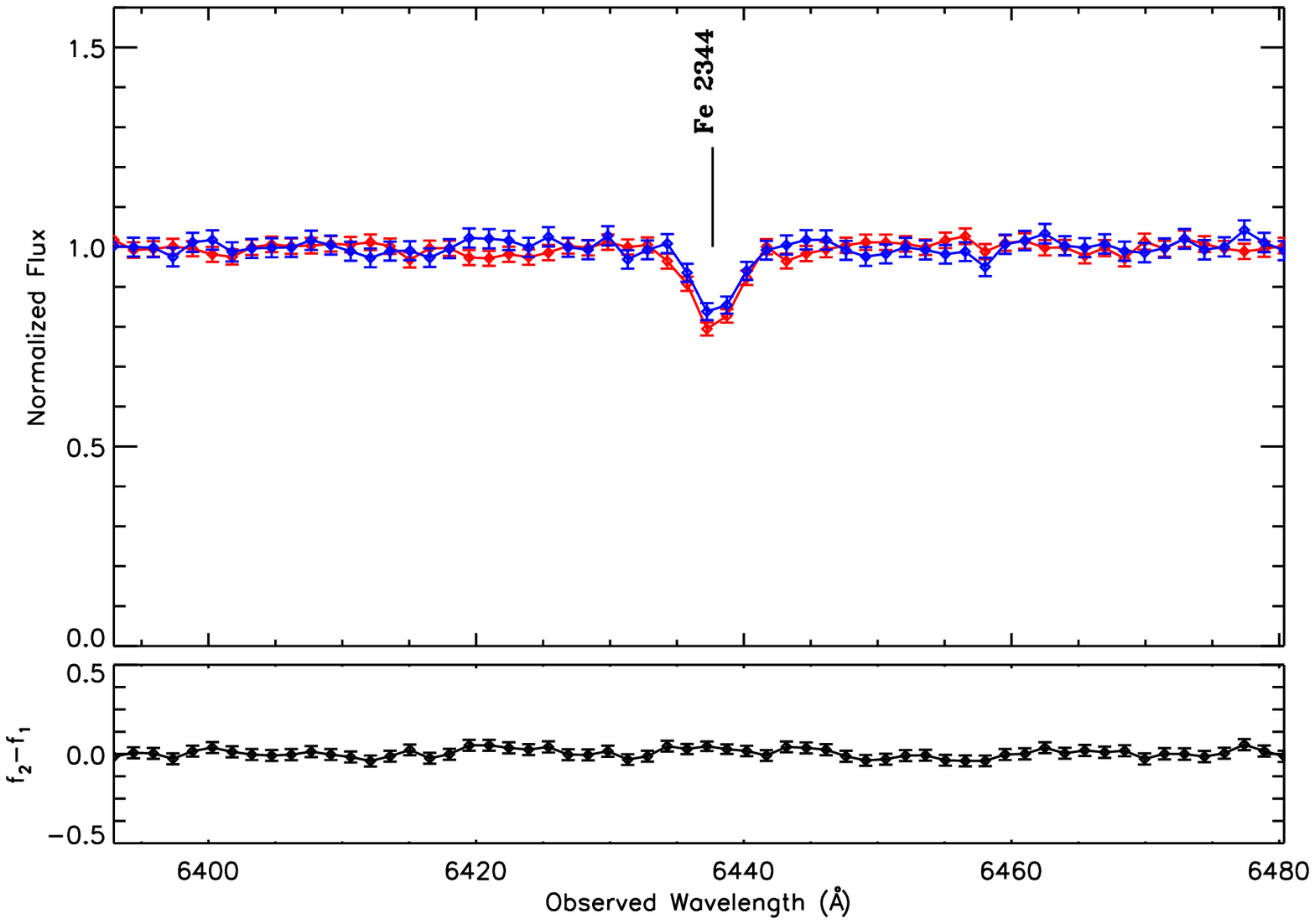}
\caption[Two-epoch normalized spectra of SDSS J170428.65+242918.0]{Two-epoch normalized spectra of the variable NAL system at $\beta$ = 0.0190 in SDSS J170428.65+242918.0.  The top panel shows the normalized pixel flux values with 1$\sigma$ error bars (first observations are red and second are blue), the bottom panel plots the difference spectrum of the two observation epochs, and shaded backgrounds identify masked pixels not included in our search for absorption line variability.  Line identifications for significantly variable absorption lines are italicised, lines detected in both observation epochs are in bold font, and undetected lines are in regular font (see Table A.1 for ion labels).  Continued in next figure.  \label{figvs32}}
\end{center}
\end{figure*}

\begin{figure*}
\ContinuedFloat
\begin{center}
\includegraphics[width=84mm]{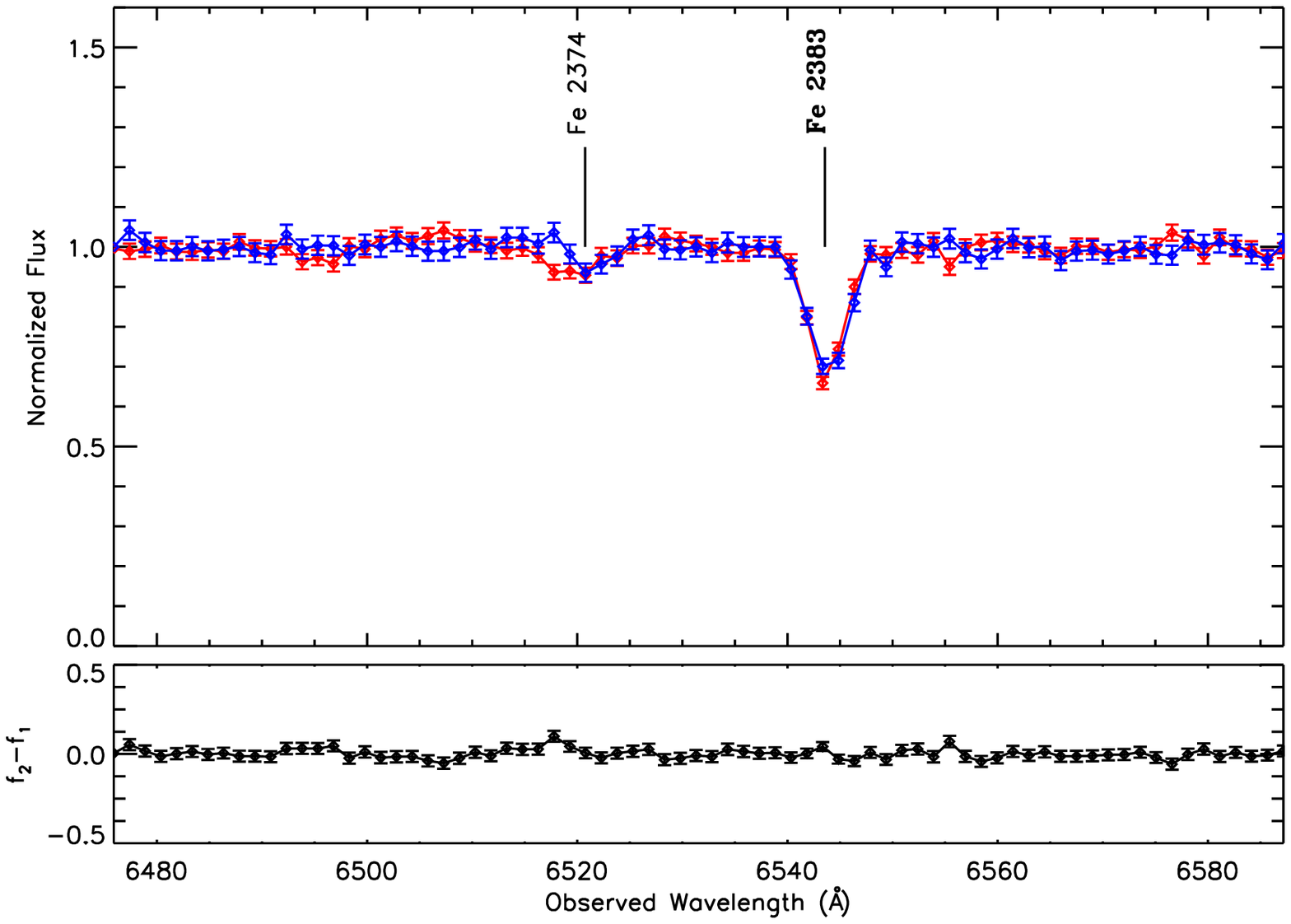}
\includegraphics[width=84mm]{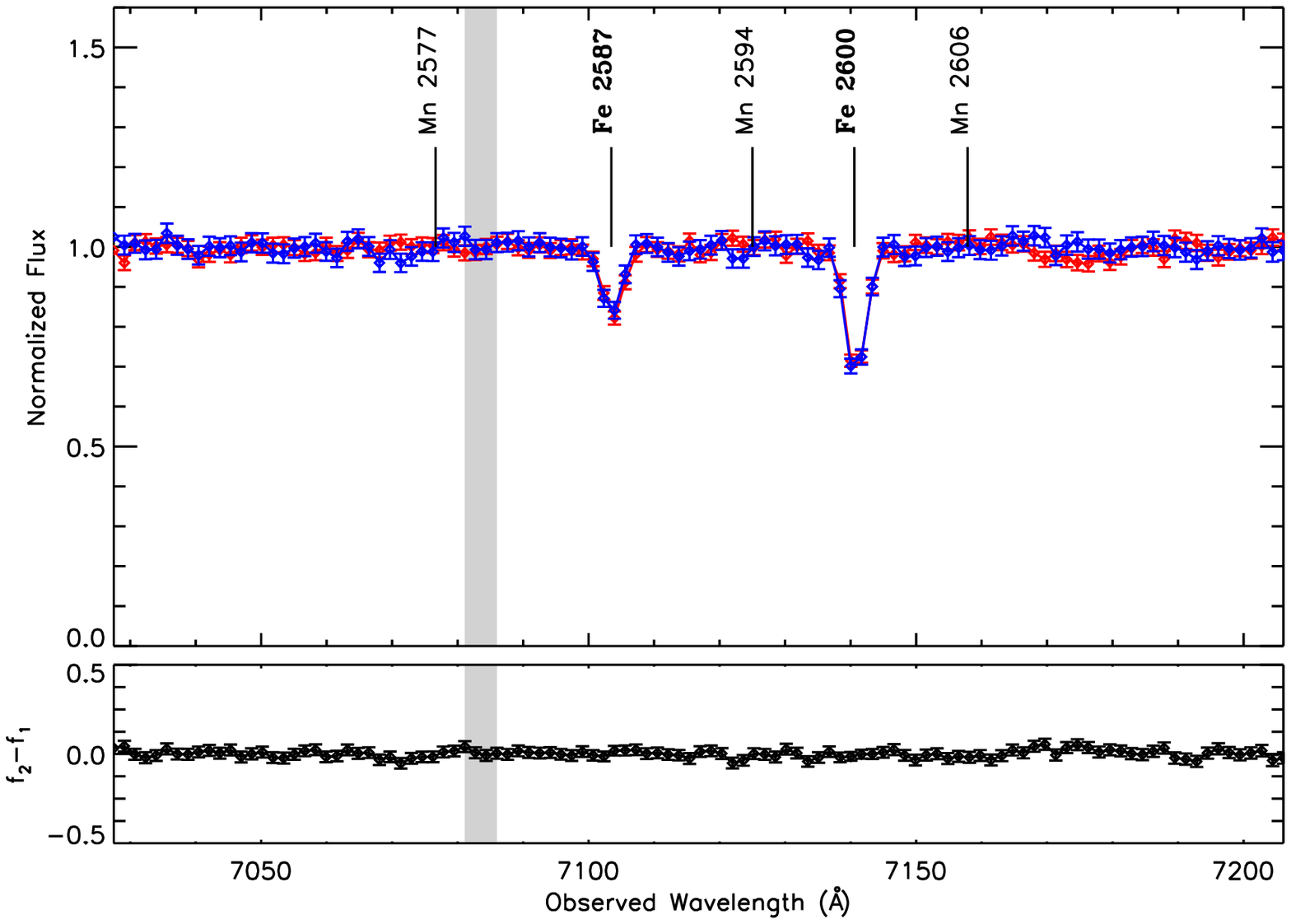}
\includegraphics[width=84mm]{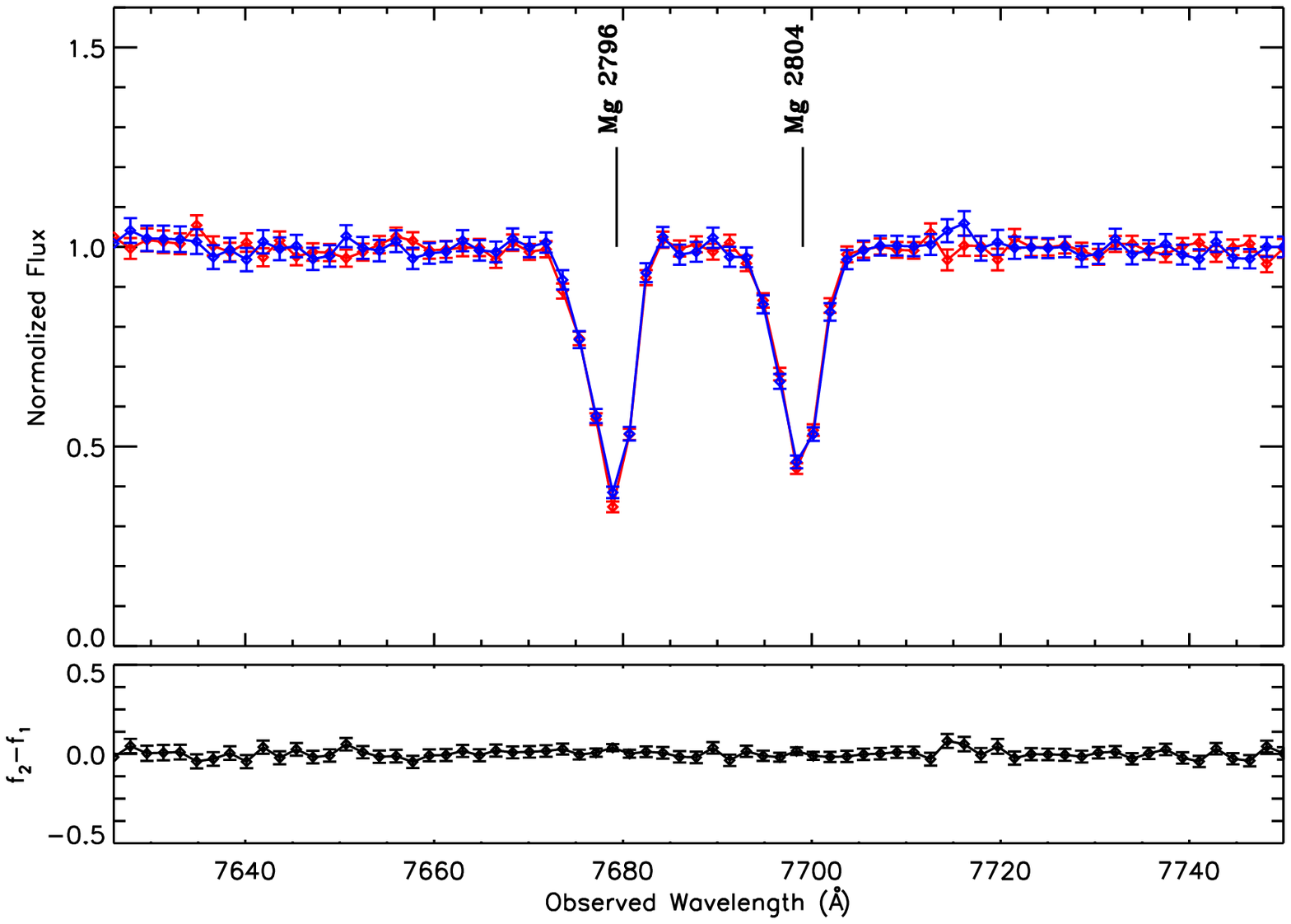}
\includegraphics[width=84mm]{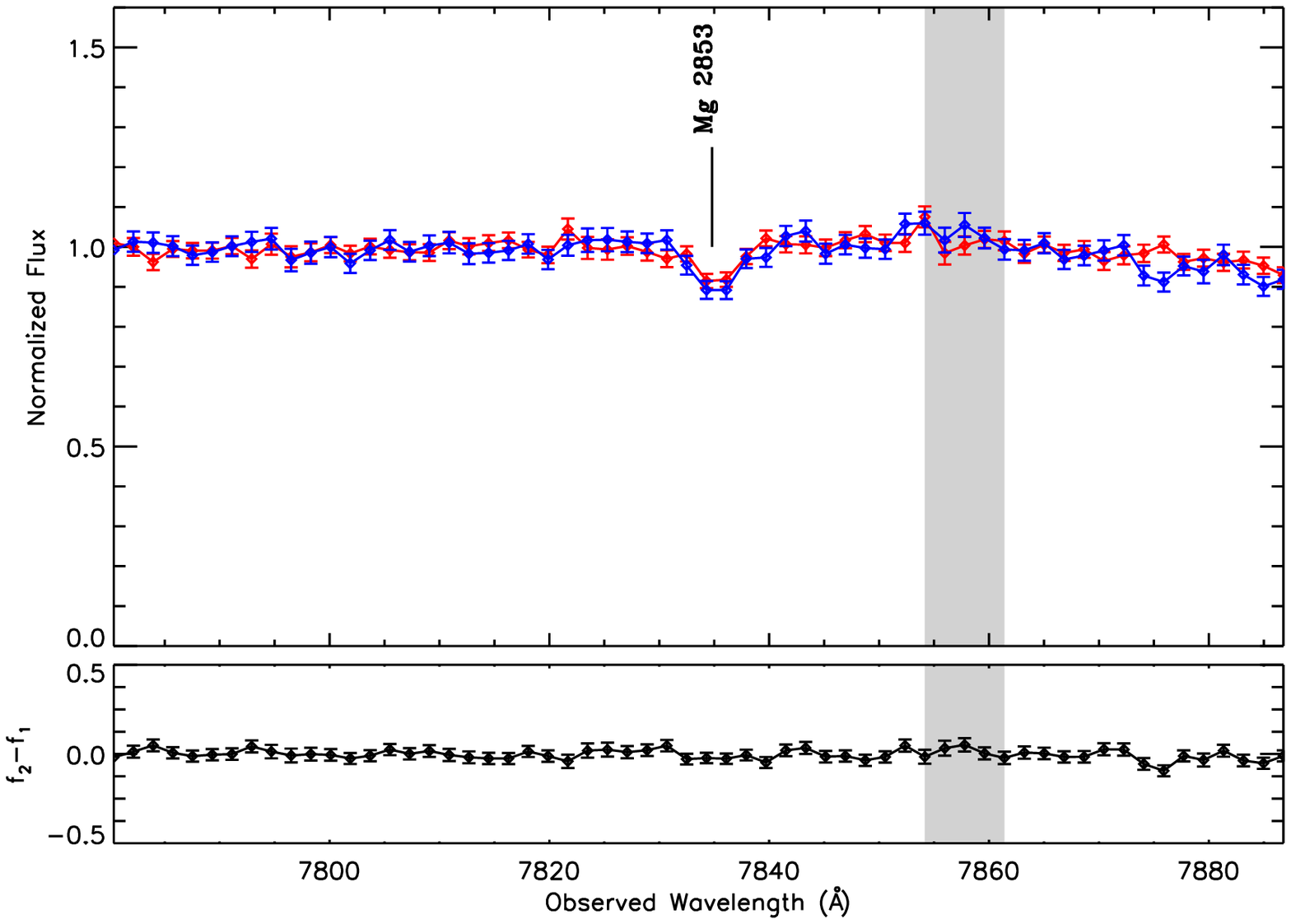}
\caption[]{Two-epoch normalized spectra of the variable NAL system at $\beta$ = 0.0190 in SDSS J170428.65+242918.0.  The top panel shows the normalized pixel flux values with 1$\sigma$ error bars (first observations are red and second are blue), the bottom panel plots the difference spectrum of the two observation epochs, and shaded backgrounds identify masked pixels not included in our search for absorption line variability.  Line identifications for significantly variable absorption lines are italicised, lines detected in both observation epochs are in bold font, and undetected lines are in regular font (see Table A.1 for ion labels).  Continued from previous figure.}
\end{center}
\end{figure*}

\begin{figure*}
\begin{center}
\includegraphics[width=84mm]{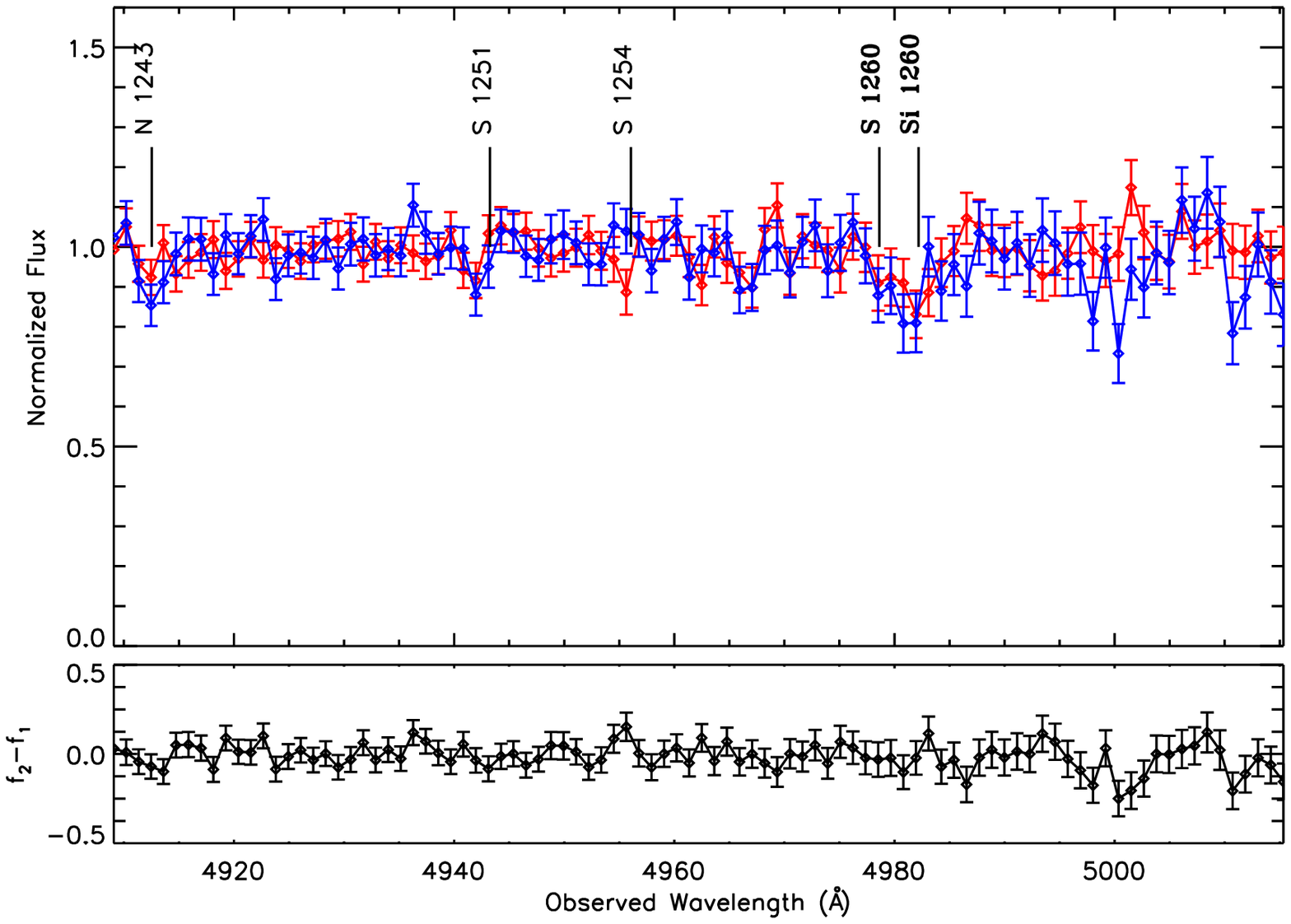}
\includegraphics[width=84mm]{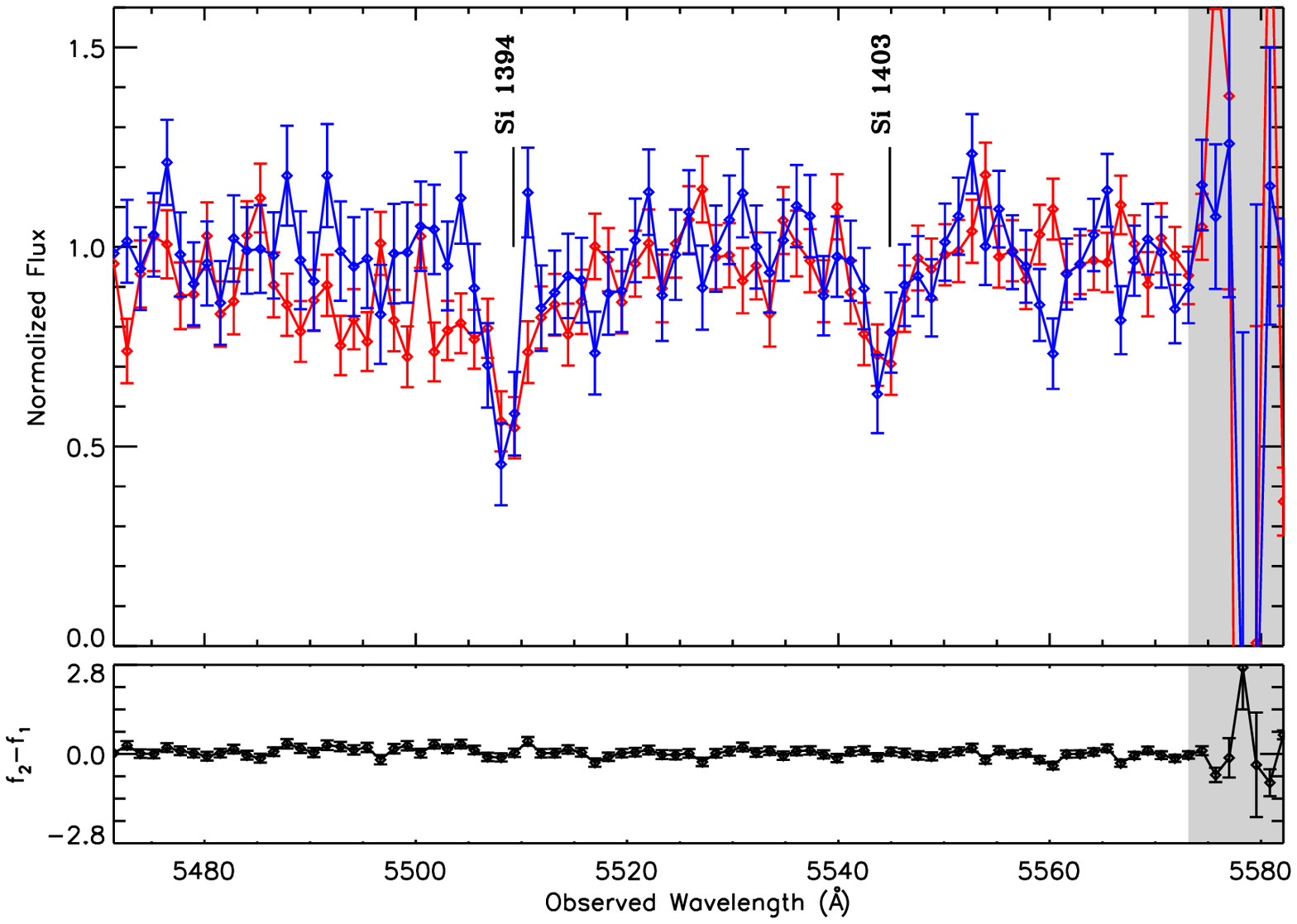}
\includegraphics[width=84mm]{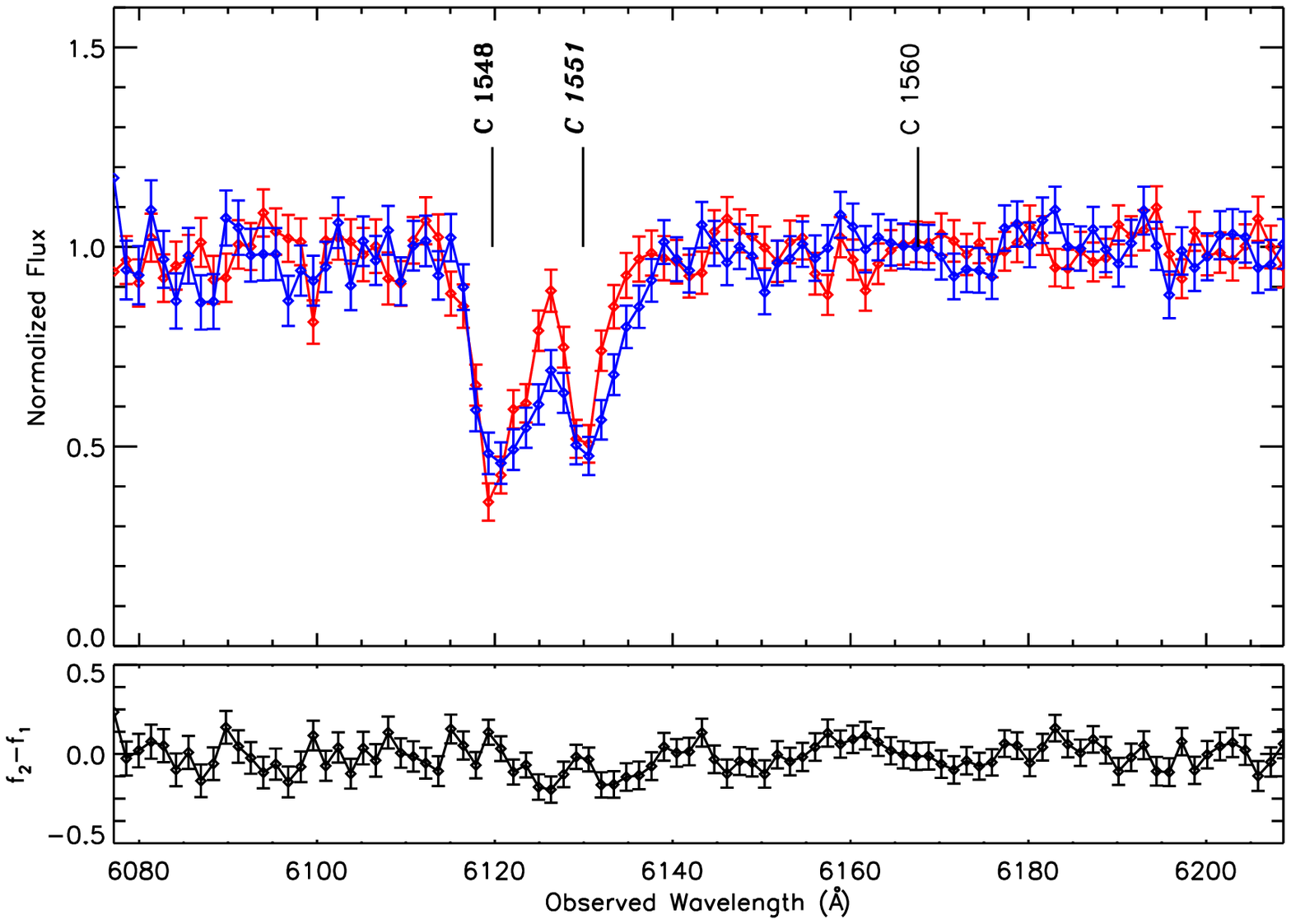}
\caption[Two-epoch normalized spectra of SDSS J231055.32+004817.1]{Two-epoch normalized spectra of the variable NAL system at $\beta$ = 0.0127 in SDSS J231055.32+004817.1.  The top panel shows the normalized pixel flux values with 1$\sigma$ error bars (first observations are red and second are blue), the bottom panel plots the difference spectrum of the two observation epochs, and shaded backgrounds identify masked pixels not included in our search for absorption line variability.  Line identifications for significantly variable absorption lines are italicised, lines detected in both observation epochs are in bold font, and undetected lines are in regular font (see Table A.1 for ion labels).  \label{figvs33}}
\end{center}
\end{figure*}

\label{lastpage}

\end{document}